\documentclass[oneside,12pt]{PhDthesisPSnPDF}

\ifpdf  
    \pdfcatalog { /PageMode (/UseOutlines)
                  /OpenAction (fitbh)  }
\fi

\title{The Self-Force Problem: \\Local Behaviour of the Detweiler-Whiting Singular Field}

\ifpdf
  \author{\href{mailto:anna.heffernan@ucd.ie}{Anna Heffernan}}
  \collegeordept{\href{http://mathsci.ucd.ie}{School of Mathematical Sciences}\\ and \\ \href{http://www.ucd.ie/casl/}{Complex and Adaptive Systems Laboratory}}
  \university{\href{http://www.ucd.ie}{University College Dublin}}
  \headofschool{\bf{Head of School:} \href{http://mathsci.ucd.ie/people/murphy_p}{\it{Dr. Patrick Murphy}}}
  \supervisor{\bf{Principal Supervisor:} \href{http://mathsci.ucd.ie/people/ottewill_a}{\it{Prof. Adrian Ottewill}}}
  \crest{\includegraphics[width=2.5cm]{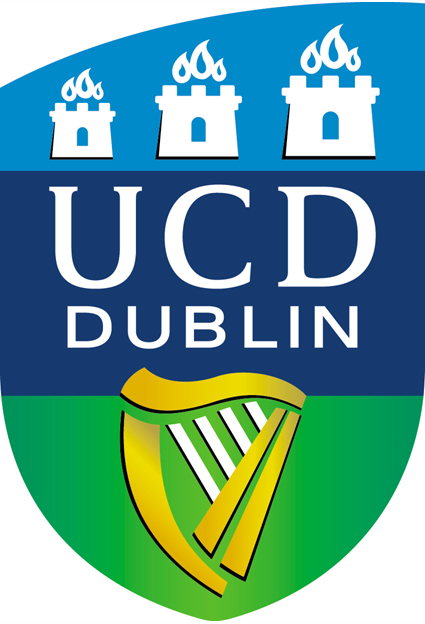}}
  
\else
  \author{YourName}
  \collegeordept{CollegeOrDept}
  \university{University}
  \crest{\includegraphics[width=4cm]{logo}}
\fi

\degree{Philosophi\ae Doctor (PhD)}
\degreedate{2012 October}

\begin{document}

\frontmatter
\renewcommand\baselinestretch{1.2}
\baselineskip=24pt plus1pt


\maketitle


\renewcommand\baselinestretch{1.5}
\baselineskip=18pt plus1pt

\setcounter{secnumdepth}{3} 
\setcounter{tocdepth}{3}    
\tableofcontents            


\listoffigures	

\listoftables  



\begin{abstracts}        

\addcontentsline{toc}{chapter}{\numberline{}Abstract}

%
%

Gravitational waves are ripples in space-time and a prediction of Einstein’s theory of relativity.  The growing reality of gravitational wave astronomy is giving age-old problems a new lease of life;
one such problem is that of the self-force.  A charged or massive particle moving in a curved background space-time \fixme{gives rise to} a field that affects its motion, pushing it off its expected geodesic.  This self-field gives rise to a so-called self-force acting on the particle.  In modelling this motion, the self-force approach uses a perturbative expansion in the mass ratio.
One of the most interesting sources of gravitational waves are extreme mass ratio inspirals. These systems 
have an extremely small mass ratio, making them perfectly suited to \fixme{perturbative,} gravitational self-force modelling.

One of the key problems that immediately arises, within the self-force model, is the divergence of the field at the particle.  To 
resolve this, the field is split into 
a singular component and a smooth regular field.  
This regular-singular split, introduced by Detweiler and Whiting, is used in most modern self-force calculations.

In this thesis, we derive high order expansions of the Detweiler-Whiting singular field, and use these to push the boundaries on current precision limits of self-force calculations.
Within the mode sum scheme, we give over 14 previously unknown regularisation parameters, almost doubling the current regularisation parameter database.  We also produce smooth effective sources to high order, and propose an application of the higher terms to improve accuracy in the $m$-mode scheme.

Finally, we investigate the status of the cosmic censorship conjecture and the role that the self-force plays.  To this end, we give regularisation parameters for non-geodesic motion.  Additionally, we show the necessity of our results in the exciting area of second order self-force calculations.  
Recently, second order self-force derivations have been developed, which benefit significantly from high-order 
coordinate expansions of the singular field, making them an immediate application of our current work.  We calculate several parameters that these schemes require, and highlight the further advancements possible from the results of this thesis. 

\end{abstracts}





\begin{declaration}        

I hereby certify that the submitted work is my own work, was completed while registered as a candidate for the degree stated on the Title Page, and I have not obtained a 
degree elsewhere on the basis of the research presented in this submitted work.

%

\vspace{10mm}


\end{declaration}


\begin{acknowledgements}      

First and foremost, I would like to offer a massive thank you to my supervisor, Adrian Ottewill, the man of never-ending patience, for his constant support, supervision, good humour, and above all, enthusiasm in sharing both his brilliance and intuition in both the complex and mundane.  There really was no question too big (or small) and I count myself very lucky to have had such a supervisor.

Throughout my PhD, I have had the good fortune of working in a fun and helpful group based in the Complex and Adaptive Systems Laboratory (CASL) at the University College of Dublin (UCD).  In particular, my colleague, collaborator and good friend, Barry Wardell, has been the absolute best in all his roles.  Switching the topic of my PhD would not have been possible without his constant support and friendship, be it in front of a desktop looking at Mathematica or in more social settings, and I am truly grateful for this.  Marc Casals and Sam Dolan also deserve a special mention, for both their enthusiasm and support.  Brien Nolan, Cliona Golden, David Brodigan, Patrick Nolan, Chris Kavanagh and all of my friends in CASL have also helped me on this journey.  

Ted Cox deserves a special mention, for his constant support throughout my education and research in Mathematical Physics, from suggesting my initial degree course in Theoretical Physics to initiating the process that became my PhD.  There are few lecturers who share his enthusiasm for teaching and research and I know how lucky both myself and UCD are to have him in our lives.  Nuria Garcia, Veronica Barker, and all of the support staff in both the School of Mathematical Sciences and CASL have been exemplary in their roles.  I know I didn't make their lives easy, but regardless, they were tireless in both their support and good humour throughout my PhD.  A quick mention to my mathematics teachers through the years - Ms. Creaney, Mr Connelly and Mr. Gunning, your teachings and enthusiasm were never wasted, thank you.

Through my research, I have had the opportunity and pleasure of meeting many helpful and fun researchers who have given me their time, in particular, Leor Barack and Niels Warburton have been especially helpful.   Roland Haas, Sarp Akcay, Jonathan Thornburg and many others from the Capra meetings were also very generous with their data, assistance and interesting discussions.  

Finally I would like to thank all my friends and family, for their support as well as keeping me sane through the years.  To all the girls, thank you.  My sister, brothers, Mum and Dad have all been amazing throughout the years, from ensuring I don't dress like a boy, to driving me around, dragging me away from my mathematics when I needed it and supporting me both financially and emotionally, I got there in the end and it was thanks to you.

This research was financially supported by the Irish Research Council for Science, Engineering and Technology (funded by the National Development Plan) as well as the School of Mathematical Sciences, UCD.


\end{acknowledgements}



\begin{dedication} 

To my Dad

\end{dedication}





\markboth{\MakeUppercase{\nomname}}{\MakeUppercase{\nomname}}


\nomenclature{GR}{General Relativity} 
\nomenclature{SNR}{Signal-to-Noise Ratio}
\nomenclature{BHB}{Black Hole Binary}
\nomenclature{EMRI}{Extreme Mass Ratio Inspiral}
\nomenclature{IMRI}{Intermediate Mass Ratio Inspiral}
\nomenclature{SMBH}{Supermassive Black Hole}
\nomenclature{IMBH}{intermediate Mass Black Hole}
\nomenclature{GC}{Clobular Cluster}
\nomenclature{NR}{Numerical Relativity}
\nomenclature{PN}{Post-Newtonian Theory}
\nomenclature{GSF}{Gravitational Self-Force}

\begin{multicols}{2} 
\begin{footnotesize} 

\printnomenclature[1.5cm] 
\label{nom} 

\end{footnotesize}
\end{multicols}


\mainmatter


\chapter{Introduction}


Every once in a while, science makes a ground-breaking discovery.  This year we were lucky enough \fixme{to witness} such an event - it will be remembered as the year in which the Higgs Boson was finally detected.  After decades of searching and non-stop research by both theorists and experimentalists of high energy particle physics, a detection was accomplished at CERN earlier this year.  As is the nature with many scientific breakthroughs, the excitement of the Higgs Boson came in two waves.  First is the theoretical wave, in this case, the production of the theory of electroweak unification \cite{Glenshow:1961, Weinberg:1967, Salam:1964} and with it, the prediction of the Higgs Boson \cite{Higgs:1964}.  As with most revolutionary theories, it took several years for people to warm to the initial idea, but after much investigation, the theory spoke for itself and became recognised as part of the standard model, to be taught to physics students globally.  As with all exciting theories, there then comes the search for physical evidence - a search which, in this case, would last almost half a decade, and result in the second wave of excitation - physical clarification that the theory is correct in the form of a direct detection of the Higgs Boson.  It was a momentous occasion for every researcher who has given their time and patience to the area.  

And while this was all happening, those of us sitting in the gravitational research area, also thrilled by the result, couldn't help but think - it's our turn next.


\section{Einstein's Theory of General Relativity} 


 \subsection{Testing the Theory}

Einstein's theory of general relativity (GR) was a revolutionary step in fundamental physics \cite{Einstein:1916}.  Like many of his era, Einstein was unsatisfied by the then accepted model of Newtonian physics, due to its inability to explain several observed effects in the world or universe around us and its unsatisfactory concept of absolute time and space.  GR successfully united Newtonian Mechanics and Special Relativity and had an immediate success as it naturally explained the precession of the perihelion of Mercury - an observation for which Newtonian theory could not completely account.  Depsite this initial success, there were many sceptics to the notion of curving space and time.  However, since there were other predictions by GR that would differ from Newtonian mechanics, it would remain only a matter of time before the theory was fully accepted.

A massive step in this direction was taken in 1919 by Sir Arthur Eddington.  Having been one of the first to receive news of the theory of GR, he organised two expeditions to observe a solar eclipse.  The reason was to measure the deflection of light by the sun, as Einstein's theory would predict a different value for this observation than that of Newtonian mechanics.  The experiment was a success \cite{Eddington:1919} and Einstein become world famous almost over night, while his theory started to overthrow its Newtonian counterpart.  Since 1919, there have been many more experiments testing the various available observables that can be used to support GR.  These have included verifying the gravitational redshift of light \cite{, Pound:1959}, gravitational lensing \cite{Walsh:1979} and time delay \cite{Shapiro:1964}, to name a few.  

One of the most exciting results to further fortify GR is the indirect detection of gravitational waves.  Gravitational waves are ripples in space-time as predicted by GR\fixme{;} they can arise from various events - compact object binaries, black hole mergers and supernovae are just a few examples.  In 1974, Hulse and Taylor discovered a new type of pulsar or radiating neutron star -  one with another pulsar in its orbit \cite{Taylor:1981}.  By observing the binary system, it was possible to calculate the orbit decay and show that the amount of energy being lost was consistent with the amount of energy that should be emitted as gravitational radiation as predicted by GR \cite{Damour:1983}.  

\begin{figure}
\begin{center}
\includegraphics[scale=0.7]{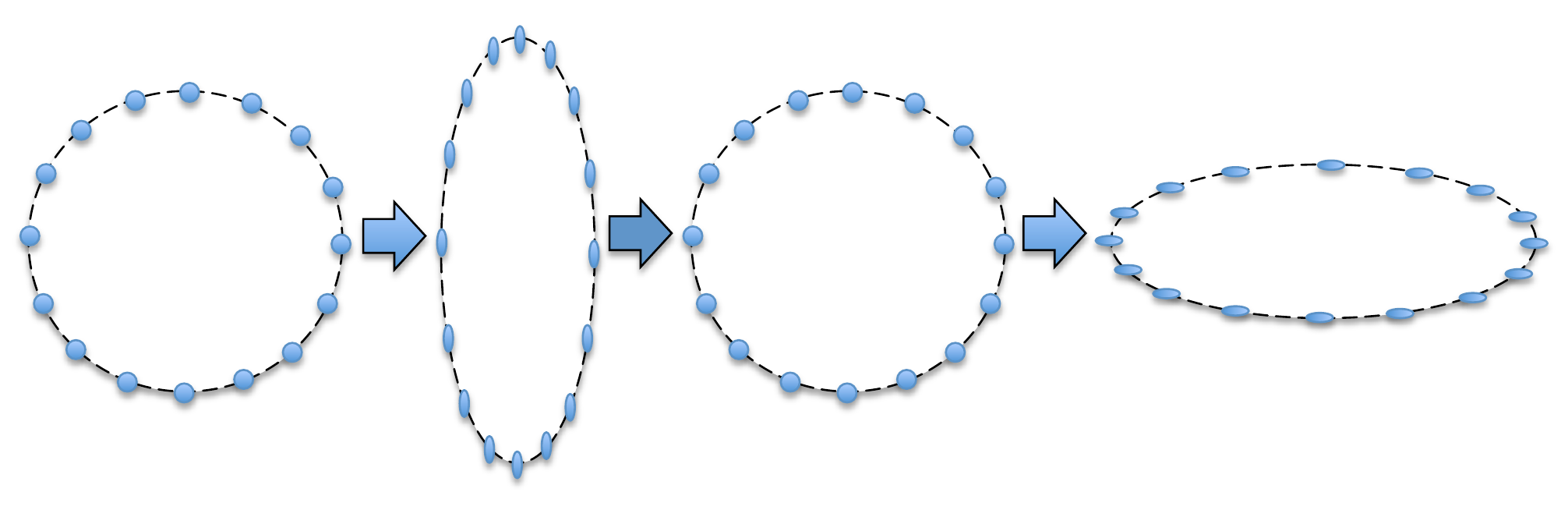}
\caption[Gravitational Wave `Stretching' Space]{If a gravitational wave were to pass through this page at a $90$ degree angle, with a plus polarisation, it would effect a circle of test masses as shown in the diagram.  There also exists the cross polarization which would have the same effect but rotated $45$ degrees.  Regardless of polarisation, the magnitude of strain to be measured is $1$ in $10^{21}$ parts.}
\label{fig: gwStretch}
\end{center}
\end{figure}

The Nobel prize winning work of Hulse and Taylor has encouraged relativists to work on the possibility of a direct detection of gravitational waves.  When a gravitational wave passes through space and time, it can be seen to \fixme{`stretch and squash'} the space it passes through, this is illustrated in Fig.~\ref{fig: gwStretch}, which shows a circle of test particles at rest being affected as a gravitational wave passes through this page.  In order to detect the waves, it is therefore necessary to be able to measure this `strain' that is placed on the test particles.  Due to the weakness of gravitational waves, however, this requires measuring a strain of $1$ in $10^{21}$ parts.  Until the 1990's, this accuracy in measurement was believed to be impossible\fixme{;} however, advances in technology and research, have now made it a possibility.

A direct detection of gravitational waves would mark a test of GR that would be the first of its kind - all previous tests of GR have measured the impact of GR on other observables in the weak regime while this would be a direct measurement of gravitational radiation predicted by GR in the strong field regime, i.e., when space and time are being strongly distorted.  Such a detection would be analogous to the recent detection of the Higgs boson, and with it would come the same thrill of accomplishment that is currently being enjoyed by our particle physics counterparts, albeit almost a century after Einstein revealed his theory.


\subsection{Gravitational Wave Astronomy}
Amazing strides have been made in Astrophysics in the last 7 decades.  We no longer rely solely on optical telescopes to inform us of the nature of our universe, instead there exists a network of satellites, antennas and telescopes that use optics, radio waves, infrared, X-ray and gamma rays to investigate the cosmos.  With each new window, came surprises that dramatically changed our understanding of the universe, some were expected but the more exciting were the unexpected, like pulsars \cite{Hewish:1968} or gamma ray bursts \cite{Klebesadel:1973}.  We are now, once again, on the verge of opening a new window onto our universe - that of gravitational wave astronomy. 

The thrill of detecting gravitational waves is not solely in the success of the detection but also in the wealth of knowledge that we can extract from the waveforms.  Gravitational waves can travel, relatively unaffected by any intervening matter, from their source to us, meaning they would carry first hand information about the violent processes that created them - processes that will often be invisible to all other types of detection available to us.  This invisibility is often due to the amount of intervening matter that would affect all other types of radiation,  but also, in some cases, such as those processes solely involving black holes, gravitational waves are the only type of classical radiation that will be emitted.  

Detection of gravitational waves is expected to occur in the next 5 years.  A network of ground-based detectors (LIGO \cite{LIGO}, VIRGO \cite{VIRGO}, GEO600 \cite{GEO}, TAMA \cite{TAMA}) have been operational for almost a decade - the first came online in 2002.  Although no detection has yet been made, hopes are high that the new advanced detectors will be successful.  This optimism is not baseless \fixme{- event rates for the gravitational wave detectors carry large error bars.  It was known that the initial detectors may not be successful, whereas the advanced detectors are expecting greater event rates than their predecessors, by a factor of  approximately 1000.  These, even with the more conservative estimates, predict that the advanced detectors should make positive detections \cite{AdvancedLigo}}.  The aim was to get an array of detectors up and running and work on reducing the noise to obtain the highest signal-to-noise ratio (SNR) possible.  Considering these detectors are required to measure strain of one part in $10^{21}$, obtaining the optimal SNR was a learning curve - some noises, although unexpected were easily removed (gunshots from hunters being such a source initially at the Louisiana LIGO site), others proved more difficult (laser shot noise).  In fact, during its final run, LIGO (Laser Interferometer Gravitational-Wave Observatory) was able to obtain a \fixme{strain sensitivity curve better} than was anticipated \cite{Pitkin:2011}.  

Gravitational wave detectors differ from their electromagnetic cousins in the sense that they have no ability to detect the direction from which the gravitational waves come.  The detector will `know' when a gravitational wave passes through it, however it has no way of telling where it came from.  For this reason, it has been crucial that there be a world network of detectors - by comparing what times each detector senses the incoming wave, we can figure out from what direction it came.  The main detectors, LIGO and VIRGO are currently offline, as they undergo major upgrades which are expected to improve the sensitivity of the detectors in strain and hence distance, by more than a factor of 10 \cite{AdvancedLigo}.  These advanced detectors are due to come online in 2015, and are fully expected to make the first gravitational wave detection.

\begin{figure}[ht]
\begin{center}
\includegraphics[scale=1.5]{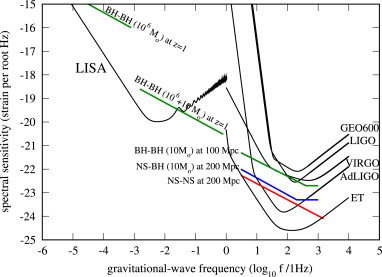}
\caption[Gravitational Wave Detectors' Sensitivity Curves]{The space (LISA) and ground based (LIGO, VIRGO, GEO600) detectors have different ranges of sensitivity and will therefore see gravitational waves from different sources.  Advanced LIGO is also shown as is ET - Einstein's Telescope - the third generation of gravitational wave detectors that will be underground.  This figure was taken from \cite{Anderson:2011}}
\label{fig: LISALIGOnoise}
\end{center}
\end{figure}

One of the unavoidable noise sources for ground-based detectors is seismic activity.  Together with other noise, this limits the range of the detectors, i.e., they can only see \fixme{gravitational waves within a certain frequency range.}  For this reason, there has been a wealth of research into the area of space-based detectors.  Such detectors, \fixme{although free from seismic noise, are still susceptible to noise sources such as detector and acceleration noise (shot noise in particular is responsible for the upward slope of all the sensitivity curves as they go towards higher frequencies as is seen in Fig.~\ref{fig: LISALIGOnoise}).  Their freedom from seismic noise opens these detectors to gravitational waves in a lower frequency range than their} ground-based counterparts. NGO/eLISA (New Gravitational-Wave Observatory/evolved Laser Interferometer Space Antenna)\cite{NGO} is such a space-based detector.  In Fig.~\ref{fig: LISALIGOnoise}, we can see the different noise curves attached to the detectors and what types of black hole binaries that they will be able to see.  It should be noted that the figure attached is for LISA and not eLISA/NGO which has a slightly higher noise curve.  We can see from the curve that EMRIs are expected to be seen by LISA.

NGO/eLISA is a modified version of the originally planned LISA which, due to cut backs in NASA, had to be redesigned on a smaller budget.  It will be up for selection as a L2 mission by the European Space Agency in 2015.  At the 2012 L1 selection process, eLISA did not get selected although it was ranked top by the  scientific review committee.  As the L2 decision will come after the launch of the LISA pathfinder \cite{LISAPath} as well as after the activation of advanced LIGO and VIRGO, the gravitational wave community are optimistic that the mission will be selected.


\section{The Two-Body Problem}

The two-body problem in Newtonian theory is readily solvable.  An isolated system of two point masses is governed by conserved integrals describing the energy and momentum resulting in periodic motion.  \begin{figure}
\begin{center}
\includegraphics[scale=0.7]{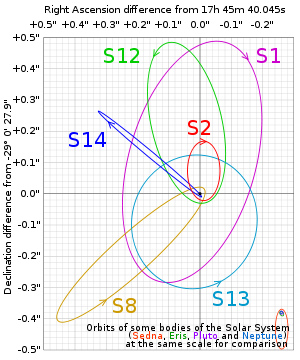}
\caption[Orbits Around Supermassive Black Hole Sagittarious A*]{The orbits of six stars that were tracked with the Very Large Telescope of the European Southern Observatory, Chile.  It is clearly seen that all 6 stars are orbiting a central mass that is invisible to the telescope.  From the motion of S2, the object's mass is estimated as $4.1 $x$ 10^6$ll solar masses, recent observations have indicated that the radius of the object is no more than $6.25$ light years - placing the object in the category of supermassive black hole.  This figure was taken from \cite{Eisenhauer:2005}}
\label{fig: BHCentre}
\end{center}
\end{figure}The two-body problem in general relativity is somewhat different - it is a longstanding open problem going back to work by Einstein himself. With recent advances in gravitational wave detector technology, this age-old problem has been given a new lease of life.  Some of the key sources expected to be seen by both space and ground based gravitational wave detectors are black hole binaries (BHBs).  These can be divided into 3 categories – extreme mass ratio inspirals (EMRIs), intermediate mass ratio inspirals (IMRIs) and comparable mass BHBs. This development is today motivating numerical, analytical and experimental relativists to work together with the prospect of bringing about the reality of gravitational wave astronomy.


\subsection{Black Hole Binary Sources}

Supermassive black holes (SMBHs - black holes with mass approximately $10^6$ times that of the sun) are believed to be located at the centre of galaxies; in fact it is known by indirect detection that one resides in the centre of our own galaxy \cite{Gillessen:2009}.  This is very clear in Fig.~\ref{fig: BHCentre} where the orbits of several stars were tracked at the centre of our galaxy.  It can be seen that they are all orbiting an `invisible' body that has dimensions that match that of a SMBH.  Near central SMBHs, there are also a disproportionately large number of stellar-mass black holes, which have sunk there through random gravitational encounters.  Every now and then, one of these stellar-mass black holes, through interactions with other bodies, will be ‘bumped’ into the grasp of the SMBH, which will initiate the start of a slow inspiral into the SMBH.  These inspirals are known as EMRIs.  EMRIs are proving to be one of the most exciting prospects for space-based detectors.  The smaller black hole can be expected to complete over $10^5$ orbits in the relativistic regime of the Kerr (rotating) black hole \cite{Barack:2009}.  The resulting emitted radiation will therefore carry information about both the inspiral parameters as well as the space-time geometry that in turn can be used to test General Relativity in the strong field regime. 

The existence of intermediate-mass black holes (IMBHs) with masses ranging from 100 and 10 000 solar masses has not yet been confirmed but there is evidence that favours their existence \cite{Farrell:2009, Farrell:2012}.  These objects are of high astrophysical interest as their existence would impact current understandings of the formation and evolution of both SMBHs and galaxies.  IMBHs are believed to reside in the centre of globular clusters (GSs), which are difficult to resolve, making detection very difficult.  Therefore, a key method of detecting an IMBH could be to detect an IMRI or comparable mass BHBs by use of gravitational wave detectors.  IMRIs can be seen as falling into two categories -– an IMBH falling into a SMBH that could be detected by space-based gravitational wave detectors or advanced ground-based detectors \cite{AmaroSeoane:2007, Knostantinidis:2012}, or a stellar-mass black hole falling into an IMBH, which is expected to be detectable by advanced ground-based detectors \cite{Smith:2013}.  IMRIs will also be interesting sources for gravitational wave detectors for similar reasons as EMRIs, they too will experience long inspirals and hence have the potential to reveal information about the space-time geometry of Kerr black holes \cite{AmaroSeoane:2007, Knostantinidis:2012}.

Comparable mass BHBs as well as comparable mass compact body binaries are also expected to be key gravitational wave sources for both ground and space based detectors.  Stellar mass BHBs are thought to form in GCs through 3 body interactions.  Their attractiveness as a source for gravitational wave detection (GWD) lies in the fact that they are not strongly bound to the cluster.  This implies the possibility of the binary being expelled from the cluster \fixme{due to} interactions with other bodies, resulting in the system evolving in isolation away from the noise of the cluster, which in turn makes them an accessible source of gravitational waves for ground based detectors.  SMBH binaries (comparable mass BHBs where both black holes are supermassive), on the other hand, are expected to be seen by space-based detectors.  SMBH binaries are of great interest to the gravitational wave detection community due to their expectantly large SNR, which should make them detectable with minimal use of data analysis.  Accurate models of the inspiral and merger will still be required for using these signals to determine source parameters.  


\subsection{Modelling Techniques}

\begin{figure}
\begin{center}
\includegraphics[scale=0.475]{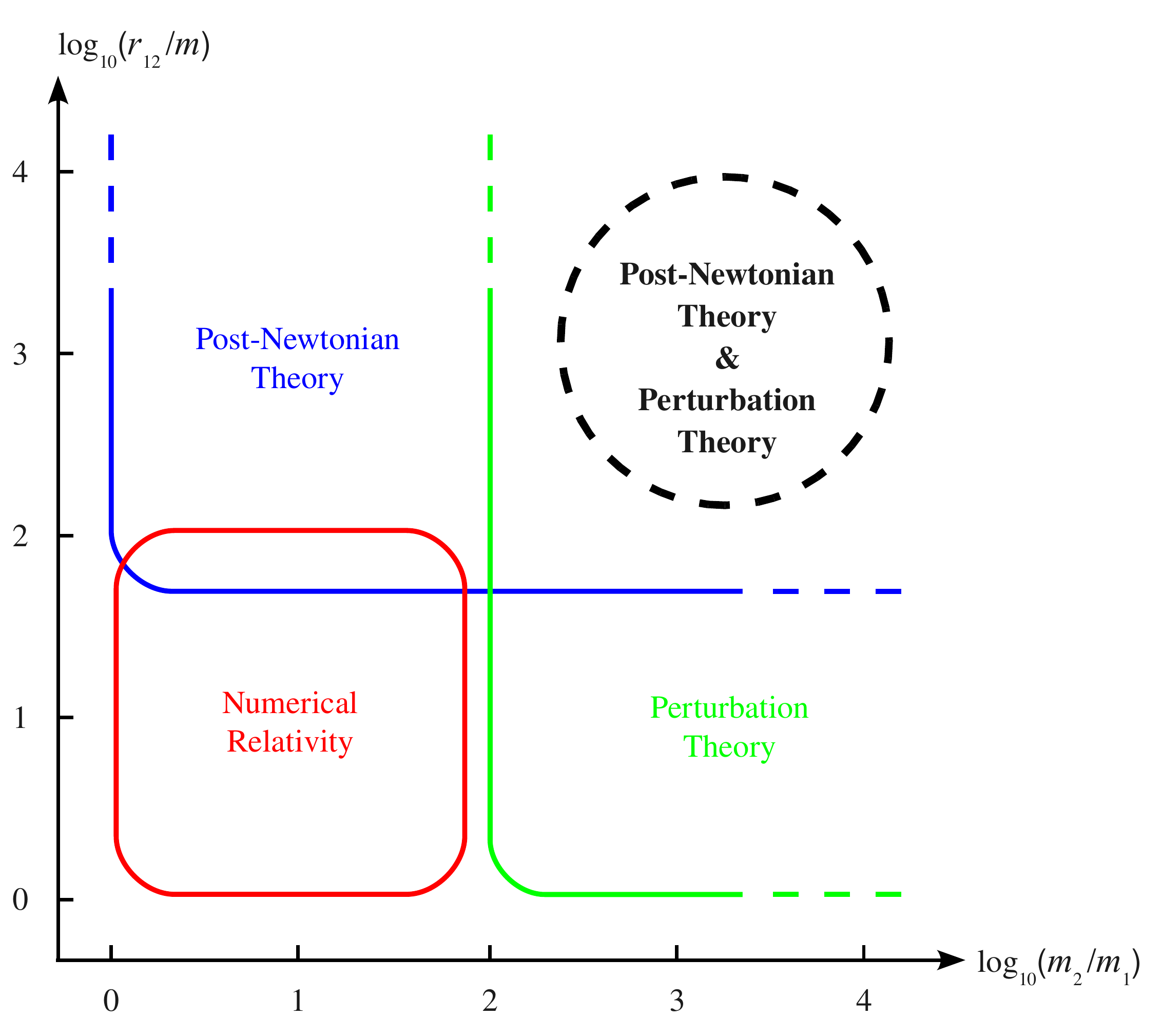}
\caption[Black Hole Parameter Space]{Application of BHB modelling methods for the parameter space of mass ratio $m_1/m_2$ and separation $r_{12}$.  This schematic figure is taken from \cite{Blanchet:2010}.}
\label{fig: parameterSpace}
\end{center}
\end{figure}

Many data analysis techniques currently being used in the search of gravitational waves are based on matched filtering; – this allows the extraction of signals buried deep in instrumental noise with significant SNR.  For successful detection, matched filtering requires accurate waveform templates.  In the case of BHBs, several methods are used to calculate the expected waveforms.  Numerical relativity (NR) has become an invaluable tool in these calculations; however, it does not come without its constraints.  It is extremely computationally expensive and is not suited to BHBs with either a large separation or large mass ratios.  In these instances, post-Newtonian (PN) and gravitational self-force (GSF) techniques are required respectively - this `sharing' of the possible parameter space between the different techniques can be visualised in Fig.~\ref{fig: parameterSpace}.

GSF theory is closely related to black hole perturbation theory and uses a perturbation of Einstein's field equations in the mass ratio to describe the motion of a point particle in a given background space-time.  At zeroth order in the small mass ratio, the point mass follows a geodesic of the background.  At first order, it deviates from this geodesic due to its interaction with its own field.  This deviation is interpreted as a force acting on the mass, the so-called GSF.  What makes these calculations difficult is that a point mass in curved space-time gives rise to a field that diverges at the particle.  It is possible to isolate that part of the physical field that is responsible for its singular behaviour.  By subtracting the singular component, the so-called Detweiler-Whiting singular field, from the retarded field, we are left with the regular part, which is (by construction) wholly responsible for the self-force.  There are three main approaches to calculating the self-force in practice, and all involve this regular-singular split of the field.

PN theory also uses a perturbation of Einstein’s field equations \fixme{by} using two parameters \fixme{–-} the typical velocity of the system (divided by the speed of light) and a measure of the deviation of the curved space-time from a flat space-time (i.e. the deviation from the flat metric).  At lowest order, PN \fixme{analysis} gives a Newtonian description and \fixme{general relativistic} effects are described as higher order perturbations.  As PN theory is a perturbation in the curvature of space-time and velocity of the system, it is effectively assuming that both are very small parameters, i.e. the theory is only applicable to slow systems in the weak field regime.  PN approximations have proven their ability to model comparable mass BHBs as well as IMRIs and EMRIs in the inspiral stage, however PN breaks down at the merger stage where NR is required for solving the final orbits of the binaries.

One can clearly see that PN and GSF by their nature are constrained to the modelling of certain systems.  GSF requires an extreme mass ratio, while PN is only applicable to slow systems with weak fields, meaning it should not be expected to be very effective in the later stages of BHB inspirals.  The word `should\fixme{'’} is intentionally used in this description as PN theory has been applied to strong-field, fast–motion systems like BHBs with remarkable success.  By going to higher orders, the PN community has shown impressive results that agree with computationally expensive NR simulations, proving the application of PN in strong fields with fast motions \cite{Baker:2007, Boyle:2007ft, Hannam:2008}.  PN theory does eventually become ineffective as the inspiral evolves in BHBs but at a much later point than previously expected.  The reason for PN’s ability to work outside its expected regime is largely unknown but welcomed by the PN community.  

GSF is also currently experiencing a similar inexplicable success outside its effective parameter space.  Recent advances \cite{LeTiec:2012, LeTiec:2011} have shown how GSF can be applied to IMRIs and comparable mass binaries with encouraging results with comparisons to PN and NR.  A great consequence of this work is extending the viability of the work from the GSF community to ground-based detector sources, which is also most welcomed by the community.

Regardless of the method used, the endgame of BHB modelling is to have a complete waveform template `bank' available for use by both ground and space-based detectors.  To this end, researchers from each of the areas are beginning to come together to compare the different methods and use them to complement each other, making it a truly global effort to assist in the detection of gravitational waves.


\section{The Self Force Problem - Thesis Outline}


\subsection{History of Theory}
This thesis will concentrate on the GSF technique, also known as the self-force problem.  As described above, the main problem in this approach lies in the singularity of the field at the particle.  Fortunately, producing expressions for such fields is nothing very new -- the self-acceleration of a charged point particle in flat space-time is given by the well known Abraham-Lorentz-Dirac formula \cite{Dirac-1938}.  In this scenario, the charge produces a field that acts as radiation, which in turn, diverts the particle from its geodesic -- for this reason, it became known as the radiation reaction.  

It was almost three decades later when DeWitt and Brehme derived the formula for the self-force of a charged particle in curved space-time \cite{DeWitt:1960}, generalizing the results of Dirac et al..  Their calculation did require a minor correction, which was provided by Hobbs several years later \cite{Hobbs:1968a}.  It was not until the late 1990s, however, that  Mino, Sasaki and Tanaka produced the most physically relevant and interesting version of the result -- that of a point mass in curved space-time \cite{Mino:Sasaki:Tanaka:1996}.  This result, also obtained by Quinn and Wald \cite{Quinn:Wald:1997} using a different approach, led to the famous MiSaTaQuWa equations, which identified the correct regularisation procedure to remove the problematic singularity.  The method they formulated, however, was not practical for calculations, and so, was `redesigned' by Barack and Ori in 2000 \cite{Barack:Ori:2000}.  Quinn was also the first to produce results in the case of a point scalar charge \cite{Quinn:2000} -- a simpler model, but one that has been used throughout the community as a test bed for new ideas and methods (it is worth noting that Barack and Ori also considered this case initially for their mode-sum scheme \cite{Barack:Ori:2000}).  There are several reviews that summarise very well all the work that has been done on this problem -- in particular those by Poisson \cite{Poisson:2003}, Detweiler \cite{Detweiler:2005} and Barack \cite{Barack:2009}.


\subsection{Main Approaches}
The three main methods of calculating the self-force are known as matched expansions, mode sum and effective source.  Like most complicated calculations, these GSF approaches are first attempted in toy-models.  In the GSF context, the complexity of the calculation increases with the spin of the field i.e., scalar is considered the simplest, followed by the electromagnetic and gravitational cases.  The space-time can also increase in complexity, with the key increase arising from going from non-rotating black holes (\fixme{Schwarzschild,} Reissner-Nordstr\"om) to rotating black holes (Kerr, Kerr-Newman).

The matched expansions concept was first suggested by Poisson and Wiseman \cite{Poisson:Wiseman:1998}.  They suggested matching together two independent expansions for the Green's function -- one in the `quasilocal' regime and one in the `distant' past regime.  The quasilocal approach was introduced by Anderson et al. \cite{Anderson:2003, Anderson:Wiseman:2005}, this method uses the MiSaTaQuWa equations to compute the relevent Green's function via an analytic Hadamard expansion.  This was built on by Ottewill and Wardell \cite{Ottewill:Wardell:2008, Ottewill:Wardell:2009}, by obtaining a very high order of accuracy from the Hadamard expansion.  Joining with Casals and Dolan, they successfully used their results to calculate the self-force on a charged particle, initially in Narai space-time (a simple toy black hole space-time) \cite{Casals:Dolan:Ottewill:Wardell:2009}, and more recently in \Sch space-time \cite{Casals:2013}.  

The effective source method was independently proposed by Barack and Goldburn \cite{Barack:Golbourn:Sago:2007, Barack:Golbourn:2007} and Detweiler and Vega \cite{Vega:Detweiler:2008}.  The methods they used were slightly different, but the concept was very much the same.  That was to solve for the fully regularised field from the homogeneous wave equation in the near neighbourhood as well as that of the retarded field outside the near neighbourhood, and uniting the results at the boundary to give that part of the field responsible for the self force.  In doing so, they were able to obtain an approximate regularised field that is fully derived from the singular field.  The difference of their methods emerged in how they separated the two regions -- Detweiler and Vega developed the window function which effectively 'smeared' the impact of the singular part of the field from full strength at the particle to zero outside the near neighbourhood; while Barack and Goldbourn introduced a world tube to separate the \fixme{two regions} and imposed boundary conditions to unite \fixme{them}.  The most exciting result from the effective source method is the production of an outline to calculate the self-force to second order - a feat that has never before been accomplished, and so, is currently receiving much attention.  This has led to another surge in excitement amongst the self force community, as second-order would no doubt lead to more accurate calculations of the self-force and resulting wave-forms \cite{Rosenthal:2005ju,Rosenthal:2006iy,Detweiler:2011tt,Pound:2012nt,Gralla:2012db}.

To date, the mode sum method has been the most successful regularisation procedure for calculating the self-force, although the effective source is very clearly catching up.  In the mode-sum method, one applies a spherical harmonic decomposition of the singular field; each of the multipole modes is then finite even at the particle, allowing to conveniently subtract the singular field mode by mode. One then numerically calculates the physical field multipoles for input into a mode-sum regularization formula -- one involving certain analytically given “regularization parameters” that characterizes the singular behaviour at large multipole numbers. The more regularization parameters one can derive, the faster the convergence of the mode sum becomes. Knowledge of high-order regularization parameters is crucial for assuring the efficiency and accuracy of the GSF calculation. 

The mode-sum was first introduced by Barack and Ori \cite{Barack:Ori:2000}, and further developed by Barack, Ori, Nakamo and Sasaki \cite{Barack:Mino:Nakano:Ori:Sasaki:2001, Barack:Ori:2002, Barack:2001, Mino:Nakano:Sasaki:2002}.  The development of the Detweiler-Whiting singular field \cite{Detweiler-Whiting-2003} furthered the approach even more, and was followed by a very clear decomposition of the scalar field into mode sums by Detweiler, Whiting and Messaritaki \cite{Detweiler:Messaritaki:Whiting:2002}.  Since its introduction, the mode-sum method has been successfully applied to the more complicated models -– including a point electric charge and point mass in \Sch space-time  \fixme{\cite{Haas:2011bt, Barack:Sago:2010}}, as well as a point scalar charge in Kerr space-time \fixme{\cite{Warburton:Barack:2010}}.  The ultimate goal is to extend this to the astrophysically interesting case of a point mass in Kerr space-time.


\subsection{Thesis Outline} \label{sec: thesisOutline}

As self-force plays \fixme{its} part in BHB modelling, and BHB modelling plays its part in the search for gravitational waves, this thesis is also aimed to assist greater goals.  We have mostly concerned ourselves with computing the singular field in the different scenarios, and using both the effective source and mode sum methods to obtain results that will assist our fellow researchers.  By specialising solely on the singular field, we were able to bring it to an accuracy not conceived possible by even the founders of some of the methods used.  To summarise, the results of this thesis \fixme{enable} more accurate and more efficient calculations of the self-force for all researchers in the field, thus making their lives a little bit easier.

Section \ref{sec: background} contains the necessary background for calculating the singular field.  This backgroud has been reviewed far more extensively in \cite{Poisson:2003}, however the scope of this section is on a `need to know basis' with respect to the rest of the thesis.

Section \ref{sec: highOrder} describes the methods used in calculating the singular field - this was done both covariantly and in coordinates, with both methods having advantages and disadvantages.  In the different methods, we also expanded around different points to introduce as much independence as possible for the two methods.  I played the the main role for the coordinate results, while my colleague, Barry Wardell, took the lead for the covariant results, which are also in this thesis for completeness.  By working in this manner, we could independently check our results and, hence, have great confidence in the results produced.  We found both methods produced the same singular field up to an order of $\epsilon^6$, where $\epsilon$ is the order of distance in the calculations.  A singular field to this accuracy has never before been calculated -- it assisted us in pushing the boundaries on both the matched expansion and mode-sum methods.

Section \ref{sec: modeSum} describes the mode sum method in detail and shows the regularisation parameters that we were able to produce in both \Sch and Kerr space-times.  These parameters have already been used by several groups and have resulted in self-force calculations to unprecedented accuracy.  This work has resulted in over ten parameters, previously unknown, and greatly appreciated by our peers.

Section \ref{sec:EffectiveSource} investigates the effective source method.  As in the mode sum, we used our high-order singular field to push the boundaries on previous results -- producing a very smooth field in both \Sch and Kerr space-times.  We also extend on the $m$-mode method, which has evolved from the effective source model, and offer up parameters in both space-times for high-order calculations.  The $m$-mode scheme is an alternative to the mode-sum scheme, introduced for Kerr black holes.  It was found that mixing of the modes occurs when calculating the retarded field using the mode-sum method for the gravitational Kerr case, therefore, an alternative that avoids this `mixing' was introduced in the form of the $m$-mode method.  Previously, researchers only used expansions of the singular field up to $\epsilon^2$ in the $m$-mode scheme, as the higher orders tend to slow the numerical calculations down. We introduce a method, whereby these higher orders can be used to further regularise the field, without slowing down the numerical calculations.

Section \ref{sec: extensions} describes further extensions of the high order expansions of the singular field. One of these is the investigation of the cosmic censorship conjecture, which involves the concept of \emph{overcharging} or \emph{overspinning} black holes.  To assist in these investigations, we produce regularisation parameters for generic motion and radial infall in a spherically symmetric space-time, as well as the motion of a charged particle in \rn space-time.  Another extension of this research is in the ongoing work towards calculating the second-order self-force.  Such calculations require regularisation parameters of the second derivative of the singular field, which we provide.

The final section summarises the results and accomplishments covered in this thesis.  We discuss the impact and importance of our results and offer several avenues, down which, this work can be continued.  

Some parts of this thesis have been in collaboration with both Barry Wardell and my thesis supervisor, Adrian Ottewill.  For clarity and completeness, that work has been included here in full.  Sections that I was not the primary contributor are indicated by an asterisk (*).

\fixme{While the primary focus of this thesis is on computing the singular field for specific
space-times, many of the expressions we give are valid in more general spacetimes. In particular, where space allows, we do not make any assumptions about the spacetime being Ricci-flat.  To make this distinction explicit, we use the Weyl tensor, $C_{abcd}$, in expressions
which are valid only in vacuum and the Riemann tensor, $R_{abcd}$ in expressions which are also
valid for non-vacuum spacetimes. Note that this is done only for space reasons\footnote{The
notable exception is the case of the gravitational singular field, as in that case the
equations of motion have not yet been derived for non-Ricci-flat spacetimes.}; our raw
calculations include all non-vacuum terms in addition to those given in this thesis and we
have made the full expressions available in electronic form \cite{BarryWardell.net}.}

Throughout this thesis, we use units in which $G=c=1$ and adopt the sign conventions of
\cite{Misner:Thorne:Wheeler:1974}. We denote symmetrization of indices using parenthesis
(e.g. $(ab)$), anti-symmetrization using square brackets (e.g. $[ab]$), and exclude indices
from (anti-)symmetrization by surrounding them by vertical
bars (e.g. $(a | b | c)$, $[a | b | c]$). We denote pairwise (anti-)symmetrization using an
overbar, e.g. $R_{(\overline{ab}\,\overline{cd})} = \frac12 (R_{abcd}+R_{cdab})$, when multiple symmetries are required.  Capital letters are used to denote the spinorial/tensorial indices appropriate to the field being considered.  \fixme{For convenience, we frequently make use of the shorthand notation of \cite{Haas:Poisson:2006} by introducing definitions such as $R_{u \sigma u \sigma | u \sigma} = R_{abcd|ef} u^a \sigma^b u^c \sigma^d u^e \sigma^f$.  As is standard practice, commas denote partial differentiation whereas semi-colans represent covariant differentiation, however, these may sometimes be omitted when they are interchangeable , i.e., covariant derivative of a scalar $\sigma_{;a} = \sigma_{,a} \equiv \sigma_a$.}



\chapter{Background} \label{sec: background} 




\section{Bitensors and Basics}

In this section, we will review the specific biscalars, bivectors and bitensors that are required to fully comprehend this thesis as well as concepts such as geodesics and Penrose diagrams.  Throughout, we are primarily dealing with two points - $x'$ which is considered to be the source or base point and $x$, which is a field point, assumed to be in the normal convex neighbourhood of $x'$ - this concept will be explained in the next sections.  


\subsection{Geodesics}
Before we look into the different categories of space-times, it is beneficial to understand how they are represented.  Space-times are described by their metric, $g_{ab}$ or line-element $ds^2$ which are related by
\begin{equation} \label{eqn: ds}
ds^2 = g_{ab} (x) dx^a dx^b,
\end{equation}
where $ds$ can be described as the infinitesimal space-time distance between two neighbouring points $x^a$ and $x^a + dx^a$.  The line element can, therefore, be seen to specify a geometry, although it should be noted that many different line elements can describe the same geometry.  The line element can be derived from the Lagrangian \cite{Chandrasekhar}, given by
\begin{equation} \label{eqn: lagrangian}
\mathcal{L} = - \frac{1}{2} g_{ab} \frac{dx^a}{d \lambda} \frac{dx^b}{d \lambda},
\end{equation}
where $\lambda$ is some affine parameter along the geodesic - for time-like geodesics, $\lambda$ may be proper time,  $\tau$.

As the line element carries information about the infinitesimal space-time distance between two points, it can be used to determine whether the two points are time-like separated, null separated or space-like separated.  If two point particles are time-like separated, it is possible for one particle (that in the past of the other), to arrive at the same point in space and time as its partner.  If they are null or light-like separated, one can only reach the position of the other in space and time if it can travel at the speed of light.  While space-like separated means that unless one particle can travel faster than the speed of light, it can never occupy the same point in  space and time as its partner.  The line-element, by its nature, can tell us how two points are separated by,
\begin{equation}
ds^2 \begin{cases}
		> 0 & \text{space-like separated} \\
		= 0 & \text{null seperated} \\
		< 0 & \text{time-like seprated}
	\end{cases}.
\end{equation}

\begin{figure}
\begin{center}
\includegraphics[scale=0.47]{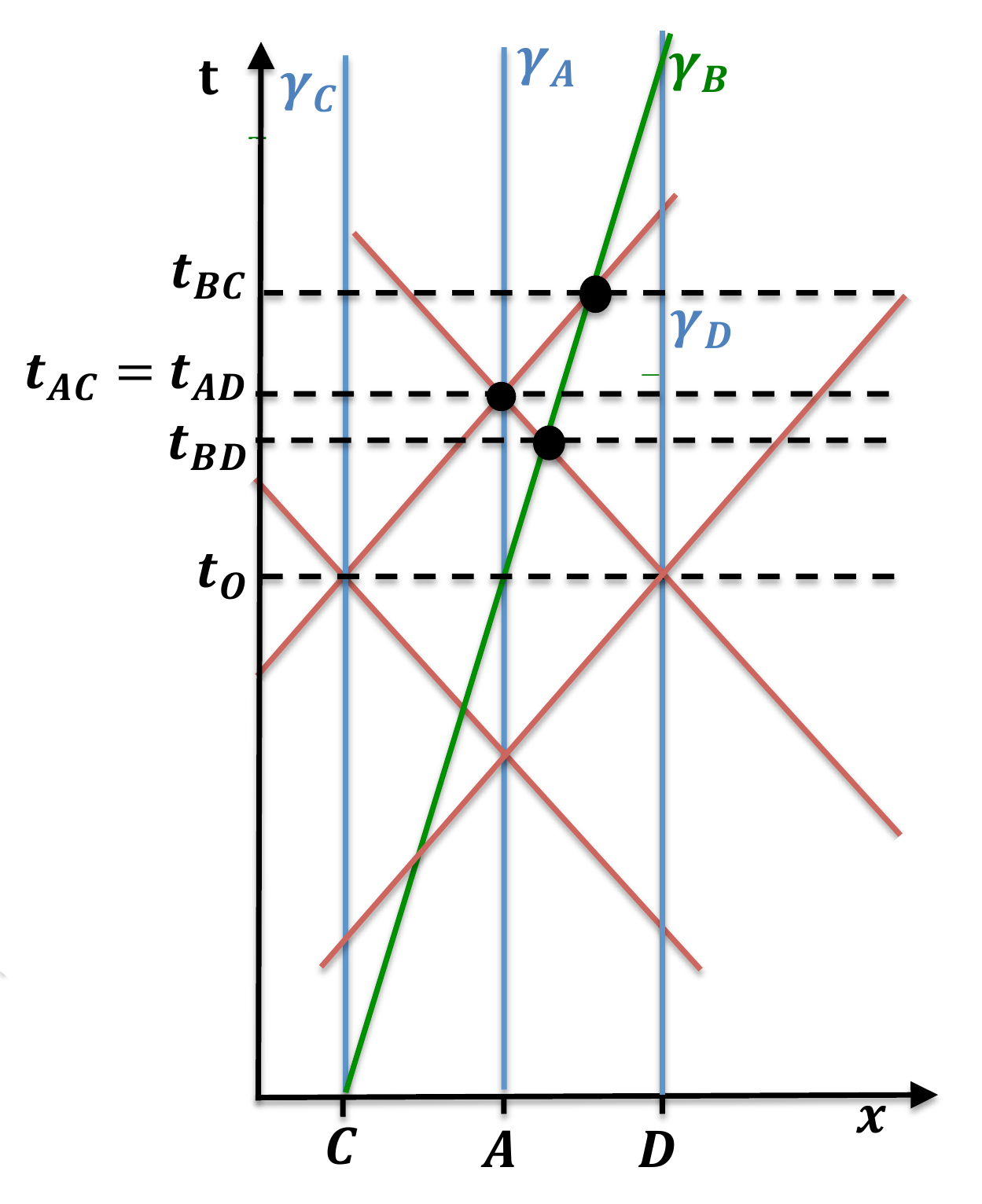}
\caption[Light-cone of Events]{ $A$, $C$ and $D$ are particles at rest and so travel along straight world lines represented in blue by $\gamma_A$, $\gamma_C$ and $\gamma_D$.  $B$ starts at $C$ and travels at a finite velocity to $D$ and is represented by the green world line, $\gamma_B$.  When $B$ passes the half way point, where $A$ resides, at $t = t_O$, light is emitted from $C$ \fixme{to} $D$ \fixme{and from $D$ to $C$.  From above,} we can see that $A$ will observe the two light rays to be emitted at the same time $t_{AC}$, while $B$ will observe light being emitted from $D$ first, then $C$.  On our time line we can clearly see that $B$ will first observe $D$ occurring, then $A$ will witness $C$ and $D$ occurring simultaneously, and then $B$ will observe $C$ happening. In this manner the concept of the light cone can be used to illustrate the relationship of events.}
\label{fig: coneExample}
\end{center}
\end{figure}

This concept of separation in space and time can also be illustrated with the use of a light cone.  Light cones are merely lines that represent the path of a particle travelling at the speed of light leaving and arriving at a point in space-time.  As we take the speed of light $c=1$, on a 2 dimensional space-time diagram this represents lines of slope $\pm1$, i.e., those that make a $45$ degree angle with the axis.  A example of their use to avoid confusion in the observation of events is illustrated in Fig.~\ref{fig: coneExample}.  With light cones, when a particle is in the future or past light cone of another, they are said to be time-like separated, if they reside on each others light cones, they are null separated and if they are outside each others light cones, they are space-like separated.

In space-time diagrams, the path a particle takes through space and time is known as a world line as it represents the points in space-time that the particle has occupied.  Geodesics are world lines that extremise proper time, that is the curve for which an infinitesimal variation in space $\delta x^a$, produces a vanishing variation in proper time.  The flat space-time equivalent of this, is a straight line connecting two points, however in four dimensional space-time, this concept, like many others, is slightly more complicated.

If two points are time-like separated, the line element in Eq.~ \eqref{eqn: ds} can be used to describe the proper time between the two points in space-time from $d\tau ^2 = -ds^2$, that is
\par \vspace{-6pt} \begin{IEEEeqnarray}{rCl}
\tau &=& 
	\int_{x(\lambda_0)}^{x(\lambda_1)} \left( -g_{ab} dx^a dx^b \right)^{1/2} \nonumber \\
&=&
	\int_{\lambda_0}^{\lambda_1} \left( -g_{ab} \frac{dx^a}{d \lambda} \frac{dx^b}{d \lambda} \right)^{1/2} d \lambda.
\end{IEEEeqnarray}
World lines that extremise the proper time between two points must satisfy Lagrange's equations,
\begin{equation}
\frac{d}{d \lambda} \frac{d \mathcal{L}}{d \dot{x}^a} - \frac{d \mathcal{L}}{d x^a} = 0,
\end{equation}
where the Lagrangian is given by Eq.~\eqref{eqn: lagrangian} and ($\dot{}$) refers to differentiation with respect to $\lambda$.  Some straight forward algebra results in the geodesic equation,
\begin{equation} \label{eqn: geodesicEqn}
\frac{d^2 x^a}{d \text{\fixme{$\lambda$}}^2} + \Gamma^a_{bc} \frac{d x^a}{d \lambda} \frac{d x^b}{d \lambda} = 0,
\end{equation}
where the $\Gamma^a_{bc}$'s are called the Christoffel symbols and are given by,
\begin{equation} \label{eqn: christoffel}
\Gamma^a_{bc} = \frac{1}{2} g^{ad} \left( g_{db,c} + g_{dc,b} - g_{bc, d} \right),
\end{equation}
where $A_{,b}$ implies $\frac{\partial A}{ \partial x^b}$.

The line element can also be used to normalise the four-velocity, which is defined to be
\begin{equation}
u^a = \frac{d x^a}{d \tau}.
\end{equation}
From the line element and $d\tau ^2 = -ds^2$, it is straight forward to show,
\begin{equation} \label{eqn: uu}
g_{ab} u^a u^b = g_{ab} \frac{dx^a}{d \tau} \frac{dx^b}{d \tau} = -1.
\end{equation}

It is now possible to define the meaning of a normal convex neighbourhood: the normal convex neighbourhood of a point is the set of points that are connected to it by a unique geodesic.  If we consider the geodesic which connects $x$ and $x'$, we can use $z(\lambda)$ to represent any point on this geodesic.


\subsection{Penrose Diagrams} \label{sec: penrose}

A Penrose diagram can be seen as a coordinate transformation that allows us to view our space-time geometry in a different light, giving us insights into the physical implications of the space-time.  In emphasising the light cone structure of space-time, it successfully maps all of space-time onto a finite space.  This is a conformal mapping that compactifies the space-time whilst preserving the light-cone structure.

\begin{figure}
\begin{center}
\includegraphics[scale=0.52]{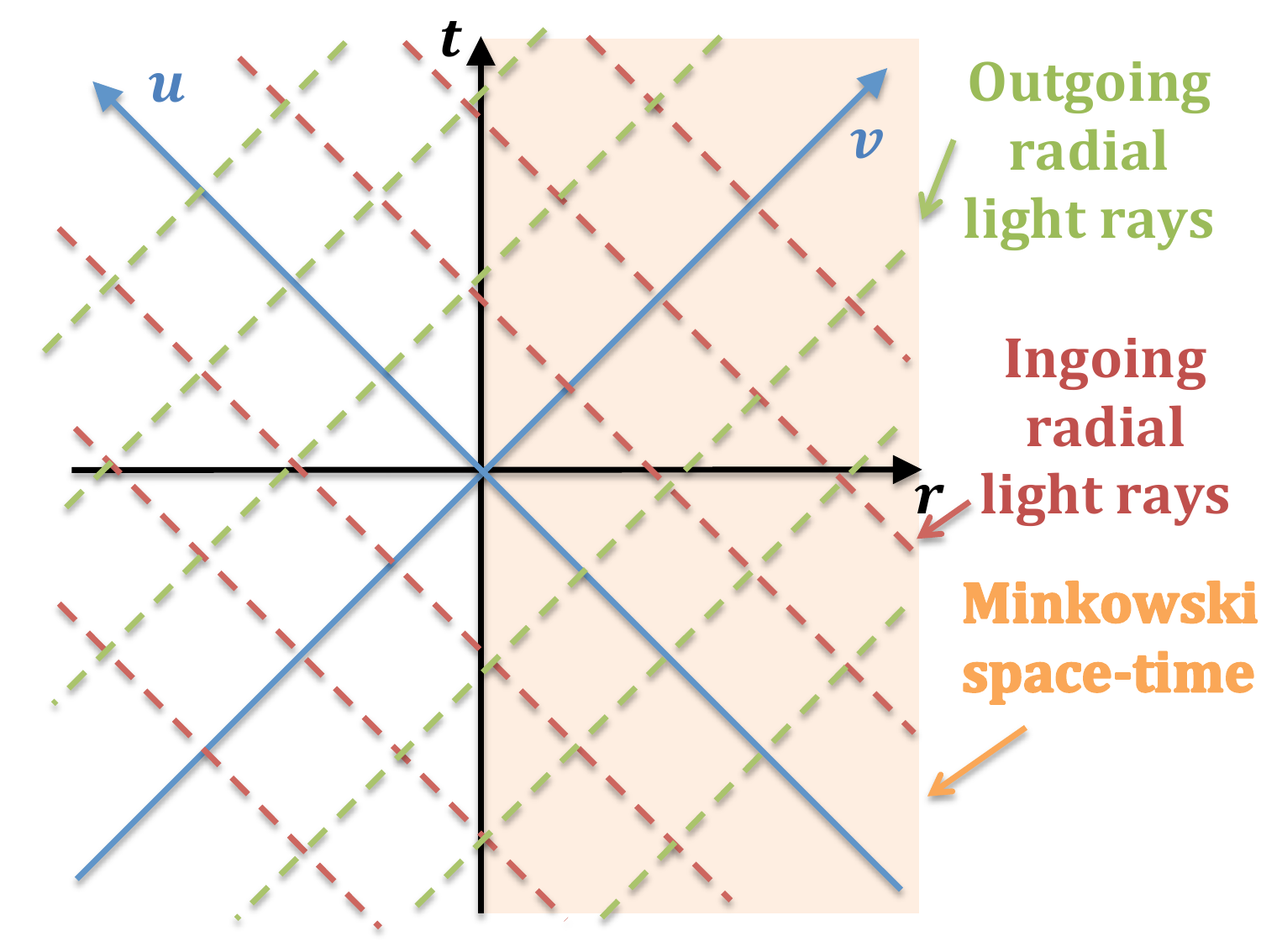}
\caption[Minkowski Space-Time]{The Minkowski (flat) space-time in null coordinates $u = t - r$ and $v = t + r$.  We note that outgoing radial light rays can be described by $u=c_1$ for \fixme{any} constant, $c_1$, while ingoing radial light rays are described by $v = c_1$.  Minkowski space-time is limited to $t \in \left( - \infty, \infty \right)$ and $r \in \left( 0, \infty \right)$, this is equivalent to $v > u$ in the above coordinate system.}
\label{fig: uvAxes}
\end{center}
\end{figure}
In Sec.~\ref{sec: BHST}, we will describe the various types of space-times that are essential for the understanding of this thesis.  However for now we will consider a flat space-time to introduce the concept of a Penrose diagrams.  Flat space-time, known as the Minkowski space-time, is described in Cartesian coordinates, by the line element
\begin{equation}
ds^2 = - dt^2 + dx^2 + dy^2 + dz^2,
\end{equation}
which can be rewritten in spherical polar coordinates as
\begin{equation} \label{eqn: minkDs}
ds^2 = - dt^2 + dr^2 + r^2 \left( d \theta^2 + \sin^2(\theta) d \phi^2 \right),
\end{equation}
where we have used the transformation
\begin{equation}
x = r \sin{\theta} \cos{\theta}, \quad y = r \sin{\theta} \sin{\phi}, \quad z = r \cos{\theta}.
\end{equation}

\begin{figure}
\begin{center}
\includegraphics[scale=0.4]{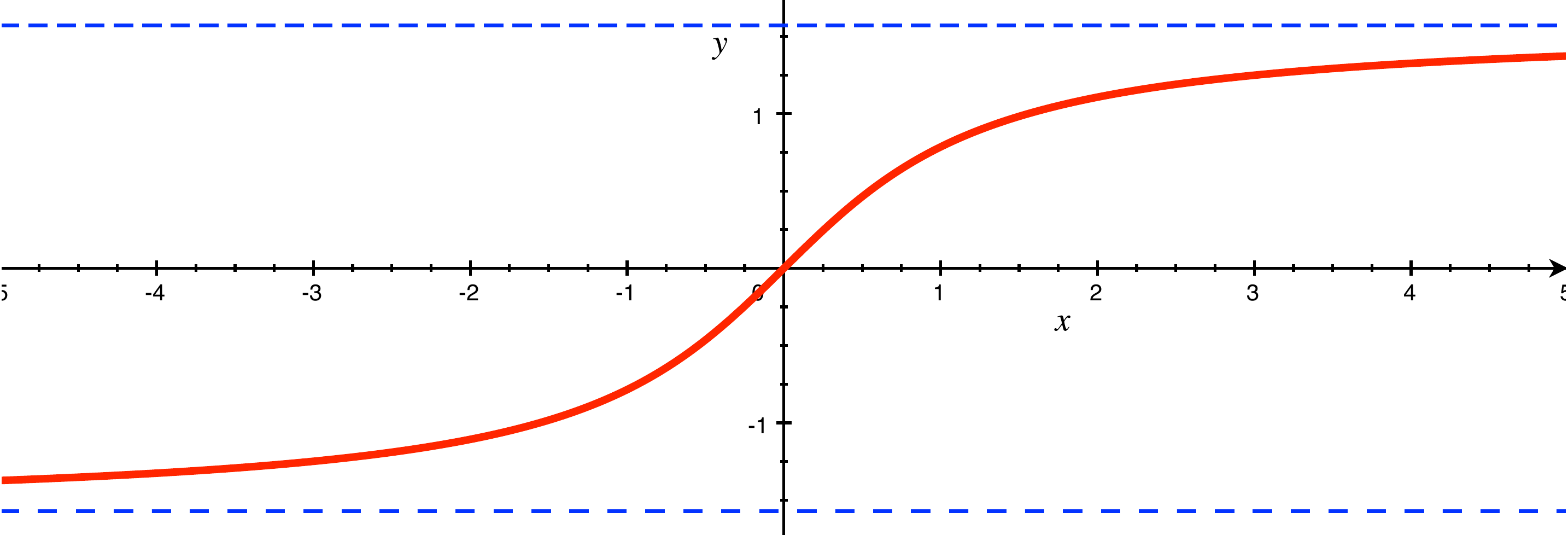}
\caption[Inverse Tangent]{The function $y = \tan^{-1} x$ maps $\left( - \infty, \infty \right)$ on to $\left( - \pi/2, \pi /2 \right)$.}
\label{fig: tanInv}
\end{center}
\end{figure}  
Introducing null coordinates in the $t$-$r$ section gives,
\begin{equation}
u \equiv t - r, \quad  v \equiv t + r, \quad \implies \quad t =\tfrac{1}{2} (v + u), \quad r = \tfrac{1}{2} (v - u).
\end{equation}
Substituting this transformation into Eq.~\eqref{eqn: minkDs} gives a new line element,
\begin{equation}
ds^2 = -du dv + \tfrac{1}{4} \left( u - v\right)^2 \left( d \theta^2 + \sin^2(\theta) d \phi^2 \right).
\end{equation}
By considering the space time of constant $\theta$ and $\phi$ it is simple to illustrate the $(u,v )$ axes with respect to the $(x, y)$ axes, as is done in Fig.~\ref{fig: uvAxes}.  

Radial light rays can be described as outgoing or ingoing, recalling that we have set $c = 1$ which implies that light rays are depicted by lines of $45$ degrees to the $x$ or $r$ axis, we note that such rays are described by
\begin{equation}
t = r + c_1, \quad \quad \text{or} \quad \quad t = -r + c_1,
\end{equation}
for any constant $c_1$.  The slope then tells us if we are dealing with outgoing radial light rays (slope $= 1$) or ingoing light rays (slope $= -1$).  Transforming these lines to our $(u,v)$ axes shows that outgoing light rays are described by $u = c_1$, while ingoing light rays are described by $v = c_1$, as is also depicted in Fig.~\ref{fig: uvAxes}.  Another way of finding the angle of radial light rays is to solve $ds^2 = 0$, which integrates to give us $u = c_1$ or $v=c_1$.

Another aspect, to consider, of the new coordinate system is its viable range and domain.  In \fixme{$(t,r)$} coordinates, we have $t \in \left(- \infty, \infty \right)$ and $r \in \left( 0, \infty \right)$, depicted as the shaded region in Fig.~\ref{fig: uvAxes}.  The equivalent of this in $(u,v)$ coordinates is the condition $v > u$ as is easily seen from Fig.~\ref{fig: uvAxes}.

\begin{figure}
\begin{center}
\includegraphics[scale=0.6]{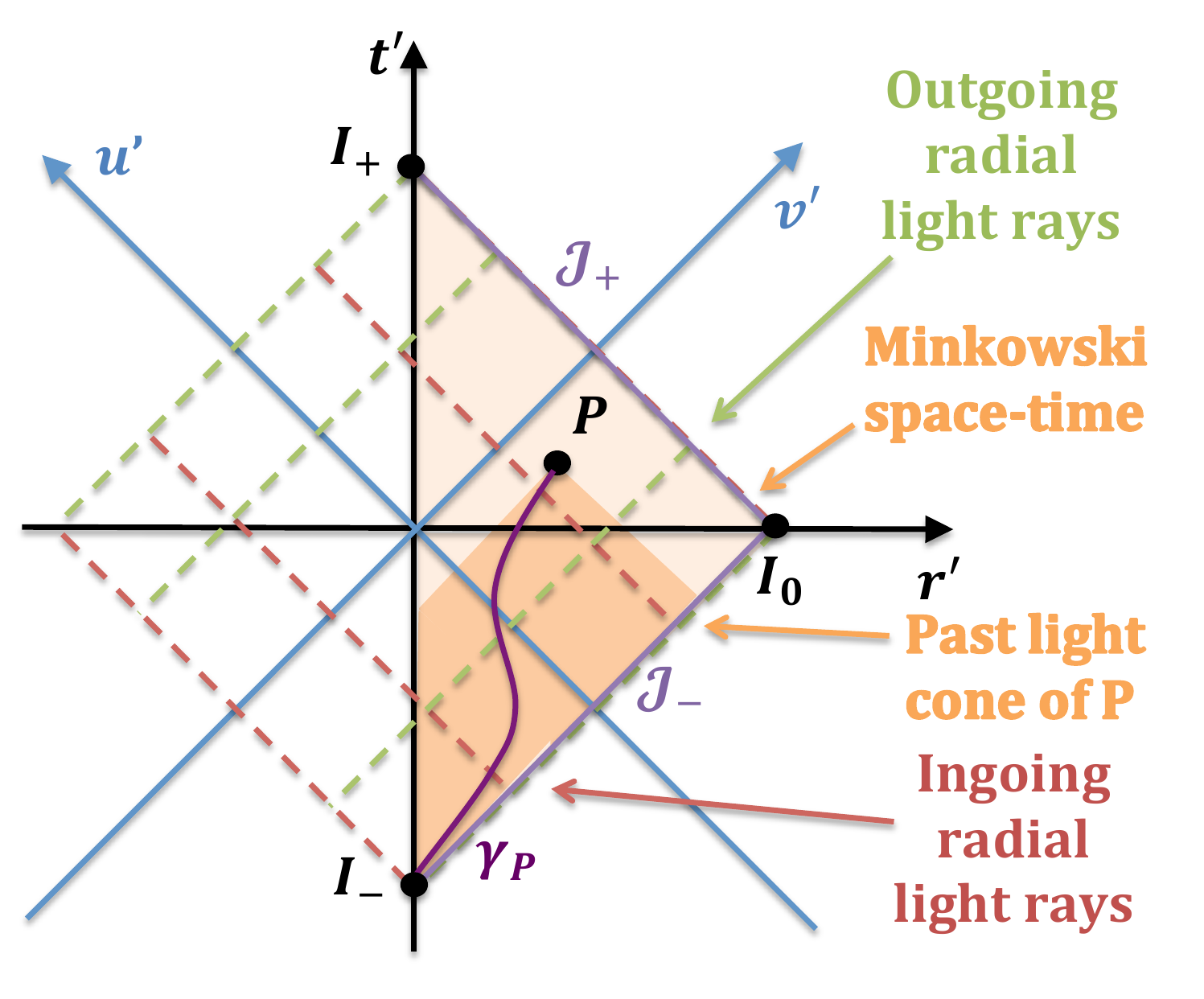}
\caption[Penrose Diagram for Minkowski Space-Time]{The Penrose diagram for Minkowski space-time is formed by $u' =\tan^{-1} u$ and $v' = \tan^{-1} v$, for Minkowski null coordinates $u$ and $v$.  Outgoing and ingoing radial light rays are described by $u'=c_1$ and $v' = c_1'$ respectively, for some constant, $c_1'$.  Minkowski space-time is limited to $t \in \left( - \infty, \infty \right)$, $r \in \left( 0, \infty \right)$, this is equivalent to $v' > u'$ and represented by the shaded region.  The darker region represents the past light cone for a particle at P, while $\gamma_P$ is its world line.  The future and past null infinities are denoted $\mathcal{J}_+$ and $\mathcal{J}_-$ respectively.  $\emph{I}_-$, $\emph{I}_+$ and $\emph{I}_0$ are the past timelike infinity, future timelike infinity and spacelike infinity respectively.}
\label{fig: uvpAxes}
\end{center}
\end{figure}

To illustrate the Penrose diagram for Minkowski space-time, we introduce another transformation,
\begin{equation}
u' \equiv \tan^{-1} u \equiv \frac{1}{2} \left( t' - r' \right), \qquad v' \equiv \tan^{-1} v \equiv \frac{1}{2} \left( t' + r' \right). 
\end{equation}
An immediate consequence of this transformation is that our coordinates now have a finite range \fixme{due to} the finite range \fixme{of} the function $\tan^{-1} x$ (illustrated in Fig.~\ref{fig: tanInv}) - all values for $u'$ and $v'$ must lie in the range $\left( - \pi / 2, \pi / 2 \right)$.  In fact, we can limit this further by recalling $v > u$ for Minkowski space-time, from Fig.~\ref{fig: tanInv}, one can clearly see that the immediate implication of this is $v' > u' $, this is illustrated as the shaded area in Fig.~\ref{fig: uvpAxes}, which is also the Penrose diagram for flat space-time.

In our null Minkowski coordinates, outgoing light rays were described by $u = c_1$ which transforms to $u' = c_1'$ in the Penrose diagram, where $c_1'$ is also a constant, similarly ingoing light rays are described by $v = c_1 \Rightarrow v' =c_1'$.  This implies that light rays are still described as lines parallel to the $(v', u')$ axes or as lines of $45$ degrees to the $(x',y')$ axes.  This preservation of angles defines the Pemrose diagram as a conformal transformation.  The direction of the light rays also gives an immediate meaning to the boundaries.  All outgoing light waves $u'=c_1'$ will end up on the boundary $v' = \pi / 2$, which can now be described as the \emph{future null infinity} and is denoted by $\mathcal{J}_+$.  Similarly all ingoing light waves, $v' = c_1'$ will originate from the boundary $u' = - \pi / 2$, known as the \emph{past null infinity} and denoted $\mathcal{J}_-$.

We can see from Fig.~\ref{fig: tanInv} that as $x \rightarrow \infty$, $\tan^{-1} x \rightarrow \pi/2$, similarly as $x \rightarrow -\infty$, $\tan^{-1} x \rightarrow -\pi/2$.  From this we can infer that as $(v,u) \rightarrow (\infty, \infty)$, $(v', u') \rightarrow (\pi/2, \pi/2)$ and as  $(u,v) \rightarrow (-\infty, -\infty)$, $(u', v') \rightarrow (-\pi/2, -\pi/2)$.  If we consider the world line of a particle, $P$, in Fig.~\ref{fig: uvpAxes}, denoted by $\gamma_P$, and follow the particle's world line into its past light-cone, it will, therefore, tend to the point $(-\pi/2, -\pi/2)$ on $(v',u')$.  This means that all (time-like) world lines originate at this point, which is known as the \emph{past time-like infinity}, $I_-$.   Similarly if we follow the particles world line into its future light cone, it will end up at $(\pi/2, \pi/2)$.  We can, therefore, conclude that all (time-like) world lines will end up at this point, known as the \emph{future time-like infinity}, $I_+$.  If we consider space-like curves, we can see that their trajectory will be forced to the point, $(\pi / 2, -\pi/2)$ in $(v',u')$, which is known as \emph{spacelike infinity}, $I_0$.

As Penrose diagrams describe a infinite space-time in a finite space, yet maintain the quality that light cones are $45$ degrees with the axes, they are very useful in comprehending from which events an observer can receive information.  This becomes extremely useful, in particular, for black-hole space-times, although these space-times are more complicated than the flat Minkowski space-time we considered here.  We will take a close look at black-hole space-times and their Penrose diagrams in Sec.~\ref{sec: BHST}.


\subsection{Synge's World Function}

\begin{figure}
\begin{center}
\includegraphics[scale=0.8]{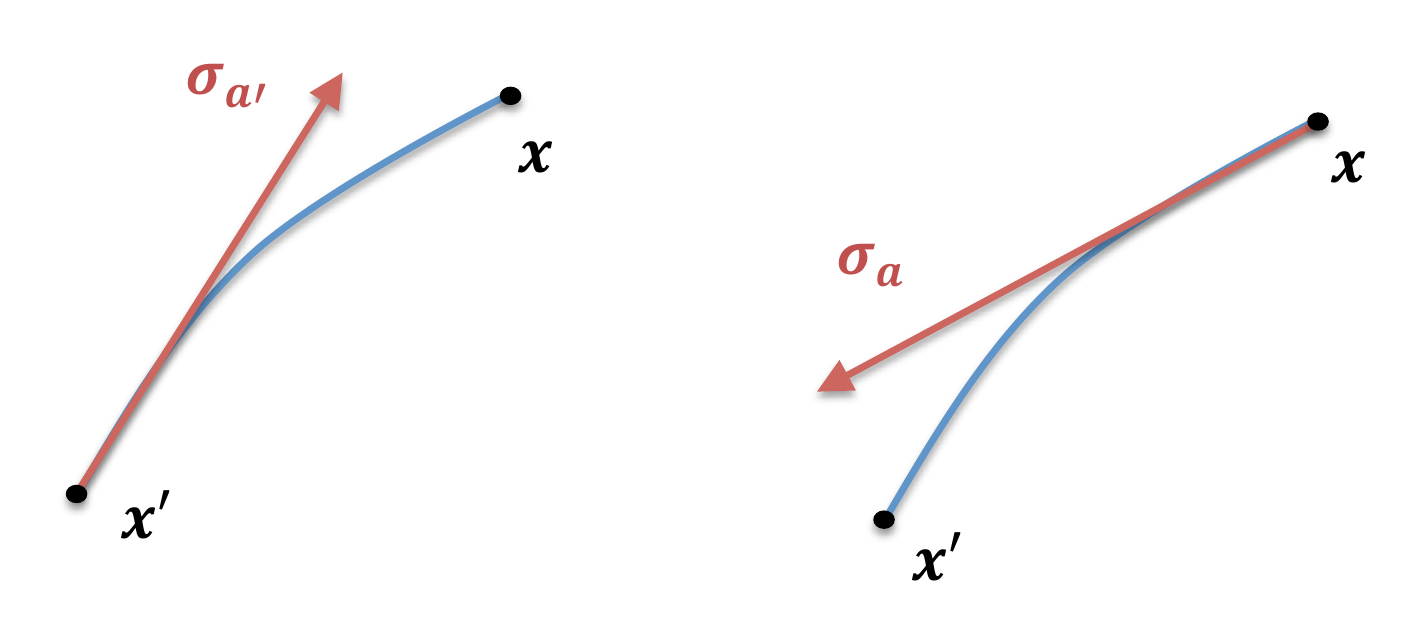}
\caption[Derivaties of Synge's World Function]{The derivatives of sigma at $x$ and $x'$, $\sigma_a$ and $\sigma_{a'}$ respectively.}
\label{fig: sigma}
\end{center}
\end{figure}

Synge's world function, $\sigma(x, x')$, is a biscalar defined as one half of the squared geodesic distance between $x$ and $x'$ \cite{Synge}.  As a biscalar, it holds the ability of a dual definition geometrically.  If one was to calculate the derivative of $\sigma(x, x')$, they could do so at either $x$ or $x'$ with the resulting vector being very different depending on where the derivative is taken.  This is clearly illustrated in Fig.~\ref{fig: sigma}.  Once differentiated, $\sigma_a$ is a vector with respect to $x$ but still a scalar with respect to $x'$.  Similarly, $\sigma_{a'}$ is a vector with respect to $x'$ and a scalar with respect to $x$.  This property leads to the ability, on taking further derivatives, of \emph{switching} the order of primed and unprimed indices with respect to each other with no change to the bitensor, i.e., $T_{ab'cd'efg'h'} = T_{acefb'd'g'h'}$ (note that the indices must stay in order with respect to indices of the same variety - except in the case of the first two due to the scalar nature before derivatives are taken).

Mathematically, Synge's world function is represented by,
\begin{equation}
\sigma \left(x, x'\right) = \frac{1}{2} \left( \lambda_1 - \lambda_0 \right) \int_{\lambda_0}^{\lambda_1} g_{ab} (z) \dot{z}^a \dot{z}^b d \lambda,
\end{equation}
where $\lambda$ affinely parameterises the geodesic connecting $x$ and $x'$, $z(\lambda_0) = x'$ and $z(\lambda_1) = x$.  The geodesic equation gives $\delta_1 = g_{ab} \dot{z}^a \dot{z}^b $, Eq.~\eqref{eqn: uu} where
\begin{equation}
\delta_1 = \begin{cases}
		 	-1 & \text{$x$ and $x'$ are timelike related} \\
			0 & \text{$x$ and $x'$ are null or lightlike related} \\
			1 & \text{$x$ and $x'$ are spacelike related}
		\end{cases}.
\end{equation}
When $x$ and $x'$ are timelike related, $\lambda$ can be taken to be proper time $\tau$, giving us
\begin{equation} \label{eqn: sigmaTime}
\sigma = \frac{1}{2} \delta_1 \Delta \lambda^2 = - \frac{1}{2} \Delta \tau^2
\end{equation}

To obtain an expression for $\sigma_a$, we define $ \delta \sigma= \sigma \left( x + \delta x, x' \right) - \sigma \left( x, x'\right)$.  In terms of $z$, the geodesic connecting $x + \delta x$ to $x'$ is written as $z (\lambda) + \delta z (\lambda)$, where $z(\lambda_0) + \delta z (\lambda_0)= x' \Rightarrow \delta z (\lambda_0) = \delta x' = 0$ and $z(\lambda_1) + \delta z (\lambda_1)= x \Rightarrow \delta z (\lambda_1) = \delta x$.  $\delta \sigma$ is now given by
\par \vspace{-6pt} \begin{IEEEeqnarray}{rCl}
\delta \sigma &=& \frac{1}{2} \Delta \lambda \left[ \int_{\lambda_0}^{\lambda_1} g_{ab} (z + 
	\delta z) \left( \dot{z}^a + \delta \dot{z}^a \right) \left( \dot{z}^b + \delta \dot{z}^b 
	\right) d \lambda - \int_{ \lambda_0 }^{ \lambda_1 } g_{ab} (z) \dot{z}^a \dot{z}^b d 
	\lambda \right] \nonumber \\
&=&
	\Delta \lambda \int_{\lambda_0}^{\lambda_1} \left[g_{ab} (z) \dot{z}^a \delta \dot{z}^b + 
	\frac{1}{2} g_{ab,c}(z)  \dot{z}^a \dot{z}^b \delta z^c  + O\left(\delta z^2\right) \right] d 
	\lambda \nonumber \\
&=&
	\Delta \lambda \left[ g_{ab} \dot{z}^a \delta z^b \right]^{\lambda_1}_{\lambda_0} - 
	\Delta \lambda \int_{\lambda_0}^{\lambda_1} \left( g_{ab} \ddot{z}^b  + \Gamma_{abc} 
	\dot{z}^b \dot{z}^c \right) \delta z^a  + O\left(\delta z^2\right) \delta \lambda 
	\nonumber \\
&=&
	\Delta \lambda g_{ab} \dot{z}^a \delta x^b,
\end{IEEEeqnarray}
where the second equality makes use of the Taylor expansion $g_{ab}(z+\delta z) = g_{ab} (z) + g_{ab,c} (z) \delta z^c + O(\delta z^2)$ and the third equality involves integration by parts on the first term and a reshuffling of indices.  The last equality makes use of the geodesic equation, Eq.~\eqref{eqn: geodesicEqn}, to make the second term disappear and recalls $\delta z (\lambda_0) = 0$ while $\delta z (\lambda_1) = dx$ and terms of order $\delta z^2$ and higher have been neglected.  This gives
\begin{equation} \label{eqn: deltaSigma}
\frac{\partial \sigma}{\partial x^a} \equiv \sigma_a \left(x, x'\right) = \Delta \lambda g_{ab} \dot{z}^b \quad \text{and} \quad 
\sigma^a \left(x, x'\right) = \Delta \lambda \dot{z}^a,
\end{equation}
where the second identity follows by simply multiplying the first by the inverse metric, $g^{ac}$. By considering $ \delta \sigma= \sigma \left( x, x' + \delta x' \right) - \sigma \left( x, x'\right)$  and the geodesic connecting $x'+\delta x'$ to $x$ in the above calculation, so that $\delta z (\lambda_0) = \delta x'$ and $\delta z (\lambda_1) = \delta x = 0$, it can also be shown that $\delta \sigma = -\Delta \lambda g_{a'b'} \dot{z}^{a'} \delta x^{b'}$ and hence,
\begin{equation} 
\frac{\partial \sigma}{\partial x^{a'}} \equiv \sigma_a' \left(x, x'\right) = -\Delta \lambda g_{a'b'} \dot{z}^{b'} \quad \text{and} \quad 
\sigma^{a'} \left(x, x'\right) = - \Delta \lambda \dot{z}^{a'}.
\end{equation}
Multiplying Eqs.~\eqref{eqn: deltaSigma} together gives
\begin{equation} \label{eqn: 2sigma}
g^{ab} \sigma_a \sigma_b = \Delta \lambda^2 \dot{z}^a \dot{z}_a = \Delta \lambda^2 \delta_1  = 2 \sigma,
\end{equation}
where we have used Eq.~\eqref{eqn: sigmaTime} in the final equality.

Taking the limit of a biscalar, bivector and bitensor as $x \rightarrow x'$ is known as the coincident limit.  Taking the coincidence limit of $\sigma$ is easy enough, as we can see directly from Eq.~\eqref{eqn: sigmaTime} that it would be zero, similarly Eq.~\eqref{eqn: deltaSigma} in the coincident limit also gives zero.  This is written as,
\begin{equation} \label{eqn: sigmaCo}
\left[ \sigma \right] = 0, \quad \quad \left[ \sigma_a \right] = \left[ \sigma_{a'} \right]  = 0.
\end{equation}
Taking a derivative of Eq.~\eqref{eqn: 2sigma} and using Eq.~\eqref{eqn: deltaSigma} gives,
\begin{equation} \label{eqn: sigmac}
\sigma_c = \sigma^b \sigma_{bc}  \quad \Rightarrow \quad \left( g_{cb} - \sigma_{bc} \right) \dot{z}^b = 0.
\end{equation}
As taking the coincident limit is independent of the geodesic, taking the coincident limit of $\sigma_{ab}$ will be independent of $\dot{z}^b$, which leaves $\left[ \sigma_{ab }\right] = g_{a'b'}$ to be a direct consequence of Eq.~\eqref{eqn: sigmac}.  Similarly it can be found that,
\begin{equation} \label{eqn: sigmaCoincidence}
\left[ \sigma_{ab }\right] = \left[ \sigma_{a'b' }\right]= g_{a'b'} \quad \text{and} \quad \left[ \sigma_{ab' }\right] = \left[ \sigma_{a'b}\right] = - g_{a'b'}
\end{equation}

Differentiating Eq.~\eqref{eqn: sigmac} twice more gives
\begin{equation} \label{eqn: sigmaAbc}
\sigma_{abc} = \sigma^d{}_{abc} \sigma_d + \sigma^d{}_{ab} \sigma_{dc} + \sigma^d{}_{ac} \sigma_{db} + \sigma^d{}_a \sigma_{dbc}.
\end{equation}
If we take the coincidence limit and use Eqs.~\eqref{eqn: sigmaCoincidence} and \eqref{eqn: sigmaCo}, we obtain
\begin{equation}
\left[ \sigma_{abc }\right]  + \left[ \sigma_{acb }\right] = 0
\end{equation}
which can be rearranged to give
\begin{equation}
\left[ \sigma_{abc }\right] = \frac{1}{2} \left[ R^d{}_{abc} \sigma_d \right] = 0.
\end{equation}
Here, we have used Ricci's identity $R^d{}_{abc} \sigma_d = \sigma_{abc} - \sigma_{acb}$, where $R^d{}_{abc}$ is the Riemann curvature tensor, a measure of the space-time curvature, defined by,
\begin{equation}
R^a{}_{bcd} = \frac{\partial \Gamma^a{}_{bd}}{\partial x^c} - \frac{\partial \Gamma^r{}_{bc}}{\partial x^d} + \Gamma^a{}_{ce} \Gamma^e_{bd} - \Gamma^a{}_{de} \Gamma^e{}_{bc}.
\end{equation}
Using Synge's rule \cite{Synge},
\begin{equation} \label{eqn: SyngeRule}
\left[\sigma_{\dots a'} \right] = \left[\sigma_{\dots} \right]_{;a'} - \left[\sigma_{\dots a} \right],
\end{equation}
it is now straight forward to also calculate
\begin{equation} \label{eqn: sigmaCoincidence3}
\left[ \sigma_{abc }\right] = \left[ \sigma_{abc' }\right] = \left[ \sigma_{ab'c' }\right] = \left[ \sigma_{a'b'c' }\right] = 0.
\end{equation}

Differentiating Eq.~\eqref{eqn: sigmaAbc} again, gives,
\par \vspace{-6pt} \begin{IEEEeqnarray}{rCl}
\sigma_{abcd} &=& 
	\sigma^e{}_{abcd}  \sigma_{e} + \sigma^e{}_{abc} \sigma_{ed} + \sigma^e{}_{abd} 
	\sigma_{ec} + \sigma^e{}_{ab} \sigma_{ecd}  \nonumber \\
&&
	+\: \sigma^e{}_{acd} \sigma_{eb} + \sigma^e{}_{ac} \sigma_{ebd} + \sigma^e{}_{ad} 
	\sigma_{ebc} + \sigma^e{}_{a} \sigma_{ebcd}.
\end{IEEEeqnarray}
If we take the coincidence limit and use Eqs.~\eqref{eqn: sigmaCoincidence}, \eqref{eqn: sigmaCo} and \eqref{eqn: sigmaCoincidence3}, we arrive at
\begin{equation} \label{eqn: sigmaRelation}
\left[ \sigma_{abcd }\right] + \left[ \sigma_{acbd }\right] + \left[ \sigma_{adbc }\right]= 0.
\end{equation}
Differentiating and taking the coincidence limit of the Ricci identity already used, $R^d{}_{abc} \sigma_d= \sigma_{abc} - \sigma_{acb}$, we arrive at,
\begin{gather}
\left[ \sigma_{acbd }\right] = \left[ \sigma_{abcd }\right] - R_{d'a'b'c'}, \\
\left[ \sigma_{adbc }\right] = \left[ \sigma_{abdc }\right]  - R_{c'a'b'd'}, \\
\left[ \sigma_{abdc }\right] = \left[ \sigma_{abcd }\right] .
\end{gather}
Substituting these into Eq.~\eqref{eqn: sigmaRelation} gives
\begin{equation} 
\left[ \sigma_{abcd }\right] = - \frac{1}{3} \left( R_{a'c'b'd'} + R_{a'd'b'c'}\right),
\end{equation}
which shows how the Riemann curvature tensor naturally comes about with taking the coincidence limit of higher derivatives of $\sigma$.  This will be useful when taking covariant Taylor series of $\sigma$ to obtain our high order covariant expansions in Sec.~\ref{sec: highOrder}.  Using Synge's rule, Eq.~\eqref{eqn: SyngeRule}, it is straight forward to also obtain,
\par \vspace{-6pt} \begin{IEEEeqnarray}{rCl}
\left[ \sigma_{abcd' }\right] &=& \frac{1}{3} \left( R_{a'c'b'd'} + R_{a'd'b'c'}\right ),\nonumber \\
\left[\sigma_{abc'd' }\right] &=& - \frac{1}{3} \left( R_{a'c'b'd'} + R_{a'd'b'c'}\right), \nonumber \\
\left[\sigma_{ab'c'd' }\right] &=& - \frac{1}{3} \left( R_{a'b'c'd'} + R_{a'c'b'd'}\right). \nonumber \\
\left[ \sigma_{a'b'c'd' }\right] &=& - \frac{1}{3} \left( R_{a'd'b'c'} + R_{a'c'b'd'}\right).
\end{IEEEeqnarray}


\subsection{Bivector of Parallel Transport}

The bivector of parallel transport, $g^a{}_{b'}$, by definition takes a tensor at $x$ and parallel transports it along the geodesic to $x'$, it can also do the opposite, parallel transport from $x'$ to $x$.  This is written as,
\begin{equation}
v^a = g^a{}_{b'} v^{b'}.
\end{equation}
Clearly as $x \rightarrow x'$, $g^a{}_{b'} \rightarrow \delta^{a'}_{b'}$, i.e.,
\begin{equation} \label{eqn: bptCoincidence}
\left[g^a{}_{b'} \right] = \delta^{a'}_{b'}.
\end{equation}
As $g^a{}_{b'}$ parallel transports along the geodesic, we immediately have,
\begin{equation} \label{eqn: gSigma}
g^a{}_{a';b} \sigma^b = g^a{}_{a';b'} \sigma^{b'} = 0
\end{equation}
If we differentiate this, we get
\begin{equation}
g^a{}_{b';cd} \sigma^c + g^a{}_{b';c} \sigma^c{}_d = 0,
\end{equation}
which at coincidence gives
\begin{equation}
\left[g^a{}_{b';c} \right] = \left[g^a{}_{b';c'} \right] = 0,
\end{equation}
where we have once again made use of Synge's rule, Eq.~\eqref{eqn: SyngeRule}.


\subsection{The Van Vleck Determinant} \label{sec: VV}

The Van Vleck determinant is biscalar defined by,
\par \vspace{-6pt} \begin{IEEEeqnarray}{rCl}  \label{eqn: vanvleck}
\Delta \left( x, x' \right) &=& \det\left[ \Delta^{a}{}_{b'} \left( x, x'\right)\right] = - \frac{\det{-\sigma_{ab'} \left(x, x' \right)}}{\sqrt{-g} \sqrt{-g'}},\nonumber \\ 
\Delta^{a'}{}_{b'} &=& -g^{a'}{}_a \left( x, x' \right) \sigma^a{}_{b'} \left(x, x' \right)
\end{IEEEeqnarray}
where the second identity, when contracted with the bivector of parallel transport, $g_{ca'}$, can be rearranged to give
\begin{equation} \label{eqn: sigmaabp}
\sigma_{ab'} = -g_{aa'} \Delta^{a'}{}_{b'}.
\end{equation}
Taking the coincidence limit of Eq.~\eqref{eqn: vanvleck} and using Eqs.~\eqref{eqn: bptCoincidence} and \eqref{eqn: sigmaCoincidence} immediately gives us $\left[\Delta^{a'}{}_{b'}\right] = \delta^{a'}{}_{b'}$ and $\left[ \Delta \right] = 1$.  Recalling Eq.~\eqref{eqn: 2sigma}, and differentiating it twice, gives
\par \vspace{-6pt} \begin{IEEEeqnarray}{rCl} \label{eqn: sigmaab}
\sigma_{ab'} &=& \sigma^c{}_a \sigma_{cb'} + \sigma_{cab'} \sigma^c \nonumber \\
&=&
	 \sigma^c{}_a \sigma_{cb'} + \sigma_{ab'c} \sigma^c,
\end{IEEEeqnarray}
where the last equality follows as sigma is a biscalar.  Combining Eqs.~\eqref{eqn: sigmaab} and \eqref{eqn: vanvleck} gives,
\par \vspace{-6pt} \begin{IEEEeqnarray}{rCl} \label{eqn: vvab}
\Delta^{a'}{}_{b'} &=& -g^{a'}{}_a \left(  \sigma^{ca} \sigma_{cb'} + \sigma^{a}{}_{b'c} \sigma^c\right) \nonumber \\
&=&
	g^{a'}{}_a g^{c}{}_{c'} \sigma^a{}_c \Delta^{c'}{}_{b'} + \Delta^{a'}{}_{b';c} \sigma^{c},
\end{IEEEeqnarray}
where the second equality follows using Eq.~\eqref{eqn: sigmaabp}.  Defining the inverse Van Vleck as $\Delta^{a'}{}_{b'} \left(\Delta^{-1} \right)^{b'}{}_{c'} = \delta^{a'}{}_{c'}$ and multiplying Eq.~\eqref{eqn: vvab} by $\left(\Delta^{-1} \right)^{b'}{}_{d'} $ gives
\begin{equation} \label{eqn: vvab2}
\delta^{a'}{}_{d'} = g^{a'}{}_a g^{c}{}_{d'} \sigma^a{}_c + \left(\Delta^{-1} \right)^{b'}{}_{d'} \Delta^{a'}{}_{b';c} \sigma^{c}.
\end{equation}
By taking the trace of Eq.~\eqref{eqn: vvab2}, i.e., setting $d' \rightarrow a' $ gives,
\begin{equation} 
4 = \sigma^a{}_{a} + \left(\Delta^{-1} \right)^{b'}{}_{a'} \Delta^{a'}{}_{b';c} \sigma^{c}
\end{equation}
which can be written as
\begin{equation}\label{eqn: 4sigma}
\left(\ln{\Delta} \right)_{,a} \sigma^a = 4 -  \sigma^a{}_{a} .
\end{equation}
From  Eq.~\eqref{eqn: 4sigma}, we can infer that $\Delta$ increases or decreases along each geodesic from $x'$ according to whether the rate of divergence of the neighbouring geodesics from $x'$ (measured by $\sigma^a{}_{a}$) is greater or lower than four.  It therefore defines a transport equation for $\Delta$.  If this divergence is largely negative, we can see that $\Delta$ blows up.


\section{Black Hole Space-Times} \label{sec: BHST}

When Einstein first published the full field equations of general relativity, often written as
\begin{equation}
G^{ab} = 8 \pi T^{ab}
\end{equation}
where $G^{ab}$ and $T^{ab}$ are the Einstein and energy-momentum tensors respectively, they were so complex that he fathomed that it would be a very long time before anybody would be able to produce an exact solution, if at all.  Therefore, you can imagine his astonishment when within months, \Sch produced such a solution to the system of equations \cite{Schwarzschild:1916}, which would become known as the \Sch solution.  Einstein had not known that an exact solution was possible as he,  himself, was only able to produce an approximation in the weak field regime (that would later become known as post-Newtonian theory) to extract values for potential observables.  What was more astonishing about this exact solution was that it contained a coordinate singularity.  This singularity can be interpreted as a region of space from which nothing can escape, a region caused by an extremely compact object that would later become known as a black hole. 

Since the success of \Sch solution, thousands more solutions have been found, however, very few are of actual physical relevance.  The concept of a black hole has grown and is now considered a reality by most in the astrophysical world.  When considering EMRIs, one can use the model of a test mass (representing the stellar mass black hole), orbiting a supermassive black hole which is responsible for the background space-time.  From astrophysical considerations, it is believed that this space-time will be either a static, spherically symmetric black hole space-time or the more interesting and more likely candidate of an axially symmetric, spinning black hole space-time.  In this section we will look at several space-times that fall into these categories.


\subsection{Spherically Symmetric Space-times}

It can be shown that spherically symmetric vacuum solutions to the Einstein's equations are also static - this is known as Birkhoff's theorem \cite{Birkhoff:1923}.  When considering space-times, stationary means that the geodesic is time independent, i.e.,
\begin{equation}
\frac{\partial g_{ab}}{\partial x^0} \equiv g_{ab,0} = 0
\end{equation}
where we are taking $x^0 = t$.  In addition, static implies that the metric must remain invariant under time reversal.  It, therefore, cannot have any $dx^{\alpha} dx^0$ terms in the line element, where $\alpha \in \{ 1, 2, 3 \}$, which in turn implies $g_{0 \alpha} = 0$.


\subsubsection{\Sch Space-Time} \label{sec: sch}

The \Sch space-time represents the space-time outside a static, spherically symmetric black hole of mass, $M$. Its line element is given by,
\begin{equation}
ds^2 = -\left( 1 - \frac{2 M}{r} \right) dt^2 + \left( 1 - \frac{2 M}{r} \right) ^{-1} dr^2 + r^2 d\Omega^2
\end{equation}
where 
\begin{equation}
d \Omega^2 = d\theta^2 + \sin ^2 \theta d \phi ^2
\end{equation}
and $\{t, r, \theta, \phi \}$ are the standard \Sch coordinates.  

The Lagrangian of a particle moving in \Sch space-time is given by Eq.~\eqref{eqn: lagrangian}, this can be written as
\begin{equation} \label{eqn: lagrangianSch}
\mathcal{L} = \frac{1}{2} \left[ \left( 1 - \frac{2 M}{r} \right) \dot{t}^2 - \left( 1 - \frac{2 M}{r} \right) ^{-1} \dot{r}^2 - r^2 \dot{\theta}^2 - r^2 \sin ^2 \theta \dot{\phi}^2\right], 
\end{equation}
where the ($\dot{}$) represents differentiation with respect to $\tau$.  From Hamiltonian mechanics we have that the canonical momenta, $p_a = \frac{\partial \mathcal{L}}{\partial \dot{q}_a}$ where $q_a=\{t, r, \theta, \phi\}$, are given by
\begin{gather}
p_t = \frac{\partial \mathcal{L}}{\partial \dot{t}} = \left( 1 - \frac{2 M}{r} \right) \dot{t} , \quad 
p_r = \frac{\partial \mathcal{L}}{\partial \dot{r}} = - \left( 1 - \frac{2 M}{r} \right)^{-1} \dot{r}, \nonumber \\
p_{\theta} = \frac{\partial \mathcal{L}}{\partial \dot{\theta}} = - r^2 \dot{\theta}, \quad
p_{\phi} =  \frac{\partial \mathcal{L}}{\partial \dot{\phi}} = - r^2 \sin ^2 \theta\dot{\phi}.
\end{gather}
The Hamiltonian itself is given by
\begin{equation} \label{eqn: HL}
\mathcal{H} = \sum_a p_a \dot{q}_a - \mathcal{L} = \mathcal{L}.
\end{equation}
As we are dealing with an isolated system, the Hamiltonian is constant, and therefore from Eq.~\eqref{eqn: HL}, so is the Lagrangian which we can set to  be equal to $\tfrac{1}{2}$, by rescaling the affine parameter, $\tau$, for timelike geodesics.  

Hamiltonian mechanics also dictates that,
\begin{equation}
\frac{\partial \mathcal{L}}{\partial q_a} = \dot{p}_a \quad \Rightarrow \quad \frac{\partial \mathcal{L}}{\partial t} = \frac{\partial p_t}{\partial \tau}, \quad \text{and} \quad \frac{\partial \mathcal{L}}{\partial \phi} = - \frac{\partial p_{\phi}}{\partial \tau},
\end{equation}
both of which we can set to zero due to the constancy of the Lagrangian.  These imply
\begin{gather}
p_t = \left( 1 - \frac{2 M}{r} \right) \dot{t}= \text{constant} \equiv E \quad \Rightarrow \quad \dot{t} = \frac{E}{\left( 1 - \frac{2 M}{r} \right)}, \nonumber \\
 p_{\phi} = - r^2 \sin^2 \theta \dot{\phi} = \text{constant} \equiv - L \quad \Rightarrow \quad \dot{\phi} = \frac{L}{r^2},
\end{gather}
where we have set $\theta = \pi/2$ in the last equality, and the constants $E$ and $L$ correspond to the energy per unit mass and angular momentum per unit mass respectively.  Taking the velocity to be in the plane of $\theta = \pi /2$ can be done without loss of generality due to the symmetry of the metric, this ensures $\dot{\theta} = 0$ so the motion stays in that plane.  Recalling that we've set $\mathcal{L}= 1/2$, Eq.~\eqref{eqn: lagrangianSch} now gives,
\begin{equation}
\frac{-E^2}{\left( 1 - \frac{2 M}{r} \right)} + \left( 1 - \frac{2 M}{r} \right)^{-1} \dot{r}^2 + \frac{L^2}{r^2} = -1
\end{equation}
which can be rearranged to give
\begin{equation}
\dot{r}^2 = E^2 - \frac{1}{r^3} \left(r - 2 M \right) \left(L^2 + r^2 \right).
\end{equation}
We now have expressions for the four velocity of a test particle in \Sch space-time.

The \Sch solution describes the space-time of a non-rotating black hole, and as such, is very useful for describing the approach of particles or light rays towards that black hole.  However, due to its singular nature at $r = 2 M$, it is not ideal for understanding the nature of the event horizon or the singularity at $r = 0$.  The \Sch solution can be rewritten in different coordinates that avoid the $r = 2 M$ singularity and give us a clearer picture of the geometry associated with the region $r < 2 M$.  The Kruskal coordinates are one such transformation - they were introduced earlier in Sec.~\ref{sec: penrose}, where we used their flat space-time equivalent to investigate the Minkowski space-time as well as to obtain a Penrose diagram for the space-time.

As with Minkowski space-time, for the \Sch solution, the Kruskal coordinates only change the $(r, t)$ components of the line element or metric.  The transformation for $r > 2 M$is given by,
\par \vspace{-6pt} \begin{IEEEeqnarray}{rCl} \label{eqn: UV1}
U &=& \left( \frac{r}{2 M} - 1\right)^{1/2} e^{r / (4 M)} \cosh \left( \frac{t}{4 M}\right), \\
V &=& \left( \frac{r}{2 M} - 1\right)^{1/2} e^{r / (4 M)} \sinh \left( \frac{t}{4 M}\right),
\end{IEEEeqnarray}
while the transformation for $r < 2 M$ is 
\par
\begin{figure}
\begin{center}
\includegraphics[scale=0.7]{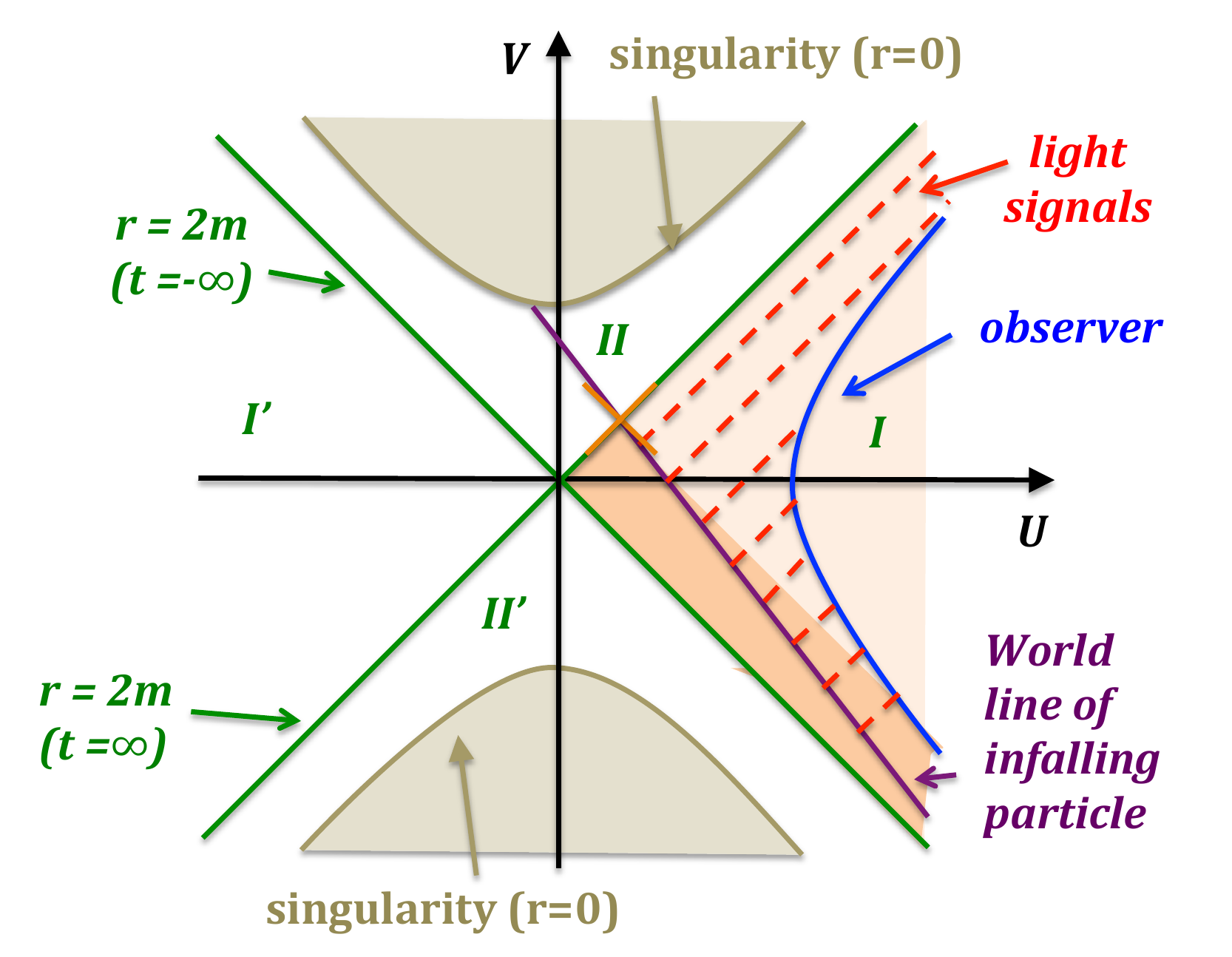}
\caption[Kruskal Diagram for Infalling Particle in \Sch Space-Time]{The Kruskal diagram showing an infalling particle for \Sch space-time. emitting radial light signals at regular intervals of proper time.  An observer receives the signals at increasing intervals until the particle crosses the horizon $r= 2M$ whereupon the observer can no longer receive any light signals.  The particle cannot return to region $I$ as this would require it travelling faster than light, i.e., travelling outside its light cone.  It will inevitably reach the singularity.}
\label{fig: UVSch2}
\end{center}
\end{figure}

 \vspace{-6pt} \begin{IEEEeqnarray}{rCl} \label{eqn: UV2}
U &=& \left(1 - \frac{r}{2 M} \right)^{1/2} e^{r / (4 M)} \sinh \left( \frac{t}{4 M}\right), \\
V &=& \left( 1 -\frac{r}{2 M} \right)^{1/2} e^{r / (4 M)} \cosh \left( \frac{t}{4 M}\right).
\end{IEEEeqnarray}
Regardless of the region, the line element transforms to,
\begin{equation} \label{eqn: lineUV}
ds^2 = \frac{32 M^3}{r} e^{-r / (2 M)} \left( -dV^2 + dU^2 \right) + r^2 \left( d \theta^2 + \sin^2 \theta d \phi^2 \right).
\end{equation}\begin{figure}
\begin{center}
\includegraphics[scale=0.71]{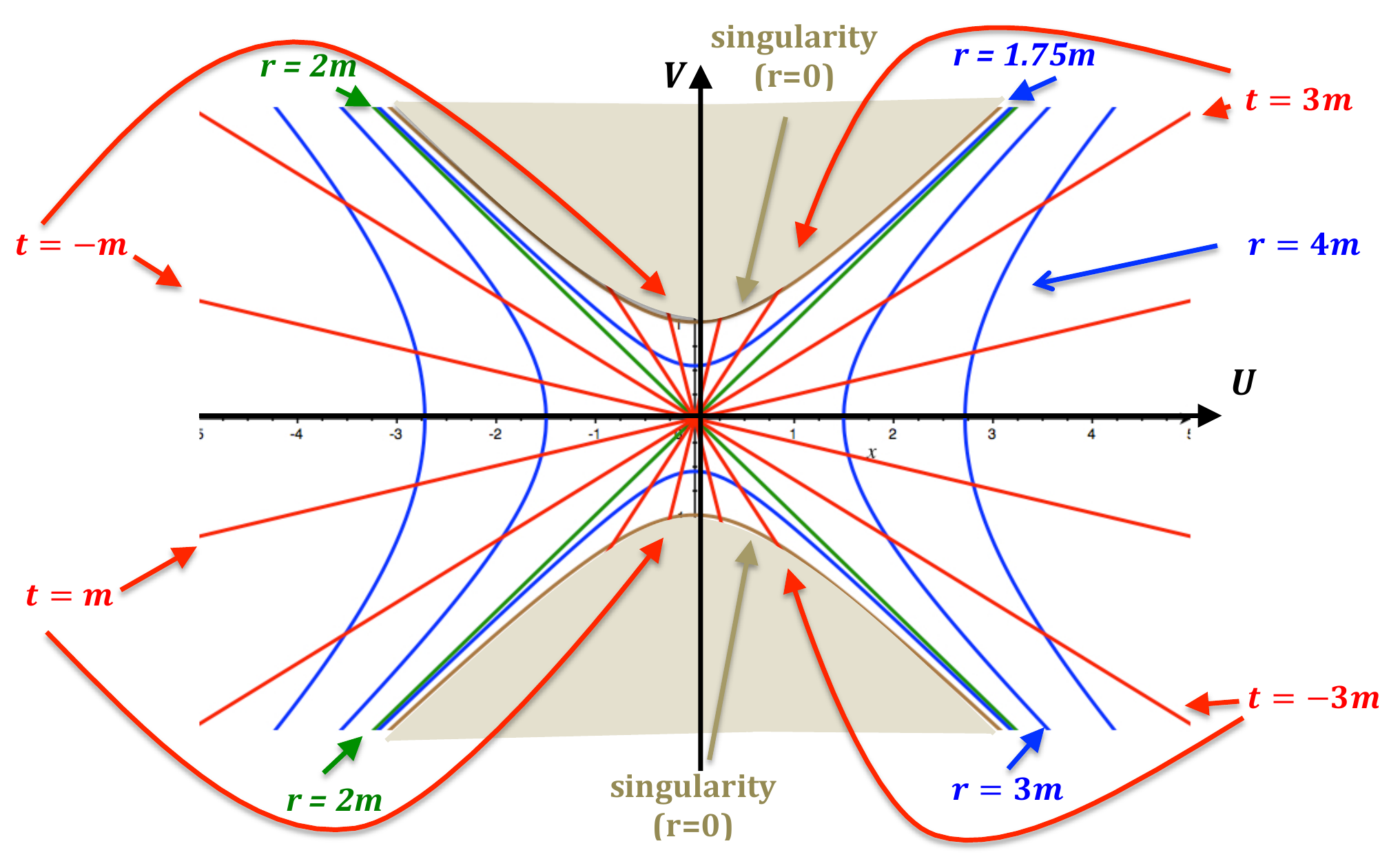}
\caption[Kruskal Diagram for \Sch Space-Time]{The Kruskal diagram for \Sch space-time is formed by the transformation in Eqs.~\eqref{eqn: UV1} and \eqref{eqn: UV2}.  Lines of constant $t$ and $r$ are highlighted in red and blue respectively.  The $r = 2 M$ horizon (green) with positive slope also represents $t=\infty$ while the horizon with negative slope also represents $t = - \infty$.  $t=0$ is the positive $U$ axis for $r>2 M$ and the positive $V$ axis for  $r < 2 M$.  The area that represents only \Sch coordinates is described by $U>V $ and $U > -V$.}
\label{fig: UVSch}
\end{center}
\end{figure}
To illustrate how $r$ is represented on a $(U,V)$ diagram, we square our $U$ and $V$ coordinates and subtract one from the other to obtain,
\begin{equation} \label{eqn: UVhyper}
U^2 - V^2 = \left( \frac{r}{2 M} - 1 \right) e^{r / (2 M)} \begin{cases} 
												> 0 & r > 2 M \\
												= 0 & r = 2 M \\
												< 0 & r < 2 M
											\end{cases}.
\end{equation}
This clearly illustrates that lines of constant $r$ are hyperbolas on the $(U,V)$ plane, in particular, $r>2M$ is represented by \emph{East-West opening hyperbola}, while $r < 2M$ will be shown as \emph{North-South opening hyperbola} as is illustrated in Fig.~\ref{fig: UVSch}.  We see by increasing $r$ in the $r > 2 M$ region that the hyperbolas move out while decreasing $r$ in the $r< 2 M$ regions has the same effect.  It can also  be seen from Eq.~\eqref{eqn: UVhyper} that $r = 2 M$ is not a singularity in Kruskal coordinates, which illustrates that $r = 2 M$ is only a coordinate singularity associated with the \Sch coordinates.  In fact, in Kruskal coordinates, it corresponds to the lines $U = \pm V$.  Eq.~\eqref{eqn: UVhyper} also tells us that $r=0$ corresponds to the hyperbola $V = + \sqrt{ U^2 + 1}$, a hyperbola in the $r < 2 M$ region.  The $(U,V)$ plane can clearly be separated into quadrants, as shown in Fig.~\ref{fig: UVSch2}, regions $I$ and $I'$ represent that area where $r > 2 M$ while areas $II$ and $II'$ are where $r < 2 M$.

To investigate the nature of $t$ with respect to the $(U,V)$ plane, we note that the above transformations give,
\begin{equation}
\frac{V}{U} = 	\begin{cases}
				\tanh \left( \frac{t}{4 M} \right) & r > 2 M \\
				\ \left[ \tanh \left( \frac{t}{4 M} \right) \right]^{-1} & r < 2 M
			\end{cases}.
\end{equation}
$t$ is, therefore, represented by a straight line in the $(U, V)$ plane, the slope of which will have the same sign as $t$ as is shown in Fig.~\ref{fig: UVSch}.  We can also see that $t=0$ corresponds to $V=0$ for $r>2M$ and $U=0$ for $r<2M$.  By taking the limit of $\tanh x$ as $x\rightarrow \infty$ and getting $1$, we note that $t = \infty$ corresponds to the line $U=V$.  Similarly taking the limit of $\tanh x$ as $x\rightarrow -\infty$ to get $-1$, gives $t = - \infty$ corresponding to the line $U=-V$.

\Sch coordinates cover the region $t \in \left( - \infty, \infty \right)$ and $r \in \left( 2 M, \infty \right) $, immediately we can see that this corresponds to the areas labelled $I$ and $I'$ in Fig.~\ref{fig: UVSch2}.  However, further investigation into the transformation given by Eq.~\eqref{eqn: UV1} for $r>2M$ shows that $U>V$, this leaves us with region $I$, in Fig.~\eqref{fig: UVSch2}, representing the \Sch coordinates.

In Sec.~\ref{sec: penrose}, we saw that we can investigate how the radial light cone is represented by looking at $ds^2 = 0$.  In our $(U,V)$ line element Eq.~\eqref{eqn: lineUV}, this corresponds to, 
\begin{equation}
dV^2 = dU^2 \qquad \implies \qquad  V = \pm U + c_1,
\end{equation}
where $c_1$ is a constant.  This means that radial light rays are depicted by straight lines at a $45$ degree angle to the $(U,V)$ axes.  If we now look at timelike particles and 
their light cones in region $I$, we can see that their light cones allow them to cross the horizon into region $II$, however once in region $II$ their future light cones never intersect with region $I$.  This means that no information about the particle can ever be received in region $I$ once the particle crosses the horizon, $r = 2 M$.  In fact once in region $II$, the particle's light cone ensures that the particle will eventually  end up at the singularity, $r=0$.  This is clearly seen in Fig.~\ref{fig: UVSch2}, where we see a particle releasing light signals to an observer, at a fixed $r$, at regular intervals of proper time.  The signals are received at increasing intervals of proper time by the observer until the particle crosses the horizon whereupon no more information or light signals can be received by the observer.

\begin{figure}
\begin{center}
\includegraphics[scale=0.64]{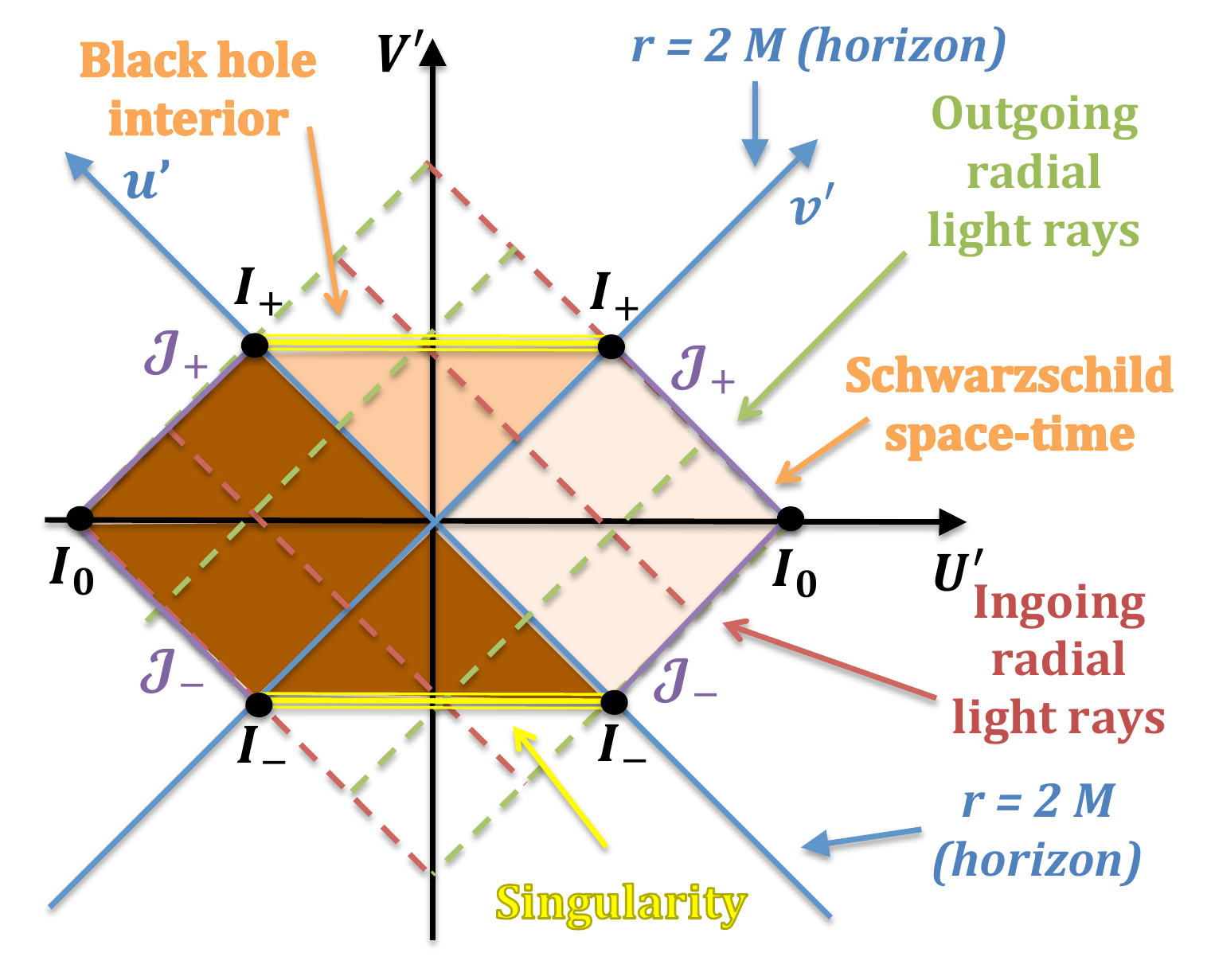}
\caption[Penrose Diagram for \Sch Space-Time]{The Penrose diagram for \Sch space-time.  The singularity at $r=0$ is represented by $V'=\pm \pi / 4$ (yellow). $I_+$ and $I_-$ are the future and past timelike infinities respectively, where all timelike world lines end up.  $\mathcal{J}_+$ and $\mathcal{J}_-$ are the future and past null infinities respectively as all light rays end and originate at these boundaries.  Spacelike infinity $I_0$, is where all space-like world lines wind up.  Once a particle crosses the horizon, the particle cannot return to the region described by \Sch coordinates as this would require it travelling faster than light, i.e., travelling outside its light cone.  It will inevitably reach the singularity.}
\label{fig: UVpSch}
\end{center}
\end{figure}
To form the Penrose diagram of \Sch space-time, we do two coordinate transformations from the Kruskal coordinates.   We introduce null coordinates by rotating our axes by $45$ degrees so light rays are parallel to our axes,
\begin{equation} \label{eqn: UVuv}
U = \frac{1}{2} \left(v - u\right), \quad V = \frac{1}{2} \left(v + u\right),
\end{equation}
and then as with the Minkowski space-time, we define our new coordinate system as
\par \vspace{-6pt} \begin{IEEEeqnarray}{rCl} \label{eqn: uvTransform}
u' \equiv \tan^{-1} u &\equiv& V' - U', \qquad v' \equiv \tan^{-1} v \equiv V' + U' , \\
 \Rightarrow U' &=& \frac{1}{2} \left( v' - u' \right), \qquad  V' = \frac{1}{2} \left(v' + u'\right).
\end{IEEEeqnarray}
As was the case for Minkowski space-time, this transformation successfully maps the infinite ranges of $u$ and $v$, $\left( - \infty, \infty \right)$, to the finite range $\left( - \pi /2, \pi /2 \right)$ for $u'$ and $v'$ and radial light rays are depicted by lines of constant $u'$ and $v'$. 

Considering $r=0$ in Eq.~\eqref{eqn: UVhyper} and using Eq.~\eqref{eqn: UVuv}, we can see that this is equivalent to setting $u=1 / v$. Using Eq.~\eqref{eqn: uvTransform} and basic trigonometry gives $V' = \pi/4$ for $v>0$, while $V'=-\pi/4$ for $v<0$ and $U' = \tfrac{1}{2} \tan^{-1} \left( v /2 - 1 / 2v \right)$ which, given the restrictions on the range of $\tan^{-1}$, implies $U' \in \left( -\pi/2, \pi / 2\right)$.  Similarly we can see that $r = 2M$, while giving us $U = \pm V$ in the Kruskal diagram, or $u=0$ and $v=0$, we consequently have $u' = 0$ and $v' = 0$ in the Penrose diagram.  This gives us enough to draw our Penrose diagram on the space where $v'$ and $u'$ both $\in \left( -\pi/2, \pi / 2\right)$, which can be seen in Fig.~\ref{fig: UVpSch}.

As in flat space we also have different types of infinity, $I_+$ and $I_-$ are the future and past timelike infinities respectively, as they are where all world lines end up.  $\mathcal{J}_+$ and $\mathcal{J}_-$ are the future and past null infinities respectively as all light rays end and originate at these boundaries.  We also have spacelike infinity $I_0$, where all space-like world lines wind up.  There are two sets of all the infinities - one for each asymptotic region.  One can clearly see from the Penrose diagram, Fig.~\ref{fig: UVpSch} that once a particle passes the horizon from the region covered by \Sch coordinates into the black hole interior, there is no returning unless it can travel faster than the speed of light, i.e., travel outside its light cone.  Similarly no light signals can leave the black hole interior so no information can ever be received from the particle to any observer remaining in the \Sch region.  The particle will eventually be forced into the singularity at $r=0$.  

It should now be obvious that diagrams such as the Kruskal and Penrose diagrams give us further insight into the physical happenings of black holes, in particular, into the space-time geometry of the black hole interior.


\subsubsection{\rn Space-Time} \label{sec: rn}

\rn space-time is a static, asymptotically
flat solution of the Einstein-Maxwell equations in
general relativity. It describes charged, non-rotating spherical
black holes or naked singularities. In general, the
\rn  space-time represents a gravitating
source which is both electrically and magnetically charged.  For the purpose of this thesis we need only concern ourselves with an electrically charged source, therefore we will take the magnetic charge to be zero.

\begin{figure}
\begin{center}
\includegraphics[scale=0.7]{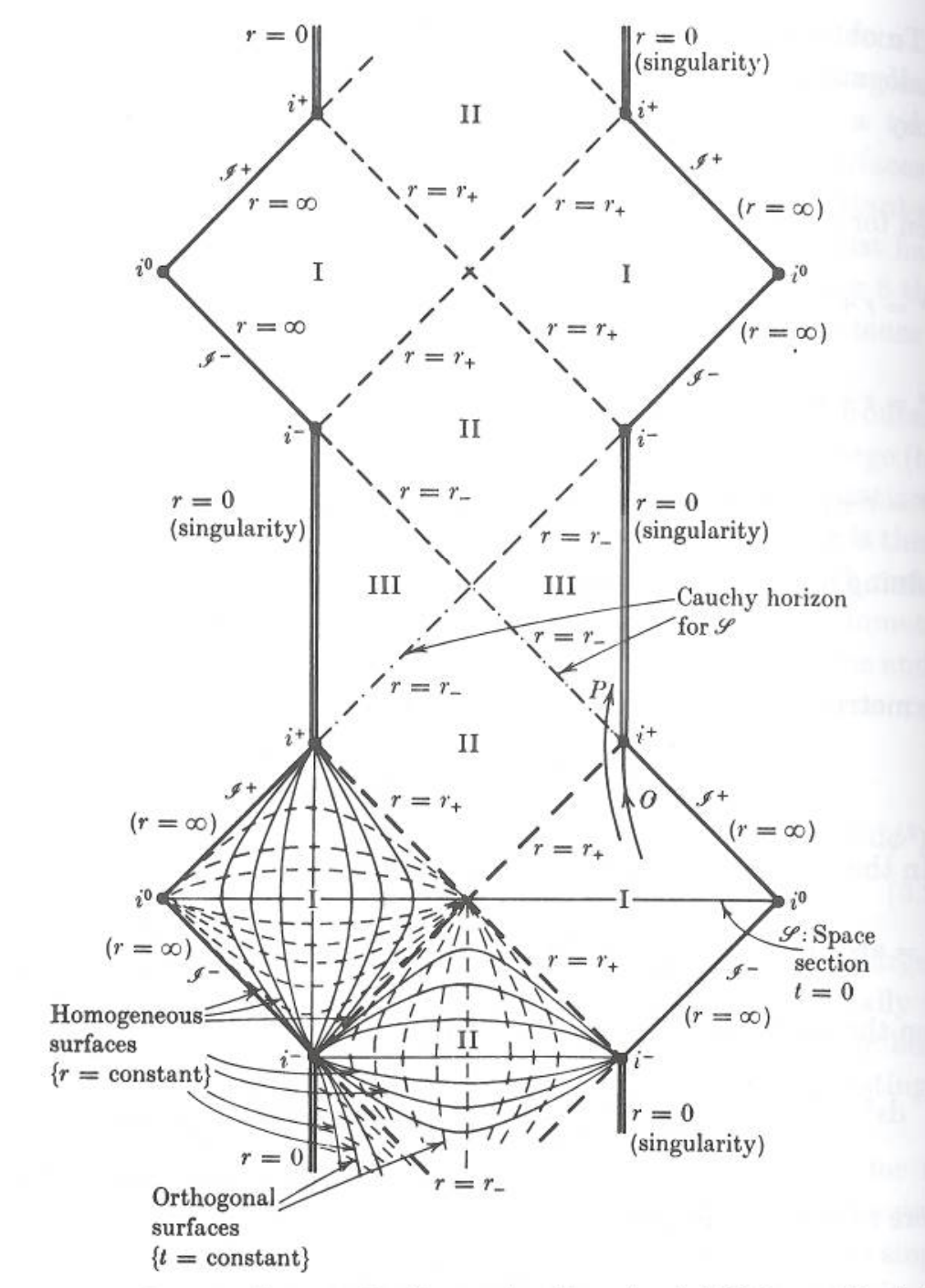}
\caption[Penrose Diagram for \rn Space-Time $Q^2 < M^2$]{The Penrose diagram for \rn space-time when $Q^2 < M^2$, has an infinite number of asymptotically flat regions where $r>r_+$, labelled $I$, connected by regions $II$ and $III$ where $r_->r>r_+$ and $0>r>r_-$ respectively.  Region $III$ contains a singularity, however it is timelike, which means it can be avoided by a timelike curve, i.e., the world line of a particle.  In fact, this singularity acts repulsive in the sense that timelike curves cannot hit them.  This Penrose diagram, therefore, hints at a very unusual scenario - it should be possible for a particle, $P$ in Fig.~\ref{fig: uvpRN} to start in region $I$, cross the horizon $r=r_+$ into region $II$, continue from this region cross another horizon $r=r_-$ into region $III$.  Here, it will avoid the singularity and can pass through another $r=r_-$ horizon in the 'next' region $II$, and again through another $r=r_+$ horizon into a 'new' region $I$.  This physically translates to a particle crossing from one universe through a 'wormhole' of sorts and arriving in a new universe.  This diagram was taken from \cite{Hawking:Ellis}.}
\label{fig: uvpRN}
\end{center}
\end{figure}

Both the \Sch and \rn space-times have singularities at their origins.  As we have seen in Sec.~\ref{sec: sch}, in the case of \Sch space-time, any particle that traverses the event horizon will inevitably be drawn into that singularity.  However, in \rn space-time, this is not to be the case.  Instead, due to the charge of the black hole, the (test) particle can and will leave the vicinity of the singularity, passing through both the event and Cauchy horizons of the \rn black hole to arrive in another universe.  In fact, the particle can continue to pass through further event and Cauchy horizons, and so, in a manner, can pass from universe to universe.  This feature of the \rn  solution is seen clearly in its Penrose diagram in Fig.~\ref{fig: uvpRN}, and will be further explained later in this section.

The \rn  solution is the unique static, spherically symmetric solution of the Einstein-Maxwell equations.
In units where the speed of light, Planck's constant, the Boltzmann constant and the Coulomb constant are set to unity, the metric is given by
\begin{equation}
ds^{2}=-\frac{\Delta}{r^{2}}dt^{2}+\frac{r^{2}}{\Delta}dr^{2}+r^{2}d\Omega_{2}^{2},
\end{equation}
where $\Delta=  r^2 - 2 M r + Q^2 \equiv (r-r_{-})(r-r_{+})$, $d\Omega_{2}^{2}$ is the metric on the two-sphere, $Q$ is the electric charge of the black hole and $M$ is the mass of the solution.  The field strength, $F_{ab} = A_{b,a} - A_{a,b}$, of the electomagnetic field is produced by the only non-zero component of the vector potential, $A_t = \frac{Q}{r}$.

There is a curvature singularity ar $r=0$ while for $0 <Q^{2} <M^2$, $\Delta(r)$ has two 
real roots given by 
\begin{equation}
r_{\pm}= M\pm \sqrt{ M^{2}-Q^{2}}.
\end{equation}
The Cauchy horizon and the event horizon are therefore defined to be at $r_{-}$  and $r_{+}$ respectively.  Between the horizons, the radial coordinate $r$ becomes
timelike and the time coordinate $t$ spacelike. When $Q^{2}=M^2$, the horizons degenerate 
and the black hole is
called extremal, while for $Q^{2} > M^2$ the solution has a
naked singularity.

A test particle with a net electric charge will not describe a geodesic in the \rn due to the
electromagnetic forces acting on it.
The motion of the particle will be determined by the Lagrangian,
\begin{align}
2 \mathcal{L} =& \frac{\Delta}{r^2} \left(\frac{dt}{d\tau}\right)^2-\frac{r^2}{\Delta} \left(\frac{dr}{d\tau}\right)^2-r^2\left(\frac{d\theta}{d\tau}\right)^2-r^2\sin^2\theta\left(\frac{d\phi}{d\tau}\right)^2 \\
&+\: 2 \frac{qQ}{r} \frac{dt}{d\tau},
\end{align}
where $q$ denotes the electric charge per unit mass of the test particle. The equations of motion which follow from this Lagrangian are
\begin{align}
  &\frac{\Delta}{r^2} \frac{dt}{d\tau} + \frac{qQ}{r} = E = \text{constant}\\
  & r^2 \sin^2\theta \frac{d\phi}{d\tau} = L = \text{constant} \label{eqn: phiRN}
\end{align}
and
\begin{align}
&\left(\frac{dr}{d\tau}\right)^2= \left(E -\frac{qQ}{r} \right)^2 - \frac{(r^2+L^2)\Delta}{r^4},\\
&r^4\left(\frac{d\theta}{d\tau}\right)^2=L^2 - \frac{1}{\sin^2 \theta} L^2. \label{eqn: thetaRN}
\end{align}
As the motion of such a particle is spherically symmetric, we can say without loss of generality that the motion takes place in a plane which we may take to be $\theta=\pi/2$.  This simplifies down Eqs.~\eqref{eqn: phiRN} and \eqref{eqn: thetaRN} to 
\begin{equation}
\frac{d\phi}{d\tau} = \frac{L}{r^2} \quad \quad \text{and} \quad \quad \frac{d\theta}{d\tau}=0.
\end{equation}

\begin{figure}
\begin{center}
\includegraphics[scale=0.67]{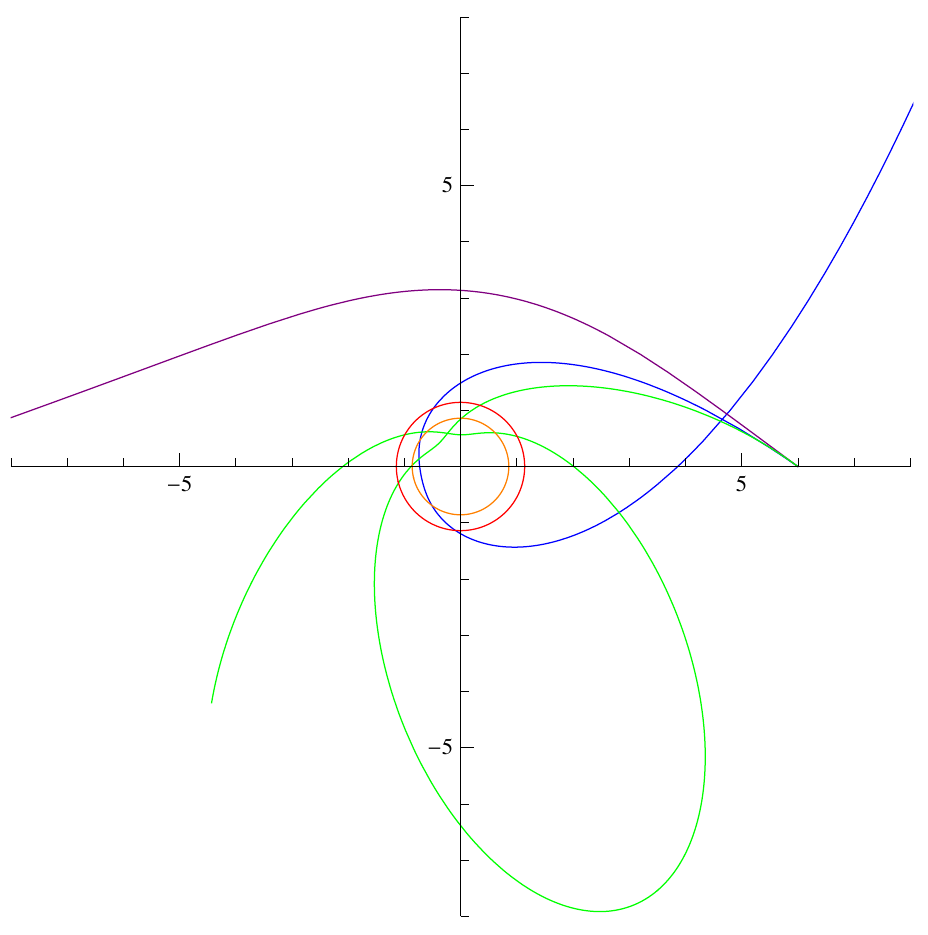}
\caption[Motion of Charged and Uncharged Particles in \rn Space-Time]{The motion of charged and uncharged particles in \rn space-time is represented by the purple, blue and green lines.  The purple line represents a charged particle carrying the same charge as the black-hole which results in it being repelled from the event horizon (red circle).  The blue line and green lines represent the geodesic motion of an uncharged particle and the motion of a charged particle of opposite charge to the black hole respectively.  These pass through both the event horizon and Cauchy horizon (orange circle) and then escape out of the black hole by traversing both horizons again.  However, the universe into which they re-emerge is not their original universe.  As is also shown in Fig.~\ref{fig: uvpRN}, this illustrates how a particle can move from universe to universe in \rn space-time.}
\label{fig: rnGeodesic}
\end{center}
\end{figure}

If we know look at how these particles behave, i.e., a charged particle in \rn space-time, we can see, with the use of mathematical software \cite{Mathematica}, that particles with the same charge as the black hole are repelled from the horizon of the black hole as illustrated in Fig.~\ref{fig: rnGeodesic} where the particle is represented by the purple line.  In a sense, the charge of the particle protects the particle from being swallowed by the black hole.  If we look at the motion of an uncharged and charged (with opposite charge to the black hole) particle in Fig.~\ref{fig: rnGeodesic}, represented by the blue and green lines respectively, we can observe something far more interesting.  The particles pass through the event horizon (red circle) and the Cauchy horizon (orange circle) and then traverses back through both horizons and essentially escape being absorbed by the black hole.  
Instead they emerge back into the universe, however the universe to which they escape is not the universe from which they came.  As described before, this is an example of particles being able to move from universe to universe in the \rn space-time.

This adventurous journey can be better explained with the use of a Penrose diagram. We, therefore, once again transform our line element into Kruskal coordinates to enable us to produce a Penrose diagram.   To do this, it is necessary to define the coordinate,
\begin{equation}
r^* = \int \frac{\Delta}{r^2} dr.
\end{equation}
For our different values of $Q^2$, this gives us different definitions after integration,
\begin{equation} \label{eqn: rstar}
r^* = 	\begin{cases}
			r + \frac{r_+^2}{r_+-r_-} \log \left(r - r_+ \right) - \frac{r_-^2}{r_+-r_-} \log 	
			\left(r - r_- \right) & Q^2 < M^2 \\
			r + M \log \left[ \left( r - M \right)^2\right] - \frac{2}{r - M} & Q^2 = M^2 \\
			r + M \log \Delta + \frac{2}{Q^2 - M^2} \tan^{-1} \left( \frac{r - M }{Q^2 - M^2 
			}\right) & Q^2 > M^2 
		\end{cases}.
\end{equation}

For now we will concentrate on the more interesting case of $Q^2 < M^2$.  We carry out the cordinate transformation,
\begin{equation}
v = t + r^*, \quad \quad u = t - r^*, 
\end{equation}
From Eq.~\eqref{eqn: rstar}, it is straight forward to calculate $dr* =dr r^2 / \Delta $, the above transformation therefore gives the line element,
\begin{equation} \label{eqn: ds2RN}
ds^2 = - \frac{\Delta} {r^2} du dv + r^2 \left(d \theta^2 + \sin^2 \theta d \phi^2 \right),
\end{equation}
which the analogue of the Kruskal solution for \Sch.

To obtain a Penrose diagram, it is now necessary to carry out another coordinate transformation, namely,
\begin{equation}
v' = \tan^{-1} \left[ \exp \left( \frac{r_+ - r_-}{4 r_+^2} v \right) \right], \qquad u' = \tan^{-1} \left[ - \exp \left( \frac{- r_+ + r_-}{4 r_+^2} u \right) \right].
\end{equation}
Taking the derivatives of either of these,  i.e., $A = \{ u, v\}$, gives, 
\par \vspace{-6pt} \begin{IEEEeqnarray}{rCl} \label{eqn: uvTransform2}
dA &=& \pm \frac{\sec^2 A'}{\tan A'} \left( \frac{4 r_+^2}{r_+ - r_-} \right) dA' \nonumber \\
&=&  \pm 2 \csc \left(2 A' \right) \left( \frac{4 r_+^2}{r_+ - r_-} \right) dA',
\end{IEEEeqnarray}
where $\pm$ is $+$ for $A=v$ and $-$ for $A=u$.   The line element, from Eq.~\eqref{eqn: ds2RN}, can now be given by
\begin{equation}
d s^2 = \frac{\Delta} {r^2} \left( \frac{64 r_+^2}{r_+ - r_-} \right) \csc \left(2 u' \right) \csc \left(2 v' \right) du' dv' + r^2 \left(d \theta^2 + \sin^2 \theta d \phi^2 \right),
\end{equation}
where $r$ is defined by
\par \vspace{-6pt} \begin{IEEEeqnarray}{rCl} \label{eqn: uvTransform3}
\tan v' \tan u' &=& - \exp \left( \frac{r_+ - r_-}{4 r_+^2} 2 r^*\right) \nonumber \\
&=& - \exp \left( \frac{r_+ - r_-}{2 r_+^2} r \right) \left( r - r_+ \right)^{1/2} \left( r - r_- \right)^{r_-^2 / 2 r_+^2}.
\end{IEEEeqnarray}

The Penrose diagram for this line element has an infinite number of asymptotically flat regions where $r>r_+$, labelled $I$, connected by regions $II$ and $III$ where $r_->r>r_+$ and $0>r>r_-$ respectively.  Region $III$ contains a singularity, however it is timelike, which means it can be avoided by a timelike curve, i.e., the world line of a particle.  In fact, this singularity acts repulsive in the sense that timelike curves cannot hit them.  This Penrose diagram, therefore, hints at a very unusual scenario - it should be possible for a particle, $P$ in Fig.~\ref{fig: uvpRN} to start in region $I$, cross the horizon $r=r_+$ into region $II$, continue from this region cross another horizon $r=r_-$ into region $III$.  Here, it will avoid the singularity and can pass through another $r=r_-$ horizon in the 'next' region $II$, and again through another $r=r_+$ horizon into a `new' region $I$.  This physically translates to a particle crossing from one universe through a `wormhole' of sorts and arriving in a new universe.  However, once the particle has left its original region $I$, like in the \Sch space-time, no information can be received from it by an observer in $I$, nor the possibility of return.   
\begin{figure}
\begin{center}
\includegraphics[scale=1.0]{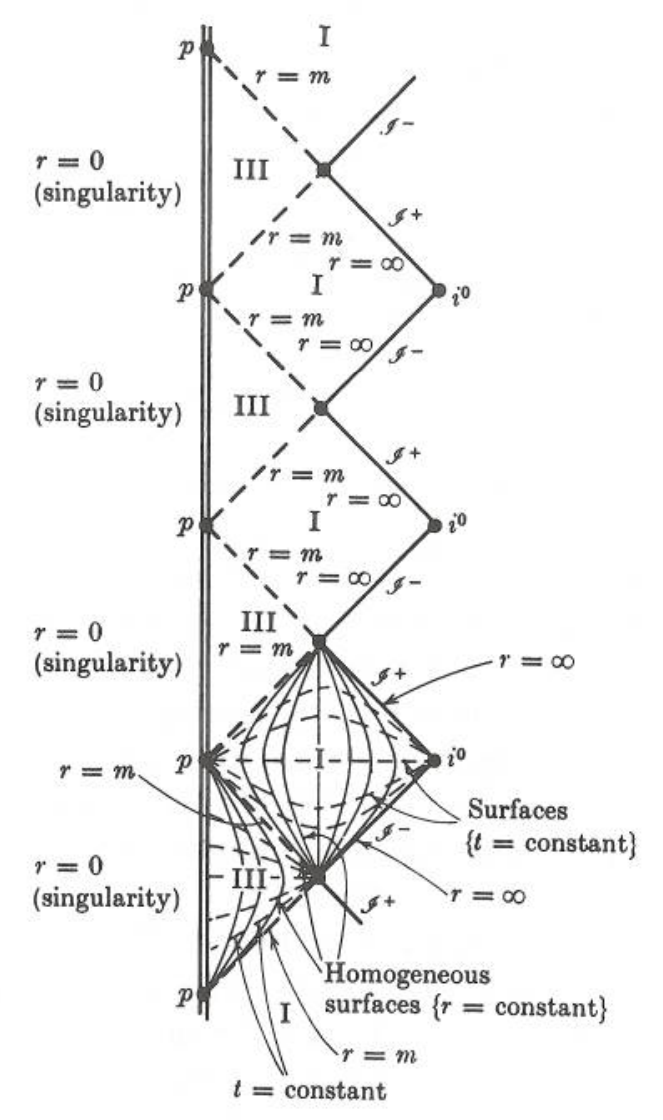}
\caption[Penrose Diagram for \rn Space-Time $Q^2 = M^2$]{The Penrose diagram for \rn space-time when $Q^2 = M^2$, has an infinite number of asymptotically flat regions where allowing the possibility of a particle crossing from one universe through a 'wormhole' of sorts and arriving in a new universe.  This diagram was taken from \cite{Hawking:Ellis}.}
\label{fig: uvpRN2}
\end{center}
\end{figure}

\begin{figure}
\begin{center}
\includegraphics[scale=1.6]{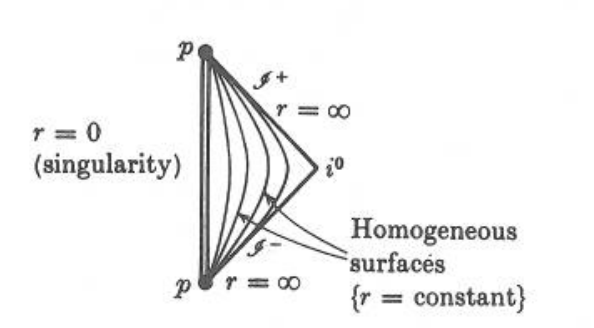}
\caption[Penrose Diagram for \rn Space-Time $Q^2 > M^2$]{The Penrose diagram for \rn space-time when the $Q^2 > M^2$, we have no horizons which suggests the concept of a naked singularity - a singularity that is not hidden behind an horizon.  Such a singularity is thought not to exist in reality (Cosmic Censorship Conjecture).  If this conjecture is to be believed, this would give us an upper limit on the possible charge of a  black hole.  This diagram was taken from \cite{Hawking:Ellis}.}
\label{fig: uvpRN3}
\end{center}
\end{figure}

The $Q^2 = M^2$ case of the \rn solution can be extended in the same manner as the $Q^2<M^2$ case to produce the Penrose diagram of Fig.~\ref{fig: uvpRN2}.  Here we see the we only have one horizon of the type $r = m$ and as in the $Q^2<M^2$, we have a timelike singularity.  For the $Q^2 > M^2$, we have no horizons which suggests the concept of a naked singularity - a singularity that is not hidden behind an horizon.  Such a singularity is thought not to exist in reality \cite{Penrose:1969} - a belief known as the Cosmic Censorship Conjecture, however we will go into this in more detail in Sec.~\ref{sec: ccc}.  If this conjecture is to be believed, this would give us an upper limit on the possible charge of a  black hole,
\begin{equation}
M^2 \geq Q^2.
\end{equation}
This naked singularity can be seen clearly with the Penrose diagam, Fig.~\ref{fig: uvpRN3}.


\subsubsection{A General Static, Spherically Symmetric Space-Time}

In Sec.~\ref{sec: nonGeodesicMotion}, we investigate the singular field and the associated regularization parameters for non-geodesic motion.  As our aim is to assist in calculations of the self-force, we tried to keep a very general space-time.  To this end we introduce the $f(r)$ space-time with the line element,
\begin{equation}
ds^2 = - f (r) dt^2 + f(r)^{-1} dr^2 + r^2 d\Omega^2.
\end{equation}
It is easily seen that this space-time is a more general version of both the \rn and \Sch space-times.


\subsection{Axially Symmetric, Stationary Space-Time}
 

\subsubsection{Kerr Space-Time}

The Kerr solution describes rotating black holes.  In Boyer-Lindquist coordinates, it has the line element,
\par \vspace{-6pt} \begin{IEEEeqnarray}{rCl}
ds^2 &=& - \left(1-\frac{2Mr}{\Sigma}\right)dt^2 
			- \frac{4aMr\sin^2\theta}{\Sigma}dtd\phi
			+ \frac{\Sigma}{\Delta}dr^2
			+ \Sigma d\theta^2 \nonumber \\
&&
			+\: \left[\Delta+\frac{2Mr(r^2+a^2)} {\Sigma}\right] \sin^2\theta d\phi^2,
\end{IEEEeqnarray}
where
\begin{eqnarray}
\Sigma &=& r^2+a^2\cos^2\theta, \\
\Delta &=& r^2-2Mr+a^2,
\end{eqnarray}
$M$ represents its mass, and $a$ its angular momentum per unit mass.  By setting $a = 0$, it is easily seen that the Kerr solution reduces to the \Sch solution.  As the metric coefficients are independent of $t$ and $\phi$, we know that it is an axially symmetric, stationary solution.  As the metric is not time invariant, it is not static, however it is invariant under the simultaneous inversion of $t$ and $\phi$, that is it remains the same under the transformation,
\begin{equation}
t \rightarrow -t, \quad \quad \phi \rightarrow - \phi.
\end{equation}
The consequence of this is that if we move forward in time with a positive spin direction, we will get the same field as if we moved backward in time with a negative spin direction - this, along with other factors, has lead to the understanding that the Kerr solution describes a rotating black hole.

As in the \Sch case, we can calculate the Lagrangian for a massive point particle moving in Kerr space-time using Eq.~\eqref{eqn: lagrangian}, looking at geodesics in the equatorial plane, i.e., setting $\theta = 0$,
\begin{equation}
\mathcal{L} = \frac{1}{2} \left[ \left(1 - \frac{2 M}{r }\right) \dot{t}^2 + \frac{4 a M}{r} \dot{t} \dot{\phi} - \frac{r^2}{\Delta} \dot{r}^2 - \left( r^2 + a^2 + \frac{2 a^2 M}{r}\right) \dot{\phi}^2 \right].
\end{equation}
As before, the canonical momenta, $p_a = \frac{\partial \mathcal{L}}{\partial \dot{q}_a}$, is given by
\begin{gather}
p_t = \left(1 - \frac{2 M}{r }\right) \dot{t} + \frac{2 a M}{r} \dot{\phi}  = E = constant, \label{eqn: ptKerr}\\
p_r = - \frac{r^2}{\Delta} \dot{r}, \\
p_{\phi} = \frac{2 a M}{r} \dot{t} - \left( r^2 + a^2 + \frac{2 a^2 M}{r}\right) \dot{\phi} = L = constant, \label{eqn: pphiKerr}
\end{gather}
where the constancy of $p_t$ and $p_{\phi}$ comes from the independence of the Lagrangian from $t$ and $\phi$.  Solving Eqs.~\eqref{eqn: ptKerr} and \eqref{eqn: pphiKerr} for $\dot{\phi}$ and $\dot{t}$ gives,
\begin{equation} \label{eqn: Kerrtphi}
\dot{t} = \frac{1}{\Delta} \left[  \left( r^2 + a^2 + \frac{2 a^2 M}{r} \right) E - \frac{2 a M}{r} L \right], \quad \dot{\phi} = \frac{1}{\Delta} \left[ \left(1 - \frac{2 M}{r} \right) L + \frac{2 a M}{r} E \right].
\end{equation}
Using the definition for the Hamiltonian in Eq.~\eqref{eqn: HL} gives
\par \vspace{-6pt} \begin{IEEEeqnarray}{rCl}
\mathcal{H} &=& \frac{1}{2} \left(1 - \frac{2 M}{r }\right) \dot{t}^2 +\frac{2 a M}{r} \dot{t} 
	\dot{\phi} - \frac{r^2}{2 \Delta} \dot{r} ^2 - \frac{1}{2} \left( r^2 + a^2 + \frac{2 a^2 M}
	{r}\right) \dot{\phi}^2 \nonumber \\
&=&
	\frac{1}{2} \left\{ \left[ \left(1 - \frac{2 M}{r }\right) \dot{t} + \frac{2 a M}{r} \dot{\phi} 
	\right] \dot{t} - \left[  \left( r^2 + a^2 + \frac{2 a^2 M}{r}  \right) \dot{t} - \frac{2 a M}{r}
	\dot{ \phi} \right] \dot{\phi}  + \frac{r^2}{\Delta} \dot{r}^2 \right\} \nonumber \\
&=&
	\frac{1}{2} \left( E \dot{t} - L \dot{\phi} -  \frac{r^2}{\Delta} \dot{r}^2 \right).
\end{IEEEeqnarray}
As $\mathcal{H}$ is independent of $t$ we are allowed to set $\mathcal{H} = 1/2$ and using Eq.~\eqref{eqn: Kerrtphi}, it is possible to solve for $\dot{r}$, to get,
\begin{equation}
\dot{r}^2 = \frac{1}{r^2} \left[ r^2 E^2 + \frac{2 M}{r} \left( a E - L \right)^2 + a^2 E^2 - L^2 - \Delta \right].
\end{equation}
As in the case of \Sch, by setting $\tau$ to be proper time, we have arrived at the four-velocity for geodesic Kerr space-time.


\subsubsection{Kerr-Newman Space-Time}

As the \Sch solution had its 'charged' counterpart in the form of the \rn solution, the Kerr solution has a similar counterpart in the Kerr-Newman solution, which describes the space-time of a charged black hole.  In Boyer-Lindquist coordinates, this has the line element, 
\begin{multline}
\label{eq:KerrNewmanMetric}
ds^2 = - \frac{\Delta - a^2 + z^2}{\rho^2} dt^2 
	+ \frac{2(\Delta-r^2-a^2)(a^2-z^2)}{a \rho^2}dtd\phi
	+ \frac{\rho^2}{\Delta}dr^2\\
	+ \rho^2  d\theta^2
	+ \frac{a^2-z^2}{a^2 \rho^2} \left( (a^2+r^2)^2 - \Delta (a^2-z^2) \right) d\phi^2, 
\end{multline}
where $\Delta=r^2-2Mr+a^2+Q^2$, $\rho^2 = r^2 + a^2 \cos^2 \theta$, $z=a \cos \theta$ and $Q$ is the charge per unit mass of the black hole.  As in the \rn case, the Kerr-Newman solution has a vanishing Ricci scalar and a non-vanishing Ricci tensor, which becomes very useful when considering non-geodesic motion.


\section{The Detweiler-Whiting Singular Field} \label{sec: DWSingularField}

When considering self-force and the potential fields involved, one is usually restricted to one of three cases, a point particle carrying a scalar or electric charge or a point mass.  In each case, a field (scalar, electromagnetic or gravitational depending on the case) is produced by the particle, which effects the motion of the particle.  This is due to the particle interacting with its own field, causing the particle or mass to deviate from the geodesic of the background space-time.  This deviation maybe interpreted as a force acting on the particle or mass, and is the so-called self-force.  Calculating the self-force, therefore, requires knowledge of the field that produces it.  

The problem in producing an expression for such a field is that it is singular at the particle or mass.  The traditional way to overcome this is to split the field into a `direct' and a `tail' part - such a decomposition originally proved very useful for describing the self-force, but as neither parts were solutions of the field equation, they did not give a meaningful explanation of the self-force.  A novel solution to the same problem, first introduced by Detweiler and Whiting \cite{Detweiler-Whiting-2003}, is to similarly split the field into two parts - a regular part and a singular part.  The singular field is designed so it will solve the same inhomogeneous wave equation as the actual field while the regular field will solve the homogeneous wave equation - in this manner the structure of the field is maintained.  This singular-regular split, by design, then gives us two fields - the singular field which has no impact on the motion of the particle but completely contains the singular structure of the original field, and the regular field which is a smooth field, that is completely responsible for the self-force.

Throughout this thesis, in the spirit of DeWitt \cite{DeWitt:1965jb}, we shall use the notation that $\varphi^A$ refers to the field in all 3 cases with $A$ specifying which case, that is
\begin{equation} \label{eqn: phiCases}
\varphi^A=\begin{cases} \Phi & \text{(scalar field)} \\
			A^a & \text{(electromagnetic field)} \\
			h^{ab} & \text{(gravitational field)}
			\end{cases}.
\end{equation}


\subsection{The wave equation}

For each of the three cases, we have a slightly different version of the inhomogeneous wave equation with a distributional source.  We will go through each of the cases here and then give a general expression which can incorporate all three scenarios.  This is to allow the reader to become familiar and confident with our more general description of the similar structures associated with the three cases.

A massless scalar field in curved space-time will satisfy the inhomogeneous wave equation
\begin{equation} \label{eqn: WaveScalar}
\left(\Box - \xi R\right) \Phi (x) = -4 \pi \mu (x),
\end{equation}
with the distributional source,
\begin{equation}
\mu(x) = q \int_{\gamma} \delta_4 \left(x, z \right) d\tau,  
\quad \text{where} \quad 
\delta_4 \left(x, x' \right) = \frac{\delta_4 \left(x - x' \right)}{\sqrt{-g}} = \frac{\delta_4 \left(x - x' \right)}{\sqrt{-g'}},
\end{equation}
$\square \equiv g^{ab}\nabla_{a}\nabla_{b}$, 
$g^{ab}$ is the (contravariant) metric tensor, $g$ is its determinant at $x$, $g'$ is its determinant at $x'$, $\nabla_{a}$ is the covariant
derivative defined by a connection $\mathcal{A}^{A}{}_{Ba}$:
$\nabla_{a}\varphi^{A}= \partial_{a}\varphi^{A}+  \mathcal{A}^{A}{}_{Ba} \varphi^{B}$, $R$ is the Ricci scalar, $\xi$ is an arbitrary coupling constant, $z(\tau)$ describes the world line, $q$ the scalar charge and $\delta_4 \left(x - x' \right) = \delta \left(x^0 - x^{0'} \right) \delta \left(x^1 - x^{1'} \right) \delta \left(x^2 - x^{2'} \right) \delta \left(x^3 - x^{3'} \right)$ is the ordinary (coordinate) Dirac delta function.  The solution of Eq.~\eqref{eqn: WaveScalar} can be written in terms of a Green's function, $G \left(x,x'\right)$,
\par \vspace{-6pt} \begin{IEEEeqnarray}{rCl}
\Phi (x) &=& \int G \left(x, x' \right) \mu(x') \sqrt{-g'} d^4 x'  \nonumber \\
&=& q\int G \left(x, z \right) d \tau.
\end{IEEEeqnarray}
Substituting this back into Eq.~\eqref{eqn: WaveScalar}, from this, it becomes clear that for Eq.~\eqref{eqn: WaveScalar} to remain true, the following condition on the Green's function must be satisfied,
\begin{equation}
\left(\Box - \xi R\right) G(x,x') = - 4 \pi \delta_4 \left(x, x' \right).
\end{equation}

An electromagnetic field in curved space-time will satisfy,
\begin{equation}\label{eqn: WaveEM}
\left(\Box \delta^a{}_b - R^a{}_b \right) A^b (x) = -4 \pi j^a (x),
\end{equation}
in the Lorentz gauge ($\nabla_a A^a = 0$), with the distributional source, the current density, given by,
\begin{equation}
j^a (x) = e \int_{\gamma} g^a{}_{c} \left(x, z \right) u^{c} \delta_4 \left( x, z \right) d \tau
\end{equation}
where $\delta^a{}_b$ is the Kronecker delta function, $R^a{}_b$ is the Ricci tensor, $u^b$ is the four velocity and $e$ is the charge of the particle.  As with the scalar case, we use a trial solution containing a Green's function of the type,
\par \vspace{-6pt} \begin{IEEEeqnarray}{rCl}
A^a (x) &=& \int G^a{}_{b'} \left(x, x' \right) j^{b'} (x') \sqrt{-g'} d^4 x' \nonumber \\
&=& e \int G^a{}_{c} \left(x, z \right) u^{c}  d \tau,
\end{IEEEeqnarray}
in Eq.~\eqref{eqn: WaveEM} to produce the required equation for the Green's function,
\begin{equation}
\left(\Box \delta^a{}_b - R^a{}_b \right) G^b{}_{b'} \left(x, x' \right) = -4 \pi g^a{}_{b'} \left(x, x'\right) \delta_4 \left(x,x'\right).
\end{equation}

The propagation of gravitational perturbations in a vacuum space-time is described by,
\begin{equation}\label{eqn: WaveG}
\left(\Box \delta^a{}_c \delta^b{}_d + 2 C_c{}^a{}_d{}^b \right) h^{cd} (x) = -16 \pi T^{ab} (x),
\end{equation}
in the Lorentz gauge ($h^{ab}{}_{;b}=0$),  and with the  energy-momentum tensor acting as the distributional source,
\begin{equation}
T^{ab} (x) = m \int_{\gamma} g^a{}_{c} \left(x, z\right) g^b{}_{d} \left(x, z\right) u^{c} u^{d} \delta_4 \left(x, z \right) d \tau,
\end{equation}
where \fixme{$C_c{}^a{}_d{}^b$} is the \fixme{Weyl} tensor and $m$ the mass of the particle.  As before we design a solution with the structure,
\par \vspace{-6pt} \begin{IEEEeqnarray}{rCl}
h^{ab} (x) &=& 4 \int G^{ab}{}_{c'd'} \left(x, x' \right) T^{c' d'} (x') \sqrt{-g'} d^4 x' \nonumber \\
&=& 4 m \int G^{ab}{}_{cd} \left(x, z \right) u^{c} u^{d} d \tau.
\end{IEEEeqnarray}
When used in Eq.~\eqref{eqn: WaveG}, we find the following constraint on the Green's function,
\begin{equation}
\left(\Box \delta^a{}_c \delta^b{}_d + 2 \text{\fixme{C}}_c{}^a{}_d{}^b \right) G^{cd}{}_{c'd'} \left(x, x' \right) = - 4 \pi g^{(a}{}_{c'} \left(x, x'\right) g^{b)}{}_{d'} \left(x, x'\right) \delta_4 \left(x, x'\right),
\end{equation}
where the symmetry brackets in $(a,b)$ are introduced to maintain the symmetry on both sides of the equation.

As one can see, all three cases are dealt with in the same manner, an attribute which applies to most calculations in this thesis.  We therefore, come back to our previous notation introduced in Eq.~\eqref{eqn: phiCases}.  That is, we now consider all of the field equations to be described by,
\begin{equation} \label{eqn: waveGen}
\mathcal{D}^{A}{}_B \varphi^{B} \equiv \left[\delta^{A}{}_B \square - P^{A}{}_B \right] \varphi^B = - 4\pi \mathcal {M}^A,
\end{equation}
with the distributional source, $\mathcal{M}$ described by,
\begin{equation} \label{eqn: MGen}
\mathcal{M}^A (x) = \mathcal{Q} \int_{\gamma} u^A \delta_4 \left(x, z\right) d \tau,
\end{equation}
where the fields here are all for the scalar, electromagnetic and gravitational cases respectively, we have
\begin{equation}
\delta^A{}_B= \begin{cases} 1 \\
					\delta^a{}_{b} \\
					\delta^{(a}{}_c \delta^{b)}{}_d
			\end{cases}, \quad
P^A{}_B = \begin{cases} \xi R \\
				R^a{}_b  \\
				-2 C_c{}^{(a}{}_d{}^{b)} 
		\end{cases},
\end{equation}
\begin{equation}
\mathcal{Q} = \begin{cases}  q \\
						e\\
						4 m
			\end{cases}, \quad
u^A = \begin{cases} 1 \\ 
				g^a{}_{b} \left(x, z \right) u^{b} \\
				g^a{}_{c} \left(x, z \right) g^b{}_{d} \left(x, z\right) u^{c} u^{d}	
	\end{cases}.		
\end{equation}
The solution of these equations can be generally written in terms of a general Green's function, $G^A{}_B \left( x, x' \right)$, as
\par \vspace{-6pt} \begin{IEEEeqnarray}{rCl}\label{eqn: varphi eqn}
\varphi^A &=& \int G^A{}_{B'} \left(x, x' \right) \mathcal{M}^{B'} (x') \sqrt{-g'} d^4 x' \nonumber \\
&=& \mathcal{Q} \int G^A{}_{B'} \left(x, x' \right) u^{B'} d \tau,
\end{IEEEeqnarray}
which gives us the general condition for our Green's function,
\begin{equation} \label{eqn: GreenGen}
\mathcal{D}^{A}{}_B G^B{}_{C'} \left(x, x'\right) = -4 \pi g^A{}_{C'} \left(x, x'\right) \delta_4 \left(x, x' \right),
\end{equation}
where we have for the scalar, electromagnetic and gravitational waves respectively,
\begin{equation} \label{eqn: GgGen}
G^A{}_{B'} \left(x, x'\right)= \begin{cases} G\left( x, x' \right) \\
					G^a{}_b' \left(x,x' \right) \\
					G^{ab}{}_{c'd'} \left(x, x' \right)
		\end{cases}, \quad
g^A{}_{C'} \left(x, x'\right)= \begin{cases} 1 \\
					g^a{}_b' \left(x, x' \right) \\
					g^{(a}{}_{c'} \left(x, x'\right) g^{b)}{}_{d'}
		\end{cases}.
\end{equation}


\subsection{Singular Field in Flat Space-time}

Before we start looking at the singular field and its associated Green's functions in curved space-time, it is beneficial to examine them in the simpler setting of flat space-time.  When we look at the physical solutions to Eq.~\eqref{eqn: waveGen}, we find there are two key solutions in the form of Eq.~\eqref{eqn: varphi eqn} - the retarded and advanced solutions given by,
\begin{equation}\label{eqn: varphiFlatRetAdv}
\varphi^A{}_{\rm \ret / \adv} = \int G^{A}{}_{B'}{}_{\rm \ret / \adv} \left(x, \xpp \right) \mathcal{M}^{B'} (x') \sqrt{-g'} d^4 \xpp'.
\end{equation}
where $ G^A{}_{B'}{}_{\rm \ret} \left(x, \xpp \right)$ and $ G^A{}_{B'}{}_{\rm \adv} \left(x, \xpp \right)$ are the retarded and advanced Green's  functions respectively.  

\begin{figure}[h]
\begin{center}
\includegraphics[scale=0.6]{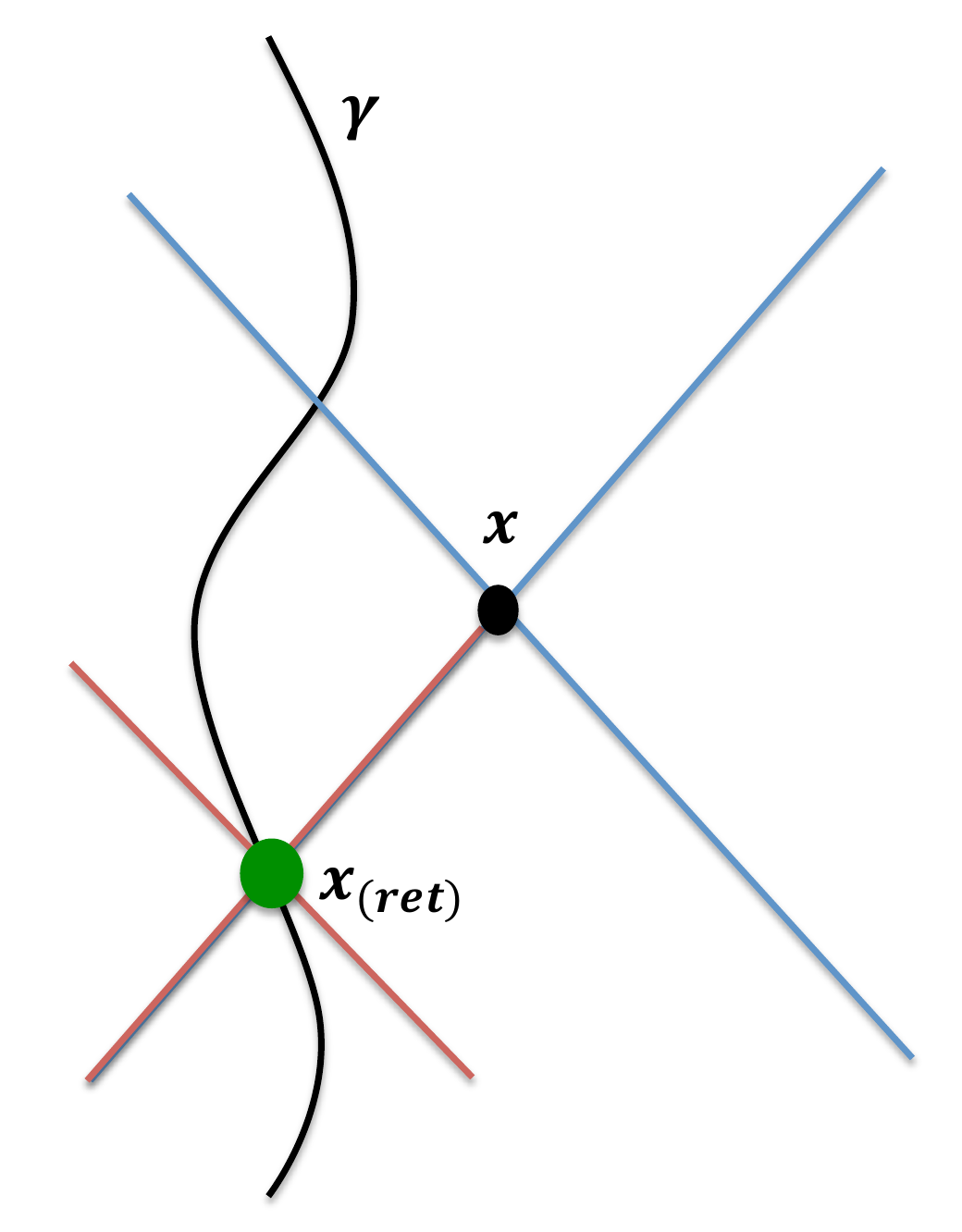}
\caption[Retarded Green's Function in Flat Space-time]{The point at which the retarded Green's function is non-zero on $\gamma$ is when $x$ is on the future light cone of $\xpp$, this point is highlighted in green and labelled $x_{\rm \ret}$. }
\label{fig: GretFlat}
\end{center}
\end{figure}

\begin{figure}
\begin{center}
\includegraphics[scale=0.6]{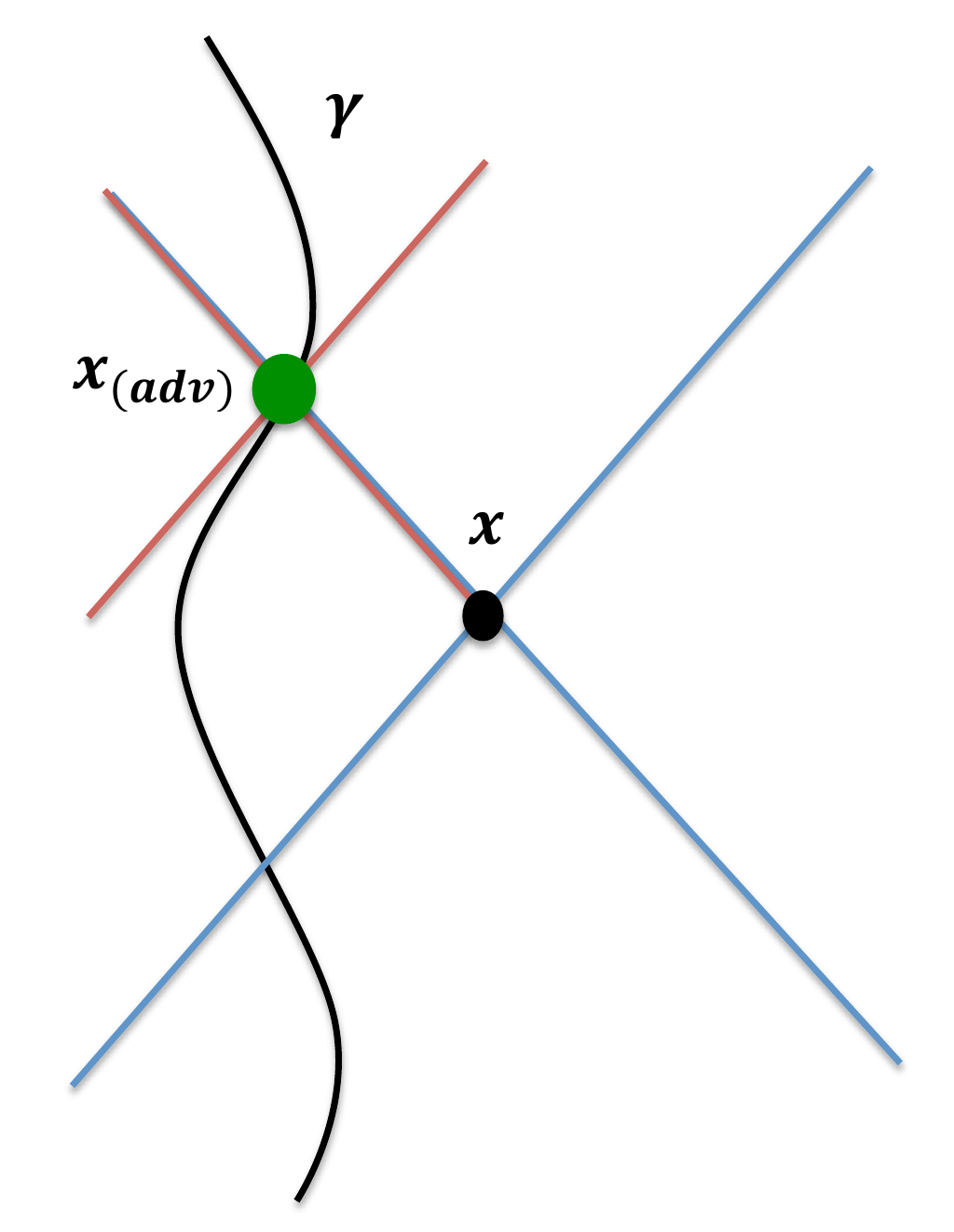}
\caption[Advanced Green's Function in Flat Space-time]{The point at which the advanced Green's function is non-zero on $\gamma$ is when $x$ is on the past light cone of $\xpp$, this point is highlighted in green and labelled $x_{\rm \adv}$. }
\label{fig: GadvFlat}
\end{center}
\end{figure}

In this chapter, we will be taking $\xpp$ to be our source point on the world line, $\gamma$, and $x$ to be a field point in the near neighbourhood of $\xpp$.   $ G^A{}_{B'}{}_{\rm \ret} \left(x, \xpp \right)$ is then only non-zero on $\gamma$ when $x$ is on the future light cone of $\xpp$, we denote this point as $x_{\rm \ret}$.  Similarly $G^A{}_{B'}{}_{\rm \adv} \left(x, \xpp \right)$ is only non-zero on $\gamma$ when $x$ is on the past light cone of $\xpp$, which we label as $x_{\rm \adv}$.  From this nature of the Green's functions, it is quite clear that they have a reciprocity nature, that is
\begin{equation}
G^A{}_{B'}{}_{\rm \ret} \left(x, \xpp \right) = G^A{}_{B'}{}_{\rm \adv} \left(\xpp, x \right).
\end{equation}
We can see from Figs.~\ref{fig: GretFlat} and \ref{fig: GadvFlat}, that $x_{\rm \ret}$ is in the past of $x$ while $x_{\rm \adv}$ is in
\newpage
\noindent its future.  When considering the physical implications of our solutions Eq.~\eqref{eqn: varphiFlatRetAdv}, we, therefore, take the retarded solution to be the physically meangingful one.  From now on we refer to the field associated with the scalar or electric charge or point mass, through the retarded Green's function, as the retarded field.

As was discussed at the beginning of Sec.~\ref{sec: DWSingularField}, the aim of this section is to obtain the desired regular-singular split of the retarded field of the form,
\begin{equation}
\varphi^A{}_{\rm \ret} = \varphi^A{}_{\rm \sing} + \varphi^A{}_{\rm \reg}.
\end{equation}
In flat space-time, this proves to be quite simple - we define the singular field to be half the sum of the retarded and advanced fields, while the regular field is half their difference.  This give us
\begin{equation} \label{eqn: varphiSingRegFlat}
\varphi^A{}_{\rm \sing} = \frac{1}{2} \left[ \varphi^A{}_{\rm \ret} + \varphi^A{}_{\rm \adv} \right], \quad \varphi^A{}_{\rm \reg} = \frac{1}{2} \left[ \varphi^A{}_{\rm \ret} - \varphi^A{}_{\rm \adv} \right].
\end{equation}
We can see by applying $\mathcal{D}^A{}_B$, that these definitions satisfy the criteria laid out earlier, i.e,, the singular field satisfies the inhomogeneous equation of Eq.~\eqref{eqn: waveGen} while the regular field satisfies the homogeneous version of the same equation.

When dealing with the singular and regular field, we often think of them in terms of their own Green's functions, therefore we make similar definitions to those in Eq.~\eqref{eqn: varphiSingRegFlat} for the corresponding Green's functions.  These are
\begin{equation} \label{eqn: GreenSingRegFlat}
G^A{}_{B'}{}_{\rm \sing} = \frac{1}{2} \left[ G^A{}_{B'}{}_{\rm \ret} + G^A{}_{B'}{}_{\rm \adv} \right]. \quad G^A{}_{B'}{}_{\rm \reg} = \frac{1}{2} \left[ G^A{}_{B'}{}_{\rm \ret} - G^A{}_{B'}{}_{\rm \adv} \right].
\end{equation} 


\subsection{Singular Green's Function in Curved Space-time}

\begin{figure}
\begin{center}
\includegraphics[scale=0.6]{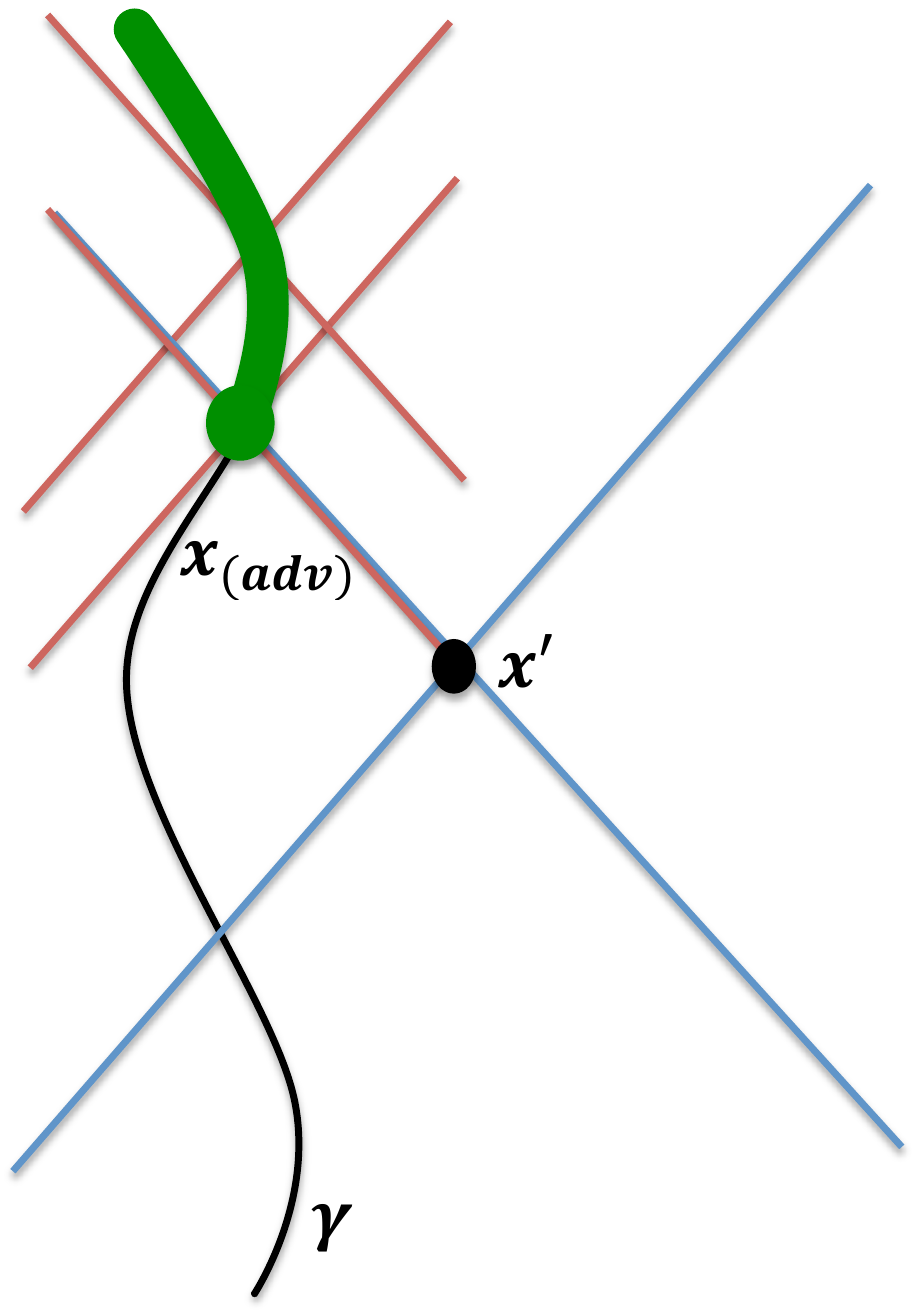}
\caption[Advanced Green's Function in Curved Space-time]{The points at which the advanced Green's function is non-zero on $\gamma$ is when $x$ is in or on the past light cone of $\xpp$, these points or that part of $\gamma$ are highlighted in green. }
\label{fig: GadvCurve}
\end{center}
\end{figure}

When we consider the regular-singular split of the retarded field in curved space-time, our lives are not as easy as they were in flat space-time.  In curved space time, energy waves, such as electromagnetic, don't just travel on the light cone (\fixme{Huygens'} Principle) as they do in flat space-time.  
\begin{figure}
\begin{center}
\includegraphics[scale=0.6]{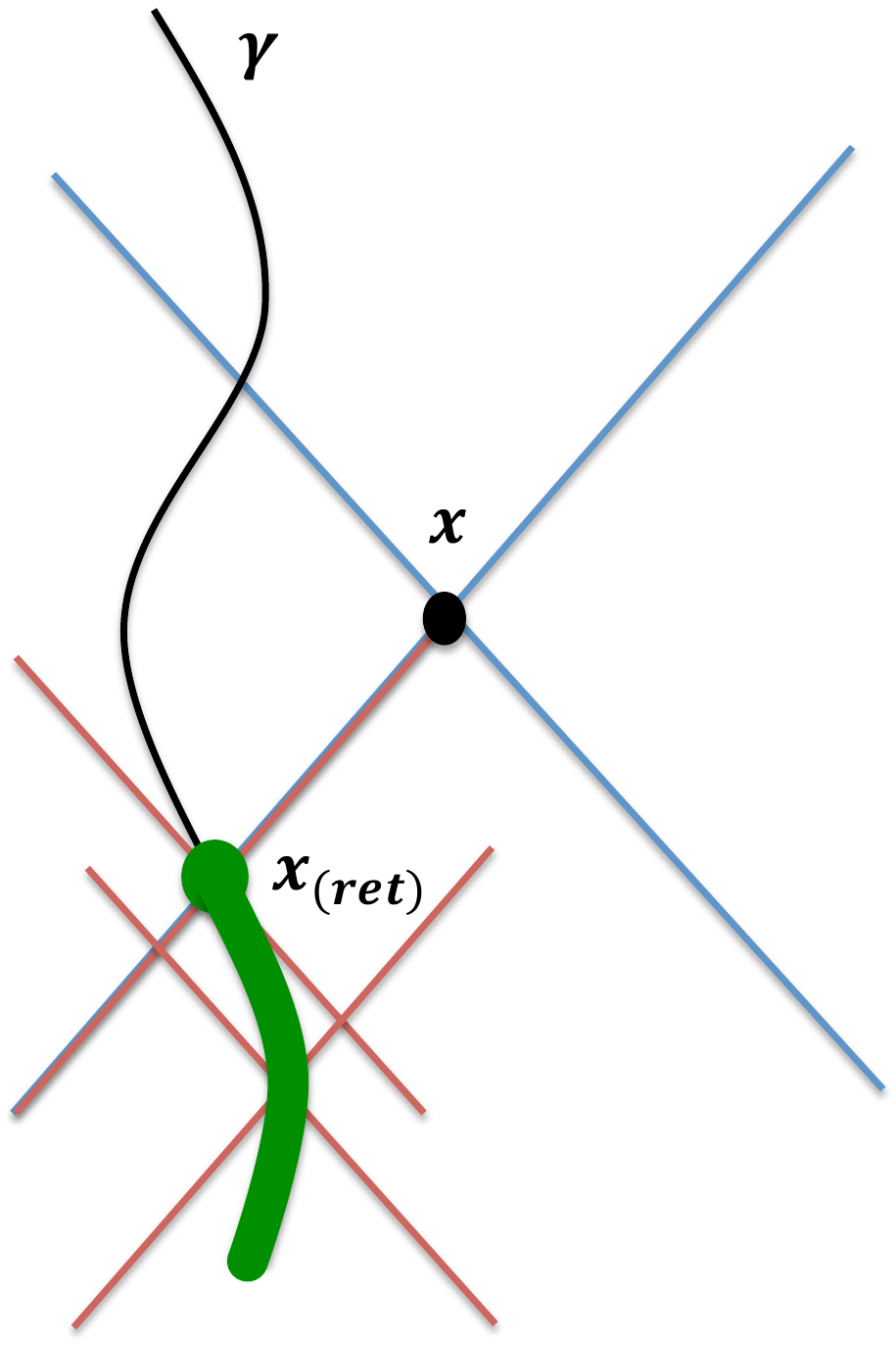}
\caption[Retarded Green's Function in Curved Space-time]{The points at which the retarded Green's function is non-zero on $\gamma$ is when $x$ is in or on the future light cone of $\xpp$, these points or that part of $\gamma$ are highlighted in green. }
\label{fig: GretCurve}
\end{center}
\end{figure}
They scatter off the space-time curvature and so reach points inside the future light-cone.  The ramifications of this for our Green's functions is that they now are not only non-zero on the light cone of $\xpp$, but also within the light cone of $\xpp$.  For the retarded Green's function, this translates to it not being zero when $x$ is on or within the future light cone of $\xpp$, in other words, when $x$ is in the chronological future of $\xpp$ or $x \in I^+ (\xpp)$.  Similarly with the \fixme{advanced} Green's function, it is non-zero on $\gamma$ when $x$ is on or within the past light cone of $\xpp$, or $x \in I^- (\xpp)$, where $I^- (\xpp)$ is the chronological past of $\xpp$.  These areas of non-zero values are shown in Figs.~\ref{fig: GretCurve} and \ref{fig: GadvCurve}.

Now that we have our Green's functions defined we re-examine our flat space-time definitions of our singular and regular fields.  If we maintain these definitions, Eq.~\eqref{eqn: varphiSingRegFlat}, and hence the associated Green's functions definitions of Eq.~\eqref{eqn: GreenSingRegFlat}, we will find that both our singular and regular definitions are now non-zero for $\xpp \in \left(- \infty, x_{\rm \ret} \right] \cup \left[ x_{\rm \adv}, \infty \right)$.  
In fact as we move our observation point, $x$, closer to the source represented by the world line, $\gamma$, we will notice that the distance between $x_{\rm \ret}$ and $x_{\rm \adv}$ tends to zero.  The implication of this on our regular and singular functions are that they will now be dependent on the entire future and past of the $\xpp$.  Not only have our expressions for the singular and regular fields become mathematically impossible to calculate, but physically the system no longer makes sense causally.  The conclusion is, therefore, that a different definition of the singular field and the resulting regular field are needed.

The required revamp of the singular-regular field split of the retarded field was introduced by Detweiler and Whiting \cite{Detweiler-Whiting-2003} and has since been called the Detweiler-Whiting singular field.  The concept is to subtract a function, $H^A{}_{B'} \left(x, \xpp \right)$, that is a biscalar in the case of the scalar field and a bitensor in the cases of the electromagnetic and gravitational fields,
\begin{equation}
H^A{}_{B'} \left(x, x'\right) = \begin{cases} H \left(x, \xpp \right) & \text{(scalar cases)} \\
								H^a{}_{b'} \left(x, \xpp \right) & \text{(electromagnetic case)} \\
								H^{ab}{}_{c'd'} \left(x, \xpp \right)  & \text{(gravitational case)}
						\end{cases}.
\end{equation}
from our singular Green's function, i.e., 
\par \vspace{-6pt} \begin{IEEEeqnarray}{rCl} \label{eqn: GreenSingRegCurve}
G^A{}_{B'}{}_{\rm \sing} \left(x, \xpp \right) &=& \frac{1}{2} \left[ G^A{}_{B'}{}_{\rm \ret} \left(x, \xpp \right) + G^A{}_{B'}{}_{\rm \adv} \left(x, \xpp \right)  - H^A{}_{B'} \left(x, \xpp \right) \right] \nonumber \\
G^A{}_{B'}{}_{\rm \reg} \left(x, \xpp \right) &=& \frac{1}{2} \left[ G^A{}_{B'}{}_{\rm \ret} \left(x, \xpp \right) - G^A{}_{B'}{}_{\rm \adv} \left(x, \xpp \right) + H^A{}_{B'} \left(x, \xpp \right)  \right].
\end{IEEEeqnarray}

\begin{figure}
\begin{center}
\includegraphics[scale=0.6]{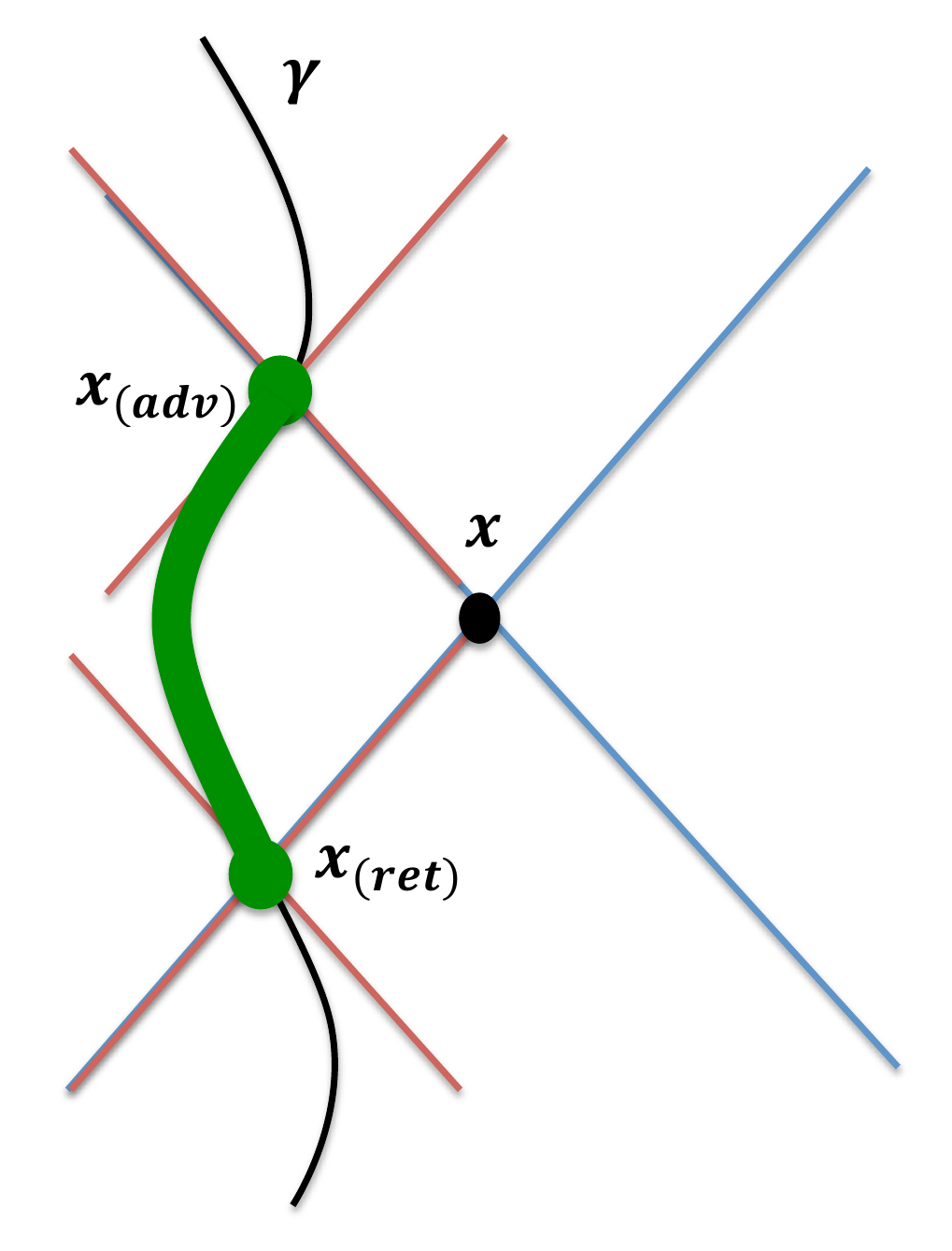}
\caption[Singular Green's Function in Curved Space-time]{The points at which the singular Green's function is non-zero on $\gamma$ is when $x$ and $\xpp$ are space and nulllike separated, that part of $\gamma$ is highlighted in green. }
\label{fig: GsingCurve}
\end{center}
\end{figure}

\begin{figure}[h]
\begin{center}
\includegraphics[scale=0.6]{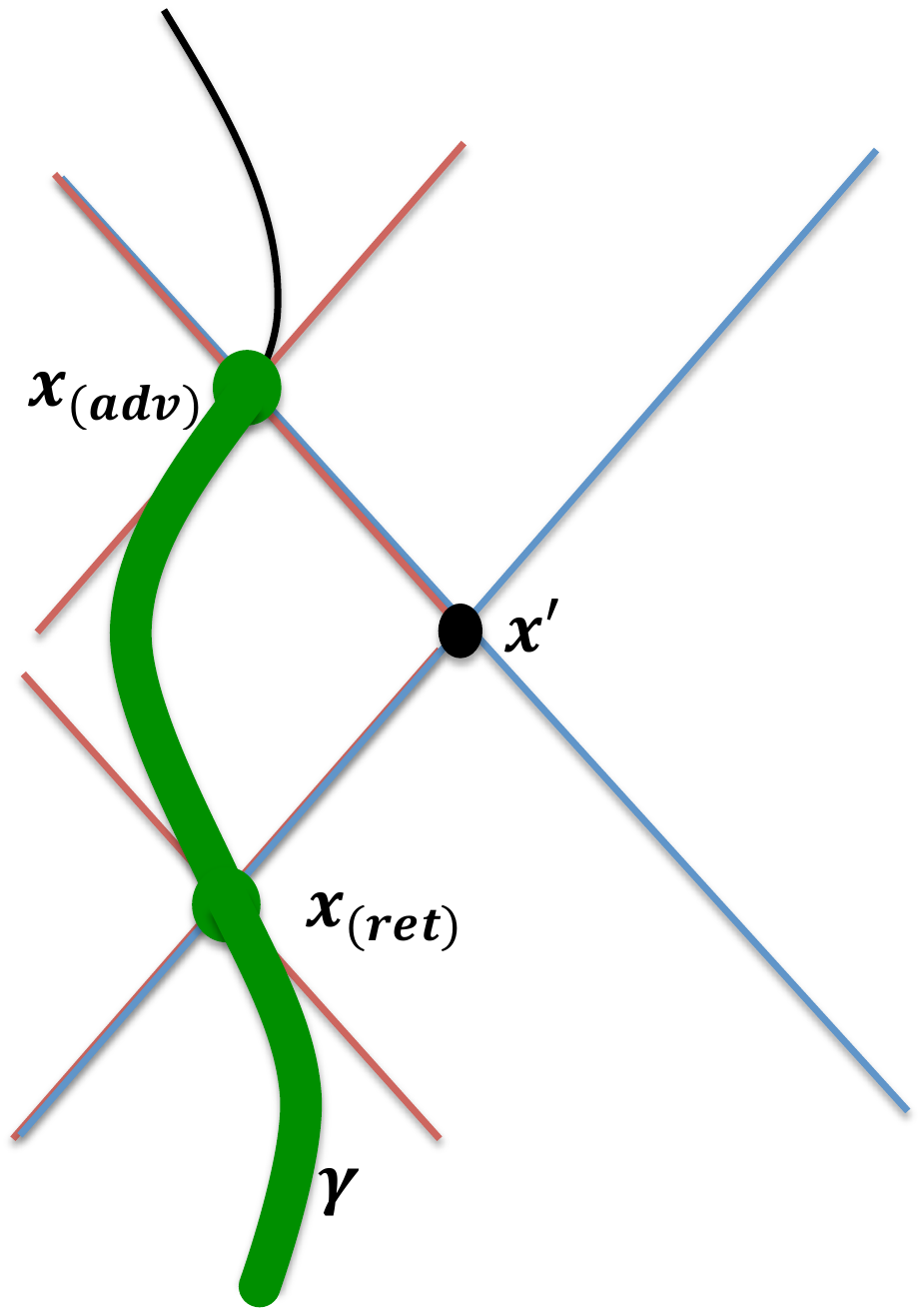}
\caption[Regular Green's Function in Curved Space-time]{The points at which the regular Green's function is non-zero on $\gamma$ is when $\xpp$ is not in the chronological past  of $x$, that part of $\gamma$ is highlighted in green. }
\label{fig: GregCurve}
\end{center}
\end{figure}

To avoid the dependence of our Green's functions on the entire past and future of $\xpp$, but maintain our previous qualities of the singular field, we demand that $H^A{}_{B'} \left(x, \xpp \right)$ has the following characteristics,
\begin{enumerate}
\item $H^A{}_{B'} \left(x, \xpp \right)$ must satisfy the homogeneous version of Eq.~\eqref{eqn: waveGen}, the wave equation. this allows the singular, retarded and regular Green's functions to still satisfy their required wave equations.
\item $H^A{}_{B'} \left(x, \xpp \right)$ must still have a reciprocity relation in respect to $x$ and $x'$ so that the other Green's functions also maintain their required reciprocity, i.e.,
$g_B{}^{B'} H^A{}_{B'} \left(x, \xpp \right) = g^A{}_{A'}H^{A'}{}_{B} \left( \xpp, x \right)$
\item $H^A{}_{B'} \left(x, \xpp \right) = G^A{}_{B'}{}_{\rm \adv} \left(x, \xpp \right)$ when $x \in I^-(x')$
\item$H^A{}_{B'} \left(x, \xpp \right) = G^A{}_{B'}{}_{\rm \ret} \left(x, \xpp \right)$ when $x \in I^+(x')$
\end{enumerate}
The resulting structure can be seen in Figs. \ref{fig: GsingCurve} and \ref{fig: GregCurve}

It will be shown in Sec.~\ref{sec: restrictions} that within a normal neighbourhood, the retarded and advanced fields can be written in the form
\begin{equation} \label{eqn: GretadvUV}
G^A{}_{B'}{}_{\rm \ret / \adv} \left(x, \xpp \right) = U^A{}_{B'} \left(x, \xpp \right) \delta_{+/-} (\sigma) \text{\fixme{$-$}} V^A{}_{B'}  \left(x, \xpp \right) \Theta_{+/-} \left( - \sigma \right)
\end{equation}
where we remind the reader of the definitions and implications of the Dirac delta function, $\delta \left (\sigma  \left(x, \xpp \right)\right)$ and the Heaviside step function $\Theta \left( - \sigma  \left(x, \xpp \right) \right)$
\begin{equation}
\delta \left (\sigma  \left(x, \xpp \right)\right) = \begin{cases} \infty & \sigma  \left(x, \xpp 
										\right) = 0 \Rightarrow \text{null geodesic 
										connecting $x$ and $\xpp$} \\
										0 & \sigma \ne 0 \Rightarrow \text{ $x$ and 
										$\xpp$ are not connected by their light-cones}
				\end{cases},
\end{equation}
\begin{equation} \label{eqn: step}
\Theta \left(- \sigma \left(x, \xpp \right) \right) = \begin{cases} 
										0 & \sigma \left(x, \xpp \right) > 0  
										\Rightarrow x  \text{ and $\xpp$ are spacelike related} \\
										1 & \sigma \left(x, \xpp \right)  \leq 0 
										\Rightarrow x \text{ and $\xpp$ are timelike related}
									\end{cases},
\end{equation}
where by $\infty$, we mean that $\delta$'s support lies purely on the light-cone.  We also introduce the definitions of $\delta_{+/-} (\sigma) $ and $\Theta_{+/-} \left( - \sigma \right)$ as
\begin{equation}
\delta_+\left(\sigma \left(x, \xpp \right)\right) = \begin{cases}
									\infty & x \text{ is on the future light-cone of } 
									\xpp \\
									0 & \text{elsewhere}
									\end{cases},
\end{equation}
\begin{equation}
\delta_-\left(\sigma \left(x, \xpp \right)\right) = \begin{cases}
									\infty & x \text{ is on the past light-cone of } \xpp 										\\
									0 & \text{elsewhere}
									\end{cases},
\end{equation}
\begin{equation}
\Theta_+ \left(- \sigma \left(x, \xpp \right) \right) = \begin{cases}
											1 & \sigma \left(x, \xpp \right) \in 
											I^+(x')\\
											0 & \text{elsewhere}
											\end{cases},
\Theta_- \left(- \sigma \left(x, \xpp \right) \right) = \begin{cases}
											1 & \sigma \left(x, \xpp \right) \in 
											I^-(x')\\
											0 & \text{elsewhere}
											\end{cases},
\end{equation}
where we can see $\delta_+\left(\sigma \left(x, \xpp \right)\right) + \delta_-\left(\sigma \left(x, \xpp \right)\right) = \delta\left(\sigma \left(x, \xpp \right)\right) $ and $\Theta_+ \left(- \sigma \left(x, \xpp \right) \right) + \Theta_- \left(- \sigma \left(x, \xpp \right) \right) = \Theta \left(- \sigma \left(x, \xpp \right) \right)$.
\newpage

Recalling the required attributes of $H^A{}_{B'} \left(x, \xpp \right)$, we examine those listed third and fourth earlier in this section, and using Eq.~\eqref{eqn: GretadvUV}, we have
\par \vspace{-6pt}\begin{IEEEeqnarray}{rCl}
H^A{}_{B'} \left(x, \xpp \right) &=& \begin{cases}
							U^A{}_{B'} \left(x, \xpp \right) \delta_{+} (\sigma) \text{\fixme{$-$}} V^A{}_{B'}  \left(x, \xpp \right) \Theta_{+} \left( - \sigma \right) & x \in I^+(x') \\
							U^A{}_{B'} \left(x, \xpp \right) \delta_{-} (\sigma) \text{\fixme{$-$}} V^A{}_{B'}  \left(x, \xpp \right) \Theta_{-} \left( - \sigma \right) & x \in I^-(x')
						\end{cases} \nonumber \\
						&=& \begin{cases}
							V^A{}_{B'}  \left(x, \xpp \right)  & x \in I^+(\xpp) \\
							V^A{}_{B'}  \left(x, \xpp \right)  & x \in I^-(\xpp) \\
						\end{cases},
\end{IEEEeqnarray}
where we note that $H^A{}_{B'} \left(x, \xpp \right) $ is also defined when $x$ and $\xpp$ are space-like related, however, there are only the first two constraints to impose in that domain.  We can therefore, set $H^A{}_{B'} \left(x, \xpp \right) = \text{\fixme{$-$}} V^A{}_{B'}  \left(x, \xpp \right)$ for all values of $\xpp$, provided $V^A{}_{B'}  \left(x, \xpp \right)$ has the required reciprocity relation and satisfies the homogeneous wave equation of Eq.~\eqref{eqn: waveGen}, which we will prove is the case in the following section.  

Substituting this result for $H^A{}_{B'} \left(x, \xpp \right) $ into our definitions for $G^A{}_{B'}{}_{\rm \sing} \left(x, \xpp \right) $ and $G^A{}_{B'}{}_{\rm \reg} \left(x, \xpp \right) $ of Eq.~\eqref{eqn: GreenSingRegCurve}, we arrive at expressions for both the regular and singlar Green's functions that are valid for all three cases, that is
\par \vspace{-6pt} \begin{IEEEeqnarray}{rCl}
G^A{}_{B'}{}_{\rm \sing} \left(x, \xpp \right) &=& \frac{1}{2} \left[U^A{}_{B'} \left(x, \xpp \right) \delta (\sigma) \text{\fixme{$+$}} V^A{}_{B'}  \left(x, \xpp \right) \Theta \left( \sigma \right) \right], \\
G^A{}_{B'}{}_{\rm \reg} \left(x, \xpp \right) &=& \frac{1}{2} U^A{}_{B'} \left(x, \xpp \right) \left[\delta_+ (\sigma) - \delta_- (\sigma) \right] \nonumber \\
&&
	\text{\fixme{$-$}}\: V^A{}_{B'}  \left(x, \xpp \right) \left[ \Theta_+ \left( - \sigma \right) + \frac{1}{2} 
	\Theta \left( \sigma \right)  \right],
\end{IEEEeqnarray}
where we have used 
$\Theta \left( - \sigma \right) - 1 = -\Theta \left( \sigma \right)$. 


\subsection{Restrictions on $U^A{}_{B'} \left(x, \xpp \right)$ and $V^A{}_{B'} \left(x, \xpp \right)$} \label{sec: restrictions}

In the previous section, we stated that the retarded and advanced Green's function can be written in the form,
\begin{equation}
G^A{}_{B'}{}_{\rm \ret / \adv} \left(x, \xpp \right) = U^A{}_{B'} \left(x, \xpp \right) \delta_{+/-} (\sigma) \text{\fixme{$-$}} V^A{}_{B'}  \left(x, \xpp \right) \Theta_{+/-} \left( - \sigma \right).
\end{equation}
By assuming this form, we can show how it is a suitable representation and deduce expressions and constraints on $U^A{}_{B'} \left(x, \xpp \right) $ and $V^A{}_{B'} \left(x, \xpp \right) $.  This is done by applying the operator $\mathcal{D}^A{}_B$ to the Green's function form and proving its viability in Eq.~\eqref{eqn: GreenGen}.

An immediate problem arises as distributions $\delta_{+/-} (\sigma)$ and $\Theta_{+/-} \left( - \sigma \right)$ are not defined at $x=\xpp$ and cannot be differentiated at this point.  Their true meaning becomes clear at the boundary value of a function in the complex plane which we may encapsulate in the prescription where we shift $\sigma \rightarrow \sigma + \epsilon $ and take the limit from the right as $\epsilon \rightarrow 0^+$.  This ensures that we are dealing with a positive $\epsilon$, which in turn demands that $x$ and $\xpp$ are time-like related, allowing us to take $\Theta_{+/-}' \left(-\sigma - \epsilon \right) = -\delta_{+/-} \left(\sigma + \epsilon \right)$, 
\fixme{where $'$ refers} to differentiating with respect to $\sigma$.  We label our shifted Green's functions as $G^{\epsilon}{}^B{}_{B'}{}_{\rm \ret / \adv}$ and notice that they satisfy $\lim_{\epsilon \to 0^+} G^{\epsilon}{}^B{}_{B'}{}_{\rm \ret / \adv} = G^B{}_{B'}{}_{\rm \ret / \adv}$.  We also note, by its definition, that $\sigma \delta \left(\sigma \right) = 0$ and by differentiating this with respect to $\sigma$ we also get the identities $\sigma \delta_{+/-}' \left(\sigma \right) = - \delta_{+/-} \left(\sigma \right)$ and $\sigma \delta_{+/-}'' \left(\sigma \right) = -2 \delta_{+/-}' \left(\sigma \right)$.  Adapting these to the shift that we have applied to  $\sigma$ gives
\par \vspace{-6pt} \begin{IEEEeqnarray}{rCl}
\sigma \delta_{+/-} \left(\sigma + \epsilon \right) &=&  - \epsilon \delta_{+/-} \left(\sigma + \epsilon \right),  \nonumber \\
\sigma \delta_{+/-}' \left(\sigma + \epsilon \right) &=& - \delta_{+/-} \left(\sigma + \epsilon \right) - \epsilon \delta_{+/-}' \left(\sigma + \epsilon \right), \nonumber \\
\sigma \delta_{+/-}'' \left(\sigma + \epsilon \right) &=& -2 \delta_{+/-}' \left(\sigma + \epsilon \right) - \epsilon \delta_{+/-}'' \left(\sigma + \epsilon \right).
\end{IEEEeqnarray}

Below for simplicity we drop the explicit $\left(x, \xpp \right)$ dependence notation and understand indices with $'$ imply a dependence on $\xpp$ and those without imply a dependence on $x$.  Applying $\mathcal{D}^A{}_B$, as defined in Eq.~\eqref{eqn: waveGen}, to the proposed shifted Green's function and recalling Eq.~\eqref{eqn: 2sigma}, gives
\par \vspace{-6pt} \begin{IEEEeqnarray}{rCl}
\mathcal{D}^A{}_B G^{\epsilon}{}^B{}_{B'}{}_{\rm \ret / \adv} &=&  \square G^{\epsilon}{}^A{}_{B'}{}_{\rm \ret / \adv}  - P^{A}{}_B  G^{\epsilon}{}^B{}_{B'}{}_{\rm \ret / \adv} \nonumber \\
&=& 
	g^{ab} \nabla_a \big[ \left(\nabla_b U^A{}_{B'} \right) \delta_{+/-} \left(\sigma + \epsilon 
	\right)  + U^A{}_{B'} \delta_{+/-}' \left(\sigma + \epsilon \right)  \sigma_b \nonumber \\
&& \quad
	\text{\fixme{$-$}}\: \nabla_b V^A{}_{B'} \Theta_{+/-} \left(-\sigma - \epsilon \right)  \text{\fixme{$+$}} V^A{}_{B'} 
	\delta_{+/-} \left(\sigma + \epsilon \right)  \sigma_b \big] \nonumber \\
&&
	-\: P^{A}{}_B \left[U^A{}_{B'} \delta_{+/-} \left(\sigma + \epsilon \right) 
	\text{\fixme{$-$}} V^A{}_{B'} \Theta_{+/-}\left(- \sigma - \epsilon \right) \right] \nonumber \\
&=&
	- 2 \epsilon  \delta_{+/-}'' \left(\sigma + \epsilon \right) U^A{}_{B'} \text{\fixme{$-$}} 2 \epsilon 
	\delta_{+/-}' \left(\sigma + \epsilon \right) V^A{}_{B'} \nonumber \\
&&
	+\: \delta_{+/-}' \left(\sigma + \epsilon \right) \left[ 2 \left(U^A{}_{B'} \right)_{;a} 
	\sigma^a + \left(\sigma^a{}_a - 4 \right) U^A{}_{B'} \right] \nonumber \\
&&
	+\: \delta_{+/-} \left(\sigma + \epsilon \right) \left[ \text{\fixme{$+$}} 2 \left(V^A{}_{B'}\right)_{;a} 
	\sigma^a \text{\fixme{$-$}} \left(2 - \sigma^a{}_a\right) V^A{}_{B'} +\mathcal{D}^A{}_B U^B{}_{B'} \right] 
	\nonumber \\
&&
	\text{\fixme{$-$}}\: \Theta_{+/-} \left( -\sigma - \epsilon \right) \mathcal{D}^A{}_{B} V^B{}_{B'}.
\end{IEEEeqnarray}
By taking the limit, we obtain $\mathcal{D}^A{}_B G^{\epsilon}{}^B{}_{B'}{}_{\rm \ret / \adv}$,
\par \vspace{-6pt} \begin{IEEEeqnarray}{rCl} \label{eqn: DG}
\mathcal{D}^A{}_B G^B{}_{B'}{}_{\rm \ret / \adv} &=&  \lim_{\epsilon \to 0^+} \mathcal{D}^A{}_B G^{\epsilon}{}^B{}_{B'}{}_{\rm \ret / \adv} \nonumber \\
&=&
	- 4 \pi \delta_4 \left(x, \xpp \right) U^A{}_{B'} + \delta_{+/-}' \left(\sigma \right) \left[ 
	2 \left(U^A{}_{B'} \right)_{;a} \sigma^a + \left(\sigma^a{}_a - 4 \right) U^A{}_{B'} \right] 
	\nonumber \\
&&
	+\:  \delta_{+/-} \left(\sigma \right) \left[ \text{\fixme{$+$}} 2 \left(V^A{}_{B'}\right)_{;a} \sigma^a \text{\fixme{$-$}} \left(2 
	- \sigma^a{}_a\right) V^A{}_{B'} +\mathcal{D}^A{}_B U^B{}_{B'} \right] \nonumber \\
&&
	\text{\fixme{$-$}}\: \Theta_{+/-} \left( -\sigma \right) \mathcal{D}^A{}_{B} V^B{}_{B'} \nonumber \\
&=&
	-4 \pi g^A{}_{B'} \left(x, x'\right) \delta_4 \left(x, x' \right),
\end{IEEEeqnarray}
where the final equality is from equating the right hand side to that of  Eq.~\eqref{eqn: GreenGen}.  We have also made use of the identities derived in \cite{Poisson:2003},
\par \vspace{-6pt} \begin{IEEEeqnarray}{rCl}
\lim_{\epsilon \to 0^+} \epsilon \delta_{+/-} \left(\sigma + \epsilon \right) &=& 0, \nonumber \\
\lim_{\epsilon \to 0^+} \epsilon \delta_{+/-}' \left(\sigma + \epsilon \right) &=& 0, \nonumber \\
\lim_{\epsilon \to 0^+} \epsilon \delta_{+/-}'' \left(\sigma + \epsilon \right) &=& 2 \pi \delta_4 \left(x, \xpp \right).
\end{IEEEeqnarray}
Comparing the final two equalities of Eq.~\eqref{eqn: DG} and equating coefficients of the distribution functions, we can immediately infer that
\begin{equation}
\delta_4 \left(x, \xpp \right) U^A{}_{B'} = g^A{}_{B'} \left(x, x'\right) \delta_4 \left(x, x' \right).
\end{equation}
As this only gives us information about $U^A{}_{B'}$ when $x = \xpp$, also known as the coincidence limit and denoted by $\left[U^A{}_{B'} \right]$, we have
\begin{equation} \label{eqn: uCoincidence}
\left[U^A{}_{B'} \right] = \left[g^A{}_{B'} \right] = \delta^A{}_{B'}.
\end{equation}

From Eqs.~\eqref{eqn: deltaSigma} and \eqref{eqn: 4sigma}, we use the identities $dx^a = (\sigma^a / \lambda ) d \lambda $ and  \\ 
$\Delta^{-1} \left(x, \xpp \right)  \left( \Delta \left(x, \xpp \right)  \sigma^a \right)_{;a} = 4$ to produce,
\begin{equation}
 \sigma^a{}_a + \Delta^{-1} \sigma^a \Delta_{;a} = 4 \Rightarrow  \sigma^a{}_a - 4 = -\Delta^{-1} \lambda \frac{d \Delta}{d \lambda} = - \lambda \frac{d}{d \lambda} \ln \Delta,
\end{equation}
where $\lambda$ affinely parameterises the geodesic connecting $x$ to $\xpp$.  The coefficient of  $\delta_{+/-}' \left(\sigma \right)$ in Eq.~\eqref{eqn: DG} is to zero to give us
\begin{equation}
2 \left(U^A{}_{B'} \right)^{-1} \left(U^A{}_{B'} \right)_{;a} \sigma^a + \left(\sigma^a{}_a - 4 \right) = \lambda \frac{d}{d \lambda} \left( 2 \ln U^A{}_{B'} - \ln \Delta \right) = 0.
\end{equation}
We, therefore, know that $\left( U^A{}_{B'} \right)^2 / \Delta$ is constant on the geodesic connecting $x$ to $\xpp$.  As it is constant we can use our values in the coincidence limit, namely Eq.~\eqref{eqn: uCoincidence} and $[\Delta] =1$ from Sec.~\ref{sec: VV}, to obtain an expression for $U^A{}_{B'}$,
\begin{equation} \label{eqn: UGen}
U^A{}_{B'} \left(x, \xpp \right) = \Delta^{\tfrac{1}{2}} \left(x, \xpp \right)  g^A{}_{B'} \left(x, \xpp \right) .
\end{equation}

Due to the nature of the delta function, the coefficient of $ \delta_{+/-} \left(\sigma \right)$ in Eq.~\eqref{eqn: DG} only has to be zero when $\sigma = 0$.  We therefore have for $\sigma = 0$
\begin{equation}\label{eqn: VR1}
\left[\text{\fixme{$+$}} 2 \left(V^A{}_{B'}\right)_{;a} \sigma^a \text{\fixme{$-$}} \left(2 - \sigma^a{}_a\right) V^A{}_{B'} +\mathcal{D}^A{}_B U^B{}_{B'} \right]_{\sigma = 0} = 0,
\end{equation}
which can be rearranged to give
\begin{equation} \label{eqn: VDerived}
2 \left(V^A{}_{B'}\right)_{;a} \sigma^a - 2 V^A{}_{B'} - \Delta^{-1} \sigma^a \Delta_{,a} V^A{}_{B'} +\mathcal{D}^A{}_B U^B{}_{B'} = 0,
\end{equation}
where we have used Eq.~\eqref{eqn: 4sigma} rearranged as $2-\sigma^a{}_a = \Delta^{-1} \sigma^a \Delta_{,a}$.  This will be later used to obtain expressions for $V^A{}_{B'}$ as described in Sec.~\ref{sec:SingularFieldExpansion} and Appendices \ref{sec:CoordinateExpansions} and \ref{sec:ExpansionCoefficients}

Similarly, due to the nature of $\Theta_{+/-} \left( -\sigma \right)$, we also have 
\begin{equation} \label{eqn: DV}
\mathcal{D}^A{}_{B} V^B{}_{B'} = 0,
\end{equation}
when $x \in I^+(\xpp) \cup I^-(\xpp)$, i.e., when $x$ and $x'$ are time-like related.  We should note that Eq.~\eqref{eqn: DV} is one of the requirements (that $H^A{}_{B'}$ satisfies the homogeneous wave equation), described in the previous section, that is  necessary for setting $H^A{}_{B'} = \text{\fixme{$-$}}V^A{}_{B'}$, which we have now shown.  As we have established an expressions for $U^A{}_{B'}$ which has reciprocity property, as does the retarded and singular Green's functions, it also follows from the definition of Eq.~\eqref{eqn: GretadvUV} that $V^A{}_{B'}$ will also have the required reciprocity, i.e., $V^A{}_{B'} \left(x, \xpp \right) = V^{A'}{}_B \left( \xpp, x \right)$.  We have therefore fulfilled all requirements that were necessary for setting $H^A{}_{B'} = \text{\fixme{$-$}} V^A{}_{B'}$, which leaves us with the singular Green's function,
\begin{equation} \label{eqn: GS}
G^A{}_{B'}{}_{\rm \sing} \left(x, \xpp \right) = \frac{1}{2} \left[U^A{}_{B'} \left(x, \xpp \right) \delta (\sigma) \text{\fixme{$+$}} V^A{}_{B'}  \left(x, \xpp \right) \Theta \left( \sigma \right) \right],
\end{equation}
where $U^A{}_{B'}$ is defined by Eq.~\eqref{eqn: UGen} and $V^A{}_{B'}$ satisfies Eqs.~\eqref{eqn: VR1} and \eqref{eqn: DV}.


\subsection{The Singular Field}
The singular field, by design, solves the inhomogeneous wave equation of \\
Eq.~\eqref{eqn: waveGen}, this means like the retarded field, it can also be written in the form of Eq.~\eqref{eqn: varphi eqn},
\begin{equation} 
\varphi^A{}_{\rm \sing} = \int_{\gamma} G^A{}_{B'}{}_{\rm \sing} \left(x, x' \right) \mathcal{M}^{B'} (x') \sqrt{-g'} d^4 x'
\end{equation}
Substituting $\mathcal{M}^{B'} (x') $ from Eq.~\eqref{eqn: MGen} and integrating over $\xpp$ gives
\begin{equation} \label{eqn: phiS eqn}
\varphi^A{}_{\rm \sing} = \mathcal{Q} \int_{\gamma} G^A{}_{B'}{}_{\rm \sing} \left(x, x' \right) u^{B'} (x') d \tau'.
\end{equation}
Recalling Fig. ~\ref{fig: GsingCurve}, we note that the singular Green's function is only supported in between the two points $x_{\rm \ret}$ and $x_{\rm \adv}$ on $\gamma$.  Combining this with our definition of the singular Green's function, Eq.~\eqref{eqn: GS}, allows us to rewrite Eq.~\eqref{eqn: phiS eqn} as
\par \vspace{-6pt} \begin{IEEEeqnarray}{rCl}
\varphi^A{}_{\rm \sing} &=& \mathcal{Q} \int_{x_{\rm \ret}}^{x_{\rm \adv}} G^A{}_{B'}{}_{\rm \sing} \left(x, x' \right) u^{B'} (x') d \tau' \nonumber \\
&=&
	\frac{\mathcal{Q}}{2} \int_{x_{\rm \ret}}^{x_{\rm \adv}} U^A{}_{B'} \left(x, \xpp \right) \delta \left( \sigma \left(x, \xpp 	\right) \right) u^{B'} (x') d \tau' \nonumber \\
&&
	\text{\fixme{+}}\: \frac{\mathcal{Q}}{2} \int_{x_{\rm \ret}}^{x_{\rm \adv}} V^A{}_{B'} \left(x, \xpp \right) 
	\Theta \left( \sigma \left(x, \xpp \right) \right) u^{B'} (x') d \tau' \nonumber \\
&=&
	\frac{\mathcal{Q}}{2} \left[ \frac{U^A{}_{B'} \left(x, \xpp \right) u^{B'} (x')}{\sigma_{a'} 
	\left(x, \xpp \right) u^{a'} (\xpp)}\right]^{\xpp = x_{\rm \adv}}_{\xpp =x_{\rm \ret}} 
	\nonumber \\
&&
	\text{\fixme{+}}\: \frac{\mathcal{Q}}{2} \int_{x_{\rm \ret}}^{x_{\rm \adv}} V^A{}_{B'} \left(x, \xpp \right) 
	 u^{B'} (x') d \tau'
\end{IEEEeqnarray}
where the integration of the first term follows from $d \sigma = \sigma_{a'} u^{a'} d \tau$ with the delta function forcing $U^A{}_{B'} \left(x, \xpp \right) u^{B'} (x')$ onto the light cone, i.e., $U^A{}_{B'} \left(x, \xpp \right) u^{B'} (x')$ is only non-zero for $\xpp = x_{\rm \ret}$ and $\xpp = x_{\rm \adv}$.  The second term simply has the $\Theta \left( \sigma \left(x, \xpp \right) \right)=1$ as its only integrating over that part of $\gamma$ that is space-like.


\subsection{The Self-Force and the Singular Field} \label{sec: selfForceSing}

We have seen that in an appropriate gauge, the retarded field, $\varphi^{A}(x)$, of an arbitrary point particle satisfies the inhomogeneous wave equation with a distributional source,
\begin{equation}
\label{eq:Wave}
\mathcal{D}^{A}{}_B \varphi^{B} = - 4\pi \mathcal{Q} \int u^{A} \delta_4 \left( x,z(\tau') \right) d\tau',
\end{equation}
where
\begin{equation}
\label{eq:Wave-operator}
\mathcal{D}^{A}{}_B  = \delta^{A}{}_B (\square - m^2) - P^{A}{}_B.
\end{equation}
The retarded solutions to this equation gives rise to a field, known as the retarded field which one might na\"ively expect to exert a self-force
\begin{equation}\label{eqn: FaCases}
F^a = p^a{}_A \varphi^{A}_{\rm{\ret}},
\end{equation}
on the particle, where $p^a{}_A(x)$ is a tensor at $x$, which depends on the type of charge.

In the previous sections we illustrated how Detweiler and Whiting \cite{Detweiler-Whiting-2003} showed how such a singular field can be
constructed through a Green function decomposition. In four spacetime dimensions and within a normal
neighbourhood, the Green function for the retarded/advanced solutions to Eq.~\eqref{eq:Wave}
may be given in Hadamard form,
\begin{eqnarray}
\label{eq:Hadamard}
G_{\mathrm{\ret/\adv}}{}^{A}{}_{B'}\left( x,x' \right) &=& \theta_{-/+} \left( x,x' \right) \big\lbrace U^{A}{}_{B'} \left( x,x' \right) \delta \left[ \sigma \left( x,x' \right) \right] \nonumber \\
&&- V^{A}{}_{B'} \left( x,x' \right) \theta \left[ - \sigma \left( x,x' \right) \right] \big\rbrace ,
\end{eqnarray}
where 
$U^{A}{}_{B'}\left( x,x' \right)$ and $V^{A}{}_{B'}\left( x,x' \right)$ are symmetric bi-spinors/tensors
which are regular for $x' \rightarrow x$, defined by Eqs.~\eqref{eqn: UGen} and \eqref{eqn: VDerived} respectively. 
The first term here, involving $U^{A}{}_{B'} \left( x,x' \right)$,
represents the \emph{direct} part of the Green function 
while the second term, involving $V^{A}{}_{B'} \left( x,x' \right)$, is known as the
\emph{tail} part of the Green function.

Detweiler and Whiting proposed to define a \emph{singular} Green function by taking the symmetric
Green function, $G_{\mathrm{(sym)}}{}^{A}{}_{B'} = \frac12(G_{\mathrm{\ret}}{}^{A}{}_{B'} + G_{\mathrm{\adv}}{}^{A}{}_{B'})$
and adding $V^{A}{}_{B'}\left( x,x' \right)$ (a homogeneous solution to Eq.~\eqref{eq:Wave}). This
leads to the previously defined singular Green function,
\begin{equation}
\label{eq:Hadamard-singular}
G_{\mathrm{\sing}}{}^{A}{}_{B'}\left( x,x' \right) =
  \frac12\left\lbrace U^{A}{}_{B'} \left( x,x' \right) \delta \left[ \sigma \left( x,x' \right) \right]
  + V^{A}{}_{B'} \left( x,x' \right) \theta \left[\sigma \left( x,x' \right) \right] \right\rbrace .
\end{equation}
Note that this has support on and \emph{outside} the past and future light-cone 
(i.e. for points $x$ and $x'$ spatially separated)
and is only uniquely defined provided $x$ and $x'$ are within a convex normal neighbourhood. 
Given this singular Green function, we may define the Detweiler-Whiting singular field,
\begin{equation} \label{eq:SingularField}
\varphi^{A}_{\rm \sing} = \int_{\tau_{\rm \adv}}^{\tau_{\rm \ret}} G_{\mathrm{\sing}}{}^{A}{}_{B'}\left( x, z(\tau') \right) u^{B'} d\tau',
\end{equation}
which also satisfies Eq.~\eqref{eq:Wave}. 
Subtracting this singular field from the retarded field, we obtain the \emph{regularized} field,
\begin{equation} \label{eqn: singRegSplitPhi}
\varphi^{A}_{\rm{\reg}} = \varphi^{A}_{\rm{\ret}} - \varphi^{A}_{\rm{\sing}},
\end{equation}
which Detweiler and Whiting showed gives the correct finite physical self-force,
\begin{equation} \label{eqn:SelfForce}
F^a = p^a{}_A \varphi^{A}_{\rm{\reg}}.
\end{equation}
Moreover, this regularized field is a solution of the homogeneous wave equation,
\begin{equation} \label{eqn: varphiReg}
\mathcal{D}^{A}{}_B \varphi^{B}_{\rm{\reg}} = 0.
\end{equation}
This holds independently of whether one is considering a scalar or electromagnetically charged
point particle or a  point mass. To make this more explicit, in the following subsections
we give the form these expressions take in each of scalar, electromagnetic and gravitational cases.

\subsubsection*{Scalar Case}
In the scalar case the singular field, $\PhiS$, is
a solution of the inhomogeneous scalar wave equation,
\begin{equation}
(\Box - \xi R - m^2) \PhiS = \text{\fixme{$- 4 \pi$}} q \int \delta_4(x,z(\tau)) d\tau,
\end{equation}
where $q$ is the scalar charge and $\xi$ is the coupling to the background scalar curvature.
An expression for $\PhiS$ may be found by considering the scalar Green function (obtained
by taking $U^{A}{}_{B'} = U(x,x')$ in Eq.~\eqref{eq:Hadamard-singular}),
\begin{equation}\label{eqn:Gs}
G^{\rm \sing} = \frac{1}{2}\left\{U(x,x') \delta[\sigma(x,x')] + V(x,x') \theta[\sigma(x,x')]\right\},
\end{equation}
with $U(x,x') = \Delta^{1/2}(x,x')$ from Eqs.~\eqref{eqn: GgGen} and \eqref{eqn: UGen}, where $\Delta^{1/2}(x,x')$ is the Van Vleck determinant as defined in Eq.~\eqref{eqn: vanvleck}.  This Green function is a solution of the equation
\begin{equation}
(\Box - \xi R - m^2)  G^{\rm \sing} = -4\pi \delta_4(x,x').
\end{equation}
Given this expression for the Green function, the scalar singular
field is
\begin{eqnarray} \label{eq:PhiS}
\PhiS(x) &=& q\int_\gamma G^{\rm \sing} (x,z(\tau)) d\tau \nonumber \\
 &=& \frac{q}{2} \Bigg[ \frac{U(x,x')}{\sigma_{c'} u^{c'}} \Bigg]_{x'=x_{\rm \ret}}^{x'=x_{\rm \adv}}
   + \frac{q}{2} \int_{\tau_{\rm \ret}}^{\tau_{\rm \adv}} V(x,z(\tau)) d\tau
\end{eqnarray}
and one computes the scalar self-force from the regularized scalar field
$\Phi^{\rm \reg} = \Phi^{\rm{\ret}} - \PhiS$ as
\begin{equation} \label{eqn:SelfForceScalar}
F^a = q \, g^{ab} \Phi^{\rm{\reg}}_{, b}.
\end{equation}

\subsubsection*{Electromagnetic Case}
In Lorenz gauge, the electromagnetic singular field satisfies the equation
\begin{equation}
\Box A_a^{\rm \sing} - R_a{}^b A_b^{\rm{\sing}}= -4\pi e \int g_{ab} \left( x, z (\tau) \right) u^{b}  \delta_4(x,z(\tau)) d\tau,
\end{equation}
where $e$ is the electric charge.
An expression for $A_a^{\rm \sing}$ may be found by considering the electromagnetic Green
function (obtained by taking $U^{A}{}_{B'} = U(x,x')_{aa'}$ in Eq.~\eqref{eq:Hadamard-singular}),
\begin{equation}
G^{\mathrm{\sing}}_{aa'}(x,x') =
\frac{1}{2} \left\lbrace U(x,x')_{aa'} \delta \left( \sigma (x,x') \right) + V(x,x')_{aa'} \theta \left( \sigma (x,x') \right) \right\rbrace,
\end{equation}
with $U^a{}_{a'} = \Delta^{1/2} g^a{}_{a'}$ from Eqs.~\eqref{eqn: GgGen} and \eqref{eqn: UGen}, where $g^a{}_{a'}$ is the bi-vector of parallel transport defined in Eq.~\eqref{eqn: gSigma}.  This Green function is a solution of the equation
\begin{equation}
\Box G_{aa'}^{\rm \sing} - R_a{}^b G_{ba'}^{\rm{\sing}}= - 4\pi g_{aa'} (x, x') \delta_4(x,x').
\end{equation}
Given this expression for the Green function, the electromagnetic singular field is
\begin{align} \label{eq:AS}
A_{a}^{\rm \sing} &= e \int_\gamma G^{\rm \sing}_{ab} (x, z(\tau)) u^{b} (z(\tau)) d\tau \nonumber \\
&= \frac{e}{2} \Bigg[\frac{u^{a'} U_{aa'}(x,x')}{\sigma_{c'} u^{c'}} \Bigg]_{x'=x_{\rm \ret}}^{x'=x_{\rm \adv}}
  + \frac{e}{2} \int_{\tau_{\rm \ret}}^{\tau_{\rm \adv}} V_{ab}(x,z(\tau)) u^{b} (z (\tau)) d\tau.
\end{align}
One computes the electromagnetic self-force from the electromagnetic regular
field,
$A_a^{\rm \reg} = A_a^{\rm{\ret}} - A_a^{\rm{\sing}}$, as
\begin{equation} \label{eqn:SelfForceEM}
F^a = e \, g^{ab} u^c A^{\rm{\reg}}_{[c, b]}.
\end{equation}

\subsubsection*{Gravitational Case}
In Lorenz gauge, the trace-reversed singular first order metric perturbation satisfies the equation
\begin{equation}
\Box \hb_{ab}^{\rm \sing} + 2 C_a{}^c{}_b{}^d \hb_{cd}^{\rm{\sing}}= -16\pi m \int g_{a'(a} u^{a'} g_{b)b'} u^{b'}  \delta_4(x,x') d\tau,
\end{equation}
where $\mu$ is the mass of the particle and the trace-reversed singular field is related to the
non-trace-reversed version by $\hb_{ab}^{\rm \sing} = h_{ab}^{\rm \sing} - \frac12 h^{\rm \sing} g_{ab}$
with $h^{\rm \sing} =  g^{ab} h^{\rm \sing}_{ab}$.
An expression for $\hb_{ab}^{\rm \sing}$ may be found by considering the gravitational Green
function (obtained by taking $U^{A}{}_{B'} = U(x,x')_{a b a' b'}$ in
Eq.~\eqref{eq:Hadamard-singular}),
\begin{equation}
G^{\mathrm{\sing}}_{a b a' b'}(x,x') =
\frac{1}{2} \left\lbrace U(x,x')_{a b a' b'} \delta \left[ \sigma (x,x') \right] + V(x,x')_{a b a' b'} \theta \left[ \sigma (x,x') \right] \right\rbrace,
\end{equation}
with $U^{ab}{}_{a'b'} = \Delta^{1/2} g^{(a}{}_{a'} g^{b)}{}_{b'}$ from Eqs.~\eqref{eqn: GgGen} and \eqref{eqn: UGen}. 
This Green function is a solution of the equation
\begin{equation}
\Box G_{aba'b'}^{\rm \sing} + 2 C_a{}^p{}_b{}^q G_{pqa'b'}^{\rm{\sing}}= - 4\pi g_{a'(a}g_{b)b'} \delta_4(x,x').
\end{equation}
Given this expression for the Green function, the trace-reversed singular first order metric
perturbation is
\begin{align} \label{eq:hS}
\hb_{a b}^{\rm \sing} &= 4 \mu \int_\gamma G^{\rm \sing}_{a b a' b'} (x, z(\tau')) u^{a'} u^{b'} d\tau' \nonumber \\
&= 2 \mu \Bigg[\frac{u^{a'} u^{b'} U_{aba'b'}(x,x')}{\sigma_{c'} u^{c'}} \Bigg]_{x'=x_{\rm \ret}}^{x'=x_{\rm \adv}}
  + 2 \mu \int_{\tau_{\rm \ret}}^{\tau_{\rm \adv}} V_{aba'b'}(x,z(\tau)) u^{a'} u^{b'} d\tau.
\end{align}
One computes the gravitational self-force from the regularized trace-reversed singular first order metric
perturbation,
$\hb_{ab}^{\rm \reg} = \hb_{ab}^{\rm{\ret}} - \hb_{ab}^{\rm{\sing}}$, as
\begin{equation}\label{eqn:SelfForceGravityBasic}
F^a = \mu \, k^{abcd} \hb^{\rm{\reg}}_{bc; d},
\end{equation}
where
\begin{equation}
k^{abcd} \equiv \frac12 g^{ad} u^b u^c - g^{ab} u^c u^d - 
  \frac12 u^a u^b u^c u^d + \frac14 u^a g^{b c} u^d + 
 \frac14 g^{a d} g^{b c}.
\end{equation}


\chapter{High-Order Expansions of the Singular Field} \label{sec: highOrder} 



\section{Coordinate and Covariant Expansions of Fundamental Bitensors} \label{sec:SingularFieldExpansion}
In the previous chapter, we gave expressions for the singular field in terms of the bitensors $U^{A}{}_{B'} \left( x,x' \right)$ and $V^{A}{}_{B'} \left( x,x' \right)$. The first of these is given by 
\begin{equation} \label{eq: U}
U^{AB'} \left( x,x' \right) = \Delta^{1/2} \left( x,x' \right) g^{AB'}\left( x,x' \right),
\end{equation}
where $\Delta \left( x,x' \right)$ is the Van Vleck determinant from Eq.~\eqref{eqn: vanvleck},
\begin{eqnarray} \label{eq:vanVleckDefinition}
\Delta \left( x,x' \right) &=& - \left[ -g \left( x \right) \right] ^{-1/2} \det \left[ -\sigma _{;		\alpha \beta '} \left( x,x' \right) \right] \left[ -g \left( x' \right) \right] ^{-1/2}  				\nonumber \\
&=& \det \left[ -g^{\alpha'}{}_\alpha \left( x,x' \right) \sigma ^{;\alpha}{}_{ \beta '} \left( x,x' \right) \right],
\end{eqnarray}
$g^{AB'}$ is the bi-tensor of parallel transport appropriate to the tensorial nature of the field defined in Eq.~\eqref{eqn: GgGen}, e.g.,
\begin{equation}
 g^{A B'} = \begin{cases}
             1 & \text{(scalar)}\\
             g^{a b'} & \text{(electromagnetic)} \\
             g^{a' (a} g^{b) b'} & \text{(gravitational)}
            \end{cases},
\end{equation}
and where the higher spin fields are taken in Lorentz gauge. Here, $g^{\alpha'}{}_\alpha\left( x,x' \right)$ is the bivector of parallel transport
defined by the transport equation
\begin{equation}
\sigma^{\alpha} g_{a b' ;\alpha}  = 0 = \sigma^{\alpha'} g_{a b' ;\alpha'}.
\end{equation}
The bitensor $V^{AB'}\left( x,x' \right)$ may be expressed in terms of a formal expansion in increasing powers of $\sigma$ \fixme{\cite{DeWitt:1960, Decanini:Folacci:2005a}}:
\begin{equation} \label{eq:V}
V^{AB'}\left( x,x' \right) = \sum_{\num =0}^{\infty} V_{\num}{}^{AB'}\left( x,x' \right) \sigma ^{\num}\left( x,x' \right),
\end{equation}
where the coefficients $V_{\num}^{AB'}\left( x,x' \right)$ satisfy the recursion relations
\begin{subequations}
\label{eq:RecursionV}
\begin{align}
\label{eq:recursionVn}
  \sigma ^{;\alpha'} (\Delta ^{-1/2} V^{AB'}_{\num})_{;\alpha'} + \left( \num +1 \right)  \Delta ^{-1/2}  V_{\num}^{AB'} + {\frac{1}{2 \num}} \Delta ^{-1/2}  \mathcal{D}^{B'}{}_{C'} V^{AC'}_{\num -1} = 0 ,
\end{align}
for $\num \in \mathbb{N}$, along with the `initial condition'
\begin{eqnarray}
\label{eq:recursionV0}
\sigma ^{;\alpha'} (\Delta ^{-1/2} V_0^{AB'}){}_{;\alpha'} + \Delta ^{-1/2} V^{AB'}_0 + {\frac{1}{2}}\Delta ^{-1/2} \mathcal{D}^{B'}{}_{C'} ( \Delta ^{1/2} g^{AC'}) &=& 0,
\end{eqnarray}
\end{subequations}
derived from Eq.~\eqref{eqn: VDerived}.

\begin{figure}
\begin{center}
\includegraphics[scale=0.6]{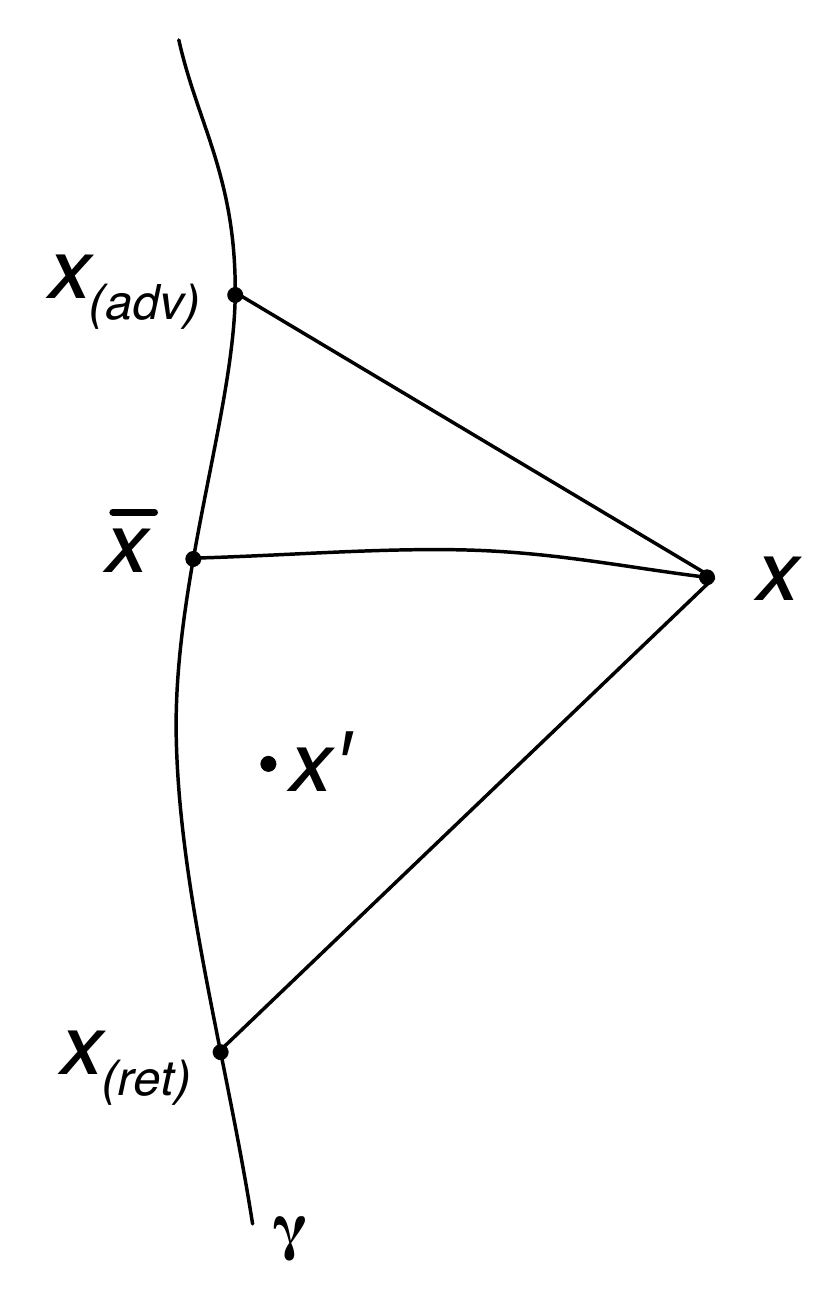}
\caption[Worldline]{We expand all bitensors (which are functions of $x$ and $x'$)
about the arbitrary point $\xb$ on the world line.}
\label{fig:SingularField}
\end{center}
\end{figure}

Looking at the above equations for $U^{AB'}\left( x,x' \right)$ and $V_{\num}{}^{AB'}\left( x,x' \right)$, we see that a key component of the present work involves the computation of several fundamental bitensors, in particular, the world function $\sigma (x, x')$, Van Vleck determinant $\Delta^{1/2}(x,x')$, four-velocity $u^{a}(x)$, and bivector of parallel transport $g_a{}^{b'}(x,x')$. This may be achieved by expressing them as expansions about some arbitrary point $\xb$ which is close to $x$ and $x'$ as shown in figure \ref{fig:SingularField}.  We derive these here using both covariant and coordinate methods, each of which has its own advantages and disadvantages.  

The covariant expression is more elegant, allowing for compact formulae; however these formulae hide complex terms such as high order derivatives of the Weyl tensor that quickly become extremely time consuming to compute, even using computer tensor algebra packages such as GRTensorII \cite{GRTensor} or xAct \cite{xTensorOnline}.  The coordinate approach is less elegant but more practical for explicit calculations and it avoids the need to use tensor algebra. Independently of the approach taken, these expansions may be used to compute expansions of $U^{AB'}\left( x,x' \right)$ and $V_{\num}{}^{AB'}\left( x,x' \right)$ (by substituting into the above equations), and hence of the singular field. In the case
of covariant expansions, for explicit calculations one must further expand the covariant
expressions in coordinates, yielding an expression which may be directly compared with
those obtained from the coordinate approach. The resulting expressions are long
but are explicit functions of the coordinates, enabling them to be transformed directly into,
for example, C functions, indeed we give them in such form online \cite{BarryWardell.net}.

\subsection{Coordinate Approach} \label{sec:Coordinate}

In this section, we describe our method for obtaining coordinate expressions of the biscalars in Eqs.~\eqref{eq:PhiS}, \eqref{eq:AS} and \eqref{eq:hS}.  We will start by considering two arbitrary points $x$ and $x'$ near $\xb$ as shown in Fig.~\ref{fig:SingularField}.  We will seek expansion where the coefficients are evaluated at $\xb$, so we introduce the notation
\begin{equation} \label{eqn:deltaxprime}
\Delta x^a = x^a - x^\ab , \qquad \delta x^{a'} = x^{a'} - x^a = x^{a'} - \Delta x^a - x^{\bar{a}},
 \end{equation}
where we use the convention that the index carries the information about the point: $\xb^a = x^\ab$.  
In the calculations below $\Delta x^{a}$ and $\delta x^{a'}$
are both assumed to be small, of order $\epsilon$.

The first item we require for our calculations is a coordinate expansion of the biscalar $\sigma \left( x,x' \right)$, the Synge world function.  We start with a standard coordinate series expansion about $x$, see for example \cite{Ottewill:Wardell:2009} (note the difference in convention for $\Delta x^a$ in that paper), to get
\begin{equation} \label{eqn:sigma}
\sigma(x,x') = \tfrac{1}{2} g_{a b}(x) \delta x^{a'} \delta x^{b'} + A_{abc}(x) \delta x^{a'} \delta x^{b'} \delta x^{c'} + B_{abcd}(x) \delta x^{a'} \delta x^{b'} \delta x^{c'} \delta x^{d'} 
+ \cdots.
\end{equation}
The coefficients are readily determined in terms of derivatives of the metric at $x$ by use of the defining identity derived in Eq.~\eqref{eqn: 2sigma}, $2 \sigma = \sigma_{a'} \sigma^{a'}$, see \cite{Ottewill:Wardell:2009}. To be explicit, the first few are given by
\begin{align*}
A_{a b c}(x) =
&
 \frac{1}{4}g_{(a b ,c)}(x), \\
B_{a b c d}(x)=
&
 \frac{1}{12}g_{(a b, c d)}(x) - \frac{1}{24}  g^{p q}(x) \big[g_{(a b ,|p|}(x) g_{c d) ,q}(x) -12   g_{(a b ,|p|} (x)g_{|q| c , d)}(x) \\
&
+ 36    g_{p (a, b} (x)g_{|q| c , d)}(x) \big] .
\end{align*}

We now go one step further by expanding the coefficients about $\xb$  to give a double expansion in $\Delta x^a$ and $\delta x^{a'}$ with coefficients at $\xb$. The first few terms are
\begin{align}
\sigma(x,x') =
&
\tfrac{1}{2} g_{\ab \bb}(\xb) \delta x^{a'}  \delta x^{b'} + \left[\tfrac{1}{2} g_{\ab \bb,\cb }(\xb) \delta x^{a'}  \delta x^{b'}  \Delta x^c+A_{\ab \bb \cb}(\xb)\delta x^{a'} \delta x^{b'}  \delta x^{c'}  \right]\nonumber \\
&
+ \Big[ \tfrac{1}{4} g_{\ab \bb,\cb\db }(\xb) \delta x^{a'}  \delta x^{b'} \Delta x^c \Delta x^d 	+A_{\ab \bb \cb,\db}(\xb)\delta x^{a'} \delta x^{b'}  \delta x^{c'}  \Delta x^{d} \nonumber \\
&
+B_{\ab \bb \cb \db}(\xb) \delta x^{a'} \delta x^{b'} \delta x^{c'}  \delta x^{d'}  \Big]+ O(\epsilon^5),
\end{align}
where now we interpret $\delta x^{a'}$ as   $x^{a'} - \Delta x^a - x^{\bar{a}}$ and we use square brackets to distinguish terms of different order in $\epsilon$.  Rather than disturb the flow here and throughout this section we just give the first few terms of each expansion for a general metric to make the structure clear and give explicit expressions in \Sch space-time to much higher order in Appendix \ref{sec:CoordinateExpansions}.

Now that as the coefficients are at the fixed point $\xb$, it is straightforward to take derivatives of $\sigma$ at $x$ and $\xp$, for example,
\begin{align}
\sigma_{{a'}} =
&   
g_{\ab \bb} \delta x^{b'}  + \left[ g_{\ab \bb,\cb } \delta x^{b'}  \Delta x^c
 +3 A_{\ab \bb \cb}\delta x^{b'}  \delta x^{c'} \right] \nonumber\\
& 
+\left[ \tfrac{1}{2} g_{\ab \bb,\cb\db }\delta x^{b'} \Delta x^c \Delta x^d +3 A_{\ab \bb \cb,\db}\delta x^{b'} \delta x^{c'} \Delta x^{d} + 4B_{\ab \bb \cb \db}\delta x^{b'} \delta x^{c'}  \delta x^{d'} \right]+O(\epsilon^4), \label{eqn: sigmaaprime} \\ 
\sigma_{a} =
&
 -\sigma_{{a'}} +\tfrac{1}{2} g_{ \bb \cb ,\ab } \delta x^{b'}  \delta x^{c'}  + \left[ \tfrac{1}{2} g_{ \bb\cb,\db\ab  } \delta x^{b'} \delta x^{c'} \delta x^{d'} +A_{ \bb \cb \db \ab}\delta x^{b'} \delta x^{c'} \delta x^{d'} \right]+O(\epsilon^4), \\
\sigma_{{a'}b} =
&
  - g_{\ab \bb} - g_{\ab\bb,\cb}  \Delta x^c +\left[ ( 3 A_{\ab \cb \db, \bb}-12 B_{\ab \bb \cb \db})\delta x^{c'}  \delta x^{d'}  -\tfrac{1}{2} g_{\ab\bb,\cb\db}\Delta x^{c}  \Delta x^{d}\right] +O(\epsilon^3).
\label{eqn: sigmaalphaprime}
\end{align}
Likewise we can calculate the Van Vleck determinant directly from its definition Eq.~\eqref{eqn: vanvleck},
\begin{equation} \label{eqn: Van Vleck}
\Delta^{\frac{1}{2}} (x,x') = \left(-\left[-g(x)\right]^{- \frac{1}{2}} \left|- \sigma_{a' b} (x,x')\right| \left[-g(x')\right]^{- \frac{1}{2}}\right)^{\frac{1}{2}},
\end{equation}
giving 
\begin{align} \label{eqn: Van Vleck 2}
\Delta^{\frac{1}{2}} (x,x') =
& 
1 + \Bigl\{\tfrac{1}{2}g^{\cb\db}(2g_{\ab\cb,\bb\db}-g_{\ab\bb,\cb\db}-g_{\cb\db,\ab\bb} )\nonumber \\
&
	+\:  \tfrac{1}{4}g^{\cb\db}g^{\eb\fb}\bigl[g_{\cb\eb,\ab}g_{\db\fb,\bb}+2g_{\cb\ab,\bb}(g_{\eb\fb,\db} 
	- 2g_{\db\eb,\fb})
\nonumber \\
&
	+\: 2g_{\ab\cb,\eb}(g_{\bb\db,\fb}-g_{\bb\fb,\db})
-g_{\ab\bb,\cb}(g_{\eb\fb,\db}-2g_{\db\eb,\fb}) \bigr]\Bigr\}\delta x^a\delta x^b + O(\epsilon^3).
\end{align}

To obtain an expressions at $x_\textrm{\adv}$ and $x_\textrm{\ret}$, we allow $x^{a'}$ to be on the world line and again give it as an expansion around the point $\bar{x}$, as shown in Fig.~\ref{fig:SingularField}.  Writing $x^{a'}$ in terms of proper time $\tau$ gives
\footnote{\noindent In principle this expression is valid for an arbitrary world line. However, later in this thesis, in our explicit calculations, we consider both geodesic and non-geodesic motion.  Often we will make the assumption that is is geodesic and derive higher derivative terms from the equations of motion.} 
\begin{equation} \label{eqn: xtilde}
x^{{a'}}(\tau) = x^{\ab} + u^{\ab} \Delta \tau + \tfrac{1}{2!} \dot{u}^{\ab} \Delta \tau^2 + \tfrac{1}{3!} \ddot{u}^{\ab} \Delta \tau^3 +  \cdots,
\end{equation}
where $u^{\ab}$ is the four velocity at the point $x^{\ab}$, $\Delta \tau = \tau - \bar{\tau}$,
and an overdot denotes differentiation with respect to $\tau$.

We are interested in determining the points on the world line that are connected to $x$ by a null geodesic, that is we want to solve
\begin{align} \label{eqn: nullseparated}
\sigma\bigl( x^a,x^{{a'}}(\tau) \bigr) &= 0  \nonumber \\
&=  \tfrac{1}{2} g_{\ab \bb}(u^{\ab} \Delta \tau - 
	\Delta x^a) (u^{\bb} \Delta \tau  -\Delta x^b) \nonumber\\
& 
	+\: \bigl[\tfrac{1}{2} g_{\ab \bb }( u^{\ab}  \Delta \tau -\Delta x^a) \dot{u}^{\bb} \Delta 
	\tau^2 +\tfrac{1}{2} g_{\ab \bb,\cb }( u^{\ab} \Delta \tau  -\Delta x^a) (u^{\bb} \Delta \tau 
	- \Delta x^b)   \Delta x^c \nonumber\\
&
	+\: \tfrac{1}{4}g_{\ab \bb , \cb}(\xb) (u^{\ab} \Delta \tau  -\Delta x^a) (u^{\bb} \Delta \tau  
	- \Delta x^b) (u^{\cb} \Delta \tau  -\Delta x^c)   \bigr]+ O(\epsilon^4).
\end{align}
By writing 
$
\Delta \tau = \tau_1 \epsilon + \tau_2 \epsilon^2 + \tau_3 \epsilon^3  + \cdots
$
and explicitly inserting an $\epsilon$ in front of $\Delta x^a$, we may equate coefficients of powers of $\epsilon$ to obtain
\begin{equation} \label{eqn: tau1}
 \tau_1{}^2  +2 g_{\ab \bb} u^{\ab} \Delta x^b\tau_1 -  g_{\ab \bb} \Delta x^a \Delta x^b =~ 
	0,
\end{equation}
which gives,
\begin{align}\label{eqn: tau2}
 g_{\ab \bb} u^{\ab}(u^{\ab} \tau_1  -\Delta x^a) \tau_2 =& -\bigl[\tfrac{1}{2} g_{\ab \bb }( u^{ 
	\ab} \tau_1 -\Delta x^a) \dot{u}^{\bb} \tau_1{}^2 \nonumber \\
&
	+\: \tfrac{1}{2} g_{\ab \bb,\cb }( u^{\ab}  \tau_1  -\Delta x^a) (u^{\bb}  \tau_1 -\Delta 
	x^b) \Delta x^c \nonumber\\
&
	+\: \tfrac{1}{4}g_{\ab \bb , \cb}(\xb) (u^{\ab}  \tau_1  -\Delta x^a) (u^{\bb}  \tau_1 -\Delta 
	x^b) (u^{\cb}  \tau_1  -\Delta x^c) \bigr].
\end{align}
Equation \eqref{eqn: tau1} is a quadratic with two real roots of opposite sign (for $x$ spacelike separated from $\xb$) corresponding to the first approximation to our points $x_\textrm{\adv}$ and $x_\textrm{\ret}$,  
\begin{align}
\tau_{1\pm} = g_{\ab \bb} u^{\ab} \Delta x^b \pm \sqrt{(g_{\ab \bb} u^{\ab} \Delta x^b)^2 + g_{\ab \bb} \Delta x^a \Delta x^b}
\equiv \rbar_{(1)} \pm \rho,
\end{align}
where $\rbar_{(1)}$ is the leading order term in the coordinate expansion of the quantity $\rbar$ that will be appearing in our covariant expansions in Section \ref{sec:Covariant}.  Equation \eqref{eqn: tau2} is typical of the higher order equations giving $\tau_n$ in terms of lower order terms.


\subsection{Covariant Approach *} \label{sec:Covariant}

In this section, we briefly discuss our method for obtaining covariant expansions for the biscalars
appearing in Eqs.~\eqref{eq:PhiS}, \eqref{eq:AS} and \eqref{eq:hS}. We eventually seek expansions about a point $\bar{x}$ on the worldline (which we may treat as fixed in the majority of this thesis).  In doing so, we follow the strategy of Haas and Poisson \cite{Poisson:2003, Haas:Poisson:2006}:
\begin{itemize}
\item For the generic biscalar $A(x,z(\tau))$, write it as $A(\tau) \equiv A(x,z(\tau))$.
\item Compute the expansion about $\tau = \bar{\tau}$. This takes the form
\begin{equation}
A(\tau) = A(\bar{\tau}) + \dot{A}(\bar{\tau}) (\tau-\bar{\tau}) + \frac{1}{2}\ddot{A}(\bar{\tau}) (\tau-\bar{\tau})^2 + \cdots,
\end{equation}
where $\dot{A}(\bar{\tau}) = A_{;\ab} u^{\ab}$,
$\ddot{A}(\bar{\tau}) = A_{;\ab \bb} u^{\ab} u^{\bb}$, $\cdots$.
\item Compute the covariant expansions of the coefficients
$\dot{A}(\bar{\tau})$, $\ddot{A}(\bar{\tau})$, $\cdots$ about $\bar{\tau}$.
\item Evaluate the expansion at the desired point, eg. $A(x') = A(x,x')$.
\item The resulting expansion depends on $\tau$ through the powers of $\tau-\bar{\tau}$.
Replace these by their expansion in $\epsilon$ (about $\bar{x}$), the distance between $x$ and
the world-line.
\end{itemize}

A key ingredient of this calculation is the expansion of $\Delta\tau \equiv \tau-\bar{\tau}$ in
$\epsilon$. The leading orders in this expansion were developed by Haas and Poisson \cite{Haas:Poisson:2006} for the particular choices $\Delta\tau_+ \equiv v-\bar{\tau}$ and $\Delta\tau_- \equiv u-\bar{\tau}$, where $u$ and $v$ are the values of $\tau$ at $x_{\rm \ret}$ and $x_{\rm \adv}$ respectively. They found
\begin{equation}
\label{eq:Delta}
\Delta\tau_\pm = (\rbar \pm \sbar) \mp \frac{(\rbar \pm \sbar)^2}{6\,\sbar} R_{u \sigma u \sigma} \mp
  \frac{(\rbar \pm \sbar)^2}{24\,\sbar} \left[
    (\rbar \pm \sbar)R_{u \sigma u \sigma | u}
    - R_{u \sigma u \sigma | \sigma}\right]
  + \mathcal{O}(\epsilon^5),
\end{equation}
where $\rbar \equiv \sigma_\ab u^\ab$ and $\sbar^\text{\fixme{$2$}} \equiv (g^{\ab \bb} + u^\ab u^\bb) \sigma_\ab \sigma_\bb$. In Appendix \ref{sec:ExpansionCoefficients} we extend their calculation to the higher orders required in the present work.  In the same Appendix, we also apply the above method to compute covariant expansions of all quantities appearing in the expression for the singular field.

In order to obtain explicit expressions, we substitute in the coordinate expansion for $\sigma_{\ab}$ (as discussed in Sec.~\ref{sec:Coordinate}) along with the metric, Riemann tensor and 4-velocity (all evaluated at $\xb$). In doing so, we only have to keep terms that contribute up to the required order and truncate any higher order terms.


\section{Expansions of the Singular Field*} \label{sec:SingularFieldExplicit}

In this section we list the covariant form of the singular field to order $\epsilon^4$, where $\epsilon$ is the fundamental scale of separation, so, for example,  $\rbar$, $\sbar$ and $\sigma(x,\xb)^{\ab}$ are all of leading order $\epsilon$.  The coordinate forms of these expansions are too long to be useful in print form so instead they are available to download~\cite{BarryWardell.net} with leading orders given in Appendix \ref{sec:ExpansionCoefficients}.  The structure of the singular field is found to be very similar for all three cases and so the scalar singular field is illustrated in Figs.~\ref{fig:SingularField1D} and \ref{fig:SingularField2D} so the reader can get a feel for this form.

\subsection{Scalar singular field}
To $\mathcal{O}(\epsilon^4)$, the scalar singular field is
\par \vspace{-6pt} \begin{IEEEeqnarray}{rCl} \label{eq:phiS-approx}
\PhiS &=& q \Big\{ \frac{1}{\sbar} + \frac{\rbar^2-\sbar^2}{6 \sbar^3} C_{u \sigma u \sigma} +
	\frac{1}{24 \sbar^3} \Big[ (\rbar^2 - 3 \sbar^2) \rbar C_{u \sigma u \sigma ; u} -
	(\rbar^2-\sbar^2) C_{u \sigma u \sigma ; \sigma} \Big] \nonumber\\
&&
	+\; \frac{1}{360 \sbar^5} \left[\PhiS\right]_{(3)} + \frac{1}{4320 \sbar^5} 
	\left[\PhiS\right]_{(4)} + \mathcal{O}(\epsilon^5) \Big\},
\end{IEEEeqnarray}
where
\par \vspace{-6pt} \begin{IEEEeqnarray}{rCl}
\left[\PhiS\right]_{(3)} &=& 15 \Big[\rbar^2 - \sbar^2 \Big]^2 C_{u\sigma u\sigma }
	C_{u\sigma u\sigma } + \sbar^2 \Big[ (\rbar^2 -  \sbar^2) (3 C_{u\sigma u\sigma; \sigma 		\sigma } + 4 C_{u\sigma \sigma \ab } C_{u\sigma \sigma }{}^{\ab }) 
	+ (\rbar^4  \nonumber \\
&&
	 -\: 6 \rbar^2 \sbar^2 - 3 \sbar^4)(4 C_{u\sigma u\ab } C_{u\sigma u}{}^{\ab } + 3 
	C_{u\sigma u\sigma; uu}) + \rbar (\rbar^2 - 3 \sbar^2) (16 C_{u\sigma u}{}^{\ab } 
	C_{u\sigma \sigma \ab } \nonumber \\
&&
	-\: 3 C_{u\sigma u\sigma; u\sigma }) \Big] + \sbar^4 \Big\{ 2 C_{u}{}^{\ab }{}_{u}{}^{\bb } [ 
	(\rbar^2 + \sbar^2) C_{\sigma \ab \sigma \bb }+ 2 \rbar (\rbar^2 + 3 \sbar^2) C_{u\ab 
	\sigma \bb }] \nonumber \\
&&
	+\: 2 C_{u}{}^{\ab }{}_{\sigma }{}^{\bb } [2 \rbar C_{\sigma \ab \sigma \bb } + (\rbar^2 + 
	\sbar^2) C_{u\bb \sigma \ab }] + (\rbar^4 + 6 \rbar^2 \sbar^2 + \sbar^4 ) C_{u\ab u\bb } 
	C_{u}{}^{\ab }{}_{u}{}^{\bb } + 2 (\rbar^2  \nonumber \\
&&
	+\: \sbar^2) C_{u\ab \sigma \bb } C_{u}{}^{\ab } {}_{\sigma }{}^{\bb } + C_{\sigma \ab 
	\sigma \bb } C_{\sigma }{}^{\ab }{}_{\sigma }{}^{\bb } \Big\}
\end{IEEEeqnarray}
and
\par \vspace{-6pt} \begin{IEEEeqnarray}{rCl}
\left[\PhiS\right]_{(4)} &=& 30 C_{u\sigma u\sigma } \Big[ \rbar (3 \rbar^4 - 10 \rbar^2 \sbar^2 	+ 15 \sbar^4) C_{u\sigma u\sigma; u} - 30 (\rbar^2 - \sbar^2)^2 C_{u\sigma u\sigma ;			\sigma } \Big] \nonumber \\
&&
    	+\: 2 \sbar^2 \Big\{ 3 C_{u\sigma u\sigma; uuu} \rbar \left (\rbar^4 - 10 \rbar^2 \sbar^2 
	- 15 \sbar^4 \right) + 3  \rbar \left(\rbar^2 - 3 \sbar^2\right) C_{u\sigma u\sigma;
	u\sigma \sigma } - 3 (\rbar^2  \nonumber \\
&&
	-\:  \sbar^2) C_{u\sigma u\sigma; \sigma \sigma \sigma } - 3 (\rbar^4 - 6 \rbar^2 
	\sbar^2 - 3 \sbar^4) C_{u\sigma u\sigma; uu\sigma } - 9 \rbar (\rbar^4 - 10 \rbar^2
	\sbar^2 \nonumber \\
&&
	-\: 15 \sbar^4) C_{u\sigma u\ab; u} - C_{u\sigma \sigma }{}^{\ab } [18 C_{u\sigma \sigma 		\ab; \sigma } (\rbar^2 - \sbar^2) - \rbar (\rbar^2 - 3 \sbar^2) (10 C_{u\sigma \sigma \ab 		;u} \nonumber \\
&&
	-\: 16 C_{u\sigma u\ab; \sigma } - 5 C_{u\sigma u\sigma; \ab }) - 30 C_{u\sigma u\ab; u}
	(\rbar^4 - 6 \rbar^2 \sbar^2 - 3 \sbar^4)] - C_{u\sigma u}{}^{\ab } [36 \rbar (\rbar^2 
	\nonumber \\
&&
	- 3 \sbar^2) C_{u\sigma \sigma \ab; \sigma } + (\rbar^4  
	- 6 \rbar^2 \sbar^2 - 3 \sbar^4) (13 C_{u\sigma u\ab; \sigma } + 5 C_{u \sigma u 				\sigma; \ab } - 25 C_{u\sigma \sigma \ab; u} )] \Big\} \nonumber \\
&&
	-\: 12 \sbar^4 \Big\{ C_{\sigma }{}^{\ab}{}_{\sigma }{}^{\bb } [ C_{\sigma \ab \sigma \bb; 
	\sigma } - \rbar ( C_{\sigma \ab \sigma \bb; u} - 2 C_{u\ab \sigma \bb ;\sigma }) -
	(\rbar^2 + \sbar^2) (2 C_{u\ab \sigma \bb; u} \nonumber \\
&&
	-\: C_{u\ab u\bb; \sigma }) - \rbar (\rbar^2 + 3 \sbar^2 )C_{u\ab u\bb; u}] + 2 C_{u}{}^{\ab }{}_{\sigma }{}^{\bb } 
	[C_{\sigma \ab \sigma \bb; \sigma } \rbar + (\rbar^2 + \sbar^2)(C_{u\ab \sigma \bb ;
	\sigma } \nonumber \\
&&
	+\: C_{u\bb \sigma \ab ;\sigma } \nonumber - C_{\sigma \ab \sigma \bb; u}) + \rbar (\rbar^2 + 3 \sbar^2)(C_{u\ab u\bb ;\sigma } 		
	- C_{u\ab \sigma \bb; u} - C_{u\bb \sigma \ab ;u}) \nonumber \\
&&
	-\: C_{u\ab u\bb u} (\rbar^4 + 6 \rbar^2 \sbar^2 + \sbar^4)] + C_{u}{}^{\ab }{}_{u}{}^{\bb } [(\rbar^2 + \sbar^2) C_{\sigma \ab \sigma \bb; 
	\sigma } \nonumber \\
&&
	-\: \rbar (\rbar^2 
	+ 3 \sbar^2) (C_{\sigma \ab \sigma \bb; u} - 2 C_{u\ab \sigma \bb; \sigma }) + (\rbar^4+ 6 \rbar^2 \sbar^2 + \sbar^4) (C_{u\ab u\bb; \sigma } - 2 C_{u\ab \sigma \bb; u})
	\nonumber \\
&&
	-\:  C_{u\ab u\bb; u} \rbar (\rbar^4 + 10 \rbar^2 \sbar^2 + 5 \sbar^4)] \Big\}.
\end{IEEEeqnarray}

\subsection{Electromagnetic singular field}
To $\mathcal{O}(\epsilon^4)$, the electromagnetic singular field is
\par \vspace{-6pt} \begin{IEEEeqnarray}{rCl} \label{eq:AS-approx}
A_{a}^{\rm \sing} &=& e g_{a}{}^\ab \Big( \frac{u_\ab}{\sbar} + \frac{1}{6 \sbar^3} \Big[3 \rbar
	\sbar^2 C_{\ab u u \sigma} + C_{u \sigma u \sigma} (\rbar^2 - \sbar^2) u_\ab \Big] \nonumber \\
&&
	+\:
	\frac{1}{24 \sbar^3} \Big\{ 4 \sbar^2 (C_{\ab uu \sigma ;u} - \rbar C_{\ab uu \sigma ;
	\sigma}) + \Big[ \rbar (\rbar^2-3\sbar^2) C_{u \sigma u \sigma ; u} - (\rbar^2 - \sbar^2) C_{u
	\sigma u \sigma \sigma}\Big] u_\ab \Big\} \nonumber \\
&&
	+\: \frac{1}{2880 \sbar^5} \left[A_\ab^{\rm
	\sing}\right]_{(3)} + \frac{1}{25920 \sbar^5} \left[A_\ab^{\rm \sing}\right]_{(4)} + \mathcal{O}(\epsilon^5)
	\Big),
\end{IEEEeqnarray}
where
\par \vspace{-6pt} \begin{IEEEeqnarray}{rCl}
[A_{\ab}^{\rm \sing}]_{(3)} &=& 120 C_{u\sigma u\ab }{}_{;}{}_{u\sigma } \sbar^4 (\rbar^2 +
	\sbar^2) - 120 C_{u\sigma u\ab }{}_{;}{}_{\sigma \sigma } \rbar \sbar^4 + 120 C_{u}{}^{\cb}
	{}_{u}{}^{\db } C_{\sigma \cb \ab \db } \rbar \sbar^6  \nonumber \\
&&
	-\: 240 C_{u\sigma u\ab } C_{u\sigma u\sigma } \rbar \sbar^2 (\rbar^2 - 3 \sbar^2)- 360
	C_{u\sigma \ab \cb } C_{u\sigma u}{}^{\cb } \sbar^4 (\rbar^2 + \sbar^2) \nonumber \\
&&
	-\: 120 C_{u\sigma u\ab }{}_{;}{}_{uu} \rbar \sbar^4 (\rbar^2 + 3 \sbar^2) + 120 \sbar^6 
	C_{u}{}^{\cb }{}_{\sigma }{}^{\db } (C_{u\db \ab \cb } \rbar + 
	C_{\sigma \db \ab \cb }) \nonumber \\
&&
	+\: 40 C_{u\cb \ab \db } C_{u}{}^{\cb }{}_{u}{}^{\db } \sbar^6 (3
	\rbar^2 + \sbar^2) - 120 C_{u\ab \sigma \cb } [C_{u\sigma \sigma }{}^{\cb } \rbar \sbar^4 
	+ 2 C_{u\sigma u}{}^{\cb } \sbar^4 (\rbar^2 + \sbar^2)] \nonumber \\
&&
	-\: 120 C_{u\ab u\cb } [ C_{u\sigma \sigma }{}^{\cb } \sbar^4 (\rbar^2 + \sbar^2) + 
	C_{u\sigma u}{}^{\cb } \rbar \sbar^4 (\rbar^2 + 3 \sbar^2)] + \{8 C_{\sigma \cb \sigma 
	\db } C_{\sigma }{}^{\cb }{}_{\sigma }{}^{\db } \sbar^4 \nonumber \\
&&
	-\: 5 C_{\cb \db \eb \fb } C^{\cb \db \eb \fb } \sbar^8 - 24 C_{u\sigma u\sigma }{}_{;}
	{}_{u\sigma } \rbar \sbar^2 (\rbar^2 - 3 \sbar^2) + 128 C_{u\sigma u}{}^{\cb } C_{u\sigma
	\sigma \cb } \rbar \sbar^2 (\rbar^2 - 3 \sbar^2) \nonumber \\
&&
	+\: 24 C_{u\sigma u\sigma }{}_{;}{}_{\sigma \sigma } \sbar^2 (\rbar^2 -  \sbar^2) + 32 
	C_{u\sigma \sigma \cb } C_{u\sigma \sigma }{}^{\cb } \sbar^2 (\rbar^2 -  \sbar^2) + 120 
	C_{u\sigma u\sigma } C_{u\sigma u\sigma } (\rbar^2 \nonumber \\
&&
	-\:  \sbar^2)^2 + 16 C_{u}{}^{\cb }{}_{u}{}^{\db } C_{\sigma \cb \sigma \db } \sbar^4 
	(\rbar^2 + \sbar^2) + 24 C_{u\sigma u\sigma }{}_{;}{}_{uu} \sbar^2 (\rbar^4 - 6 \rbar^2 
	\sbar^2 - 3 \sbar^4)  \nonumber \\
&&
	+\: 32 C_{u\sigma u\cb } C_{u\sigma u}{}^{\cb } \sbar^2 (\rbar^4 - 6 \rbar^2 \sbar^2 - 3 
	\sbar^4) + 8 C_{u\cb u\db } C_{u}{}^{\cb }{}_{u}{}^{\db } \sbar^4 (\rbar^4 + 6 \rbar^2 
	\sbar^2 + \sbar^4) \nonumber \\
&&
	+\: 16 \sbar^4 C_{u}{}^{\cb }{}_{\sigma }{}^{\db } [2 C_{\sigma \cb \sigma \db } \rbar + 
	C_{u\db \sigma \cb } (\rbar^2 + \sbar^2)] + 16 \sbar^4 C_{u\cb \sigma \db } [C_{u}{}^{\cb
	}{}_{\sigma }{}^{\db } (\rbar^2 + \sbar^2) \nonumber \\
&&
	+\: 2 C_{u}{}^{\cb }{}_{u}{}^{\db } \rbar (\rbar^2 + 3 \sbar^2)]\} u_{\ab }
\end{IEEEeqnarray}
and 
\par \vspace{-6pt} \begin{IEEEeqnarray}{rCl}
[A_\ab^{\rm \sing}]_{(4)} &=& 216 C_{u\sigma u\ab }{}_{;}{}_{\sigma \sigma \sigma } \rbar
	\sbar^4 + 540 C_{u\sigma u\sigma }{}_{;}{}_{\sigma } C_{u\sigma u\ab } \rbar \sbar^2
	(\rbar^2 - 3 \sbar^2) \nonumber \\
&&
	+\: 720 C_{u\sigma u\ab }{}_{;}{}_{\sigma } C_{u\sigma u\sigma } \rbar
	\sbar^2 (\rbar^2 -  3 \sbar^2) - 216 C_{u\sigma u\ab }{}_{;}{}_{u\sigma \sigma } \sbar^4 (\rbar^2 + 
	\sbar^2) \nonumber \\
&&
	+\: 1512 C_{u\sigma \ab }{}^{\cb }{}_{;}{}_{\sigma } C_{u\sigma u\cb } \sbar^4 
	(\rbar^2 + \sbar^2) + 216 C_{u\sigma u\ab }{}_{;}{}_{uu\sigma } \rbar \sbar^4 (\rbar^2 + 3 \sbar^2) \nonumber \\
&&
	-\: 864 
	C_{u\sigma \ab }{}^{\cb }{}_{;}{}_{u} C_{u\sigma u\cb } \rbar \sbar^4 (\rbar^2 + 3 \sbar^2) 
	- 216 C_{u\sigma u\ab }{}_{;}{}_{uuu} \sbar^4 (\rbar^4 + 6 \rbar^2 \sbar^2 + \sbar^4) \nonumber \\
&&
	-\: 
	540 C_{u\sigma u\sigma }{}_{;}{}_{u} C_{u\sigma u\ab } \sbar^2 (\rbar^4 - 6 \rbar^2 
	\sbar^2 - 3 \sbar^4) + 720 C_{u\sigma u\ab }{}_{;}{}_{u} C_{u\sigma u\sigma } \sbar^2 (- \rbar^4 + 6 \rbar^2 
	\sbar^2 \nonumber \\
&&
	+\: 3 \sbar^4) - 432 \sbar^6 C_{\sigma }{}^{\cb }{}_{\ab }{}^{\db }{}_{;}{}_{\sigma } (
	C_{u\cb u\db } \rbar + C_{u\db \sigma \cb }) - 648 \sbar^6 C_{u}{}^{\cb }{}_{\sigma }{}^{\db }{}_{;}{}_{\sigma } ( C_{u\db \ab \cb } \rbar 
	+ C_{\sigma \db \ab \cb } ) \nonumber \\
&&
	+\: 16 \sbar^8 C_{u\ab }{}^{\cb \db }{}^{;}{}^{\eb } ( C_{u\cb \db 
	\eb } \rbar - C_{u\eb \cb \db } \rbar + C_{\sigma \cb \db \eb } - C_{\sigma \eb \cb \db }) + 432 \sbar^4 C_{u\sigma \sigma }{}^{\cb }{}_{;}{}_{\sigma } [ 
	C_{u\ab \sigma \cb } \rbar \nonumber \\
&&
	+\: C_{u\ab u\cb } (\rbar^2 + \sbar^2)] + 504 \sbar^4 C_{u\ab 
	\sigma }{}^{\cb }{}_{;}{}_{\sigma } [ C_{u\sigma \sigma \cb } \rbar + 2 C_{u\sigma u\cb } (\rbar^2 + \sbar^2)]  \nonumber \\
&&
	-\: 324 \sbar^4 C_{u\sigma \sigma }{}^{\cb }
	{}_{;}{}_{u} [ C_{u\ab \sigma \cb } (\rbar^2 + \sbar^2) + C_{u\ab u\cb } \rbar (\rbar^2 + 3 
	\sbar^2)] \nonumber \\
&&
	+\: 108 \sbar^4 C_{u\sigma u}{}^{\cb }{}_{;}{}_{\sigma } [5 C_{u\ab \sigma \cb } (\rbar^2 + 
	\sbar^2) + 6 C_{u\sigma \ab \cb } (\rbar^2 + \sbar^2) + 3 C_{u\ab u\cb } \rbar (\rbar^2 + 
	3 \sbar^2)] \nonumber \\
&&
	-\: 96 \sbar^4 C_{u\ab \sigma }{}^{\cb }{}_{;}{}_{u} [3 C_{u\sigma \sigma \cb } (\rbar^2 + 
	\sbar^2) + 8 C_{u\sigma u\cb } \rbar (\rbar^2 + 3 \sbar^2)] \nonumber \\
&&
	+\: 96 \sbar^4 C_{u\sigma 
	u\ab }{}^{;}{}^{\cb } [3 C_{u\sigma \sigma \cb } (\rbar^2 + \sbar^2) + 4 C_{u\sigma u\cb } \rbar (\rbar^2 + 3 \sbar^2)] \nonumber \\
&&
	+\: 24 \sbar^4 C_{u\ab u}
	{}^{\cb }{}_{;}{}_{\sigma } [9 C_{u\sigma \sigma \cb } (\rbar^2 + \sbar^2) + 17 C_{u\sigma 
	u\cb } \rbar (\rbar^2 + 3 \sbar^2)] \nonumber \\
&&
	-\: 216 \sbar^4 C_{u\sigma u}{}^{\cb }{}_{;}{}_{u} [3 C_{u\ab \sigma \cb } 
	\rbar (\rbar^2  + 3 \sbar^2) + 6 C_{u\sigma \ab \cb } \rbar (\rbar^2 + 3 \sbar^2) + 2 
	C_{u\ab u\cb } \sbar^4 (\rbar^4 \nonumber \\
&&
	+\: 6 \rbar^2 \sbar^2 + \sbar^4)]- 72 \sbar^4 C_{u\ab u}{}^{\cb }{}_{;}{}_{u} [8 C_{u\sigma 
	\sigma \cb } \rbar (\rbar^2 + 3 \sbar^2) + 7 C_{u\sigma u\cb } (\rbar^4 + 6 \rbar^2 
	\sbar^2 + \sbar^4)] \nonumber \\
&&
	-\: 216 \sbar^6 C_{u}{}^{\cb }{}_{u}{}^{\db }{}_{;}{}_{\sigma } [3 C_{\sigma \cb \ab \db } \rbar 
	+ C_{u\cb \ab \db } (3 \rbar^2 + \sbar^2)] - 144 \sbar^6 C_{u}{}^{\cb }{}_{\ab }{}^{\db }
	{}_{;}{}_{\sigma } [3 C_{u\db \sigma \cb } \rbar \nonumber \\
&&
	+\:  C_{u\cb u\db } (3 \rbar^2 + \sbar^2)] + 48 \sbar^6 C_{u\ab \sigma }{}^{\cb }{}^{;}
	{}^{\db } [3 C_{u\cb \sigma \db } \rbar + 3 C_{u\db \sigma \cb } \rbar + 3 C_{\sigma \cb 
	\sigma \db } + C_{u\cb u\db } (3 \rbar^2 \nonumber \\
&&
	+\: \sbar^2)] + 216 \sbar^6 C_{\sigma }{}^{\cb }{}_{\ab }{}^{\db }{}_{;}{}_{u} [3 C_{u\db 
	\sigma \cb } \rbar + C_{u\cb u\db } (3 \rbar^2 + \sbar^2)] + 144 \sbar^6 C_{u}{}^{\cb }
	{}_{\sigma }{}^{\db }{}_{;}{}_{u} [3 C_{\sigma \db \ab \cb } \rbar \nonumber \\
&&
	+\: C_{u\db \ab \cb } (3 \rbar^2 + \sbar^2)] + 48 \sbar^6 C_{u\ab u}{}^{\cb }{}^{;}
	{}^{\db } [3 \rbar C_{\sigma \cb \sigma \db } + 3 \rbar C_{u\cb u\db } (\rbar^2 + \sbar^2) 
	+ C_{u\cb \sigma \db } (3 \rbar^2 \nonumber \\
&&
	+\: \sbar^2) + C_{u\db \sigma \cb } (3 \rbar^2 + \sbar^2)] + 216 \sbar^6 C_{u}{}^{\cb }{}_{\ab}{}^{\db 
	}{}_{;}{}_{u} [3 C_{u\cb u\db } \rbar (\rbar^2 + \sbar^2) + C_{u\db \sigma \cb } (3 \rbar^2 \nonumber \\
&&
	+\: 
	\sbar^2)] + 144 \sbar^6 C_{u}{}^{\cb }{}_{u}{}^{\db }{}_{;}{}_{u} [3 C_{u\cb \ab \db } \rbar (\rbar^2 + 
	\sbar^2) + C_{\sigma \cb \ab \db } (3 \rbar^2 + \sbar^2)] \nonumber \\
&&
	+\: \{45 C^{\cb \db \eb \fb }{}_{ 
	\sigma } C_{\cb \db \eb \fb } \sbar^8 - 45 C^{\cb \db \eb \fb }{}_{u} C_{\cb \db \eb \fb} \rbar \sbar^8 + 36 C_{u\sigma 
	u\sigma }{}_{;}{}_{u\sigma \sigma } \rbar \sbar^2 (\rbar^2 - 3 \sbar^2) \nonumber \\
&&
	-\: 540 C_{u\sigma u 
	\sigma }{}_{;}{}_{\sigma } C_{u\sigma u\sigma } (\rbar^2 - \sbar^2)^2 - 36 C_{u\sigma u\sigma }{}_{;}{}_{\sigma \sigma \sigma } \sbar^2 (\rbar^2 - \sbar^2) \nonumber \\
&&
	+\: 
	36 C_{u\sigma u\sigma }{}_{;}{}_{uuu} \rbar \sbar^2 (\rbar^4 - 10 \rbar^2 \sbar^2 - 15 
	\sbar^4) - 36 C_{u\sigma u\sigma }{}_{;}{}_{uu\sigma } \sbar^2 (\rbar^4 - 6 \rbar^2 \sbar^2 - 3 
	\sbar^4) \nonumber \\
&&
	+\: 180 C_{u\sigma u\sigma }{}_{;}{}_{u} C_{u\sigma u\sigma } (3 \rbar^5 - 10 
	\rbar^3 \sbar^2 + 15 \rbar \sbar^4) - 216 \sbar^2 C_{u\sigma \sigma }{}^{\cb }{}_{;}{}_{\sigma } [2 \rbar C_{u\sigma u\cb } 
	(\rbar^2 \nonumber \\
&&
	-\: 3 \sbar^2) - 2 C_{u\sigma \sigma \cb } (- \rbar^2 + \sbar^2)] - 72 \sbar^4 
	C_{\sigma }{}^{\cb }{}_{\sigma }{}^{\db }{}_{;}{}_{\sigma } [2 C_{u\cb \sigma \db } \rbar 
	\nonumber + C_{\sigma \cb \sigma \db } + C_{u\cb u\db } (\rbar^2 \nonumber \\
&&
	+\: \sbar^2)] - 144 \sbar^4 C_{u}
	{}^{\cb }{}_{\sigma }{}^{\db }{}_{;}{}_{\sigma } [C_{\sigma \cb \sigma \db } \rbar + C_{u\cb 
	\sigma \db } (\rbar^2 + \sbar^2) + C_{u\db \sigma \cb } (\rbar^2 + \sbar^2) \nonumber \\
&&
	+\: C_{u\cb u\db } \rbar (\rbar^2 + 3 \sbar^2)] + 72 \sbar^4 C_{\sigma }{}^{\cb 
	}{}_{\sigma }{}^{\db }{}_{;}{}_{u} [C_{\sigma \cb \sigma \db} \rbar + 2 C_{u\cb \sigma \db } 
	(\rbar^2 + \sbar^2) + C_{u\cb u\db } \rbar (\rbar^2 \nonumber \\
&&
	+\: 3 \sbar^2)] + 60 \sbar^2 C_{u\sigma \sigma }{}^{\cb }{}_{;}{}_{u} [2 C_{u\sigma \sigma 
	\cb } \rbar (\rbar^2 - 3 \sbar^2) + 5 C_{u\sigma u\cb } (\rbar^4 - 6 \rbar^2 \sbar^2 - 3 
	\sbar^4)] \nonumber \\
&&
	+\: 72 \sbar^2 C_{u\sigma u}{}^{\cb }{}_{;}{}_{u} [3 C_{u\sigma u\cb } \rbar (\rbar^4 - 10
	\rbar^2 \sbar^2 - 15 \sbar^4) + 5 C_{u\sigma \sigma \cb } (\rbar^4 - 6 \rbar^2 \sbar^2 -
	3 \sbar^4)] \nonumber \\
&&
	-\: 72 \sbar^4 C_{u}{}^{\cb }{}_{u}{}^{\db }{}_{;}{}_{\sigma } [C_{\sigma \cb \sigma \db } 
	(\rbar^2 + \sbar^2) + 2 C_{u\cb \sigma \db } \rbar (\rbar^2 + 3 \sbar^2) + C_{u\cb u\db} 	
	(\rbar^4 + 6 \rbar^2 \sbar^2 \nonumber \\
&&
	+\: \sbar^4)] + 144 \sbar^4 C_{u}{}^{\cb }{}_{\sigma }{}^{\db }{}_{;}{}_{u} [C_{\sigma \cb \sigma \db } 
	(\rbar^2 + \sbar^2) + C_{u\cb \sigma \db } \rbar (\rbar^2 + 3 \sbar^2) + C_{u\db \sigma
	\cb } \rbar (\rbar^2 \nonumber \\
&&
	+\: 3 \sbar^2) + C_{u\cb u\db } (\rbar^4 + 6 \rbar^2 \sbar^2 + \sbar^4)] - 60 \sbar^2 C_{u\sigma
	u\sigma }{}^{;}{}^{\cb } [C_{u\sigma \sigma \cb } \rbar (\rbar^2 - 3 \sbar^2) \nonumber \\
&&
	+\: C_{u\sigma
	u\cb } (\rbar^4 - 6 \rbar^2 \sbar^2 - 3 \sbar^4)] - 12 \sbar^2 C_{u\sigma u}{}^{\cb }{}_{;}{}_{\sigma } [16 C_{u\sigma \sigma
	\cb } \rbar (\rbar^2 - 3 \sbar^2) \nonumber \\
&&
	+\: 13 C_{u\sigma u\cb } (\rbar^4 - 6 \rbar^2 \sbar^2 - 3
	\sbar^4)] + 72 \sbar^4 C_{u}{}^{\cb }{}_{u}{}^{\db }{}_{;}{}_{u} [C_{\sigma \cb \sigma \db } \rbar
	(\rbar^2 + 3 \sbar^2) + 2 C_{u\cb \sigma \db } (\rbar^4 \nonumber \\
&&
	+\: 6 \rbar^2 \sbar^2 + \sbar^4) + C_{u\cb u\db } \rbar (\rbar^4 + 10 \rbar^2 \sbar^2 + 5 \sbar^4)]\} u_{\ab }.
\end{IEEEeqnarray}
\begin{figure}
\begin{center}
\includegraphics[width=14cm]{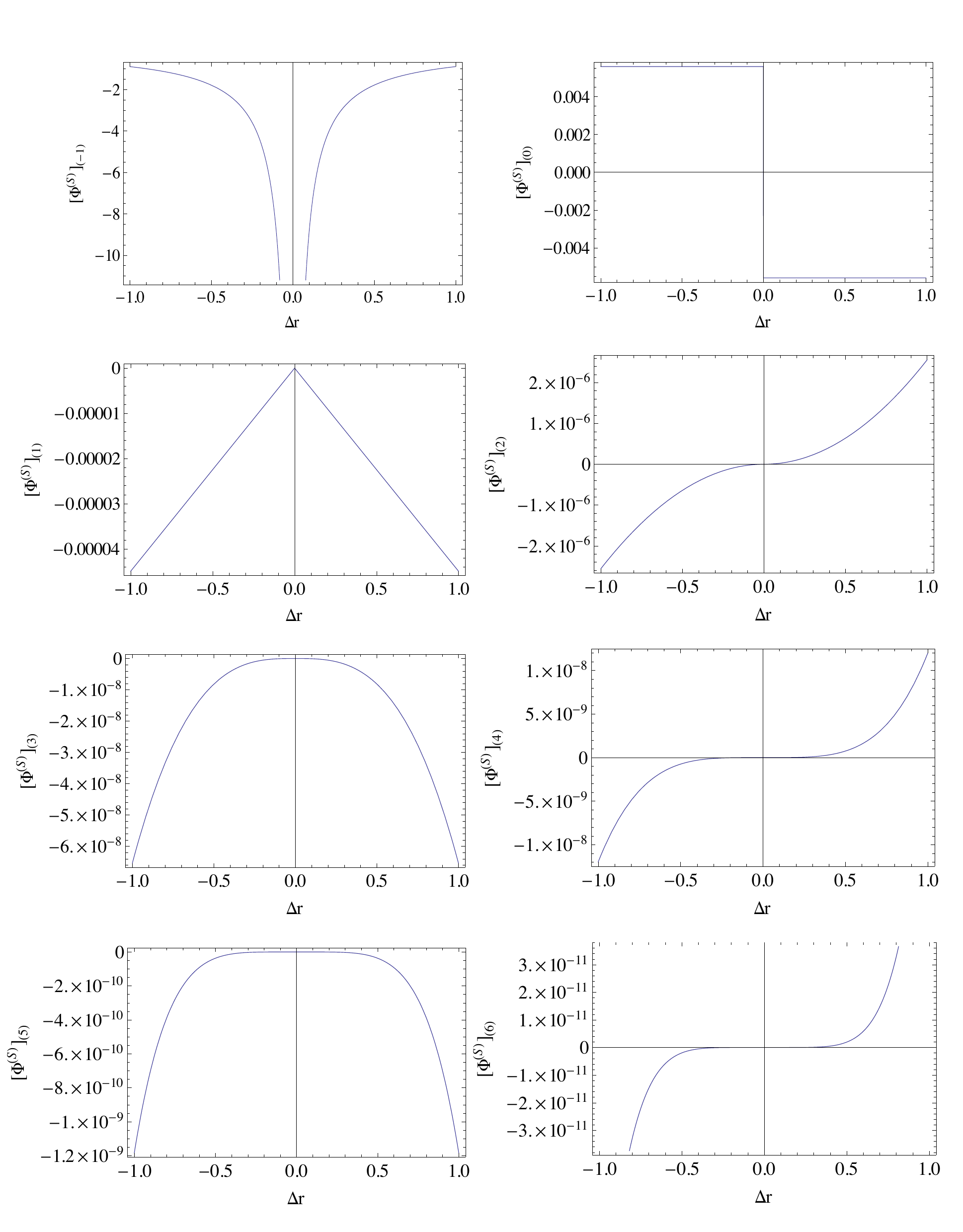}
\caption[Singular Field Components in 2D]{Terms in the coordinate expansion of the singular field
for $\mathcal{O}(\epsilon^{-1})$ (top left) to $\mathcal{O}(\epsilon^{6})$ (bottom right).
Shown is the scalar case of a circular geodesic of radius $r_0 = 10M$ in \Sch space-time.}
\label{fig:SingularField1D}
\end{center}
\end{figure}
\begin{figure}
\begin{center}
\includegraphics[width=15cm]{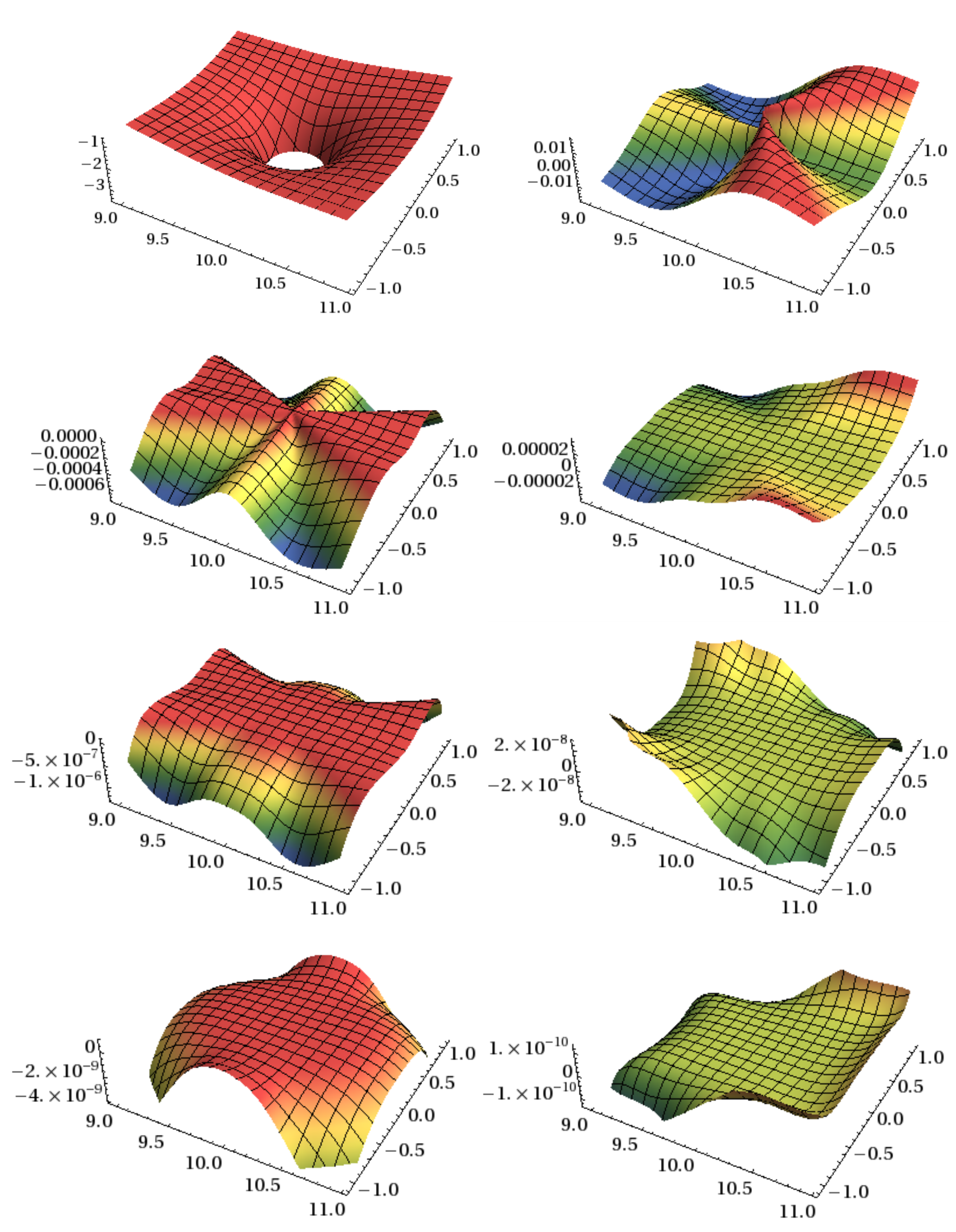}
\caption[Singular Field Components in 3D]{Terms in the coordinate expansion of the singular field, $[\Phi^{\sing}]_{(n)}$, in the region of the particle
for $\mathcal{O}(\epsilon^{-1})$ (top left) to $\mathcal{O}(\epsilon^{6})$ (bottom right).
Shown is the scalar case of a circular geodesic of radius $r_0 = 10M$ in \Sch space-time.}
\label{fig:SingularField2D}
\end{center}
\end{figure}
\subsection{Gravitational singular field}
To $\mathcal{O}(\epsilon^4)$, the gravitational singular field is
\par \vspace{-6pt} \begin{IEEEeqnarray}{rCl} \label{eq:hS-approx}
\hb_{ab}^{\rm \sing} &=& \mu g_{a}{}^\ab g_b{}^\bb \Big( \frac{u_\ab u_\bb}{\sbar} + \frac{1}
	{6 \sbar^3} \Big[ (\rbar^2 - \sbar^2) C_{u \sigma u \sigma} u_\ab u_\bb - 6 \rbar \sbar^2 
	C_{ u \sigma u (\ab} u_{\bb)} - 6 \sbar^4 C_{\ab u \bb u} \Big] \nonumber \\
&& 
	+\: \frac{1}{24 \sbar^3} \Big\{ 12 \sbar^4 (C_{\ab u \bb u ;\sigma} - \rbar C_{\ab u \bb u ; 
	u}) + 4 \sbar^2 \Big[u_{(\ab}  C_{\bb) uu \sigma; u} (\rbar^2 + \sbar^2) - \rbar u_{(\ab}  
	C_{\bb) u u \sigma ; \sigma}\Big] \nonumber \\
&&
	+\: u_\ab u_\bb \Big[ \rbar (\rbar^2 - 3\sbar^2) C_{u \sigma u \sigma ; u}  - (\rbar^2 - 
	\sbar^2) C_{u \sigma u \sigma ; \sigma} \Big] \Big\} + \frac{1}{1440 \sbar^5} 	
	\left[\hb_{\ab \bb}^{\rm \sing}\right]_{(3)} \nonumber \\
&&
	+\: \frac{1}{6480 \sbar^5} \left[\hb_{\ab \bb}^{\rm \sing}\right]_{(4)} + \mathcal{O}
	(\epsilon^5) \Big),
\end{IEEEeqnarray}
where
\par \vspace{-6pt} \begin{IEEEeqnarray}{rCl}
\Big[\hb_{\ab \bb}^{\rm \sing}&&\Big]_{(3)} = -240 C_{\ab u\bb u}{}_{;}{}_{\sigma \sigma } 
	\sbar^6 + 240 C_{\ab u\bb u}{}_{;}{}_{u\sigma } \rbar \sbar^6 + 480  C_{\ab }{}^{\cb }
	{}_{u\sigma } C_{\bb uu\cb } \rbar \sbar^6 - 120 C_{\ab u\sigma }{}^{\cb } C_{\bb u\sigma 
	\cb } \sbar^6 \nonumber \\
&&
	+\: 960 C_{\ab u\bb }{}^{\cb } C_{u\sigma u\cb } \rbar  \sbar^6 + 40 C_{\ab u\bb u}{}^{;}
	{}^{\cb }{}_{\cb } \sbar^8 + 20 C_{\ab u}{}^{\cb \db } C_{\bb u\cb \db } \sbar^8 + 240 
	C_{\ab }{}^{\cb }{}_{\bb }{}^{\db } C_{u\cb u\db } \sbar^8 \nonumber \\
&&
	+\: 360 C_{\ab uu\sigma } C_{\bb uu\sigma } \sbar^4 (\rbar^2 + \sbar^2) + 240 C_{\ab u 
	\bb u} C_{u\sigma u\sigma } \sbar^4 (\rbar^2 + \sbar^2) - 80 C_{\ab u\bb u}{}_{;}{}_{uu}  
	\sbar^6 (3 \rbar^2  \nonumber \\
&&
	+\: \sbar^2) - 40 C_{\ab uu}{}^{\cb } \sbar^6 [6  C_{\bb u\sigma \cb } \rbar + C_{\bb uu\cb } (3 
	\rbar^2 + \sbar^2)] + \{120 C_{\bb uu\sigma }{}_{;}{}_{\sigma \sigma } \rbar \sbar^4 \nonumber \\
&&
	+\: 
	240  C_{\bb uu\sigma } C_{u\sigma u\sigma } \rbar \sbar^2 (\rbar^2 - 3 \sbar^2) - 120 C_{\bb uu\sigma }{}_{;}{}_{u\sigma } \sbar^4 (\rbar^2 + \sbar^2) - 
	360 C_{\bb }{}^{\cb }{}_{u\sigma } C_{u\sigma u\cb } \sbar^4 (\rbar^2 \nonumber \\
&&
	+\: \sbar^2) + 120 
	C_{\bb uu\sigma }{}_{;}{}_{uu} \rbar \sbar^4 (\rbar^2 + 3 \sbar^2) + 120 C_{\bb }{}^{\cb }{}_{\sigma }{}^{\db } ( C_{u\cb u\db } \rbar +  C_{u\cb 	\sigma \db } ) \sbar^6  \nonumber \\
&&
	+\: 120 C_{\bb u\sigma }{}^{\cb } [ C_{u\sigma \sigma \cb }\rbar + 2 
	C_{u\sigma u\cb }  (\rbar^2 + \sbar^2)] \sbar^4 + 40 C_{\bb }{}^{\cb }{}_{u}{}^{\db } [3  C_{u\cb \sigma \db } \rbar  +  C_{u\cb u\db } (3 
	\rbar^2 + \sbar^2)] \sbar^6 \nonumber \\
&&
	+\: 120 C_{\bb uu}{}^{\cb } [C_{u\sigma \sigma \cb }  (\rbar^2 
	+ \sbar^2) +  C_{u\sigma u\cb } \rbar (\rbar^2 + 3 \sbar^2)]\sbar^4\} u_{\ab } + \{4 C_{\sigma \cb \sigma \db } C_{\sigma }{}^{\cb }
	{}_{\sigma }{}^{\db } \sbar^4  \nonumber \\
&&
	-\: 5 C_{\cb \db \eb \fb} C^{\cb \db \eb \fb} \sbar^8 + 12 
	C_{u\sigma u\sigma }{}_{;}{}_{\sigma \sigma } \sbar^2 (\rbar^2 - \sbar^2) + 16 C_{u\sigma \sigma \cb } C_{u\sigma \sigma }{}^{\cb } \sbar^2  (\rbar^2 - \sbar^2) 
	 \nonumber \\
&&
	-\: 12 C_{u\sigma u\sigma }{}_{;}{}_{u\sigma } \rbar \sbar^2 (\rbar^2 - 3 \sbar^2) + 64 
	C_{u\sigma u}{}^{\cb } C_{u\sigma \sigma \cb } \rbar\sbar^2 (\rbar^2 - 3 \sbar^2) 
	+ 60 C_{u\sigma u\sigma }^2 (\rbar^2 -  \sbar^2)^2  \nonumber \\
&&
	+\: 8 C_{u}{}^{\cb }{}_{u}{}^{\db } 
	C_{\sigma \cb \sigma \db } \sbar^4 (\rbar^2 + \sbar^2) + 12 C_{u\sigma u\sigma }{}_{;}
	{}_{uu} \sbar^2 (\rbar^4 - 6 \rbar^2 \sbar^2 - 3 \sbar^4)   \nonumber \\
&&
	+\: 16 C_{u\sigma u\cb } C_{u\sigma u}{}^{\cb } \sbar^2 (\rbar^4 - 6 \rbar^2 \sbar^2 - 3 
	\sbar^4) + 4 C_{u\cb u\db } C_{u}{}^{\cb }{}_{u}{}^{\db } \sbar^4 (\rbar^4 + 6 \rbar^2 
	\sbar^2 + \sbar^4) \nonumber \\
&&
	+\: 8 C_{u}{}^{\cb }{}_{\sigma }{}^{\db } [2  C_{\sigma \cb \sigma \db } \rbar+  C_{u\db 
	\sigma \cb }  (\rbar^2 + \sbar^2)] \sbar^4+ 8 C_{u\cb \sigma \db } [C_{u}{}^{\cb }
	{}_{\sigma }{}^{\db }  (\rbar^2 + \sbar^2) \nonumber \\
&&
	+\: 2 C_{u}{}^{\cb }{}_{u}{}^{\db } \rbar  (\rbar^2 + 3 \sbar^2)]\sbar^4\} u_{\ab } u_{\bb }
\end{IEEEeqnarray}
and
\par \vspace{-6pt} \begin{IEEEeqnarray}{rCl}
\Big[\hb_{\ab \bb}^{\rm \sing}&&\Big]_{(4)} = 270 (C_{\ab u\bb u}{}_{;}{}_{\sigma \sigma \sigma 
	} -  C_{\ab u\bb u}{}_{;}{}_{u\sigma \sigma } \rbar )\sbar^6 - 1080 (C_{u\sigma u}{}^{\cb }
	{}_{;}{}_{\sigma }  C_{\ab u\bb \cb } + C_{\ab }{}^{\cb }{}_{u\sigma }{}_{;}{}_{\sigma } C_{\bb 
	uu\cb } ) \rbar\sbar^6 \nonumber \\
&&
	-\: 2700 C_{\ab u\bb }{}^{\cb }{}_{;}{}_{\sigma } \rbar C_{u\sigma u\cb } \sbar^6 - 90 
	C_{\ab u\bb u}{}^{;}{}^{\cb }{}_{\cb \sigma } \sbar^8+ 90 C_{\ab u\bb u}{}^{;}{}^{\cb }{}_{\cb 
	u} \rbar \sbar^8 - 360 C_{u}{}^{\cb }{}_{u}{}^{\db }{}_{;}{}_{\sigma } C_{\ab \cb \bb \db } 
	\sbar^8 \nonumber \\
&&
	+\: 720 C_{u}{}^{\cb }{}_{u}{}^{\db }{}_{;}{}_{u} \rbar C_{\ab \cb \bb \db } \sbar^8 + 180 
	C_{u}{}^{\cb }{}_{\sigma }{}^{\db }{}_{;}{}_{\db } C_{\ab u\bb \cb } \sbar^8 + 180 C_{u}{}^{\cb}
	{}_{u}{}^{\db }{}_{;}{}_{\cb } \rbar C_{\ab u\bb \db } \sbar^8 \nonumber \\
&&
	-\: 120 C_{\ab u}{}^{\cb \db }{}_{;}{}_{\sigma } C_{\bb u\cb \db } \sbar^8  - 60 C_{\ab u 
	\sigma }{}^{\cb }{}^{;}{}^{\db } C_{\bb u\cb \db } \sbar^8 + 60 C_{\ab u}{}^{\cb \db }{}_{;}
	{}_{u} \rbar C_{\bb u\cb \db } \sbar^8  \nonumber \\
&&
	-\: 180 C_{\ab }{}^{\cb }{}_{\sigma }{}^{\db }{}_{;}{}_{\db 
	} C_{\bb uu\cb } \sbar^8 - 180 C_{\ab }{}^{\cb }{}_{u}{}^{\db }{}_{;}{}_{\db } \rbar C_{\bb uu\cb } \sbar^8 - 720 
	C_{\ab }{}^{\cb }{}_{\bb }{}^{\db }{}_{;}{}_{\sigma } C_{u\cb u\db } \sbar^8  \nonumber \\
&&
	+\: 360 C_{\ab }
	{}^{\cb }{}_{\bb }{}^{\db }{}_{;}{}_{u} \rbar C_{u\cb u\db } \sbar^8 - 270 C_{u\sigma u\sigma }{}_{;}{}_{\sigma } C_{\ab u\bb u} \sbar^4 (\rbar^2 + \sbar^2) 
	 - 1080 C_{\ab uu\sigma }{}_{;}{}_{\sigma } C_{\bb uu\sigma } \sbar^4 (\rbar^2  \nonumber \\
&&
	+\: \sbar^2) 
	- 540 C_{\ab u\bb u}{}_{;}{}_{\sigma } C_{u\sigma u\sigma } \sbar^4 (\rbar^2 + \sbar^2) 
	- 270 C_{\ab u\bb u}{}_{;}{}_{uuu} \rbar \sbar^6 (\rbar^2 + \sbar^2)  \nonumber \\
&&
	+\: 270 (C_{u\sigma u 
	\sigma }{}_{;}{}_{u}  C_{\ab u\bb u}  + 4 C_{\ab uu\sigma }{}_{;}{}_{u}  C_{\bb uu\sigma }   + 2 C_{\ab u\bb u}{}_{;}{}_{u} 
	C_{u\sigma u\sigma } )\rbar \sbar^4 (\rbar^2 + 3 \sbar^2)   \nonumber \\
&&
	+\: 540 C_{\ab u\sigma }{}^{\cb 
	}{}_{;}{}_{\sigma } ( C_{\bb uu\cb } \rbar  + C_{\bb u\sigma \cb } )\sbar^6 + 90( C_{\ab u\bb u}{}_{;}{}_{uu\sigma }  + 6 C_{u\sigma u}{}^{\cb }{}_{;}{}_{u} C_{\ab u\bb 
	\cb } + 2 C_{\ab }{}^{\cb }{}_{u\sigma }{}_{;}{}_{u} C_{\bb uu\cb }  \nonumber \\
&&
	+\: 6 C_{\ab u\bb }{}^{\cb }
	{}_{;}{}_{u} C_{u\sigma u\cb }) (3 \rbar^2  + \sbar^2) \sbar^6 + 60 C_{\ab uu}{}^{\cb }{}^{;}{}^{\db } (3  C_{\bb \cb u\db } \rbar + 3 C_{\bb \cb \sigma 
	\db }  -   C_{\bb u\cb \db } \rbar ) \sbar^8 \nonumber \\
&&
	+\: 180 C_{\ab u\bb }{}^{\cb }{}^{;}{}^{\db } ( C_{u 
	\cb u \db } \rbar + C_{u\cb \sigma \db }) \sbar^8 - 180 C_{\ab u\sigma }{}^{\cb }{}_{;}{}_{u} [ 3  C_{\bb u\sigma \cb }  \rbar+ C_{\bb uu\cb } 
	(3 \rbar^2  + \sbar^2)] \sbar^6  \nonumber \\
&&
	+\: 180 C_{\ab uu}{}^{\cb }{}_{;}{}_{\sigma } [-3 \rbar C_{\bb 
	\cb u\sigma } + 3 \rbar C_{\bb u\sigma \cb } + C_{\bb uu\cb } (3 \rbar^2  + \sbar^2)]\sbar^6+ 90 C_{\ab u\bb u}{}^{;}{}^{\cb } [3 C_{u 
	\sigma \sigma \cb } \rbar  \nonumber \\
&&
	+\: 2 C_{u\sigma u\cb } (3 \rbar^2 + \sbar^2)]\sbar^6 
	+ 180 C_{\ab u u}{}^{\cb }{}_{;}{}_{u} [3 C_{\bb \cb u\sigma } (3 \rbar^2 + \sbar^2)-3 
	C_{\bb uu\cb }  \rbar (\rbar^2 + \sbar^2)  \nonumber \\
&&
	-\: C_{\bb u\sigma \cb } (3 \rbar^2 + \sbar^2)] 
	\sbar^6 + \big\{36 C_{u}{}^{\cb }{}_{\sigma }{}^{\db }{}^{;}{}^{\eb } C_{\bb \cb \db \eb } \sbar^8 - 
	108 C_{\bb uu\sigma }{}_{;}{}_{\sigma \sigma \sigma } \rbar \sbar^4 + 36 C_{u}{}^{\cb }
	{}_{u}{}^{\db }{}^{;}{}^{\eb } \rbar C_{\bb \cb \db \eb } \sbar^8   \nonumber \\
&&
	+\: 12 C_{\bb }{}^{\cb \db \eb 
	}{}_{;}{}_{\sigma } C_{u\cb \db \eb } \sbar^8 + 24 C_{\bb }{}^{\cb }{}_{\sigma }{}^{\db }{}^{;}{}^{\eb } C_{u\cb \db \eb } \sbar^8 + 12 
	C_{\bb }{}^{\cb \db \eb }{}_{;}{}_{u}  C_{u\cb \db \eb } \rbar \sbar^8 + 24 C_{\bb }{}^{\cb }
	{}_{u}{}^{\db }{}^{;}{}^{\eb }  C_{u\cb \db \eb } \rbar \sbar^8  \nonumber \\
&&
	-\: 90(3 C_{u\sigma u\sigma }
	{}_{;}{}_{\sigma }  C_{\bb uu\sigma } + 4 C_{\bb uu\sigma }{}_{;}{}_{\sigma } C_{u\sigma u\sigma }) \rbar\sbar^2 (\rbar^2 - 3 
	\sbar^2) + 108 (C_{\bb uu\sigma }{}_{;}{}_{u\sigma \sigma }   \nonumber \\
&&
	+\: 7 C_{\bb }{}^{\cb }{}_{u 
	\sigma }{}_{;}{}_{\sigma } C_{u\sigma u\cb }) \sbar^4 (\rbar^2 + \sbar^2) - 108 (C_{\bb uu\sigma }{}_{;}{}_{uu\sigma } +4  C_{\bb }{}^{\cb }{}_{u\sigma }{}_{;}{}_{u} 
	C_{u\sigma u\cb } )\rbar \sbar^4 (\rbar^2 + 3 \sbar^2)  \nonumber \\
&&
	+\: 90  \sbar^2 (3 C_{u\sigma u 
	\sigma }{}_{;}{}_{u} C_{\bb uu\sigma }  + 4 C_{\bb uu\sigma }{}_{;}{}_{u} C_{u\sigma u\sigma }) (\rbar^4 - 6 \rbar^2 \sbar^2 - 3 
	\sbar^4) + 108 C_{\bb uu\sigma }{}_{;}{}_{uuu} \sbar^4 (\rbar^4  \nonumber \\
&&
	+\: 6 \rbar^2 \sbar^2 + 
	\sbar^4) - 216 C_{u}{}^{\cb }{}_{\sigma }{}^{\db }{}_{;}{}_{\sigma } ( C_{\bb \cb u\db } \rbar + C_{\bb 
	\cb \sigma \db }) \sbar^6 -324 C_{\bb }{}^{\cb }{}_{\sigma }{}^{\db }{}_{;}{}_{\sigma } (C_{u 
	\cb u \db }  \rbar  + C_{u\cb \sigma \db } )\sbar^6  \nonumber \\
&&
	+ 8 C_{\bb u}{}^{\cb \db }{}^{;}{}^{\eb } 
	( C_{u\cb \db \eb } \rbar -  C_{u\eb \cb \db } \rbar +  C_{\sigma \cb \db \eb }  -  C_{\sigma \eb \cb \db } ) 
	\sbar^8-216 C_{u\sigma \sigma }{}^{\cb }{}_{;}{}_{\sigma } [  C_{\bb u\sigma \cb } \rbar+ 
	C_{\bb u u\cb }  (\rbar^2  \nonumber \\
&&
	+\: \sbar^2)] \sbar^4 - 252 C_{\bb u\sigma }{}^{\cb }{}_{;}{}_{\sigma } [ \rbar C_{u\sigma \sigma \cb }  +2 
	C_{u\sigma u\cb } (\rbar^2 + \sbar^2)]\sbar^4 +54  C_{u\sigma u}{}^{\cb }{}_{;}{}_{\sigma } 
	[6 C_{\bb \cb u\sigma }  (\rbar^2 + \sbar^2)  \nonumber \\
&&
	-\: 5 C_{\bb u\sigma \cb }  (\rbar^2 
	+ \sbar^2) - 3 \rbar C_{\bb uu\cb }  (\rbar^2 + 3 \sbar^2)]\sbar^4   + 162 C_{u\sigma 
	\sigma }{}^{\cb }{}_{;}{}_{u} [ C_{\bb u\sigma \cb }  (\rbar^2 + \sbar^2)  \nonumber \\
&&
	+\: C_{\bb uu\cb }
	\rbar (\rbar^2  + 3 \sbar^2)] \sbar^4 + 12 C_{\bb uu}{}^{\cb }{}_{;}{}_{\sigma } [-9 C_{u\sigma \sigma \cb }  (\rbar^2 + \sbar^2) 
	- 17 \rbar C_{u\sigma u\cb }  (\rbar^2 + 3 \sbar^2)]\sbar^4  \nonumber \\
&&
	-\:48  C_{\bb uu\sigma }{}^{;}
	{}^{\cb } [3 C_{u\sigma \sigma \cb }  (\rbar^2+ \sbar^2) + 4 \rbar C_{u\sigma u\cb }  (\rbar^2 + 3 \sbar^2)] \sbar^4   + 48 C_{\bb u\sigma }{}^{ 
	\cb }{}_{;}{}_{u} [3 C_{u\sigma \sigma \cb } (\rbar^2 + \sbar^2)  \nonumber \\
&&
	+\: 8  C_{u\sigma u\cb } \rbar 
	(\rbar^2+ 3 \sbar^2)] \sbar^4 + 108 C_{u\sigma u}{}^{\cb }{}_{;}{}_{u} [  3 \rbar C_{\bb u\sigma \cb }  (\rbar^2 + 3 
	\sbar^2) -6 C_{\bb \cb u\sigma } \rbar (\rbar^2 + 3 \sbar^2)  \nonumber \\
&&
	+\: 2 C_{\bb uu\cb } (\rbar^4 
	+ 6 \rbar^2 \sbar^2 + \sbar^4)] \sbar^4 + 36  C_{\bb uu}{}^{\cb }{}_{;}{}_{u} [8  C_{u\sigma \sigma \cb } \rbar (\rbar^2 + 3 
	\sbar^2) + 7 C_{u\sigma u\cb }  (\rbar^4 + 6 \rbar^2 \sbar^2 \nonumber \\
&&
	+\: \sbar^4)] \sbar^4 - 72 
	C_{u}{}^{\cb }{}_{u}{}^{\db }{}_{;}{}_{\sigma } [3 C_{\bb \cb \sigma \db } \rbar + C_{\bb \cb u\db } (3 \rbar^2  + \sbar^2)] \sbar^6+ 108 C_{u}{}^{\cb }{}_{\sigma }{}^{ 
	\db }{}_{;}{}_{u} [3 C_{\bb \cb \sigma \db } \rbar  \nonumber \\
&&
	+\: C_{\bb \cb u\db } (3 \rbar^2 + \sbar^2)] 
	\sbar^6 + 108 C_{u}{}^{\cb }{}_{u}{}^{\db }{}_{;}{}_{u} [3 C_{\bb \cb u\db } \rbar (\rbar^2 + \sbar^2) 
	+ C_{\bb \cb \sigma \db } (3 \rbar^2 + \sbar^2)]\sbar^6   \nonumber \\
&&
	-\:108 C_{\bb }{}^{\cb }{}_{u}{}^{ 
	\db }{}_{;}{}_{\sigma } [3  C_{u\cb \sigma \db }  \rbar + C_{u\cb u\db } (3 \rbar^2 + \sbar^2)] \sbar^6 + 24 C_{\bb u\sigma }{}^{\cb }{}^{;}
	{}^{\db } [3  C_{u\cb \sigma \db } \rbar + 3  C_{u\db \sigma \cb }\rbar   \nonumber \\
&&
	+\: 3 C_{\sigma \cb 
	\sigma \db }  +  C_{u\cb u\db } (3 \rbar^2  + \sbar^2)]\sbar^6 + 72 C_{\bb }{}^{\cb }{}_{\sigma }{}^{\db }{}_{;}{}_{u} [3 \rbar C_{u\cb \sigma \db }+ C_{u 
	\cb u \db } (3 \rbar^2  + \sbar^2)] \sbar^6  \nonumber \\
&&
	+\: 72 C_{\bb }{}^{\cb }{}_{u}{}^{\db }{}_{;}{}_{u} [3 
	C_{u \cb u\db} \rbar (\rbar^2 + \sbar^2) + C_{u \cb \sigma \db } (3 \rbar^2 + \sbar^2)] \sbar^6 + 24 C_{\bb uu}{}^{\cb }{}^{;}{}^{\db } [3  C_{\sigma \cb \sigma \db } 
	\rbar  \nonumber \\
&&
	+\: 3  C_{u\cb u\db } \rbar (\rbar^2 + \sbar^2) +  C_{u\cb \sigma \db } (3 \rbar^2 + 
	\sbar^2) +  C_{u\db \sigma \cb } (3 \rbar^2 + \sbar^2)] \sbar^6\big\} u_{\ab } + \{-18 (C_{\sigma }{}^{\cb }{}_{\sigma }{}^{\db }{}_{;}
	{}_{\sigma }  \nonumber \\
&&
	-\:  C_{\sigma }{}^{\cb }{}_{\sigma }{}^{\db }{}_{;}{}_{u} \rbar) C_{\sigma \cb 
	\sigma \db } \sbar^4 + 27 C^{\cb \db \eb \fb }{}_{\sigma } C_{\cb \db \eb \fb } \sbar^8 
	- 18 C^{\cb \db \eb \fb }{}_{u} \rbar C_{\cb \db \eb \fb } \sbar^8 + 9 C_{u\sigma u 
	\sigma }{}_{;}{}_{u\sigma \sigma } \rbar \sbar^2 (\rbar^2  \nonumber \\
&&
	-\: 3 \sbar^2) - 15 C_{u\sigma u 
	\sigma }{}^{;}{}^{\cb } \rbar C_{u\sigma \sigma \cb } \sbar^2 (\rbar^2 - 3 \sbar^2) 
	- 135 C_{u\sigma u\sigma }{}_{;}{}_{\sigma } C_{u\sigma u\sigma } (\rbar^2 -  \sbar^2)^2 
	 \nonumber \\
&&
	+\: 9 C_{u\sigma u\sigma }{}_{;}{}_{uuu} \rbar \sbar^2 (\rbar^4 - 10 \rbar^2 \sbar^2 - 15 
	\sbar^4) + 9 C_{u\sigma u\sigma }{}_{;}{}_{uu\sigma } \sbar^2 (3 \sbar^4 + 6 \rbar^2 \sbar^2 - 
	\rbar^4 )  \nonumber \\
&&
	+\: 15 C_{u\sigma u\sigma }{}^{;}{}^{\cb } C_{u\sigma u\cb } \sbar^2 (- \rbar^4 + 
	6 \rbar^2 \sbar^2 + 3 \sbar^4) + C_{u\sigma u\sigma }{}_{;}{}_{\sigma \sigma \sigma } (-9 \rbar^2 \sbar^2 + 9 \sbar^4) 
	 \nonumber \\
&&
	+\: 45 C_{u\sigma u\sigma }{}_{;}{}_{u} C_{u\sigma u\sigma } (3 \rbar^5 - 10 \rbar^3 
	\sbar^2 + 15 \rbar \sbar^4) - 54 C_{u\sigma \sigma }{}^{\cb }{}_{;}{}_{\sigma } [2  C_{u\sigma u\cb } \rbar  (\rbar^2 - 
	3 \sbar^2)  \nonumber \\
&&
	+\:  C_{u\sigma \sigma \cb }  (\rbar^2 - \sbar^2)]\sbar^2 + 36 C_{u\cb \sigma 
	\db} \sbar^4 [- C_{\sigma }{}^{\cb }{}_{\sigma }{}^{\db }{}_{;}{}_{\sigma } \rbar + C_{\sigma }{}^{\cb }{}_{\sigma }{}^{\db }{}_{;}{}_{u} (\rbar^2 + \sbar^2)]  \nonumber \\
&&
	-\: 18 C_{u\cb u 
	\db } \sbar^4 [C_{\sigma }{}^{\cb }{}_{\sigma }{}^{\db }{}_{;}{}_{\sigma } (\rbar^2 + \sbar^2)-  
	C_{\sigma }{}^{\cb}{}_{\sigma }{}^{\db }{}_{;}{}_{u} \rbar (\rbar^2 + 3 \sbar^2)] - 36 C_{u}{}^{\cb }{}_{\sigma }{}^{\db }{}_{;}{}_{\sigma } [ C_{\sigma \cb \sigma \db } \rbar 
	 \nonumber \\
&&
	+\: C_{u\cb \sigma \db }  (\rbar^2 + \sbar^2) + C_{u\db \sigma \cb }  (\rbar^2+ \sbar^2) + 
	C_{u \cb u\db } \rbar (\rbar^2 + 3 \sbar^2)] \sbar^4 + 15 C_{u\sigma \sigma }{}^{\cb }{}_{;}{}_{u} [2  C_{u\sigma \sigma \cb } \rbar (\rbar^2  \nonumber \\
&&
	-\: 
	3 \sbar^2) + 5 C_{u\sigma u\cb }  (\rbar^4 - 6 \rbar^2 \sbar^2 - 3 \sbar^4)] \sbar^2 
	+ 18 C_{u\sigma u}{}^{\cb }{}_{;}{}_{u} [3 \rbar C_{u\sigma u\cb }  (\rbar^4- 10 \rbar^2 
	\sbar^2 - 15 \sbar^4)  \nonumber \\
&&
	+\: 5 C_{u\sigma \sigma \cb }  (\rbar^4  - 6 \rbar^2 \sbar^2 - 3 
	\sbar^4)]\sbar^2 - 18 C_{u}{}^{\cb }{}_{u}{}^{\db }{}_{;}{}_{\sigma } [ C_{\sigma \cb \sigma \db }  (\rbar^2 + 
	\sbar^2) + 2  C_{u\cb \sigma \db }  \rbar (\rbar^2 + 3 \sbar^2)   \nonumber \\
&&
	+\: 18 C_{u\cb u\db }  ( 
	\rbar^4 + 6 \rbar^2 \sbar^2 + \sbar^4)]\sbar^4 + 36 C_{u}{}^{\cb }{}_{\sigma }{}^{\db }{}_{;}{}_{u} [C_{\sigma \cb \sigma \db }  (\rbar^2 + 
	\sbar^2) + C_{u\cb \sigma \db }\rbar (\rbar^2  + 3 \sbar^2)  \nonumber \\
&&
	+\:   C_{u\db \sigma \cb } 
	\rbar ( \rbar^2 + 3 \sbar^2) +  C_{u\cb u\db } (\rbar^4 + 6 \rbar^2 \sbar^2 + \sbar^4)]  \sbar^4 + C_{u\sigma u}{}^{\cb }{}_{;}{}_{\sigma } [-48 
	C_{ u\sigma \sigma \cb }  \rbar (\rbar^2 - 3 \sbar^2)  \nonumber \\
&&
	+\: 39 C_{u\sigma u\cb }  (- \rbar^4 + 
	6 \rbar^2 \sbar^2 + 3 \sbar^4)] \sbar^2 + 18 C_{u}{}^{\cb }{}_{u}{}^{\db }{}_{;}{}_{u} [ C_{\sigma \cb \sigma \db } \rbar (\rbar^2 + 3 
	\sbar^2) + 2 C_{u\cb \sigma \db }  (\rbar^4  \nonumber \\
&&
	+\: 6 \rbar^2 \sbar^2 + \sbar^4) + C_{u \cb u \db } \rbar (\rbar^4 + 10 \rbar^2 \sbar^2  + 5 \sbar^4)]\sbar^4\big\} 
	u_{\ab } u_{\bb }.
\end{IEEEeqnarray}


\chapter{Mode-Sum Decomposition of the Singular Field} \label{sec: modeSum}



The singular field expansions derived in the previous sections have several applications in
explicit self-force calculations. One of the most successful computational approaches to date
is the \emph{mode-sum} scheme of Barack and Ori
\cite{Barack:Ori:2000, Barack:Mino:Nakano:Ori:Sasaki:2001}; the majority of existing calculations
are based on it in one form or another
\cite{Barack:Burko:2000, Burko:2000b,Detweiler:Messaritaki:Whiting:2002, DiazRivera:2004, Haas:Poisson:2006, Haas:2007, Canizares:Sopuerta:2009, Canizares:Sopuerta:Jaramillo:2010, Barack:Sago:2007, Barack:Lousto:2002, Sago:Barack:Detweiler:2008, Detweiler:2008, Sago:2009, Barack:Sago:2010, Warburton:Barack:2009, Warburton:Barack:2010, Thornburg:2010, Haas:2011bt, Warburton:2011fk, Hopper:2010uv, Keidl:2010pm,Shah:2010bi}.  The basic idea is to decompose the singular retarded field into spherical harmonic modes which are continuous and finite in general for the scalar case and in Lorenz gauge for the electromagnetic and gravitational cases. A key component of the calculation involves the subtraction of so-called \emph{regularization parameters} - analytically derived expressions which render the formally divergent sum over spherical harmonic modes finite.  In this section, we derive these parameters from our singular field expressions and show how they may be used to compute the self-force with unprecedented accuracy.


\section{Mode Sum Concept}
The self-force, for each case can be represented by Eq.~\eqref{eqn:SelfForce}, or alternatively as
\begin{equation} 
F_a = p_a{}_A \varphi^{A}_{\rm{\reg}}, \quad \text{where} \quad \varphi^{A}_{\rm{\reg}} = \varphi^{A}_{\rm{\ret}} - \varphi^{A}_{\rm{\sing}},
\end{equation}
is the regularised field and $p_a{}_A(x)$ is a tensor at $x$ and depends on the type of charge.  We can therefore rewrite the self-force as
\begin{equation}
F_a = p_a{}_A \varphi^{A}_{\rm{\ret}} - p_a{}_A \varphi^{A}_{\rm{\sing}}.
\end{equation}
\fixme{Carrying} out a spherical harmonic decomposition on the field terms, i.e.,
\begin{equation}
\varphi^{A}_{\rm{\ret / \sing}} = \sum_{lm}^{\infty} \varphi^{lm}{}^{A}_{\rm{\ret / \sing}} Y^{lm},
\end{equation}
allows the self-force to be rewritten as,
\begin{equation}
F_a = \sum_{lm}^{\infty} \left(p_a{}_A \varphi^{lm}{}^{A}_{\rm{\ret}} - p_a{}_A \varphi^{lm}{}^{A}_{\rm{\sing}} \right).
\end{equation}
By defining the $l$ component of the retarded or singular self-force to be,
\begin{equation} \label{eqn: FretSing}
F_a^l{}_{\rm \ret / \sing} = p_a{}_A  \sum _{m=-l}^{l} \varphi^{lm}{}^{A}_{\rm{\ret / \sing}},
\end{equation}
the self-force can be expressed as
\begin{equation} \label{eqn: FaSplit}
F_a = \sum^l \left( F_a^l{}_{\rm \ret}  -  F_a^l{}_{\rm \sing}  \right).
\end{equation}
It is the last term on the right that we calculate in this chapter for each of the 3 cases, in both Kerr and \Sch space-times.  

As our singular field is an expansion, it is written in terms of order $\epsilon$,  and evaluated at $x'$, that is
\par \vspace{-6pt} \begin{IEEEeqnarray}{rCl}
 F^a_l{}_{\rm \sing} &=& F^l_{a\lnpow{1}}\left(r_0,t_0\right) + F^l_{a[0]}\left(r_0,t_0\right) + F^l_{a\lpow{2}}\left(r_0,t_0\right)  \nonumber \\
&&
	+\: F^l_{a\lpow{4}}\left(r_0,t_0\right) + F^l_{a\lpow{6}}\left(r_0,t_0\right) +  \dots,
\end{IEEEeqnarray}
where we are missing odd \fixme{orders above $-1$}, as these are zero - this will be shown to be the case later in this chapter.  When summed over $l$, the contribution of $ F^l_{a\lpow{2}}\left(r_0,t_0\right)$ and higher terms to the self-force is \fixme{also} zero.  However, if we ignore these higher terms in the approximation of $\varphi^{lm}{}^{A}_{\rm{\sing}}$, \fixme{our resulting expression for} $\varphi^{lm}{}^{A}_{\rm{\reg}}$ is only $C^1$, meaning we can only differentiate it once, which is not \fixme{sufficient} for $\varphi^{lm}{}^{A}_{\rm{\reg}}$ to be a solution of the homogeneous wave equation, Eq.~\eqref{eqn: varphiReg}.  
When it comes to numerically calculating the self-force using the mode-sum method, the inclusion of the higher order terms dramatically speeds up computation times.   For this reason, every extra term or regularisation parameter that can be calculated is of great benefit to the self-force community.


\section{Rotated Coordinates}
In order to obtain expressions which are readily written as mode-sums, previous calculations
\cite{Barack:Ori:2000,Detweiler:Messaritaki:Whiting:2002,Haas:Poisson:2006} found it useful
to work in a rotated coordinate frame.  We found it most efficient to carry out this rotation
prior to doing any calculations.  To this end, we introduce Riemann normal coordinates on the 2-sphere at $\xb$ in the
form
\begin{equation}
w_{1} = 2 \sin\left(\frac{\alpha}{2}\right) \cos\beta, \quad\quad w_2 = 2 \sin\left(\frac{\alpha}{2}\right) \sin\beta,
\end{equation}
where $\alpha$ and $\beta$ are rotated angular coordinates given by
\begin{eqnarray}
\sin \theta \cos \phi &=& \cos \alpha, \\
\sin \theta \sin \phi &=& \sin \alpha \cos \beta, \\
\cos \theta &=& \sin \alpha \sin \beta.
\end{eqnarray}
In these coordinates, the \Sch metric is given by the line element
\begin{align}
ds^2 =& -\left(\frac{r - 2 M}{r}\right) dt^2 + \left(\frac{r}{r - 2 M}\right) dr^2 +
r^2 \Bigg\{\left[\frac{16 - w_2^2 \left(8 - w_1^2 - w_2^2\right)}{4 \left(4 - w_1^2 - w_2^2 \right)}\right] dw_1^2 \nonumber \\ 
& \qquad + 2 dw_1 dw_2 \left[ \frac{w_1 w_2 \left( 8 - w_1^2 - w_2^2 \right)}{4 \left( 4 - w_1^2 - w_2^2\right)} \right]
+ \left[ \frac{16 - w_1^2 \left(8 - w_1^2 - w_2^2\right)}{4 \left(4 - w_1^2 - w_2^2 \right)}\right] dw_2^2\Bigg\}.
\end{align}
The algebraic form of the metric makes it very suitable for using with computer algebra programmes such as Mathematica. 
The apparent complexity of having a non-diagonal metric on $S^2$ is in fact minimal since the determinant of that
metric is simply $1$.  

The Kerr metric in these coordinates is given by the line element
\begin{align}
ds^2 = & \left[\frac{8 M r}{ 4 r^2 +a^2 w_2^2 \left(4 - w_1^2 - w_2^2 \right)} - 1 \right] dt^2
	+  \left[\frac{4 r^2 + a^2 w_2^2 \left(4 - w_1^2 - w_2^2 \right)}{4 \left(r^2 - 2 M r + a^2
	\right)} \right] dr^2 \nonumber \\
&
	+\: 2 dt dw_1 \left\{\frac{-2 a M r \left[8 - w_2^2 \left(6 - w_1^2 - w_2^2 \right) \right]}
	{\sqrt{4 - w_1^2 - w_2^2} \left[4 r^2 +a^2 w_2^2 \left(4 - w_1^2 - w_2^2 \right) \right]} 
	\right\}  \nonumber \\
&
	+\:2 dt dw_2 \left\{ \frac{2 a M r w_1 w_2 \left(6 - w_1^2 - w_2^2 \right)}{\sqrt{4 - w_1^2 
	- w_2^2} \left[4 r^2 +a^2 w_2^2 \left(4 - w_1^2 - w_2^2 \right) \right]} \right\} 
	\nonumber \\
&	
	+\: g_{w_1 w_1} dw_1^2 + 2 g_{w_1 w_2} dw_1 dw_2 + g_{w_2 w_2} dw_2^2,
\end{align}
where
\par \vspace{-6pt} \begin{IEEEeqnarray}{rCl}
g_{w_1 w_1} &=& \frac{1}{4 \left(4 - w_1^2 - w_2 ^2 \right) \left[4 - w_2^2 \left(4 - w_1^2 - 
	w_2^2 \right) \right]} \Bigg(w_1^2 w_2^2 \left[4 r^2 +a^2 w_2^2 \left(4 - w_1^2 - w_2^2 
	\right) \right] \nonumber \\
&&
	+\: \left[8 - w_2^2 \left(6 - w_1^2 - w_2^2 \right) \right]^2 \Bigg\{r^2 + a^2  \nonumber \\
&&
	+\: 2 M a^2 r 
	\left[ \frac{4 - w_2^2 \left(4 - w_1^2 - w_2^2 \right) }{4 r^2 + a^2 w_2^2 \left(4 - w_1^2 
	- w_2^2 \right) } \right] \Bigg\} \Bigg), \nonumber \\
g_{w_1 w_2} &=& \frac{1}{4 \left(4 - w_1^2 - w_2 ^2 \right) \left[4 - w_2^2 \left(4 - w_1^2 - 
	w_2^2 \right) \right]} \Bigg(w_1 w_2 \left(w_1^2 + 2 w_2^2 - 4 \right) \big[4 r^2 +a^2 
	w_2^2 \big(4 \nonumber \\
&&
	-\: w_1^2 - w_2^2 \big) \big] + w_1 w_2 \left(6 - w_1^2 - w_2^2 \right) \left[8 - w_2^2 
	\left(6 - w_1^2 - w_2^2 \right) \right] \Bigg\{r^2 + a^2 \nonumber \\
&&
	+\: 2 M a^2 r \left[ \frac{4 - w_2^2 \left(4 - w_1^2 - w_2^2 \right) }{4 r^2 + a^2 w_2^2 
	\left(4 - w_1^2 - w_2^2 \right) } \right] \Bigg\} \Bigg), \nonumber \\
g_{w_2 w_2} & = & \frac{1}{4 \left(4 - w_1^2 - w_2 ^2 \right) \left[4 - w_2^2 \left(4 - w_1^2 - 
	w_2^2 \right) \right]} \Bigg( \left(4 - w_1^2 - w_2^2 \right)^2 \big[4 r^2 +a^2 
	w_2^2 \big(4 - w_1^2 \nonumber \\
&&
	-\: w_2^2 \big) \big] + w_1^2 w_2^2 \left(6 - w_1^2 - w_2^2 \right)^2 \Bigg\{r^2 + a^2 
	\nonumber \\
&&
	+\: 2 M a^2 r \left[ \frac{4 - w_2^2 \left(4 - w_1^2 - w_2^2 \right) }{4 r^2 + a^2 w_2^2 
	\left(4 - w_1^2 - w_2^2 \right) } \right] \Bigg\} \Bigg).
\end{IEEEeqnarray}
As in the \Sch case, this algebraic form has an advantage over its trigonometric counterpart in computer algebraic programmes where trigonometric functions tend to slow calculations down.  The complexity of the Kerr metric does slow down the calculation of the singular field.  However, despite this, rotating the metric and then calculating the singular field and its resulting regularization parameters still remains faster than calculating the singular field in regular Kerr co-ordinates (such as Boyer-Lindquist) and then rotating the resulting complicated expression to obtain the desired regularisation parameters.


\section{Mode decomposition} \label{sec: modeDecomp}
The method of regularization of the self force through $l$-mode decomposition is by now
standard, see, for example, \cite{Barack:Ori:2000}, \cite{Detweiler:Messaritaki:Whiting:2002}
and \cite{Haas:Poisson:2006}.  Having calculated the singular field, it is straightforward
to calculate the component of the self-force that arises from the singular field\footnote{In this section, for notational convenience
we drop the implied $\sing$ superscript denoting ``singular'' as we are always referring
to the singular component.}, $F_a$, for scalar, electromagnetic and gravitational cases using Eqs.~\eqref{eqn:SelfForceScalar}, \eqref{eqn:SelfForceEM} and \eqref{eqn:SelfForceGravityBasic} with the singular field substituted for the regular field.  We study the multipole decomposition of $F_a$ by writing
\begin{equation}\label{eqn:falpha}
F_a \left(r, t, \alpha, \beta \right) = \sum_{lm} F_a^{lm} \left(r, t \right) Y^{lm} \left( \alpha, \beta \right),
\end{equation}
where $Y^{lm} \left( \theta, \phi \right)$ are scalar spherical harmonics, and accordingly
\begin{equation} \label{eqn:flm}
F_a^{lm} \left(r, t\right) = \int F_a \left(r, t, \alpha, \beta \right) Y^{lm*} \left( \alpha, \beta \right) d \Omega.
\end{equation}
To calculate the $l$-mode contribution at $\xb = \left( t_0, r_0, \alpha_0, \beta_0 \right)$, we have
\begin{equation} \label{eqn:falphal}
F_a^l \left(r_0, t_0 \right) = \lim_{\Delta r \rightarrow 0} \sum_{m} F_a^{lm} \left(r_0+\Delta r, t_0 \right) Y^{lm} \left( \alpha_0, \beta_0 \right).
\end{equation}

In previous calculations, Eq. (\ref{eqn:falpha}) has naturally arisen in \Sch coordinates with $\theta_0 = \frac{\pi}{2}$, and it was necessary to perform a rotation to move the coordinate location of the particle from the equatorial plane to a pole in the new coordinate system.  However, by choosing to work in an $S^2$ Riemann normal coordinate system from the start, our particle is already located on the pole.  This saves us from further transformation and expansions at this stage.  With the particle on the pole, $Y^{lm} \left( \alpha_0 = 0, \beta_0 \right) = 0$  for all $m \neq 0$.  This also allows us, without loss of generality, to take $\beta_0 = 0$.  Taking $\alpha_0$, $\beta_0$ and $m$ all to be equal to zero in Eq.~(\ref{eqn:falphal}) gives us
\par \vspace{-6pt} \begin{IEEEeqnarray}{rCl} \label{eqn:fla}
F_a^l \left(r_0, t_0\right) &=& \lim_{\Delta r \rightarrow 0} \sqrt{\frac{2 l + 1}{4 \pi}} F_a^{l, m = 
	0} \left(r_0+\Delta r, t_0 \right) \nonumber \\
&=& 
	\frac{2 l + 1}{4 \pi} \lim_{\Delta r \rightarrow 0} \int F_a \left( r_0+\Delta r, t_0, \alpha, 
	\beta \right) P_l \left( \cos \alpha \right) d \Omega .
\end{IEEEeqnarray}

For each spin field, the singular self-force, $F_a \left( r, t, \alpha, \beta \right)$, has the form
\begin{equation}
\label{eqn:fasum} 
F_a \left( r, t, \alpha, \beta \right) = \sum_{n=1}^{\infty} \frac{B_a^{(3 n -2)}}{ \zrho^{2 n + 1}} \epsilon^{n-3},
\end{equation}
where $B_a^{(k)} = b_{a_1 a_2 \dots a_k}(\bar{x}) \Delta x^{a_1} \Delta x^{a_2} \dots \Delta x^{a_k}$.  On identifying $\tau_1 = \rbar_{(1)} \pm \zrho$, this form can be easily seen to follow from the coordinate representation of the above expressions for the singular field. In using Eq.~(\ref{eqn:fasum}) to determine the regularization parameters, we only need to take the sum to the appropriate order: $n=1$ for $A_a$, $n=2$ for $B_a$, etc.

Explicitly, in our coordinates 
$\zrho = \sqrt{(g_{\alphab \betab} u^{\alphab} \Delta x^b)^2 +g_{\alphab \betab} \Delta x^a \Delta x^b}$ takes the form
\begin{align} \label{eqn: rhoSch}
\zrho \left(r, t, \alpha, \beta \right)^2 =& \frac{\left(E^2 \rb^3-L^2 (\rb-2M)\right)}{\rb (\rb-2 
	M) ^2}\Delta r^2 +\left(L^2+\rb^2\right)\Delta w_1^2  \nonumber \\
&
	-\:  \Bigg(\frac{2 E \rb \rbdot}{\rb - 
	2M} \Delta r 
	+ 2 E L\Delta w_1 \Bigg)\Delta t +\frac{2  L \rb \rbdot}{\rb-2M} \Delta r \Delta w_1  \nonumber \\
&
	+\:  
	\left( E^2+\frac{2 M}{\rb}-1\right)\Delta t^2 +\rb^2\Delta w_2^2 ,
\end{align}
in Schwarzschild space-time, and
\begin{align}
\zrho \left(r, t, \alpha, \beta \right)^2 =&\frac{\Delta r^2 \rb \left[\rb \left(a^2 E^2-L^2 \right) 
	+ 2 M (L-a E)^2+E^2 \rb^3\right]}{\left(a^2-2 M \rb +\rb^2\right)^2} \nonumber \\
&
	+\: \Delta t \Bigg[ 
	\Delta w_1 \Big(-\frac{4 a M}{\rb} - 2 E L\Big)-\frac{2 \Delta r E \rb^2 \rbdot}{a^2-2 M \rb+\rb^2}\Bigg] \nonumber \\
&
	+\: \Delta w_1^2 
	\left( \frac{2 a^2 M}{\rb}+a^2+L^2+\rb^2\right) 
	+ \frac{2 \Delta r \Delta w_1 L \rb^2 \rbdot}{a^2 - 2 M \rb+\rb^2} \nonumber \\
&
	+\: \Delta t^2 
	\left(E^2+\frac{2 M}{\rb}-1\right)+\Delta w_2^2 \rb^2,
\end{align}
in Kerr space-time, where the $\alpha$, $\beta$ dependence is contained exclusively in $\Delta w_1$ and $\Delta w_2$, and $E = -u_t$ and $L = u_{\phi}$ are the energy per unit mass and angular momentum along the axis of symmetry, respectively.  In particular, taking $t = t_0$ ($\Delta t = 0$) allows us to write
\par \vspace{-6pt} \begin{IEEEeqnarray}{rCl}\label{eqn: rhotSchwar}
\zrho \left(r, t_0, \alpha, \beta \right)^2 &&= \frac{E^2  \rb^4}{\left(L^2+\rb^2\right) (\rb-2 
	M)^2}\Delta r^2\nonumber \\
&&
	 +\: \left(L^2+\rb^2\right)\left(\Delta w_1 + \frac{L \rb \rbdot}{\left( \rb -2M\right)\left( 
	L^2 + \rb^2 \right)} \Delta r\right)^2+\rb^2\Delta w_2^2,
\end{IEEEeqnarray}
in Schwarzschild space-time, and
\par \vspace{-6pt} \begin{IEEEeqnarray}{rCl}\label{eqn: rhotKerr}
\zrho \left(r, t_0, \alpha, \beta \right)^2 &=& \frac{\Delta r^2 \rb \left[E \rb \left( a^2 + \rb^2 
	\right) +2 a M (a E-L)\right]^2}{\left(a^2-2 M \rb+\rb^2\right)^2 \left[\rb \left( a^2 + L^2 
	\right) +2 a^2 M+\rb^3\right]}+\Big(\frac{2 a^2 M}{\rb}+a^2+L^2 \nonumber\\
&&
	+\: \rb^2\Big) \left[\Delta w_1 + \frac{ \Delta r L \rb^3 \rbdot}{\left(a^2-2 M \rb +\rb^2 
	\right) \left( 2 a^2 M+a^2 \rb+L^2 \rb + \rb^3 \right)} \right]^2 \nonumber \\
&&
	+\: \Delta w_2^2 \rb^2,
\end{IEEEeqnarray}
in Kerr space-time.

For mode-sum decomposition, it is favourable to get $\zrhoz\left(\alpha, \beta \right)^2 \equiv  \zrho\left(r_0, t_0, \alpha, \beta \right)^2$ in the form
\begin{equation} \label{eqn: rhoz}
\zrhoz\left(\alpha, \beta \right)^2 = 2 \left(1 - \cos{\alpha}\right) \zeta^2 \left(1 - k \sin^2{\beta} \right)
\end{equation}
This is can be done by rewriting Eqs.~\eqref{eqn: rhotSchwar} and \eqref{eqn: rhotKerr} with $\Delta r \rightarrow 0$ as
\begin{equation}
\zrhoz\left(\alpha, \beta \right)^2  = \zeta^2 \Delta w_1 ^2+\rb^2\Delta w_2^2,
\end{equation}
where
\begin{equation}
\zeta^2 = L^2 + \rb^2 \quad \quad \text{and} \quad \quad \zeta^2 = L^2 + \rb^2 +\frac{2 a^2 M}{\rb} + a^2 
\end{equation}
in Schwarzschild and Kerr space-times respectively, and rearranging to give
\begin{equation}
\zrhoz\left(\alpha, \beta \right)^2  = 2 \left(1 - \cos{\alpha}\right) \zeta^2 \left[1 - \left(\frac{\zeta^2 - \rb^2}{\zeta^2}\right) \sin^2{\beta} \right].
\end{equation}
which is equivalent to Eq.~\eqref{eqn: rhoz} with $k = \frac{\zeta^2-\rb^2}{\zeta^2}$.  Defining $\chi(\beta) \equiv 1 - k \sin^2 \beta$, we can rewrite our $\Delta w$'s in the alternate forms

\begin{align}
\Delta w_1 ^2 &= 2 \left( 1-\cos{\alpha}\right) \cos^2 \beta = \frac{\zrhoz^2}{\zeta^2 \chi}\cos^2\beta= \frac{\zrhoz^2}{\left(\zeta^2-\rb^2\right)\chi}\left[k-(1-\chi)\right], \\
 \Delta w_2^2 &= 2 \left( 1-\cos{\alpha}\right) \sin^2 \beta = \frac{\zrhoz^2}{\zeta^2 \chi}\sin^2\beta= \frac{\zrhoz^2}{\left(\zeta ^2 - \rb^2 \right)\chi} (1 - \chi),
\end{align}

Suppose, for the moment, that we may take the limit in Eq.~(\ref{eqn:fla}) through the integral sign, then using our alternate forms we have
\begin{equation}
\lim_{\Delta r \to 0} \frac{B_a^{(3 n -2)}}{ \zrho^{2 n + 1}} \epsilon^{n-3} = \frac{b_{i_1 i_2 \dots i_{3n-2}}(\rb) \Delta w^{i_1} \Delta w^{i_2} \dots \Delta w^{i_{3n-2}}} { \zrhoz^{2 n + 1}} \epsilon^{n-3} = \zrhoz^{n -3 } \epsilon^{n-3} c_{a(n)}(\rb,\chi).
\end{equation}
In \cite{Barack:Ori:2002}, it was shown that the integral and limit in Eq.(\ref{eqn:fla}) are indeed interchangeable for all orders except the leading order, $n=1$ term, where the limiting $\zrhoz^{-3 }$ would not be integrable. Thus we
find the singular self-force now has the form
\begin{align} \label{eqn:fa}
F_a^l \left(r_0, t_0 \right) =& \frac{2 l + 1}{4 \pi}\Bigg[\epsilon^{-2}  \lim_{\Delta r \rightarrow 0} \int\frac{B_a^{(1)} \left(r, t_0, \alpha, \beta \right)}{\zrho^3 \left(r, t_0, \alpha, \beta \right)} P_l \left( \cos \alpha \right) d \Omega \nonumber \\
& + \sum_{n=2}^{\infty} \epsilon^{n-3} \int \zrhoz^{n-3} c_{a (n)} \left(r_0, \chi \right) P_l \left( \cos \alpha \right) d \Omega \Bigg]\nonumber\\
\equiv & F^l_{a\lnpow{1}}\left(r_0,t_0\right) \epsilon^{-2} + F^l_{a[0]}\left(r_0,t_0\right) \epsilon^{-1} + F^l_{a\lpow{2}}\left(r_0,t_0\right) \epsilon^1 + F^l_{a\lpow{4}}\left(r_0,t_0\right) \epsilon^3 \nonumber \\
& + F^l_{a\lpow{6}}\left(r_0,t_0\right) \epsilon^5 +  \dots,
\end{align}
where the $\beta$ dependence in the $c_n$'s are hidden in $\chi$, and the $\alpha$, $\beta$ dependence of $F_a \left(r, t_0, \alpha, \beta \right)$ is hidden in both the $\zrho$'s and $c_n$'s. Note here that we use the convention that
a subscript in square brackets denotes the term which will contribute at that order in $1/l$. Furthermore the integrand in the summation is odd or even
under $\Delta w_i \to - \Delta w_i$ according to whether $n$ (and so $3n-2$) is odd or even.  This means only the even terms are non-vanishing, while
$F^l_{a\lpow{1}}\left(r_0,t_0\right) = F^l_{a\lpow{3}}\left(r_0,t_0\right)= F^l_{a\lpow{5}}\left(r_0,t_0\right)=0$ etc.   

Some care is required in dealing with taking the limit in the first term. This has been addressed previously in \Sch space-time
\cite{Detweiler:Messaritaki:Whiting:2002, Barack:Ori:2002,Mino:Nakano:Sasaki:2002, Haas:Poisson:2006} which we will now extend to Kerr space-time.  As is standard for the \Sch first order, we shift our $\Delta w_1$ coordinate to enable us to remove the cross-terms $\Delta r \Delta w_1$, by setting $\Delta w_1 \rightarrow \Delta w_1 + \mu \Delta r$, where $\mu$ can easily be read off from Eqs.~\eqref{eqn: rhotSchwar} and \eqref{eqn: rhotKerr}, to be
\par \vspace{-6pt} \begin{IEEEeqnarray}{rCl}
\mu_{Schwar} &=& \frac{- L \rb \rbdot}{\left( \rb -2M\right)\left( L^2 + \rb^2\right)}, \\
\mu_{Kerr} &=& \frac{ - L \rb^3 \rbdot}{\left(a^2-2 M \rb +\rb^2 \right) \left( 2 a^2 M+a^2 
	\rb+L^2 \rb + \rb^3 \right)}
\end{IEEEeqnarray}
for \Sch and Kerr space-times respectively.  This allows us to write
\begin{align}
\zrho \left(r, t_0, \alpha, \beta \right)^2 &= \nu^2 \Delta r^2 + \zeta^2 \Delta w_1^2 + \rb^2 \Delta w_2^2 \nonumber \\
&=  \nu^2 \Delta r^2 + 2 \chi \zeta^2 \left(1 - \cos{\alpha}\right)
\end{align}
where the expressions for $\nu$ for \Sch and Kerr space-times can be read off from Eqs.~\eqref{eqn: rhotSchwar} and\eqref{eqn: rhotKerr}  respectively.  This can be easily rearranged to give
\par \vspace{-6pt} \begin{IEEEeqnarray}{rCl}
\zrho \left(r, t_0, \alpha, \beta \right)^{-3} &=& \zeta^{-3} \left(2 \chi \right)^{-3/2} \left(\delta^2 + 
	1 - \cos{\alpha} \right)^{-3/2} \nonumber \\
&=&
	\zeta^{-3} \left(2\chi \right)^{-3/2} \sum_{l=0} \mathcal{A}_l^{{-3}/{2}} (\delta) P_l \left( \cos \alpha \right),
\end{IEEEeqnarray}
where 
\begin{equation}
\delta^2 = \frac{\nu^2 \Delta r^2}{2 \zeta^2 \chi}   \quad \quad \text{and} \quad \quad  \mathcal{A}_l^{-\frac{3}{2}} (\delta) = \frac{2 l + 1}{\delta}
\end{equation} 
Here $\mathcal{A}_l^{-\frac{3}{2}} (\delta)$ is derived from the generating function of the Legendre polynomials as shown in Eq.~(D12) of \cite{Detweiler:Messaritaki:Whiting:2002}.  $\zrho \left(r, t_0, \alpha, \beta \right)^{-3} $ can now be expressed as
\begin{equation}
\zrho \left(r, t_0, \alpha, \beta \right)^{-3} =  \frac{1}{\zeta^2 \nu \chi \sqrt{\Delta r^2}} \sum_{l=0} \left( l + \tfrac{1}{2} \right) P_l \left( \cos \alpha \right)
\end{equation}
Bringing this result into our expression for $F^l_{a\lnpow{1}}\left(r_0,t_0\right)$ from Eq.~\eqref{eqn:fa} and integrating over $\alpha$ gives
\par \vspace{-6pt} \begin{IEEEeqnarray}{rCl} 
F^l_{a\lnpow{1}}\left(r_0,t_0\right) &=& \frac{1}{2 \pi} \left(l + \tfrac{1}{2} \right) \lim_{\Delta 
	r \rightarrow 0} \frac{1}{\zeta^2 \nu \sqrt{\Delta r^2}} \int \frac{\tilde{B}_a^{(1)} P_l \left( 
	\cos \alpha \right)}{\chi} d\Omega \nonumber \\
&=&
	\left(l + \tfrac{1}{2} \right) \lim_{\Delta r \rightarrow 0} \frac{\tilde{b}_{a_r} \Delta r}
	{\zeta^2 \nu \sqrt{\Delta r^2}} \frac{1}{2 \pi} \int \chi^{-1} d\beta \nonumber \\
&=&
	\left(l + \tfrac{1}{2} \right) \lim_{\Delta r \rightarrow 0} \frac{\tilde{b}_{a_r} \Delta r}
	{\zeta^2 \nu \sqrt{\Delta r^2}} \left<\chi^{-1} \right> \nonumber \\
&=&
	\left(l + \tfrac{1}{2} \right) \frac{\tilde{b}_{a_r} \sgn \left(\Delta r\right)} {\zeta \nu \rb} \label{eqn: Aterm}
\end{IEEEeqnarray}
where the first equality takes advantage of the orthogonal nature of the $P_l \left( \cos \alpha \right)$ and last equality comes from taking the limit as $\Delta r \rightarrow  0$ and noting from Appendix C of \cite{Detweiler:Messaritaki:Whiting:2002} that $\left<\chi^{-1} \right>$ is a special type of hypergeometric function given by
\begin{equation}
\left<\chi^{-1} \right> = F \left( 1, \tfrac{1}{2};1;k \right) = \frac{1}{\sqrt{1-k}} = \frac{\zeta}{\rb}
\end{equation}
$B_a^{(1)}$ and $b_{a_r}$ also now carry a tilde to signify that they are not the same $B_a^{(1)}$ and $b_{a_r}$ as Eq.~\eqref{eqn:fasum}, but rather the tilde represents that they have also undergone the coordinate shift $\Delta w_1 \rightarrow \Delta w_1 + \mu \Delta r$.

In the higher order tems in Eq.~(\ref{eqn:fa}),
we may immediately work with  $\zrhoz^2= 2\chi \zeta^2 (1-\cos\alpha)$ so,
\par \vspace{-6pt} \begin{IEEEeqnarray}{rCl} \label{eqn: rhoAl}
\zrhoz \left(r_0, t_0, \alpha, \beta \right)^n &=& \zeta^n \left[2\chi \left( 1 - 
	\cos \alpha \right)\right]^{{n}/{2}} \nonumber \\
&=&
	\zeta^n \left(2\chi \right)^{n/2} \sum_{l=0} \mathcal{A}_l^{{n}/{2}} (0) P_l 
	\left( \cos \alpha \right),
\end{IEEEeqnarray}
where $\mathcal{A}_l^{-\frac{1}{2}} (0) = \sqrt{2}$ from the generating function of the Legendre polynomials
and, as derived  in Appendix~D of \cite{Detweiler:Messaritaki:Whiting:2002}, for $(n+1)/2\in\mathbb{N}$
\begin{align}
\mathcal{A}^{n/2}_l \left(0 \right) =& 
		 \frac{\mathcal{P}_{n/2} \left( 2 l + 1\right) }{\left(2 l - n\right)\left(2 l - n +2\right) \dots \left(2 l + n\right) \left(2 l + n +2 \right)}, \\
\text{where} \quad \quad \quad \quad \mathcal{P}_{n/2} = &\left(-1\right)^{(n+1)/2} 2^{1 + n/2} \left( n!! \right)^2, \nonumber \\
\end{align}
In this case the angular integrals involve
\begin{eqnarray} \label{eqn:zeta minus n}
\frac{1}{2 \pi} \int  \frac{d \beta}{\chi(\beta)^{n/2}} &=& 
\left<\chi^{-{n/2}}(\beta) \right> 
={}_2 F_1 \left(\frac{n}{2}, \frac{1}{2}, 1, k \right)
\end{eqnarray}
where $(n+1)/2\in\mathbb{N}\cup \{0\}$. The resulting equations can then be tidied up using the following special cases of hypergeometric functions
\begin{IEEEeqnarray}{lClClCl}
\left<\chi^{-\frac{1}{2}}\right> &=& \mathcal{F}_{\frac{1}{2}}(k) &=& {}_2F_1 \left(\frac{1}{2}, \frac{1}{2};1;k \right) &=& \frac{2}{\pi} \mathcal{K} (k), \\
\left<\chi^{\frac{1}{2}}\right> &=& \mathcal{F}_{-\frac{1}{2}}(k) &=& {}_2F_1 \left(-\frac{1}{2}, \frac{1}{2};1;k \right) &=& \frac{2}{\pi} \mathcal{E}(k) ,
\end{IEEEeqnarray}
where
\begin{equation}
\mathcal{K }(k) \equiv \int_0^{\pi/2} (1 - k \sin^2 \beta)^{-1/2} d\beta, \quad
\mathcal{E}(k) \equiv \int_0^{\pi/2} (1 - k \sin^2 \beta)^{1/2} d\beta
\end{equation}
are complete elliptic integrals of the first and second kinds, respectively.  
\begin{table}[htb]
\begin{center}
 \begin{tabular}{|c|c|c|}
\hline
  RP & BO & DMW \\
\hline
 $F_{a\lnpow{1}}$ & $A_a$ & $A_a$ \\
 $F_{a[0]}$ & $B_a$ & $B_a$ \\
 $F_{a\lpow{1}}$ & $C_a$ & $C_a$ \\
 $F_{a\lpow{2}}$ & --- & $D_a$ \\
 $F_{a\lpow{4}}$ & --- & $E^1_a$ \\
 $F_{a\lpow{6}}$ & --- & $E^2_a$ \\
\hline
 \end{tabular}
\caption[Regularization Parameters' Notation]{Relation between notational choices for the regularization parameters (RPs). The 
  most common choices are those of either Barack and Ori \cite{Barack:Ori:2000} or
  Detweiler, Messaritaki and Whiting \cite{Detweiler:Messaritaki:Whiting:2002}.}
\label{table:rp}
\end{center}
\end{table}
All other powers of $\chi$ can be integrated to hypergeometric functions that can then be manipulated to be one of the above by the use of the recurrence relation in Eq. (15.2.10) of \cite{Abramowitz:Stegun};  that is
\begin{equation}
\mathcal{F}_{p+1} (k) = \frac{p-1}{p \left(k - 1\right)} \mathcal{F}_{p-1}(k) + \frac{1 - 2p + \left(p - \frac{1}{2}\right) k}{p \left(k - 1\right)} \mathcal{F}_p(k).
\end{equation}

In the next sections, we give the results of applying this calculation to each of scalar,
electromagnetic and gravitational cases in turn. In doing so, we omit the explicit dependence on $l$ which in each case is
\begin{gather}
F^l_{a\lnpow{1}} = (2l+1) F_{a\lnpow{1}}, \quad F^l_{a[0]} = F_{a[0]}, \quad F^l_{a\lpow{2}} = 
	\frac{F_{a\lpow{2}}}{(2l-1)(2l+3)}, \nonumber \\
F^l_{a\lpow{4}} = \frac{F_{a\lpow{4}}}{(2l-3)(2l-1)(2l+3)(2l+5)}, \nonumber \\
F^l_{a\lpow{6}} = \frac{F_{a\lpow{6}}}{(2l-5)(2l-3)(2l-1)(2l+3)(2l+5)(2l+7)}.
\end{gather}
It is \fixme{also} worth pointing out that there exists in the literature several different
notations for the regularization parameters. We have adopted a notation which is readily
extensible to other orders and which makes the dependence on $l$ explicit. To avoid
confusion, in Table \ref{table:rp} we give the relation between our notation and
other common notations.


\section{\Sch Space-time} \label{sec: schRPs}
\subsection{Scalar case} \label{sec:scalar-regularization}
In the \Sch scalar case, the regularization parameters for the self-force, as described in Eq.~\eqref{eqn:SelfForceScalar}, are given by 
\begin{gather}
F_{t\lnpow{1}} = \frac{\rbdot \sgn(\Delta r)}{2 (L^2+\rb^2)}, \quad
F_{r\lnpow{1}} = -\frac{E\rb \sgn(\Delta r)}{2 (\rb-2M) (L^2+\rb^2)}, \quad
F_{\theta\lnpow{1}} = 0, \quad
F_{\phi\lnpow{1}} = 0,
\end{gather}
\begin{gather}
F_{t[0]} = - \frac{ E \rb \rbdot}{\pi (L^2+\rb^2)^{3/2}} (2\mathcal{E} - \mathcal{K}), \\
F_{r[0]} = \frac{1}{\pi \rb (\rb-2 M) (L^2+\rb^2)^{3/2}} \Big\{ F^{\mathcal{E}}_{r\lpow{0}} 
	\mathcal{E} + F^{\mathcal{K}}_{r\lpow{0}} \mathcal{K} \Big\},
\end{gather}
where
\par \vspace{-6pt} \par \vspace{-6pt} \begin{IEEEeqnarray}{rCl}
F^{\mathcal{E}}_{r\lpow{0}} &=&  [2 E^2 \rb^3 - (\rb - 2 M) (L^2 + \rb^2)],  \nonumber \\
F^{\mathcal{K}}_{r\lpow{0}} &=& - [E^2 \rb^3 + (\rb - 2 M) (L^2 + \rb^2)], \nonumber
\end{IEEEeqnarray}

\begin{gather}
F_{\theta[0]} = 0, \quad
F_{\phi[0]} = - \frac{ \rb \rbdot}{L \pi (L^2+\rb^2)^{1/2}} (\mathcal{E} - \mathcal{K}),
\end{gather}

\begin{equation}
F_{t\lpow{2}} = \frac{E \rbdot}{2 \pi \rb^4 (L^2+\rb^2)^{7/2}} (F^{\mathcal{E}}_{t\lpow{2}} \mathcal{E} + F^{\mathcal{K}}_{t\lpow{2}} \mathcal{K}),
\end{equation}
where
\par \vspace{-6pt} \begin{IEEEeqnarray*}{rCl}
F^{\mathcal{E}}_{t\lpow{2}} &=& 8 E^2 (L^2 - \rb^2 ) \rb^7 \nonumber \\
&&
	-\:  (L^2 + \rb^2) (36 L^6 M + 104 L^4 M \rb^2 + 98 L^2 M \rb^4 + L^2 \rb^5 + 46 M 
	\rb^6 - 7 \rb^7), \nonumber \\
F^{\mathcal{K}}_{t\lpow{2}} &=& - E^2 \rb^7 (3 L^2 - 5 \rb^2) + 2 \rb^2 (L^2 + \rb^2) (9 L^4 M 
	+ 18 L^2 M \rb^2 + 13 M \rb^4 - 2 \rb^5),
\end{IEEEeqnarray*}

\begin{equation}
F_{r\lpow{2}} = \frac{1}{2 \pi \rb^6 (\rb - 2M) (L^2+\rb^2)^{7/2}} (F^{\mathcal{E}}_{r\lpow{2}} \mathcal{E} + F^{\mathcal{K}}_{r\lpow{2}} \mathcal{K}),
\end{equation}
where
\par \vspace{-6pt} \begin{IEEEeqnarray*}{rCl}
F^{\mathcal{E}}_{r\lpow{2}} &=& - 8 E^4 \rb^{10} (L^2 -\rb^2 ) \nonumber \\
&&
	+\: 4 E^2 \rb^3 (L^2 + \rb^2) (9 L^6 M + 26 L^4 M \rb^2 + 23 L^2 M \rb^4 + L^2 \rb^5 
	+ 14 M \rb^6 - 3 \rb^7) \nonumber \\ 
&&
	-\: (\rb - 2 M) (L^2 + \rb^2)^2 (28 L^6 M + 82 L^4 M \rb^2 + 82 L^2 M \rb^4 -  L^2\rb^5 
	+ 32 M \rb^6  \nonumber \\
&&	
	\quad -\:  3 \rb^7), \nonumber \\
F^{\mathcal{K}}_{r\lpow{2}} &=& E^4 \rb^{10} (3 L^2 - 5 \rb^2) \\
&&
	-\:  E^2 \rb^5 (L^2 + \rb^2) (18 
	L^4 M + 34 L^2 M \rb^2 + L^2 \rb^3 + 32 M \rb^4 - 7 \rb^5) \nonumber \\ 
&&
	+\:  (\rb - 2 M ) \rb^2 (L^2 + \rb^2)^2 (14 L^4 M + 28 L^2 M \rb^2 + 16 M \rb^4 - 
	\rb^5), 
\end{IEEEeqnarray*}

\begin{equation}
F_{\theta\lpow{2}} = 0,
\end{equation}

\begin{equation}
F_{\phi\lpow{2}} = \frac{\rbdot}{2 \pi L \rb^4 (L^2+\rb^2)^{5/2}} (F^{\mathcal{E}}_{\phi\lpow{2}} \mathcal{E} + F^{\mathcal{K}}_{\phi\lpow{2}} \mathcal{K}),
\end{equation}
where
\par \vspace{-6pt} \begin{IEEEeqnarray*}{rCl}
F^{\mathcal{E}}_{\phi\lpow{2}} &=& E^2 \rb^7 (7 L^2 - \rb^2)
  + (L^2 + \rb^2) (28 L^6 M + 58 L^4 M \rb^2 + 34 L^2 M \rb^4 -  L^2 \rb^5 + \rb^7),
\nonumber \\
F^{\mathcal{K}}_{\phi\lpow{2}} &=& - E^2 \rb^7 (3 L^2 - \rb^2)
  - \rb^2 (L^2 + \rb^2) (14 L^4 M + 16 L^2 M \rb^2 + \rb^5),
\end{IEEEeqnarray*}

\begin{equation}
F_{t\lpow{4}} = \frac{3 E \rbdot}{40 \pi \rb^{11} (L^2+\rb^2)^{11/2}} (F^{\mathcal{E}}_{t\lpow{4}} \mathcal{E} + F^{\mathcal{K}}_{t\lpow{4}} \mathcal{K}),
\end{equation}
where
\par \vspace{-6pt} \begin{IEEEeqnarray*}{rCl}
F^{\mathcal{E}}_{t\lpow{4}} &=& -30 E^4 \rb^{16} (23 L^4 - 82 L^2 \rb^2 + 23 \rb^4) 
	\nonumber \\ 
&&
  	+\: 2 E^2 \rb^5 (L^2 + \rb^2) (44800 L^{12} M + 219136 L^{10} M \rb^2 + 428252 L^8 M 
	\rb^4 \\
&&
	\quad +\:  418776 L^6 M \rb^6 + 206374 L^4 M \rb^8 + 45 L^4 \rb^9 + 45188 L^2 M \rb^{10} - 1230 L^2 
	\rb^{11}  \\
&&
	\quad -\:  166 M \rb^{12} + 645 \rb^{13}) \nonumber \\ 
&&
  	-\: 2 (L^2 + \rb^2)^2 (20480 L^{14} M^2 - 97280 L^{12} M^2 \rb^2 + 85120 L^{12} M 
	\rb^3  \\
&&
	\quad -\:  700832 L^{10} M^2 \rb^4 + 388480 L^{10} M \rb^5 - 1426472 L^8 M^2 \rb^6 + 704552 L^8 M \rb^7  \\
&&
	\quad -\:  
	1358276 L^6 M^2 \rb^8 + 635226 L^6 M \rb^9 - 635180 L^4 M^2 \rb^{10} + 286498 L^4 M \rb^{11}  \\
&&
	\quad -\:  15 
	L^4 \rb^{12}  - 124540 L^2 M^2 \rb^{12} + 54086 L^2 M \rb^{13} - 90 L^2 \rb^{14} - 2796 
	M^2 \rb^{14}  \\
&&
	\quad +\:  182 M \rb^{15} + 285 \rb^{16}), \nonumber \\
F^{\mathcal{K}}_{t\lpow{4}} &=& 15 E^4 \rb^{16} (15 L^4 - 82 L^2 \rb^2 + 31 \rb^4) 
	\nonumber \\ 
&&
  	-\: 4 E^2 \rb^7 (L^2 + \rb^2) (11200 L^{10} M + 44984 L^8 M \rb^2 + 68227 L^6 M 
	\rb^4 \nonumber \\ 
&&
  	\quad +\: 46849 L^4 M \rb^6 + 13493 L^2 M \rb^8 - 270 L^2 \rb^9 + 127 M \rb^{10} + 
	210 \rb^{11}) \nonumber \\ 
&&
  	+\: \rb^2 (L^2 + \rb^2)^2 (20480 L^{12} M^2 - 115200 L^{10} M^2 \rb^2 + 85120 L^{10} 
	M \rb^3  \\
&&
	\quad -\:  599072 L^8 M^2 \rb^4 + 314000 L^8 M \rb^5 - 908104 L^6 M^2 \rb^6 + 433792 L^6 M \rb^7  \\
&&
	\quad -\:  589164 
	L^4 M^2 \rb^8 + 268648 L^4 M \rb^9 - 151484 L^2 M^2 \rb^{10} + 66380 L^2 M \rb^{11}  \\
&&
	\quad -\:  15 
	L^2 \rb^{12} - 5592 M^2 \rb^{12} + 1204 M \rb^{13} + 345 \rb^{14}),
\end{IEEEeqnarray*}

\begin{equation}
F_{r\lpow{4}} = \frac{3}{40 \pi \rb^{13} (\rb - 2M) (L^2+\rb^2)^{11/2}} (F^{\mathcal{E}}_{r\lpow{4}} \mathcal{E} + F^{\mathcal{K}}_{r\lpow{4}} \mathcal{K}),
\end{equation}
where
\par \vspace{-6pt} \begin{IEEEeqnarray*}{rCl}
F^{\mathcal{E}}_{r\lpow{4}} &=& 30 E^6 \rb^{19} (23 L^4 - 82 L^2 \rb^2 + 23 \rb^4) 
	\nonumber \\ 
&&
  	-\: E^4 \rb^8 (L^2 + \rb^2) (89600 L^{12} M + 438272 L^{10} M \rb^2 + 856504 L^8 M 
	\rb^4  \\
&&
	\quad +\:  837552 L^6 M \rb^6 + 411938 L^4 M \rb^8 + 495 L^4 \rb^9 + 92836 L^2 M \rb^{10} - 3690 L^2 
	\rb^{11}  \\
&&
	\quad -\:  902 M \rb^{12} + 1575 \rb^{13}) \nonumber \\ 
&&
  	+\: 8 E^2 \rb^3 (L^2 + \rb^2)^2 (5120 L^{14} M^2 - 35200 L^{12} M^2 \rb^2 + 26720 
	L^{12} M \rb^3 \nonumber \\
&&
	\quad -\: 227368 L^{10} M^2 \rb^4 + 123200 L^{10} M \rb^5 - 456300 L^8 M^2 \rb^6 + 
	225979 L^8 M \rb^7 \nonumber \\
&&
	\quad -\: 434510 L^6 M^2 \rb^8 + 206277 L^6 M \rb^9 - 203983 L^4 M^2 \rb^{10} + 
	94211 L^4 M \rb^{11} \nonumber \\
&&
	\quad -\: 40376 L^2 M^2 \rb^{12} + 18367 L^2 M \rb^{13} - 135 L^2 
	\rb^{14} - 571 M^2 \rb^{14} - 146 M \rb^{15}  \\
&&
	\quad +\:  135 \rb^{16}) \nonumber \\ 
&&
  	-\: (\rb - 2 M ) (L^2 + \rb^2)^3 (40960 L^{14} M^2 - 86016 L^{12} M^2 \rb^2 + 116480 
	L^{12} M \rb^3 \nonumber \\
&&
	\quad -\: 860224 L^{10} M^2 \rb^4 + 510080 L^{10} M \rb^5 - 1780112 L^8 M^2 \rb^6 
	+ 882400 L^8 M \rb^7 \nonumber \\
&&
	\quad -\: 1657392 L^6 M^2 \rb^8 + 752340 L^6 M \rb^9 - 743164 L^4 M^2 \rb^{10} + 
	316100 L^4 M \rb^{11} \nonumber \\
&&
	\quad +\: 30 L^4 \rb^{12} - 136236 L^2 M^2 \rb^{12} + 53200 L^2 M \rb^{13} + 75 L^2 
	\rb^{14} - 3120 M^2 \rb^{14} \nonumber \\
&&
	\quad +\: 160 M \rb^{15} + 165 \rb^{16}), \nonumber \\
F^{\mathcal{K}}_{r\lpow{4}} &=&-15 E^6 \rb^{19} (15 L^4 - 82 L^2 \rb^2 + 31 \rb^4) 
	\nonumber \\ 
&&
  	+\: E^4 \rb^{10} (L^2 + \rb^2) (44800 L^{10} M + 179936 L^8 M \rb^2 + 272908 L^6 M 
	\rb^4 \nonumber \\ 
&&
  	\quad + 187126 L^4 M \rb^6 + 135 L^4 \rb^7 + 55232 L^2 M \rb^8 - 1710 L^2 \rb^9 + 
	118 M \rb^{10}  \\
&&
	\quad +\:  1035 \rb^{11}) \nonumber \\ 
&&
  	-\:  E^2 \rb^5 (L^2 + \rb^2)^2 (20480 L^{12} M^2 - 158720 L^{10} M^2 \rb^2 + 106880 
	L^{10} M \rb^3 \nonumber \\
&&
	\quad -\: 769632 L^8 M^2 \rb^4 + 399280 L^8 M \rb^5 - 1159632 L^6 M^2 \rb^6 + 
	559556 L^6 M \rb^7 \nonumber \\
&&
	\quad -\: 755876 L^4 M^2 \rb^8 + 352004 L^4 M \rb^9 - 196524 L^2 M^2 \rb^{10} + 
	89680 L^2 M \rb^{11} \nonumber \\
&&
	\quad -\: 405 L^2 \rb^{12} - 5528 M^2 \rb^{12} + 512 M \rb^{13} + 675 \rb^{14}) 
	\nonumber \\ 
&&
  	+\:  (\rb - 2 M) \rb^2 (L^2 + \rb^2)^3 (20480 L^{12} M^2 - 60928 L^{10} M^2 \rb^2 + 
	58240 L^{10} M \rb^3 \nonumber \\
&&
	\quad -\: 375840 L^8 M^2 \rb^4 + 204080 L^8 M \rb^5 - 564472 L^6 M^2 \rb^6 + 
	265360 L^6 M \rb^7 \nonumber \\
&&
	\quad -\: 350956 L^4 M^2 \rb^8 + 152380 L^4 M \rb^9 - 84276 L^2 M^2 \rb^{10} + 
	33740 L^2 M \rb^{11}  \\
&&
	\quad +\:  15 L^2 \rb^{12} - 3120 M^2 \rb^{12} + 640 M \rb^{13} + 75 \rb^{14}),
\end{IEEEeqnarray*}

\begin{equation}
F_{\theta\lpow{4}} = 0,
\end{equation}

\begin{equation}
F_{\phi\lpow{4}} = \frac{3 \rbdot}{40 \pi L \rb^{11} (L^2+\rb^2)^{9/2}}  (F^{\mathcal{E}}_{\phi\lpow{4}} \mathcal{E} + F^{\mathcal{K}}_{\phi\lpow{4}} \mathcal{K}),
\end{equation}
where
\par \vspace{-6pt} \begin{IEEEeqnarray*}{rCl}
F^{\mathcal{E}}_{\! \phi\lpow{4}} &=& -15 E^4 \rb^{16} (43 L^4 - 82 L^2 \rb^2 + 3 \rb^4) 
	\nonumber \\ 
&&
  -\: 10 E^2 \rb^5 (L^2 + \rb^2) (4352 L^{12} M + 16512 L^{10} M \rb^2 + 22948 L^8 M 
	\rb^4 \nonumber \\ 
&&
  	\quad +\: 13346 L^6 M \rb^6 + 2136 L^4 M \rb^8 - 9 L^4 \rb^9 - 710 L^2 M \rb^{10} + 
	126 L^2 \rb^{11} - 9 \rb^{13}) \nonumber \\ 
&&
  +\: (L^2 + \rb^2)^2 (40960 L^{14} M^2 - 96256 L^{12} M^2 \rb^2 + 116480 L^{12} M 
	\rb^3  \\
&&
	\quad -\:  704064 L^{10} M^2 \rb^4 + 429440 L^{10} M \rb^5 - 1134992 L^8 M^2 \rb^6 + 595040 L^8 M \rb^7  \\
&&
	\quad -\:  
	755632 L^6 M^2 \rb^8 + 372500 L^6 M \rb^9 - 194724 L^4 M^2 \rb^{10} + 94940 L^4 M \rb^{11}  \\
&&
	\quad +\:  30 
	L^4 \rb^{12} - 6276 L^2 M^2 \rb^{12} + 4040 L^2 M \rb^{13} + 105 L^2 \rb^{14} + 480 M^2 \rb^{14}  \\
&&
	\quad -\:  45 \rb^{16}),
	\nonumber \\
F^{\mathcal{K}}_{\! \phi\lpow{4}} &=& 15 E^4 \rb^{16} (L^2 - 3 \rb^2) (15 L^2 -  \rb^2) 
	\nonumber \\
&&
  	+\: 10 E^2 \rb^7 (L^2 + \rb^2) (2176 L^{10} M + 6352 L^8 M \rb^2 + 6018 L^6 M \rb^4 
	\nonumber \\ 
&&
  	\quad +\: 1666 L^4 M \rb^6 - 320 L^2 M \rb^8 + 63 L^2 \rb^9 - 9 \rb^{11}) \nonumber 
	\\ 
&&
  -\:  \rb^2 (L^2 + \rb^2)^2 (20480 L^{12} M^2 - 66048 L^{10} M^2 \rb^2 + 58240 L^{10} 
	M \rb^3  \\
&&
	\quad -\:  293280 L^8 M^2 \rb^4 + 163760 L^8 M \rb^5 - 314392 L^6 M^2 \rb^6 + 156960 L^6 M \rb^7  \\
&&
	\quad -\:  114876 
	L^4 M^2 \rb^8 + 55420 L^4 M \rb^9 - 6516 L^2 M^2 \rb^{10} + 3740 L^2 M \rb^{11}  \\
&&
	\quad +\:  15 L^2 
	\rb^{12} + 480 M^2 \rb^{12} - 45 \rb^{14}),
\end{IEEEeqnarray*}

\begin{equation}
F_{t\lpow{6}} = \frac{-3 E \rbdot}{560 \pi  \rb^{16} \left(L^2+\rb^2\right)^{15/2}} \left(F^{\mathcal{E}}_{t\lpow{6}} \mathcal{E} + F^{\mathcal{K}}_{t\lpow{6}} \mathcal{K} \right),
\end{equation}
where
\par \vspace{-6pt} \begin{IEEEeqnarray*}{rCl}
F^{\mathcal{E}}_{t\lpow{6}} &=& 28000 E^6 \rb^{23} (\rb-L) (L+\rb) \left(11 L^4-74 L^2 \rb^2 + 
	11 \rb^4\right) \nonumber \\
&& 
  -\: 25 E^4 \rb^8 \left(L^2+\rb^2\right) \big(-16056320 L^{18} M-107151360 L^{16} M 
	\rb^2  \\
&&
	\quad -\: 302586880 L^{14} M \rb^4 - 464979968 L^{12} M \rb^6 - 412568652 L^{10} M \rb^8 \\
&&
	\quad -\: 201055024 L^8 M 
	\rb^{10} - 39268410 L^6 M \rb^{12}-1575 L^6 \rb^{13}+5226426 L^4 M \rb^{14}  \\
&&
	\quad +\:  
	99435 L^4 \rb^{15} + 3185118 L^2 M \rb^{16}-186165 L^2 \rb^{17}+19662 M \rb^{18}+35385 
	\rb^{19}\big) \nonumber \\
&& 
  +\: 2 E^2 \rb^3 \left(L^2+\rb^2\right)^2 \big(-1007616000 L^{20} M^2-2885324800 
	L^{18} M^2 \rb^2 \nonumber \\
&&
	\quad -\: 1548288000 L^{18} M \rb^3 + 5271990272 L^{16} M^2 \rb^4 - 9940582400 
	L^{16} M \rb^5 \nonumber \\
&&
	\quad +\: 35832487264 L^{14} M^2 \rb^6-27145052800 L^{14} M \rb^7 + 69571689904 
	L^{12} M^2 \rb^8 \nonumber \\
&&
	\quad -\: 40793731200 L^{12} M \rb^9+69887626312 L^{10} M^2 \rb^{10} - 
	36329433800 L^{10} M \rb^{11} \nonumber \\
&&
	\quad +\: 39015325900 L^8 M^2 \rb^{12}-19063343950 L^8 M \rb^{13}+11166709052 
	L^6 M^2 \rb^{14} \nonumber \\
&&
	\quad -\: 5373108900 L^6 M \rb^{15} + 7875 L^6 \rb^{16}+1046817944 L^4 M^2 
	\rb^{16}  \\
&&
	\quad -\:  575985300 L^4 M \rb^{17} + 94500 L^4 \rb^{18}-118281276 L^2 M^2 \rb^{18}  \\
&&
	\quad +\:  29440500 L^2 M \rb^{19} 
	- 1178625 L^2 \rb^{20} - 8271468 M^2 \rb^{20}+479850 M \rb^{21} \\
&&
	\quad +\: 414750 \rb^{22} \big) \nonumber 
	\\
&& 
  +\: \left(L^2+\rb^2\right)^3 \big(-5775360000 L^{20} M^3+2580480000 L^{20} M^2 \rb 
	\nonumber \\
&&
	\quad -\: 18980904960 L^{18} M^3 \rb^2+750796800 L^{18} M^2 \rb^3 + 3429888000 
	L^{18} M \rb^4 \nonumber \\
&&
	\quad +\: 10876463104 L^{16} M^3 \rb^4-53063915520 L^{16} M^2 \rb^5 + 
	21396480000 L^{16} M \rb^6 \nonumber \\
&&
	\quad +\: 143196789568 L^{14} M^3 \rb^6-191859546624 L^{14} M^2 \rb^7 + 
	56793312800 L^{14} M \rb^8 \nonumber \\
&&
	\quad +\:292841560608 L^{12} M^3 \rb^8 - 317782413664 L^{12} M^2 \rb^9  \\
&&
	\quad +\:  
	83096000800 L^{12} M \rb^{10} + 297880915104 L^{10} M^3 \rb^{10} \\
&&
	\quad -\: 295661821784 L^{10} M^2 \rb^{11} + 
	72364880400 L^{10} M \rb^{12}  \\
&&
	\quad +\:  168534399040 L^8 M^3 \rb^{12}-159472848000 L^8 M^2 \rb^{13} + 
	37560515600 L^8 M \rb^{14}  \\
&&
	\quad +\:  50707761864 L^6 M^3 \rb^{14}-46826640820 L^6 M^2 \rb^{15} + 
	10839698800 L^6 M \rb^{16}  \\
&&
	\quad +\:  7000 L^6 \rb^{17} + 6272875728 L^4 M^3 \rb^{16} - 5878415984 L^4 M^2 
	\rb^{17}  \\
&&
	\quad +\:  1398754050 L^4 M \rb^{18} + 41125 L^4 \rb^{19}-86931352 L^2 M^3 
	\rb^{18}  \\
&&
	\quad +\:  212436 L^2 M^2 \rb^{19} + 21357300 L^2 M \rb^{20} + 124250 L^2 \rb^{21}-29620256 M^3 \rb^{20}  \\
&&
	\quad +\:  
	16081176 M^2 \rb^{21}  - 597150 M \rb^{22}-245875 \rb^{23}\big), 
\end{IEEEeqnarray*}
\par \vspace{-6pt} \begin{IEEEeqnarray*}{rCl}
F^{\mathcal{K}}_{t\lpow{6}} &=& -875 E^6 \rb^{23} \left(-105 L^6+1189 L^4 \rb^2-1531 L^2 
	\rb^4 +247 \rb^6\right) \nonumber \\
&& 
  +\: 50 E^4 \rb^{10} \left(L^2+\rb^2\right) \big(-4014080 L^{16} M-23275520 L^{14} M 
	\rb^2  \\
&&
	\quad -\: 55468800 L^{12} M \rb^4 - 68718512 L^{10} M \rb^6 - 45183275 L^8 M \rb^8 \\
&&
	\quad -\: 13052460 L^6 M \rb^{10} 
	+ 362358 L^4 M \rb^{12} + 18900 L^4 \rb^{13}+919476 L^2 M \rb^{14} \\
&&
	\quad -\: 49560 L^2 \rb^{15} +15501 M 
	\rb^{16}+12180 \rb^{17}\big) \nonumber \\
&& 
  -\: E^2 \rb^5 \left(L^2+\rb^2\right)^2 \big(-1007616000 L^{18} M^2-2003660800 
	L^{16} M^2 \rb^2 \nonumber \\
&& 
	\quad -\: 1548288000 L^{16} M \rb^3 + 6977961472 L^{14} M^2 \rb^4-8585830400 
	L^{14} M \rb^5 \nonumber \\
&&
	\quad +\: 29653513376 L^{12} M^2 \rb^6-19705027200 L^{12} M \rb^7 + 43981240344 
	L^{10} M^2 \rb^8 \nonumber \\
&&
	\quad -\: 23922541200 L^{10} M \rb^9+32635169200 L^8 M^2 \rb^{10}-16162950800 
	L^8 M \rb^{11} \nonumber \\
&&
	\quad +\: 12004692860 L^6 M^2 \rb^{12}-5726827800 L^6 M \rb^{13}+1578230992 
	L^4 M^2 \rb^{14} \nonumber \\
&&
	\quad -\: 803087700 L^4 M \rb^{15}+7875 L^4 \rb^{16} - 113069964 L^2 M^2 \rb^{16} 
	 \\
&&
	\quad +\:  24092400 L^2 M \rb^{17} - 1118250 L^2 \rb^{18}-13324056 M^2 \rb^{18}+1526700 M \rb^{19} \\
&&
	\quad +\: 553875 
	\rb^{20} \big)\nonumber \\
&& 
  +\: 2 \rb^2 \left(L^2+\rb^2\right)^3 \big(1443840000 L^{18} M^3-645120000 L^{18} 
	M^2 \rb \\
&&
	\quad +\: 3481866240 L^{16} M^3 \rb^2 + 376780800 L^{16} M^2 \rb^3 - 857472000 L^{16} M \rb^4 \\
&&
	\quad -\: 5698068736 L^{14} 
	M^3 \rb^4 + 12906055680 L^{14} M^2 \rb^5-4598832000 L^{14} M \rb^6  \\
&&
	\quad -\:  30679784768 
	L^{12} M^3 \rb^6 + 36702979536 L^{12} M^2 \rb^7-10214544200 L^{12} M \rb^8 \\
&&
	\quad -\: 46687160912 
	L^{10} M^3 \rb^8 + 47920180732 L^{10} M^2 \rb^9-12034259400 L^{10} M \rb^{10}  \\
&&
	\quad -\:  
	34910457500 L^8 M^3 \rb^{10} + 33450421620 L^8 M^2 \rb^{11} - 7955786400 L^8 M \rb^{12} \\
&&
	\quad -\: 13231827540 
	L^6 M^3 \rb^{12} + 12237529680 L^6 M^2 \rb^{13}-2830778675 L^6 M \rb^{14}  \\
&&
	\quad -\:  2081061396 
	L^4 M^3 \rb^{14} + 1920152080 L^4 M^2 \rb^{15}-448289925 L^4 M \rb^{16} \\
&&
	\quad -\: 1750 L^4 \rb^{17} 
	- 4481148 L^2 M^3 \rb^{16} + 24248796 L^2 M^2 \rb^{17}-11278125 L^2 M \rb^{18} \\
&&
	\quad -\: 8750 L^2 \rb^{19} + 
	11067088 M^3 \rb^{18} - 6431148 M^2 \rb^{19} + 440325 M \rb^{20} \\
&&
	\quad +\: 77000 \rb^{21}\big),
\end{IEEEeqnarray*}

\begin{equation}
F_{r\lpow{6}} = \frac{-3}{560 \pi  \rb^{18} \left(L^2+\rb^2\right)^{15/2} (\rb-2 M)} \left(F^{\mathcal{E}}_{r\lpow{6}} \mathcal{E} + F^{\mathcal{K}}_{r\lpow{6}} \mathcal{K} \right),
\end{equation}
where
\par \vspace{-6pt} \begin{IEEEeqnarray*}{rCl}
F^{\mathcal{E}}_{r\lpow{6}} &=& -28000 E^8 (\rb-L) (L+\rb) \left(11 L^4-74 \rb^2 L^2+11 
	\rb^4 \right) \rb^{26} \nonumber \\
&&
  +\: 50 E^6 \left(L^2+\rb^2\right) \big(19565 \rb^{19}+6086 M \rb^{18}-113785 L^2 
	\rb^{17} + 1633964 L^2 M \rb^{16} \nonumber \\
&&
	\quad +\: 76615 L^4 \rb^{15}+2559418 L^4 M \rb^{14} - 5075 L^6 \rb^{13}-19625630 
	L^6 M \rb^{12} \nonumber \\
&&
	\quad -\: 100527512 L^8 M \rb^{10}-206284326 L^{10} M \rb^8-232489984 L^{12} M 
	\rb^6 \nonumber \\
&&
	\quad -\: 151293440 L^{14} M \rb^4-53575680 L^{16} M \rb^2-8028160 L^{18} M\big) 
	\rb^{11} \nonumber \\
&& 
  -\: E^4 \left(L^2+\rb^2\right)^2 \big(1094625 \rb^{22}+545450 M \rb^{21}-4202625 L^2 
	\rb^{20}  \\
&&
	\quad -\: 16774936 M^2 \rb^{20} + 110101000 L^2 M \rb^{19} + 1414875 L^4 \rb^{18} \\
&&
	\quad -\: 331621052 L^2 M^2 
	\rb^{18}  - 822447000 L^4 M \rb^{17} - 7875 L^6 \rb^{16} \\
&&
	\quad +\: 1429685188 L^4 M^2 \rb^{16} 
	- 9480504900 L^6 M \rb^{15} + 19802086804 L^6 M^2 \rb^{14}  \\
&&
	\quad -\:  35145688050 L^8 M \rb^{13}+72068652100 
	L^8 M^2 \rb^{12} - 68030273700 L^{10} M \rb^{11}  \\
&&
	\quad +\:  130518064824 L^{10} M^2 \rb^{10} - 
	76873222400 L^{12} M \rb^9  \\
&&
	\quad +\:  129714899808 L^{12} M^2 \rb^8  -  51274604800 L^{14} M \rb^7 + 
	65633972928 L^{14} M^2 \rb^6  \\
&&
	\quad -\:  18785894400 L^{16} M \rb^5 + 8353439744 L^{16} M^2 \rb^4 - 2924544000 
	L^{18} M \rb^3  \\
&&
	\quad -\:  6114713600 L^{18} M^2 \rb^2 - 2015232000 L^{20} M^2\big) \rb^6 \nonumber 
	\\
&& 
  +\: 2 E^2 \left(L^2+\rb^2\right)^3 \big(242375 \rb^{23}+138500 M \rb^{22}-419125 L^2 
	\rb^{21}  \\
&&
	\quad -\: 10992820 M^2 \rb^{21} + 20399392 M^3 \rb^{20} + 6959650 L^2 M \rb^{20}+9625 L^4 \rb^{19}  \\
&&
	\quad -\:  
	81665756 L^2 M^2 \rb^{19} + 138887552 L^2 M^3 \rb^{18}-822884550 L^4 M \rb^{18}  \\
&&
	\quad -\:  875 L^6 \rb^{17} + 
	3518901164 L^4 M^2 \rb^{17} - 3802035608 L^4 M^3 \rb^{16} \\
&&
	\quad -\: 7067551050 L^6 M \rb^{16}+30648837404 L^6 
	M^2 \rb^{15} - 33234129320 L^6 M^3 \rb^{14} \\
&&
	\quad -\: 25352340750 L^8 M \rb^{14}+106617027800 
	L^8 M^2 \rb^{13} - 111740075320 L^8 M^3 \rb^{12}  \\
&&
	\quad -\:  49660036200 L^{10} M 
	\rb^{12} + 197830827680 L^{10} M^2 \rb^{11}  \\
&&
	\quad -\:  195029907128 L^{10} M^3 \rb^{10}-57553509200 L^{12} M \rb^{10}  \\
&&
	\quad +\:  
	209550665904 L^{12} M^2 \rb^9 - 183717663248 L^{12} M^3 \rb^8-39556896400 L^{14} M \rb^8  \\
&&
	\quad +\:  
	121346652064 L^{14} M^2 \rb^7 - 77791192288 L^{14} M^3 \rb^6-14956032000 L^{16} M \rb^6 \\
&&
	\quad +\: 28755231744 
	L^{16} M^2 \rb^5 + 7146388480 L^{16} M^3 \rb^4 - 2403072000 L^{18} M \rb^4 \\
&&
	\quad -\: 3619737600 
	L^{18} M^2 \rb^3 + 18731642880 L^{18} M^3 \rb^2-2297856000 L^{20} M^2 \rb  \\
&&
	\quad +\:  4902912000 
	L^{20} M^3\big) \rb^3 \nonumber \\
&& 
  +\: (2 M-\rb) \left(L^2+\rb^2\right)^4 \big(53375 \rb^{23}+173600 M \rb^{22}+29750 L^2 
	\rb^{21}  \\
&&
	\quad -\: 5212944 M^2 \rb^{21} + 10317184 M^3 \rb^{20} - 14183750 L^2 M \rb^{20}+25375 L^4 \rb^{19}  \\
&&
	\quad +\:  
	62499436 L^2 M^2 \rb^{19} - 76153328 L^2 M^3 \rb^{18}-679641900 L^4 M \rb^{18}  \\
&&
	\quad +\:  7000 L^6 \rb^{17} + 
	3345821696 L^4 M^2 \rb^{17} - 4111469784 L^4 M^3 \rb^{16} \\
&&
	\quad -\: 5518261350 L^6 M \rb^{16}+25823273900 L^6 
	M^2 \rb^{15} - 30032966752 L^6 M^3 \rb^{14} \\
&&
	\quad -\: 20150264400 L^8 M \rb^{14} + 88797134680 
	L^8 M^2 \rb^{13} - 96877958296 L^8 M^3 \rb^{12}  \\
&&
	\quad -\:  40738068000 L^{10} M \rb^{12} + 
	166165277616 L^{10} M^2 \rb^{11}  \\
&&
	\quad -\:  165473393280 L^{10} M^3 \rb^{10} - 48857572400 L^{12} M \rb^{10}  \\
&&
	\quad +\:  
	177451524416 L^{12} M^2 \rb^9 - 149256330784 L^{12} M^3 \rb^8-34733020000 L^{14} M \rb^8  \\
&&
	\quad +\:  
	101711045376 L^{14} M^2 \rb^7 - 51263318208 L^{14} M^3 \rb^6-13563648000 L^{16} M \rb^6 \\
&&
	\quad +\: 21232814080 
	L^{16} M^2 \rb^5 +\: 21391316992 L^{16} M^3 \rb^4 - 2247168000 L^{18} M \rb^4 \\
&&
	\quad -\: 5520998400 
	L^{18} M^2 \rb^3 +\: 23777402880 L^{18} M^3 \rb^2-2580480000 L^{20} M^2 \rb  \\
&&
	\quad +\:  5775360000 
	L^{20} M^3\big),
\end{IEEEeqnarray*}
\par \vspace{-6pt} \begin{IEEEeqnarray*}{rCl}
F^{\mathcal{K}}_{r\lpow{6}} &=& 875 E^8 \left(-105 L^6+1189 \rb^2 L^4-1531 \rb^4 L^2+247 
	\rb^6 \right) \rb^{26} \nonumber \\
&& 
  -\: 25 E^6 \left(L^2+\rb^2\right) \big(27055 \rb^{17}+25612 M \rb^{16}-123235 L^2 
	\rb^{15} +1887182 L^2 M \rb^{14} \nonumber \\
&&
	\quad +\: 62125 L^4 \rb^{13}+676066 L^4 M \rb^{12} - 2625 L^6 \rb^{11}-26099670 
	L^6 M \rb^{10} \nonumber \\
&&
	\quad -\: 90366550 L^8 M \rb^8-137437024 L^{10} M \rb^6-110937600 L^{12} M \rb^4 
	\nonumber \\
&&
	\quad -\: 46551040 L^{14} M \rb^2 - 8028160 L^{16} M\big) \rb^{13} \nonumber \\
&& 
  +\: 2 E^4 \left(L^2+\rb^2\right)^2 \big(370125 \rb^{20}+697975 M \rb^{19}-1065750 L^2 
	\rb^{18}  \\
&&
	\quad -\: 6904028 M^2 \rb^{18} + 27977700 L^2 M \rb^{17} + 244125 L^4 \rb^{16}-86371482 L^2 M^2 \rb^{16} 
	 \\
&&
	\quad -\:  315079050 L^4 M \rb^{15} + 615225146 L^4 M^2 \rb^{14}-2582807100 L^6 M \rb^{13}  \\
&&
	\quad +\:  5441132830 L^6 
	M^2 \rb^{12} - 7522573725 L^8 M \rb^{11}+15199781250 L^8 M^2 \rb^{10} \\
&&
	\quad -\: 11253096600 
	L^{10} M \rb^9 + 20574272172 L^{10} M^2 \rb^8-9303284800 L^{12} M \rb^7 \\
&&
	\quad +\: 13728299088 
	L^{12} M^2 \rb^6 - 4056729600 L^{14} M \rb^5 + 3016609536 L^{14} M^2 \rb^4 \\
&&
	\quad -\: 731136000 
	L^{16} M \rb^3 - 1087846400 L^{16} M^2 \rb^2-503808000 L^{18} M^2\big) \rb^8 \nonumber 
	\\
&& 
  -\: E^2 \left(L^2+\rb^2\right)^3 \big(314125 \rb^{21}+970000 M \rb^{20}-358750 L^2 
	\rb^{19} -18222440 M^2 \rb^{19} \nonumber \\
&& 
	\quad +\: 31216064 M^3 \rb^{18} - 3780450 L^2 M \rb^{18}-875 L^4 \rb^{17} - 
	36108732 L^2 M^2 \rb^{17} \nonumber \\
&&
	\quad +\: 87877152 L^2 M^3 \rb^{16}-1092352500 L^4 M \rb^{16} + 4756636984 L^4 
	M^2 \rb^{15} \nonumber \\
&&
	\quad -\: 5211718840 L^4 M^3 \rb^{14}-7494234450 L^6 M \rb^{14}+32418083300 L^6 
	M^2 \rb^{13} \nonumber \\
&&
	\quad -\: 35018994560 L^6 M^3 \rb^{12}-21674793600 L^8 M \rb^{12}+89821855320 
	L^8 M^2 \rb^{11} \nonumber \\
&&
	\quad -\: 92610055960 L^8 M^3 \rb^{10} - 33236721600 L^{10} M \rb^{10} + 
	127823271040 L^{10} M^2 \rb^9 \nonumber \\
&&
	\quad -\: 120667329776 L^{10} M^3 \rb^8-28422864400 L^{12} M \rb^8 + 
	95019791488 L^{12} M^2 \rb^7 \nonumber \\
&&
	\quad -\: 72612130368 L^{12} M^3 \rb^6-12853344000 L^{14} M \rb^6+30055494144 
	L^{14} M^2 \rb^5 \nonumber \\
&&
	\quad -\: 5260183040 L^{14} M^3 \rb^4-2403072000 L^{16} M \rb^4-1609113600 
	L^{16} M^2 \rb^3 \nonumber \\
&&
	\quad +\: 14441594880 L^{16} M^3 \rb^2 - 2297856000 L^{18} M^2 \rb+4902912000 
	L^{18} M^3\big) \rb^5 \nonumber \\
&& 
  +\: (\rb-2 M) \left(L^2+\rb^2\right)^4 \big(27125 \rb^{21}+299600 M \rb^{20}+9625 L^2 
	\rb^{19}  \\
&&
	\quad -\:  4164624 M^2 \rb^{19} + 7521664 M^3 \rb^{18} - 12312300 L^2 M \rb^{18}+3500 L^4 \rb^{17}  \\
&&
	\quad +\:  
	60136468 L^2 M^2 \rb^{17} - 77649520 L^2 M^3 \rb^{16}-435265950 L^4 M \rb^{16}  \\
&&
	\quad +\:  2136130980 L^4 
	M^2 \rb^{15} - 2610250432 L^4 M^3 \rb^{14}-2918350050 L^6 M \rb^{14} \\
&&
	\quad +\: 13485739400 L^6 
	M^2 \rb^{13}  - 15471992072 L^6 M^3 \rb^{12}-8700997200 L^8 M \rb^{12} \\
&&
	\quad +\: 37444962000 
	L^8 M^2 \rb^{11} - 39767055768 L^8 M^3 \rb^{10} - 13875397200 L^{10} M \rb^{10}  \\
&&
	\quad +\:  
	54025898376 L^{10} M^2 \rb^9 - 50546368608 L^{10} M^3 \rb^8-12345326000 L^{12} M \rb^8  \\
&&
	\quad +\:  40319920928 
	L^{12} M^2 \rb^7 - 27561410368 L^{12} M^3 \rb^6-5798688000 L^{14} M \rb^6 \\
&&
	\quad +\: 11983523840 
	L^{14} M^2 \rb^5 + 2639284736 L^{14} M^3 \rb^4-1123584000 L^{16} M \rb^4 \\
&&
	\quad -\: 1631539200 
	L^{16} M^2 \rb^3 + 9361981440 L^{16} M^3 \rb^2 - 1290240000 L^{18} M^2 \rb \\
&&
	\quad +\: 2887680000 
	L^{18} M^3\big) \rb^2,
\end{IEEEeqnarray*}

\begin{equation}
F_{\theta\lpow{6}} = 0,
\end{equation}

\begin{equation}
F_{\phi\lpow{6}} = \frac{-3\rbdot}{560 \pi  L \rb^{16} \left(L^2+\rb^2\right)^{13/2}} \left(F^{\mathcal{E}}_{\phi\lpow{6}} \mathcal{E} + F^{\mathcal{K}}_{\phi\lpow{6}} \mathcal{K} \right),
\end{equation}
where
\par \vspace{-6pt} \begin{IEEEeqnarray*}{rCl}
F^{\mathcal{E}}_{\phi\lpow{6}} &=& 875 E^6 \rb^{23} \left(1773 L^4 \rb^2 -337 L^6-947 L^2 
	\rb^4 +15 \rb^6\right) \nonumber \\
&& 
  +\: 175 E^4 \rb^8 \left(L^2+\rb^2\right) \big(983040 L^{18} M + 7045120 L^{16} M \rb^2 + 
	21570560 L^{14} M \rb^4 \nonumber \\
&&
	\quad +\: 36582912 L^{12} M \rb^6 + 37139572 L^{10} M \rb^8 + 22617566 L^8 M 
	\rb^{10}  \\
&&
	\quad +\:  7708830 L^6 M \rb^{12} + 225 L^6 \rb^{13} + 1199730 L^4 M \rb^{14} - 9375 L^4 \rb^{15}  \\
&&
	\quad +\:  4926 L^2 M 
	\rb^{16} +  9375 L^2 \rb^{17} - 225 \rb^{19}\big) \nonumber \\
&& 
  +\: E^2 \rb^3 \left(L^2+\rb^2\right)^2 \big(2015232000 L^{20} M^2+8400691200  L^{18} 
	M^2 \rb^2 \nonumber \\
&&
	\quad +\: 1376256000 L^{18} M \rb^3+11063078912 L^{16} M^2 \rb^4 + 7276953600 
	L^{16} M \rb^5 \nonumber \\
&&
	\quad +\: 82626752 L^{14} M^2 \rb^6+15631481600 L^{14} M \rb^7-12481032128 
	L^{12} M^2 \rb^8 \nonumber \\
&& 
	\quad +\: 17141443200 L^{12} M \rb^9-10632497080 L^{10} M^2 \rb^{10}+9613116800 
	L^{10} M \rb^{11} \nonumber \\
&&
	\quad -\: 2120762600 L^8 M^2 \rb^{12} + 2046335900 L^8 M \rb^{13}+1035946292 
	L^6 M^2 \rb^{14} \nonumber \\
&&
	\quad -\: 316454600 L^6 M \rb^{15}+15750 L^6 \rb^{16}+431941952 L^4 M^2 \rb^{16} 
	 \\
&&
	\quad -\:  166655300 L^4 M \rb^{17} + 133875 L^4 \rb^{18}+17346492 L^2 M^2 \rb^{18} \\
&&
	\quad -\: 2290400 L^2 M \rb^{19} - 
	850500 L^2 \rb^{20} - 312480 M^2 \rb^{20}+39375 \rb^{22}\big) \nonumber \\
&& 
  -\: \left(L^2+\rb^2\right)^3 \big(2580480000 L^{20} M^2 \rb -5775360000 L^{20} M^3  \\
&&
	\quad -\:  
	21381242880 L^{18} M^3 \rb^2 + 4448665600 L^{18} M^2 \rb^3 + 2247168000 L^{18} M \rb^4 \\
&&
	\quad -\: 16436058112 
	L^{16} M^3 \rb^4 - 20228515840 L^{16} M^2 \rb^5+12144384000 L^{16} M \rb^6  \\
&&
	\quad +\:  37967372736 
	L^{14} M^3 \rb^6 - 78865174016 L^{14} M^2 \rb^7+27288335200 L^{14} M \rb^8 \\
&&
	\quad +\: 91117248928 
	L^{12} M^3 \rb^8 - 114866020480 L^{12} M^2 \rb^9+32738062000 L^{12} M \rb^{10}  \\
&&
	\quad +\:  
	80974789248 L^{10} M^3 \rb^{10} - 86565871136  L^{10} M^2 \rb^{11}  \\
&&
	\quad +\:  22304895200 L^{10} M \rb^{12}  + 
	35112838392  L^8 M^3 \rb^{12} - 34540540744 L^8 M^2 \rb^{13} \\
&&
	\quad +\: 8392988800 L^8 M \rb^{14} + 6757023136 
	L^6 M^3 \rb^{14} - 6373891596  L^6 M^2 \rb^{15} \\
&&
	\quad +\; 1516716950 L^6 M \rb^{16}-7000 L^6 
	\rb^{17} + 273916728 L^4 M^3 \rb^{16}  \\
&&
	\quad - 286552800 L^4 M^2 \rb^{17}+80573500 L^4 M 
	\rb^{18} -28875 L^4 \rb^{19}  \\
&&
	\quad -\:  24711184 L^2 M^3 \rb^{18}+14372004 L^2 M^2 \rb^{19} - 855050 L^2 M 
	\rb^{20} -50750 L^2 \rb^{21} \nonumber \\
&&
	\quad +\: 309120 M^3 \rb^{20}-245280 M^2 \rb^{21}+13125 \rb^{23}\big),
\end{IEEEeqnarray*}
\par \vspace{-6pt} \begin{IEEEeqnarray*}{rCl}
F^{\mathcal{K}}_{\phi\lpow{6}} &=& -875 E^6 \rb^{23} \left(-105 L^6+829 L^4 \rb^2-587 L^2 
	\rb^4 +15 \rb^6\right) \nonumber \\
&& 
  +\: 175 E^4 \rb^{10} \left(L^2+\rb^2\right) \big(-491520 L^{16} M-3092480 L^{14} M \rb^2 
	- 8102400 L^{12} M \rb^4 \nonumber \\
&&
	\quad -\: 11336736 L^{10} M \rb^6 - 8972282 L^8 M \rb^8-3855372 L^6 M \rb^{10} - 
	750282 L^4 M \rb^{12} \nonumber \\
&&
	\quad +\: 3825 L^4 \rb^{13}-12696 L^2 M \rb^{14} - 5550 L^2 \rb^{15}+225 \rb^{17} 
	\big) \nonumber \\
&& 
  -\: E^2 \rb^5 \left(L^2+\rb^2\right)^2 \big(1007616000 L^{18} M^2+3318681600 L^{16} 
	M^2 \rb^2 \nonumber \\
&&
	\quad +\: 688128000 L^{16} M \rb^3 + 2674925056 L^{14} M^2 \rb^4 + 3036364800 
	L^{14} M \rb^5 \nonumber \\
&&
	\quad -\: 2164346848 L^{12} M^2 \rb^6 + 5191177600 L^{12} M \rb^7-4277576360 
	L^{10} M^2 \rb^8 \nonumber \\
&&
	\quad +\: 4156658800 L^{10} M \rb^9-1698491920 L^8 M^2 \rb^{10}+1358613200 L^8 
	M \rb^{11} \nonumber \\
&&
	\quad +\:281096500 L^6 M^2 \rb^{12} - 46924500 L^6 M \rb^{13}+248366216 L^4 M^2 
	\rb^{14} \nonumber \\
&&
	\quad -\: 97522600 L^4 M \rb^{15}+7875 L^4 \rb^{16}+15819372 L^2 M^2 \rb^{16} - 
	3686900 L^2 M \rb^{17} \nonumber \\
&&
	\quad -\: 456750 L^2 \rb^{18}-312480 M^2 \rb^{18}+39375 \rb^{20}\big) \nonumber \\
&& 
  +\: \rb^2 \left(L^2+\rb^2\right)^3 \big(-2887680000 L^{18} M^3+1290240000 L^{18} M^2 
	\rb \nonumber \\
&&
	\quad -\: 8163901440 L^{16} M^3 \rb^2+1095372800 L^{16} M^2 \rb^3 + 1123584000 
	L^{16} M \rb^4 \nonumber \\
&&
	\quad -\: 1209975296 L^{14} M^3 \rb^4-11012229120 L^{14} M^2 \rb^5+5089056000 
	L^{14} M \rb^6\nonumber \\
&& 
	\quad +\: 19718951872 L^{12} M^3 \rb^6-29772000928  L^{12} M^2 \rb^7 + 
	9243911600 L^{12} M \rb^8 \nonumber \\
&&
	\quad +\: 28381407744 L^{10} M^3 \rb^8 - 31906051368 L^{10} M^2 \rb^9 + 
	8496115600 L^{10} M \rb^{10} \nonumber \\
&&
	\quad +\: 16529207256 L^8 M^3 \rb^{10}-16531893472 L^8 M^2 \rb^{11} + 
	4060456400 L^8 M \rb^{12} \nonumber \\
&&
	\quad +\: 4051974744 L^6 M^3 \rb^{12}-3820351368 L^6 M^2 \rb^{13}+903318850 
	L^6 M \rb^{14} \nonumber \\
&&
	\quad +\: 233659760 L^4 M^3 \rb^{14}-228732252 L^4 M^2 \rb^{15}+59520650 L^4 M 
	\rb^{16}  \\
&&
	\quad -\: 3500 L^4 \rb^{17} - 17332624 L^2 M^3 \rb^{16} + 10640724 L^2 M^2 \rb^{17}-891800 L^2 M 
	\rb^{18}  \\
&&
	\quad -\: 11375 L^2 \rb^{19} + 309120 M^3 \rb^{18}-245280 M^2 \rb^{19}+13125 \rb^{21}\big).
\end{IEEEeqnarray*}


\subsection{Electromagnetic case} \label{sec: emRPsch}
In the electromagnetic case, an ambiguity arises in the definition of $u^a$ in the angular
directions \emph{away} from the world-line. In \eqref{eqn:SelfForceEM} one is free to define
$u^a(x)$ as they wish provided $\lim_{x \rightarrow \xb} u^a(x) = u^\ab$. A natural covariant choice
would be to define this through parallel transport, $u^a(x) = g^a{}_\bb u^\bb$. However,
in reality it is more practical in numerical calculations to define $u^a$ such that its
components in \Sch
coordinates are equal to the components of $u^\ab$ in \Sch coordinates \cite{Barack:Sago:2010}. Doing so,
the regularization parameters are given by
\begin{gather}
F_{t\lnpow{1}} = -\frac{\rbdot \sgn(\Delta r)}{2 (L^2+\rb^2)}, \quad
F_{r\lnpow{1}} = \frac{E\rb \sgn(\Delta r)}{2 (\rb-2M) (L^2+\rb^2)}, \quad
F_{\theta\lnpow{1}} =0, \quad
F_{\phi\lnpow{1}} = 0,
\end{gather}

\begin{equation}
F_{t[0]} = -\frac{E \rbdot}{\pi  \rb \left(\rb^2+L^2\right)^{3/2}} \left( \rb^2 \mathcal{K} +2 L^2 \mathcal{E}\right),
\end{equation}

\begin{equation}
F_{r[0]} = \frac{1}{\pi  \rb^3 \left(\rb^2+L^2\right)^{3/2} \left(\rb-2 M\right)} \left(F^{\mathcal{E}}_{r[0]} \mathcal{E} + F^{\mathcal{K}}_{r[0]} \mathcal{K} \right),
\end{equation}
where
\par \vspace{-6pt} \begin{IEEEeqnarray*}{rCl}
F^{\mathcal{E}}_{r[0]} &=& 2 E^2 L^2 \rb^3 + \left(L^2+\rb^2\right) \left(2 L^2+\rb^2\right) (2 
	M - \rb), \nonumber \\
F^{\mathcal{K}}_{r[0]} &=& E^2 \rb^5 + \rb^2 \left(L^2+\rb^2\right) (\rb-2 M),
\end{IEEEeqnarray*}

\begin{equation}
F_{\theta[0]} = 0,
\end{equation}

\begin{equation} 
F_{\phi[0]} = \frac{ \rbdot}{\pi  L \rb \sqrt{L^2+\rb^2}} \left[\mathcal{E} \left(2 
	L^2+\rb^2\right)-\mathcal{K} \rb^2 \right],
\end{equation}

\begin{equation}
F_{t\lpow{2}} = -\frac{E \rbdot}{2 \pi  \rb^4 \left(L^2+\rb^2\right)^{7/2}} \left( F^{\mathcal{E}}_{t\lpow{2}} \mathcal{E} + F^{\mathcal{K}}_{t\lpow{2}} \mathcal{K} \right),
\end{equation}
where
\par \vspace{-6pt} \begin{IEEEeqnarray*}{rCl}
F^{\mathcal{E}}_{t\lpow{2}} &=& 2 E^2 \rb^5 \left(-L^4+10 L^2 \rb^2+3 \rb^4\right) 
	\nonumber \\
&&
  +\: \left(L^2+\rb^2\right) \left(60 L^6 M+168 L^4 M \rb^2+182 L^2 M \rb^4 -13 L^2 
	\rb^5 +58 M \rb^6-5 \rb^7\right), \nonumber \\
F^{\mathcal{K}}_{t\lpow{2}} &=& - E^2 \rb^7 \left(11 L^2+3\rb^2\right)\\
&&
	+\: 2 \rb^2 \left(L^2+\rb^2\right) \left(-21 L^4 M-48 L^2 M \rb^2 
	+ 3 L^2 \rb^3-23 M \rb^4+\rb^5\right),
\end{IEEEeqnarray*}

\begin{equation}
F_{r\lpow{2}} = -\frac{1}{2 \pi  \rb^6 \left(L^2+\rb^2\right)^{7/2} (\rb-2 M)} \left( F^{ 
	\mathcal{E}}_{r\lpow{2}} \mathcal{E} + F^{\mathcal{K}}_{r\lpow{2}} \mathcal{K} \right),
\end{equation}
where
\par \vspace{-6pt} \begin{IEEEeqnarray*}{rCl}
F^{\mathcal{E}}_{r\lpow{2}} &=& \: -2 E^4 \rb^8 \left(-L^4+10 L^2 \rb^2+3 \rb^4\right) \nonumber \\
&& 
  +\: 2 E^2 \rb^3 \left(L^2+\rb^2\right) \Big(-30 L^6 M-86 L^4 M \rb^2+L^4 \rb^3-98 L^2 
	M \rb^4+10 L^2 \rb^5 \\
&&
	\quad -\: 26 M \rb^6 + \rb^7\Big) \nonumber \\
&&
  -\: \left(L^2+\rb^2\right)^2 (2 M-\rb) \left(44 L^6 M+94 L^4 M \rb^2+ 54 L^2 M \rb^4 + 
	L^2 \rb^5+3 \rb^7\right), \nonumber \\
F^{\mathcal{K}}_{r\lpow{2}} &=& \: E^4 \rb^{10} \left(11 L^2+3 \rb^2\right) \nonumber \\
&& 
  +\: E^2 \rb^5 \left(L^2+\rb^2\right) \left(42 L^4 M+98 L^2 M \rb^2-7 L^2 \rb^3+40 M 
	\rb^4 +\rb^5\right) \nonumber \\
&&
  -\: \rb^2 \left(L^2+\rb^2\right)^2 (\rb-2 M) \left(22 L^4 M+24 L^2 M \rb^2+2 L^2 \rb^3 
	+ 3 \rb^5\right),
\end{IEEEeqnarray*}

\begin{equation}
F_{\theta\lpow{2}} = 0,
\end{equation}

\begin{equation}
F_{\phi\lpow{2}} = -\frac{\rbdot}{2 \pi  L \rb^4 \left(L^2+\rb^2\right)^{5/2}} \left( F^{ 
	\mathcal{E}}_{\phi\lpow{2}} \mathcal{E} + F^{\mathcal{K}}_{\phi\lpow{2}} \mathcal{K}\right),
\end{equation}
where
\par \vspace{-6pt} \begin{IEEEeqnarray*}{rCl}
F^{\mathcal{E}}_{\phi\lpow{2}} &=& E^2 \rb^5 \left(-2 L^4-7 L^2 \rb^2+3 \rb^4\right) \nonumber \\
&&
  -\: \left(L^2+\rb^2\right) \left(44 L^6 M+94 L^4 M \rb^2+54 L^2 M \rb^4+L^2 \rb^5+3 
	\rb^7 \right), \nonumber \\
F^{\mathcal{K}}_{\phi\lpow{2}} &=& -E^2 \rb^7 \left(3 \rb^2-L^2\right) + \rb^2 \left(L^2+\rb^2\right) \left(22 L^4 M+24 L^2 M 
	\rb^2 +2 L^2 \rb^3+3 \rb^5\right),
\end{IEEEeqnarray*}

\begin{equation}
F_{t\lpow{4}} = \frac{3 E \rbdot}{40 \pi  \rb^{11} \left(L^2+\rb^2\right)^{11/2}} \left( 
	F^{\mathcal{E}}_{t\lpow{4}} \mathcal{E} + F^{\mathcal{K}}_{t\lpow{4}} \mathcal{K} \right),
\end{equation}
where
\par \vspace{-6pt} \begin{IEEEeqnarray*}{rCl}
F^{\mathcal{E}}_{t\lpow{4}} &=& -30 E^4 \rb^{14} \left(3 L^6-102 L^4 \rb^2+43 L^2 \rb^4+20 
	\rb^6 \right) \nonumber \\
&& 
  +\: 2 E^2 \rb^5 \left(L^2+\rb^2\right) \big(34560 L^{12} M+169728 L^{10} M \rb^2 + 
	333564 L^8 M \rb^4 \nonumber \\
&&
	\quad +\: 328912 L^6 M \rb^6+167074 L^4 M \rb^8 - 1245 L^4 \rb^9+32948 L^2 M 
	\rb^{10} +1230 L^2 \rb^{11} \nonumber \\
&&
	\quad +\: 230 M \rb^{12} + 555 \rb^{13}\big) \nonumber \\
&& 
  -\: 4 \left(L^2+\rb^2\right)^2 \big(11520 L^{14} M^2-18240 L^{12} M^2 \rb^2+33600 L^{12} 
	M \rb^3 \\
&&
	\quad -\: 226624 L^{10} M^2\rb^4 + 153960 L^{10} M \rb^5 - 496164 L^8 M^2 \rb^6+280764 L^8 M \rb^7 \\
&&
	\quad -\: 485652 
	L^6 M^2 \rb^8 + 255197 L^6 M \rb^9-230930 L^4 M^2 \rb^{10} + 116771 L^4 M \rb^{11}   \\
&&
	\quad -\:  30 
	L^4 \rb^{12}  - 45200 L^2 M^2 \rb^{12}+21487 L^2 M \rb^{13}+270 L^2 \rb^{14}-1342 M^2 
	\rb^{14}  \\
&&
	\quad +\: 229 M \rb^{15} + 120 \rb^{16}\big), \nonumber \\
F^{\mathcal{K}}_{t\lpow{4}} &=& 15 E^4 \rb^{16} \left(-87 L^4+66 L^2 \rb^2+25 \rb^4\right) 
	\nonumber \\
&& 
  -\: 4 E^2 \rb^7 \left(L^2+\rb^2\right) \big(8640 L^{10} M+34872 L^8 M \rb^2+53283 L^6 M 
	\rb^4  \\
&&
	\quad +\:  37555 L^4 M \rb^6 - 225 L^4 \rb^7+9809 L^2 M \rb^8 + 420 L^2 \rb^9+265 M \rb^{10}+165 
	\rb^{11} \big) \nonumber \\
&& 
  +\: \rb^2 \left(L^2+\rb^2\right)^2 \big(23040 L^{12} M^2 - 56640 L^{10} M^2 \rb^2+67200 
	L^{10} M \rb^3 \\
&&
	\quad -\: 402608 L^8 M^2 \rb^4 + 249120 L^8 M \rb^5 - 643736 L^6 M^2 \rb^6+346608 L^6 M \rb^7  \\
&&
	\quad -\:  427796 
	L^4 M^2 \rb^8 + 217192 L^4 M \rb^9-110916 L^2 M^2 \rb^{10} + 52580 L^2 M \rb^{11} \\
&&
	\quad +\: 615 
	L^2 \rb^{12}-5368 M^2 \rb^{12} + 1516 M \rb^{13}+255 \rb^{14}\big),
\end{IEEEeqnarray*}

\begin{equation}
F_{r\lpow{4}} = \frac{3}{40 \pi  \rb^{13} \left(L^2+\rb^2\right)^{11/2} (\rb-2 M)} \left( 
	F^{\mathcal{E}}_{r\lpow{4}} \mathcal{E} + F^{\mathcal{K}}_{r\lpow{4}} \mathcal{K} \right),
\end{equation}
where
\par \vspace{-6pt} \begin{IEEEeqnarray*}{rCl}
F^{\mathcal{E}}_{r\lpow{4}} &=& 30 E^6 \rb^{17} \left(3 L^6-102 L^4 \rb^2+43 L^2 \rb^4+20 
	\rb^6 \right) \nonumber \\
&& 
  -\: E^4 \rb^8 \left(L^2+\rb^2\right) \big(69120 L^{12} M+339456 L^{10} M \rb^2+667128 
	L^8 M \rb^4  \\
&&
	\quad +\:  657884 L^6 M \rb^6 - 30 L^6 \rb^7+334658 L^4 M \rb^8 - 2745 L^4 \rb^9+62656 L^2 M \rb^{10}  \\
&&
	\quad +\:  
	4080 L^2 \rb^{11}+610 M \rb^{12} + 1035 \rb^{13}\big) \nonumber \\
&&
  +\: 2 E^2 \rb^3 \left(L^2+\rb^2\right)^2 \big(23040 L^{14} M^2 - 36480 L^{12} M^2 \rb^2 + 
	67200 L^{12} M \rb^3 \nonumber \\
&& 
	\quad -\: 445056 L^{10} M^2 \rb^4+303824 L^{10} M \rb^5 - 959952 L^8 M^2 \rb^6 + 
	545340 L^8 M \rb^7 \nonumber \\
&&
	\quad -\: 922996 L^6 M^2 \rb^8+486240 L^6 M \rb^9-428644 L^4 M^2 \rb^{10} + 
	216604 L^4 M \rb^{11} \nonumber \\
&&
	\quad +\: 105 L^4 \rb^{12}-78428 L^2 M^2 \rb^{12} + 35368 L^2 M \rb^{13}+1350 L^2 
	\rb^{14} -2684 M^2 \rb^{14} \nonumber \\
&&
	\quad +\: 608 M \rb^{15}+165 \rb^{16}\big) \nonumber \\
&& 
  +\: \left(L^2+\rb^2\right)^3 (2 M-\rb) \big(46080 L^{14} M^2+89856 L^{12} M^2 \rb^2 + 
	53760 L^{12} M \rb^3 \nonumber \\
&& 
	\quad -\: 86336 L^{10} M^2 \rb^4+211520 L^{10} M \rb^5 - 344128 L^8 M^2 \rb^6 + 
	317600 L^8 M \rb^7\nonumber \\
&&
	\quad -\: 306808 L^6 M^2 \rb^8+221140 L^6 M \rb^9-98676 L^4 M^2 \rb^{10}+66220 
	L^4 M \rb^{11}  \\
&&
	\quad +\:  60 L^4 \rb^{12} - 5244 L^2 M^2 \rb^{12}+4440 L^2 M \rb^{13}+105 L^2 \rb^{14}+160 M^2 
	\rb^{14}  \\
&&
	\quad -\:  75 \rb^{16}\big),\\
F^{\mathcal{K}}_{r\lpow{4}} &=& -15 E^6 \rb^{19} \left(-87 L^4+66 L^2 \rb^2+25 \rb^4\right) 
	\nonumber \\
&& 
  +\: E^4 \rb^{10} \left(L^2+\rb^2\right) \big(34560 L^{10} M+139488 L^8 M \rb^2+213132 
	L^6 M \rb^4  \\
&&
	\quad +\:  150250 L^4 M \rb^6 - 915 L^4 \rb^7+37496 L^2 M \rb^8 + 2550 L^2 \rb^9+1210 M \rb^{10} \\
&&
	\quad +\: 585 
	\rb^{11} \big) \nonumber \\
&& 
  -\: E^2 \rb^5 \left(L^2+\rb^2\right)^2 \big(23040 L^{12} M^2-56640 L^{10} M^2 \rb^2 + 
	67200 L^{10} M \rb^3 \nonumber \\
&&
	\quad -\: 394416 L^8 M^2 \rb^4 + 245024 L^8 M \rb^5 - 618528 L^6 M^2 \rb^6 + 
	334004 L^6 M \rb^7 \nonumber \\
&&
	\quad -\: 400636 L^4 M^2 \rb^8+203252 L^4 M \rb^9+180 L^4 \rb^{10} - 97892 L^2 
	M^2 \rb^{10}  \\
&&
	\quad +\:  44568 L^2 M \rb^{11} + 1365 L^2 \rb^{12}-5368 M^2 \rb^{12}+1816 M \rb^{13}+105 \rb^{14}\big) \\
&& 
  -\: \rb^2 \left(L^2+\rb^2\right)^3 \left(\rb-2 M \right) \big(63760 L^8 M^2 \rb^4 -23040 
	L^{12} M^2-24768 L^{10} M^2 \rb^2 \nonumber \\
&&
	\quad -\: 26880 L^{10} M \rb^3-82240 L^8 M \rb^5 + 115848 L^6 M^2 \rb^6-87920 L^6 
	M \rb^7 \nonumber \\
&&
	\quad +\: 56084 L^4 M^2 \rb^8-36340 L^4 M \rb^9+5324 L^2 M^2 \rb^{10}-3540 L^2 
	M \rb^{11} + 15 L^2 \rb^{12} \nonumber \\
&&
	\quad -\: 160 M^2 \rb^{12}+75 \rb^{14}\big),
\end{IEEEeqnarray*}

\begin{equation}
F_{\theta\lpow{4}} = 0,
\end{equation}

\begin{equation}
F_{\phi\lpow{4}} = \frac{3\rbdot}{40 \pi  L \rb^{11} \left(L^2+\rb^2\right)^{9/2}} \left( F^{ 
	\mathcal{E}}_{\phi\lpow{4}} \mathcal{E} + F^{\mathcal{K}}_{\phi\lpow{4}} \mathcal{K} 
	\right),
\end{equation}
where
\par \vspace{-6pt} \begin{IEEEeqnarray*}{rCl}
F^{\mathcal{E}}_{\phi\lpow{4}} &=& -15 E^4 \rb^{14} \left(2 L^6+17 L^4 \rb^2-108 L^2 \rb^4 
	+ 5 \rb^6\right) \nonumber \\
&& 
  +\: 2 E^2 \rb^7 \left(L^2+\rb^2\right) \big(4096 L^{10} M+16188 L^8 M \rb^2+24154 L^6 M 
	\rb^4  \\
&&
	\quad +\:  16608 L^4 M \rb^6 - 165 L^4 \rb^7+5986 L^2 M \rb^8 - 810 L^2 \rb^9+75 \rb^{11}\big) 
	\nonumber \\
&& 
  +\: \left(L^2+\rb^2\right)^2 \big(46080 L^{14} M^2+89856 L^{12} M^2 \rb^2 + 53760 
	L^{12} M \rb^3 \\
&&
	\quad -\: 86336 L^{10} M^2 \rb^4 + 211520 L^{10} M \rb^5 - 344128 L^8 M^2 \rb^6+317600 L^8 M \rb^7 \\
&&
	\quad -\: 306808 
	L^6 M^2 \rb^8 + 221140 L^6 M \rb^9-98676 L^4 M^2 \rb^{10} + 66220 L^4 M \rb^{11} \\
&&
	\quad +\: 60 L^4 
	\rb^{12} -5244 L^2 M^2 \rb^{12} + 4440 L^2 M \rb^{13}+105 L^2 \rb^{14} + 160 M^2 \rb^{14} \\
&&
	\quad -\: 75 \rb^{16}\big), 
	\nonumber \\
F^{\mathcal{K}}_{\phi\lpow{4}} &=& 15 E^4 \rb^{16} \left(L^4-58 L^2 \rb^2+5 \rb^4\right) 
	\nonumber \\
&& 
  -\: 2 E^2 \rb^9 \left(L^2+\rb^2\right) \big(2048 L^8 M+6302 L^6 M \rb^2+6790 L^4 M 
	\rb^4 - 90 L^4 \rb^5 \nonumber \\
&&
	\quad +\: 3256 L^2 M \rb^6-375 L^2 \rb^7+75 \rb^9\big) \nonumber \\
&& 
  +\: \rb^2 \left(L^2+\rb^2\right)^2 \big(-23040 L^{12} M^2-24768 L^{10} M^2 \rb^2-26880 
	L^{10} M \rb^3 \nonumber \\
&&
	\quad +\: 63760 L^8 M^2 \rb^4-82240 L^8 M \rb^5 + 115848 L^6 M^2 \rb^6-87920 
	L^6 M \rb^7 \nonumber \\
&&
	\quad +\: 56084 L^4 M^2 \rb^8-36340 L^4 M \rb^9 + 5324 L^2 M^2 \rb^{10}-3540 L^2 
	M \rb^{11} + 15 L^2 \rb^{12} \nonumber \\
&&
	\quad -\: 160 M^2 \rb^{12}+75 \rb^{14}\big).
\end{IEEEeqnarray*}


\subsection{Gravitational case}


\subsubsection{Self-force regularization} \label{sec: gravityRPsch}
The self force on a gravitational particle is given by
\begin{equation}
F^a = k^{abcd} \hb^{\rm R}_{bc;d},
\end{equation}
where
\begin{equation}
k^{abcd} \equiv \frac12 g^{ad} u^b u^c - g^{ab} u^c u^d - 
  \frac12 u^a u^b u^c u^d + \frac14 u^a g^{b c} u^d + 
 \frac14 g^{a d} g^{b c}.
\end{equation}
Note that, as in the electromagnetic case, an ambiguity arises here due to the presence of
terms involving the four-velocity at $x$. One is free to arbitrarily choose how to define this
provided $\lim_{x\to\bar{x}} u^a = u^{\ab}$. Following Barack and Sago~\cite{Barack:Sago:2010},
we choose to take the \Sch
components of the four velocity at $x$ to be exactly those at $\xb$.  The regularisation parameters in the gravitational case are given by

\begin{gather}
F^t_{\lnpow{1}} = \mp\frac{\rb \rbdot}{2 (L^2+\rb^2) (\rb-2M)}, \quad
F^r_{\lnpow{1}} = \mp\frac{E}{2 (L^2+\rb^2)}, \quad
F^\theta_{\lnpow{1}} = 0, \quad
F^\phi_{\lnpow{1}} = 0,
\end{gather}
\begin{gather}
F^t_{[0]} = -\frac{\rbdot E}{\pi (\rb-2 M) (L^2+\rb^2)^{3/2}} (2 L^2 \mathcal{E}+ \rb^2 
	\mathcal{K}), \\
F^r_{[0]} = \frac{1}{\pi \rb^4 (L^2+\rb^2)^{3/2}} \Big(F_{\mathcal{E}\lpow{0}}^r \mathcal{E}
 	+ F_{\mathcal{K}\lpow{0}}^r \mathcal{K} \Big),
\end{gather}
where
\par \vspace{-6pt} \begin{IEEEeqnarray*}{rCl}
F_{\mathcal{E}\lpow{0}}^r &=& -2 E^2 L^2 \rb^3 + (\rb-2 M) (L^2 + \rb^2) (2 L^2 + \rb^2), 
	\nonumber \\
F_{\mathcal{K}\lpow{0}}^r &=&  - \rb^2 \big[E^2 \rb^3 + ( \rb - 2 M ) (L^2 + \rb^2)\big], 
	\nonumber
\end{IEEEeqnarray*}
\begin{gather}
F^\theta_{[0]} = 0, \quad
F^\phi_{[0]} = - \frac{\rbdot }{\pi L \rb^3 (L^2+\rb^2)^{1/2}}[(2L^2+\rb^2) \mathcal{E}-\rb^2 
	\mathcal{K}],
\end{gather}

\begin{equation}
F^t_{\lpow{2}} = \frac{E \rbdot}{2 \pi \rb^3 (\rb-2 M) (L^2+r^2)^{7/2}}  (F_{ \mathcal{E} 
	\lpow{2}} ^t \mathcal{E} + D_{\mathcal{K}\lpow{2}}^t \mathcal{K}),
\end{equation}
where
\par \vspace{-6pt} \begin{IEEEeqnarray*}{rCl}
F_{\mathcal{E}\lpow{2}}^t &=& - 2 E^2 \rb^5 (11 L^4 + 34 L^2 \rb^2 + 15 \rb^4) \nonumber \\
&& 
  -\: (L^2 + \rb^2) (276 L^6 M + 768 L^4 M \rb^2 + 782 L^2 M \rb^4 - 37 L^2 \rb^5 + 274 M 
	\rb^6 \\
&&
	\quad -\:  29 \rb^7), \\
F_{\mathcal{K}\lpow{2}}^t &=& E^2 \rb^5 (12 L^4 + 35 L^2 \rb^2 + 15 \rb^4) \nonumber \\
&&
  +\: 2 \rb^2 (L^2 + \rb^2) (93 L^4 M + 204 L^2 M \rb^2 - 9 L^2 \rb^3 + 107 M \rb^4 - 7 
	\rb^5),
\end{IEEEeqnarray*}

\begin{equation}
F^r_{\lpow{2}} = \frac{1}{2 \pi \rb^7 (L^2+\rb^2)^{7/2}}(F_{\mathcal{E}\lpow{2}}^r \mathcal{E} + 
	F_{ \mathcal{K}\lpow{2}}^r \mathcal{K}),
\end{equation}
where
\par \vspace{-6pt} \begin{IEEEeqnarray*}{rCl}
F_{\mathcal{E}\lpow{2}}^r &=& - 2 E^4 \rb^8 (11 L^4 + 34 L^2 \rb^2 + 15 \rb^4) \nonumber \\
&& 
  -\: 2 E^2 \rb^3 (L^2 + \rb^2) (138 L^6 M + 422 L^4 M \rb^2 - 19 L^4 \rb^3 + 422 L^2 M 
	\rb^4 - 34 L^2 \rb^5 \nonumber \\
&&
	\quad +\: 122 M \rb^6 - 7 \rb^7), \nonumber\\
&&
  +\: (\rb-2M) (L^2 + \rb^2)^2 (188 L^6 M + 406 L^4 M \rb^2 + 222 L^2 M \rb^4 + 13 L^2 
	\rb^5  \\
&&
	\quad +\: 15 \rb^7), \nonumber \\
F_{\mathcal{K}\lpow{2}}^r &=& E^4 \rb^8 (12 L^4 + 35 L^2 \rb^2 + 15 \rb^4)\nonumber \\
&&
  +\: E^2 \rb^5 (L^2 + \rb^2) (210 L^4 M - 12 L^4 \rb + 410 L^2 M \rb^2 - 19 L^2 \rb^3 + 
	184 M \rb^4 + \rb^5)] \nonumber \\
&&
  -\: \rb^2 (\rb-2 M) (L^2 + \rb^2)^2 (94 L^4 M + 96 L^2 M \rb^2 + 14 L^2 \rb^3 + 15 \rb^5), 
	\nonumber
\end{IEEEeqnarray*}

\begin{equation}
F^\theta_{\lpow{2}} = 0,
\end{equation}

\begin{equation}
F^\phi_{\lpow{2}} = \frac{\rbdot}{2 \pi  L^3 \rb^6 (L^2+\rb^2)^{5/2}}(F_{ \mathcal{E} \lpow{2}} 
	^\phi \mathcal{E} + F_{\mathcal{K}\lpow{2}}^\phi \mathcal{K}),
\end{equation}
where
\par \vspace{-6pt} \begin{IEEEeqnarray*}{rCl}
F_{\mathcal{E}\lpow{2}}^\phi &=& - E^2 L^2 \rb^5 (38 L^4 + 31 L^2 \rb^2 - 15 \rb^4), 
	\nonumber \\
&& 
  -\: (L^2 + \rb^2) (188 L^8 M + 406 L^6 M \rb^2 - 64 L^6 \rb^3 + 222 L^4 M \rb^4 - 163 L^4 
	\rb^5  \\
&&
	\quad -\: 145 L^2 \rb^7  - 48 \rb^9) \nonumber \\
F_{\mathcal{K}\lpow{2}}^\phi &=& E^2 L^2 \rb^5 (12 L^4 + L^2 \rb^2 - 15 \rb^4) \nonumber 
	\\
&&
	+\: \rb^2 (L^2 + \rb^2) (94 L^6 M + 96 L^4 M \rb^2 - 74 L^4 \rb^3 - 121 L^2 \rb^5 - 48 
	\rb^7),
\end{IEEEeqnarray*}

\begin{equation}
F^t_{\lpow{4}} = \frac{3 E \rbdot }{40 \pi \rb^{10} (\rb-2 M) (L^2+r^2)^{11/2}}(F_{ \mathcal{E} 
	\lpow{4}} ^t \mathcal{E} + F_{\mathcal{K}\lpow{4}}^t \mathcal{K}),
\end{equation}
where
\par \vspace{-6pt} \begin{IEEEeqnarray*}{rCl}
F_{\mathcal{E}\lpow{4}}^t &=& 30 E^4 \rb^{10} (64 L^{10} + 384 L^8 \rb^2 + 989 L^6 \rb^4 + 
	1222 L^4 \rb^6 + 437 L^2 \rb^8 + 12 \rb^{10}) \nonumber \\ 
&&
  +\: 2 E^2 \rb^5 (L^2 + \rb^2) (92160 L^{12} M + 445008 L^{10} M \rb^2 + 859044 L^8 M 
	\rb^4  \\
&&
	\quad -\: 1920 L^8 \rb^5 + 838312 L^6 M \rb^6 - 9780 L^6 \rb^7 + 433114 L^4 M \rb^8 - 19065 L^4 
	\rb^9  \\
&&
	\quad +\: 102188 L^2 M \rb^{10} - 9870 L^2 \rb^{11} + 5030 M \rb^{12} - 585 \rb^{13}) \nonumber \\ 
&&
  +\: 4 (L^2 + \rb^2)^2 (46080 L^{14} M^2 + 403200 L^{12} M^2 \rb^2 - 48000 L^{12} M 
	\rb^3 \nonumber \\
&&
	\quad +\: 1231984 L^{10} M^2 \rb^4 - 219840 L^{10} M \rb^5 + 1841004 L^8 M^2 \rb^6 - 
	411324 L^8 M \rb^7 \nonumber \\
&&
	\quad +\: 1490772 L^6 M^2 \rb^8 - 406397 L^6 M \rb^9 + 480 L^6 \rb^{10} + 668810 
	L^4 M^2 \rb^{10} \nonumber \\
&&
	\quad -\: 232331 L^4 M \rb^{11} + 2040 L^4 \rb^{12} + 161840 L^2 M^2 \rb^{12} - 
	75367 L^2 M \rb^{13}  \\
&&
	\quad +\: 1590 L^2 \rb^{14} + 18382 M^2 \rb^{14} - 10669 M \rb^{15} + 210 \rb^{16}), \\
F_{\mathcal{K}\lpow{4}}^t &=& -15 E^4 \rb^{12} (64 L^8 + 328 L^6 \rb^2 + 495 L^4 \rb^4 + 
	22 L^2 \rb^6 - 81 \rb^8) \nonumber \\ 
&&
  -\: 4 E^2 \rb^7 (L^2 + \rb^2) (25920 L^{10} M + 102372 L^8 M \rb^2 + 152523 L^6 M \rb^4 
	- 480 L^6 \rb^5 \nonumber \\
&&
	\quad +\: 103375 L^4 M \rb^6 - 1575 L^4 \rb^7 + 25889 L^2 M \rb^8 - 120 L^2 \rb^9 - 
	455 M \rb^{10}  \\
&&
	\quad +\: 495 \rb^{11}) \nonumber \\ 
&&
  -\: \rb^2 (L^2 + \rb^2)^2 (92160 L^{12} M^2 + 766080 L^{10} M^2 \rb^2 - 130560 L^{10} M 
	\rb^3 \nonumber \\
&&
	\quad +\: 2014208 L^8 M^2 \rb^4 - 497760 L^8 M \rb^5 + 2443016 L^6 M^2 \rb^6 - 
	743088 L^6 M \rb^7 \nonumber \\
&&
	\quad +\: 1527236 L^4 M^2 \rb^8 - 552712 L^4 M \rb^9 + 960 L^4 \rb^{10} + 496596 
	L^2 M^2 \rb^{10} \nonumber \\
&&
	\quad -\: 206180 L^2 M \rb^{11} - 135 L^2 \rb^{12} + 73528 M^2 \rb^{12} - 30796 M 
	\rb^{13} - 735 \rb^{14}),
\end{IEEEeqnarray*}

\begin{equation}
F^r_{\lpow{4}} = \frac{3 }{40 \pi \rb^{14} (L^2+\rb^2)^{11/2}}(F_{\mathcal{E}\lpow{4}}^r 
	\mathcal{E} + F_{\mathcal{K}\lpow{4}}^r \mathcal{K}),
\end{equation}
where
\par \vspace{-6pt} \begin{IEEEeqnarray*}{rCl}
F_{\mathcal{E}\lpow{4}}^r &=& 30 E^6 \rb^{13} (64 L^{10} + 384 L^8 \rb^2 + 989 L^6 \rb^4 + 
	1222 L^4 \rb^6 + 437 L^2 \rb^8 + 12 \rb^{10}) \nonumber \\ 
&&
  +\: E^4 \rb^8 (L^2 + \rb^2) (184320 L^{12} M + 893856 L^{10} M \rb^2 - 1920 L^{10} \rb^3 
	 \\
&&
	\quad +\: 1746888 L^8 M \rb^4- 18240 L^8 \rb^5 + 1744604 L^6 M \rb^6 - 53550 L^6 \rb^7  \\
&&
	\quad +\: 907298 L^4 M 
	\rb^8 - 58665 L^4 \rb^9 + 197536 L^2 M \rb^{10} - 16320 L^2 \rb^{11}  \\
&&
	\quad +\: 9010 M \rb^{12}  - 645 
	\rb^{13}) \nonumber \\ 
&&
  +\: 2 E^2 \rb^3 (L^2 + \rb^2)^2 (92160 L^{14} M^2 + 1036800 L^{12} M^2 \rb^2 - 211200 
	L^{12} M \rb^3 \nonumber \\ 
&&
	\quad +\: 3363456 L^{10} M^2 \rb^4 - 889424 L^{10} M \rb^5 + 5003472 L^8 M^2 \rb^6 
	- 1487220 L^8 M \rb^7 \nonumber \\
&&
	\quad +\: 1920 L^8 \rb^8 + 3866356 L^6 M^2 \rb^8 - 1272840 L^6 M \rb^9 + 9780 L^6 
	\rb^{10} \nonumber \\
&&
	\quad +\: 1570564 L^4 M^2 \rb^{10} - 594724 L^4 M \rb^{11} + 10875 L^4 \rb^{12} + 
	321308 L^2 M^2 \rb^{12} \nonumber \\
&&
	\quad -\: 146848 L^2 M \rb^{13} + 1830 L^2 \rb^{14} + 36764 M^2 \rb^{14} - 20288 M 
	\rb^{15} - 105 \rb^{16}) \nonumber \\ 
&&
  - \:(\rb - 2 M) (L^2 + \rb^2)^3 (184320 L^{14} M^2 + 1711104 L^{12} M^2 \rb^2 - 245760 
	L^{12} M \rb^3 \nonumber \\ 
&&
	\quad +\: 4872896 L^{10} M^2 \rb^4 - 884480 L^{10} M \rb^5 + 6311728 L^8 M^2 \rb^6 
	- 1185120 L^8 M \rb^7 \nonumber \\
&&
	\quad +\: 4083688 L^6 M^2 \rb^8 - 721620 L^6 M \rb^9 + 1920 L^6 \rb^{10} + 
	1299396 L^4 M^2 \rb^{10} \nonumber \\
&&
	\quad -\: 198700 L^4 M \rb^{11} + 2460 L^4 \rb^{12} + 209484 L^2 M^2 \rb^{12} - 
	28120 L^2 M \rb^{13}  \\
&&
	\quad -\: 105 L^2 \rb^{14} + 28640 M^2 \rb^{14} - 5120 M \rb^{15} - 525 \rb^{16}), \nonumber \\
F_{\mathcal{K}\lpow{4}}^r &=& -15 E^6 \rb^{15} (64 L^8 + 328 L^6 \rb^2 + 495 L^4 \rb^4 + 
	22 L^2 \rb^6 - 81 \rb^8) \nonumber \\ 
&&
  -\: E^4 \rb^{10} (L^2 + \rb^2) (103680 L^{10} M + 415248 L^8 M \rb^2 - 2880 L^8 \rb^3  \\
&&
	\quad +\: 
	627612 L^6 M \rb^4 - 10680 L^6 \rb^5 + 418570 L^4 M \rb^6 - 8835 L^4 \rb^7  \\
&&
	\quad +\: 93896 L^2 M 
	\rb^8 + 4350 L^2 \rb^9  - 2870 M \rb^{10} + 2505 \rb^{11}) \nonumber \\ 
&&
  -\: E^2 \rb^5 (L^2 + \rb^2)^2 (92160 L^{12} M^2 + 1019520 L^{10} M^2 \rb^2 - 257280 
	L^{10} M \rb^3 \nonumber \\ 
&&
	\quad +\: 2806176 L^8 M^2 \rb^4 - 893744 L^8 M \rb^5 + 3309168 L^6 M^2 \rb^6 - 
	1180004 L^6 M \rb^7 \nonumber \\
&&
	\quad +\: 1920 L^6 \rb^8 + 1878796 L^4 M^2 \rb^8 - 724772 L^4 M \rb^9 - 900 L^4 
	\rb^{10}  \\
&&
	\quad +\: 517652 L^2 M^2 \rb^{10} - 205608 L^2 M \rb^{11} - 5685 L^2 \rb^{12} + 73528 M^2 \rb^{12}  \\
&&
	\quad -\: 28696 M 
	\rb^{13}  - 1785 \rb^{14}) \nonumber \\ 
&&
  +\: \rb^2 (\rb - 2 M) (L^2 + \rb^2)^3 (92160 L^{12} M^2 + 815232 L^{10} M^2 \rb^2 - 
	157440 L^{10} M \rb^3 \nonumber \\ 
&&
	\quad +\: 1939840 L^8 M^2 \rb^4 - 468160 L^8 M \rb^5 + 1942488 L^6 M^2 \rb^6 - 
	494560 L^6 M \rb^7 \nonumber \\
&&
	\quad +\: 892484 L^4 M^2 \rb^8 - 217780 L^4 M \rb^9 - 960 L^4 \rb^{10} + 195164 
	L^2 M^2 \rb^{10} \nonumber \\
&&
	\quad -\: 38820 L^2 M \rb^{11} - 1545 L^2 \rb^{12} + 28640 M^2 \rb^{12} - 5120 M 
	\rb^{13} - 525 \rb^{14}),
\end{IEEEeqnarray*}

\begin{equation}
F^\theta_{\lpow{4}} = 0,
\end{equation}

\begin{equation}
F^\phi_{\lpow{4}} = \frac{3 \rbdot }{40 \pi \rb^{13} L^5 (L^2+r^2)^{9/2}} (F_{ \mathcal{E} 
	\lpow{4}} ^\phi \mathcal{E} + F_{\mathcal{K}\lpow{4}}^\phi  \mathcal{K}) ,
\end{equation}
where
\par \vspace{-6pt} \begin{IEEEeqnarray*}{rCl}
F_{\mathcal{E}\lpow{4}}^\phi  &=& 15 E^4 L^4 \rb^{10} (128 L^{10} + 960 L^8 \rb^2 + 2266 
	L^6 \rb^4 + 1369 L^4 \rb^6 - 228 L^2 \rb^8 - 35 \rb^{10}) \nonumber \\ 
&&
  +\: 2 E^2 L^2 \rb^5 (L^2 + \rb^2) (115200 L^{14} M + 449744 L^{12} M \rb^2 + 660732 
	L^{10} M \rb^4 \nonumber \\ 
&&
	\quad -\: 5760 L^{10} \rb^5 + 442406 L^8 M \rb^6 - 21300 L^8 \rb^7 + 116472 L^6 M 
	\rb^8 - 18475 L^6 \rb^9 \nonumber \\
&&
	\quad -\: 1186 L^4 M \rb^{10} + 4310 L^4 \rb^{11} + 10765 L^2 \rb^{13} + 4240 
	\rb^{15}) \nonumber \\ 
&&
  +\: (L^2 + \rb^2)^2 (184320 L^{18} M^2 + 1711104 L^{16} M^2 \rb^2 - 245760 L^{16} M 
	\rb^3 \nonumber \\
&&
	\quad +\: 4872896 L^{14} M^2 \rb^4 - 884480 L^{14} M \rb^5 + 6311728 L^{12} M^2 
	\rb^6  \\
&&
	\quad -\: 1090720 L^{12} M \rb^7 + 4083688 L^{10} M^2 \rb^8 - 436180 L^{10} M \rb^9 - 5760 L^{10} \rb^{10}  \\
&&
	\quad +\: 
	1299396 L^8 M^2 \rb^{10} + 73140 L^8 M \rb^{11} - 83700 L^8 \rb^{12} + 209484 L^6 M^2 \rb^{12}  \\
&&
	\quad +\: 
	15720 L^6 M \rb^{13} - 236585 L^6 \rb^{14} + 28640 L^4 M^2 \rb^{14} - 63200 L^4 M \rb^{15}  \\
&&
	\quad -\: 
	271325 L^4 \rb^{16} - 21120 L^2 M \rb^{17} - 138400 L^2 \rb^{18} - 25600 \rb^{20}), \nonumber \\
F_{\mathcal{K}\lpow{4}}^\phi  &=& -15 E^4 L^4 \rb^{12} (192 L^8 + 584 L^6 \rb^2 + 169 L^4 
	\rb^4 - 322 L^2 \rb^6 - 35 \rb^8) \nonumber \\ 
&&
  -\: 2 E^2 L^2 \rb^7 (L^2 + \rb^2) (63360 L^{12} M + 197992 L^{10} M \rb^2 + 216538 L^8 M 
	\rb^4  \\
&&
	\quad -\: 4800 L^8 \rb^5 + 87890 L^6 M \rb^6 - 7110 L^6 \rb^7 + 5264 L^4 M \rb^8 + 2455 L^4 \rb^9  \\
&&
	\quad +\: 
	8645 L^2 \rb^{11} + 4240 \rb^{13}) \nonumber \\ 
&&
  -\: \rb^2 (L^2 + \rb^2)^2 (92160 L^{16} M^2 + 815232 L^{14} M^2 \rb^2 - 157440 L^{14} M 
	\rb^3 \nonumber \\ 
&&
	\quad +\: 1939840 L^{12} M^2 \rb^4 - 422080 L^{12} M \rb^5 + 1942488 L^{10} M^2 
	\rb^6  \\
&&
	\quad -\: 309120 L^{10} M \rb^7 + 892484 L^8 M^2 \rb^8 + 9580 L^8 M \rb^9 - 39360 L^8 \rb^{10}  \\
&&
	\quad +\: 195164 
	L^6 M^2 \rb^{10} + 22780 L^6 M \rb^{11} - 152745 L^6 \rb^{12} + 28640 L^4 M^2 \rb^{12}  \\
&&
	\quad -\: 
	52640 L^4 M \rb^{13} - 213325 L^4 \rb^{14} - 21120 L^2 M \rb^{15} -\: 125600 L^2 \rb^{16}  \\
&&
	\quad -\: 25600 
	\rb^{18}).
\end{IEEEeqnarray*}


\subsubsection{$huu$ regularization} \label{sec: huu}
The quantity
\begin{equation}
H^{\reg} = \frac12 h^\reg_{ab} u^a u^b
\end{equation}
was first proposed by Detweiler \cite{Detweiler:2005} as a tool for constructing gauge invariant
measurements from self-force calculations. It has since proven invaluable in extracting gauge invariant
results from gauge dependent self-force calculations \cite{Detweiler:2008,Barack:2011ed}.

Much the same as with self-force calculations, the calculation of $H^{\reg}$ requires
the subtraction of the appropriate singular piece, $H^{\sing} = \frac12 h^{\sing}_{ab} u^a u^b$
from the full retarded field. In this section, we give this subtraction in the form of
mode-sum regularization parameters.
In doing so, we keep with our convention that the term proportional to
$l+\tfrac12$ is denoted by $H_{\lnpow{1}}$ ($=0$ in this case), the constant term is denoted by $H_{[0]}$,
and so on.

Note that, as in the self-force case, an ambiguity arises here due to the presence of
terms involving the four-velocity at $x$. One is free to arbitrarily choose how to define this
provided $\lim_{x\to\bar{x}} u^a = u^\ab$. As before, we choose this in such a way that the \Sch components of the four velocity at $x$ are exactly those at $\xb$.  The regularisation parameters are then given by

\begin{equation}
H_{[0]} = \frac{2 \mathcal{K}}{\pi \sqrt{L^2+\rb^2}},
\end{equation}
\begin{equation}
H_{\lpow{2}} = \frac{H^{\mathcal{E}}_{\lpow{2}} \mathcal{E} + H^{\mathcal{K}}_{\lpow{2}} 
	\mathcal{K}}{\pi \rb^3 (L^2+\rb^2)^{3/2}},
\end{equation}
where
\par \vspace{-6pt} \begin{IEEEeqnarray*}{rCl}
H^{\mathcal{E}}_{\lpow{2}} &=& 2 E^2 \rb^5 + (L^2 + \rb^2) (36 L^2 M - 8 L^2 \rb + 38 M \
	rb^2 - 9 \rb^3), \\
H^{\mathcal{K}}_{\lpow{2}} &=& - E^2 \rb^3 (16 L^2 + 17 \rb^2) - 2 (L^2 + \rb^2) (16 L^2 M - 
	4 L^2 \rb + 33 M \rb^2 - 12 \rb^3),
\end{IEEEeqnarray*}
\begin{equation}
H_{\lpow{4}} = \frac{3(H^{\mathcal{E}}_{\lpow{4}} \mathcal{E} + H^{\mathcal{K}}_{\lpow{4}} 
	\mathcal{K})}{20 \pi  r^{10} (L^2+r^2)^{7/2}},
\end{equation}
where
\par \vspace{-6pt} \begin{IEEEeqnarray*}{rCl}
H^{\mathcal{E}}_{\lpow{4}} &=& -120 E^4 \rb^{12} (8 L^4 + 17 L^2 \rb^2 + 7 \rb^4) \nonumber 
	\\
&&
  +\: 2 E^2 \rb^5 (L^2 + \rb^2) (3584 L^8 M + 12712 L^6 M \rb^2 + 15516 L^4 M \rb^4 + 120 
	L^4 \rb^5 \nonumber \\
&&
	\quad +\: 6182 L^2 M \rb^6 + 735 L^2 \rb^7 + 34 M \rb^8 + 495 \rb^9) \nonumber \\ 
&&
  +\: 2 (L^2 + \rb^2)^2 (1536 L^{10} M^2 + 13888 L^8 M^2 \rb^2 - 1600 L^8 M \rb^3 + 
	40584 L^6 M^2 \rb^4 \nonumber \\ 
&&
	\quad -\: 9440 L^6 M \rb^5 + 46888 L^4 M^2 \rb^6 - 14100 L^4 M \rb^7 + 120 L^4 
	\rb^8 + 18936 L^2 M^2 \rb^8 \nonumber \\
&&
	\quad -\: 5350 L^2 M \rb^9 + 15 L^2 \rb^{10} + 340 M^2 \rb^{10} + 850 M \rb^{11} - 90 
	\rb^{12}), \\
H^{\mathcal{K}}_{\lpow{4}} &=& 15 E^4 \rb^{10} (64 L^6 + 224 L^4 \rb^2 + 259 L^2 \rb^4 + 
	91 \rb^6) \nonumber \\ 
&&
  -\: 4 E^2 \rb^7 (L^2 + \rb^2) (1376 L^6 M + 3174 L^4 M \rb^2 + 420 L^4 \rb^3 + 1965 L^2 
	M \rb^4  \\
&&
	\quad +\: 960 L^2 \rb^5 + 227 M \rb^6 + 510 \rb^7) \nonumber \\ 
&&
  -\: \rb^2 (L^2 + \rb^2)^2 (1536 L^8 M^2 + 15904 L^6 M^2 \rb^2 - 7360 L^6 M \rb^3 + 
	36160 L^4 M^2 \rb^4 \nonumber \\ 
&&
	\quad -\: 19320 L^4 M \rb^5 + 22412 L^2 M^2 \rb^6 - 11040 L^2 M \rb^7 - 720 L^2 
	\rb^8 + 680 M^2 \rb^8 \nonumber \\
&&
	\quad +\: 860 M \rb^9 - 705 \rb^{10}).
\end{IEEEeqnarray*}


\subsection{Example - Scalar Self-Force} \label{sec: example}
As an example application of our high order regularization parameters, we consider the case
of a scalar particle on a circular geodesic of the \Sch space-time. In this case, the
retarded field may be computed using the frequency domain method
described in \cite{Detweiler:Messaritaki:Whiting:2002}, along with improved asymptotics for the
boundary conditions (by expanding inside the exponential rather than outside) and with the use of the arbitrary precision differential equation solving support in
\emph{Mathematica} \cite{Mathematica}.  These improvements allowed us to substantially increase the accuracy of the computed retarded field. We found this to be necessary to get the full benefit from the the higher order regularization parameters.

If we consider the the scalar wave equation in Eq.~\eqref{eqn: WaveScalar} with zero Ricci scalar as is the case for \Sch space-time,
\begin{equation} \label{eqn: boxPhi}
\Box\Phi (x) = -4 \pi \mu (x),
\end{equation}
where the distributional source,
\begin{equation}
\mu(x) = q \int_{\gamma} \frac{\delta_4 \left(x - x' \right)}{\sqrt{-g}} d\tau,  
\end{equation}
is representing a point charge, $q$ moving along a world line $\gamma$ described by $z^a (\tau)$, where $\tau$ is proper time.  If we consider a circular orbit, i.e., $x' = \{t(\tau), \rb, \pi/2, \Omega t(\tau)\}$ with $\Omega = \left(M / \rb^3 \right)^{1/2}$, we can rewrite the distributional source accordingly,
\par \vspace{-6pt} \begin{IEEEeqnarray}{rCl} \label{eqn: scalarSource}
\mu &=& q \int_{\gamma} \left(- g \right)^{-\tfrac{1}{2}} \delta \left(t - t (\tau) \right) \delta 
	\left( r -  \rb \right) \delta \left(\theta - \pi/2 \right) \delta \left(\phi - \Omega t(\tau) 
	\right) d\tau \nonumber \\
&=&
	q r^{-2} \delta \left( r -  \rb \right) \delta \left(\theta - \pi/2 \right) \delta \left(\phi - 
	\Omega t \right) \left(\frac{dt}{d \tau}\right)^{-1}
\end{IEEEeqnarray}
where we have used $d\tau = \left(\frac{dt}{d \tau}\right)^{-1} dt$.  It is beneficial to now decompose \\
$\delta \left(\theta - \pi/2 \right) \delta \left(\phi - \Omega t \right)$ into spherical harmonics,
\begin{equation}
\delta \left(\theta - \pi/2 \right) \delta \left(\phi - \Omega t \right) = \sum_{lm} D_{lm} Y_{lm} \left(\theta, \phi \right),
\end{equation}
where
\par \vspace{-6pt} \begin{IEEEeqnarray}{rCl} 
D_{lm} &=& \sqrt{\frac{\left(2l + 1\right) }{4 \pi} \frac{ \left(l -m \right) !}{ \left(l + m \right) ! } 
	}\int_{-\pi}^{\pi} e^{- i m \phi} \delta \left(\phi - \Omega t \right) \int_{-\pi/2}^{\pi/2} 
	\delta \left(\theta - \frac{\pi}{2} \right)  P_{lm} \left( \cos{\theta} \right) \sin{\theta} d \theta d 
	\phi \nonumber \\
&=&
	\sqrt{\frac{\left(2l + 1\right) }{4 \pi} \frac{ \left(l -m \right) !}{ \left(l + m \right) ! } } e^{-i 
	m \Omega t} P_{lm} \left( \cos{\pi/2} \right) \nonumber \\
&=&
	e^{-i m \Omega t} Y_{lm}^* \left( \pi/2, 0 \right).
\end{IEEEeqnarray}
Incorporating this into Eq.~\eqref{eqn: scalarSource} gives,
\begin{equation} \label{eqn: muMode}
\mu = \sum_{lm} \frac{q_{lm}}{4 \pi r} \delta \left( r -  \rb \right) e^{ i \omega_m  t} Y_{lm} \left( \theta, \phi \right),
\end{equation}
where
\begin{equation}
\omega_m = -m \Omega, \quad q_{lm} = \frac{4 \pi q}{ r} \frac{ Y_{lm}^* \left( \pi/2, 0 \right)}{ \frac{dt}{d \tau}}, \quad \frac{dt}{d \tau} = \sqrt{\frac{\rb}{\rb - 3 M}},
\end{equation}
and $\frac{dt}{d \tau} $ can be drived from $g_{ab} \dot{x}^a \dot{x}^b = -1$

Expanding out $\Box \Phi = g^{ab} \nabla_a \nabla_b \Phi$ in \Sch space-time gives
\par \vspace{-6pt} \begin{IEEEeqnarray}{rCl} \label{eqn: boxPhiExp}
\frac{r}{\left( r - 2 M \right)} \Box \Phi &=& \frac{\partial ^2 \Phi}{\partial r^2} + \frac{2 \left( r - M \right)}{ 
	\left( r - 2 M \right)} \frac{\partial  \Phi}{\partial  r} - \frac{r^2}{\left(r - 2 M\right)^2} \frac{\partial ^2 \Phi}{\partial  
	t^2} \nonumber \\
&&
	+\: \frac{1}{r \left(r - 2 M\right)} \left[ \frac{\partial ^2 \Phi}{\partial  \theta^2} + \frac{\cos{\theta}}{  
	\sin{ \theta}} \frac{\partial  \Phi}{\partial  \theta} +\frac{1}{\sin^2 \theta} \frac{\partial ^2 \Phi}{\partial  \phi^2} 
	\right],
\end{IEEEeqnarray}
while decomposing the retarded field gives,
\begin{equation} \label{eqn: phiMode}
\Phi = \sum_{lm} \Phi_{lm} e^{i \omega_m t} Y_{lm} \left( \theta, \phi \right). 
\end{equation}
Using Eqs.~\eqref{eqn: muMode}, \eqref{eqn: boxPhiExp} and \eqref{eqn: phiMode} in Eq.~\eqref{eqn: boxPhi} allows us to write the $lm$ part of the scalar wave equation as,
\begin{equation} \label{eqn: phiLm}
\frac{d^2 \Phi_{lm}}{d r^2} + \frac{2 \left( r - M \right)}{\left(r - 2 M \right)} \frac{d \Phi_{lm}}{d r} + \left[ \frac{\omega^2 r^2}{ \left( r - 2 M \right)^2} - \frac{l \left(l + 1 \right)}{r \left( r - 2 M \right)} \right] \Phi_{lm} = - \frac{q_{lm}}{ r - 2 M} \delta \left( r - \rb \right).
\end{equation}

The tortoise coordinate was first introduced by Wheeler \cite{Wheeler:1955}, it is designed to remove the single derivative, $d/dr$ and is given by,
\begin{equation}
\frac{d}{d r_*} = \left( 1 - \frac{2 M}{r} \right) \frac{d}{d r}, \quad \quad \text{or} \quad \quad r_* = r + 2 M \log{ \left(\frac{r}{2 M} - 1 \right)}.
\end{equation}
To obtain appropriate boundary conditions, we consider ingoing waves at the horizon,
\begin{equation} \label{eqn: phiIn}
\Phi_{lm} = \frac{e^{i \omega r_*}}{r}, \quad \quad \text{for} \quad r \rightarrow 2 M,
\end{equation}
and outgoing waves at infinity,
\begin{equation} \label{eqn: phiOut}
\Phi_{lm} = \frac{e^{- i \omega r_*}}{r}, \quad \quad \text{for} \quad r \rightarrow \infty.
\end{equation}

Our aim is now to solve the inhomogeneous equation of Eq.~\eqref{eqn: phiLm} using the above boundary conditions.  To agree with these conditions, we assume that $\Phi_{lm}$ admits an asymptotic expansion in $1 / r$ at $r \rightarrow \infty$ and an asymptotic expansion in $ \left( r - 2 M \right)$ as $r \rightarrow 2 M$.  We know what the leading order behaviour should look like from Eqs.~\eqref{eqn: phiIn} and \eqref{eqn: phiOut}, keeping this in mind, we assume expansions of $\Phi_{lm}$ for $\Phi_{lm}^{(in)}$ and $\Phi_{lm}^{(out)}$ to be,
\begin{equation} \label{eqn: phiOutIn}
\Phi_{lm}^{(out)}(r) = \frac{\exp \left(i \omega r_* \sum_{n=0} \frac{a_n}{r^n} \right)}{r}, \quad \quad \text{and} \quad \quad \Phi_{lm}^{(in)}(r) =\frac{ e^{- i \omega r_*}}{r} \sum_{n=0} b_n \left( r - 2 M\right)^n
\end{equation}
To determine the coefficients $a_n$ and $b_n$, we use Eq.~\eqref{eqn: phiLm}.  For the outgoing waves, hence the $a_n$'s, we can use mathematical packages like Mathematica to substitute  $\Phi_{lm}^{(out)}$ into the homogeneous equation of Eq.~\eqref{eqn: phiLm} and solve for the coefficients using initial conditions, $a_0 = 1$, $a_{n<0} = 0$.  For the ingoing wave, it is possible to analytically obtain a recursion relation for the $b_n$'s by substituting in $\Phi_{lm}^{(in)}$ into Eq.~\eqref{eqn: phiLm}, that is,
\par \vspace{-6pt} \begin{IEEEeqnarray}{rCl}
b_n &=& - \frac{-12 i \omega M \left( n - 1 \right) + \left( 2 n - 3 \right) \left( n - 1 \right) - 
	\left( l^2 + l + 1 \right)}{ 2 M \left( - 4 i n \omega M + n^2 \right)} b_{n-1} \nonumber \\
&&
	-\: \frac{ - 12 i \omega M \left( n - 2 \right) + \left( n - 2 \right) \left( n - 3 \right) - l 
	\left( l + 1 \right) }{4 M^2 \left( - 4 i n \omega M + n^2 \right)} b_{n-2} \nonumber \\
&&
	+\: \frac{ i \omega \left(n - 3 \right) }{2 M^2 \left( - 4 i n \omega M + n^2 \right)} b_{n-3},
\end{IEEEeqnarray}
and use starting values, $b_0 = 1$ and $b_{n<0} = 0$ to determine the required $b_n$'s.

Once values for the coefficients $a_n$ and $b_n$ have been obtained, we use our expressions 
for $\Phi_{lm}^{(in)}$ and $\Phi_{lm}^{(out)}$ from Eqs.~\eqref{eqn: phiOutIn} in Eq.~\eqref{eqn: phiLm} to derive initial conitions away from the singular parts so we can numerically integrate to obtain homogeneous solutions, with the appropriate boundary conditions.  The inhomogeneous solution to Eq.~\eqref{eqn: phiLm} is now of the type,
\par \vspace{-6pt} \begin{IEEEeqnarray}{rCl} \label{eqn: phiFull}
\Phi_{lm} &=& \mathcal{A}_{lm} \Phi_{lm}^{(in)} (r) \Theta \left( \rb - r \right) + \mathcal{B}_{lm} \Phi_{lm}^{(out)} (r) \Theta 
	\left( r - \rb \right) \nonumber \\
&=&
	\Phi_{lm}^{-} (r) \Theta \left( \rb - r \right) + \Phi_{lm}^{+} (r) \Theta \left( r - \rb \right),
\end{IEEEeqnarray}
where $\Theta\left( \rb - r \right)$ is the Heaviside step function previously introduced in Eq.~\eqref{eqn: step} and we have introduced the notation $\Phi_{lm}^{-} (r) = \mathcal{A}_{lm} \Phi_{lm}^{(in)} (r)$ and  $\Phi_{lm}^{+} (r) = \mathcal{B}_{lm} \Phi_{lm}^{(out)} (r)$.  $\mathcal{A}_{lm}$ and $\mathcal{B}_{lm}$ can be determined by imposing suitable matching conditions.  These are found by substituting $\Phi_{lm} $ from Eq.~\eqref{eqn: phiFull} into the radial equation, Eq.~\eqref{eqn: phiLm}\fixme{, and gathering like terms;}
\par \vspace{-6pt} \begin{IEEEeqnarray}{rCl} \label{eqn: phiPlusMinus}
0&=&\delta ' \left( r - \rb \right) \left( \Phi_{lm}^{+} - \Phi_{lm}^{-} \right) \nonumber \\
&&
	+\: \delta \left( r - \rb \right) \left[ 2 \left( \Phi_{lm}^+{} ' - \Phi_{lm}^- {} ' \right) + \frac{2 
	\left( r - M \right)}{r - 2 M} \left( \Phi_{lm}^{+} - \Phi_{lm}^{-} \right) + \frac{q_{lm}}{r - 2 
	M} \right] \nonumber \\
&&
	+\: \Theta \left( \rb - r\right) \left[ \Phi_{lm}^{-}{} '' + \frac{2 \left( r - M \right)}{r - 2 M} 
	\Phi_{lm}^{-} {}' + W \Phi_{lm}^{-} \right] \nonumber \\
&&
	+\: \Theta \left( r - \rb \right) \left[ \Phi_{lm}^{+}{} '' + \frac{2 \left( r - M \right)}{r - 2 M} 
	\Phi_{lm}^{+} {}' + W \Phi_{lm}^{+} \right] ,
\end{IEEEeqnarray}
where
\begin{equation}
W = \frac{\omega^2 r^2}{ \left( r - 2 M \right)^2} - \frac{l \left(l + 1 \right)}{r \left( r - 2 M 
	\right)},
\end{equation}
and $'$ refers to differentiation with respect to $r$.  It can be clearly seen that the last two terms of Eq.~\eqref{eqn: phiPlusMinus} are zero as they are the homogeneous radial equation, which $\Phi_{lm}^{-} (r)$ and $\Phi_{lm}^{+} (r)$ solve by design.  Our other two terms give,
\par \vspace{-6pt} \begin{IEEEeqnarray}{rCl} \label{eqn: Philms}
\left[\Phi_{lm}^{+} - \Phi_{lm}^{-} \right]_{r=\rb} &=& 0, \nonumber \\
\left[\Phi_{lm}^{+}{} ' - \Phi_{lm}^{-} {} ' \right]_{r=\rb} &=& \frac{1}{2} \left[\frac{- q_{lm}}{r - 2 M} \right]_{r = \rb}.
\end{IEEEeqnarray}
It is now possible to substitute $\Phi_{lm}^{-} (r) = \mathcal{A}_{lm} \Phi_{lm}^{(in)} (r)$ and  $\Phi_{lm}^{+} (r) = \mathcal{B}_{lm} \Phi_{lm}^{(out)} (r)$ back into Eq.~\eqref{eqn: Philms}, to obtain a simple system of simultaneous equations in $\mathcal{A}_{lm}$ and $\mathcal{B}_{lm}$.  These are easily solved to give,
\par \vspace{-6pt} \begin{IEEEeqnarray}{rCl} \label{eqn: AB}
\mathcal{A}_{lm} &=& \frac{ - q_{lm} \Phi_{lm}^{(in)} (\rb)}{ 2 \left( r - 2 M \right) \left[ \Phi_{lm}^{(out)} (\rb) \Phi_{lm}^{(in)}{} ' (\rb) - \Phi_{lm}^{(in)} (\rb) \Phi_{lm}^{(out)} {}' (\rb) \right]}, \nonumber \\
\mathcal{B}_{lm} &=& \frac{ - q_{lm} \Phi_{lm}^{(out)} (\rb)}{ 2 \left( r - 2 M \right) \left[ \Phi_{lm}^{(out)} (\rb) \Phi_{lm}^{(in)} {}' (\rb) - \Phi_{lm}^{(in)} (\rb) \Phi_{lm}^{(out)}{} ' (\rb) \right]}.
\end{IEEEeqnarray}

The $l$ component of the retarded self-force \begin{figure}
\begin{center}
\includegraphics[width=14cm]{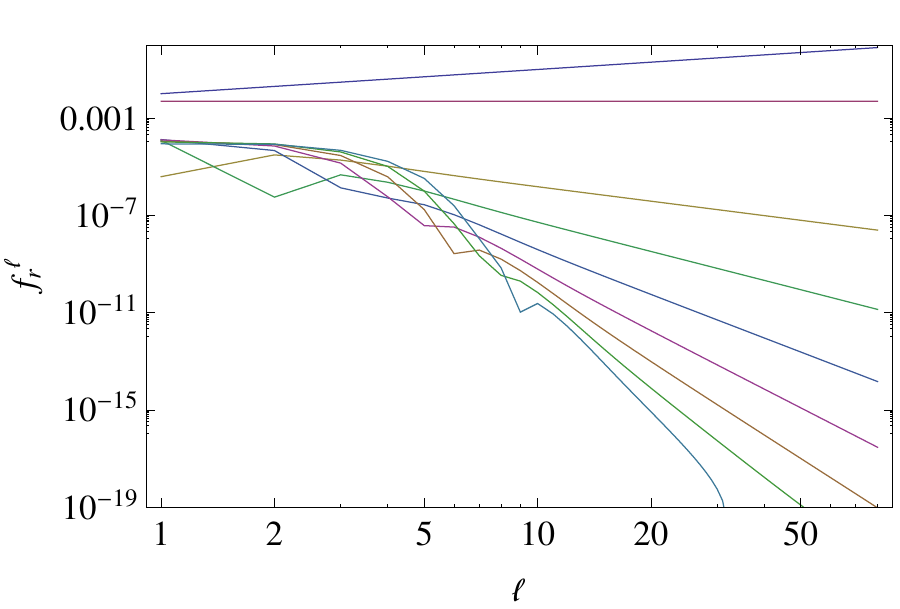}
\caption[Regularization of the Scalar Self-Force Radial Component in \Sch Space-Time]{Regularization of the radial component of the scalar self-force in \Sch space-time for the case of a scalar particle on a circular geodesic of radius $r_0 = 10M$ in \Sch space-time.  In decreasing slope the above lines represent the unregularised self-force, self-force regularised by subtracting from it in turn the cumulative sum of $F^l_{r[-1]}$, $F^l_{r[0]}$, $F^l_{r\lpow{2}}$, $F^l_{r\lpow{4}}$, $F^l_{r\lpow{6}}$, $F^l_{r\lpow{8}}$, $F^l_{r\lpow{10}}$, $F^l_{r\lpow{12}}$.}
\label{fig:scalar-circular-reg}
\end{center}
\end{figure} with the help of Eq.~\eqref{eqn: FretSing} is now given by
\begin{equation}
F_a^l{}_{\rm \ret} = \frac{\partial}{\partial x^a} \sum _{m=-l}^{l}  \left[\mathcal{A}_{lm} 
	\Phi_{lm}^{(in)} (r) \Theta \left( \rb - r \right) + \mathcal{B}_{lm} \Phi_{lm}^{(out)} (r) \Theta 
	\left( r - \rb \right) \right] \nonumber \\
\end{equation}
where we have numerically solved for $\Phi_{lm}^{in} (r)$ and $\Phi_{lm}^{out} (r)$ and $\mathcal{A}_{lm}$ and $\mathcal{B}_{lm}$ are given by Eq.~\eqref{eqn: AB}.  We have also replaced $p_a{}_A$ with its scalar operator $\frac{\partial}{\partial x^a}$.  The $l$ mode of the full self force can now be calculated from Eq.~\eqref{eqn: FaSplit},
\begin{equation} \label{eqn: FaSplit}
F_a = \sum^l \left( F_a^l{}_{\rm \ret}  -  F_a^l{}_{\rm \sing}  \right),
\end{equation}
where we have shown the calculation of both $F_a^l{}_{\rm \ret}$ and $F_a^l{}_{\rm \sing}$ in the last two sections.


\subsection{Impact of Regularisation Parameters in \Sch Space-Time}

\begin{figure}[t]
\begin{center}
\includegraphics[width=14cm]{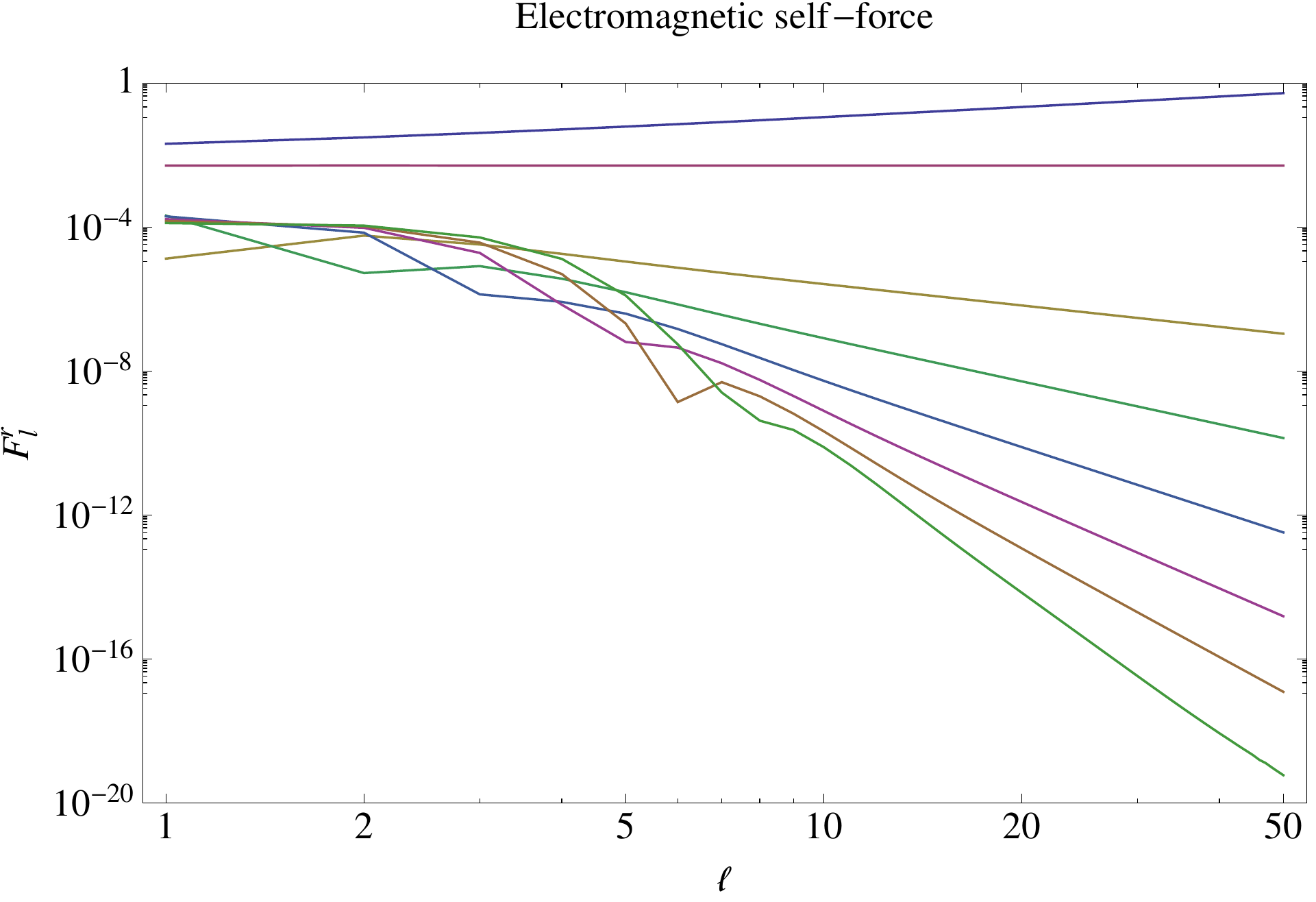}
\caption[Regularization of the Electromagnetic Self-Force Radial Component in \Sch Space-Time]{Regularization of the radial component of the self-force for the case of a electromagnetic particle on a circular geodesic of radius $r_0 = 10M$ in \Sch space-time.  In decreasing slope the above lines represent the unregularised self-force and the self-force regularised by subtracting from it in turn the cumulative sum of $F^l_{r[-1]}$, $F^l_{r[0]}$, $F^l_{r\lpow{2}}$, $F^l_{r\lpow{4}}$, $F^l_{r\lpow{6}}$, $F^l_{r\lpow{8}}$, $F^l_{r\lpow{10}}$.}
\label{fig: emSchw}
\end{center}
\end{figure}

\begin{figure} [ht]
\begin{center}
\includegraphics[width=14cm]{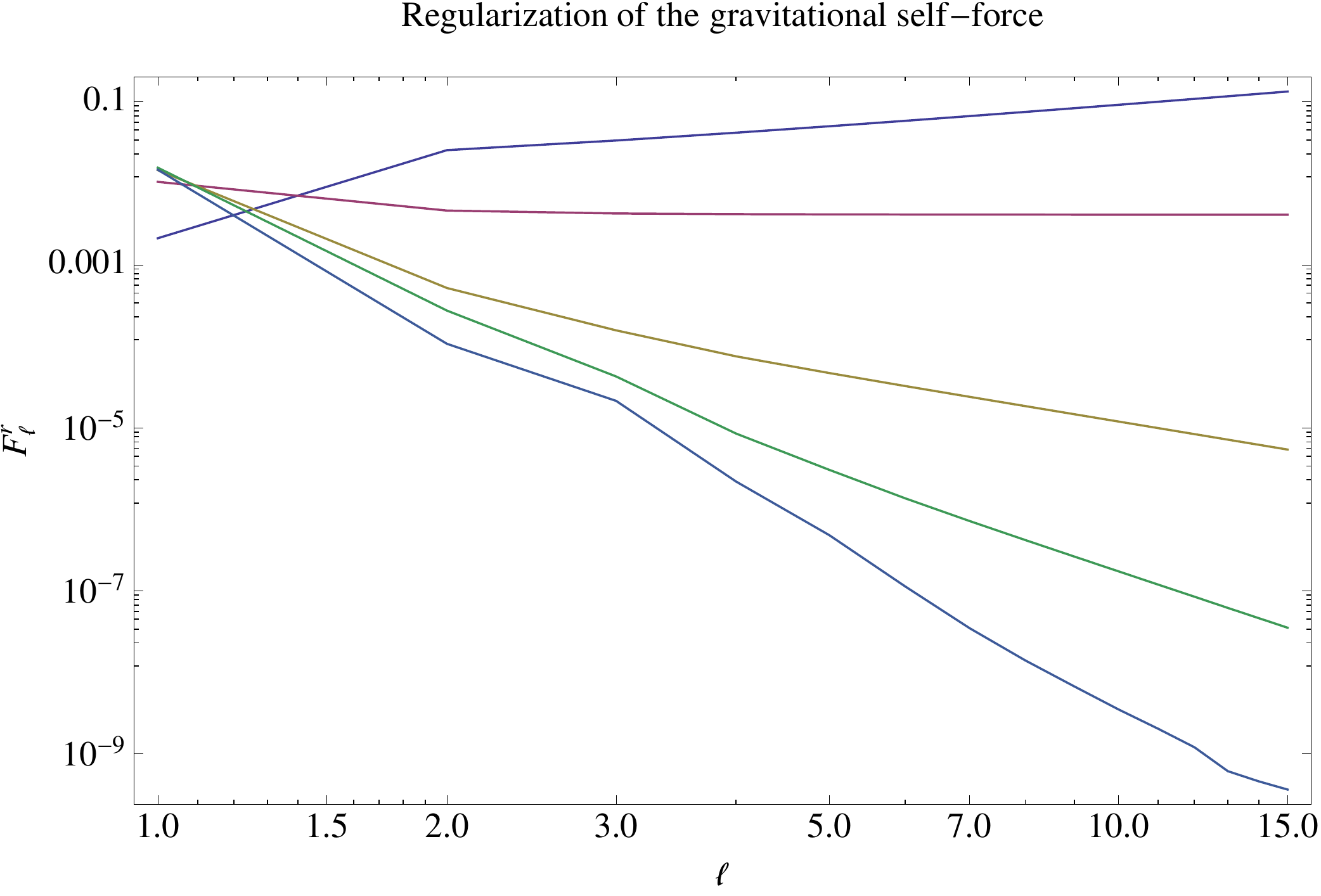}
\caption[Regularization of the Gravitational Self-Force Radial Component in \Sch Space-Time]{Regularization of the radial component of the self-force for the case of a gravitational particle on a elliptic geodesic of radius $r_0 = 10M$ in \Sch space-time.  The graph is plotting $F^l_{r}$ against $l$.  In decreasing slope the above lines represent the unregularised self-force and the self-force regularised by subtracting from it in turn the cumulative sum of $F^l_{r[-1]}$, $F^l_{r[0]}$, $F^l_{r\lpow{2}}$, $F^l_{r\lpow{4}}$.}
\label{fig: gravitySch}
\end{center}
\end{figure}

The results (and benefits) of the calculations in Sec.~
\ref{sec: schRPs} are illustrated in Figs.~\ref{fig:scalar-circular-reg}, \ref{fig: emSchw} and \ref{fig: gravitySch}.  There we show the effect of subtracting in turn the cumulative sums of the regularization
parameters from the full retarded field.

In Fig.~\ref{fig:scalar-circular-reg}, in order from top to bottom are $F_r^{\rm ret}$ and the result of subtracting
from it in turn the cumulative sum of the regularization terms $F^l_{r\lnpow{1}}$, $F^l_{r[0]}$, $F^l_{r\lpow{2}}$, $F^l_{r\lpow{4}}$, $F^l_{r\lpow{6}}$, $F^l_{r\lpow{8}}$, $F^l_{r\lpow{10}}$ and $F^l_{r\lpow{12}}$.
The parameters $F_{r\lnpow{1}}$, $F_{r[0]}$, $F_{r\lpow{2}}$, $F_{r\lpow{4}}$ and $F_{r\lpow{6}}$ are analytically derived in
Sec.~\ref{sec:scalar-regularization}, while $F_{r\lpow{8}}$, $F_{r\lpow{10}}$ and $F_{r\lpow{12}}$ were determined
through a numerical fit to the data. The resulting rapid convergence with $l$ enables the
calculation of an extremely accurate value for the self-force. Summing over $l$, we find
$F_r = 0.000013784482575667959(3)$, where the uncertainty in the last digit is estimated by
assuming that the only error comes from limiting the sum to a finite $l_{\rm max} = 80$.

In addition to providing a highly accurate benchmark, the example in the previous section may be used to assess the benefits which can be obtained from the use of higher-order regularization parameters.  The most obvious benefit is that with fixed computational resources (i.e. fixed number of
spherical harmonic modes) one can obtain a much more accurate value for the self-force.
This is highlighted by comparison of our value for $F_r$ with that of the previous
benchmark given in \cite{Detweiler:Messaritaki:Whiting:2002}, \fixme{$F_r = 0.0000137844828(2)$}.  Both calculations consider the same case of a scalar charge in a circular orbit of radius $10M$ around a
\Sch black hole. 
 \begin{table} [ht]
\centering
\scalebox{0.8}{\begin{tabular}{ | l | l l | l l | }
\hline
 			& \multicolumn{2}{c|}{$l_\text{max}=25,n=12$} 						& \multicolumn{2}{c|}{$l_\text{max}=80,n=50$} \\
  RPs used	& abs.	& rel.			& abs.	& rel.		\\
  \hline                        
  $AB$ 		& $1.3784482573\times10^{-5}$		& $1.2\times10^{-10}$ 	& $1.37844825756674\times10^{-5}$			& $3.7\times10^{-14}$\\
  $ABD$ 		& $1.37844825757\times10^{-5}$		& $5.0\times10^{-12}$ 	& $1.378448257566791\times10^{-5}$			& $3.3\times10^{-15}$\\
  $ABDF$ 		& $1.378448257567\times10^{-5}$		& $4.2\times10^{-13}$ 	& $1.378448257566793\times10^{-5}$			& $1.7\times10^{-15}$\\
  $ABDFH$ 	& $1.37844825756675\times10^{-5}$	& $3.0\times10^{-14}$ 	& $1.3784482575667951\times10^{-5}$			& $5.5\times10^{-16}$\\    
  \hline\hline
  CPU time	&  \multicolumn{2}{c|}{155s} 						& \multicolumn{2}{c|}{4247s}\\
  \hline
\end{tabular}}
\caption[Impact of Regularisation Parameters for Scalar Self-Force]{Table demonstrating the usefulness of the analytically derived higher order regularization parameters in practical self-force calculations. In this example we show the regularization of the radial component of the scalar self-force for a circular orbit at $r_0=10M$ about a \Sch black hole. In the left most column we list the analytically derived regularization parameters employed in each calculation. For each calculation we numerically fit the higher order regularization parameters up to $F_{r\lpow{15}}$. The next two wide columns show the result of computing the scalar self-force for $25$ and $80$ $l$-modes respectively ($l_{max}$ = number of modes) and numerically fitting the unknown regularization parameters.
We show the resulting absolute value of the self-force and its relative difference verses the highly accurate value provided in the main text. The CPU time taken to compute the $l$-modes of the retarded field using a code running on 12 cores of a machine with a 3GHz clock speed is also given.  This shows the improvement in run-time is over a factor of 36}
\end{table}\label{tab: schRP}
Using $40$ $l$-modes and regularization parameters up
to $F^l_{r[2]}$, \cite{Detweiler:Messaritaki:Whiting:2002}
obtained a value for the self-force with a fractional accuracy of $10^{-9}$. The inclusion
of the next two regularization parameters improves this to a fractional error of $10^{-12}$
which increases to a fractional error of $10^{-17}$ when $80$ modes are used.  

To illustrate this further we have included Table~\ref{tab: schRP}, where the physical time taken to run numerical code is recorded as is the accuracy obtained.  The results in this table show that, by using the regularization parameters derived in this thesis, it is possible to calculate a scalar self-force using only 25 $l$-modes which is as accurate as a calculation made using 80 $l$-modes and regularizing with only the $A$ and $B$ parameters. As the high $l$-modes are computationally expensive to calculate, using the higher order regularization parameters offers a substantial improvement in code run-time for a fixed level of accuracy in the final result. In this example the improvement in run-time is over a factor of 36.

This example represents a somewhat extreme case: it uses highly accurate frequency domain
methods combined with high-precision numerical integration and a relatively large number
of spherical harmonic modes. In more typical time-domain calculations, numerical
data up to $l \sim 15$ is used and it is common that the dominant source of error comes
from the tail fit. While it may seem that one merely needs to compute more modes to reduce
this error, this is not a realistic solution. 
In a mode-sum
calculation, the number of spherical harmonic modes required for each $l$ scales as $l^2$,
meaning that simply running simulations for larger and larger $l$ rapidly becomes
prohibitively expensive in terms of computational cost.  Additionally, the improvement
with each additional mode falls off as an inverse power in $l$, meaning that many more
$l$ modes are required for an increasingly small benefit. In this case, the inclusion of
higher order regularization parameters essentially eliminates this problem: without them
the tail fit is the dominant source of error, with them sufficiently accurate results
may be obtained without even fitting for a tail. 

In the electromagnetic and gravitational cases, data for the retarded field, the ansatz of what was calculated in  Sec.~\ref{sec: example}, is increasingly more complicated with higher spins. As outlined in Sec.~\ref{sec: thesisOutline}, the aim of this thesis is to concentrate on the singular component of the self-force.  We therefore did not derive the retarded field for the higher spin cases.  However, with the assistance  of Roland Haas, Patrick Nolan and the Southampton group (Sarp Ackay, Niels Warburton, Leor Barack), who gave us access to their data for the retarded field in these cases, we are able to show the success of our regularisation parameters.  Figs.~\ref{fig: emSchw} and \ref{fig: gravitySch} depict the retarded field in the electromagnetic and gravitational cases respectively; unregularised and then regularised by the sum of the parameters for different $n$, where $n$ can be considered as the order of $\epsilon$ to which we calculated the singular field - this is currently being increased with ongoing work from the Dublin Self-Force group \cite{Heffernan:2012b}.  

In all three cases, we can clearly see that with the regularisation parameters comes a jump in fractional accuracy with the same number of $l$-modes - meaning it is possible to calculate more accurate data with the same number of $l$ modes.  The importance of this lies with the exponential increase in computation time with the higher $l$ modes, making the higher regularisation parameters invaluable to those numerically calculating the self-force.

It should be pointed out that there is one caveat to our conclusions. The use of high
order regularization parameters requires the subtraction of increasingly
(relatively) large numbers to obtain a small regularized remainder. It is therefore
essential that any numerically provided data for the retarded field must be of sufficient
accuracy for the subtraction to yield meaningful results. As a result, calculations which were
previously deemed sufficient would not necessarily gain an immediate benefit from higher order
regularization parameters.


\section{Kerr Space-time}

The Kerr cases follow the same necessary calculations, for the self-force, as their \Sch counterparts.  I therefore won't go through the calculations again but remind the reader that the necessary calculations for the scalar, electromagnetic, gravitational and $h u u$ regularisation parameters is explained in detail in Secs.~\ref{sec:scalar-regularization}, \ref{sec: emRPsch}, \ref{sec: gravityRPsch} and \ref{sec: huu} respectively.  The regularisation parameters for Kerr at the higher orders prove to be too large for paper format, in these instances they have been made available in electronic form \cite{BarryWardell.net}


\subsection{Scalar Case}
The regularisation parameters in the Kerr scalar case are given by,
\begin{gather}
F_{t\lnpow{1}} = \frac{\rb \rbdot \text{sgn$\Delta $r}}{\rb \left(a^2+L^2\right)+2
   a^2 M+\rb^3}, \nonumber \\
F_{r\lnpow{1}} = -\frac{\text{sgn$\Delta $r} \left(E \rb \left(a^2+\rb^2\right)+2
   a M (a E-L)\right)}{\left(a^2-2 M \rb+\rb^2\right)
   \left(\rb \left(a^2+L^2\right)+2 a^2 M+\rb^3\right)}, \nonumber \\
F_{\theta\lnpow{1}} = 0, \quad
F_{\phi\lnpow{1}} = 0,
\end{gather}

\begin{equation}
F_{t[0]} = \frac{\rbdot}{\pi  \rb^2 \left(\rb^2+L^2 + \frac{2 a^2 M}{\rb} + a^2 \right)^{3/2}} \left(F^{\mathcal{E}}_{t[0]} \mathcal{E} + F^{\mathcal{K}}_{t[0]} \mathcal{K} \right),
\end{equation}
where
\par \vspace{-6pt} \begin{IEEEeqnarray*}{rCl}
F^{\mathcal{E}}_{t[0]} &=& 4 a L M \left(4 a^4 M^2+2 a^4 M \rb+2 a^2 L^2 M \rb-a^2 M \rb^3-a^2
   \rb^4-L^2 \rb^4\right) \nonumber \\
&&
	+\: E \big(-12 a^6 M^3-16 a^6 M^2
   \rb-7 a^6 M \rb^2-a^6 \rb^3-4 a^4 L^2 M^2 \rb-6 a^4 L^2 M
   \rb^2 \nonumber \\
&& \quad
	-\: 2 a^4 L^2 \rb^3-6 a^4 M^2 \rb^3-5 a^4 M \rb^4-a^4
   \rb^5+a^2 L^4 M \rb^2-a^2 L^4 \rb^3 \\
&& \quad
	-\:5 a^2 L^2 M \rb^4-3
   a^2 L^2 \rb^5-2 L^4 \rb^5\big), \\
F^{\mathcal{K}}_{t[0]} &=& -2 a
   L M \big(2 a^4 M^2-a^4 M \rb-a^4 \rb^2-a^2 L^2 M \rb-2 a^2 L^2
   \rb^2-2 a^2 M \rb^3-2 a^2 \rb^4 \\
&& \quad
	-\:L^4 \rb^2-2 L^2
   \rb^4\big) \nonumber \\
&&
	+\: E \big(4 a^6 M^3+4 a^6 M^2 \rb+a^6 M \rb^2-2 a^4 L^2 M^2
   \rb-a^4 L^2 M \rb^2+2 a^4 M^2 \rb^3 \\
&& \quad
	+\:a^4 M \rb^4-2 a^2 L^4
   M \rb^2+a^2 L^2 M \rb^4+a^2 L^2 \rb^5+L^4 \rb^5\big),
\end{IEEEeqnarray*}

\begin{equation}
F_{r[0]} = \frac{F^{\mathcal{E}}_{\text{\fixme{$r$}}[0]} \mathcal{E} + F^{\mathcal{K}}_{\text{\fixme{$r$}}[0]} \mathcal{K}}{\pi  \rb^3 \left( 2 a^2 M + a^2 \rb + L^2 \rb \right)^2 \left(\rb^2+L^2 + \frac{2 a^2 M}{\rb} + a^2 \right)^{3/2} \left(\rb^2-2 M \rb +a^2 \right)},
\end{equation}
where
\par \vspace{-6pt} \begin{IEEEeqnarray*}{rCl}
F^{\mathcal{E}}_{\text{\fixme{$r$}}[0]} &=& \big(-24 a^8 M^3 \rb-32 a^8 M^2 \rb^2-14 a^8 M \rb^3-2 a^8
   \rb^4+24 a^6 L^2 M^4+12 a^6 L^2 M^3 \rb \\
&& \quad
	-\: 30 a^6 L^2 M^2 \rb^2-27
   a^6 L^2 M \rb^3-6 a^6 L^2 \rb^4+48 a^6 M^4 \rb^2+40 a^6 M^3
   \rb^3 \\
&& \quad
	-\:16 a^6 M^2 \rb^4-20 a^6 M \rb^5-4 a^6 \rb^6+8 a^4
   L^4 M^3 \rb-12 a^4 L^4 M \rb^3-6 a^4 L^4 \rb^4 \\
&& \quad
	+\: 36 a^4 L^2 M^3
   \rb^3+12 a^4 L^2 M^2 \rb^4-21 a^4 L^2 M \rb^5-9 a^4 L^2
   \rb^6+24 a^4 M^3 \rb^5 \\
&& \quad 
	+\: 8 a^4 M^2 \rb^6-6 a^4 M \rb^7-2 a^4
   \rb^8-2 a^2 L^6 M^2 \rb^2+a^2 L^6 M \rb^3-2 a^2 L^6
   \rb^4 \\
&& \quad
	+\: 6 a^2 L^4 M^2 \rb^4+a^2 L^4 M \rb^5-6 a^2 L^4
   \rb^6+12 a^2 L^2 M^2 \rb^6-3 a^2 L^2 \rb^8  \\
&& \quad
	+\: 2 L^6
   M \rb^5-L^6 \rb^6+2 L^4 M \rb^7-L^4 \rb^8\big) \\
&&
	-\:2 a E L M \big(24 a^6 M^3+28 a^6 M^2
   \rb+10 a^6 M \rb^2+a^6 \rb^3
	+ 8 a^4 L^2 M^2 \rb \\
&& \quad
	+\: 8 a^4 L^2 M
   \rb^2+2 a^4 L^2 \rb^3-4 a^4 M \rb^4-2 a^4 \rb^5-2 a^2 L^4
   M \rb^2+a^2 L^4 \rb^3 \\
&& \quad
	-\: 2 a^2 L^2 M \rb^4-a^2 L^2 \rb^5-6
   a^2 M \rb^6-3 a^2 \rb^7+L^4 \rb^5-3 L^2 \rb^7\big)\\
&&
	+\: E^2 \left(2 a^2 M+a^2 \rb+\rb^3\right) \big(12 a^6 M^3+16 a^6 M^2 \rb+7 a^6 M 
	\rb^2+a^6 \rb^3\\
&& \quad
	+\:4 a^4 L^2 M^2 \rb+6 a^4 L^2 M \rb^2+2 a^4 L^2 \rb^3+6 a^4 M^2 \rb^3+5 a^4 M 
	\rb^4+a^4 \rb^5 \\
&& \quad
	-\: a^2 L^4 M \rb^2+a^2 L^4 \rb^3+5 a^2 L^2 M \rb^4+3 a^2 L^2 \rb^5+2 L^4 \rb^5 
	\big), \\
F^{\mathcal{K}}_{\text{\fixme{$r$}}[0]} &=& \big(8 a^8 M^3 \rb+8 a^8 M^2 \rb^2+2 a^8 M \rb^3-8 a^6 L^2 M^4+4 a^6 L^2
   M^3 \rb+12 a^6 L^2 M^2 \rb^2 \\
&& \quad
	+\:4 a^6 L^2 M \rb^3-16 a^6 M^4
   \rb^2-8 a^6 M^3 \rb^3+8 a^6 M^2 \rb^4+4 a^6 M \rb^5+4 a^4
   L^4 M^3 \rb \\
&& \quad 
	+\: 8 a^4 L^4 M^2 \rb^2+2 a^4 L^4 M \rb^3
	+ 8 a^4 L^2 M^3
   \rb^3+12 a^4 L^2 M^2 \rb^4+2 a^4 L^2 M \rb^5 \\
&& \quad
	-\: a^4 L^2
   \rb^6-8 a^4 M^3 \rb^5+2 a^4 M \rb^7+4 a^2 L^6 M^2 \rb^2+16
   a^2 L^4 M^2 \rb^4-2 a^2 L^4 \rb^6 \\
&& \quad 
	+\: 4 a^2 L^2 M^2 \rb^6-a^2 L^2
   \rb^8 +2 L^6 M \rb^5-L^6
   \rb^6+2 L^4 M \rb^7-L^4 \rb^8 \big)\\
&&
	+\: 2 a E L M
   \big(8 a^6 M^3+4 a^6 M^2 \rb-2 a^6 M \rb^2-a^6 \rb^3-4 a^4 L^2
   M^2 \rb-6 a^4 L^2 M \rb^2 \\
&& \quad
	-\:2 a^4 L^2 \rb^3-8 a^4 M^2
   \rb^3-12 a^4 M \rb^4-4 a^4 \rb^5-4 a^2 L^4 M \rb^2-a^2 L^4
   \rb^3 \\
&& \quad 
	-\: 10 a^2 L^2 M \rb^4-5 a^2 L^2 \rb^5-6 a^2 M \rb^6-3
   a^2 \rb^7-L^4 \rb^5-3 L^2 \rb^7\big) \\
&&
	-\: E^2 \left(2 a^2 M+a^2 \rb+\rb^3\right) \big(4 a^6
   M^3+4 a^6 M^2 \rb+a^6 M \rb^2-2 a^4 L^2 M^2 \rb \\
&&
	-\: a^4 L^2 M
   \rb^2+2 a^4 M^2 \rb^3+a^4 M \rb^4-2 a^2 L^4 M \rb^2+a^2
   L^2 M \rb^4+a^2 L^2 \rb^5+L^4 \rb^5\big), 
\end{IEEEeqnarray*}

\begin{equation}
F_{\theta[0]} = 0, 
\end{equation}

\begin{equation}
F_{\phi[0]} = \frac{L \rbdot}{\pi  \rb \left( 2 a^2 M + a^2 \rb + L^2 \rb \right)^2 \left(\rb^2+L^2 + \frac{2 a^2 M}{\rb} + a^2 \right)^{1/2} } \left(F^{\mathcal{E}}_{\text{\fixme{$\phi$}}[0]} \mathcal{E} + F^{\mathcal{K}}_{\text{\fixme{$\phi$}}[0]} \mathcal{K} \right),
\end{equation}
where
\par \vspace{-6pt} \begin{IEEEeqnarray*}{rCl}
F^{\mathcal{E}}_{\text{\fixme{$\phi$}}[0]} &=& -2 a^4 M^2-a^4 M \rb-a^2 L^2 M \rb-4 a^2 M \rb^3-a^2
   \rb^4-L^2 \rb^4, \nonumber \\
F^{\mathcal{K}}_{\text{\fixme{$\phi$}}[0]} &=& \rb^3 \left(4 a^2 M+a^2 \rb+L^2 \rb\right).
\end{IEEEeqnarray*}

The regularisation parameters for $F_{a [2]}$ and $F_{a [4]}$ are too large for paper format and have instead been made available electronically \cite{BarryWardell.net}.  For the reader to get an understanding of the form and size of these expressions, we have included $F_{t [2]}$ for eccentric orbits, and $F_{r [2]}$ for a circular orbit.  Expressions online are for eccentric orbits  (all expressions in Kerr space-time are for the case equatorial plane).  To condense $F_{t [2]}$, the resulting expression we have used the notation,
\begin{equation}
\mathcal{L}^2 = L^2 + a^2 + \rb^2 + \frac{2 M a^2}{\rb}.
\end{equation}
$F_{t [2]}$ is described by,
\begin{equation}
F_{t [2]} = \frac{\rbdot}{6 \pi \rb^7 \left( 2 a^2 M + a^2 \rb + L^2 \rb \right)^5 \left(\rb^2+L^2 + \frac{2 a^2 M}{\rb} + a^2 \right)^{7/2}}\left(F^{\mathcal{E}}_{t[2]} \mathcal{E} + F^{\mathcal{K}}_{t[2]} \mathcal{K} \right),
\end{equation}
where
\par \vspace{-6pt} \begin{IEEEeqnarray*}{rCl}
F^{\mathcal{E}}_{t[2]} &=&  \big[120 a L M (2 M-\rb) \rb^5 \mathcal{L}^{18}-12 a L M \rb^4 \big(-42 \rb^4+91 M \rb^3+5 a^2 \rb^2 \\
&& \qquad 
	+\: 20 a^2 M^2\big) \mathcal{L}^{16}-2 a L M \rb^3 \big(399 \rb^7-984 M \rb^6-120 a^2 \rb^5-82 a^2 M
   \rb^4 \\
&& \qquad
	+\: 240 a^2 M^2 \rb^3-138 a^4 M \rb^2+170 a^4 M^2 \rb+1068 a^4 M^3\big) \mathcal{L}^{14} \\
&& \quad
	+\: 2 a L M \rb^2 \big(273 \rb^{10}-870 M \rb^9-181 a^2 \rb^8-273 a^2 M \rb^7+1386 a^2 M^2 \rb^6 \\
&& \qquad
	-\: 236 a^4 M
   \rb^5+1165 a^4 M^2 \rb^4+4608 a^4 M^3 \rb^3-150 a^6 M^2 \rb^2+840 a^6 M^3 \rb \\
&& \qquad
	+\: 2676 a^6 M^4\big) \mathcal{L}^{12}-2 a L M \rb \big(18 \rb^{13}-312 M \rb^{12}-152 a^2 \rb^{11}-290 a^2 M \rb^{10} \\
&& \qquad
	+\: 1584 a^2
   M^2 \rb^9-21 a^4 M \rb^8+1832 a^4 M^2 \rb^7+5418 a^4 M^3 \rb^6+299 a^6 M^2 \rb^5 \\
&& \qquad
	+\: 2393 a^6 M^3 \rb^4+3234 a^6 M^4 \rb^3-60 a^8 M^3 \rb^2+1020 a^8 M^4 \rb \\
&& \qquad
	+\: 2460 a^8 M^5\big) \mathcal{L}^{10}+2 a L M \big(-138
   \rb^{16}+162 M \rb^{15}-144 a^2 \rb^{14}+20 a^2 M \rb^{13} \\
&& \qquad
	+\: 780 a^2 M^2 \rb^{12}+229 a^4 M \rb^{11}+806 a^4 M^2 \rb^{10}+1110 a^4 M^3 \rb^9+341 a^6 M^2 \rb^8 \\
&& \qquad
	+\: 1563 a^6 M^3 \rb^7-978 a^6 M^4 \rb^6+453
   a^8 M^3 \rb^5-812 a^8 M^4 \rb^4-4656 a^8 M^5 \rb^3 \\
&& \qquad
	+\: 420 a^{10} M^5 \rb+840 a^{10} M^6\big) \mathcal{L}^8+2 a L M \rb^2 \big(165 \rb^{16}-312 M \rb^{15} \\
&& \qquad
	+\: 160 a^2 \rb^{14}-300 a^2 M \rb^{13}-768 a^2 M^2
   \rb^{12}-339 a^4 M \rb^{11}-174 a^4 M^2 \rb^{10} \\
&& \qquad
	+\: 1470 a^4 M^3 \rb^9+311 a^6 M^2 \rb^8+541 a^6 M^3 \rb^7+714 a^6 M^4 \rb^6-222 a^8 M^3 \rb^5 \\
&& \qquad
	+\: 908 a^8 M^4 \rb^4+5556 a^8 M^5 \rb^3+2328 a^{10} M^5
   \rb+4656 a^{10} M^6\big) \mathcal{L}^6 \\
&& \quad
	-\: 2 a L M \rb^4 \big(99 \rb^{16}-198 M \rb^{15}+101 a^2 \rb^{14}-253 a^2 M \rb^{13}-1074 a^2 M^2 \rb^{12} \\
&& \qquad
	-\: 259 a^4 M \rb^{11}-645 a^4 M^2 \rb^{10}+1422 a^4 M^3
   \rb^9+275 a^6 M^2 \rb^8+1935 a^6 M^3 \rb^7 \\
&& \qquad
	-\: 102 a^6 M^4 \rb^6-117 a^8 M^3 \rb^5-2588 a^8 M^4 \rb^4-2424 a^8 M^5 \rb^3 \\
&& \qquad
	+\: 1980 a^{10} M^5 \rb+3960 a^{10} M^6\big) \mathcal{L}^4+16 a L M \rb^6 \left(a^2
   M-\rb^3\right) \big(-3 \rb^{13} \\
&& \qquad
	+\: 6 M \rb^{12}-3 a^2 \rb^{11}+6 a^2 M \rb^{10}+96 a^2 M^2 \rb^9+6 a^4 M \rb^8+85 a^4 M^2 \rb^7 \\
&& \qquad
	-\: 18 a^4 M^3 \rb^6-3 a^6 M^2 \rb^5-160 a^6 M^3 \rb^4-180 a^6 M^4
   \rb^3+120 a^8 M^4 \rb \\
&& \qquad
	+\: 240 a^8 M^5\big) \mathcal{L}^2-384 a^3 L M^3 \rb^8 \left(a^2 M-\rb^3\right)^3 \left(\rb^3+a^2 \rb+2 a^2 M\right) \big]\\
&&
	+\: E \big[-12 M \rb^5 \left(9 \rb^3+40 a^2 M\right) \mathcal{L}^{18}+6 M \rb^4 \big(92 \rb^6+15 a^2 \rb^4+402 a^2 M \rb^3 \\
&& \qquad
	+\: 40 a^4 M \rb+80
   a^4 M^2\big) \mathcal{L}^{16}+\rb^3 \big(-3 \rb^{10}-1134 M \rb^9-402 a^2 M \rb^7 \\
&& \qquad
	-\:5010 a^2 M^2 \rb^6-60 a^4 M \rb^5+118 a^4 M^2 \rb^4+3192 a^4 M^3 \rb^3-90 a^6 M \rb^3 \\
&& \qquad 
	+\: 48 a^6 M^2 \rb^2+3040 a^6 M^3
   \rb+5616 a^6 M^4\big) \mathcal{L}^{14}-\rb^2 \big(-39 \rb^{13}-1122 M \rb^{12}  \\
&& \qquad
	-\: 42 a^2 \rb^{11}-681 a^2 M \rb^{10}-9 a^4 \rb^9-5238 a^2 M^2 \rb^9-192 a^4 M \rb^8 \\
&& \qquad
	+\: 1685 a^4 M^2 \rb^7+11310 a^4 M^3
   \rb^6-198 a^6 M \rb^6+1341 a^6 M^2 \rb^5 \\
&& \qquad
	+\: 16509 a^6 M^3 \rb^4+27846 a^6 M^4 \rb^3-126 a^8 M^2 \rb^3+1704 a^8 M^3 \rb^2 \\
&& \qquad
	+\: 11400 a^8 M^4 \rb+15312 a^8 M^5\big) \mathcal{L}^{12}+\rb \big(-150 \rb^{16}-360 M
   \rb^{15}-216 a^2 \rb^{14} \\
&& \qquad
	-\: 432 a^2 M \rb^{13}-66 a^4 \rb^{12}-1668 a^2 M^2 \rb^{12}-132 a^4 M \rb^{11}+2564 a^4 M^2 \rb^{10} \\
&& \qquad
	+\: 10404 a^4 M^3 \rb^9-63 a^6 M \rb^9+2525 a^6 M^2 \rb^8+22422 a^6 M^3
   \rb^7 \\
&& \qquad
	+\: 32616 a^6 M^4 \rb^6+177 a^8 M^2 \rb^6+6010 a^8 M^3 \rb^5+23048 a^8 M^4 \rb^4 \\
&& \qquad
	+\: 20880 a^8 M^5 \rb^3-60 a^{10} M^3 \rb^3+2640 a^{10} M^4 \rb^2+12780 a^{10} M^5 \rb \\
&& \qquad
	+\: 14520 a^{10} M^6\big)
   \mathcal{L}^{10}+\big(270 \rb^{19}-372 M \rb^{18}+444 a^2 \rb^{17}-192 a^2 M \rb^{16} \\
&& \qquad
	+\: 168 a^4 \rb^{15}-3624 a^2 M^2 \rb^{15}-252 a^4 M \rb^{14}-4082 a^4 M^2 \rb^{13} \\
&& \qquad
	-\: 2712 a^4 M^3 \rb^{12}-243 a^6 M
   \rb^{12}-1633 a^6 M^2 \rb^{11}-5708 a^6 M^3 \rb^{10} \\
&& \qquad
	-\: 3684 a^6 M^4 \rb^9-159 a^8 M^2 \rb^9-3990 a^8 M^3 \rb^8-6152 a^8 M^4 \rb^7 \\
&& \qquad
	+\: 12336 a^8 M^5 \rb^6-453 a^{10} M^3 \rb^6+670 a^{10} M^4 \rb^5+16640 a^{10}
   M^5 \rb^4 \\
&& \qquad
	+\: 26976 a^{10} M^6 \rb^3-1260 a^{12} M^5 \rb^2-5040 a^{12} M^6 \rb-5040 a^{12} M^7\big) \mathcal{L}^8 \\
&& \quad
	-\: \rb^2 \big(255 \rb^{19}-498 M \rb^{18}+456 a^2 \rb^{17}-498 a^2 M \rb^{16}+198 a^4
   \rb^{15} \\
&& \qquad
	-\: 7014 a^2 M^2 \rb^{15}-528 a^4 M \rb^{14}-8054 a^4 M^2 \rb^{13}-1428 a^4 M^3 \rb^{12} \\
&& \qquad
	-\: 405 a^6 M \rb^{12}-1935 a^6 M^2 \rb^{11}+8254 a^6 M^3 \rb^{10}+12408 a^6 M^4 \rb^9 \\
&& \qquad
	+\: 357 a^8 M^2 \rb^9+3314 a^8 M^3 \rb^8+3368 a^8 M^4 \rb^7+9456 a^8 M^5 \rb^6 \\
&& \qquad
	-\: 222 a^{10} M^3 \rb^6+460 a^{10} M^4 \rb^5+24164 a^{10} M^5 \rb^4+44712 a^{10} 
	M^6 \rb^3 \\
&& \qquad
	+\: 6984 a^{12} M^5 \rb^2+27936 a^{12} M^6 \rb+27936 a^{12} M^7\big)
   \mathcal{L}^6+\rb^4 \big(123 \rb^{19} \\
&& \qquad
	-\: 246 M \rb^{18}+234 a^2 \rb^{17}-315 a^2 M \rb^{16}+111 a^4 \rb^{15}-6234 a^2 M^2 \rb^{15} \\
&& \qquad
	-\: 372 a^4 M \rb^{14}-9233 a^4 M^2 \rb^{13}-2874 a^4 M^3 \rb^{12}-279 a^6 M
   \rb^{12} \\
&& \qquad
	-\: 3230 a^6 M^2 \rb^{11}+9401 a^6 M^3 \rb^{10}+13818 a^6 M^4 \rb^9+285 a^8 M^2 \rb^9 \\
&& \qquad
	+\: 7638 a^8 M^3 \rb^8+3632 a^8 M^4 \rb^7-12000 a^8 M^5 \rb^6-117 a^{10} M^3 \rb^6 \\
&& \qquad
	-\: 8818 a^{10} M^4 \rb^5-19016 a^{10}
   M^5 \rb^4-3696 a^{10} M^6 \rb^3+5940 a^{12} M^5 \rb^2 \\
&& \qquad
	+\: 23760 a^{12} M^6 \rb+23760 a^{12} M^7\big) \mathcal{L}^4-8 \rb^6 \left(a^2 M-\rb^3\right) \big(\rb^3+a^2 \rb \\
&& \qquad
	+\: 2 a^2 M\big) \big(-3 \rb^{13}+6 M
   \rb^{12}-3 a^2 \rb^{11}+6 a^2 M \rb^{10}+360 a^2 M^2 \rb^9+6 a^4 M \rb^8 \\
&& \qquad
	+\: 281 a^4 M^2 \rb^7-102 a^4 M^3 \rb^6-3 a^6 M^2 \rb^5-500 a^6 M^3 \rb^4-552 a^6 M^4 \rb^3 \\
&& \qquad
	+\: 360 a^8 M^4 \rb+720 a^8 M^5\big)
   \mathcal{L}^2 \\
&& \qquad
	+\: 576 a^2 M^2 \rb^8 \left(a^2 M-\rb^3\right)^3 \left(\rb^3+a^2 \rb+2 a^2 M\right)^2\big] \\
&&
	+\: E^2 \big[24 a L M \rb^5 \left(9 \rb^3+5 a^2 \rb+10 a^2 M\right)
   \mathcal{L}^{16}+12 a L M \rb^4 \big(-77 \rb^6-32 a^2 \rb^4 \\
&& \qquad
	-\: 49 a^2 M \rb^3+5 a^4 \rb^2+20 a^4 M \rb+20 a^4 M^2\big) \mathcal{L}^{14}-2 a L M \rb^3 \big(-732 \rb^9 \\
&& \qquad
	-\: 217 a^2 \rb^7-276 a^2 M \rb^6+147 a^4
   \rb^5+1540 a^4 M \rb^4+2946 a^4 M^2 \rb^3 \\
&& \qquad
	+\: 378 a^6 M \rb^2+1830 a^6 M^2 \rb+2172 a^6 M^3\big) \mathcal{L}^{12}+2 a L M \rb^2 \big(-348 \rb^{12} \\
&& \qquad
	+\: 56 a^2 \rb^{10}-54 a^2 M \rb^9+200 a^4 \rb^8+2551 a^4 M
   \rb^7+4554 a^4 M^2 \rb^6 \\
&& \qquad
	+\:949 a^6 M \rb^5+4995 a^6 M^2 \rb^4+5940 a^6 M^3 \rb^3+615 a^8 M^2 \rb^2+2700 a^8 M^3 \rb \\
&& \qquad
	+\: 2940 a^8 M^4\big) \mathcal{L}^{10}-2 a L M \rb \big(612 \rb^{15}+890 a^2 \rb^{13}+618 a^2
   M \rb^{12} \\
&& \qquad
	+\: 258 a^4 \rb^{11}+845 a^4 M \rb^{10}+522 a^4 M^2 \rb^9+443 a^6 M \rb^8+1845 a^6 M^2 \rb^7 \\
&& \qquad
	+\: 372 a^6 M^3 \rb^6-179 a^8 M^2 \rb^5-2179 a^8 M^3 \rb^4-3642 a^8 M^4 \rb^3+315 a^{10} M^3
   \rb^2 \\
&& \qquad
	+\: 1260 a^{10} M^4 \rb+1260 a^{10} M^5\big) \mathcal{L}^8-2 a L M \rb^3 \big(-1410 \rb^{15}-2072 a^2 \rb^{13} \\
&& \qquad
	-\: 1452 a^2 M \rb^{12}-674 a^4 \rb^{11}+955 a^4 M \rb^{10}+2658 a^4 M^2 \rb^9+1041 a^6 M
   \rb^8 \\
&& \qquad
	+\: 2079 a^6 M^2 \rb^7+2508 a^6 M^3 \rb^6-229 a^8 M^2 \rb^5+3322 a^8 M^3 \rb^4 \\
&& \qquad
	+\: 7560 a^8 M^4 \rb^3+1746 a^{10} M^3 \rb^2+6984 a^{10} M^4 \rb+6984 a^{10} M^5\big) \mathcal{L}^6 \\
&& \quad
	+\: 2 a L M \rb^5 \big(-1380
   \rb^{15}-2339 a^2 \rb^{13}-1434 a^2 M \rb^{12}-963 a^4 \rb^{11} \\
&& \qquad
	+\: 1509 a^4 M \rb^{10}+3396 a^4 M^2 \rb^9+2129 a^6 M \rb^8+2703 a^6 M^2 \rb^7 \\
&& \qquad
	-\: 1200 a^6 M^3 \rb^6-2351 a^8 M^2 \rb^5-6013 a^8 M^3
   \rb^4-2622 a^8 M^4 \rb^3 \\
&& \qquad
	+\: 1485 a^{10} M^3 \rb^2+5940 a^{10} M^4 \rb+5940 a^{10} M^5\big) \mathcal{L}^4-16 a L M \rb^7 \big(a^2 M \\
&& \qquad
	-\: \rb^3\big) \left(\rb^3+a^2 \rb+2 a^2 M\right) \big(87 \rb^9+76 a^2
   \rb^7-130 a^4 M \rb^4-159 a^4 M^2 \rb^3 \\
&& \qquad
	+\: 90 a^6 M^2 \rb+180 a^6 M^3\big) \mathcal{L}^2+288 a L M \rb^9 \left(a^2 M-\rb^3\right)^3 \big(\rb^3+a^2 \rb \\
&& \qquad
	+\: 2 a^2 M\big)^2\big] \\
&&
	+\: E^3 \big[-60 a^2 M \rb^8
   \mathcal{L}^{16}-6 a^2 M \rb^7 \left(-44 \rb^3+9 a^2 \rb+22 a^2 M\right) \mathcal{L}^{14}+3 \rb^3 \big(8 \rb^{12} \\
&& \qquad
	+\: 10 a^2 \rb^{10}-131 a^2 M \rb^9+53 a^4 M \rb^7-2 a^6 \rb^6+80 a^4 M^2 \rb^6+72 a^6 M
   \rb^5 \\
&& \qquad
	+\: 470 a^6 M^2 \rb^4+644 a^6 M^3 \rb^3+30 a^8 M \rb^3+228 a^8 M^2 \rb^2+560 a^8 M^3 \rb \\
&& \qquad
	+\: 448 a^8 M^4\big) \mathcal{L}^{12}-3 \rb^2 \big(56 \rb^{15}+96 a^2 \rb^{13}+20 a^2 M \rb^{12}+38 a^4
   \rb^{11} \\
&& \qquad
	+\:132 a^4 M \rb^{10}-2 a^6 \rb^9+26 a^4 M^2 \rb^9+145 a^6 M \rb^8+788 a^6 M^2 \rb^7 \\
&& \qquad
	+\: 911 a^6 M^3 \rb^6+67 a^8 M \rb^6+628 a^8 M^2 \rb^5+1679 a^8 M^3 \rb^4+1382 a^8 M^4 \rb^3 \\
&& \qquad
	+\: 60 a^{10} M^2
   \rb^3+400 a^{10} M^3 \rb^2+880 a^{10} M^4 \rb+640 a^{10} M^5\big) \mathcal{L}^{10} \\
&& \quad
	+\: 3 \rb \big(160 \rb^{18}+332 a^2 \rb^{16}+378 a^2 M \rb^{15}+202 a^4 \rb^{14}+370 a^4 M \rb^{13}+30 a^6 \rb^{12} \\
&& \qquad
	+\: 14 a^4
   M^2 \rb^{12}+43 a^6 M \rb^{11}+16 a^6 M^2 \rb^{10}-163 a^6 M^3 \rb^9+13 a^8 M \rb^9 \\
&& \qquad
	+\: 244 a^8 M^2 \rb^8+325 a^8 M^3 \rb^7-222 a^8 M^4 \rb^6+4 a^{10} M^2 \rb^6-173 a^{10} M^3 \rb^5 \\
&& \qquad
	-\: 740 a^{10} M^4
   \rb^4-756 a^{10} M^5 \rb^3+35 a^{12} M^3 \rb^3+210 a^{12} M^4 \rb^2+420 a^{12} M^5 \rb \\
&& \qquad
	+\: 280 a^{12} M^6\big) \mathcal{L}^8+3 \rb^3 \left(\rb^3+a^2 \rb+2 a^2 M\right) \big(-240 \rb^{15}-328 a^2 \rb^{13} \\
&& \qquad
	-\: 172
   a^2 M \rb^{12}-98 a^4 \rb^{11}+202 a^4 M \rb^{10}+366 a^4 M^2 \rb^9+151 a^6 M \rb^8 \\
&& \qquad
	+\: 138 a^6 M^2 \rb^7+23 a^6 M^3 \rb^6-76 a^8 M^2 \rb^5+232 a^8 M^3 \rb^4+768 a^8 M^4 \rb^3 \\
&& \qquad
	+\: 194 a^{10} M^3 \rb^2+776
   a^{10} M^4 \rb+776 a^{10} M^5\big) \mathcal{L}^6-3 \rb^5 \big(\rb^3+a^2 \rb \\
&& \qquad
	+\: 2 a^2 M\big) \big(-200 \rb^{15}-322 a^2 \rb^{13}-165 a^2 M \rb^{12}-124 a^4 \rb^{11}+248 a^4 M \rb^{10} \\
&& \qquad
	+\: 496 a^4 M^2
   \rb^9+271 a^6 M \rb^8+310 a^6 M^2 \rb^7-219 a^6 M^3 \rb^6-284 a^8 M^2 \rb^5 \\
&& \qquad
	-\: 728 a^8 M^3 \rb^4-320 a^8 M^4 \rb^3+165 a^{10} M^3 \rb^2 +660 a^{10} M^4 \rb \\
&& \qquad
	+\: 660 a^{10} M^5\big) \mathcal{L}^4+24 \rb^7 \left(a^2
   M-\rb^3\right) \left(\rb^3+a^2 \rb+2 a^2 M\right)^2 \big(11 \rb^9+9 a^2 \rb^7 \\
&& \qquad
	-\: a^2 M \rb^6-15 a^4 M \rb^4-18 a^4 M^2 \rb^3+10 a^6 M^2 \rb+20 a^6 M^3\big) \mathcal{L}^2 \\
&& \quad
	-\: 48 \rb^9 \left(a^2
   M-\rb^3\right)^3 \left(\rb^3+a^2 \rb+2 a^2 M\right)^3\big],
\end{IEEEeqnarray*}

\par \vspace{-6pt} \begin{IEEEeqnarray*}{rCl}
F^{\mathcal{K}}_{t[2]} &=&  \big[-60 a L M (2 M-\rb) \rb^7 \mathcal{L}^{16}+24 a L M \rb^6 \big(-10 \rb^4+22 M \rb^3+2 a^2 \rb^2+a^2 M \rb \\
&& \quad \quad
	+\: 20 a^2 M^2\big) \mathcal{L}^{14}+2 a L M \rb^5 \big(180 \rb^7-456 M \rb^6-99 a^2 
	\rb^5-116 a^2M \rb^4 \\
&& \qquad
	-\: 138 a^2 M^2 \rb^3-234 a^4 M \rb^2-209 a^4 M^2 \rb+894 a^4 M^3\big) \mathcal{L} 
	^{12} \\
&& \quad
	-\: 2 a L M \rb^4 \big(132 \rb^{10}-408 M \rb^9-140 a^2 \rb^8-290 a^2 M \rb^7+996 
	a^2 M^2 \rb^6 \\
&& \qquad
	-\:535 a^4 M \rb^5-36 a^4 M^2 \rb^4+4746 a^4 M^3 \rb^3-495 a^6 M^2 \rb^2+855 a^6 M^3 \rb \\
&& \qquad
	+\: 5298 a^6 M^4\big) \mathcal{L}^{10}+2 a L M \rb^3 \big(78 \rb^{13}-252 M \rb^{12}-54 a^2 \rb^{11}-330 a^2 M \rb^{10} \\
&& \qquad
	+\: 1356 a^2
   M^2 \rb^9-421 a^4 M \rb^8+586 a^4 M^2 \rb^7+6870 a^4 M^3 \rb^6-353 a^6 M^2 \rb^5 \\
&& \qquad
	+\: 2509 a^6 M^3 \rb^4+10506 a^6 M^4 \rb^3-345 a^8 M^3 \rb^2+2820 a^8 M^4 \rb \\
&& \qquad
	+\: 8160 a^8 M^5\big) \mathcal{L}^8-2 a L M \rb^2
   \big(72 \rb^{16}-168 M \rb^{15}+48 a^2 \rb^{14}-230 a^2 M \rb^{13} \\
&& \qquad
	+\: 216 a^2 M^2 \rb^{12}-209 a^4 M \rb^{11}+332 a^4 M^2 \rb^{10}+3330 a^4 M^3 \rb^9+239 a^6 M^2 \rb^8 \\
&& \qquad
	+\: 1861 a^6 M^3 \rb^7+4770 a^6 M^4
   \rb^6-6 a^8 M^3 \rb^5+2444 a^8 M^4 \rb^4+6732 a^8 M^5 \rb^3 \\
&& \qquad
	+\: 2100 a^{10} M^5 \rb+4200 a^{10} M^6\big) \mathcal{L}^6+2 a L M \rb^4 \big(48 \rb^{16}-96 M \rb^{15}+49 a^2 \rb^{14} \\
&& \qquad
	-\: 122 a^2 M \rb^{13}-498 a^2
   M^2 \rb^{12}-125 a^4 M \rb^{11}-289 a^4 M^2 \rb^{10}+660 a^4 M^3 \rb^9 \\
&& \qquad
	+\: 133 a^6 M^2 \rb^8+863 a^6 M^3 \rb^7-114 a^6 M^4 \rb^6-57 a^8 M^3 \rb^5-1172 a^8 M^4 \rb^4 \\
&& \qquad
	-\: 1032 a^8 M^5 \rb^3+936 a^{10} M^5
   \rb+1872 a^{10} M^6\big) \mathcal{L}^4-8 a L M \rb^6 \big(a^2 M \\
&& \qquad
	-\: \rb^3\big) \big(-3 \rb^{13}+6 M \rb^{12}-3 a^2 \rb^{11}+6 a^2 M \rb^{10}+93 a^2 M^2 \rb^9+6 a^4 M \rb^8 \\
&& \qquad
	+\: 82 a^4 M^2 \rb^7-18 a^4 M^3
   \rb^6-3 a^6 M^2 \rb^5-154 a^6 M^3 \rb^4-171 a^6 M^4 \rb^3 \\
&& \qquad
	+\: 117 a^8 M^4 \rb+234 a^8 M^5\big) \mathcal{L}^2+192 a^3 L M^3 \rb^8 \left(a^2 M-\rb^3\right)^3 \big(\rb^3+a^2 \rb \\
&& \qquad
	+\: 2 a^2 M\big) \big] \\
&&
	+\: E \big[6 M \rb^7 \left(9 \rb^3+40 a^2 M\right) \mathcal{L}^{16}-6 M
   \rb^6 \big(45 \rb^6+11 a^2 \rb^4+158 a^2 M \rb^3+3 a^4 \rb^2 \\
&& \quad \quad
	+\: 76 a^4 M \rb+220 a^4 M^2\big) \mathcal{L}^{14}-\rb^5 \big(12 \rb^{10}-564 M \rb^9+18 a^2 \rb^8 \\
&& \qquad
	-\: 273 a^2 M \rb^7-1512 a^2 M^2
   \rb^6-120 a^4 M \rb^5-613 a^4 M^2 \rb^4+1320 a^4 M^3 \rb^3 \\
&& \qquad 
	-\: 126 a^6 M \rb^3-678 a^6 M^2 \rb^2+848 a^6 M^3 \rb+4968 a^6 M^4\big) \mathcal{L}^{12}+\rb^4 \big(60 \rb^{13} \\
&& \qquad
	-\: 660 M \rb^{12}+96 a^2
   \rb^{11}-468 a^2 M \rb^{10}+18 a^4 \rb^9-1644 a^2 M^2 \rb^9 \\
&& \qquad
	-\: 288 a^4 M \rb^8+660 a^4 M^2 \rb^7+11664 a^4 M^3 \rb^6-333 a^6 M \rb^6-873 a^6 M^2 \rb^5 \\
&& \qquad
	+\: 11474 a^6 M^3 \rb^4+33012 a^6 M^4 \rb^3-378 a^8
   M^2 \rb^3+1047 a^8 M^3 \rb^2 \\
&& \qquad
	+\: 18060 a^8 M^4 \rb+31416 a^8 M^5\big) \mathcal{L}^{10}-\rb^3 \big(120 \rb^{16}-510 M \rb^{15}+204 a^2 \rb^{14} \\
&& \qquad
	-\: 462 a^2 M \rb^{13}+66 a^4 \rb^{12}-2304 a^2 M^2 \rb^{12}-372 a^4
   M \rb^{11}+310 a^4 M^2 \rb^{10} \\
&& \qquad
	+\: 13788 a^4 M^3 \rb^9-351 a^6 M \rb^9+133 a^6 M^2 \rb^8+21122 a^6 M^3 \rb^7 \\
&& \qquad
	+\: 51120 a^6 M^4 \rb^6-258 a^8 M^2 \rb^6+4841 a^8 M^3 \rb^5+44044 a^8 M^4 \rb^4 \\
&& \qquad
	+\: 73428 a^8 M^5
   \rb^3-345 a^{10} M^3 \rb^3+6570 a^{10} M^4 \rb^2+38940 a^{10} M^5 \rb \\
&& \qquad
	+\: 48840 a^{10} M^6\big) \mathcal{L}^8+\rb^2 \big(120 \rb^{19}-294 M \rb^{18}+216 a^2 \rb^{17}-318 a^2 M \rb^{16} \\
&& \qquad
	+\: 90 a^4 \rb^{15}-3276
   a^2 M^2 \rb^{15}-318 a^4 M \rb^{14}-2896 a^4 M^2 \rb^{13}+4092 a^4 M^3 \rb^{12} \\
&& \qquad
	-\: 243 a^6 M \rb^{12}-351 a^6 M^2 \rb^{11}+12762 a^6 M^3 \rb^{10}+25512 a^6 M^4 \rb^9 \\
&& \qquad
	+\: 222 a^8 M^2 \rb^9+5057 a^8 M^3
   \rb^8+24668 a^8 M^4 \rb^7+38028 a^8 M^5 \rb^6 \\
&& \qquad
	-\: 6 a^{10} M^3 \rb^6+5800 a^{10} M^4 \rb^5+36320 a^{10} M^5 \rb^4+49392 a^{10} M^6 \rb^3 \\
&& \qquad
	+\: 6300 a^{12} M^5 \rb^2+25200 a^{12} M^6 \rb+25200 a^{12} M^7\big)
   \mathcal{L}^6-\rb^4 \big(60 \rb^{19} \\
&& \qquad
	-\: 120 M \rb^{18}+114 a^2 \rb^{17}-153 a^2 M \rb^{16}+54 a^4 \rb^{15}-2952 a^2 M^2 \rb^{15} \\
&&\qquad
	-\: 180 a^4 M \rb^{14}-4329 a^4 M^2 \rb^{13}-1332 a^4 M^3 \rb^{12}-135 a^6 M
   \rb^{12} \\
&& \qquad
	-\: 1491 a^6 M^2 \rb^{11}+4390 a^6 M^3 \rb^{10}+6312 a^6 M^4 \rb^9+138 a^8 M^2 \rb^9 \\
&& \qquad
	+\: 3475 a^8 M^3 \rb^8+1348 a^8 M^4 \rb^7-5796 a^8 M^5 \rb^6-57 a^{10} M^3 \rb^6 \\
&& \qquad
	-\: 4030 a^{10} M^4 \rb^5-8372 a^{10} M^5
   \rb^4-1080 a^{10} M^6 \rb^3+2808 a^{12} M^5 \rb^2 \\
&& \qquad
	+\: 11232 a^{12} M^6 \rb+11232 a^{12} M^7\big) \mathcal{L}^4+4 \rb^6 \left(a^2 M-\rb^3\right) \big(\rb^3+a^2 \rb \\
&& \qquad
	+\: 2 a^2 M\big) \big(-3 \rb^{13}+6 M
   \rb^{12}-3 a^2 \rb^{11}+6 a^2 M \rb^{10}+351 a^2 M^2 \rb^9+6 a^4 M \rb^8 \\
&& \qquad
	+\: 272 a^4 M^2 \rb^7-102 a^4 M^3 \rb^6-3 a^6 M^2 \rb^5-482 a^6 M^3 \rb^4-525 a^6 M^4 \rb^3 \\
&& \qquad
	+\: 351 a^8 M^4 \rb+702 a^8 M^5\big)
   \mathcal{L}^2 \\
&& \quad
	-\: 288 a^2 M^2 \rb^8 \left(a^2 M-\rb^3\right)^3 \left(\rb^3+a^2 \rb+2 a^2 M\right)^2\big] \\
&&
	+\: E^2 \big[-12 a
   L M \rb^7 \left(9 \rb^3+5 a^2 \rb+10 a^2 M\right) \mathcal{L}^{14}+6 a L M \rb^6 \big(77 \rb^6+23 a^2 \rb^4 \\
&&
 \qquad -36 a^2 M \rb^3+4 a^4 \rb^2+80 a^4 M \rb+160 a^4 M^2\big) \mathcal{L}^{12}+2 a L M \rb^5
   \big(-435 \rb^9 \\
&& \qquad
	-\: 86 a^2 \rb^7+591 a^2 M \rb^6+33 a^4 \rb^5+695 a^4 M \rb^4+2253 a^4 M^2 \rb^3 \\
&& \qquad
	+\: 384 a^6 M \rb^2+2265 a^6 M^2 \rb+3246 a^6 M^3\big) \mathcal{L}^{10}-2 a L M \rb^4 \big(-582 \rb^{12} \\
&& \qquad
	-\: 346 a^2
   \rb^{10}+402 a^2 M \rb^9+28 a^4 \rb^8+2143 a^4 M \rb^7+6006 a^4 M^2 \rb^6 \\
&& \qquad
	+\: 989 a^6 M \rb^5+7305 a^6 M^2 \rb^4+11733 a^6 M^3 \rb^3+1515 a^8 M^2 \rb^2 \\
&& \qquad
	+\: 7545 a^8 M^3 \rb+9030 a^8 M^4\big) \mathcal{L}^8+2 a L M
   \rb^3 \big(-708 \rb^{15}-906 a^2 \rb^{13} \\
&& \qquad
	-\: 432 a^2 M \rb^{12}-242 a^4 \rb^{11}+1499 a^4 M \rb^{10}+3900 a^4 M^2 \rb^9+999 a^6 M \rb^8 \\
&& \qquad
	+\: 5325 a^6 M^2 \rb^7+8067 a^6 M^3 \rb^6+1073 a^8 M^2 \rb^5+7168
   a^8 M^3 \rb^4 \\
&& \qquad
	+\: 10044 a^8 M^4 \rb^3+1575 a^{10} M^3 \rb^2+6300 a^{10} M^4 \rb+6300 a^{10} M^5\big) \mathcal{L}^6 \\
&& \quad
	-\: 2 a L M \rb^5 \big(-651 \rb^{15}-1097 a^2 \rb^{13}-678 a^2 M \rb^{12}-448 a^4 \rb^{11}+693 a^4 M
   \rb^{10} \\
&& \qquad
	+\: 1554 a^4 M^2 \rb^9+975 a^6 M \rb^8+1203 a^6 M^2 \rb^7-585 a^6 M^3 \rb^6-1079 a^8 M^2 \rb^5 \\
&& \qquad
	-\: 2707 a^8 M^3 \rb^4-1098 a^8 M^4 \rb^3+702 a^{10} M^3 \rb^2+2808 a^{10} M^4 \rb \\
&& \qquad
	+\: 2808 a^{10} M^5\big)
   \mathcal{L}^4+2 a L M \rb^7 \left(a^2 M-\rb^3\right) \left(\rb^3+a^2 \rb+2 a^2 M\right) \big(339 \rb^9 \\
&& \qquad
	+\: 295 a^2 \rb^7-502 a^4 M \rb^4-609 a^4 M^2 \rb^3+351 a^6 M^2 \rb+702 a^6 M^3\big) \mathcal{L}^2 \\
&& \quad
	-\: 144 a L M
   \rb^9 \left(a^2 M-\rb^3\right)^3 \left(\rb^3+a^2 \rb+2 a^2 M\right)^2\big] \\
&&
	+\: E^3 \big[30 a^2 M \rb^{10} \mathcal{L}^{14}-3 \rb^6 \big(3 \rb^9+3 a^2 \rb^7+48 
	a^2 M \rb^6-30 a^4 M \rb^4-112 a^4 M^2\rb^3 \\
&& \quad \quad
	+\: 18 a^6 M \rb^2+96 a^6 M^2 \rb+120 a^6 M^3\big) \mathcal{L}^{12}-3 \rb^5 \big( 
	-23 \rb^{12}-37 a^2 \rb^{10} \\
&& \qquad
	-\: 121 a^2 M \rb^9-15 a^4 \rb^8+55 a^4 M \rb^7-a^6 \rb^6+311 a^4 M^2 \rb^6+24 
	a^6 M \rb^5 \\
&& \qquad
	+\: 361 a^6 M^2 \rb^4+682 a^6 M^3 \rb^3+30 a^8 M \rb^3+264 a^8 M^2 \rb^2+760 
	a^8 M^3 \rb \\
&& \qquad
	+\: 704 a^8 M^4\big) \mathcal{L}^{10}+3 \rb^4 \big(-70 \rb^{15}-142 a^2 \rb^{13}-228 a^2 M \rb^{12}-86 a^4
   \rb^{11} \\
&& \qquad
	-\: 68 a^4 M \rb^{10}-14 a^6 \rb^9+270 a^4 M^2 \rb^9+113 a^6 M \rb^8+1064 a^6 M^2 \rb^7 \\
&& \qquad
	+\: 1739 a^6 M^3 \rb^6+71 a^8 M \rb^6+851 a^8 M^2 \rb^5+2957 a^8 M^3 \rb^4+3078 a^8 M^4 \rb^3 \\
&& \qquad
	+\: 135 a^{10} M^2
   \rb^3+1040 a^{10} M^3 \rb^2+2540 a^{10} M^4 \rb+2000 a^{10} M^5\big) \mathcal{L}^8 \\
&& \quad
	-\: 3 \rb^3 \big(-110 \rb^{18}-258 a^2 \rb^{16}-312 a^2 M \rb^{15}-192 a^4 \rb^{14}-258 a^4 M \rb^{13} \\
&& \qquad
	-\: 44 a^6 \rb^{12}+88
   a^4 M^2 \rb^{12}+127 a^6 M \rb^{11}+874 a^6 M^2 \rb^{10}+1077 a^6 M^3 \rb^9 \\
&& \qquad
	+\: 97 a^8 M \rb^9+673 a^8 M^2 \rb^8+2071 a^8 M^3 \rb^7+2226 a^8 M^4 \rb^6+77 a^{10} M^2 \rb^6 \\
&& \qquad
	+\: 991 a^{10} M^3 \rb^5+3040 a^{10} M^4
   \rb^4+2732 a^{10} M^5 \rb^3+175 a^{12} M^3 \rb^3 \\
&& \qquad
	+\: 1050 a^{12} M^4 \rb^2+2100 a^{12} M^5 \rb+1400 a^{12} M^6\big) \mathcal{L}^6+3 \rb^5 \big(\rb^3+a^2 \rb \\
&& \qquad
	+\: 2 a^2 M\big) \big(-95 \rb^{15}-152 a^2
   \rb^{13}-78 a^2 M \rb^{12}-58 a^4 \rb^{11}+116 a^4 M \rb^{10} \\
&& \qquad
	+\: 230 a^4 M^2 \rb^9+125 a^6 M \rb^8+138 a^6 M^2 \rb^7-107 a^6 M^3 \rb^6-131 a^8 M^2 \rb^5 \\
&& \qquad
	-\: 330 a^8 M^3 \rb^4-136 a^8 M^4 \rb^3+78 a^{10}
   M^3 \rb^2+312 a^{10} M^4 \rb+312 a^{10} M^5\big) \mathcal{L}^4 \\
&& \quad
	-\: 3 \rb^7 \left(a^2 M-\rb^3\right) \left(\rb^3+a^2 \rb+2 a^2 M\right)^2 \big(43 \rb^9+35 a^2 \rb^7-4 a^2 M \rb^6 \\
&& \qquad
	-\: 58 a^4 M \rb^4-69 a^4 M^2
   \rb^3+39 a^6 M^2 \rb+78 a^6 M^3\big) \mathcal{L}^2 \\
&& \quad
	+\: 24 \rb^9 \left(a^2 M-\rb^3\right)^3 \left(\rb^3+a^2 \rb+2 a^2 M\right)^3\big],
\end{IEEEeqnarray*}

\begin{IEEEeqnarray}{rCl}
F_{r[2]} &=& \frac{1}{6 \pi  \rb^6} \sqrt{\frac{2 a \sqrt{M
   \rb^5}-3 M \rb^3+\rb^4}{\left(a
   \sqrt{M}+\rb^{3/2}\right)^2 \left[a^2+\rb (\rb-2 M)\right]}} \\
&& \times \: \frac{\left[2 a \sqrt{M}+\sqrt{\rb} (\rb-3 M)\right]^{-2} \left(F^{\mathcal{E}}_{r[2]} \mathcal{E} + F^{\mathcal{K}}_{ r[2]} \mathcal{K} \right)}{  \left[a^4 M+2 a^3 \sqrt{M \rb^3}+a^2 \rb
   \left(M \rb+\rb^2-2 M^2 \right)-4 a M^{3/2} \rb^{5/2}+M \rb^4\right]^3}, \nonumber
\end{IEEEeqnarray}
where
\begin{IEEEeqnarray*}{rCl}
F^{\mathcal{E}}_{r[2]}&=&-\left(\rb^{3/2}+a \sqrt{M}\right)^2 \big(3 M^3 \rb^{33/2}-36 M^4 \rb^{31/2}+9
   a^2 M^2 \rb^{31/2}-78 a M^{7/2} \rb^{15}\\
&& \quad
	+\: 93 M^5 \rb^{29/2}+27 a^2 M^3
   \rb^{29/2} + 9 a^4 M \rb^{29/2}+744 a M^{9/2} \rb^{14}+44 a^3 M^{5/2}
   \rb^{14}\\
&& \quad
	+\: 3 a^6 \rb^{27/2}-60 M^6 \rb^{27/2}-1068 a^2 M^4 \rb^{27/2}-120
   a^4 M^2 \rb^{27/2} \\
&& \quad
	-\:  1890 a M^{11/2} \rb^{13}+60 a^3 M^{7/2} \rb^{13}+18 a^5
   M^{3/2} \rb^{13}+3681 a^2 M^5 \rb^{25/2}\\
&& \quad
	+\: 913 a^4 M^3 \rb^{25/2}+15 a^6 M
   \rb^{25/2} + 1332 a M^{13/2} \rb^{12}-704 a^3 M^{9/2} \rb^{12} \\
&& \quad
	-\: 404 a^5 M^{5/2}
   \rb^{12}+24 a^7 \sqrt{M} \rb^{12}+6 a^8 \rb^{23/2}-4251 a^2 M^6
   \rb^{23/2} \\
&& \quad
	-\:  3846 a^4 M^4 \rb^{23/2}-357 a^6 M^2 \rb^{23/2}+2472 a^3 M^{11/2}
   \rb^{11}+3434 a^5 M^{7/2} \rb^{11}\\
&& \quad
	+\: 24 a^7 M^{3/2} \rb^{11}+2178 a^2 M^7
   \rb^{21/2} + 4559 a^4 M^5 \rb^{21/2}+732 a^6 M^3 \rb^{21/2}\\
&& \quad
	+\: 81 a^8 M
   \rb^{21/2}-3336 a^3 M^{13/2} \rb^{10}-7502 a^5 M^{9/2} \rb^{10}-1328 a^7
   M^{5/2} \rb^{10} \\
&& \quad
	+\:  48 a^9 \sqrt{M} \rb^{10}-786 a^4 M^6 \rb^{19/2}+1421 a^6 M^4
   \rb^{19/2}-366 a^8 M^2 \rb^{19/2}\\
&& \quad
	-\: 1272 a^3 M^{15/2} \rb^9+5324 a^5 M^{11/2}
   \rb^9 + 4322 a^7 M^{7/2} \rb^9+174 a^9 M^{3/2} \rb^9\\
&& \quad
	+\: 3501 a^4 M^7
   \rb^{17/2}-5631 a^6 M^5 \rb^{17/2}-1651 a^8 M^3 \rb^{17/2}+156 a^{10} M
   \rb^{17/2} \\
&& \quad
	+\:  942 a^5 M^{13/2} \rb^8-10 a^7 M^{9/2} \rb^8-1360 a^9 M^{5/2}
   \rb^8-3798 a^4 M^8 \rb^{15/2}\\
&& \quad
	-\: 5611 a^6 M^6 \rb^{15/2}+4900 a^8 M^4
   \rb^{15/2} + 288 a^{10} M^2 \rb^{15/2}+6648 a^5 M^{15/2} \rb^7\\
&& \quad
	-\: 2186 a^7 M^{11/2}
   \rb^7-1512 a^9 M^{7/2} \rb^7+264 a^{11} M^{3/2} \rb^7+4740 a^6 M^7
   \rb^{13/2} \\
&& \quad
	+\:  5292 a^8 M^5 \rb^{13/2}-1686 a^{10} M^3 \rb^{13/2}-2088 a^5
   M^{17/2} \rb^6-11734 a^7 M^{13/2} \rb^6\\
&& \quad
	+\: 2090 a^9 M^{9/2} \rb^6 + 384 a^{11}
   M^{5/2} \rb^6+2844 a^6 M^8 \rb^{11/2}-2100 a^8 M^6 \rb^{11/2}\\
&& \quad
	-\: 2490 a^{10} M^4
   \rb^{11/2}+246 a^{12} M^2 \rb^{11/2} + 3528 a^7 M^{15/2} \rb^5+8106 a^9 M^{11/2}
   \rb^5\\
&& \quad
	-\: 790 a^{11} M^{7/2} \rb^5-348 a^6 M^9 \rb^{9/2}-5398 a^8 M^7
   \rb^{9/2}+495 a^{10} M^5 \rb^{9/2} \\
&& \quad
	+\:  414 a^{12} M^3 \rb^{9/2}+360 a^7 M^{17/2}
   \rb^4-2174 a^9 M^{13/2} \rb^4-2598 a^{11} M^{9/2} \rb^4 \\
&& \quad
	+\: 120 a^{13} M^{5/2}
   \rb^4+708 a^8 M^8 \rb^{7/2} + 3844 a^{10} M^6 \rb^{7/2}-126 a^{12} M^4
   \rb^{7/2}\\
&& \quad
	-\: 728 a^9 M^{15/2} \rb^3+570 a^{11} M^{11/2} \rb^3+324 a^{13} M^{7/2}
   \rb^3-543 a^{10} M^7 \rb^{5/2} \\
&& \quad
	-\:  1216 a^{12} M^5 \rb^{5/2}+24 a^{14} M^3
   \rb^{5/2}+554 a^{11} M^{13/2} \rb^2-52 a^{13} M^{9/2} \rb^2\\
&& \quad
	+\: 186 a^{12} M^6
   \rb^{3/2}+144 a^{14} M^4 \rb^{3/2} - 188 a^{13} M^{11/2} \rb-24 a^{14} M^5
   \sqrt{\rb}\\
&& \quad
	+\: 24 a^{15} M^{9/2}\big), \\
F^{\mathcal{K}}_{r[2]} &=&\rb^{7/2} \left(\rb^{3/2}-3 M \sqrt{\rb}+2 a \sqrt{M}\right) \big(3 M^3
   \rb^{29/2}-24 M^4 \rb^{27/2}+9 a^2 M^2 \rb^{27/2}\\
&& \quad
	-\: 78 a M^{7/2}
   \rb^{13}+33 M^5 \rb^{25/2} + 57 a^2 M^3 \rb^{25/2}+9 a^4 M \rb^{25/2}+486
   a M^{9/2} \rb^{12}\\
&& \quad
	+\: 44 a^3 M^{5/2} \rb^{12}+3 a^6 \rb^{23/2}-885 a^2 M^4
   \rb^{23/2}-102 a^4 M^2 \rb^{23/2} \\
&& \quad
	-\:  612 a M^{11/2} \rb^{11}+146 a^3 M^{7/2}
   \rb^{11}+18 a^5 M^{3/2} \rb^{11}+1845 a^2 M^5 \rb^{21/2}\\
&& \quad
	+\: 688 a^4 M^3
   \rb^{21/2}+9 a^6 M \rb^{21/2} - 1086 a^3 M^{9/2} \rb^{10}-354 a^5 M^{5/2}
   \rb^{10}\\
&& \quad
	+\: 24 a^7 \sqrt{M} \rb^{10}-1206 a^2 M^6 \rb^{19/2}-1805 a^4 M^4
   \rb^{19/2}-252 a^6 M^2 \rb^{19/2} \\
&& \quad
	+\:  2580 a^3 M^{11/2} \rb^9+2546 a^5 M^{7/2}
   \rb^9+66 a^7 M^{3/2} \rb^9-528 a^4 M^5 \rb^{17/2}\\
&& \quad
	-\: 158 a^6 M^3
   \rb^{17/2}+78 a^8 M \rb^{17/2} + 12 a^3 M^{13/2} \rb^8-3176 a^5 M^{9/2}
   \rb^8\\
&& \quad
	-\: 998 a^7 M^{5/2} \rb^8-189 a^4 M^6 \rb^{15/2}+3058 a^6 M^4
   \rb^{15/2}+171 a^8 M^2 \rb^{15/2} \\
&& \quad
	-\:  24 a^5 M^{11/2} \rb^7+200 a^7 M^{7/2}
   \rb^7+132 a^9 M^{3/2} \rb^7+1143 a^4 M^7 \rb^{13/2}\\
&& \quad
	-\: 368 a^6 M^5
   \rb^{13/2}-1263 a^8 M^3 \rb^{13/2} - 1962 a^5 M^{13/2} \rb^6+1250 a^7 M^{9/2}
   \rb^6\\
&& \quad
	+\: 216 a^9 M^{5/2} \rb^6-639 a^6 M^6 \rb^{11/2}-347 a^8 M^4
   \rb^{11/2}+123 a^{10} M^2 \rb^{11/2} \\
&& \quad
	+\:  684 a^5 M^{15/2} \rb^5+1978 a^7 M^{11/2}
   \rb^5-590 a^9 M^{7/2} \rb^5-879 a^6 M^7 \rb^{9/2}\\
&& \quad
	+\: 232 a^8 M^5
   \rb^{9/2}+186 a^{10} M^3 \rb^{9/2} - 684 a^7 M^{13/2} \rb^4-832 a^9 M^{9/2}
   \rb^4\\
&& \quad
	+\: 60 a^{11} M^{5/2} \rb^4+114 a^6 M^8 \rb^{7/2}+1048 a^8 M^6
   \rb^{7/2}-90 a^{10} M^4 \rb^{7/2} \\
&& \quad
	-\:  108 a^7 M^{15/2} \rb^3+252 a^9 M^{11/2}
   \rb^3+144 a^{11} M^{7/2} \rb^3-147 a^8 M^7 \rb^{5/2}\\
&& \quad
	-\: 452 a^{10} M^5
   \rb^{5/2}+12 a^{12} M^3 \rb^{5/2} + 146 a^9 M^{13/2} \rb^2-38 a^{11} M^{9/2}
   \rb^2\\
&& \quad
	+\: 69 a^{10} M^6 \rb^{3/2}+72 a^{12} M^4 \rb^{3/2}-70 a^{11} M^{11/2}
   \rb-12 a^{12} M^5 \sqrt{\rb} \\
&& \quad
	+\:  12 a^{13} M^{9/2}\big).
\end{IEEEeqnarray*}


\subsection{Electromagnetic Case}
The regularisation parameters for Kerr space-time in the electromagnetic case are given by
\begin{gather}
F_{t\lnpow{1}} =-\frac{\rb \rbdot \sgn \Delta r}{\rb \left(a^2+L^2\right)+2
   a^2 M+\rb^3}, \nonumber \\
F_{r\lnpow{1}} = \frac{\sgn \Delta r \left(E \rb \left(a^2+\rb^2\right)+2 a
   M (a E-L)\right)}{\left(a^2-2 M \rb+\rb^2\right)
   \left(\rb \left(a^2+L^2\right)+2 a^2 M+\rb^3\right)}, \nonumber \\
F_{\theta\lnpow{1}} = 0, \quad
F_{\phi\lnpow{1}} = 0,
\end{gather}

\begin{equation}
F_{t[0]} = \frac{\rbdot}{\pi  \rb^2 \left(\rb^2+L^2 + \frac{2 a^2 M}{\rb} + a^2 \right)^{3/2} \left( 2 a^2 M + a^2 \rb +L^2 \rb \right)^2} \left(F^{\mathcal{E}}_{t[0]} \mathcal{E} + F^{\mathcal{K}}_{t[0]} \mathcal{K} \right),
\end{equation}
where
\par \vspace{-6pt} \begin{IEEEeqnarray*}{rCl}
F^{\mathcal{E}}_{t[0]} &=& 
	-4 a L M \left(4 a^4 M^2+2 a^4 M \rb+2 a^2 L^2 M
   \rb-a^2 M \rb^3-a^2 \rb^4-L^2 \rb^4\right) \\
&&
	+\: E \big(-12 a^6 M^3-16 a^6 M^2 \rb-7 a^6 M \rb^2-a^6
   \rb^3-28 a^4 L^2 M^2 \rb-22 a^4 L^2 M \rb^2 \\
&& \quad
	-\: 4 a^4 L^2
   \rb^3-6 a^4 M^2 \rb^3-5 a^4 M \rb^4-a^4 \rb^5-15 a^2 L^4 M
   \rb^2-5 a^2 L^4 \rb^3 \\
&& \quad
	-\: 5 a^2 L^2 M \rb^4-a^2 L^2 \rb^5-2 L^6
   \rb^3\big), \\
F^{\mathcal{K}}_{t[0]} &=& 2 a L M \big(2 a^4 M^2-a^4 M \rb-a^4 \rb^2-a^2 L^2 M \rb-2 a^2 
	L^2\rb^2-2 a^2 M \rb^3-2 a^2 \rb^4 \\
&& \quad
	-\: L^4 \rb^2-2 L^2 \rb^4\big) \\
&&
	+\: E \big(4 a^6 M^3+4 a^6 M^2 \rb+a^6 M
   \rb^2+10 a^4 L^2 M^2 \rb+5 a^4 L^2 M \rb^2+2 a^4 M^2
   \rb^3 \\
&& \quad
	+\: a^4 M \rb^4+4 a^2 L^4 M \rb^2+a^2 L^2 M \rb^4-a^2 L^2
   \rb^5-L^4 \rb^5\big),
\end{IEEEeqnarray*}

\begin{equation}
F_{r[0]} = \frac{\left(F^{\mathcal{E}}_{\text{\fixme{$r$}}[0]} \mathcal{E} + F^{\mathcal{K}}_{\text{\fixme{$r$}}[0]} \mathcal{K} \right)}{\pi  \rb^3 \left(\rb^2+L^2 + \frac{2 a^2 M}{\rb} + a^2 \right)^{3/2} \left( 2 a^2 M + a^2 \rb + L^2 \rb \right)^2 \left( a^2 - 2 M \rb + \rb^2\right)} ,
\end{equation}
where
\par \vspace{-6pt} \begin{IEEEeqnarray*}{rCl}
F^{\mathcal{E}}_{r[0]} &=& 
L^2 \big(24 a^6 M^4+28 a^6 M^3 \rb-6 a^6 M^2 \rb^2-11 a^6 M \rb^3-2
   a^6 \rb^4+56 a^4 L^2 M^3 \rb \\
&& \quad
	+\: 24 a^4 L^2 M^2 \rb^2-18 a^4 L^2 M
   \rb^3-6 a^4 L^2 \rb^4+52 a^4 M^3 \rb^3+20 a^4 M^2 \rb^4 \\
&& \quad
	-\: 11 a^4 M \rb^5-3 a^4 \rb^6+30 a^2 L^4 M^2 \rb^2-3 a^2 L^4 M
   \rb^3-6 a^2 L^4 \rb^4+42 a^2 L^2 M^2 \rb^4 \\
&& \quad
	-\: 5 a^2 L^2 M
   \rb^5-6 a^2 L^2 \rb^6+8 a^2 M^2 \rb^6-2 a^2 M \rb^7-a^2
   \rb^8+4 L^6 M \rb^3-2 L^6 \rb^4 \\
&& \quad
	+\: 6 L^4 M \rb^5-3 L^4
   \rb^6+2 L^2 M \rb^7-L^2 \rb^8\big) \\
&& 
	-\: 2 a E L M \big(24
   a^6 M^3+36 a^6 M^2 \rb+18 a^6 M \rb^2+3 a^6 \rb^3+56 a^4 L^2 M^2
   \rb \\
&& \quad
	+\: 48 a^4 L^2 M \rb^2+10 a^4 L^2 \rb^3+24 a^4 M^2 \rb^3+24
   a^4 M \rb^4+6 a^4 \rb^5+30 a^2 L^4 M \rb^2 \\
&& \quad
	+\: 11 a^2 L^4
   \rb^3+22 a^2 L^2 M \rb^4+9 a^2 L^2 \rb^5+6 a^2 M \rb^6+3
   a^2 \rb^7+4 L^6 \rb^3+3 L^4 \rb^5 \\
&& \quad
	+\: 3 L^2 \rb^7\big) \\
&&
	+\: E^2 \left(2 a^2 M+a^2 \rb+\rb^3\right) \big(12 a^6 M^3+16 a^6 M^2
   \rb+7 a^6 M \rb^2+a^6 \rb^3 \\
&& \quad
	+\: 28 a^4 L^2 M^2 \rb+22 a^4 L^2 M
   \rb^2+4 a^4 L^2 \rb^3+6 a^4 M^2 \rb^3+5 a^4 M \rb^4+a^4
   \rb^5 \\
&& \quad
	+\: 15 a^2 L^4 M \rb^2+5 a^2 L^4 \rb^3+5 a^2 L^2 M
   \rb^4+a^2 L^2 \rb^5+2 L^6 \rb^3\big), \\
F^{\mathcal{K}}_{r[0]} &=& 
	-L^2 \big(8 a^6 M^4+12 a^6 M^3
   \rb-2 a^6 M \rb^3+20 a^4 L^2 M^3 \rb+8 a^4 L^2 M^2 \rb^2 \\
&& \quad
	-\: 4 a^4 L^2 M \rb^3+32 a^4 M^3 \rb^3+12 a^4 M^2 \rb^4-6 a^4 M
   \rb^5-a^4 \rb^6+8 a^2 L^4 M^2 \rb^2 \\
&& \quad
	-\: 2 a^2 L^4 M \rb^3+24
   a^2 L^2 M^2 \rb^4-4 a^2 L^2 M \rb^5-2 a^2 L^2 \rb^6+8 a^2 M^2
   \rb^6-2 a^2 M \rb^7 \\
&& \quad
	-\: a^2 \rb^8+2 L^4 M \rb^5-L^4
   \rb^6+2 L^2 M \rb^7-L^2 \rb^8\big)\\
&&
	+\: 2 a E L M \big(8
   a^6 M^3+12 a^6 M^2 \rb+6 a^6 M \rb^2+a^6 \rb^3+20 a^4 L^2 M^2
   \rb \\
&& \quad
	+\: 14 a^4 L^2 M \rb^2+2 a^4 L^2 \rb^3+16 a^4 M^2 \rb^3+16
   a^4 M \rb^4+4 a^4 \rb^5+8 a^2 L^4 M \rb^2 \\
&& \quad
	+\: a^2 L^4 \rb^3+14
   a^2 L^2 M \rb^4+5 a^2 L^2 \rb^5+6 a^2 M \rb^6+3 a^2
   \rb^7+L^4 \rb^5+3 L^2 \rb^7\big) \\
&&
	-\: E^2 \left(2 a^2 M+a^2 \rb+\rb^3\right) \big(4 a^6
   M^3+4 a^6 M^2 \rb+a^6 M \rb^2+10 a^4 L^2 M^2 \rb \\
&& \quad
	+\: 5 a^4 L^2 M
   \rb^2+2 a^4 M^2 \rb^3+a^4 M \rb^4+4 a^2 L^4 M \rb^2+a^2 L^2
   M \rb^4-a^2 L^2 \rb^5 \\
&& \quad
	-\: L^4 \rb^5\big),
\end{IEEEeqnarray*}

\begin{equation}
F_{\theta[0]}  = 0
\end{equation}

\begin{equation}
F_{\phi[0]} = \frac{ L \rbdot \left(F^{\mathcal{E}}_{\phi[0]} \mathcal{E} + F^{\mathcal{K}}_{ \phi[0]} \mathcal{K} \right)}{\pi  \rb \left(\rb^2+L^2 + \frac{2 a^2 M}{\rb} + a^2 \right)^{1/2} \left( 2 a^2 M + a^2 \rb + L^2 \rb \right)^2} ,
\end{equation}
where
\par \vspace{-6pt} \begin{IEEEeqnarray*}{rCl}
F^{\mathcal{E}}_{\phi[0]} &=& 14 a^4 M^2+11 a^4 M \rb+2 a^4 \rb^2+11 a^2 L^2 M \rb+4 a^2 L^2
   \rb^2+4 a^2 M \rb^3+a^2 \rb^4 \\
&&
	+\: 2 L^4 \rb^2+L^2 \rb^4, \\
F^{\mathcal{K}}_{\phi[0]} &=& -4 a^4 M^2-2 a^4 M \rb-2 a^2 L^2 M \rb-4 a^2 M \rb^3-a^2
   \rb^4-L^2 \rb^4.
\end{IEEEeqnarray*}
As with the Scalar case, $F_{a[2]}$ proves too large to include in paper format and so is available electronically \cite{BarryWardell.net}, although we provide $F_{r[2]}$ for circular orbits below.

\par \vspace{-6pt} \begin{IEEEeqnarray}{rCl}
F_{r[2]} &=&  \frac{1}{6 \pi  M \rb^{11/2}}\sqrt{\frac{2 a M^{5/2} \rb^{3/2}+M^2 \rb^2 (\rb-3 M)}{\left(a \sqrt{M}+\rb^{3/2}\right)^2 \left[a^2+\rb
   (\rb-2 M)\right]}}  \\
&& 
	\times \:
\frac{ \left(2 a \sqrt{M}+\sqrt{\rb} (\rb-3 M)\right)^{-2} \left(F^{\mathcal{E}}_{r[2]} \mathcal{E} + F^{\mathcal{K}}_{ r[2]} \mathcal{K} \right)}{   \left[a^4 M+2 a^3 \sqrt{M \rb^3}+a^2 \rb \left(-2 M^2+M 
   \rb+\rb^2\right)-4 a M^{3/2} \rb^{5/2}+M \rb^4\right]^3} , \nonumber
\end{IEEEeqnarray}
where
\par \vspace{-6pt} \begin{IEEEeqnarray*}{rCl}
F^{\mathcal{E}}_{r[2]} &=&
	-\left(\rb^{3/2}+a \sqrt{M}\right)^2 \big(-3 M^3 \rb^{33/2}-96 M^4 \rb^{31/2}-9 a^2 M^2 \rb^{31/2}+438 a M^{7/2} \rb^{15} \\
&& \quad
	+\: 495 M^5
   \rb^{29/2}-399 a^2 M^3 \rb^{29/2}-9 a^4 M \rb^{29/2}-2088 a M^{9/2} \rb^{14} \\
&& \quad
	+\: 388 a^3 M^{5/2} \rb^{14}-3 a^6 \rb^{27/2}-588
   M^6 \rb^{27/2}+1080 a^2 M^4 \rb^{27/2} \\
&& \quad
	-\: 228 a^4 M^2 \rb^{27/2}+2058 a M^{11/2} \rb^{13}-372 a^3 M^{7/2} \rb^{13}+54 a^5 M^{3/2}
   \rb^{13} \\
&& \quad
	+\: 4467 a^2 M^5 \rb^{25/2}+599 a^4 M^3 \rb^{25/2}-123 a^6 M \rb^{25/2}+828 a M^{13/2} \rb^{12} \\
&& \quad
	-\: 7504 a^3 M^{9/2}
   \rb^{12}+644 a^5 M^{5/2} \rb^{12}-24 a^7 \sqrt{M} \rb^{12}-6 a^8 \rb^{23/2} \\
&& \quad
	-\: 11265 a^2 M^6 \rb^{23/2}+1578 a^4 M^4
   \rb^{23/2}+609 a^6 M^2 \rb^{23/2}+9576 a^3 M^{11/2} \rb^{11} \\
&& \quad
	-\: 5882 a^5 M^{7/2} \rb^{11}-528 a^7 M^{3/2} \rb^{11}+2358 a^2 M^7
   \rb^{21/2}+14389 a^4 M^5 \rb^{21/2} \\
&& \quad
	+\: 3516 a^6 M^3 \rb^{21/2}-189 a^8 M \rb^{21/2}+7608 a^3 M^{13/2} \rb^{10}-2650 a^5 M^{9/2}
   \rb^{10} \\
&& \quad
	+\: 3296 a^7 M^{5/2} \rb^{10}-48 a^9 \sqrt{M} \rb^{10}-39138 a^4 M^6 \rb^{19/2}-16229 a^6 M^4 \rb^{19/2} \\
&& \quad
	-\: 330 a^8 M^2
   \rb^{19/2}-2184 a^3 M^{15/2} \rb^9+21172 a^5 M^{11/2} \rb^9-3530 a^7 M^{7/2} \rb^9 \\
&& \quad
	-\: 750 a^9 M^{3/2} \rb^9+21603 a^4 M^7
   \rb^{17/2}+39699 a^6 M^5 \rb^{17/2}+8911 a^8 M^3 \rb^{17/2} \\
&& \quad
	-\: 156 a^{10} M \rb^{17/2}-29982 a^5 M^{13/2} \rb^8-26366 a^7 M^{9/2}
   \rb^8+1576 a^9 M^{5/2} \rb^8 \\
&& \quad
	-\: 5706 a^4 M^8 \rb^{15/2}-16697 a^6 M^6 \rb^{15/2}-16348 a^8 M^4 \rb^{15/2}-1368 a^{10} M^2
   \rb^{15/2} \\
&& \quad
	+\: 13800 a^5 M^{15/2} \rb^7+39338 a^7 M^{11/2} \rb^7+9096 a^9 M^{7/2} \rb^7-264 a^{11} M^{3/2} \rb^7 \\
&& \quad
	+\: 2436 a^6 M^7
   \rb^{13/2}+1368 a^8 M^5 \rb^{13/2}+3426 a^{10} M^3 \rb^{13/2}-3096 a^5 M^{17/2} \rb^6 \\
&& \quad
	-|: 23546 a^7 M^{13/2} \rb^6-16754 a^9 M^{9/2}
   \rb^6-1032 a^{11} M^{5/2} \rb^6+4404 a^6 M^8 \rb^{11/2} \\
&& \quad
	+\: 4464 a^8 M^6 \rb^{11/2}+654 a^{10} M^4 \rb^{11/2}-246 a^{12} M^2
   \rb^{11/2}+5880 a^7 M^{15/2} \rb^5 \\
&& \quad
	+\: 15222 a^9 M^{11/2} \rb^5+2686 a^{11} M^{7/2} \rb^5-516 a^6 M^9 \rb^{9/2}-9146 a^8 M^7
   \rb^{9/2} \\
&& \quad
	-\: 3435 a^{10} M^5 \rb^{9/2}+54 a^{12} M^3 \rb^{9/2}+600 a^7 M^{17/2} \rb^4-4018 a^9 M^{13/2} \rb^4 \\
&& \quad
	-\: 4530 a^{11} M^{9/2}
   \rb^4-120 a^{13} M^{5/2} \rb^4+1164 a^8 M^8 \rb^{7/2}+6992 a^{10} M^6 \rb^{7/2} \\
&&\quad
	+\: 702 a^{12} M^4 \rb^{7/2}-1288 a^9 M^{15/2}
   \rb^3+1158 a^{11} M^{11/2} \rb^3+540 a^{13} M^{7/2} \rb^3 \\
&& \quad
	-\: 969 a^{10} M^7 \rb^{5/2}-2336 a^{12} M^5 \rb^{5/2}-24 a^{14} M^3
   \rb^{5/2}+1030 a^{11} M^{13/2} \rb^2 \\
&& \quad
	-\: 116 a^{13} M^{9/2} \rb^2+354 a^{12} M^6 \rb^{3/2}+288 a^{14} M^4 \rb^{3/2}-364 a^{13}
   M^{11/2} \rb \\
&& \quad
	-\: 48 a^{14} M^5 \sqrt{\rb}+48 a^{15} M^{9/2}\big), \\
F^{\mathcal{E}}_{r[2]} &=&
-\rb^3 \left(\rb^{3/2}-3 M \sqrt{\rb}+2 a \sqrt{M}\right) \big(3 M^3 \rb^{15}+96 M^4 \rb^{14}+9 a^2 M^2 \rb^{14} \\
&& \quad
	-\: 438 a
   M^{7/2} \rb^{27/2}-255 M^5 \rb^{13}+393 a^2 M^3 \rb^{13}+9 a^4 M \rb^{13} \\
&& \quad
	+\: 1014 a M^{9/2} \rb^{25/2}-388 a^3 M^{5/2}
   \rb^{25/2}+3 a^6 \rb^{12}-69 a^2 M^4 \rb^{12}+210 a^4 M^2 \rb^{12} \\
&& \quad
	+\: 252 a M^{11/2} \rb^{23/2}-406 a^3 M^{7/2} \rb^{23/2}-54
   a^5 M^{3/2} \rb^{23/2}-2763 a^2 M^5 \rb^{11} \\
&& \quad
	-\: 32 a^4 M^3 \rb^{11}+105 a^6 M \rb^{11}+3594 a^3 M^{9/2} \rb^{21/2}-738 a^5 M^{5/2}
   \rb^{21/2} \\
&& \quad
	+\: 954 a^2 M^6 \rb^{10}+763 a^4 M^4 \rb^{10}-204 a^6 M^2 \rb^{10}-2940 a^3 M^{11/2} \rb^{19/2} \\
&& \quad
	+\: 1610 a^5 M^{7/2}
   \rb^{19/2}+474 a^7 M^{3/2} \rb^{19/2}-3648 a^4 M^5 \rb^9-3014 a^6 M^3 \rb^9 \\
&& \quad
	+\: 222 a^8 M \rb^9-564 a^3 M^{13/2}
   \rb^{17/2}+2656 a^5 M^{9/2} \rb^{17/2}-2366 a^7 M^{5/2} \rb^{17/2} \\
&& \quad
	+\: 10731 a^4 M^6 \rb^8+12274 a^6 M^4 \rb^8+963 a^8 M^2
   \rb^8-21096 a^5 M^{11/2} \rb^{15/2} \\
&& \quad
	-\: 3328 a^7 M^{7/2} \rb^{15/2}+996 a^9 M^{3/2} \rb^{15/2}-1881 a^4 M^7 \rb^7-7640 a^6 M^5
   \rb^7 \\
&& \quad
	-\: 9303 a^8 M^3 \rb^7+17622 a^5 M^{13/2} \rb^{13/2}+30698 a^7 M^{9/2} \rb^{13/2} \\
&& \quad
	+\: 1056 a^9 M^{5/2} \rb^{13/2}-21807 a^6 M^6
   \rb^6+517 a^8 M^4 \rb^6+2283 a^{10} M^2 \rb^6 \\
&& \quad
	-\: 1044 a^5 M^{15/2} \rb^{11/2}-16022 a^9 M^{7/2} \rb^{11/2}+8817 a^6 M^7
   \rb^5+30400 a^8 M^5 \rb^5 \\
&& \quad
	+\: 426 a^{10} M^3 \rb^5-8268 a^7 M^{13/2} \rb^{9/2}+4904 a^9 M^{9/2} \rb^{9/2}+2940 a^{11} M^{5/2}
   \rb^{9/2} \\
&& \quad
	-\: 174 a^6 M^8 \rb^4-10712 a^8 M^6 \rb^4-13842 a^{10} M^4 \rb^4+1428 a^7 M^{15/2} \rb^{7/2} \\
&& \quad
	+\: 12492 a^9 M^{11/2}
   \rb^{7/2}-360 a^{11} M^{7/2} \rb^{7/2}-1011 a^8 M^7 \rb^3+4108 a^{10} M^5 \rb^3 \\
&& \quad
	+\: 2172 a^{12} M^3 \rb^3-2062 a^9 M^{13/2}
   \rb^{5/2}-5774 a^{11} M^{9/2} \rb^{5/2}+1725 a^{10} M^6 \rb^2 \\
&& \quad
	-\: 480 a^{12} M^4 \rb^2+986 a^{11} M^{11/2} \rb^{3/2}+864 a^{13}
   M^{7/2} \rb^{3/2}-876 a^{12} M^5 \rb \\
&& \quad
	-\: 17222 a^7 (M \rb)^{11/2}+144 a^{14} M^4-156 a^{13} \sqrt{M^9 \rb}+24 a^7 \sqrt{M
   \rb^{21}}\big)
\end{IEEEeqnarray*}


\subsection{Gravitational Case}

\subsubsection{Self-Force Regularisation}

\begin{gather}
F_{t\lnpow{1}} = \frac{\rb \rbdot \text{sgn$\Delta $r}}{\rb
   \left(a^2+L^2\right)+2 a^2 M+\rb^3} , \\
F_{r\lnpow{1}} = -\frac{\text{sgn$\Delta $r} \left(E \rb
   \left(a^2+\rb^2\right)+2 a M (a
   E-L)\right)}{\left(a^2-2 M \rb+\rb^2\right)
   \left(\rb \left(a^2+L^2\right)+2 a^2 M+\rb^3\right)}, \\
F_{\theta \lnpow{1}} = 0,  \qquad
F_{\phi \lnpow{1}} =0,
\end{gather}

\begin{equation}
F_{t[0]} = \frac{\rbdot \left(F^{\mathcal{E}}_{t[0]} \mathcal{E} + F^{\mathcal{K}}_{ t [0]} \mathcal{K} \right)}{\pi  \rb^2 \left(\rb^2+L^2 + \frac{2 a^2 M}{\rb} + a^2 \right)^{3/2} \left( 2 a^2 M + a^2 \rb + L^2 \rb \right)^2} ,
\end{equation}
where
\par \vspace{-6pt} \begin{IEEEeqnarray*}{rCl}
F^{\mathcal{E}}_{t[0]} &=& 
	4 a L M \left(4 a^4 M^2+2 a^4 M \rb+2 a^2 L^2 M \rb-a^2 M \rb^3-a^2\rb^4-L^2 \rb^4 
	\right) \\
&&
	+\: E \big(12 a^6 M^3+16 a^6 M^2 \rb+7 a^6
   M \rb^2+a^6 \rb^3+28 a^4 L^2 M^2 \rb+22 a^4 L^2 M \rb^2 \\
&& \quad
	+\: 4 a^4 L^2
   \rb^3+6 a^4 M^2 \rb^3+5 a^4 M \rb^4+a^4 \rb^5+15 a^2 L^4 M
   \rb^2+5 a^2 L^4 \rb^3 \\
&& \quad
	+\: 5 a^2 L^2 M \rb^4+a^2 L^2 \rb^5+2 L^6
   \rb^3\big), \\
F^{\mathcal{K}}_{t[0]} &=&
-2 a L M \big(2 a^4 M^2-a^4
   M \rb-a^4 \rb^2-a^2 L^2 M \rb-2 a^2 L^2 \rb^2-2 a^2 M
   \rb^3-2 a^2 \rb^4 \\
&& \quad
	-\: L^4 \rb^2-2 L^2 \rb^4\big) \\
&&
	+\: E \big(-4 a^6 M^3-4 a^6 M^2 \rb-a^6 M \rb^2-10 a^4 L^2 M^2 \rb-5
   a^4 L^2 M \rb^2-2 a^4 M^2 \rb^3 \\
&& \quad
	-\: a^4 M \rb^4-4 a^2 L^4 M \rb^2-a^2
   L^2 M \rb^4+a^2 L^2 \rb^5+L^4 \rb^5\big),
\end{IEEEeqnarray*}

\begin{equation}
F_{r[0]} = \frac{ \left(F^{\mathcal{E}}_{r[0]} \mathcal{E} + F^{\mathcal{K}}_{ r [0]} \mathcal{K} \right)}{\pi  \rb^6 \left(\rb^2+L^2 + \frac{2 a^2 M}{\rb} + a^2 \right)^{3/2} \left( 2 a^2 M + a^2 \rb + L^2 \rb \right)^2 \left( a^2 - 2 M \rb  + \rb^2 \right)} ,
\end{equation}
where
\par \vspace{-6pt} \begin{IEEEeqnarray*}{rCl}
F^{\mathcal{E}}_{r[0]} &=& 
+L^2 \rb^3 \big(-24 a^6
   M^4-28 a^6 M^3 \rb+6 a^6 M^2 \rb^2+11 a^6 M \rb^3+2 a^6 \rb^4 \\
&& \quad
	-\: 56
   a^4 L^2 M^3 \rb-24 a^4 L^2 M^2 \rb^2+18 a^4 L^2 M \rb^3+6 a^4 L^2
   \rb^4-52 a^4 M^3 \rb^3 \\
&& \quad
	-\: 20 a^4 M^2 \rb^4+11 a^4 M \rb^5+3 a^4
   \rb^6-30 a^2 L^4 M^2 \rb^2+3 a^2 L^4 M \rb^3+6 a^2 L^4 \rb^4 \\
&& \quad
	-\: 42 a^2 L^2 M^2 \rb^4+5 a^2 L^2 M \rb^5+6 a^2 L^2 \rb^6-8 a^2 M^2
   \rb^6+2 a^2 M \rb^7+a^2 \rb^8 \\
&&\quad
	-\: 4 L^6 M \rb^3+2 L^6 \rb^4-6
   L^4 M \rb^5+3 L^4 \rb^6-2 L^2 M \rb^7+L^2 \rb^8\big) \\
&& 
	+\: 2 a E L M \rb^3 \big(24 a^6
   M^3+36 a^6 M^2 \rb+18 a^6 M \rb^2+3 a^6 \rb^3+56 a^4 L^2 M^2
   \rb \\
&& \quad
	+\: 48 a^4 L^2 M \rb^2+10 a^4 L^2 \rb^3+24 a^4 M^2 \rb^3+24 a^4 M
   \rb^4+6 a^4 \rb^5+30 a^2 L^4 M \rb^2 \\
&& \quad
	+\: 11 a^2 L^4 \rb^3+22 a^2 L^2
   M \rb^4+9 a^2 L^2 \rb^5+6 a^2 M \rb^6+3 a^2 \rb^7+4 L^6
   \rb^3+3 L^4 \rb^5 \\
&& \quad
	+\: 3 L^2 \rb^7\big) \\
&&
	-\: E^2 \rb^3 \left(2 a^2 M+a^2 \rb+\rb^3\right) \big(12 a^6 M^3+16
   a^6 M^2 \rb+7 a^6 M \rb^2+a^6 \rb^3 \\
&& \quad
	+\: 28 a^4 L^2 M^2 \rb+22 a^4 L^2
   M \rb^2+4 a^4 L^2 \rb^3+6 a^4 M^2 \rb^3+5 a^4 M \rb^4+a^4
   \rb^5 \\
&& \quad
	+\: 15 a^2 L^4 M \rb^2+5 a^2 L^4 \rb^3+5 a^2 L^2 M \rb^4+a^2
   L^2 \rb^5+2 L^6 \rb^3\big), \\
F^{\mathcal{K}}_{r[0]} &=& 
	-L^2 \rb^3 \big(-8 a^6 M^4-12 a^6 M^3 \rb+2 a^6 M
   \rb^3-20 a^4 L^2 M^3 \rb-8 a^4 L^2 M^2 \rb^2 \\
&& \quad
	+\: 4 a^4 L^2 M \rb^3-32
   a^4 M^3 \rb^3-12 a^4 M^2 \rb^4+6 a^4 M \rb^5+a^4 \rb^6-8 a^2 L^4
   M^2 \rb^2 \\
&& \quad
	+\: 2 a^2 L^4 M \rb^3-24 a^2 L^2 M^2 \rb^4+4 a^2 L^2 M
   \rb^5+2 a^2 L^2 \rb^6-8 a^2 M^2 \rb^6+2 a^2 M \rb^7 \\
&& \quad
	+\: a^2
   \rb^8-2 L^4 M \rb^5+L^4 \rb^6-2 L^2 M \rb^7+L^2
   \rb^8\big) \\
&&
	-\: 2 a E L M \rb^3 \big(8 a^6 M^3+12
   a^6 M^2 \rb+6 a^6 M \rb^2+a^6 \rb^3+20 a^4 L^2 M^2 \rb \\
&& \quad
	+\: 14 a^4 L^2
   M \rb^2+2 a^4 L^2 \rb^3+16 a^4 M^2 \rb^3+16 a^4 M \rb^4+4 a^4
   \rb^5+8 a^2 L^4 M \rb^2 \\
&& \quad
	+\: a^2 L^4 \rb^3+14 a^2 L^2 M \rb^4+5 a^2
   L^2 \rb^5+6 a^2 M \rb^6+3 a^2 \rb^7+L^4 \rb^5+3 L^2
   \rb^7\big) \\
&&
	-\: E^2 \rb^3 \left(2 a^2 M+a^2 \rb+\rb^3\right) \big(-4 a^6 M^3-4
   a^6 M^2 \rb-a^6 M \rb^2-10 a^4 L^2 M^2 \rb \\
&& \quad
	-\: 5 a^4 L^2 M \rb^2-2
   a^4 M^2 \rb^3-a^4 M \rb^4-4 a^2 L^4 M \rb^2-a^2 L^2 M \rb^4+a^2
   L^2 \rb^5 \\
&& \quad
	+\: L^4 \rb^5\big),
\end{IEEEeqnarray*}

\begin{equation}
F_{\theta[0]}  = 0,
\end{equation}

\begin{equation}
F_{\phi[0]} = \frac{ \left(F^{\mathcal{E}}_{\phi[0]} \mathcal{E} + F^{\mathcal{K}}_{ \phi [0]} \mathcal{K} \right)}{\pi  \rb^6 \left(\rb^2+L^2 + \frac{2 a^2 M}{\rb} + a^2 \right)^{3/2} \left( 2 a^2 M + a^2 \rb + L^2 \rb \right)^2 \left( a^2 - 2 M \rb  + \rb^2 \right)} ,
\end{equation}
where
\par \vspace{-6pt} \begin{IEEEeqnarray*}{rCl}
F^{\mathcal{E}}_{\phi[0]} &=& -14 a^4 M^2-11 a^4 M \rb-2 a^4 \rb^2-11 a^2 L^2 M \rb-4 a^2 L^2
   \rb^2-4 a^2 M \rb^3-a^2 \rb^4 \\
&&
	-\: 2 L^4 \rb^2-L^2 \rb^4, \\
F^{\mathcal{K}}_{\phi[0]} &=& 4 a^4 M^2+2 a^4 M \rb+2 a^2 L^2 M \rb+4 a^2 M \rb^3+a^2 \rb^4+L^2
   \rb^4.
\end{IEEEeqnarray*}
As with the scalar and electromagnetic cases, $F_{a[\text{\fixme{$2$}}]} $ is too large for paper format and so is available electronically \cite{BarryWardell.net}.
As outlined above, we do give $F_{\text{\fixme{$r$}} [2]}$ for circular orbits,
\par \vspace{-6pt} \begin{IEEEeqnarray}{rCl}
F_{r [2]} &=& \frac{1}{\pi  \rb^{11/2}} \sqrt{\frac{2 a M^{5/2} \rb^{3/2}+M^2 \rb^2 (\rb-3 M)}{\left(a
   \sqrt{M}+\rb^{3/2}\right)^2 \left[a^2+\rb (\rb-2 M)\right]}} \left[2 a \sqrt{M}+\sqrt{\rb} (\rb-3 M)\right]^{-2}  \nonumber \\
&&
	\times \: \frac{ \left[a^2+\rb (\rb-2 M)\right]^{-1} \left(F^{\mathcal{E}}_{r[2]} \mathcal{E} + F^{\mathcal{K}}_{r[2]} \mathcal{K} \right)}{  \left[a^4 M+2 a^3
   \sqrt{M \rb^3}+a^2 \rb \left(-2 M^2+M \rb+\rb^2\right)-4 a M^{3/2} \rb^{5/2}+M \rb^4\right]^2}, \nonumber
\end{IEEEeqnarray}
where
\par \vspace{-6pt} \begin{IEEEeqnarray*}{rCl}
F^{\mathcal{E}}_{r[2]} &=& 
-\left(a \sqrt{M}+\rb^{3/2}\right)^2 \left(a^2-2 M \rb+\rb^2\right) \big(32 a^{11} M^{5/2}-48 a^{10} M^3 \sqrt{\rb} \\
&& \quad
	+\: 144 a^{10} M^2
   \rb^{3/2}+478 a^9 M^{3/2} \rb^3-104 a^9 M^{5/2} \rb^2-192 a^9 M^{7/2} \rb \\
&& \quad
	+\: 288 a^8 M^4 \rb^{3/2}-992 a^8 M^3 \rb^{5/2}-473
   a^8 M^2 \rb^{7/2}+779 a^8 M \rb^{9/2} \\
&& \quad
	-\: 292 a^7 M^{3/2} \rb^5-2496 a^7 M^{5/2} \rb^4+800 a^7 M^{7/2} \rb^3+384 a^7 M^{9/2}
   \rb^2 \\
&& \quad
	+\: 540 a^7 \sqrt{M} \rb^6-576 a^6 M^5 \rb^{5/2}+2240 a^6 M^4 \rb^{7/2}+2489 a^6 M^3 \rb^{9/2} \\
&& \quad
	-\: 4623 a^6 M^2
   \rb^{11/2}+227 a^6 M \rb^{13/2}+127 a^6 \rb^{15/2}-1846 a^5 M^{3/2} \rb^7 \\
&& \quad
	+\: 3574 a^5 M^{5/2} \rb^6+4018 a^5 M^{7/2}
   \rb^5-1952 a^5 M^{9/2} \rb^4-256 a^5 M^{11/2} \rb^3 \\
&& \quad
	+\: 386 a^5 \sqrt{M} \rb^8+384 a^4 M^6 \rb^{7/2}-1664 a^4 M^5
   \rb^{9/2}-3950 a^4 M^4 \rb^{11/2} \\
&& \quad
	+\: 6038 a^4 M^3 \rb^{13/2}-2275 a^4 M^2 \rb^{15/2}-462 a^4 M \rb^{17/2}+105 a^4
   \rb^{19/2} \\
&& \quad
	-\: 14 a^3 M^{3/2} \rb^9+4444 a^3 M^{5/2} \rb^8-6514 a^3 M^{7/2} \rb^7-1876 a^3 M^{9/2} \rb^6 \\
&& \quad
	+\: 1536 a^3 M^{11/2}
   \rb^5-144 a^3 \sqrt{M} \rb^{10}+1728 a^2 M^5 \rb^{13/2}+674 a^2 M^4 \rb^{15/2} \\
&& \quad
	+\: 588 a^2 M^3 \rb^{17/2}-1312 a^2 M^2
   \rb^{19/2}+348 a^2 M \rb^{21/2}-22 a^2 \rb^{23/2} \\
&& \quad
	-\: 148 a M^{3/2} \rb^{11}+884 a M^{5/2} \rb^{10}-1370 a M^{7/2} \rb^9+384 a
   M^{9/2} \rb^8 \\
&& \quad
	-\: 192 M^4 \rb^{19/2}+287 M^3 \rb^{21/2}-131 M^2 \rb^{23/2}+18 M \rb^{25/2}\big), \\
F^{\mathcal{K}}_{r[2]} &=&
\rb^3 \left(2 a \sqrt{M}-3 M \sqrt{\rb}+\rb^{3/2}\right) \big(144 a^{12} M^2-96 a^{11} \sqrt{M^5 \rb}+576 a^{11} (M \rb)^{3/2} \\
&& \quad
	-\: 921
   a^{10} M^3 \rb+81 a^{10} M^2 \rb^2+864 a^{10} M \rb^3+862 a^9 M^{3/2} \rb^{7/2} \\
&& \quad
	+\: 582 a^9 M^{7/2} \rb^{3/2}+576 a^9 \sqrt{M
   \rb^9}-4010 a^9 (M \rb)^{5/2}+2052 a^8 M^4 \rb^2 \\
&& \quad
	+\: 313 a^8 M^3 \rb^3-5342 a^8 M^2 \rb^4+1351 a^8 M \rb^5+144 a^8
   \rb^6-2308 a^7 M^{3/2} \rb^{11/2} \\
&& \quad
	-\: 3816 a^7 M^{5/2} \rb^{9/2}-1176 a^7 M^{9/2} \rb^{5/2}+920 a^7 \sqrt{M \rb^{13}}+9524 a^7 (M
   \rb)^{7/2} \\
&& \quad
	-\: 1764 a^6 M^5 \rb^3-2260 a^6 M^4 \rb^4+12726 a^6 M^3 \rb^5-6148 a^6 M^2 \rb^6+60 a^6 M \rb^7 \\
&& \quad
	+\: 254 a^6
   \rb^8-2580 a^5 M^{3/2} \rb^{15/2}+4204 a^5 M^{5/2} \rb^{13/2}+3504 a^5 M^{7/2} \rb^{11/2} \\
&&\quad
	+\: 792 a^5 M^{11/2} \rb^{7/2}+204 a^5
   \sqrt{M \rb^{17}}-8384 a^5 (M \rb)^{9/2}+384 a^4 M^6 \rb^4 \\
&& \quad
	+\: 2620 a^4 M^5 \rb^5-11878 a^4 M^4 \rb^6+8425 a^4 M^3 \rb^7-1656
   a^4 M^2 \rb^8-111 a^4 M \rb^9 \\
&& \quad
	+\: 88 a^4 \rb^{10}-242 a^3 M^{3/2} \rb^{19/2}+2928 a^3 M^{5/2} \rb^{17/2}-4866 a^3 M^{7/2}
   \rb^{15/2} \\
&& \quad
	+\: 1372 a^3 M^{9/2} \rb^{13/2}-144 a^3 \sqrt{M \rb^{21}}+1536 a^3 (M \rb)^{11/2}+1728 a^2 M^5 \rb^7 \\
&& \quad
	-\: 1614 a^2 M^4
   \rb^8+1373 a^2 M^3 \rb^9-1075 a^2 M^2 \rb^{10}+334 a^2 M \rb^{11}-22 a^2 \rb^{12} \\
&& \quad
	-\: 148 a M^{3/2} \rb^{23/2}+718 a M^{5/2}
   \rb^{21/2}-1040 a M^{7/2} \rb^{19/2}+384 a M^{9/2} \rb^{17/2} \\
&& \quad
	-\: 192 M^4 \rb^{10}+280 M^3 \rb^{11}-128 M^2 \rb^{12}+18 M
   \rb^{13}\big).
\end{IEEEeqnarray*}

\subsubsection{$h u u$ Regularisation}

\begin{equation}
H_{[0]} = \frac{2 \mathcal{K}}{\pi \left(\rb^2+L^2 + \frac{2 a^2 M}{\rb} + a^2 \right)^{1/2} },
\end{equation}

\begin{equation}
H_{[1]} = 0,
\end{equation}

\begin{equation}
H_{[2]} = \frac{ \left(H^{\mathcal{E}}_{[2]} \mathcal{E} + H^{\mathcal{K}}_{[2]} \mathcal{K} \right)}{3 \pi  \rb^7 \left(\rb^2+L^2 + \frac{2 a^2 M}{\rb} + a^2 \right)^{3/2} \left( 2 a^2 M + a^2 \rb + L^2 \rb \right)^3 } ,
\end{equation}
where
\par \vspace{-6pt} \begin{IEEEeqnarray*}{rCl}
H^{\mathcal{E}}_{[2]} &=& 
	\big(12 M \rb^5 a^{12}+92 M^2 \rb^4 a^{12}+264 M^3 \rb^3 a^{12}+336 M^4
   \rb^2 a^{12}+160 M^5 \rb a^{12} \\
&& \quad
	-\: 24 \rb^8 a^{10}-240 M \rb^7
   a^{10}-1104 L^2 M^6 a^{10}-1104 M^2 \rb^6 a^{10}-2880 M^3 \rb^5 a^{10} \\
&& \quad
	+\: 48 L^2
   M \rb^5 a^{10}-4272 M^4 \rb^4 a^{10}+230 L^2 M^2 \rb^4 a^{10}-3264 M^5
   \rb^3 a^{10} \\
&& \quad
	+\: 96 L^2 M^3 \rb^3 a^{10}-960 M^6 \rb^2 a^{10}-1116 L^2 M^4
   \rb^2 a^{10}-2096 L^2 M^5 \rb a^{10} \\
&& \quad
	-\: 48 \rb^{10} a^8-420 M \rb^9
   a^8-120 L^2 \rb^8 a^8-1556 M^2 \rb^8 a^8-2872 M^3 \rb^7 a^8 \\
&& \quad
	-\: 882 L^2 M
   \rb^7 a^8-2448 M^4 \rb^6 a^8-2781 L^2 M^2 \rb^6 a^8-672 M^5 \rb^5
   a^8 \\
&& \quad
	-\: 4770 L^2 M^3 \rb^5 a^8+72 L^4 M \rb^5 a^8-4272 L^2 M^4 \rb^4 a^8+90
   L^4 M^2 \rb^4 a^8 \\
&& \quad
	-\: 1440 L^2 M^5 \rb^3 a^8-1044 L^4 M^3 \rb^3 a^8-2928
   L^4 M^4 \rb^2 a^8-2112 L^4 M^5 \rb a^8 \\
&& \quad
	-\: 24 \rb^{12} a^6-168 M
   \rb^{11} a^6-195 L^2 \rb^{10} a^6-456 M^2 \rb^{10} a^6-480 M^3
   \rb^9 a^6 \\
&& \quad
	-\: 1086 L^2 M \rb^9 a^6-240 L^4 \rb^8 a^6-96 M^4 \rb^8
   a^6-2119 L^2 M^2 \rb^8 a^6 \\
&& \quad
	-\: 1528 L^2 M^3 \rb^7 a^6-1098 L^4 M \rb^7
   a^6-84 L^2 M^4 \rb^6 a^6-1578 L^4 M^2 \rb^6 a^6 \\
&& \quad
	-\: 696 L^4 M^3 \rb^5
   a^6+48 L^6 M \rb^5 a^6+84 L^4 M^4 \rb^4 a^6-190 L^6 M^2 \rb^4 a^6 \\
&& \quad
	-\: 1320
   L^6 M^3 \rb^3 a^6-1476 L^6 M^4 \rb^2 a^6-75 L^2 \rb^{12} a^4-246 L^2 M
   \rb^{11} a^4 \\
&& \quad
	-\: 297 L^4 \rb^{10} a^4-84 L^2 M^2 \rb^{10} a^4+216 L^2 M^3
   \rb^9 a^4-690 L^4 M \rb^9 a^4-240 L^6 \rb^8 a^4 \\
&& \quad
	+\: 529 L^4 M^2 \rb^8
   a^4+1374 L^4 M^3 \rb^7 a^4-402 L^6 M \rb^7 a^4+771 L^6 M^2 \rb^6
   a^4 \\
&& \quad
	+\: 1194 L^6 M^3 \rb^5 a^4+12 L^8 M \rb^5 a^4-190 L^8 M^2 \rb^4 a^4-444
   L^8 M^3 \rb^3 a^4 \\
&& \quad
	-\: 78 L^4 \rb^{12} a^2+36 L^4 M \rb^{11} a^2-201 L^6
   \rb^{10} a^2+384 L^4 M^2 \rb^{10} a^2+198 L^6 M \rb^9 a^2 \\
&& \quad
	-\: 120 L^8
   \rb^8 a^2+1092 L^6 M^2 \rb^8 a^2+162 L^8 M \rb^7 a^2+672 L^8 M^2
   \rb^6 a^2 \\
&& \quad
	-\: 48 L^{10} M^2 \rb^4 a^2 -27 L^6 \rb^{12}+114 L^6 M \rb^{11}-51 L^8
   \rb^{10}+222 L^8 M \rb^9 \\
&& \quad
	-\: 24 L^{10} \rb^8+108 L^{10} M \rb^7 \big) \\
&&
	+\: 4 a E L M \big(-71 a^4
   \rb^{11}-75 L^4 \rb^{11}-146 a^2 L^2 \rb^{11}-262 a^4 M
   \rb^{10}-270 a^2 L^2 M \rb^{10} \\
&& \quad
	-\: 147 a^6 \rb^9-159 L^6 \rb^9-465
   a^2 L^4 \rb^9-453 a^4 L^2 \rb^9-240 a^4 M^2 \rb^9-811 a^6 M
   \rb^8 \\
&& \quad
	-\: 873 a^2 L^4 M \rb^8-1684 a^4 L^2 M \rb^8-76 a^8 \rb^7-78
   L^8 \rb^7-310 a^2 L^6 \rb^7 \\
&& \quad
	-\: 462 a^4 L^4 \rb^7-306 a^6 L^2
   \rb^7-1490 a^6 M^2 \rb^7-1542 a^4 L^2 M^2 \rb^7-912 a^6 M^3
   \rb^6 \\
&& \quad
	-\: 522 a^8 M \rb^6-549 a^2 L^6 M \rb^6-1620 a^4 L^4 M
   \rb^6-1593 a^6 L^2 M \rb^6 \\
&& \quad
	-\: 1323 a^8 M^2 \rb^5-1368 a^4 L^4 M^2
   \rb^5-2691 a^6 L^2 M^2 \rb^5-1460 a^8 M^3 \rb^4 \\
&& \quad
	-\: 1458 a^6 L^2 M^3
   \rb^4+37 a^{10} M \rb^4+24 a^2 L^8 M \rb^4+109 a^4 L^6 M
   \rb^4+183 a^6 L^4 M \rb^4 \\
&& \quad
	+\: 135 a^8 L^2 M \rb^4-588 a^8 M^4
   \rb^3+291 a^{10} M^2 \rb^3+222 a^4 L^6 M^2 \rb^3 \\
&& \quad
	+\: 735 a^6 L^4 M^2
   \rb^3+804 a^8 L^2 M^2 \rb^3+858 a^{10} M^3 \rb^2+738 a^6 L^4 M^3
   \rb^2 \\
&& \quad
	+\: 1596 a^8 L^2 M^3 \rb^2+1124 a^{10} M^4 \rb+1056 a^8 L^2 M^4
   \rb+552 a^{10} M^5\big) \\
&& 
	-\: 3 E^2 \big(368 M^6 a^{12}+4 M \rb^5 a^{12}+55 M^2 \rb^4 a^{12}+280 M^3
   \rb^3 a^{12}+680 M^4 \rb^2 a^{12} \\
&& \quad
	+\: 800 M^5 \rb a^{12}-9 \rb^8
   a^{10}-124 M \rb^7 a^{10}-604 M^2 \rb^6 a^{10}-1368 M^3 \rb^5 a^{10} \\
&& \quad
	+\: 12
   L^2 M \rb^5 a^{10}-1472 M^4 \rb^4 a^{10}+160 L^2 M^2 \rb^4 a^{10}-608
   M^5 \rb^3 a^{10} \\
&& \quad
	+\: 672 L^2 M^3 \rb^3 a^{10}+1152 L^2 M^4 \rb^2 a^{10}+704
   L^2 M^5 \rb a^{10}-18 \rb^{10} a^8 \\
&& \quad
	-\: 224 M \rb^9 a^8-35 L^2 \rb^8
   a^8-901 M^2 \rb^8 a^8-1492 M^3 \rb^7 a^8-438 L^2 M \rb^7 a^8 \\
&& \quad
	-\: 884 M^4
   \rb^6 a^8-1685 L^2 M^2 \rb^6 a^8-2604 L^2 M^3 \rb^5 a^8+12 L^4 M
   \rb^5 a^8 \\
&& \quad
	- \: 1412 L^2 M^4 \rb^4 a^8+171 L^4 M^2 \rb^4 a^8+540 L^4 M^3
   \rb^3 a^8+492 L^4 M^4 \rb^2 a^8 \\
&& \quad
	-\: 9 \rb^{12} a^6-96 M \rb^{11}
   a^6-55 L^2 \rb^{10} a^6-278 M^2 \rb^{10} a^6-244 M^3 \rb^9 a^6 \\
&& \quad
	-\: 596 L^2
   M \rb^9 a^6-51 L^4 \rb^8 a^6-1679 L^2 M^2 \rb^8 a^6-1414 L^2 M^3
   \rb^7 a^6 \\
&& \quad
	-\: 572 L^4 M \rb^7 a^6-1567 L^4 M^2 \rb^6 a^6-1254 L^4 M^3
   \rb^5 a^6+4 L^6 M \rb^5 a^6 \\
&& \quad
	+\: 82 L^6 M^2 \rb^4 a^6+148 L^6 M^3
   \rb^3 a^6-20 L^2 \rb^{12} a^4-176 L^2 M \rb^{11} a^4-56 L^4
   \rb^{10} a^4 \\
&& \quad
	-\: 266 L^2 M^2 \rb^{10} a^4-512 L^4 M \rb^9 a^4-33 L^6
   \rb^8 a^4-774 L^4 M^2 \rb^8 a^4 \\
&& \quad
	-\: 326 L^6 M \rb^7 a^4-486 L^6 M^2
   \rb^6 a^4+16 L^8 M^2 \rb^4 a^4-13 L^4 \rb^{12} a^2-80 L^4 M
   \rb^{11} a^2 \\
&& \quad
	-\: 19 L^6 \rb^{10} a^2-140 L^6 M \rb^9 a^2-8 L^8 \rb^8
   a^2-68 L^8 M \rb^7 a^2-2 L^6 \rb^{12}\big),
\end{IEEEeqnarray*}
\par \vspace{-6pt} \begin{IEEEeqnarray*}{rCl}
H^{\mathcal{K}}_{[2]} &=& 
	\rb^3 \big(24 \rb^5 a^{10}+180 M \rb^4 a^{10}+512
   M^2 \rb^3 a^{10}+656 M^3 \rb^2 a^{10}+320 M^4 \rb a^{10} \\
&&\quad
	+\: 93 \rb^7
   a^8+624 M \rb^6 a^8-912 L^2 M^5 a^8+120 L^2 \rb^5 a^8+1448 M^2 \rb^5
   a^8 \\
&& \quad
	+\: 976 M^3 \rb^4 a^8+636 L^2 M \rb^4 a^8-912 M^4 \rb^3 a^8+891 L^2 M^2
   \rb^3 a^8-1152 M^5 \rb^2 a^8 \\
&& \quad
	-\: 434 L^2 M^3 \rb^2 a^8-1720 L^2 M^4
   \rb a^8+69 \rb^9 a^6+354 M \rb^8 a^6+375 L^2 \rb^7 a^6 \\
&& \quad
	+\: 492 M^2
   \rb^7 a^6-168 M^3 \rb^6 a^6+1620 L^2 M \rb^6 a^6+240 L^4 \rb^5
   a^6-576 M^4 \rb^5 a^6 \\
&& \quad
	+\: 1207 L^2 M^2 \rb^5 a^6-2720 L^2 M^3 \rb^4 a^6+732
   L^4 M \rb^4 a^6-3372 L^2 M^4 \rb^3 a^6 \\
&& \quad
	-\: 450 L^4 M^2 \rb^3 a^6-2896 L^4
   M^3 \rb^2 a^6-2052 L^4 M^4 \rb a^6+210 L^2 \rb^9 a^4 \\
&& \quad
	+\: 534 L^2 M
   \rb^8 a^4+567 L^4 \rb^7 a^4-384 L^2 M^2 \rb^7 a^4-1224 L^2 M^3
   \rb^6 a^4 \\
&& \quad
	+\:  1074 L^4 M \rb^6 a^4+240 L^6 \rb^5 a^4-2017 L^4 M^2
   \rb^5 a^4-3726 L^4 M^3 \rb^4 a^4 \\
&& \quad
	+\: 180 L^6 M \rb^4 a^4-1525 L^6 M^2
   \rb^3 a^4-1806 L^6 M^3 \rb^2 a^4+213 L^4 \rb^9 a^2 \\
&& \quad
	-\: 18 L^4 M \rb^8
   a^2+381 L^6 \rb^7 a^2-888 L^4 M^2 \rb^7 a^2-216 L^6 M \rb^6 a^2+120 L^8
   \rb^5 a^2 \\
&& \quad
	-\: 1776 L^6 M^2 \rb^5 a^2-192 L^8 M \rb^4 a^2-696 L^8 M^2
   \rb^3 a^2+72 L^6 \rb^9-198 L^6 M \rb^8 \\
&& \quad
	+\: 96 L^8 \rb^7-294 L^8 M
   \rb^6+24 L^{10} \rb^5-96 L^{10} M \rb^4\big) \\
&&
	+\: 4 a  \rb^3 E L M \big(456 M^4 a^8+45 \rb^4 a^8+322 M \rb^3 a^8+862 M^2
   \rb^2 a^8+1024 M^3 \rb a^8 \\
&& \quad
	+\: 116 \rb^6 a^6+636 M \rb^5 a^6+183 L^2
   \rb^4 a^6+1162 M^2 \rb^4 a^6+708 M^3 \rb^3 a^6 \\
&& \quad
	+\: 992 L^2 M \rb^3
   a^6+1765 L^2 M^2 \rb^2 a^6+1026 L^2 M^3 \rb a^6+71 \rb^8 a^4+262 M
   \rb^7 a^4 \\
&& \quad
	+\: 358 L^2 \rb^6 a^4+240 M^2 \rb^6 a^4+1326 L^2 M \rb^5
   a^4+279 L^4 \rb^4 a^4+1206 L^2 M^2 \rb^4 a^4 \\
&& \quad
	+\: 1018 L^4 M \rb^3 a^4+903
   L^4 M^2 \rb^2 a^4+146 L^2 \rb^8 a^2+270 L^2 M \rb^7 a^2+368 L^4
   \rb^6 a^2 \\
&& \quad
	+\: 690 L^4 M \rb^5 a^2+189 L^6 \rb^4 a^2+348 L^6 M \rb^3
   a^2+75 L^4 \rb^8+126 L^6 \rb^6+48 L^8 \rb^4\big) \\
&&
	-\: 3 \rb^3 E^2 \big(304 M^5 a^{10}+8 \rb^5 a^{10}+90 M \rb^4 a^{10}+386 M^2
   \rb^3 a^{10}+796 M^3 \rb^2 a^{10} \\
&& \quad
	+\: 792 M^4 \rb a^{10}+32 \rb^7
   a^8+308 M \rb^6 a^8+32 L^2 \rb^5 a^8+1068 M^2 \rb^5 a^8 \\
&& \quad
	+\: 1600 M^3
   \rb^4 a^8+304 L^2 M \rb^4 a^8+880 M^4 \rb^3 a^8+1003 L^2 M^2
   \rb^3 a^8 \\
&& \quad
	+\: 1388 L^2 M^3 \rb^2 a^8+684 L^2 M^4 \rb a^8+24 \rb^9
   a^6+186 M \rb^8 a^6+113 L^2 \rb^7 a^6 \\
&& \quad
	+\: 458 M^2 \rb^7 a^6+364 M^3
   \rb^6 a^6+854 L^2 M \rb^6 a^6+48 L^4 \rb^5 a^6+2017 L^2 M^2 \rb^5
   a^6 \\
&& \quad
	+\: 1522 L^2 M^3 \rb^4 a^6+370 L^4 M \rb^4 a^6+849 L^4 M^2 \rb^3
   a^6+602 L^4 M^3 \rb^2 a^6 \\
&& \quad
	+\: 65 L^2 \rb^9 a^4+356 L^2 M \rb^8 a^4+146 L^4
   \rb^7 a^4+446 L^2 M^2 \rb^7 a^4+776 L^4 M \rb^6 a^4 \\
&& \quad
	+\: 32 L^6 \rb^5
   a^4+945 L^4 M^2 \rb^5 a^4+188 L^6 M \rb^4 a^4+232 L^6 M^2 \rb^3 a^4+58
   L^4 \rb^9 a^2 \\
&& \quad
	+\: 170 L^4 M \rb^8 a^2+81 L^6 \rb^7 a^2+230 L^6 M
   \rb^6 a^2+8 L^8 \rb^5 a^2+32 L^8 M \rb^4 a^2 \\
&& \quad
	+\: 17 L^6 \rb^9+16 L^8
   \rb^7\big) .
\end{IEEEeqnarray*}

\subsection{Results}

Using the mode sum, to date, researchers have only been able to produce values for the Kerr retarded field and resulting self-force in the scalar case.  As there is no known decomposition of the metric perturbation into tensor harmonic modes that separates the field equations in Kerr Lorentz gauge metric perturbation, higher spins are yet to be calculated.  Therefore it is not possible to demonstrate the validity of our electromagnetic and gravitational parameters.  However, as with our \Sch parameters, deriving the expressions by independent methods gives us confidence in our results.  The only other `check' we can do is to set the spin of the black hole to be zero, i.e., $a=0$, and compare with our \Sch results which we know to be correct.  All of the parameters given have passed these tests.
\begin{figure} [ht]
\begin{center}
\includegraphics[scale=0.66]{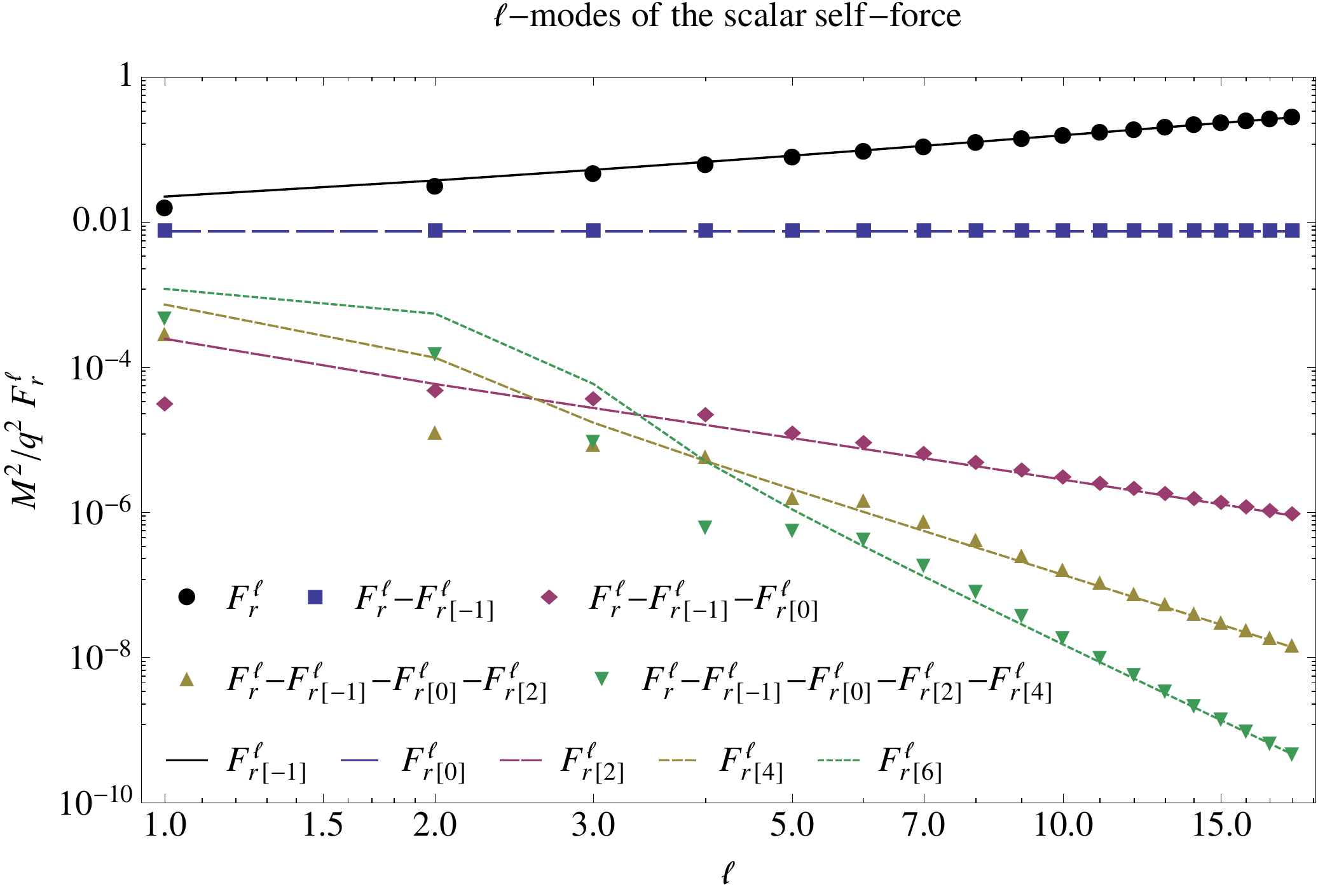}
\caption[Regularisation of the Scalar Self-Force Radial Component in Kerr Space-Time]{Regularization of the radial component of the scalar self-force in Kerr space-time for the case of a scalar particle as $F_{r}^l$ against $l$.  This data is for an eccentric geodesic with energy $E = 0.955492$, spin $a=1/2M$ and angular momentum $L=3.59656$.  In decreasing slope the above lines represent the unregularised self-force (black), self-force regularised by subtracting from it in turn the cumulative sum of $F^l_{r[-1]}$ (blue), $F^l_{r[0]}$ (red), $F^l_{r\lpow{2}}$ (yellow), $F^l_{r\lpow{4}}$ (green). }
\label{fig: kerrRP}
\end{center}
\end{figure}

For the scalar case we were able to use the numerical data from Barack and Warburton \cite{Warburton:Barack:2010}, for calculating the retarded field for Kerr scalar eccentric orbits in the equatorial plane.  We were able to show that our parameters successfully regularised their data as illustrated in Fig.~\ref{fig: kerrRP}.
\begin{table}[tb] 
\begin{center}
 \begin{tabular}{|c|c|c|c|c|c|}
\hline
  Cases & $F_{a\lnpow{1}}$ & $F_{a\fixme{\text{$\lpow{0}$}}}$ & $F_{a\fixme{\text{$\lpow{2}$}}}$ & $F_{a\fixme{\text{$\lpow{4}$}}}$ & $F_{a\fixme{\text{$\lpow{6}$}}}$\\
\hline
\Sch scalar & BO& BO& DMW / HP & HOW & HOW\\
\Sch electromagnetic & BO& BO& HP & HOW & HOW\\
\Sch gravity & BO& BO& HOW  & HOW &---\\
\Sch $h u u$ &---& BO& HOW & HOW &---\\
Kerr scalar & BO& BO & HOW & HOW &---\\
Kerr electromagnetic& BO& BO& HOW &--- &---\\
Kerr gravity & BO& BO & HOW &---  &--- \\
Kerr $h u u$ &---& HOW & HOW &--- &--- \\
\hline
 \end{tabular}
\caption[Regularization Parameters to Date]{This table represents the current regularisation parameters that are known for each case in \Sch and Kerr space-times.  We have indicated which authors first derived the regularisation parameters.  BO is Barack and Ori,  DMW is Detweiler, Messaritaki and Whiting, HP is Haas and Poisson and HOW is Heffernan, Ottewill and Wardell.    DMW and HP share the authorship for \Sch scalar $F_{a\lnpow{2}}$ as DMW produced the first expressions for circular orbits while HP extended this to the elliptic orbits. The HOW results were produced as part of this thesis.}
\label{tab: RPs}
\end{center}
\end{table}

\section{Regularisation Parameters to Date}

In Table~\ref{tab: RPs}, we summarise all the regularisation parameters that are known to date and the authors of each parameter.  We can see that the work of this thesis has greatly added to the already existing data base of parameters, which in turn is dramatically reducing the computation time necessary for accurate predictions of the self-force.

By deriving our expressions using two independent methods we were able to be confident in our results\fixme{.  In} particular cases, this was boosted by the success of the parameters in regularising numerical data for the retarded field.  The alternative approaches used both came with their advantages and disadvantages.  For the lower orders, computation time as well as the personal time of the researcher were very much on par, however, the covariant method produces more `elegant' expressions.   However, when we increase the order of our expansions, the covariant technique becomes more time consuming, both computationally and personally for the researcher.  In comparison, working in coordinates has a `sense' of automation attached to it for the higher orders\fixme{.  S}ome finesse is required for the very high orders when the calculation may get `stuck', but this usually only requires minutes of the researchers time.  Such `blips' usually occur due to the high number of unknown parameters, memory issues and certain formalisms that sufficiently slow our software (Mathematica) down, most of which can be circumnavigated with experience and knowledge of the software.



\chapter{Effective Source} \label{sec:EffectiveSource} 


As another application of our high-order expansions of the Detweiler-Whiting singular field,
we consider its use in the effective source approach to calculating the self-force. The effective
source approach -- independently proposed by Barack and Golbourn
\cite{Barack:Golbourn:2007} and by Vega and Detweiler \cite{Vega:Detweiler:2008} -- relies on
knowledge of the singular field to derive an equation for a regularized field that gives
the self-force without any need for post-processed regularization.  Where the mode-sum method solves for the retarded field, which is in turn regularised by using the singular field after it is numerically calculated, the effective source method numerically calculates the regularised field directly from the wave equation.

There are numerous advantages for this method.  
\begin{itemize}
\item By design, there are no delta functions or singularities - a desirable attribute for numerical calculations.
\item There is no post-processed construction of the regularized field - that is we don't have to cancel two large quantities $\varphi^{A}_{\rm{\ret}}$ and $\varphi^{A}_{\rm{\sing}}$ which when carried out numerically can lead to large round-off errors.
\item It does not rely on the separability of the perturbation equations, an advantage for time-domain calculations in Kerr space-time where the perturbation equations are not fully separable.
\end{itemize}

\section{Methods}

Following directly from Eq.~\eqref{eqn: singRegSplitPhi}, we have
\begin{equation} 
\mathcal{D^A{}_B} \varphi^{B}_{\rm{\reg}} = \mathcal{D^A{}_B}\varphi^{B}_{\rm{\ret}} - \mathcal{D^A{}_B} \varphi ^{B}_{ \rm{\sing}},
\end{equation}
where the definition of the singular field and Eq.~\eqref{eq:Wave} gives us
\begin{equation}\label{eqn: EffectiveSourceWave}
\mathcal{D}^{A}{}_B \varphi^{B}_{\rm{\reg}}  = - 4\pi \mathcal{Q} \int u^{A} \delta_4 \left( x,z(\tau') \right) d\tau' - \mathcal{D} \varphi ^{A}_{ \rm{\sing}} = 0
\end{equation}
However, as the singular field is only well-defined in the neighbourhood of the particle, it should be noted that Eq.~\eqref{eqn: EffectiveSourceWave} is also only valid in this neighbourhood.  Therefore a similar equation is required for when we are not in this region.  The two methods introduced to do this are the window function as first described by Vega and Detweiler \cite{Vega:Detweiler:2008} and the world tube, which was proposed by Barack and Golbourn \cite{Barack:Golbourn:2007}.


\subsection{World Tube Method}

The world tube method resolves the problem of lack of global definition of the singular field by introducing a world tube - inside the world tube one solves for the regular field, $\varphi^{B}_{\rm{\reg}}$, and outside the world tube one solves for the retarded field, $\varphi^{B}_{\rm{\reg}}$. If we consider the boundary of this world tube to be at $\tilde{x}$ and world line at $x_0$, we then have
\begin{align} \label{eqn: SeffRegions}
\mathcal{D}^{A}{}_B \varphi^{B}_{\rm{\reg}} =& - 4\pi \mathcal{Q} \int u^{A} \delta_4 \left( x,z(\tau') \right) d\tau' - \mathcal{D} \varphi ^{A}_{ \rm{\sing}} = 0, \quad \quad |x-x_0| < \tilde{x} \nonumber \\
\mathcal{D}^{A}{}_B \varphi^{B}_{\rm{\ret}} = & 0, \quad \quad |x-x_0| > \tilde {x} 
\end{align}
where the matching condition $\varphi^{A}_{\rm{\reg}} = \varphi^{A}_{\rm{\ret}} -\varphi ^{A}_{ \rm{\sing}}$ is imposed on the boundary.  As the use of the singular field is now restricted to the region near the particle, its non-global definition is no longer a problem.

\subsection{Window Function Method}
The window function method involves the use of a globally defined \emph{approximate} singular field, $\tilde{\varphi} ^{A}_{ \rm{\sing}}$.  This is obtained by the use of a smooth window function, $W$, that is chosen \fixme{so} that at the particle, $\tilde{\varphi} ^{A}_{ \rm{\sing}}$ is given by the exact singular field, while away from the particle $\tilde{\varphi} ^{A}_{ \rm{\sing}}$ quickly diminishes and only the retarded field remains on and past the boundary, $\tilde{x}$, i.e., $\tilde{\varphi} ^{A}_{ \rm{\sing}} = W \varphi ^{A}_{ \rm{\sing}}$.  In this manner, the world function method unites the two regions of the world tube method avoiding the use of two seperate computational domains.  The definition of an approximate singular field, gives a new definition of an approximate regular field, $\tilde{\varphi}^{A}_{\rm{\reg}} = \varphi^{A}_{\rm{\ret}} - \tilde{\varphi} ^{A}_{ \rm{\sing}}$, and with it, a slightly altered wave equation,
\begin{equation} \label{eqn: SeffWF}
\mathcal{D}^{A}{}_B \tilde{\varphi}^{B}_{\rm{\reg}}  = - 4\pi \mathcal{Q} \int u^{A} \delta_4 \left( x,z(\tau') \right) d\tau' - \mathcal{D^A{}_B} \left( W \varphi ^{B}_{ \rm{\sing}} \right)
\end{equation}
Due to the nature of the window function, Eq.~\eqref{eqn: SeffWF} is not restricted to a certain region of space unlike Eqs.~\eqref{eqn: EffectiveSourceWave} and \eqref{eqn: SeffRegions}.

By design, the window function will have certain restrictions -  the resulting approximate regular field must give the correct self-force, the approximate singular field must be equal to the exact singular field at the particle and equal to zero far from the particle.  These give
\begin{itemize}
\item $W = 1 + f$, where $f = \mathcal{O} \left(\epsilon^n\right)$ and $p^a{}_A f = \mathcal{O} \left( \epsilon^{n-1} \right)$  where $p^a{}_A$ is that of Eq.~\eqref{eqn: FaCases}.
\item $W$ is smooth
\item $W= 0$ for $x > \tilde{x}$
\end{itemize}
where $n$ is an integer $> 2$.  This lower bound comes from the equation defining the self-force as
\par \vspace{-6pt} \begin{IEEEeqnarray}{rCl}
F^a  &=& p^{a}{}_A \tilde{\varphi}^{A}_{\rm{\reg}} \nonumber \\
&=& p^{a}{}_A \left( \varphi^{A}_{\rm{\ret}}  - \varphi^{A}_{\rm{\sing}} \right) - \left( p^{a}{}_A \fixme{\text{$f$}} \right)\varphi^{\text{\fixme{$A$}}}_{\rm{\sing}} - \fixme{\text{$f$}} \left(p^{a}{}_A \varphi^{\text{\fixme{$A$}}}_{\rm{\sing}}  \right) \nonumber \\
&\rightarrow&  p^{a}{}_A \left( \varphi^{A}_{\rm{\ret}}  - \varphi^{A}_{\rm{\sing}} \right) \quad \quad \text{as} \quad \quad x\rightarrow x_0
\end{IEEEeqnarray}
where the last equality holds at the particle which requires the last two terms of the second equality $\rightarrow 0$ as $\epsilon \rightarrow 0$.

As stated before, one of the key advantages of the effective source is its avoidance of $\delta$ functions  - this can easily be shown when considering the window function approach.   The effective source is defined by,
\par \vspace{-6pt} \begin{IEEEeqnarray}{rCl} \label{eqn: effSource}
S_{\rm eff}^{A} &=& \mathcal{D}^{A}{}_B \tilde{\varphi}^{B}_{\rm{\reg}} \nonumber \\
&=& -\mathcal{D}^{A}{}_B \tilde{\varphi}^{B}_{\rm \sing} - 4\pi \mathcal{Q} \int u^{A} \delta_4 \left( x,z(\tau') \right) d\tau' \nonumber \\
&=& -\mathcal{D}^{A}{}_B \left(\fixme{\text{$f$}}\varphi^{\fixme{\text{$B$}}}_{\rm \sing} \right) \nonumber \\
&=& -\left(\Box\fixme{\text{$f$}} \right) \varphi^{\fixme{\text{$A$}}}_{\rm \sing} - 2 \nabla_a\fixme{\text{$f$}} \nabla^a \varphi^{\fixme{\text{$A$}}}_{\rm \sing}  - \fixme{\text{$f$}} \Box \varphi^{\fixme{\text{$A$}}}_{\rm \sing} \nonumber \\
&=& - \left(\Box \fixme{\text{$f$}} \right) \varphi^{\fixme{\text{$A$}}}_{\rm \sing} - 2 \nabla_a \fixme{\text{$f$}} \nabla^a \varphi^{\fixme{\text{$A$}}}_{\rm \sing} 
\end{IEEEeqnarray}
where the last two equalities are for a vacuum space-time, i.e., $\mathcal{D}^{A}{}_B = \delta^A{}_B \Box$ and we take advantage of $\fixme{\text{$f$}}$ being zero at the particle and $\Box \varphi^{A}_{\rm \sing}$ begin zero away from the particle.


\section{Effective Source from the Singular Field}

The effective source, defined by Eq.~\eqref{eqn: effSource},
 for sufficiently good approximations gives
\begin{equation} \label{eqn: SeffEqn}
S_{\rm eff}^{A} = \begin{cases}
             		0 & \text{(at the particle)}\\
             		-\mathcal{D}^{A}{}_B \tilde{\varphi}^{B}_{\rm \sing}  & \text{(away from the 		
			particle)} 
            		\end{cases},
\end{equation}
where \fixme{`}at the particle\fixme{'} follows from $\fixme{\text{$f$}}=0$ \fixme{in Eq.~\eqref{eqn: effSource}} (and the fact that $\varphi^{B}_{\rm \sing}$ solves the same inhomogeneous wave equation as $\varphi^{B}_{\rm \ret} $) \fixme{. `A}way from the particle\fixme{'} holds as $\mathcal{D}^{A}{}_B \varphi^{B}_{\rm \ret} =0$ \fixme{in the first line of Eq.~\eqref{eqn: effSource}}.

If the singular field is known exactly, then the regularized field is totally regular and is a solution of the homogeneous wave equation. In reality, exact expressions for the singular field can only be obtained for very simple space-times. More generally, the best one can do is an approximation such as that given in Sec.~\ref{sec:SingularFieldExpansion}, which we now define as $\tilde{\varphi}^{A}_{\rm \sing}$, so we have
\begin{equation}
\tilde{\varphi}^{A}_{\rm \sing} = \left[ 1 + \mathcal{O} \left( \epsilon \right)^n\right] \varphi^{A}_{\rm \sing} 
\end{equation}
which satisfies the above conditions for $W$.  Our calculated high-order expansion of the singular field can, therefore, be placed into Eq.~\eqref{eqn: SeffEqn} as the approximate singular field.

\begin{figure}
\begin{center}
\includegraphics[width=15cm]{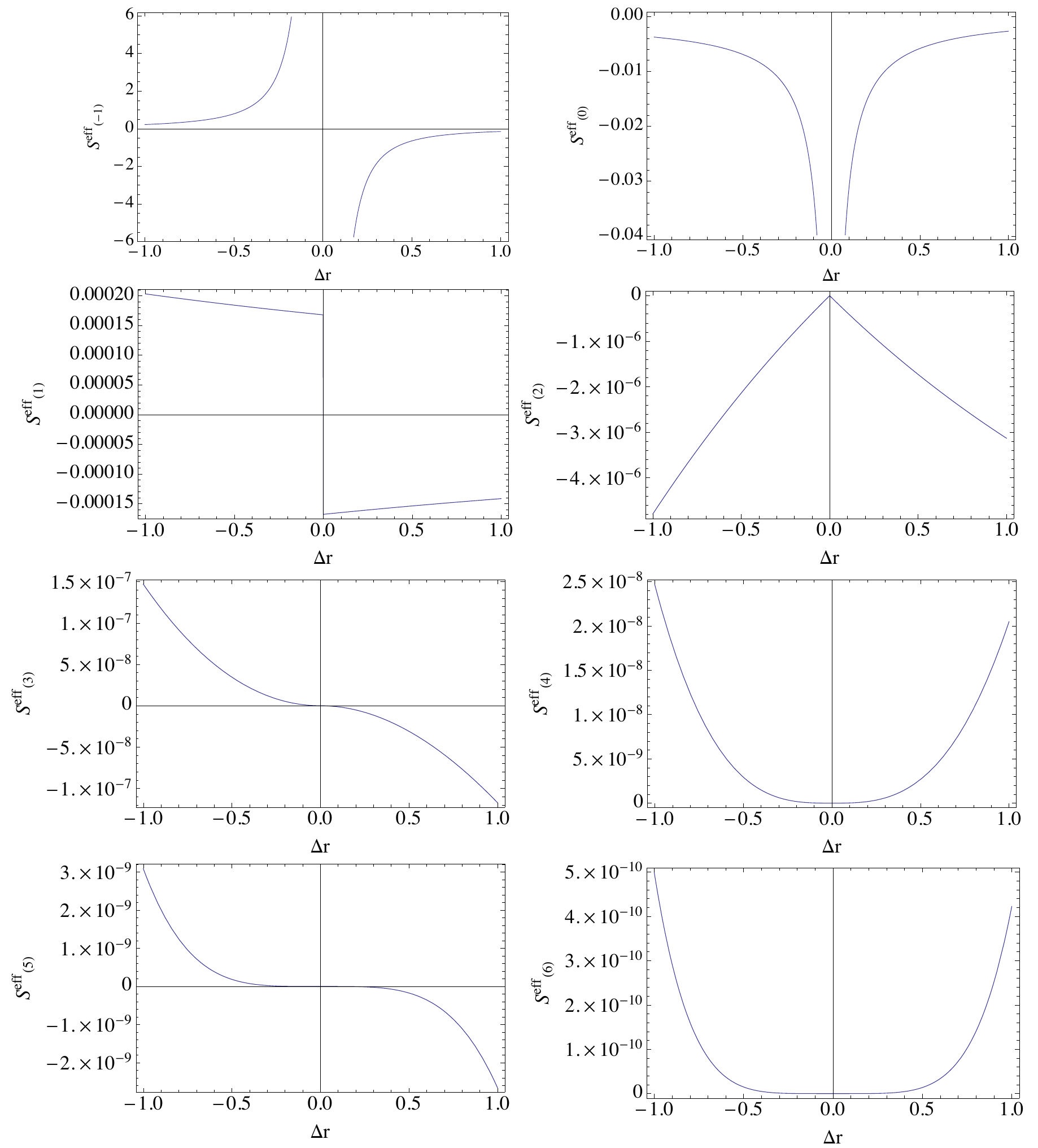}
\caption[Effective Source Components in 2D]{Effective source for the approximate singular field of order
$\mathcal{O}(\epsilon^{-1})$ (top left) to $\mathcal{O}(\epsilon^{6})$ (bottom right).
Shown is the scalar case of a circular geodesic of radius $r_0 = 10M$ in \Sch space-time.}
\label{fig:EffectiveSource1D}
\end{center}
\end{figure}
\begin{figure}
\begin{center}
\includegraphics[width=15cm]{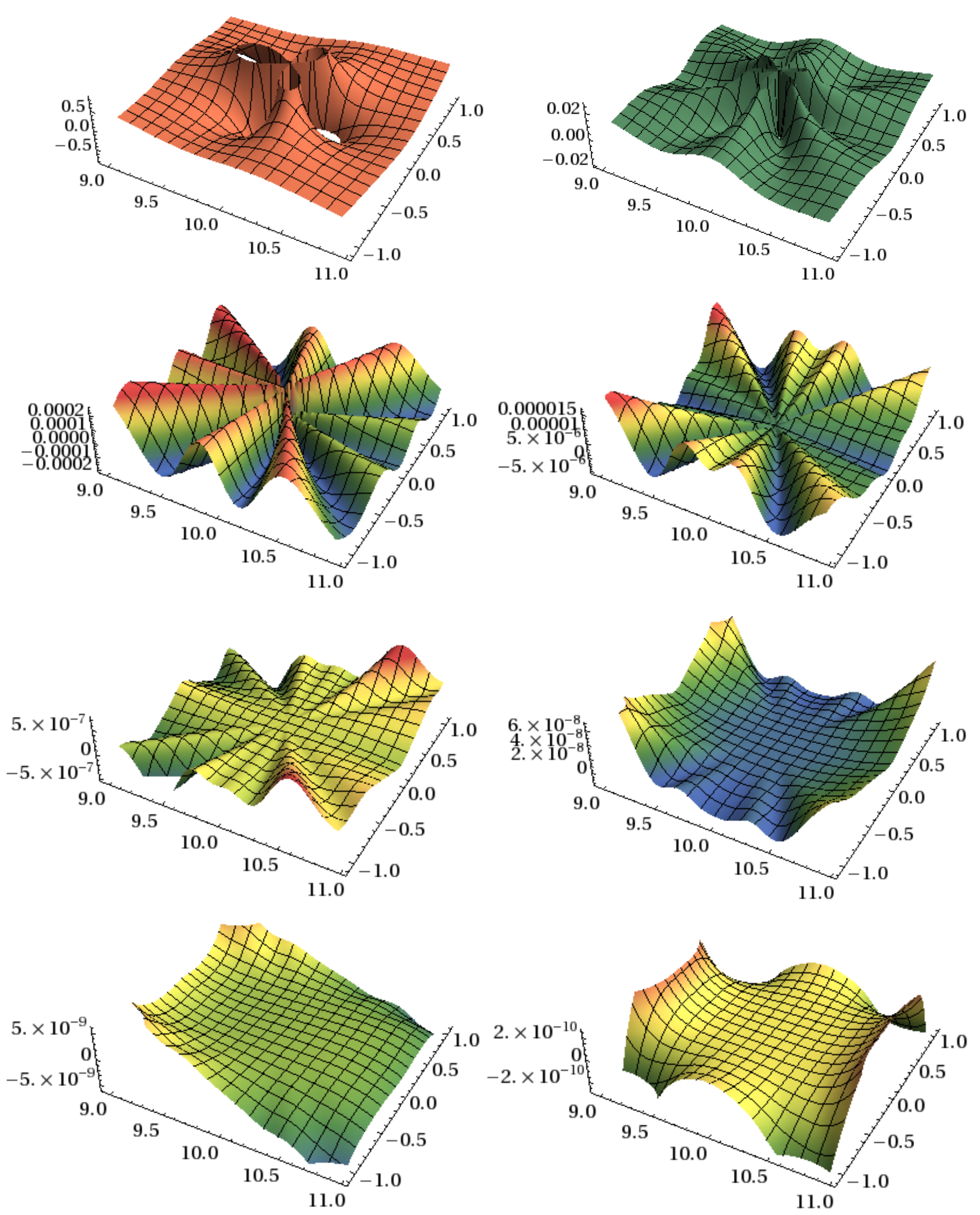}
\caption[Effecitive Source Components in 3D]{Effective source for the approximate singular field, $S^{\rm eff}{}_{(n)}$, in the region of the particle of order
$\mathcal{O}(\epsilon^{-1})$ (top left) to $\mathcal{O}(\epsilon^{6})$ (bottom right).
Shown is the scalar case of a circular geodesic of radius $r_0 = 10M$ in \Sch space-time.}
\label{fig:EffectiveSource2D}
\end{center}
\end{figure}
In Figs.~\ref{fig:SingularField1D},\ref{fig:SingularField2D}, \ref{fig:EffectiveSource1D}
and \ref{fig:EffectiveSource2D} we show the result of applying our expansions
to the case of a scalar particle on a circular geodesic of radius $r_0 = 10M$ in
\Sch space-time. Similar plots can be obtained for the electromagnetic case, gravitational case
and for more generic motion. However, the general structure does not change and is best illustrated
by this simple example.


\section{$m$-mode Scheme}
One disadvantage of the effective source approach stems from the fact that the
source must be evaluated in an extended region around the world-line. Since the
source is derived from a complicated expansion approximating the singular field,
its evaluation can dominate the run time of a numerical code. This problem is
exasperated as increasingly good approximations to the singular field --- using
increasingly high order series expansions (such as those in Sec.~\ref{sec:SingularFieldExplicit}) --- are used, placing a practical
upper limit on the order of singular field approximation which may be used in
effective source calculations. Existing calculations
\cite{Dolan:Barack,Dolan:Barack:Wardell,Diener:Vega:Wardell:Detweiler} settled
on what appears to be a ``sweet spot'', using an approximation accurate to
$\mathcal{O}(\epsilon^2)$.

This may appear to rule out the usefulness of high-order expansions of the
singular field in effective source calculations, particularly in the case of the
Kerr space-time where even an order $\mathcal{O}(\epsilon^2)$ approximation to
the singular field is quite unwieldy. However, it turns out that high-order
expansions can, in fact, be put to good use in effective source calculations. In
this section, we show how this may be achieved in the case of the $m$-mode
approach to effective source calculations. In this approach, one first performs
a decomposition into $m$-modes
\begin{equation}
\label{eq:m-decomposition}
\Phi^{(m)} = \frac{1}{2\pi}\int_{-\pi}^{\pi} \Phi e^{-i m \phi} d\phi,
\end{equation}
and independently evolves the $m$-decomposed form of the wave equation for each
$m$-mode. These equations have an $m$ dependent effective source which is
derived from the particular choice of approximation to the singular field. The
full field is then given as a sum of these individual modes:
\begin{equation}
\Phi = \sum\limits_{m=-\infty}^{\infty} \Phi^{(m)} e^{i m \phi_0}.
\end{equation}

For an approximation accurate to $\mathcal{O}(\epsilon^n)$, the numerical
solutions for the field fall off as $m^{-(n+2)}$ for $m$ even and as
$m^{-(n+3)}$ for $m$ odd. Obviously, only finitely many $m$-modes (typically
$\sim10-20$) can ever be computed numerically; with the error from truncating
the sum at a finite $m$ putting an upper limit on the accuracy of the self-force
which can be computed. This may be mitigated somewhat by fitting for a large-$m$
tail, but that fit itself requires more modes and is only ever approximate.
Here, we propose a much better solution, using the higher-order terms in the
singular field (those which have not been used in computing the effective
source) to analytically deriving expressions for the tail. In many ways, this is
analogous to the $l$-mode regularization scheme where there is a large-$l$ tail
and one can compute $l$-mode regularization parameters.

\subsection{Derivation of $m$-mode regularization parameters}
To derive analytic expressions for the large-$m$ tail, we first note that an
approximation to the singular field accurate to $\mathcal{O}(\epsilon^n)$ can be
written in the form
\begin{align}
\label{eq:full-phis-m-mode}
\Phi^{\rm S} (x)
  =& \frac{1}{\rho_0^{2n+3}} \Big[ \sum\limits_{i=0\atop i~\text{even}}^{3(n+1)} A_{n,i} \sin^i(\Delta\phi/2) + \sum\limits_{i=0\atop i~\text{odd}}^{3(n+1)} A_{n,i} \sin^{i-1}(\Delta\phi/2) \sin(\Delta\phi) \Big] + \mathcal{O}(\epsilon^{n+1})
\nonumber \\
  =& \frac{1}{\rho_0^{2n+3}} \Big[ \sum\limits_{i=0\atop i~\text{even}}^{3(n+1)} A_{n,i} \sin^i(\Delta\phi/2) + 2 \sum\limits_{i=0\atop i~\text{odd}}^{3(n+1)} A_{n,i} \sin^{i}(\Delta\phi/2) \cos(\Delta\phi/2) \Big]  \nonumber \\
&+ \mathcal{O}(\epsilon^{n+1})
\end{align}
where the coefficients $A_{n,i}$ are functions of position $r_0$ and $\theta_0$,
of the constants of motion $E$, $L$, $C$, and also of $\Delta r$ and $\Delta
\theta$. This form has the benefit of ensuring that the approximation is regular
everywhere except on the world-line, while still being amenable to analytic
integration in the $\phi$ direction. This makes it particularly appropriate for
use in $m$-mode effective source calculations \cite{Thornburg:Wardell}.

Using the leading orders (say, to $\mathcal{O}(\epsilon^p)$) in this expansion
to compute an effective source, one is left with a singular field remainder
which is finite, but of limited differentiability on the world-line. Since it is
finite, we can safely set $\Delta r = \Delta \theta = 0$ in
Eq.~\eqref{eq:full-phis-m-mode}, leading to a singular field remainder which has
the form
\begin{align}
\label{eq:remainder-phis-m-mode}
\Phi^{\rm S} (x)
  =& \Big[2\Theta(\Delta\phi)-1\Big] \Big[ \sum\limits_{i=p+1\atop i~\text{odd}}^{n} B_{n,i} \sin^i(\Delta\phi/2) + 2 \sum\limits_{i=p+1\atop i~\text{even}}^{n} B_{n,i} \sin^{i}(\Delta\phi/2) \cos(\Delta\phi/2) \Big]  \nonumber \\
&+ \mathcal{O}(\epsilon^{n+1}),
\end{align}
where $\Theta(\Delta\phi)$ is the Heaviside step function. Substituting this
into Eq.~\eqref{eq:m-decomposition} and noting that for even $j$
\begin{eqnarray}
\int_{-\pi}^{\pi} \Big[2\Theta(\Delta\phi)-1\Big] \sin^j(\Delta\phi/2) \cos(\Delta\phi/2)  e^{-i m \phi} d\phi  \qquad \qquad \qquad \nonumber \\
 =\frac{2 i m}{j+1} \int_{-\pi}^{\pi} \Big[2\Theta(\Delta\phi)-1\Big] \sin^{j+1}(\Delta\phi/2) e^{-i m \phi} d \phi
\end{eqnarray}
we are left with trivial integrals of the form
\begin{eqnarray}
\int_{-\pi}^{\pi} \Big[2\Theta(\Delta\phi)-1\Big] \sin^{j+1}(\Delta\phi/2) e^{-i m \phi} d \phi  \qquad \qquad \qquad \nonumber \\
= \int_{-\pi}^{\pi} \Big[2\Theta(\Delta\phi)-1\Big] \sin^{j+1}(\Delta\phi/2) \cos(m \phi) d \phi.
\end{eqnarray}
As a result, we see that the real regularization parameters are given by the odd
terms in the expansion of the singular field and the imaginary parameters are
given by the the even terms. Furthermore, we see that the falloff with $m$ is
always an even power of $1/m$ in the real part and an odd power of $1/m$ in the
imaginary part.

While this analysis was done for the field, it should be noted that it equally
well applies to the self-force. The only modification necessary is to compute
the self-force from the singular field before setting $\Delta r = \Delta \theta
= 0$; the remainder of the calculation proceeds in exactly the same way.

Finally, we note the $m$-mode regularization parameters derived in this way are
dependent on the singular field being written in the form given in
Eq.~\eqref{eq:full-phis-m-mode}. Effective source calculations may use some other
form for the approximation to the singular field (while still being accurate to
the same order), in which case there is no guarantee that the regularization
parameters given here are appropriate.


\subsection{$m$-mode regularization parameters}
Below, we give the results of applying this calculation to both scalar and
gravitational cases in \Sch and Kerr space-times. In doing so, we omit the
explicit dependence on $m$ which in each case is
\begin{gather}
F^m_{a[2]} = \frac{-4 F_{a[2]}}{\pi(2m-1)(2m+1)}, \quad
F^m_{a[4]} = \frac{24 F_{a[4]}}{\pi(2m-3)(2m-1)(2m+1)(2m+3)}, \nonumber \\
F^m_{a[6]} = \frac{-480 F_{a[6]}}{\pi(2m-5)(2m-3)(2m-1)(2m+1)(2m+3)(2m+5)}, \nonumber \\
F^m_{a[8]} = \frac{20160F_{a[8]}}{\pi(2m-7)(2m-5)(2m-3)(2m-1)(2m+1)(2m+3)(2m+5)(2m+7)}.
\end{gather}

\subsubsection*{\Sch $m$-modes}
For \Sch eccentric orbits, the $m$-modes of the radial component of the self-force is given by
\par \vspace{-6pt}\begin{IEEEeqnarray}{rCl}
F_{r [2]} &=& \frac{1}{ 24 \rb^6 \left( \rb - 2 M \right) \left( L^2 + \rb^2\right)^{7/2}} \bigg[
	2
   \left(L^2+\rb^2\right) (\rb-2 M) \big(12 L^8 M+47 L^6 M \rb^2 \nonumber \\
&& \qquad
	+\: 67
   L^4 M \rb^4+12 L^4 \rb^5 
	+ 92 L^2 M \rb^6-9 L^2 \rb^7+24 M
   \rb^8-3 \rb^9\big) \nonumber \\
&& \quad
	+\: E^2 \rb^3
   \big(-48 L^8 M-178 L^6 M \rb^2-140 L^4 M \rb^4-66 L^4 \rb^5-172
   L^2 M \rb^6 \nonumber \\
&& \qquad
	-\: 6 L^2 \rb^7 
	-72 M \rb^8+15 \rb^9\big) \nonumber \\
&& \quad
	\:  -9 E^4 \rb^{10} (\rb-2 L) (2 L+\rb) \bigg],
\end{IEEEeqnarray}

\par \vspace{-6pt}\begin{IEEEeqnarray}{rCl}
F_{r [4]} &=& \frac{1}{ 480 \rb^13 \left( \rb - 2 M \right) \left( L^2 + \rb^2\right)^{11/2}} \bigg[ 
-2 \left(L^2+\rb^2\right)^2
   (\rb-2 M) \big(19200 L^{16} M^2 \nonumber \\
&& \qquad
	+\: 52096 L^{14} M^2 \rb^2+24640 L^{14} M
   \rb^3-35744 L^{12} M^2 \rb^4+126560 L^{12} M \rb^5 \nonumber \\
&& \qquad
	-\: 294960 L^{10}
   M^2 \rb^6+266320 L^{10} M \rb^7-452392 L^8 M^2 \rb^8+292420 L^8 M
   \rb^9 \nonumber \\
&& \qquad
	-\: 323040 L^6 M^2 \rb^{10}+175360 L^6 M \rb^{11}-112088 L^4 M^2
   \rb^{12}+55100 L^4 M \rb^{13} \nonumber \\
&& \qquad
	-\: 675 L^4 \rb^{14}-11520 L^2 M^2
   \rb^{14}+3920 L^2 M \rb^{15}+270 L^2 \rb^{16}-240 M
   \rb^{17} \nonumber \\
&& \qquad
	+\: 45 \rb^{18}\big) \nonumber \\
&& \quad
	+\: E^2 \rb^3
   \left(L^2+\rb^2\right) \big(38400 L^{16} M^2-30464 L^{14} M^2
   \rb^2+115840 L^{14} M \rb^3 \nonumber \\
&& \qquad
	-\: 820032 L^{12} M^2 \rb^4+622784 L^{12}
   M \rb^5-2307616 L^{10} M^2 \rb^6 \nonumber \\
&& \quad
	+\: 1379728 L^{10} M \rb^7-2983600
   L^8 M^2 \rb^8+1607760 L^8 M \rb^9 \nonumber \\
&& \qquad
	-\: 2040472 L^6 M^2 \rb^{10}+1031608
   L^6 M \rb^{11}+1350 L^6 \rb^{12}-736872 L^4 M^2 \rb^{12} \nonumber \\
&& \qquad
	+\: 369188
   L^4 M \rb^{13}-5895 L^4 \rb^{14}-97392 L^2 M^2 \rb^{14}+45676 L^2
   M \rb^{15} \nonumber \\
&& \qquad
	-\: 540 L^2 \rb^{16}+960 M^2 \rb^{16}-1680 M
   \rb^{17}+405 \rb^{18}\big) \nonumber \\
&& \quad
	-\: E^4 \rb^8 \big(55040 L^{14} M+310912
   L^{12} M \rb^2+726176 L^{10} M \rb^4+896032 L^8 M \rb^6 \nonumber \\
&& \qquad
	+\: 609740 L^6
   M \rb^8+3240 L^6 \rb^9+233848 L^4 M \rb^{10}-5580 L^4
   \rb^{11} \nonumber \\
&& \qquad
	+\: 39366 L^2 M \rb^{12}-3555 L^2 \rb^{13}-1020 M
   \rb^{14}+540 \rb^{15}\big) \nonumber \\
&& \quad
	+\: 225 E^6 \rb^{19} \left(8 L^4-12 L^2
   \rb^2+\rb^4\right)
\bigg].
\end{IEEEeqnarray}

\subsubsection*{Kerr $m$-modes}
As the expressions for generic orbits of the Kerr spacetime are too large to be
of use in printed form we give here only the results for the case of a circular
geodesic orbit (in which case only the parameter for the $r$ component of the
self-force is non-zero) and direct the reader online \cite{m-mode-online} for
more generic expressions in electronic form.

For circular orbits, the scalar $m$-mode regularization parameters are:
\par \vspace{-6pt}\begin{IEEEeqnarray}{rCl}
F_{r [2]} &=& 
   \frac{M}{24 \rb^4 [a M+\rb \sqrt{M \rb}] [2 a \sqrt{M \rb}+\rb (\rb-3 M)]^{3/2} [a^2+\rb (\rb-2 M)]^{3/2}} \nonumber \\ 
&& 
   \times \: \bigg[24 a^7 M^2 - 24 a^6 M \sqrt{M \rb} (M-2 \rb) -4 a^5 M \rb (23 M^2+M \rb-6 \rb^2) \nonumber \\
&& \quad
	+\: 2 a^4 M \rb \sqrt{M \rb} (45 M^2-112 M \rb+31\rb^2)
   +2 a^3 M \rb^2 (45 M^3+45 M^2 \rb \nonumber \\
&& \qquad
	-\: 73 M \rb^2+19 \rb^3) -3 a^2 \rb^2 \sqrt{M \rb} (29 M^4-88 M^3 \rb+38 M^2 \rb^2-4 M \rb^3 \nonumber \\
&& \qquad
	+\: \rb^4)
    -6 a M \rb^4 (29 M^3-43 M^2 \rb+21 M \rb^2-3 \rb^3) -3 \rb^5 \sqrt{M \rb} (29 M^3 \nonumber \\
&& \qquad
	-\: 25 M^2\rb+3 M \rb^2+\rb^3) \bigg],
\end{IEEEeqnarray}
\par \vspace{-6pt}\begin{IEEEeqnarray}{rCl}
F_{r [4]} &=& 
   \frac{M^2}{1440 \rb^{9} [a M + \rb \sqrt{M \rb}][2 a \sqrt{M \rb}+\rb (\rb-3 M)]^{7/2} [a^2+\rb (\rb-2 M)]^{3/2} } \nonumber \\ 
&& 
   	\times \: \bigg[-23040 a^{14} M^2 \sqrt{M\rb}
    +\:11520 a^{13} M^2 \rb (M-8 \rb) \nonumber \\
&& \quad
	+\: 384 a^{12} M \rb \sqrt{M \rb}(461 M^2-81 M \rb-360 \rb^2) -192 a^{11} M \rb^2(307 M^3 \nonumber \\
&& \qquad
    	-\: 3780 M^2 \rb+1233 M \rb^2+480 \rb^3)-64 a^{10} \rb^2 \sqrt{M \rb} (8549 M^4 \nonumber \\
&& \qquad
	-\: 3593 M^3 \rb-16212 M^2 \rb^2
    + 6336 M\rb^3+360 \rb^4) +32 a^9 M \rb^3 (2835 M^4 \nonumber \\
&& \qquad
	-\: 69401 M^3 \rb+46565 M^2 \rb^2+13779 M \rb^3-8748 \rb^4) \nonumber \\
&& \quad
    	+\: 192 a^8 \rb^3 \sqrt{M \rb}(4470 M^5-3621 M^4 \rb-15645 M^3 \rb^2+12662 M^2 \rb^3 \nonumber \\
&& \qquad
	-\: 1529 M \rb^4-342 \rb^5)
    + 16 a^7 \rb^4 (-1479 M^6+210966 M^5\rb-224760 M^4 \rb^2 \nonumber \\
&& \qquad
	-\: 49213 M^3 \rb^3+93619 M^2 \rb^4-19953 M \rb^5+180 \rb^6) \nonumber \\
&& \quad
    	-\:  16 a^6 \rb^4 \sqrt{M \rb} (43101 M^6-61443 M^5\rb-271980 M^4 \rb^2+343776 M^3 \rb^3 \nonumber \\
&& \quad
	-\: 100489 M^2 \rb^4-7763 M \rb^5
    +4158 \rb^6)+12 a^5 \rb^5 (-2367 M^7-221220 M^6\rb \nonumber \\
&& \qquad
	+\: 337457 M^5 \rb^2+71894 M^4 \rb^3-262111 M^3 \rb^4+111498 M^2 \rb^5 \nonumber \\
&& \qquad
    - \: 14459 M \rb^6+588 \rb^7) +3 a^4 \rb^5 \sqrt{M \rb}(76125 M^7-176307 M^6 \rb \nonumber \\
&& \qquad
	-\: 1157559 M^5 \rb^2+1949709 M^4 \rb^3
    -855873 M^3 \rb^4-26505 M^2 \rb^5 \nonumber \\
&& \qquad
	+\: 76235 M \rb^6-8065 \rb^7)+12 a^3\rb^7 (76125 M^7-152637 M^6 \rb-93174 M^5 \rb^2 \nonumber \\
&& \qquad
    	+\:  281414 M^4 \rb^3-166063 M^3 \rb^4+36555 M^2 \rb^5-3480 M \rb^6+460\rb^7) \nonumber \\
&& \quad
	 +\: 18 a^2 \rb^8 \sqrt{M \rb} (76125 M^6
    -145182 M^5 \rb+70771 M^4 \rb^2+16696 M^3 \rb^3 \nonumber \\
&& \qquad
	-\: 19905 M^2 \rb^4+3190 M \rb^5+225 \rb^6) +36 a \rb^{10} (25375 M^6
    - 47369 M^5 \rb \nonumber \\
&& \qquad
	+\: 31856 M^4 \rb^2-8692 M^3 \rb^3+705 M^2 \rb^4-75 M \rb^5+40 \rb^6) \nonumber \\
&& \qquad
    	+\:  9 \rb^{11} \sqrt{M\rb} (25375 M^5-47015 M^4 \rb+29014 M^3 \rb^2-4814 M^2 \rb^3 \nonumber \\
&& \qquad
	-\: 1365 M \rb^4+405 \rb^5)
   \bigg].
\end{IEEEeqnarray}

The gravitational, $m$-mode parameters for $H$ are
\begin{IEEEeqnarray}{rCl}
H_{[2]} &=&
  \frac{M^{3/2} \left[2 a \sqrt{M \rb}+\rb (\rb-3 M)\right]^{-1/2}}{12
   \rb^{7/2} \left(a M+\sqrt{M\rb^3}\right) \left[a^2+\rb (\rb-2 M)\right]^{1/2}} 
\bigg[44
   a^4 M+88 a^3 \sqrt{M \rb^3} \nonumber \\ 
&& \quad -\: 3 a^2 \rb (M-\rb) (29 M+15 \rb)+6 a \sqrt{M \rb^5} (14 \rb-29 M)-87 M \rb^4 \nonumber \\
&& \quad +\: 45 \rb^5\bigg],
\end{IEEEeqnarray}
and
\begin{IEEEeqnarray}{rCl}
  H_{[4]} &=& \frac{M^{3/2}}{720 \rb^{15/2} (a
   M^{1/2}+\rb^{3/2}) (2 a (M \rb)^{1/2}+\rb (\rb-3 M))^{7/2} (a^2+\rb (\rb-2 M))^{1/2}} 
\nonumber \\
&&
  \times\:\Big[13824 a^{12} M^{5/2}+6912 a^{11} M^2 \sqrt{\rb} (3 M+8 \rb)-64 a^{10} M^{3/2} \rb (2249 M^2 \nonumber \\
&&
	\qquad -\: 2484 M \rb-1296 \rb^2)-64 a^9 M \rb^{3/2} (1920 M^3+8383 M^2 \rb-6777 M \rb^2 \nonumber \\
&&
	\qquad -\: 864 \rb^3)+48 a^8 M^{1/2} \rb^2 (11005 M^4-21094 M^3 \rb-12019 M^2 \rb^2\nonumber \\
&&
	\qquad +\: 11664 M \rb^3+288 \rb^4)+64 a^7 M \rb^{5/2} (3879 M^4+30408 M^3 \rb\nonumber \\
&&
	\qquad-\: 44007 M^2 \rb^2+2981 M \rb^3+5562 \rb^4)+4 a^6 M^{1/2} \rb^3 (-208989 M^5\nonumber \\
&&
	\qquad +\: 544428 M^4 \rb+483978 M^3 \rb^2-880476 M^2 \rb^3+209395 M \rb^4+24192 \rb^5) \nonumber \\
&&
	\quad -\: 12 a^5 \rb^{7/2} (14247 M^6+263427 M^5 \rb-515490 M^4 \rb^2+95446 M^3 \rb^3\nonumber \\
&&
	\qquad +\: 154187 M^2 \rb^4-52041 M \rb^5-432 \rb^6)+3 a^4 M^{1/2} \rb^4 (163125 M^6\nonumber \\
&&
	\qquad -\: 528642 M^5 \rb-1218021 M^4 \rb^2+2583348 M^3 \rb^3-1176005 M^2 \rb^4\nonumber \\
&&
	\qquad +\: 7790 M \rb^5 + 57445 \rb^6)+12 a^3 \rb^{11/2} (163125 M^6-386172 M^5 \rb \nonumber \\
&&
	\qquad +\: 19074 M^4 \rb^2+335148 M^3 \rb^3-201235 M^2 \rb^4+31920 M \rb^5 + 860 \rb^6) \nonumber \\
&&
	\quad +\: 18 a^2 M^{1/2} \rb^7 (163125 M^5-337377 M^4 \rb+197294 M^3 \rb^2-2450 M^2 \rb^3 \nonumber \\
&&
	\qquad -\: 28315 M \rb^4+6075 \rb^5)+36 a \rb^{17/2} (54375 M^5-103674 M^4 \rb\nonumber \\
&&
	\qquad +\: 71932 M^3 \rb^2-20850 M^2 \rb^3+1725 M \rb^4+140 \rb^5) + 9 \sqrt{M} \rb^{10} (54375 M^4\nonumber \\
&&
	\qquad -\: 97620 M^3 \rb+66074 M^2 \rb^2-20020 M \rb^3+2295 \rb^4)\Big].
\end{IEEEeqnarray}


\subsection{Example - Kerr Scalar Self-Force}

As an example application of these $m$-mode regularization parameters, we consider the case of a scalar charge, on a circular geodesic orbit of radius $10M$, in the Kerr space-time with $a=0.6M$. The self-force, in this case, was computed in \cite{Thornburg:Wardell} using the $m$-mode effective source approach, with an effective source derived from an approximation to the singular field of the form \eqref{eq:full-phis-m-mode} accurate to $\mathcal{O}(\epsilon^2)$.   As expected, this gave numerical results for the $m$-modes of the self-force which asymptotically fall off as $m^{-4}$. In this case, the $F^m_{r[2]}$ parameter is not needed as it has already been subtracted through the effective source calculation. However, the $F^m_{r[4]}$ parameter has not been subtracted and asymptotically gives the leading order behaviour (in $1/m$) of the modes. Subtracting this from the numerical results, therefore, leaves a remainder which falls off as $m^{-6}$. Furthermore, a numerical fit of this remainder can be done to numerically determine the next two parameters, in this case, giving $F_{r[6]}=0.108797 q^2 / M^2 $ and $F_{r[8]}=11.3398 q^2 / M^2$.

\begin{figure}
\includegraphics[width=14.5cm]{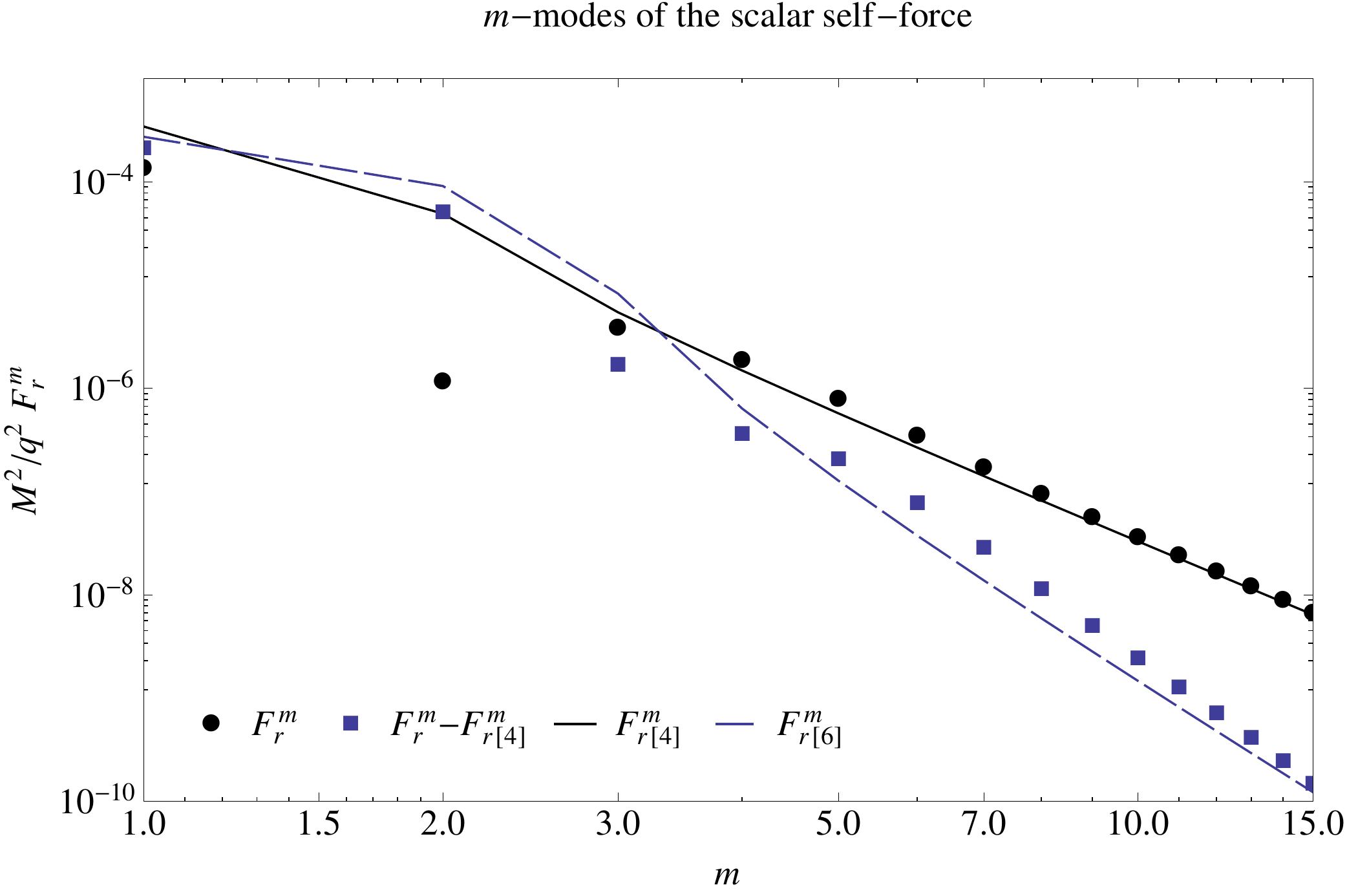}
\caption[$m$-mode Regularization of the Radial Component of the Self-Force]{Regularization of the radial component of the self-force for the case of
a scalar particle on a circular geodesic of radius $r_0 = 10M$ in Kerr space-time with $a=0.6M$.  The numerical self-force modes
asymptotically match the $F^m_{r[4]}$ regularization parameter for
large $m$. After regularization, the remainder fall off as
$m^{-6}$, as expected.}
\label{fig:m-modes}
\end{figure}

In Fig.~\ref{fig:m-modes}, we plot the results of subtracting the $F^m_{r[4]}$ and numerically fitted $F^m_{r[6]}$ regularization parameters, in turn, from the raw numerical data. For large $m$, the numerical data (black dots) falls of as $m^{-4}$, with the coefficient matching our analytic prediction given by $F^m_{r[4]}$ (black line). Subtracting this leading order behaviour, we find that the remainder falls off as $m^{-6}$ (purple squares and line), as expected.

\chapter{Further Extensions}  \label{sec: extensions}

\section{Non-geodesic Motion} \label{sec: nonGeodesicMotion}
Thus far, we have only considered calculating the singular field, and the resulting regularisation parameters, for geodesic motions.  However, as not all situations that occur in the universe will be geodesic, it is, therefore, beneficial to examine other types of motion.

\subsection{Cosmic Censorship Conjecture} \label{sec: ccc}
One of the truely interesting applications of non-geodesic motion self-force calculations is using the back reaction to investigate the possibility of cosmic censorship.  It has long been known that if a sufficient large amount of mass is contained in a small enough region, that gravitational collapse to singularity will occur.  The cosmic censorship conjecture was first proposed by Penrose, \cite{Penrose:1969}: The complete gravitational collapse of a body always results in a black hole rather than a naked singularity; i.e., all singularities of gravitational collapse are `hidden' within black holes, where they cannot be `seen' by distant observers.  Another way of looking at this is that any observer who is sufficiently far away from a singularity, will never encounter a singularity nor see any effect arising from singularities.  The issue of cosmic censorship is a major unresolved issue in the understanding of gravitational collapse.  

We saw earlier in Sec.~\ref{sec: rn}, how \rn space-time (static, spherically symmetric charged space-time) can be shown to have a \emph{naked singularity} - one not hidden behind an horizon.  However, this only occurred in the situation where $Q^2>M^2$, where $Q$ is the charge of the black hole and $M$ its mass. Therefore, we can say that the cosmic censorship conjecture demands an upper limit on the charge, that is
\begin{equation} \label{eqn: cccRN}
Q^2 \leq M^2.
\end{equation}  
Similarly it can be shown that the the Kerr extension of the \rn solution, the Kerr-Newman solution (axially symmetric, stationary, charged space-time), also has a naked singularity in the event that $Q^2 + a^2 > M^2$ where $a$ is the the angular momentum per unit mass.  Again, for cosmic censorship to hold, this implies an upper limit on the charge (and angular momentum), that is 
\begin{equation} \label{eqn: cccKN}
Q^2 + a^2 \leq M^2.
\end{equation}

 As of yet, there is no decisive evidence for or against the validity of cosmic censorship.  To prove such a conjecture, would no doubt demand some rigour, that is if anybody can think of a method to do so.  However, to prove it incorrect, all that is required is one example.  The method of investigation to date, therefore, has involved attempts of producing a \rn or Kerr-Newman black hole that is \emph{overcharged} or \emph{overspun} (obviously this only pertains to the Kerr-Newman black hole), that is one that disobeys the upper limit on its charge as described in Eqs.~\eqref{eqn: cccRN} and \eqref{eqn: cccKN} for \rn \cite{Hubeny:1998, Zimmerman:2010} and Kerr-Newman respectively \cite{Wald:1974, Jacobson:2009, Barausse:2011}, where Kerr is considered a special case of the Kerr-Newman ($Q=0$).

When a black hole satisfies the above (relevant) condition, with an equal sign instead of the less than equal, the black hole is said to be extremal.  Therefore, one avenue of investigation has been to take an extremal black hole and get it to `absorb' an object whose charge and/or angular momentum is sufficiently large to break the above conditions.  There would then be no final black hole state for the system to settle down to and a naked singularity would be born.  This was first attempted by Wald \cite{Wald:1974}, it was not a success as it appeared to be not possible to get an extremal black hole to `swallow' a particle with sufficient charge or angular momentum.  It appears that the parameters that would allow us to overcharge a black hole actually protect the particle from being drawn into an extremal black hole.

A slightly different approach to this was taken by Hubeny in 1999 \cite{Hubeny:1998}, where a black hole that was non-extremal, but very close to being extremal, was considered in \rn space-time. As in previous research, they allowed a test particle with sufficient charge to drop into the black hole, to push it over its upper charge limit.  In this paper, they assumed that the impact of the back-reaction was negligible compared to the rest of the system.  As long as this assumption holds, they were successfully able to overcharge the black hole.  This was followed up by Jacobson \cite{Jacobson:2009} who also considered a near extremal black hole, but instead of over charging, he attempted to overspin a Kerr black hole.  Like Hubeny, he was successful, but his investigation neglected the self-force effects.

Recent reanalysis, however, \cite{Zimmerman:2010, Barausse:2011}, is now showing that neglecting the electromagnetic self-force is not justified, and that for a clear analysis, it will be required to be included in the calculation.  This is now leading researchers to investigate the possibility that it is the self-force that `protects' these particles (particles with the ability to overcharge) from being swallowed by the black holes.

To this end, it is now becoming increasingly important, to calculate the self-force for non-geodesic motions, particularly for charged black hole space-times. This \fixme{chapter,} therefore, investigates several different scenarios with calculations of the singular field for these different scenarios as well as the resulting regularisation parameters.

\subsection{Singular Field}

For calculating non-geodesic motion, I have so far solely concentrated on the scalar case, with the aim to use this as a toy model to ensure efficient coding which we can then apply to the more complicated cases of electromagnetic and gravitational particles.  

This chapter concentrates on three different space-times with accelerated motion, that is Schwarzschild, \rn and the generic $f(r)$ space-time, as described in Sec.~\ref{sec: BHST} - all are static, spherically symmetric space-times.  Several choices of motion are then considered - motion with arbitrary four-velocity in the equatorial plane, radial infall and that of a charged particle.  A lot of the expressions, proved too large for paper format, therefore, some regularisation parameters have been omitted but are available electronically \cite{BarryWardell.net} instead.

To calculate the singular field, the technique remains the same as described in Sec.~\ref{sec:Coordinate}.  It differs only in the expressions introduced for the four velocity, $u^{\ab}$ when obtaining an expressions for $x^{a'}$ in Eq.~\eqref{eqn: xtilde}.  The four velocity of the particle at $x'$ is then easily obtained by differentiating the resulting expression with respect to proper $\tau$.  For generic motion in the equatorial plane, one can use
\begin{equation}
u^{\bar{t}} = \dot{t_0}, \quad u^{\bar{r}} = \rbdot, \quad u^{\bar{w_1}} = \dot{\phi}_0, \quad u^{\bar{w_2}} =0,
\end{equation}
where I have taken motion to be in the $\theta_0 = \pi/2$ plane without loss of generality due to the spherical symmetry of the space-times.  For radial infall, one can use,
\begin{equation}
u^{\bar{t}} = \dot{t_0} , \quad u^{\bar{r}} = \rbdot, \quad u^{\bar{\theta}} = 0, \quad u^{\bar{\phi}} = 0,
\end{equation}
and for a charged particle of charge per unit mass $q$, the standard four-velocity, previously derived in Sec.~\ref{sec: rn}, is
\begin{gather}
u^{\bar{t}} = \frac{E \rb^2 - q Q \rb}{\rb^2 - 2 m \rb + Q^2}, \nonumber \\
 u^{\bar{r}} = \sqrt{\left( E - \frac{q Q}{\rb} \right)^2 - \frac{1}{\rb^2}\left( \rb^2 - 2 m \rb + Q^2 \right) \left( 1 + \frac{L^2}{\rb^2} \right)}, \nonumber \\
 u^{\bar{\theta}} = 0, \qquad u^{\bar{\phi}} = \frac{L}{\rb^2}.
\end{gather}
For $f(r)$, we can also make use of the fact that 
\begin{equation}
\rbdot^2  = -f(\rb) + f(\rb)^2 \dot{t_0}^2 - f(\rb) \rb^2 \dot{\phi}_0^2.
\end{equation}


\subsection{Mode Sum Decomposition}

As with the singular field, the regularisation parameters follow much the same method as that described in Sec.~\ref{sec: modeSum} for geodesic motions.  The rotation is the same as with the \Sch metric as all the space-times that we use here are identical to the \Sch space-time in respect to $(\theta, \phi)$ which is where the rotation changes the metric or line element.

The reader is reminded that the self force for the scalar field is given by Eq.~\eqref{eqn:fasum}, that is
\begin{equation}
F_a \left( r, t, \alpha, \beta \right) = \sum_{n=1}^{\infty} \frac{B_a^{(3 n -2)}}{ \zrho^{2 n + 1}} \epsilon^{n-3},
\end{equation}
where $B_a^{(k)} = b_{a_1 a_2 \dots a_k}(\bar{x}) \Delta x^{a_1} \Delta x^{a_2} \dots \Delta x^{a_k}$.  In coordinates, 
\begin{equation}
\zrho = \sqrt{(g_{\alphab \betab} u^{\alphab} \Delta x^b)^2 +g_{\alphab \betab} \Delta x^a \Delta x^b},
\end{equation}
 will take a different form according to the metric used.  
For the $f(r)$ metric, this is explicitly given by,
\par \vspace{-6pt} \begin{IEEEeqnarray}{rCl}
\zrho \left( t, r, \alpha, \beta \right)^\text{\fixme{$2$}} &=& \Delta r^2 \left[\dot{t}_0^2-\frac{\rb^2
   \dot{\phi}_0^2}{f(\rb)}\right]+\Delta t \left[-2 \Delta
   w_1 \rb^2 \dot{t}_0 \dot{\phi}_0 f(\rb)-2 \Delta r \dot{t}_0
   \dot{\rb}_0\right] \nonumber \\
&&
	+\: \frac{2 \Delta r \Delta w_1 \rb^2
   \dot{\rb}_0 \dot{\phi}_0}{f(\rb)}+\Delta t^2 f(\rb) \left[\dot{t}_0^2
   f(\rb)-1\right]+\Delta w_2^2 \rb^2 \nonumber \\
&&
	+\: \Delta
   w_1^2 \left(\rb^4 \dot{\phi}_0^2+\rb^2\right)
\end{IEEEeqnarray}
with $\dot{\phi}_0$ set to zero for radial motion, while for \rn space-time, it becomes
\par \vspace{-6pt} \begin{IEEEeqnarray}{rCl}
\zrho \left( t, r, \alpha, \beta \right)^\text{\fixme{$2$}} &=& \frac{\Delta r^2 \left[E^2 \rb^4-2
   E q Q \rb^3+2 L^2 M \rb+\rb^2
   \left(q^2 Q^2-L^2\right)-L^2 Q^2\right]}{\left(-2 M
   \rb+Q^2+\rb^2\right)^2} \nonumber \\
&&
	+\: \frac{\Delta
   t^2 \left[\left(E^2-1\right) \rb^2+2
   \rb (M-E q Q)+\left(q^2-1\right)
   Q^2\right]}{\rb^2} \nonumber \\
&&
	+\: \Delta t
   \left[\Delta w_1 \left(\frac{2 L q
   Q}{\rb}-2 E L\right)+\frac{2 \Delta
   r \rb \rbdot (q Q-E \rb)}{-2
   M \rb+Q^2+\rb^2}\right] \nonumber \\
&&
	+\: \Delta w_1^2
   \left(L^2+\rb^2\right)+\frac{2 \Delta r
  \Delta w_1 L \rb^2 \rbdot}{-2 M
   \rb+Q^2+\rb^2}+\Delta w_2^2
   \rb^2,
\end{IEEEeqnarray}
where $E$, $L$ and $M$ are the same as those previously defined for \Sch and Kerr space-times in Sec.~\ref{sec: modeDecomp}.

In Sec.~\ref{sec: modeDecomp}, it was also beneficial to have an expression for $\zrho\left( t, r, \alpha, \beta \right)$ when $\Delta t = 0$, particularly in the form
\begin{equation}
 \zrho \left(r, t_0, \alpha, \beta \right)^\text{\fixme{$2$}} = \nu^2 \Delta r^\text{\fixme{$2$}} + \zeta^2 \left(\Delta w_1 - \mu \Delta r \right)^\text{\fixme{$2$}} + \rb^2 \Delta w_2^2
\end{equation} 
This allows us to write,
\par \vspace{-6pt} \begin{IEEEeqnarray}{rCl}
\zrho \left(r, t_0, \alpha, \beta \right) ^\text{\fixme{$2$}} &=& \frac{\dot{t}_0^2}{1 + \rb^2 \dot{\phi}_0^2} \Delta r^2 + \left( \rb^2 + \rb^4 \dot{\phi}_0^2 \right) \left[ \Delta w_1^2 + \frac{\rbdot \dot{\phi}_0}{f(\rb) \left( 1 + \rb^2 \dot{\phi}_0 \right)} \Delta r\right]^2  \nonumber \\
&&
	+\:  \rb^2 \Delta w_2^2,
\end{IEEEeqnarray}
in the case of the $f(r)$ metric and
\par \vspace{-6pt} \begin{IEEEeqnarray}{rCl}
\zrho \left(r, t_0, \alpha, \beta \right)^\text{\fixme{$2$}} &=& \frac{\Delta r^2 \rb^4 (q Q-E
   \rb)^2}{\left(L^2+\rb^2\right) \left(-2 M
   \rb+Q^2+\rb^2\right)^2} +\Delta
   w_2^2 \rb^2 \nonumber \\
&&
	+\: \left(L^2+\rb^
   2\right) \left(\Delta w_1+\frac{\Delta
   r L \rb^2
   \rbdot}{\left(L^2+\rb^2\right) \left(-2 M
   \rb+Q^2+\rb^2\right)}\right)^2
\end{IEEEeqnarray}
for \rn space-time.  From this, it is possible to read off the following expressions for the $f(r)$ and \rn metrics,
\begin{gather}
\zeta^2_{f(r)}  = \rb^2 + \rb^4 \dot{\phi}_0^2, \quad \quad \zeta^2_{RN} = L^2+\rb^2,\nonumber \\
\nu^2_{f(r)} = \frac{\dot{t}_0^2}{1 + \rb^2 \dot{\phi}_0^2}, \quad \nu^2_{RN} = \frac{\rb^4 (q Q-E
   \rb)^2}{\left(L^2+\rb^2\right) \left(-2 M
   \rb+Q^2+\rb^2\right)^2}, \nonumber \\
\mu_{f(r)}  = -\frac{\rbdot \dot{\phi}_0}{f(\rb) \left( 1 + \rb^2 \dot{\phi}_0 \right)}, \quad \mu_{RN} =  -\frac{L \rb^2
   \rbdot}{\left(L^2+\rb^2\right) \left(-2 M
   \rb+Q^2+\rb^2\right)}.
\end{gather}
We also recall from Sec.~\ref{sec: modeDecomp}, our definitions for $k$ and $\chi$,
\begin{equation}
k = \frac{\zeta^2-\rb^2}{\zeta^2}, \qquad \chi(\beta) \equiv 1 - k \sin^2 \beta,
\end{equation}
We now have everything necessary to calculate the mode sum decomposition.  The first term can be calculated from Eq.~\eqref{eqn: Aterm} while the higher terms are required to follow the full method previously outlined in Sec.~\ref{sec: modeDecomp}.


\subsection{$f(r)$ Regularisation Parameters}

For generic motion, described above we calculate the following parameters,
\begin{gather}
F_{t\lnpow{1}} = \frac{\sgn \Delta r \rbdot}{2 \left(\rb^4
   \dot{\phi}_0^2+\rb^2\right)}, \quad
F_{r\lnpow{1}} = -\frac{\sgn \Delta r \dot{t}_0}{2 \left(\rb^4
   \dot{\phi}_0^2+\rb^2\right)}, \quad
F_{\theta\lnpow{1}} = 0, \quad
F_{\phi\lnpow{1}} = 0,
\end{gather}

\begin{equation}
F_{t\lpow{0}} = \frac{1}{2 \pi \rb^2 \dot{\phi}_0 \left(1+\rb^2 \dot{\phi}_0^2 \right)^{3/2}} (F^{\mathcal{E}}_{t\lpow{0}} \mathcal{E} + F^{\mathcal{K}}_{t\lpow{0}} \mathcal{K}),
\end{equation}
where
\par \vspace{-6pt} \begin{IEEEeqnarray*}{rCl}
F^{\mathcal{E}}_{t\lpow{0}} &=&
2 \rb \dot{t}_0 \Big[2 \rb^4 \dot{\phi}_0^4 \ddot{\phi}_0f(\rb)+3 \rb^2 \dot{\phi}_0^2
   \ddot{\phi}_0f(\rb)+\ddot{\phi}_0f(\rb)+2 \rb^2 \rbdot \dot{\phi}_0^3
   \ddot{rb} 
	+ 4 \rbdot \dot{\phi}_0 \ddot{\rb} \\
&& \quad
	+\: \rb^4 \rbdot \dot{\phi}_0^5 f'(\rb)+4
   \rb^2 \rbdot \dot{\phi}_0^3 f'(\rb)+3 \rbdot \dot{\phi}_0 f'(\rb) 
	+ 2 \rb^3
   \rbdot \dot{\phi}_0^5 f(\rb) \\
&& \quad
	+\: 2 \rb \rbdot \dot{\phi}_0^3 f(\rb)\Big] \\
&&
	-\: 4 \rb
   \dot{\phi}_0 \dot{t}_0^2 \ddot{t}_0f(\rb)^2 \left(\rb^2 \dot{\phi}_0^2+2\right)+2 \rb \dot{\phi}_0
   \ddot{t}_0f(\rb) \left(\rb^2 \dot{\phi}_0^2+1\right)  \\
&& 
	-\:   4 \rb \rbdot \dot{\phi}_0 \dot{t}_0^3
   f(\rb) \left(\rb^2 \dot{\phi}_0^2+2\right) f'(\rb), \\
F^{\mathcal{K}}_{t\lpow{0}} &=&
\dot{t}_0 \Big[-2 \rb^3 \dot{\phi}_0^2 \ddot{\phi}_0f(\rb)-2 \rb \ddot{\phi}_0f(\rb)-2
   \rb \rbdot \dot{\phi}_0 \ddot{\rb}-\rb^3 \rbdot \dot{\phi}_0^3 f'(\rb)
	- \rb
   \rbdot \dot{\phi}_0 f'(\rb) \\
&& \quad
	-\: 2 \rb^2 \rbdot \dot{\phi}_0^3 f(\rb)-2 \rbdot \dot{\phi}_0
   f(\rb)\Big] \\
&&
	+\: 2 \rb \dot{\phi}_0 \dot{t}_0^2 \ddot{t}_0f(\rb)^2+2 \rb \rbdot
   \dot{\phi}_0 \dot{t}_0^3 f(\rb) f'(\rb),
\end{IEEEeqnarray*}

\begin{equation}
F_{r\lpow{0}} = \frac{1}{2 \pi f(\rb) \rb^2 \dot{\phi}_0 \left(1+\rb^2 \dot{\phi}_0^2 \right)^{3/2}} (F^{\mathcal{E}}_{r\lpow{0}} \mathcal{E} + F^{\mathcal{K}}_{r\lpow{0}} \mathcal{K}),
\end{equation}
where
\par \vspace{-6pt} \begin{IEEEeqnarray*}{rCl}
F^{\mathcal{E}}_{r\lpow{0}} &=&
	-\left(\rb^2 \dot{\phi}_0^2+1\right)
   \Big[4 \rb^3 \rbdot \dot{\phi}_0^2 \ddot{\phi}_0-4 \rb^3 \dot{\phi}_0^3 \ddot{\rb}+2
   \rb \rbdot \ddot{\phi}_0-6 \rb \dot{\phi}_0 \ddot{rb} 
	- 2 \rb^5 \dot{\phi}_0^5
   f'(\rb) \\
&& \quad
	-\: 5 \rb^3 \dot{\phi}_0^3 f'(\rb)-3 \rb \dot{\phi}_0 f'(\rb)-4 \rb^4
   \dot{\phi}_0^5 f(\rb)
	- 6 \rb^2 \dot{\phi}_0^3 f(\rb)-2 \dot{\phi}_0 f(\rb)\Big] \\
&&
	-\: 2 \rb \dot{\phi}_0 \dot{t}_0^2 f(\rb) \Big[2 \rb^2 \dot{\phi}_0^2 \ddot{\rb}+4 \ddot{rb}+3
   \rb^4 \dot{\phi}_0^4 f'(\rb)+9 \rb^2 \dot{\phi}_0^2 f'(\rb)+6 f'(\rb) 
	 \\
&& \quad
	+\:  2 \rb^3
   \dot{\phi}_0^4 f(\rb)+2 \rb \dot{\phi}_0^2 f(\rb)\Big]\\
&&
	+\: 4 \rb
   \rbdot \dot{\phi}_0 \dot{t}_0 \ddot{t}_0 f(\rb) \left(\rb^2 \dot{\phi}_0^2+2\right)+4 \rb
   \dot{\phi}_0 \dot{t}_0^4 f(\rb)^2 \left(\rb^2 \dot{\phi}_0^2+2\right) f'(\rb), \\
F^{\mathcal{K}}_{r\lpow{0}} &=&
\left(\rb^2 \dot{\phi}_0^2+1\right) \Big[2 \rb \rbdot \ddot{\phi}_0-2 \rb \dot{\phi}_0
   \ddot{rb}-\rb^3 \dot{\phi}_0^3 f'(\rb)-\rb \dot{\phi}_0 f'(\rb)
	- 2 \rb^2 \dot{\phi}_0^3
   f(\rb) \\
&& \quad
	-\: 6 \dot{\phi}_0 f(\rb)\Big]\\
&&
	+\: \dot{\phi}_0 \dot{t}_0^2 f(\rb) \left[2 \rb \ddot{rb}+3
   \rb^3 \dot{\phi}_0^2 f'(\rb)+3 \rb f'(\rb)+2 \rb^2 \dot{\phi}_0^2 f(\rb)+2
   f(\rb)\right] \\
&&
	-\: 2 \rb \rbdot \dot{\phi}_0 \dot{t}_0 \ddot{t}_0 f(\rb)-2 \rb \dot{\phi}_0
   \dot{t}_0^4 f(\rb)^2 f'(\rb),
\end{IEEEeqnarray*}

\begin{equation}
F_{\theta \lpow{0}} = 0,
\end{equation}

\begin{equation}
F_{\phi \lpow{0}} = \frac{1}{2 \pi f(\rb) \rb^2 \dot{\phi}_0^2 \left(1+\rb^2 \dot{\phi}_0^2 \right)^{1/2}} (F^{\mathcal{E}}_{\phi \lpow{0}} \mathcal{E} + F^{\mathcal{K}}_{\phi \lpow{0}} \mathcal{K}),
\end{equation}
where
\par \vspace{-6pt} \begin{IEEEeqnarray*}{rCl}
F^{\mathcal{E}}_{ \phi \lpow{0}} &=&
-4 \rb^5 \dot{\phi}_0^4 \ddot{\phi}_0 f(\rb)-2 \rb^3 \dot{\phi}_0^2 \ddot{\phi}_0 f(\rb)+2
   \rb \dot{t}_0 \dot{\phi}_0 \ddot{t}_0 f(\rb)^2 \left(2 \rb^2 \dot{\phi}_0^2+1\right)+2 \rb
   \ddot{\phi}_0 f(\rb) \\
&&
	-\: 4 \rb^3 \rbdot \dot{\phi}_0^3 \ddot{\rb}-2 \rb
   \rbdot \dot{\phi}_0 \ddot{\rb}-2 \rb^5 \rbdot \dot{\phi}_0^5 f'(\rb)-3 \rb^3
   \rbdot \dot{\phi}_0^3 f'(\rb) \\
&&
	+\: 2 \rb \rbdot \dot{t}_0^2 \dot{\phi}_0 f(\rb) \left(2
   \rb^2 \dot{\phi}_0^2+1\right) f'(\rb)-\rb \rbdot \dot{\phi}_0 f'(\rb)-4 \rb^4
   \rbdot \dot{\phi}_0^5 f(\rb) \\
&&
	-\: 2 \rb^2 \rbdot \dot{\phi}_0^3 f(\rb)+2 \rbdot \dot{\phi}_0
   f(\rb),\\
F^{\mathcal{K}}_{ \phi \lpow{0}} &=&
2 \rb^3 \dot{\phi}_0^2 \ddot{\phi}_0 f(\rb)-2 \rb \dot{t}_0 \dot{\phi}_0 \ddot{t}_0 f(\rb)^2-2
   \rb \ddot{\phi}_0 f(\rb)+2 \rb \rbdot \dot{\phi}_0 \ddot{\rb}+\rb^3
   \rbdot \dot{\phi}_0^3 f'(\rb) \\
&& 
	-\: 2 \rb \rbdot \dot{t}_0^2 \dot{\phi}_0 f(\rb)
   f'(\rb)+\rb \rbdot \dot{\phi}_0 f'(\rb)+2 \rb^2 \rbdot \dot{\phi}_0^3
   f(\rb)-2 \rbdot \dot{\phi}_0 f(\rb).
\end{IEEEeqnarray*}

Due to the increasing complexity of the resulting regularisation parameters, for orders higher than $F_{a \lpow{0}}$, we needed to simplify the system.  We proceeded in two ways, completely generic motion in \Sch space-time and radial infall with our $f(r)$ metric.  The \Sch space-time regularisation parameters proved to be quite unwieldy, therefore, all powers up to $F_{a \lpow{2}}$ have been made electronically available \cite{BarryWardell.net}.  For radial infall we set $\dot{\phi}_0 = 0$, and we also allow our coupling constant $\xi = \xi_{1/6} + 1/6$ so that $\xi_{1/6} = -1/6$ represents minimal coupling and $\xi_{1/6}  = 0$ depicts conformal coupling. These simplify our expressions immensely to give the following parameters,  

\par \vspace{-6pt} \begin{IEEEeqnarray}{rCl}
F_{t\lpow{2}} &=& \frac{1}{24 \rb} \Big( 3 \big\{39 \ddot{t}_0 f'(\rb)^2+28 \left(6 \ddot{t}_0
   \ddot{r}_0-\rbdot \dddot{t}_0 \right) f'(\rb)+20 \big[\ddot{t}_0 \left(9 \ddot{r}_0^2-2 \rbdot
   \dddot{r}_0\right) \nonumber \\
&&
	\qquad \quad -\: 2 \rbdot \dddot{t}_0 \ddot{r}_0\big]\big\} \rb^2+2 \ddot{t}_0 \left(45
   \rb^2 \ddot{t}_0^2+24 \xi_{1/6}-8\right) f(\rb)^2+2 f(\rb) \big\{[ (12 \xi_{1/6} \nonumber \\
&&
	\qquad \qquad +\: 41)
   \ddot{t}_0 f''(\rb)-6 \ddddot{t}_0] \rb^2+2 \big[-6 \rbdot \dddot{t}_0+36 \ddot{t}_0
   \ddot{r}_0+(24 \xi_{1/6} \nonumber \\
&&
	\qquad \qquad +\: 31 ) \ddot{t}_0 f'(\rb)\big] \rb-\left(24 \xi_{1/6}+1\right)
   \ddot{t}_0\big\}\Big) \nonumber \\
&&
	+\: \frac{ 1}{24 \rb^2 f(\rb)} \dot{t}_0 \Big(9 \rbdot \left[140 \ddot{r}_0^3+140 f'(\rb) \ddot{r}_0^2+39
   f'(\rb)^2 \ddot{r}_0+2 f'(\rb)^3\right] \rb^3 \nonumber \\
&&
	\quad +\: f(\rb) \big(2 \big\{360 \ddot{r}_0
   \dddot{r}_0-30 \rbdot \ddddot{r}_0+3 \rbdot (12 \xi_{1/6}+41) \ddot{r}_0 f''(\rb) \nonumber \\
&&
	\quad \qquad +\: 2
   f'(\rb) \left[84 \dddot{r}_0+\rbdot \left(15 \xi_{1/6}+31\right) f''(\rb)\right]\big\} \rb^2+4
   \rbdot \big[90 \ddot{r}_0^2+3 (24 \xi_{1/6} \nonumber \\
&& 
	\qquad \qquad +\: 31) f'(\rb) \ddot{r}_0+4 (15 \xi_{1/6}+7)
   f'(\rb)^2\big] \rb-\rbdot (24 \xi_{1/6}+1) \big[6 \ddot{r}_0 \nonumber \\
&&
	\qquad \qquad +\: 5 f'(\rb)\big]\big)
   \rb+2 \left[180 \ddot{t}_0 \dddot{t}_0 \rb^3+135 \rbdot \ddot{t}_0^2 \rb^2+8
   \rbdot (1-3 \xi_{1/6})\right] f(\rb)^3 \nonumber \\
&&
	\qquad +\: f(\rb)^2 \big\{\rbdot \left[1890 \ddot{r}_0
   \ddot{t}_0^2+1485 f'(\rb) \ddot{t}_0^2+8 (3 \xi_{1/6}+5) f^{(3)}(\rb)\right] \rb^3 \nonumber \\
&&
	\quad \qquad +\: 8
   \left[12 \dddot{r}_0+5 \rbdot (3 \xi_{1/6}+2) f''(\rb)\right] \rb^2 \nonumber \\
&&	
	\quad \qquad +\: 8 \rbdot (3 \xi_{1/6}-1) \left[6 \ddot{r}_0+7 f'(\rb)\right] \rb+2 \rbdot (24 \xi_{1/6}+1)\big\} \Big)
   \nonumber \\
&&
	-\: \frac{ 1}{16 \rb} f(\rb) \dot{t}_0^2 \Big(5 \big\{357
   \ddot{t}_0 f'(\rb)^2+4 \left(255 \ddot{t}_0 \ddot{r}_0-26 \rbdot \dddot{t}_0\right) f'(\rb)+28
   \big[\ddot{t}_0 \big(27 \ddot{r}_0^2 \nonumber \\
&&
	\qquad \qquad -\: 4 \rbdot \dddot{r}_0\big)-4 \rbdot \dddot{t}_0
   \ddot{r}_0\big]\big\} \rb^2+4 \ddot{t}_0 \left(210 \rb^2 \ddot{t}_0^2+24 \xi_{1/6}-17\right) f(\rb)^2 \nonumber \\
&&
	\quad +\: 4 f(\rb) \big\{\left[(12 \xi_{1/6}+143) \ddot{t}_0 f''(\rb)-10
   \ddddot{t}_0\right] \rb^2+\big[-20 \rbdot \dddot{t}_0+210 \ddot{t}_0 \ddot{r}_0 \nonumber \\
&&
	\quad \qquad +\: (48 \xi_{1/6}+215) \ddot{t}_0 f'(\rb)\big] \rb-(24 \xi_{1/6}+1) \ddot{t}_0\big\}\Big)
  \nonumber \\
&&
	-\: \frac{1}{32 \rb^2} \dot{t}_0^3 \Big[\rbdot \left[2520 \ddot{r}_0^3+7980 f'(\rb) \ddot{r}_0^2+5170
   f'(\rb)^2 \ddot{r}_0+937 f'(\rb)^3\right] \rb^3 \nonumber \\
&&
	\quad +\: 2 f(\rb) \big(8 \big\{5 \ddot{r}_0
   \left[14 \dddot{r}_0+17 \rbdot f''(\rb)\right]+f'(\rb) \big[95 \dddot{r}_0+2 \rbdot (3
   \xi_{1/6} \nonumber \\
&&
	\quad \qquad \qquad+\: 40) f''(\rb)\big]\big\} \rb^2+3 \rbdot \big[140 \ddot{r}_0^2+340 f'(\rb)
   \ddot{r}_0+(64 \xi_{1/6} \nonumber \\
&&
	\qquad \qquad+\: 171) f'(\rb)^2\big] \rb-4 \rbdot (24 \xi_{1/6}+1)
   f'(\rb)\big) \rb+4 f(\rb)^2 \big\{\rbdot \big[1890 \ddot{r}_0 \ddot{t}_0^2 \nonumber \\
&&
	\quad \qquad +\: 2205
   f'(\rb) \ddot{t}_0^2+34 f^{(3)}(\rb)\big] \rb^2+\left[40 \dddot{r}_0+68 \rbdot
   f''(\rb)\right] \rb+\rbdot \big[(48 \xi_{1/6} \nonumber \\
&&
	\qquad \qquad -\: 49) f'(\rb)-30 \ddot{r}_0\big]\big\}
   \rb+8 \left(140 \ddot{t}_0 \dddot{t}_0 \rb^3+105 \rbdot \ddot{t}_0^2 \rb^2+3
   \rbdot\right) f(\rb)^3 \Big]  \nonumber \\
&& 
	+\: \frac{5 }{16 \rb} f(\rb)^2  \dot{t}_0^4 \Big\{\left[756 \ddot{t}_0
   \ddot{r}_0^2+1377 \ddot{t}_0 f'(\rb)^2+28 \left(87 \ddot{t}_0 \ddot{r}_0-4 \rbdot \dddot{t}_0 \right)
   f'(\rb)\right] \rb^2 \nonumber \\
&&
	\quad +\: 12 \ddot{t}_0 f(\rb) \left[14 \ddot{r}_0+27 f'(\rb)+18 \rb
   f''(\rb)\right] \rb+12 \ddot{t}_0 \left(21 \rb^2 \ddot{t}_0^2-1\right) f(\rb)^2\Big\}
   \nonumber \\
&&
	+\: \frac{5 }{16 \rb} f(\rb) \dot{t}_0^5 \Big[7 \rbdot f'(\rb) \left[108 \ddot{r}_0^2+172 f'(\rb)
   \ddot{r}_0+59 f'(\rb)^2\right] \rb^2 \nonumber \\
&&
	\quad +\: 4 f(\rb) \big(\rbdot f'(\rb) \left[42
   \ddot{r}_0+43 f'(\rb)\right]+28 \rb \big\{\rbdot \ddot{r}_0 f''(\rb)+f'(\rb)
   \big[\dddot{r}_0 \nonumber \\
&&
	\qquad \qquad+\: 2 \rbdot f''(\rb)\big]\big\}\big) \rb+4 \rbdot f(\rb)^2 \big\{3
   \left(63 \rb^2 \ddot{t}_0^2-1\right) f'(\rb) \nonumber \\
&&
	\qquad +\: 2 \rb \left[2 f''(\rb)+\rb
   f^{(3)}(\rb)\right]\big\}\Big]  \nonumber \\
&&
	-\: \frac{35}{2} \ddot{t}_0 f(\rb)^3 \dot{t}_0^6 \left\{\rb f'(\rb)
   \left[27 \ddot{r}_0+32 f'(\rb)\right]+f(\rb) \left[3 f'(\rb)+2 \rb f''(\rb)\right]\right\}
    \nonumber \\
&&
	-\: \frac{35}{8} \rbdot f(\rb)^2 f'(\rb) \dot{t}_0^7 \left\{\rb
   f'(\rb) \left[54 \ddot{r}_0+41 f'(\rb)\right]+f(\rb) \left[6 f'(\rb)+8 \rb
   f''(\rb)\right]\right\}  \nonumber \\
&&
	+\: \frac{945}{4} \rb \ddot{t}_0
   f(\rb)^4 f'(\rb)^2 \dot{t}_0^8 + \frac{315}{4} \rb \rbdot f(\rb)^3 f'(\rb)^3 \dot{t}_0^9, 
\end{IEEEeqnarray}

\par \vspace{-6pt} \begin{IEEEeqnarray}{rCl}
F_{r\lpow{2}} &=& \frac{1}{12 \rb^2 f(\rb)^2} \Big(6 \left[60 \ddot{r}_0^3+3 f'(\rb)^2
   \ddot{r}_0-40 \rbdot \dddot{r}_0 \ddot{r}_0+12 \left(3 \ddot{r}_0^2-\rbdot \dddot{r}_0\right) f'(\rb)\right]
   \rb^3 \nonumber \\
&&
	\quad +\: f(\rb) \big(\left\{\left[2 (12 \xi_{1/6}+23) \ddot{r}_0+(12 \xi_{1/6}+5) f'(\rb)\right] f''(\rb)-24
   \ddddot{r}_0\right\} \rb^2 \nonumber \\
&&
	\qquad +\: 8 \left[18 \ddot{r}_0^2+(12 \xi_{1/6}+11) f'(\rb) \ddot{r}_0+(6 \xi_{1/6}+1)
   f'(\rb)^2-6 \rbdot \dddot{r}_0\right] \rb \nonumber \\
&&
	\qquad -\: (24 \xi_{1/6}+1) \left[2 \ddot{r}_0+f'(\rb)\right]\big)
   \rb+\left(54 \rb^2 \ddot{t}_0^2-48 \xi_{1/6}+4\right) f(\rb)^3 \nonumber \\
&&
	\quad +\: f(\rb)^2 \big\{\left[189 f'(\rb)
   \ddot{t}_0^2+30 (9 \ddot{t}_0 \ddot{r}_0-2 \rbdot \dddot{t}_0) \ddot{t}_0+2 (12 \xi_{1/6}+5)
   f^{(3)}(\rb)\right] \rb^3 \nonumber \\
&&
	\qquad +\: 2 (60 \xi_{1/6}+13) f''(\rb) \rb^2+2 (12 \xi_{1/6}-1) \left[2 \ddot{r}_0+3
   f'(\rb)\right] \rb \nonumber \\
&&
	\qquad +\: 48 \xi_{1/6}+2\big\}\Big) \nonumber \\
&&
	+\: \frac{1}{8 \rb f(\rb)} \dot{t}_0 \Big(3 \rbdot \ddot{t}_0 \left[420 \ddot{r}_0^2+420 f'(\rb) \ddot{r}_0+109
   f'(\rb)^2\right] \rb^2+2 \big[105 \rb^2 \rbdot \ddot{t}_0^3 \nonumber \\
&&
	\qquad+\: 8 \rbdot (3 \xi_{1/6}-1) \ddot{t}_0+24
   \rb \dddot{t}_0\big] f(\rb)^2+2 f(\rb) \big\{\big[10 \big(-\rbdot \ddddot{t}_0+12 \dddot{t}_0 \ddot{r}_0 \nonumber \\
&&
	\qquad \qquad +\: 12
   \ddot{t}_0 \dddot{r}_0\big)+84 \dddot{t}_0 f'(\rb)+\rbdot (12 \xi_{1/6}+83) \ddot{t}_0 f''(\rb)\big]
   \rb^2 \nonumber \\
&&
	\qquad +\: 2 \rbdot \ddot{t}_0 \left[90 \ddot{r}_0+(24 \xi_{1/6}+73) f'(\rb)\right] \rb-\rbdot (24
   \xi_{1/6}+1) \ddot{t}_0\big\}\Big) \nonumber \\
&&
	-\: \frac{1}{48
   \rb^2 f(\rb)} \dot{t}_0^2 \Big[3 \big[129 f'(\rb)^3+1272 \ddot{r}_0 f'(\rb)^2+60 \left(51
   \ddot{r}_0^2-10 \rbdot \dddot{r}_0\right) f'(\rb) \nonumber \\
&&
	\qquad +\: 560 \left(3 \ddot{r}_0^3-\rbdot \ddot{r}_0
   \dddot{r}_0\right)\big] \rb^3+6 f(\rb) \big(\big\{2 \big[(12 \xi_{1/6}+143) \ddot{r}_0+3 (6 \xi_{1/6} \nonumber \\
&&
	\quad \qquad \qquad +\: 31)
   f'(\rb)\big] f''(\rb)-20 \ddddot{r}_0\big\} \rb^2+\big[300 \ddot{r}_0^2+16 (6 \xi_{1/6}+31) f'(\rb)
   \ddot{r}_0 \nonumber \\
&&
	\quad \qquad +\: 3 (48 \xi_{1/6}+59) f'(\rb)^2-40 \rbdot \dddot{r}_0\big] \rb-(24 \xi_{1/6}+1) \left[2
   \ddot{r}_0+3 f'(\rb)\right]\big) \rb \nonumber \\
&&
	\quad -\: 4 \left(-405 \rb^2 \ddot{t}_0^2+24 \xi_{1/6}-14\right) f(\rb)^3+2
   f(\rb)^2 \big(\big\{4995 f'(\rb) \ddot{t}_0^2+210 \big[27 \ddot{t}_0 \ddot{r}_0 \nonumber \\
&&
	\qquad \qquad -\: 4 \rbdot \dddot{t}_0 \big]
   \ddot{t}_0+8 (3 \xi_{1/6}+14) f^{(3)}(\rb)\big\} \rb^3+20 (6 \xi_{1/6}+13) f''(\rb) \rb^2 \nonumber \\
&&
	\qquad+\: 2 \left[6 (12
   \xi_{1/6}-7) \ddot{r}_0+(132 \xi_{1/6}-53) f'(\rb)\right] \rb+48 \xi_{1/6}+2\big)\Big] \nonumber \\
&&
	-\: \frac{5}{16 \rb} \dot{t}_0^3 
   \Big(\rbdot \ddot{t}_0 \left[756 \ddot{r}_0^2+1932 f'(\rb) \ddot{r}_0+901 f'(\rb)^2\right] \rb^2+4
   f(\rb) \big\{3 \rbdot \ddot{t}_0 \big[14 \ddot{r}_0 \nonumber \\
&&
	\quad \qquad +\: 25 f'(\rb)\big]+\rb \left[28 \dddot{t}_0
   \ddot{r}_0+28 \ddot{t}_0 \dddot{r}_0+46 \dddot{t}_0 f'(\rb)+46 \rbdot \ddot{t}_0 f''(\rb)\right]\big\}
   \rb \nonumber \\
&&
	\quad +\: 4 \left(63 \rb^2 \rbdot \ddot{t}_0^3-3 \rbdot \ddot{t}_0+4 \rb \dddot{t}_0\right)
   f(\rb)^2\Big) \nonumber \\
&&
	+\: \frac{1}{32 \rb^2} \dot{t}_0^4 \Big(5 \big[504
   \ddot{r}_0^3+2394 f'(\rb)^2 \ddot{r}_0+561 f'(\rb)^3+28 \big(93 \ddot{r}_0^2 \nonumber \\
&&
	\quad \qquad -\: 8 \rbdot \dddot{r}_0\big)
   f'(\rb)\big] \rb^3+2 f(\rb) \big\{8 \left[135 \ddot{r}_0+2 (3 \xi_{1/6}+83) f'(\rb)\right] f''(\rb)
   \rb^2 \nonumber \\
&&
	\qquad +\: \left[420 \ddot{r}_0^2+1740 f'(\rb) \ddot{r}_0+(192 \xi_{1/6}+1145) f'(\rb)^2\right] \rb-4 (24
   \xi_{1/6} \nonumber \\
&&
	\quad \qquad +\: 1) f'(\rb)\big\} \rb+4 f(\rb)^2 \big\{5 \left[378 \ddot{r}_0 \ddot{t}_0^2+693 f'(\rb)
   \ddot{t}_0^2+10 f^{(3)}(\rb)\right] \rb^2 \nonumber \\
&&
	\qquad +\: 108 f''(\rb) \rb-30 \ddot{r}_0+(48 \xi_{1/6}-61)
   f'(\rb)\big\} \rb+24 \left(35 \rb^2 \ddot{t}_0^2+1\right) f(\rb)^3\Big) \nonumber \\
&&
	+\: \frac{35}{4} f(\rb) \dot{t}_0^5 \Big(\rb \rbdot \ddot{t}_0
   f'(\rb) \left[54 \ddot{r}_0+55 f'(\rb)\right]+f(\rb) \big\{6 \rbdot \ddot{t}_0 f'(\rb) \nonumber \\
&&
	\qquad +\: 4 \rb
   \left[\dddot{t}_0 f'(\rb)+\rbdot \ddot{t}_0 f''(\rb)\right]\big\}\Big) \nonumber \\
&&
	-\: \frac{5}{16 \rb} f(\rb) \dot{t}_0^6
   \Big(f'(\rb) \left[756 \ddot{r}_0^2+1708 f'(\rb) \ddot{r}_0+745 f'(\rb)^2\right] \rb^2 \nonumber \\
&&
	\quad +\: 4 f(\rb) \left\{61
   f'(\rb)^2+\left[42 \ddot{r}_0+76 \rb f''(\rb)\right] f'(\rb)+28 \rb \ddot{r}_0 f''(\rb)\right\}
   \rb \nonumber \\
&&
	\quad +\: 4 f(\rb)^2 \left\{3 \left(63 \rb^2 \ddot{t}_0^2-1\right) f'(\rb)+2 \rb \left[2 f''(\rb)+\rb
   f^{(3)}(\rb)\right]\right\}\Big)  \nonumber \\
&&
	-\: \frac{945}{4} \rb \rbdot \ddot{t}_0 f(\rb)^2 f'(\rb)^2 \dot{t}_0^7 \nonumber \\
&&
	+\: \frac{35}{8} f(\rb)^2 f'(\rb) \dot{t}_0^8 \left\{\rb f'(\rb)
   \left[54 \ddot{r}_0+53 f'(\rb)\right]+f(\rb) \left[6 f'(\rb)+8 \rb f''(\rb)\right]\right\}
    \nonumber \\
&&
	-\: \frac{315}{4} \rb f(\rb)^3 f'(\rb)^3 \dot{t}_0^{10},
\end{IEEEeqnarray}

\begin{equation}
F_{\theta \lpow{2}} = 0, \qquad \qquad F_{\phi \lpow{2}} = 0.
\end{equation}

We have also obtained $F_{a \lpow{4}} $, however, once again, these are not suitable for printed format, and so, have been made available online \cite{BarryWardell.net}.  Unfortunately, we have not been able to obtain numerical data to test the regularising capabilities of these regularisation parameters.  However, Roland Haas, has also calculated the generic \Sch parameters, so we were able to confirm our results matched his up to $F_{a \lpow{0}}$ \cite{Haas:Heffernan:2012}.  We also verified that our parameters agreed with the results of Casals and collaborators \cite{Casals:Poisson:Vega:2012}, when given the required $f(r)$ and four-velocity.


\subsection{\rn Regularisation Parameters}

For motion of a charged particle of charge per unit mass $q$, as described above we calculate the following parameters,
\begin{gather}
F_{t\lnpow{1}} = -\frac{\rbdot \sgn \Delta r}{2
   \left(L^2+\rb^2\right)}, \quad
F_{r\lnpow{1}} = \frac{\rb \sgn \Delta r (E \rb-q
   Q)}{2 \left(L^2+\rb^2\right) \left(Q^2+\rb^2-2 M
   \rb\right)}, \quad
F_{\theta\lnpow{1}} = 
F_{\phi\lnpow{1}} = 0,
\end{gather}

\begin{equation}
F_{t\lpow{0}} = \frac{\rbdot}{\pi  \rb^2 \left(L^2+\rb^2\right)^{3/2}} \left\{ \left[q Q \left(L^2+3 \rb^2\right)-2
   E\rb^3\right] \mathcal{E} + \left(E \rb^3-q Q \rb^2\right) \mathcal{K} \right\},
\end{equation}

\begin{equation}
F_{r\lpow{0}} = \frac{1}{\pi  \rb\left(L^2+\rb^2\right)^{3/2} \left(-2
   M \rb+Q^2+\rb^2\right)} (F^{\mathcal{E}}_{r\lpow{0}} \mathcal{E} + F^{\mathcal{K}}_{r\lpow{0}} \mathcal{K}),
\end{equation}
where
\par \vspace{-6pt} \begin{IEEEeqnarray*}{rCl}
F^{\mathcal{E}}_{r\lpow{0}} &=& 2 E^2 \rb^4-E q Q \rb
   \left(L^2+5 \rb^2\right)+2 L^2 M
   \rb-\rb^2 \left[L^2+\left(1-3 q^2\right)
   Q^2\right] \\
&&
	+\: L^2 \left(q^2-1\right) Q^2+2 M
   \rb^3-\rb^4, \\
F^{\mathcal{K}}_{r\lpow{0}} &=& -E^2 \rb^4+2 E q Q \rb^3+2 L^2
   M \rb-\rb^2 \left[L^2+\left(q^2+1\right)
   Q^2\right]-L^2 Q^2+2 M \rb^3-\rb^4,
\end{IEEEeqnarray*}

\begin{equation}
F_{\theta\lpow{0}} =0,
\end{equation}

\begin{equation}
F_{\phi \lpow{0}} =\frac{-\rb \rbdot}{\pi  L \sqrt{L^2+\rb^2}} \left( \mathcal{E}-\mathcal{K}\right),
\end{equation}

\begin{equation}
F_{t\lpow{2}} = \frac{\rbdot}{6 \pi  \rb^8 \left(L^2+\rb^2\right)^{7/2}} (F^{\mathcal{E}}_{t\lpow{2}} \mathcal{E} + F^{\mathcal{K}}_{t\lpow{2}} \mathcal{K}),
\end{equation}
where
\par \vspace{-6pt} \begin{IEEEeqnarray*}{rCl}
F^{\mathcal{E}}_{t\lpow{2}} &=& 24 E^3 \rb^{11} (L-\rb)
   (L+\rb)+E^2 q Q \rb^6 \left(16
   L^6+59 L^4 \rb^2+38 L^2 \rb^4+139
   \rb^6\right) \\
&&
	+\: E \rb \big(-288 L^{10}
   q^2 Q^2+192 L^{10} Q^2-108 L^8 M \rb^3-1224 L^8
   q^2 Q^2 \rb^2 \\
&&
	 \quad +\: 912 L^8 Q^2 \rb^2-420 L^6 M
   \rb^5-2065 L^6 q^2 Q^2 \rb^4+1728 L^6 Q^2
   \rb^4-606 L^4 M \rb^7 \\
&&
	\quad -\: 1761 L^4 q^2 Q^2
   \rb^6+1635 L^4 Q^2 \rb^6-3 L^4
   \rb^8-432 L^2 M \rb^9-791 L^2 q^2 Q^2
   \rb^8 \\
&&
	\quad +\: 798 L^2 Q^2 \rb^8+18 L^2
   \rb^{10}-138 M \rb^{11}-303 q^2 Q^2
   \rb^{10}+171 Q^2 \rb^{10}+21
   \rb^{12}\big) \\
&&
	+\: q Q \big(-480 L^{10} M
   \rb+336 L^{10} q^2 Q^2+240 L^{10} \rb^2-1856
   L^8 M \rb^3 \\
&&
	 \quad +\: 1440 L^8 q^2 Q^2 \rb^2-160 L^8
   Q^2 \rb^2+1002 L^8 \rb^4-2642 L^6 M
   \rb^5+2433 L^6 q^2 Q^2 \rb^4 \\
&&
	\quad -\: 639 L^6 Q^2
   \rb^4+1610 L^6 \rb^6-1576 L^4 M
   \rb^7+2038 L^4 q^2 Q^2 \rb^6-970 L^4 Q^2
   \rb^6 \\
&&
	\quad +\: 1211 L^4 \rb^8-202 L^2 M
   \rb^9+873 L^2 q^2 Q^2 \rb^8-687 L^2 Q^2
   \rb^8+376 L^2 \rb^{10} \\
&&
	\quad +\: 108 M
   \rb^{11}+212 q^2 Q^2 \rb^{10}-196 Q^2
   \rb^{10}+13 \rb^{12}\big), \\
F^{\mathcal{K}}_{t\lpow{2}} &=& 3 E^3 \rb^{11} \left(5 \rb^2-3
   L^2\right)-E^2 q Q \rb^8 \left(8 L^4+9 L^2
   \rb^2+73 \rb^4\right) \\
&&
	-\: E \rb^3
   \big(-144 L^8 q^2 Q^2+96 L^8 Q^2-54 L^6 M
   \rb^3-486 L^6 q^2 Q^2 \rb^2+372 L^6 Q^2
   \rb^2 \\
&&
	\quad -\: 162 L^4 M \rb^5-614 L^4 q^2 Q^2
   \rb^4+543 L^4 Q^2 \rb^4-186 L^2 M
   \rb^7-359 L^2 q^2 Q^2 \rb^6 \\
&&
	\quad +\: 366 L^2 Q^2
   \rb^6+12 L^2 \rb^8-78 M \rb^9-159 q^2
   Q^2 \rb^8+99 Q^2 \rb^8+12
   \rb^{10}\big) \\
&&
	-\: q Q \rb^2 \big(-240 L^8 M
   \rb+168 L^8 q^2 Q^2+120 L^8 \rb^2-718 L^6 M
   \rb^3+573 L^6 q^2 Q^2 \rb^2 \\
&&
	\quad -\: 80 L^6 Q^2
   \rb^2+396 L^6 \rb^4-704 L^4 M
   \rb^5+723 L^4 q^2 Q^2 \rb^4-248 L^4 Q^2
   \rb^4 \\
&&
	\quad +\: 463 L^4 \rb^6-190 L^2 M
   \rb^7+410 L^2 q^2 Q^2 \rb^6-268 L^2 Q^2
   \rb^6+206 L^2 \rb^8 \\
&&
	\quad +\: 36 M \rb^9+116 q^2
   Q^2 \rb^8-100 Q^2 \rb^8+19
   \rb^{10}\big),
\end{IEEEeqnarray*}

\begin{equation}
F_{r\lpow{2}} = \frac{1}{6 \pi  \rb^9 \left(L^2+\rb^2\right)^{7/2}
   \left(-2 M \rb+Q^2+\rb^2\right)} (F^{\mathcal{E}}_{r\lpow{2}} \mathcal{E} + F^{\mathcal{K}}_{r\lpow{2}} \mathcal{K}),
\end{equation}
where
\par \vspace{-6pt} \begin{IEEEeqnarray*}{rCl}
F^{\mathcal{E}}_{r\lpow{2}} &=& -24 E^4
   (L-\rb) (L+\rb) \rb^{14} -E^3 q Q
   \left(16 L^6+59 \rb^2 L^4+14 \rb^4 L^2+163
   \rb^6\right) \rb^9 \\
&&
	-\: E^2 \big(36 \rb^{12}-168 M
   \rb^{11}+24 L^2 \rb^{10}-442 q^2 Q^2
   \rb^{10}+186 Q^2 \rb^{10}-444 L^2 M
   \rb^9 \\
&&
	\quad -\: 12 L^4 \rb^8+804 L^2 Q^2
   \rb^8-829 L^2 q^2 Q^2 \rb^8-588 L^4 M
   \rb^7+1626 L^4 Q^2 \rb^6 \\
&&
	\quad -\: 1820 L^4 q^2 Q^2
   \rb^6-420 L^6 M \rb^5+1728 L^6 Q^2
   \rb^4-2081 L^6 q^2 Q^2 \rb^4 \\
&&
	\quad \: -108 L^8 M
   \rb^3+912 L^8 Q^2 \rb^2-1224 L^8 q^2 Q^2
   \rb^2+192 L^{10} Q^2-288 L^{10} q^2 Q^2\big)
   \rb^4 \\
&&
	-\: E q Q \big(-120
   \rb^{12}+470 M \rb^{11}-120 L^2
   \rb^{10}+515 q^2 Q^2 \rb^{10}-479 Q^2
   \rb^{10} \\
&&
	\quad +\: 1186 L^2 M \rb^9+114 L^4
   \rb^8-1963 L^2 Q^2 \rb^8+1664 L^2 q^2 Q^2
   \rb^8+1230 L^4 M \rb^7 \\
&&
	\quad +\: 252 L^6
   \rb^6-3705 L^4 Q^2 \rb^6+3799 L^4 q^2 Q^2
   \rb^6+494 L^6 M \rb^5+186 L^8
   \rb^4 \\
&&
	\quad -\: 3725 L^6 Q^2 \rb^4+4498 L^6 q^2 Q^2
   \rb^4-116 L^8 M \rb^3+48 L^{10}
   \rb^2-1888 L^8 Q^2 \rb^2 \\
&&
	\quad +\: 2664 L^8 q^2 Q^2
   \rb^2-96 L^{10} M \rb-384 L^{10} Q^2+624
   L^{10} q^2 Q^2\big) \rb^3 \\
&&
	+\:  \big(9 \rb^{16}-114 M \rb^{15}+21 L^2
   \rb^{14}+192 M^2 \rb^{14}-163 q^2 Q^2
   \rb^{14}+161 Q^2 \rb^{14} \\
&&
	\quad +\: 460 M q^2 Q^2
   \rb^{13}-400 M Q^2 \rb^{13}-480 L^2 M
   \rb^{13}+15 L^4 \rb^{12}+212 q^4 Q^4
   \rb^{12} \\
&&
	\quad -\: 372 q^2 Q^4 \rb^{12}+152 Q^4
   \rb^{12}+876 L^2 M^2 \rb^{12}+915 L^2 Q^2
   \rb^{12}-736 L^2 q^2 Q^2 \rb^{12} \\
&&
	\quad +\: 2022 L^2 M
   q^2 Q^2 \rb^{11}-2226 L^2 M Q^2 \rb^{11}-864
   L^4 M \rb^{11}+3 L^6 \rb^{10} \\
&&
	\quad +\: 873 L^2 q^4
   Q^4 \rb^{10}+894 L^2 Q^4 \rb^{10}-1799 L^2
   q^2 Q^4 \rb^{10}+1668 L^4 M^2 \rb^{10} \\
&& 
	\quad +\: 2287
   L^4 Q^2 \rb^{10}-1960 L^4 q^2 Q^2
   \rb^{10}+4766 L^4 M q^2 Q^2 \rb^9-5378 L^4 M
   Q^2 \rb^9 \\
&&
	\quad -\: 828 L^6 M \rb^9+2272 L^4 Q^4
   \rb^8+2038 L^4 q^4 Q^4 \rb^8-4141 L^4 q^2
   Q^4 \rb^8 \\
&&
	\quad +\: 1644 L^6 M^2 \rb^8+3161 L^6 Q^2
   \rb^8-3172 L^6 q^2 Q^2 \rb^8+6922 L^6 M q^2
   Q^2 \rb^7 \\
&& 
	\quad -\: 7138 L^6 M Q^2 \rb^7-414 L^8 M
   \rb^7+3158 L^6 Q^4 \rb^6+2433 L^6 q^4 Q^4
   \rb^6 \\
&& \quad -\: 5421 L^6 q^2 Q^4 \rb^6+828 L^8 M^2
   \rb^6+2508 L^8 Q^2 \rb^6-2929 L^8 q^2 Q^2
   \rb^6 \\
&&
	\quad +\: 6006 L^8 M q^2 Q^2 \rb^5-5430 L^8 M
   Q^2 \rb^5-84 L^{10} M \rb^5+2508 L^8 Q^4
   \rb^4 \\
&& 
	\quad +\: 1440 L^8 q^4 Q^4 \rb^4-4091 L^8 q^2
   Q^4 \rb^4+168 L^{10} M^2 \rb^4+1072 L^{10}
   Q^2 \rb^4 \\
&&
	\quad -\: 1432 L^{10} q^2 Q^2
   \rb^4+2864 L^{10} M q^2 Q^2 \rb^3-2228
   L^{10} M Q^2 \rb^3+1072 L^{10} Q^4
   \rb^2 \\
&& 
	\quad +\: 336 L^{10} q^4 Q^4 \rb^2-1672 L^{10}
   q^2 Q^4 \rb^2+192 L^{12} Q^2 \rb^2-288
   L^{12} q^2 Q^2 \rb^2 \\
&&
	\quad +\: 576 L^{12} M q^2 Q^2
   \rb-384 L^{12} M Q^2 \rb+192 L^{12} Q^4-288
   L^{12} q^2 Q^4 \big), \\
F^{\mathcal{K}}_{r\lpow{2}} &=& -3 E^4 \left(5 \rb^2-3 L^2\right)
   \rb^{14}+8 E^3 q Q \left(L^4+11
   \rb^4\right) \rb^{11} \\
&&
	+\: E^2 \big(21
   \rb^{10}-96 M \rb^9+18 L^2 \rb^8-232
   q^2 Q^2 \rb^8+108 Q^2 \rb^8-198 L^2 M
   \rb^7 \\
&&
	\quad -\: 3 L^4 \rb^6+372 L^2 Q^2
   \rb^6-368 L^2 q^2 Q^2 \rb^6-156 L^4 M
   \rb^5+540 L^4 Q^2 \rb^4 \\
&&
	\quad -\: 622 L^4 q^2 Q^2
   \rb^4-54 L^6 M \rb^3+372 L^6 Q^2
   \rb^2-486 L^6 q^2 Q^2 \rb^2+96 L^8 Q^2 \\
&&
	\quad -\: 144
   L^8 q^2 Q^2\big) \rb^6 \\
&&
	+\: E q Q \big(-57
   \rb^{10}+242 M \rb^9-42 L^2 \rb^8+275
   q^2 Q^2 \rb^8-263 Q^2 \rb^8+468 L^2 M
   \rb^7\\
&&
	\quad +\: 63 L^4 \rb^6-870 L^2 Q^2
   \rb^6+769 L^2 q^2 Q^2 \rb^6+258 L^4 M
   \rb^5+72 L^6 \rb^4 \\
&&
	\quad -\: 1191 L^4 Q^2
   \rb^4+1337 L^4 q^2 Q^2 \rb^4-16 L^6 M
   \rb^3+24 L^8 \rb^2-776 L^6 Q^2
   \rb^2 \\
&&
	\quad +\: 1059 L^6 q^2 Q^2 \rb^2-48 L^8 M
   \rb-192 L^8 Q^2+312 L^8 q^2 Q^2\big)
   \rb^5 \\
&&
	-\: \big(3 \rb^{14}-54 M \rb^{13}+6
   L^2 \rb^{12}+96 M^2 \rb^{12}-85 q^2 Q^2
   \rb^{12}+83 Q^2 \rb^{12} \\
&&
	\quad +\: 244 M q^2 Q^2
   \rb^{11}-208 M Q^2 \rb^{11}-192 L^2 M
   \rb^{11}+3 L^4 \rb^{10}+116 q^4 Q^4
   \rb^{10} \\
&&
	\quad -\: 204 q^2 Q^4 \rb^{10}+80 Q^4
   \rb^{10}+360 L^2 M^2 \rb^{10}+407 L^2 Q^2
   \rb^{10}-353 L^2 q^2 Q^2 \rb^{10} \\
&&
	\quad +\: 928 L^2 M
   q^2 Q^2 \rb^9-982 L^2 M Q^2 \rb^9-264 L^4 M
   \rb^9+410 L^2 q^4 Q^4 \rb^8 \\
&&
	\quad +\: 401 L^2 Q^4
   \rb^8-827 L^2 q^2 Q^4 \rb^8+516 L^4 M^2
   \rb^8+825 L^4 Q^2 \rb^8-777 L^4 q^2 Q^2
   \rb^8 \\
&&
	\quad +\: 1776 L^4 M q^2 Q^2 \rb^7-1902 L^4 M
   Q^2 \rb^7-168 L^6 M \rb^7+822 L^4 Q^4
   \rb^6 \\
&&
	\quad +\: 723 L^4 q^4 Q^4 \rb^6-1488 L^4 q^2 Q^4
   \rb^6+336 L^6 M^2 \rb^6+857 L^6 Q^2
   \rb^6 \\
&&
	\quad -\: 955 L^6 q^2 Q^2 \rb^6+1984 L^6 M q^2
   Q^2 \rb^5-1882 L^6 M Q^2 \rb^5-42 L^8 M
   \rb^5 \\
&&
	\quad +\: 857 L^6 Q^4 \rb^4+573 L^6 q^4 Q^4
   \rb^4-1431 L^6 q^2 Q^4 \rb^4+84 L^8 M^2
   \rb^4+452 L^8 Q^2 \rb^4 \\
&& \quad -\: 590 L^8 q^2 Q^2
   \rb^4+1180 L^8 M q^2 Q^2 \rb^3-946 L^8 M Q^2
   \rb^3+452 L^8 Q^4 \rb^2 \\
&& 
	\quad +\: 168 L^8 q^4 Q^4
   \rb^2-710 L^8 q^2 Q^4 \rb^2+96 L^{10} Q^2
   \rb^2-144 L^{10} q^2 Q^2 \rb^2 \\
&&
	\quad +\: 288 L^{10} M
   q^2 Q^2 \rb-192 L^{10} M Q^2 \rb+96 L^{10}
   Q^4-144 L^{10} q^2 Q^4\big) \rb^2,
\end{IEEEeqnarray*}

\begin{equation}
F_{\theta \lpow{2}} =0,
\end{equation}

\begin{equation}
F_{\phi \lpow{2}} = \frac{\rbdot}{6 \pi  L \rb^7 \left(L^2+\rb^2\right)^{5/2}} (F^{\mathcal{E}}_{\phi \lpow{2}} \mathcal{E} + F^{\mathcal{K}}_{\phi \lpow{2}} \mathcal{K}),
\end{equation}
where
\par \vspace{-6pt} \begin{IEEEeqnarray*}{rCl}
F^{\mathcal{E}}_{\phi \lpow{2}} &=& -3 E^2 \rb^{10} \left(\rb^2-7
   L^2\right)-6 E L^2 q Q \rb^3 \left(32
   L^6+96 L^4 \rb^2+99 L^2 \rb^4+43
   \rb^6\right) \\
&&
	+\: \big(288 L^{10} q^2 Q^2-192 L^{10} Q^2+84
   L^8 M \rb^3+1288 L^8 q^2 Q^2 \rb^2-784 L^8
   Q^2 \rb^2 \\
&&
	\quad +\: 258 L^6 M \rb^5+2147 L^6 q^2 Q^2
   \rb^4-1220 L^6 Q^2 \rb^4+276 L^4 M
   \rb^7 \\
&&
	\quad +\: 1611 L^4 q^2 Q^2 \rb^6-876 L^4 Q^2
   \rb^6-3 L^4 \rb^8+102 L^2 M \rb^9+502
   L^2 q^2 Q^2 \rb^8 \\
&&
	\quad -\: 262 L^2 Q^2 \rb^8+14 q^2
   Q^2 \rb^{10}-14 Q^2 \rb^{10}+3
   \rb^{12} \big), \\
F^{\mathcal{K}}_{\phi \lpow{2}} &=& 3 E^2 \rb^{10} \left(\rb^2-3
   L^2\right)+12 E L^2 q Q \rb^5 \left(8
   L^4+17 L^2 \rb^2+11 \rb^4\right) \\
&&
	-\: \rb^2
   \big(144 L^8 q^2 Q^2-96 L^8 Q^2+42 L^6 M
   \rb^3+518 L^6 q^2 Q^2 \rb^2-308 L^6 Q^2
   \rb^2 \\
&&
	\quad +\: 90 L^4 M \rb^5+627 L^4 q^2 Q^2
   \rb^4-345 L^4 Q^2 \rb^4+48 L^2 M
   \rb^7+279 L^2 q^2 Q^2 \rb^6 \\
&&
	\quad -\: 147 L^2 Q^2
   \rb^6+3 L^2 \rb^8+14 q^2 Q^2 \rb^8-14
   Q^2 \rb^8+3 \rb^{10}\big),
\end{IEEEeqnarray*}

\begin{equation}
F_{t \lpow{4}} = \frac{\rbdot}{120 \pi  \rb^{14} \left(L^2+\rb^2\right)^{11/2}} (F^{\mathcal{E}}_{t \lpow{4}} \mathcal{E} + F^{\mathcal{K}}_{t \lpow{4}} \mathcal{K}),
\end{equation}
where
\par \vspace{-6pt} \begin{IEEEeqnarray*}{rCl}
F^{\mathcal{E}}_{t \lpow{4}} &=& -270 E^5 \left(23 L^4-82 \rb^2 L^2+23 \rb^4\right) \rb^{19} \\
&&
	-\: 3 E^4 q Q \big(409600 L^{14}+2428928 \rb^2
   L^{12}+6009984 \rb^4 L^{10}+7960704 \rb^6 L^8 \\
&&
	\quad +\:  5988165 \rb^8 L^6+2462841 \rb^{10} L^4+504611 \rb^{12} L^2+2287
   \rb^{14}\big) \rb^8 \\
&& 	
	+\: 6 E^3 \big(1935 \rb^{18}-498 M \rb^{17}-1755 L^2 \rb^{16}+73233 q^2 Q^2
   \rb^{16}-23793 Q^2 \rb^{16} \\
&&
	\quad +\:  135066 L^2 M \rb^{15}-3555 L^4 \rb^{14}-503331 L^2 Q^2 \rb^{14}+2026086 L^2 q^2 Q^2
   \rb^{14} \\
&&
	\quad +\: 754686 L^4 M \rb^{13}+135 L^6 \rb^{12}-2928267 L^4 Q^2 \rb^{12}+11024742 L^4 q^2 Q^2 \rb^{12} \\
&&
	\quad +\: 1875450
   L^6 M \rb^{11}-8467761 L^6 Q^2 \rb^{10}+29703734 L^6 q^2 Q^2 \rb^{10} \\
&&
	\quad +\: 2541084 L^8 M \rb^9-14208728 L^8 Q^2
   \rb^8+45990045 L^8 q^2 Q^2 \rb^8 \\
&&
	\quad +\: 1942164 L^{10} M \rb^7-14542528 L^{10} Q^2 \rb^6+43192984 L^{10} q^2 Q^2
   \rb^6 \\
&&
	\quad +\: 791808 L^{12} M \rb^5-8995008 L^{12} Q^2 \rb^4+24402464 L^{12} q^2 Q^2 \rb^4 \\
&&
	\quad +\: 134400 L^{14} M \rb^3-3100928
   L^{14} Q^2 \rb^2+7649280 L^{14} q^2 Q^2 \rb^2 \\
&&
	\quad -\: 458752 L^{16} Q^2+1024000 L^{16} q^2 Q^2\big) \rb^5 \\
&&
	-\: 6 E^2 q Q
   \big(-2599 \rb^{18}+43620 M \rb^{17}-264463 L^2 \rb^{16}+253092 q^2 Q^2 \rb^{16} \\
&&
	\quad -\:  147788 Q^2 \rb^{16}+1316950
   L^2 M \rb^{15}-1500961 L^4 \rb^{14}-2663357 L^2 Q^2 \rb^{14} \\
&&
	\quad +\: 5343633 L^2 q^2 Q^2 \rb^{14}+7105430 L^4 M
   \rb^{13}-4045767 L^6 \rb^{12} \\
&&
	\quad -\: 14657063 L^4 Q^2 \rb^{12}+29673969 L^4 q^2 Q^2 \rb^{12}+18183158 L^6 M
   \rb^{11} \\
&&
	\quad -\: 6290580 L^8 \rb^{10}-40769095 L^6 Q^2 \rb^{10}+82391819 L^6 q^2 Q^2 \rb^{10} \\
&&
	\quad +\: 26404942 L^8 M
   \rb^9-5934582 L^{10} \rb^8-66318465 L^8 Q^2 \rb^8 \\
&&
	\quad +\: 132771523 L^8 q^2 Q^2 \rb^8+22886076 L^{10} M \rb^7-3370352
   L^{12} \rb^6 \\
&&
	\quad -\: 66125024 L^{10} Q^2 \rb^6+130657452 L^{10} q^2 Q^2 \rb^6+11685056 L^{12} M \rb^5 \\
&&
	\quad -\: 1063040 L^{14}
   \rb^4-39974000 L^{12} Q^2 \rb^4+77788400 L^{12} q^2 Q^2 \rb^4 \\
&&
	\quad +\: 3213312 L^{14} M \rb^3-143360 L^{16}
   \rb^2-13499776 L^{14} Q^2 \rb^2 \\
&&
	\quad +\: 25838080 L^{14} q^2 Q^2 \rb^2+360448 L^{16} M \rb-1959936 L^{16} Q^2 \\
&&
	\quad +\: 3686400 L^{16}
   q^2 Q^2\big) \rb^4 \\
&& 	
	-\: 2 E \big(2565 \rb^{22}+1638 M \rb^{21}+4320 L^2 \rb^{20}-25164 M^2 \rb^{20}+191439
   q^2 Q^2 \rb^{20}  \\
&&
	\quad -\:  74259 Q^2 \rb^{20}-646632 M q^2 Q^2 \rb^{19}+222360 M Q^2 \rb^{19}+490050 L^2 M \rb^{19} \\
&&
	\quad +\: 810
   L^4 \rb^{18}-890175 q^4 Q^4 \rb^{18}+894150 q^2 Q^4 \rb^{18}-125415 Q^4 \rb^{18} \\
&&
	\quad -\: 1171188 L^2 M^2
   \rb^{18}-1749168 L^2 Q^2 \rb^{18}+4755813 L^2 q^2 Q^2 \rb^{18} \\
&&
	\quad -\: 13980438 L^2 M q^2 Q^2 \rb^{17}+4871790 L^2 M Q^2
   \rb^{17}+3553668 L^4 M \rb^{17} \\
&&
	\quad -\: 1080 L^6 \rb^{16}-16716890 L^2 q^4 Q^4 \rb^{16}-2411564 L^2 Q^4
   \rb^{16} \\
&&
	\quad +\: 15435190 L^2 q^2 Q^4 \rb^{16}-7983504 L^4 M^2 \rb^{16}-12068802 L^4 Q^2 \rb^{16} \\
&&
	\quad +\: 30721971 L^4 q^2 Q^2
   \rb^{16}-84853512 L^4 M q^2 Q^2 \rb^{15}+32445084 L^4 M Q^2 \rb^{15} \\
&&
	\quad +\: 11360772 L^6 M \rb^{15}-135 L^8
   \rb^{14}-15718486 L^4 Q^4 \rb^{14}-92623570 L^4 q^4 Q^4 \rb^{14} \\
&&
	\quad +\: 87621761 L^4 q^2 Q^4 \rb^{14}-24778584 L^6 M^2
   \rb^{14}-41946804 L^6 Q^2 \rb^{14} \\
&&
	\quad +\: 101241717 L^6 q^2 Q^2 \rb^{14}-264445050 L^6 M q^2 Q^2 \rb^{13}+109297668 L^6 M Q^2
   \rb^{13} \\
&&
	\quad +\: 20353518 L^8 M \rb^{13}-53123352 L^6 Q^4 \rb^{12}-258359070 L^6 q^4 Q^4 \rb^{12} \\
&&
	\quad +\: 258395799 L^6 q^2 Q^4
   \rb^{12}-43003836 L^8 M^2 \rb^{12}-86589699 L^8 Q^2 \rb^{12} \\
&&
	\quad +\: 201164550 L^8 q^2 Q^2 \rb^{12}-500092170 L^8 M q^2 Q^2
   \rb^{11}+218747784 L^8 M Q^2 \rb^{11} \\
&&
	\quad +\: 21895290 L^{10} M \rb^{11}-107756587 L^8 Q^4 \rb^{10}-418949260 L^8 q^4 Q^4
   \rb^{10} \\
&&	
	\quad +\: 458466857 L^8 q^2 Q^4 \rb^{10}-44208468 L^{10} M^2 \rb^{10}-113210100 L^{10} Q^2 \rb^{10} \\
&&
	\quad +\: 256457802 L^{10}
   q^2 Q^2 \rb^{10}-611954976 L^{10} M q^2 Q^2 \rb^9+277125054 L^{10} M Q^2 \rb^9 \\
&&
	\quad +\: 14099688 L^{12} M \rb^9-139148756
   L^{10} Q^4 \rb^8-413709950 L^{10} q^4 Q^4 \rb^8 \\
&&
	\quad +\: 518343847 L^{10} q^2 Q^4 \rb^8-26328744 L^{12} M^2 \rb^8-95140992
   L^{12} Q^2 \rb^8 \\
&&
	\quad +\: 212582412 L^{12} q^2 Q^2 \rb^8-491696910 L^{12} M q^2 Q^2 \rb^7+225447636 L^{12} M Q^2 \rb^7 \\
&&
	\quad +\: 5028480
   L^{14} M \rb^7-115837968 L^{12} Q^4 \rb^6-244956825 L^{12} q^4 Q^4 \rb^6 \\
&&
	\quad +\: 377634564 L^{12} q^2 Q^4 \rb^6-7874208 L^{14}
   M^2 \rb^6-50005248 L^{14} Q^2 \rb^6 \\
&&
	\quad +\: 111319920 L^{14} q^2 Q^2 \rb^6-252175704 L^{14} M q^2 Q^2 \rb^5+114559968 L^{14} M
   Q^2 \rb^5 \\
&&
	\quad +\: 766080 L^{16} M \rb^5-60419552 L^{14} Q^4 \rb^4-78854720 L^{14} q^4 Q^4 \rb^4 \\
&&
	\quad +\: 172053376 L^{14} q^2 Q^4
   \rb^4-506880 L^{16} M^2 \rb^4-15003648 L^{16} Q^2 \rb^4 \\
&&
	\quad +\: 33569280 L^{16} q^2 Q^2 \rb^4-75235584 L^{16} M q^2 Q^2
   \rb^3+33181824 L^{16} M Q^2 \rb^3 \\
&&
	\quad -\: 18011776 L^{16} Q^4 \rb^2-9673600 L^{16} q^4 Q^4 \rb^2+44648960 L^{16} q^2 Q^4
   \rb^2 \\
&&
	\quad +\: 184320 L^{18} M^2 \rb^2-1966080 L^{18} Q^2 \rb^2+4454400 L^{18} q^2 Q^2 \rb^2 \\
&&
	\quad -\: 9968640 L^{18} M q^2 Q^2
   \rb+4190208 L^{18} M Q^2 \rb-2347008 L^{18} Q^4 \\
&&
	\quad +\: 460800 L^{18} q^4 Q^4+5038080 L^{18} q^2 Q^4\big) \rb \\
&&
	+\: q Q \big(-3873
   \rb^{22}+151704 M \rb^{21}+104220 L^2 \rb^{20}-318480 M^2 \rb^{20} \\
&&
	\quad +\:  537432 q^2 Q^2 \rb^{20}-437400 Q^2
   \rb^{20}-1400880 M q^2 Q^2 \rb^{19}+1158000 M Q^2 \rb^{19} \\
&&
	\quad +\: 2209068 L^2 M \rb^{19}+2385642 L^4 \rb^{18}-681540
   q^4 Q^4 \rb^{18}+1222536 q^2 Q^4 \rb^{18} \\
&&
	\quad -\: 540996 Q^4 \rb^{18}-5259780 L^2 M^2 \rb^{18}-7254060 L^2 Q^2
   \rb^{18} \\
&&
	\quad +\: 11070048 L^2 q^2 Q^2 \rb^{18}-27285760 L^2 M q^2 Q^2 \rb^{17}+19146976 L^2 M Q^2 \rb^{17} \\
&&
	\quad +\: 5896800 L^4 M
   \rb^{17}+13693068 L^6 \rb^{16}-11937935 L^2 q^4 Q^4 \rb^{16} \\
&&
	\quad -\: 8816919 L^2 Q^4 \rb^{16}+21030902 L^2 q^2 Q^4
   \rb^{16}-23413728 L^4 M^2 \rb^{16} \\
&&
	\quad -\: 41586612 L^4 Q^2 \rb^{16}+72518934 L^4 q^2 Q^2 \rb^{16}-172668464 L^4 M q^2 Q^2
   \rb^{15} \\
&&
	\quad +\: 109539944 L^4 M Q^2 \rb^{15}-9832056 L^6 M \rb^{15}+38639727 L^8 \rb^{14} \\
&&
	\quad -\: 52066836 L^4 Q^4
   \rb^{14}-65374750 L^4 q^4 Q^4 \rb^{14}+124997662 L^4 q^2 Q^4 \rb^{14} \\
&&
	\quad -\: 39974136 L^6 M^2 \rb^{14}-124129464 L^6 Q^2
   \rb^{14}+246341034 L^6 q^2 Q^2 \rb^{14} \\
&&
	\quad -\: 570585876 L^6 M q^2 Q^2 \rb^{13}+325325568 L^6 M Q^2 \rb^{13}-79257528 L^8 M
   \rb^{13} \\
&&
	\quad +\: 63371520 L^{10} \rb^{12}-162622650 L^6 Q^4 \rb^{12}-180762075 L^6 q^4 Q^4 \rb^{12} \\
&&
	\quad +\: 391693206 L^6 q^2 Q^4
   \rb^{12}-1531008 L^8 M^2 \rb^{12}-221987472 L^8 Q^2 \rb^{12} \\
&&
	\quad +\: 504774618 L^8 q^2 Q^2 \rb^{12}-1144260908 L^8 M q^2 Q^2
   \rb^{11}+576749888 L^8 M Q^2 \rb^{11} \\
&&
	\quad -\: 177359988 L^{10} M \rb^{11}+63661152 L^{12} \rb^{10}-307579872 L^8 Q^4
   \rb^{10} \\
&&
	\quad -\: 289670365 L^8 q^4 Q^4 \rb^{10}+746170690 L^8 q^2 Q^4 \rb^{10}+99954492 L^{10} M^2 \rb^{10} \\
&&
	\quad -\: 251169132 L^{10}
   Q^2 \rb^{10}+661812822 L^{10} q^2 Q^2 \rb^{10}-1476501628 L^{10} M q^2 Q^2 \rb^9 \\
&&
	\quad +\: 644010784 L^{10} M Q^2
   \rb^9-209116320 L^{12} M \rb^9+38790000 L^{14} \rb^8 \\
&&
	\quad -\: 372049431 L^{10} Q^4 \rb^8-279974030 L^{10} q^4 Q^4
   \rb^8+914474396 L^{10} q^2 Q^4 \rb^8 \\
&&
	\quad +\: 167517648 L^{12} M^2 \rb^8-181248228 L^{12} Q^2 \rb^8+561925680 L^{12} q^2 Q^2
   \rb^8 \\
&&
	\quad -\: 1240022916 L^{12} M q^2 Q^2 \rb^7+455921832 L^{12} M Q^2 \rb^7-140817312 L^{14} M \rb^7 \\
&&
	\quad +\: 13211520 L^{16}
   \rb^6-290766144 L^{12} Q^4 \rb^6-158502105 L^{12} q^4 Q^4 \rb^6 \\
&&
	\quad +\: 729106752 L^{12} q^2 Q^4 \rb^6+131222304 L^{14} M^2
   \rb^6-80627184 L^{14} Q^2 \rb^6 \\
&&
	\quad +\: 300097560 L^{14} q^2 Q^2 \rb^6-657913984 L^{14} M q^2 Q^2 \rb^5+197253568 L^{14} M Q^2
   \rb^5 \\
&&
	\quad -\: 51400704 L^{16} M \rb^5+1935360 L^{18} \rb^4-142472976 L^{14} Q^4 \rb^4 \\
&&
	\quad -45372320 L^{14} q^4 Q^4
   \rb^4+367169792 L^{14} q^2 Q^4 \rb^4+52204032 L^{16} M^2 \rb^4 \\
&&
	\quad -\: 20015232 L^{16} Q^2 \rb^4+91898880 L^{16} q^2 Q^2
   \rb^4-200906240 L^{16} M q^2 Q^2 \rb^3 \\
&&
	\quad +\: 46962944 L^{16} M Q^2 \rb^3-7925760 L^{18} M \rb^3-39880320 L^{16} Q^4
   \rb^2 \\
&&
	\quad -2784640 L^{16} q^4 Q^4 \rb^2+106448896 L^{16} q^2 Q^4 \rb^2+8515584 L^{18} M^2 \rb^2 \\
&&
	\quad -\: 2095104 L^{18} Q^2
   \rb^2+12334080 L^{18} q^2 Q^2 \rb^2-26972160 L^{18} M q^2 Q^2 \rb \\
&&
	\quad +\: 4595712 L^{18} M Q^2 \rb-4872192 L^{18} Q^4+998400
   L^{18} q^4 Q^4 \\
&&
	\quad +\:  13578240 L^{18} q^2 Q^4\big),
\end{IEEEeqnarray*}
\par \vspace{-6pt} \begin{IEEEeqnarray*}{rCl}
F^{\mathcal{K}}_{t \lpow{4}} &=& 135 E^5 \left(15 L^4-82 \rb^2 L^2+31 \rb^4\right) \rb^{19} \\
&&
	+\: 3 E^4 q Q \big(204800 L^{12}+1035264 \rb^2
   L^{10}+2108736 \rb^4 L^8+2179536 \rb^6 L^6 \\
&&
	\quad +\:  1165317 \rb^8 L^4+297930 \rb^{10} L^2+5077 \rb^{12}\big)
   \rb^{10} \\
&&
	-\: 6 E^3 \big(1260 \rb^{16}+762 M \rb^{15}-360 L^2 \rb^{14}+60393 q^2 Q^2 \rb^{14}-18033 Q^2
   \rb^{14} \\
&&
	\quad +\:  81720 L^2 M \rb^{13}-1620 L^4 \rb^{12}-300507 L^2 Q^2 \rb^{12}+1205172 L^2 q^2 Q^2 \rb^{12} \\
&&
	\quad +\: 362052 L^4
   M \rb^{11}-1439895 L^4 Q^2 \rb^{10}+5327535 L^4 q^2 Q^2 \rb^{10}+690456 L^6 M \rb^9 \\
&&
	\quad -\: 3369105 L^6 Q^2
   \rb^8+11437958 L^6 q^2 Q^2 \rb^8+679266 L^8 M \rb^7-4418516 L^8 Q^2 \rb^6 \\
&&
	\quad +\: 13632434 L^8 q^2 Q^2 \rb^6+337104
   L^{10} M \rb^5-3327216 L^{10} Q^2 \rb^4 \\
&&
	\quad +\: 9270672 L^{10} q^2 Q^2 \rb^4+67200 L^{12} M \rb^3-1349760 L^{12} Q^2
   \rb^2 \\
&&
	\quad +\: 3376640 L^{12} q^2 Q^2 \rb^2-229376 L^{14} Q^2+512000 L^{14} q^2 Q^2\big) \rb^7 \\
&&
	+\: 6 E^2 q Q \big(-4129
   \rb^{16}+38700 M \rb^{15}-157571 L^2 \rb^{14}+194532 q^2 Q^2 \rb^{14}  \\
&&
	\quad -\:  109868 Q^2 \rb^{14}+785110 L^2 M
   \rb^{13}-726375 L^4 \rb^{12}-1571128 L^2 Q^2 \rb^{12} \\
&&
	\quad +\: 3173052 L^2 q^2 Q^2 \rb^{12}+3393570 L^4 M
   \rb^{11}-1560969 L^6 \rb^{10} \\
&&
	\quad -\: 7108428 L^4 Q^2 \rb^{10}+14421009 L^4 q^2 Q^2 \rb^{10}+6829778 L^6 M \rb^9-1869252
   L^8 \rb^8 \\
&&
	\quad -\: 15983728 L^6 Q^2 \rb^8+32215910 L^6 q^2 Q^2 \rb^8+7497018 L^8 M \rb^7-1278336 L^{10} \rb^6 \\
&&
	\quad -\: 20301184
   L^8 Q^2 \rb^6+40369221 L^8 q^2 Q^2 \rb^6+4583136 L^{10} M \rb^5-468800 L^{12} \rb^4 \\
&&
	\quad -\: 14877072 L^{10} Q^2
   \rb^4+29087640 L^{10} q^2 Q^2 \rb^4+1448960 L^{12} M \rb^3 \\
&&
	\quad -\: 71680 L^{14} \rb^2-5892416 L^{12} Q^2 \rb^2+11306240
   L^{12} q^2 Q^2 \rb^2+180224 L^{14} M \rb \\
&&
	\quad -\: 979968 L^{14} Q^2+1843200 L^{14} q^2 Q^2\big) \rb^6 \\
&&
	+\: E \big(3105
   \rb^{20}+10836 M \rb^{19}+6075 L^2 \rb^{18}-50328 M^2 \rb^{18}+305838 q^2 Q^2 \rb^{18}   \\
&&
	\quad -\:  113958 Q^2
   \rb^{18}-1005264 M q^2 Q^2 \rb^{17}+335280 M Q^2 \rb^{17}+619092 L^2 M \rb^{17} \\
&&
	\quad +\: 2835 L^4 \rb^{16}-1330350 q^4
   Q^4 \rb^{16}+1313100 q^2 Q^4 \rb^{16}-182430 Q^4 \rb^{16} \\
&&
	\quad -\: 1464012 L^2 M^2 \rb^{16}-2142162 L^2 Q^2
   \rb^{16}+5759652 L^2 q^2 Q^2 \rb^{16} \\
&&
	\quad -\: 16783704 L^2 M q^2 Q^2 \rb^{15}+5929320 L^2 M Q^2 \rb^{15}+3623508 L^4 M
   \rb^{15} \\
&&
	\quad -\: 135 L^6 \rb^{14}-19788205 L^2 q^4 Q^4 \rb^{14}-2906833 L^2 Q^4 \rb^{14} \\
&&
	\quad +\: 18237230 L^2 q^2 Q^4
   \rb^{14}-8079516 L^4 M^2 \rb^{14}-12410082 L^4 Q^2 \rb^{14} \\
&&
	\quad +\: 31087080 L^4 q^2 Q^2 \rb^{14}-84537504 L^4 M q^2 Q^2
   \rb^{13}+33093468 L^4 M Q^2 \rb^{13} \\
&&
	\quad +\: 9337212 L^6 M \rb^{13}-15995343 L^4 Q^4 \rb^{12}-90035325 L^4 q^4 Q^4
   \rb^{12} \\
&&
	\quad +\: 85965222 L^4 q^2 Q^4 \rb^{12}-20141244 L^6 M^2 \rb^{12}-35944326 L^6 Q^2 \rb^{12} \\
&&
	\quad +\: 85247508 L^6 q^2 Q^2
   \rb^{12}-218096664 L^6 M q^2 Q^2 \rb^{11}+92595732 L^6 M Q^2 \rb^{11} \\
&&
	\quad +\: 13052088 L^8 M \rb^{11}-45146379 L^6 Q^4
   \rb^{10}-202594530 L^6 q^4 Q^4 \rb^{10} \\
&&
	\quad +\: 208408062 L^6 q^2 Q^4 \rb^{10}-27039996 L^8 M^2 \rb^{10}-60463344 L^8 Q^2
   \rb^{10} \\
&&
	\quad +\: 138277194 L^8 q^2 Q^2 \rb^{10}-335909592 L^8 M q^2 Q^2 \rb^9+150422196 L^8 M Q^2 \rb^9 \\
&&
	\quad +\: 10322208 L^{10} M
   \rb^9-74712839 L^8 Q^4 \rb^8-255627035 L^8 q^4 Q^4 \rb^8 \\
&&
	\quad +\: 296963974 L^8 q^2 Q^4 \rb^8-19993032 L^{10} M^2
   \rb^8-62058144 L^{10} Q^2 \rb^8 \\
&&
	\quad +\: 139011672 L^{10} q^2 Q^2 \rb^8-324471696 L^{10} M q^2 Q^2 \rb^7+148931124 L^{10} M Q^2
   \rb^7 \\
&&
	\quad +\: 4358160 L^{12} M \rb^7-75792456 L^{10} Q^4 \rb^6-184174695 L^{10} q^4 Q^4 \rb^6 \\
&&
	\quad +\: 259473420 L^{10} q^2 Q^4
   \rb^6-7280928 L^{12} M^2 \rb^6-38474496 L^{12} Q^2 \rb^6 \\
&&
	\quad +\: 85566000 L^{12} q^2 Q^2 \rb^6-194444088 L^{12} M q^2 Q^2
   \rb^5+88930416 L^{12} M Q^2 \rb^5 \\
&&
	\quad +\: 766080 L^{14} M \rb^5-46566192 L^{12} Q^4 \rb^4-70015920 L^{12} q^4 Q^4
   \rb^4 \\
&&
	\quad +\: 137078976 L^{12} q^2 Q^4 \rb^4-668160 L^{14} M^2 \rb^4-13283328 L^{14} Q^2 \rb^4 \\
&&
	\quad +\: 29671680 L^{14} q^2 Q^2
   \rb^4-66513024 L^{14} M q^2 Q^2 \rb^3+29515392 L^{14} M Q^2 \rb^3 \\
&&
	\quad -\: 15958144 L^{14} Q^4 \rb^2-10076800 L^{14} q^4 Q^4
   \rb^2+40240640 L^{14} q^2 Q^4 \rb^2 \\
&&
	\quad +\: 184320 L^{16} M^2 \rb^2-1966080 L^{16} Q^2 \rb^2+4454400 L^{16} q^2 Q^2
   \rb^2 \\
&&
	\quad -\: 9968640 L^{16} M q^2 Q^2 \rb+4190208 L^{16} M Q^2 \rb-2347008 L^{16} Q^4 \\
&&
	\quad +\: 460800 L^{16} q^4 Q^4+5038080 L^{16} q^2
   Q^4\big) \rb^3 \\
&&
	+\: q Q \big(4683 \rb^{20}-122184 M \rb^{19}-85299 L^2 \rb^{18}+246480 M^2 \rb^{18}  - 402072 q^2 Q^2
   \rb^{18} \\
&&
	\quad +\: 319320 Q^2 \rb^{18}+1040880 M q^2 Q^2 \rb^{17}-841200 M Q^2 \rb^{17}-1240236 L^2 M \rb^{17} \\
&&
	\quad -\: 1440711 L^4
   \rb^{16}+501540 q^4 Q^4 \rb^{16}-888456 q^2 Q^4 \rb^{16}+386916 Q^4 \rb^{16} \\
&&
	\quad +\: 3065100 L^2 M^2 \rb^{16}+4292790
   L^2 Q^2 \rb^{16}-6691122 L^2 q^2 Q^2 \rb^{16} \\
&&
	\quad +\: 16400320 L^2 M q^2 Q^2 \rb^{15}-11326336 L^2 M Q^2 \rb^{15}-1857564 L^4
   M \rb^{15} \\
&&
	\quad -\: 6690429 L^6 \rb^{14}+7040165 L^2 q^4 Q^4 \rb^{14}+5218941 L^2 Q^4 \rb^{14} \\
&&
	\quad -\: 12483314 L^2 q^2 Q^4
   \rb^{14}+10448748 L^4 M^2 \rb^{14}+20474022 L^4 Q^2 \rb^{14} \\
&&
	\quad -\: 36932034 L^4 q^2 Q^2 \rb^{14}+87218124 L^4 M q^2 Q^2
   \rb^{13}-53889036 L^4 M Q^2 \rb^{13} \\
&&
	\quad +\: 8853396 L^6 M \rb^{13}-15209496 L^8 \rb^{12}+25888791 L^4 Q^4
   \rb^{12} \\
&&
	\quad +\: 31686225 L^4 q^4 Q^4 \rb^{12}-62199108 L^4 q^2 Q^4 \rb^{12}+10784652 L^6 M^2 \rb^{12} \\
&&
	\quad +\: 50285358 L^6 Q^2
   \rb^{12}-104940750 L^6 q^2 Q^2 \rb^{12}+240786324 L^6 M q^2 Q^2 \rb^{11} \\
&&
	\quad -\: 131456148 L^6 M Q^2 \rb^{11}+37493124 L^8 M
   \rb^{11}-19479636 L^{10} \rb^{10} \\
&&
	\quad +\: 67184151 L^6 Q^4 \rb^{10}+70609410 L^6 q^4 Q^4 \rb^{10}-162072600 L^6 q^2 Q^4
   \rb^{10} \\
&&
	\quad -\: 13000764 L^8 M^2 \rb^{10}+72091914 L^8 Q^2 \rb^{10}-176070546 L^8 q^2 Q^2 \rb^{10} \\
&&
	\quad +\: 395451188 L^8 M q^2 Q^2
   \rb^9-186243908 L^8 M Q^2 \rb^9+60770952 L^{10} M \rb^9 \\
&&
	\quad -\: 14401200 L^{12} \rb^8+103102425 L^8 Q^4 \rb^8+87628105
   L^8 q^4 Q^4 \rb^8 \\
&&
	\quad -\: 251350660 L^8 q^2 Q^4 \rb^8-44267976 L^{10} M^2 \rb^8+62671716 L^{10} Q^2 \rb^8 \\
&&
	\quad -\: 182240280 L^{10} q^2
   Q^2 \rb^8+403374780 L^{10} M q^2 Q^2 \rb^7-158966172 L^{10} M Q^2 \rb^7 \\
&&
	\quad +\: 51140688 L^{12} M \rb^7-5759040 L^{14}
   \rb^6+97370736 L^{10} Q^4 \rb^6 \\
&&
	\quad +\: 60917535 L^{10} q^4 Q^4 \rb^6-241917126 L^{10} q^2 Q^4 \rb^6-46231344 L^{12} M^2
   \rb^6 \\
&&
	\quad +\: 32408064 L^{12} Q^2 \rb^6-114853740 L^{12} q^2 Q^2 \rb^6+252017952 L^{12} M q^2 Q^2 \rb^5 \\
&&
	\quad -\: 79947504 L^{12} M Q^2
   \rb^5+22232832 L^{14} M \rb^5-967680 L^{16} \rb^4 \\
&&
	\quad +\: 55768176 L^{12} Q^4 \rb^4+21062280 L^{12} q^4 Q^4
   \rb^4-142529664 L^{12} q^2 Q^4 \rb^4 \\
&&
	\quad -\: 22376448 L^{14} M^2 \rb^4+9091008 L^{14} Q^2 \rb^4-40553280 L^{14} q^2 Q^2
   \rb^4 \\
&&
	\quad +\: 88652800 L^{14} M q^2 Q^2 \rb^3-21470848 L^{14} M Q^2 \rb^3+3962880 L^{16} M \rb^3 \\
&&
	\quad +\: 17808576 L^{14} Q^4
   \rb^2+1829120 L^{14} q^4 Q^4 \rb^2-47283968 L^{14} q^2 Q^4 \rb^2 \\
&&
	\quad -\: 4257792 L^{16} M^2 \rb^2+1047552 L^{16} Q^2
   \rb^2-6167040 L^{16} q^2 Q^2 \rb^2 \\
&&
	\quad +\: 13486080 L^{16} M q^2 Q^2 \rb-2297856 L^{16} M Q^2 \rb+2436096 L^{16} Q^4 \\
&&
	\quad -\: 499200
   L^{16} q^4 Q^4-6789120 L^{16} q^2 Q^4\big) \rb^2,
\end{IEEEeqnarray*}

\begin{equation}
F_{r \lpow{4}} = \frac{1}{120 \pi  \rb^{15} \left(L^2+\rb^2\right)^{11/2} \left(-2 M
   \rb+Q^2+\rb^2\right)} (F^{\mathcal{E}}_{r \lpow{4}} \mathcal{E} + F^{\mathcal{K}}_{r \lpow{4}} \mathcal{K}),
\end{equation}
where
\par \vspace{-6pt} \begin{IEEEeqnarray*}{rCl}
F^{\mathcal{E}}_{r \lpow{4}} &=& 270 E^6 \left(23
   L^4-82 \rb^2 L^2+23 \rb^4\right) \rb^{22} \\
&&
	+\: 3 E^5 q Q \big(409600 L^{14}+2428928
   \rb^2 L^{12}+6009984 \rb^4 L^{10}+7960704 \rb^6 L^8 \\
&&
	\quad +\: 5988165
   \rb^8 L^6+2460771 \rb^{10} L^4+511991 \rb^{12} L^2+217
   \rb^{14}\big) \rb^{11} \\
&& 
	-\: 3 E^4 \big(4725 \rb^{18}-2706 M \rb^{17}-6345
   L^2 \rb^{16}+148753 q^2 Q^2 \rb^{16}-46731 Q^2 \rb^{16} \\
&&
	\quad +\: 275802
   L^2 M \rb^{15}-9585 L^4 \rb^{14}-1009497 L^2 Q^2
   \rb^{14}+4556783 L^2 q^2 Q^2 \rb^{14} \\
&&
	\quad +\: 1514322 L^4 M
   \rb^{13}+1485 L^6 \rb^{12}-5859009 L^4 Q^2 \rb^{12}+24512325
   L^4 q^2 Q^2 \rb^{12} \\
&&
	\quad +\: 3748470 L^6 M \rb^{11}-16934307 L^6 Q^2
   \rb^{10}+65395633 L^6 q^2 Q^2 \rb^{10} \\
&&
	\quad +\: 5082168 L^8 M
   \rb^9-28417456 L^8 Q^2 \rb^8+99940794 L^8 q^2 Q^2
   \rb^8 \\
&&
	\quad +\: 3884328 L^{10} M \rb^7-29085056 L^{10} Q^2
   \rb^6+92395952 L^{10} q^2 Q^2 \rb^6 \\
&&
	\quad +\; 1583616 L^{12} M
   \rb^5-17990016 L^{12} Q^2 \rb^4+51233856 L^{12} q^2 Q^2
   \rb^4 \\
&&
	\quad +\: 268800 L^{14} M \rb^3-6201856 L^{14} Q^2 \rb^2+15708160
   L^{14} q^2 Q^2 \rb^2 \\
&&
	\quad -\: 917504 L^{16} Q^2+2048000 L^{16} q^2 Q^2\big)
   \rb^8 \\
&& 
	+\: 6 E^3 q Q \big(-1000
   \rb^{18}+43794 M \rb^{17}-422808 L^2 \rb^{16}+326325 q^2 Q^2
   \rb^{16} \\
&&
	\quad -\: 171917 Q^2 \rb^{16}+1765196 L^2 M \rb^{15}-2660942
   L^4 \rb^{14}-3323278 L^2 Q^2 \rb^{14} \\
&&
	\quad +\: 7369719 L^2 q^2 Q^2
   \rb^{14}+10172968 L^4 M \rb^{13}-7966266 L^6 \rb^{12} \\
&&
	\quad -\: 18741756
   L^4 Q^2 \rb^{12}+40698711 L^4 q^2 Q^2 \rb^{12}+27899876 L^6 M
   \rb^{11} \\
&&
	\quad -\: 13726866 L^8 \rb^{10}-53157490 L^6 Q^2
   \rb^{10}+112095553 L^6 q^2 Q^2 \rb^{10} \\
&&
	\quad +\: 43818598 L^8 M
   \rb^9-14330022 L^{10} \rb^8-87963479 L^8 Q^2 \rb^8 \\
&&
	\quad +\: 178761568
   L^8 q^2 Q^2 \rb^8+41619120 L^{10} M \rb^7-9006128 L^{12}
   \rb^6 \\
&&
	\quad -\: 89062992 L^{10} Q^2 \rb^6+173850436 L^{10} q^2 Q^2
   \rb^6+23748416 L^{12} M \rb^5 \\
&&
	\quad -\: 3146880 L^{14} \rb^4-54604784
   L^{12} Q^2 \rb^4+102190864 L^{12} q^2 Q^2 \rb^4 \\
&&
	\quad +\: 7515392 L^{14} M
   \rb^3-471040 L^{16} \rb^2-18684544 L^{14} Q^2 \rb^2 \\
&&
	\quad +\: 33487360
   L^{14} q^2 Q^2 \rb^2+1015808 L^{16} M \rb-2746368 L^{16} Q^2 \\
&&
	\quad +\: 4710400
   L^{16} q^2 Q^2\big) \rb^7 \\
&&
	+\: 2 E^2 \big(4860 \rb^{22}-5256 M \rb^{21}+4860
   L^2 \rb^{20}-20556 M^2 \rb^{20}+287988 q^2 Q^2 \rb^{20} \\
&&
	\quad -\: 113820
   Q^2 \rb^{20}-954996 M q^2 Q^2 \rb^{19}+303768 M Q^2
   \rb^{19}+650700 L^2 M \rb^{19} \\
&&
	\quad -\: 4860 L^4 \rb^{18}-1649451 q^4
   Q^4 \rb^{18}+1426266 q^2 Q^4 \rb^{18}-167271 Q^4
   \rb^{18} \\
&&
	\quad -\: 1494648 L^2 M^2 \rb^{18}-2700336 L^2 Q^2
   \rb^{18}+8760606 L^2 q^2 Q^2 \rb^{18} \\
&&
	\quad -\: 24354096 L^2 M q^2 Q^2
   \rb^{17}+6936936 L^2 M Q^2 \rb^{17}+4708764 L^4 M
   \rb^{17} \\
&&
	\quad -\: 4860 L^6 \rb^{16}-32747789 L^2 q^4 Q^4
   \rb^{16}-3363272 L^2 Q^4 \rb^{16} \\
&&
	\quad +\:  26636665 L^2 q^2 Q^4
   \rb^{16}-10271016 L^4 M^2 \rb^{16}-18674064 L^4 Q^2
   \rb^{16} \\
&&
	\quad +\: 58880031 L^4 q^2 Q^2 \rb^{16}-153480156 L^4 M q^2 Q^2
   \rb^{15}+46788024 L^4 M Q^2 \rb^{15} \\
&&
	\quad +\: 14870376 L^6 M
   \rb^{15}-22318078 L^4 Q^4 \rb^{14}-181645477 L^4 q^4 Q^4
   \rb^{14} \\
&&
	\quad +\: 155248127 L^4 q^2 Q^4 \rb^{14}-31782672 L^6 M^2
   \rb^{14}-65376552 L^6 Q^2 \rb^{14} \\
&&
	\quad +\: 197390685 L^6 q^2 Q^2
   \rb^{14}-487017858 L^6 M q^2 Q^2 \rb^{13}+159651648 L^6 M Q^2
   \rb^{13} \\
&&
	\quad +\: 26378784 L^8 M \rb^{13}-76549320 L^6 Q^4
   \rb^{12}-505534527 L^6 q^4 Q^4 \rb^{12} \\
&&
	\quad +\: 464714751 L^6 q^2 Q^4
   \rb^{12}-55054908 L^8 M^2 \rb^{12}-136397988 L^8 Q^2
   \rb^{12} \\
&&
	\quad +\: 393051093 L^8 q^2 Q^2 \rb^{12}-925336602 L^8 M q^2 Q^2
   \rb^{11}+324390168 L^8 M Q^2 \rb^{11} \\
&&
	\quad +\: 28131660 L^{10} M
   \rb^{11}-157565011 L^8 Q^4 \rb^{10}-817263829 L^8 q^4 Q^4
   \rb^{10} \\
&&
	\quad +\: 830437055 L^8 q^2 Q^4 \rb^{10}-56681208 L^{10} M^2
   \rb^{10}-180622728 L^{10} Q^2 \rb^{10} \\
&&
	\quad +\: 495894657 L^{10} q^2 Q^2
   \rb^{10}-1123879422 L^{10} M q^2 Q^2 \rb^9 \\
&&
	\quad +\: 418186680 L^{10} M Q^2
   \rb^9 + 17967564 L^{12} M \rb^9-206561384 L^{10} Q^4
   \rb^8 \\
&&
	\quad -\: 805682306 L^{10} q^4 Q^4 \rb^8 + 938352028 L^{10} q^2 Q^4
   \rb^8-34064496 L^{12} M^2 \rb^8 \\
&&
	\quad -\: 153977808 L^{12} Q^2
   \rb^8 + 402481380 L^{12} q^2 Q^2 \rb^8-886327902 L^{12} M q^2 Q^2
   \rb^7 \\
&&
	\quad +\: 346989144 L^{12} M Q^2 \rb^7 + 6359040 L^{14} M
   \rb^7-174674784 L^{12} Q^4 \rb^6 \\
&&
	\quad -\: 478322025 L^{12} q^4 Q^4
   \rb^6+677344476 L^{12} q^2 Q^4 \rb^6-10535328 L^{14} M^2
   \rb^6 \\
&&
	\quad -\: 82191552 L^{14} Q^2 \rb^6+204471504 L^{14} q^2 Q^2
   \rb^6-441740568 L^{14} M q^2 Q^2 \rb^5 \\
&&
	\quad +\: 180263136 L^{14} M Q^2
   \rb^5+961920 L^{16} M \rb^5-92605856 L^{14} Q^4
   \rb^4 \\
&&
	\quad -\: 156368960 L^{14} q^4 Q^4 \rb^4+302515168 L^{14} q^2 Q^4
   \rb^4-898560 L^{16} M^2 \rb^4 \\
&&
	\quad -\: 25069824 L^{16} Q^2
   \rb^4+59343360 L^{16} q^2 Q^2 \rb^4-127004928 L^{16} M q^2 Q^2
   \rb^3 \\
&&
	\quad +\; 53510016 L^{16} M Q^2 \rb^3-28077952 L^{16} Q^4
   \rb^2-20732800 L^{16} q^4 Q^4 \rb^2 \\
&&
	\quad +\: 75872768 L^{16} q^2 Q^4
   \rb^2+184320 L^{18} M^2 \rb^2-3342336 L^{18} Q^2 \rb^2 \\
&&
	\quad +\: 7526400
   L^{18} q^2 Q^2 \rb^2-16112640 L^{18} M q^2 Q^2 \rb+6942720 L^{18} M
   Q^2 \rb \\
&&
	\quad -\: 3723264 L^{18} Q^4+460800 L^{18} q^4 Q^4+8110080 L^{18} q^2
   Q^4\big) \rb^4 \\
&&
	+\: E q Q \big(5079 \rb^{22}-235620 M
   \rb^{21}+992100 L^2 \rb^{20}+504744 M^2 \rb^{20} \\
&&
	\quad -\: 1350310 q^2
   Q^2 \rb^{20}+888686 Q^2 \rb^{20}+3554144 M q^2 Q^2
   \rb^{19}-2263552 M Q^2 \rb^{19} \\
&&
	\quad -\: 9084768 L^2 M \rb^{19}+7718154
   L^4 \rb^{18}+2461890 q^4 Q^4 \rb^{18}-3440836 q^2 Q^4
   \rb^{18} \\
&&
	\quad +\: 1088258 Q^4 \rb^{18}+14973516 L^2 M^2
   \rb^{18}+18405168 L^2 Q^2 \rb^{18} \\
&&
	\quad -\: 31970494 L^2 q^2 Q^2
   \rb^{18}+78024276 L^2 M q^2 Q^2 \rb^{17}-45671860 L^2 M Q^2
   \rb^{17} \\
&&
	\quad -\: 62892456 L^4 M \rb^{17}+27980196 L^6
   \rb^{16}+45371715 L^2 q^4 Q^4 \rb^{16} \\
&&
	\quad +\: 20187859 L^2 Q^4
   \rb^{16}-63290102 L^2 q^2 Q^4 \rb^{16}+98735712 L^4 M^2
   \rb^{16} \\
&&
	\quad +\: 121291696 L^4 Q^2 \rb^{16}-215221456 L^4 q^2 Q^2
   \rb^{16}+504892648 L^4 M q^2 Q^2 \rb^{15} \\
&&
	\quad -\: 295031728 L^4 M Q^2
   \rb^{15}-209019312 L^6 M \rb^{15}+59451759 L^8
   \rb^{14} \\
&&
	\quad +\: 128965872 L^4 Q^4 \rb^{14}+250621890 L^4 q^4 Q^4
   \rb^{14}-381499764 L^4 q^2 Q^4 \rb^{14} \\
&&
	\quad +\: 315106536 L^6 M^2
   \rb^{14}+411774712 L^6 Q^2 \rb^{14}-736675088 L^6 q^2 Q^2
   \rb^{14} \\
&&
	\quad +\: 1675177216 L^6 M q^2 Q^2 \rb^{13}-980869592 L^6 M Q^2
   \rb^{13}-407418468 L^8 M \rb^{13} \\
&&
	\quad +\: 79660800 L^{10}
   \rb^{12}+430949890 L^6 Q^4 \rb^{12}+697480215 L^6 q^4 Q^4
   \rb^{12} \\
&&
	\quad -\: 1196335424 L^6 q^2 Q^4 \rb^{12}+587111736 L^8 M^2
   \rb^{12}+839497638 L^8 Q^2 \rb^{12} \\
&&
	\quad -\: 1506597818 L^8 q^2 Q^2
   \rb^{12}+3343433448 L^8 M q^2 Q^2 \rb^{11} \\
&&
	\quad -\: 1956511088 L^8 M Q^2
   \rb^{11}-499890096 L^{10} M \rb^{11}+68459784 L^{12}
   \rb^{10} \\
&&
	\quad +\: 869332598 L^8 Q^4 \rb^{10}+1127568885 L^8 q^4 Q^4
   \rb^{10}-2262598504 L^8 q^2 Q^4 \rb^{10} \\
&&
	\quad +\: 683252172 L^{10} M^2
   \rb^{10}+1091330168 L^{10} Q^2 \rb^{10}-1959126506 L^{10} q^2 Q^2
   \rb^{10} \\
&&
	\quad +\: 4269207740 L^{10} M q^2 Q^2 \rb^9-2487072788 L^{10} M Q^2
   \rb^9-392419632 L^{12} M \rb^9 \\
&&
	\quad +\: 36753360 L^{14}
   \rb^8+1121055459 L^{10} Q^4 \rb^8+1107393930 L^{10} q^4 Q^4
   \rb^8 \\
&&
	\quad -\: 2735560170 L^{10} q^2 Q^4 \rb^8+503329248 L^{12} M^2
   \rb^8+916123916 L^{12} Q^2 \rb^8 \\
&&
	\quad -\: 1642504384 L^{12} q^2 Q^2
   \rb^8+3534244496 L^{12} M q^2 Q^2 \rb^7 \\
&&
	\quad -\: 2040857344 L^{12} M Q^2
   \rb^7 - 191819232 L^{14} M \rb^7+11256960 L^{16} \rb^6 \\
&&
	\quad +\: 934914848
   L^{12} Q^4 \rb^6 + 648415755 L^{12} q^4 Q^4 \rb^6 \\
&&
	\quad -\: 2139789760 L^{12}
   q^2 Q^4 \rb^6+227511840 L^{14} M^2 \rb^6 + 482597808 L^{14} Q^2
   \rb^6 \\
&&
	\quad -\: 863335480 L^{14} q^2 Q^2 \rb^6+1843461552 L^{14} M q^2 Q^2
   \rb^5  \\
&&
	\quad -\:  1050699904 L^{14} M Q^2 \rb^5  - 53273856 L^{16} M
   \rb^5+1505280 L^{18} \rb^4  \\
&&
	\quad +\:  489728848 L^{14} Q^4
   \rb^4  + 203081760 L^{14} q^4 Q^4 \rb^4-1051874624 L^{14} q^2 Q^4
   \rb^4 \\
&&
	\quad +\: 57220608 L^{16} M^2 \rb^4 + 145505920 L^{16} Q^2
   \rb^4-259578880 L^{16} q^2 Q^2 \rb^4 \\
&&
	\quad +\: 552460288 L^{16} M q^2 Q^2
   \rb^3-309561856 L^{16} M Q^2 \rb^3-6426624 L^{18} M
   \rb^3 \\
&&
	\quad +\: 146918784 L^{16} Q^4 \rb^2+22131840 L^{16} q^4 Q^4
   \rb^2-296288256 L^{16} q^2 Q^4 \rb^2 \\
&&
	\quad +\: 6057984 L^{18} M^2
   \rb^2+19200000 L^{18} Q^2 \rb^2-34145280 L^{18} q^2 Q^2
   \rb^2 \\
&&
	\quad +\: 72714240 L^{18} M q^2 Q^2 \rb-39911424 L^{18} M Q^2
   \rb+19298304 L^{18} Q^4 \\
&&
	\quad -\: 1920000 L^{18} q^4 Q^4-36556800 L^{18} q^2
   Q^4\big) \rb^3 \\
&&
	+\: \big(- 1485 \rb^{26}+1530 M \rb^{25}-5130 L^2 \rb^{24}+30960 M^2
   \rb^{24}-143937 q^2 Q^2 \rb^{24} \\
&&
	\quad +\: 87267 Q^2 \rb^{24}-56160 M^3
   \rb^{23}+852232 M q^2 Q^2 \rb^{23}-465776 M Q^2
   \rb^{23} \\
&&
	\quad -\: 472860 L^2 M \rb^{23}-6750 L^4 \rb^{22}+782620 q^4
   Q^4 \rb^{22}-988320 q^2 Q^4 \rb^{22} \\
&&
	\quad +\: 254420 Q^4
   \rb^{22}+2276604 L^2 M^2 \rb^{22}+2285088 L^2 Q^2
   \rb^{22}+601744 M^2 Q^2 \rb^{22} \\
&&
	\quad -\: 4501074 L^2 q^2 Q^2
   \rb^{22}-1159280 M^2 q^2 Q^2 \rb^{22}-1891256 M q^4 Q^4
   \rb^{21} \\
&&
	\quad +\: 2400112 M q^2 Q^4 \rb^{21}-618168 M Q^4
   \rb^{21}-2620728 L^2 M^3 \rb^{21} \\
&&
	\quad +\: 23322076 L^2 M q^2 Q^2
   \rb^{21}-11761256 L^2 M Q^2 \rb^{21}-4272120 L^4 M
   \rb^{21} \\
&&
	\quad -\: 4320 L^6 \rb^{20}-681540 q^6 Q^6 \rb^{20}+1467724 q^4
   Q^6 \rb^{20}-951852 q^2 Q^6 \rb^{20} \\
&&
	\quad +\: 165668 Q^6
   \rb^{20}+16609959 L^2 q^4 Q^4 \rb^{20}+5777113 L^2 Q^4
   \rb^{20} \\
&&
	\quad -\: 21087008 L^2 q^2 Q^4 \rb^{20}+19022328 L^4 M^2
   \rb^{20}+18528642 L^4 Q^2 \rb^{20} \\
&&
	\quad +\: 14705764 L^2 M^2 Q^2
   \rb^{20}-37951554 L^4 q^2 Q^2 \rb^{20}-29064620 L^2 M^2 q^2 Q^2
   \rb^{20} \\
&&
	\quad -\: 38365582 L^2 M q^4 Q^4 \rb^{19}+49504704 L^2 M q^2 Q^4
   \rb^{19}-13671490 L^2 M Q^4 \rb^{19} \\
&&
	\quad -\: 20902176 L^4 M^3
   \rb^{19}+185208580 L^4 M q^2 Q^2 \rb^{19}-92789276 L^4 M Q^2
   \rb^{19} \\
&&
	\quad -\: 16734960 L^6 M \rb^{19}-1485 L^8 \rb^{18}-11937935
   L^2 q^6 Q^6 \rb^{18} \\
&&
	\quad +\: 26570813 L^2 q^4 Q^6 \rb^{18}+3486895 L^2 Q^6
   \rb^{18}-18044573 L^2 q^2 Q^6 \rb^{18} \\
&&
	\quad +\: 44676012 L^4 Q^4
   \rb^{18}+110210600 L^4 q^4 Q^4 \rb^{18}-151530634 L^4 q^2 Q^4
   \rb^{18} \\
&&
	\quad +\: 72175608 L^6 M^2 \rb^{18}+77056620 L^6 Q^2
   \rb^{18}+113316832 L^4 M^2 Q^2 \rb^{18}  \\
&&
	\quad -\:  161379888 L^6 q^2 Q^2
   \rb^{18}-220688504 L^4 M^2 q^2 Q^2 \rb^{18}  \\
&&
	\quad -\:  248051796 L^4 M q^4 Q^4
   \rb^{17} + 347390984 L^4 M q^2 Q^4 \rb^{17}  \\
&&
	\quad -\:  103714112 L^4 M Q^4
   \rb^{17}-77376816 L^6 M^3 \rb^{17} + 754691492 L^6 M q^2 Q^2
   \rb^{17} \\
&&
	\quad -\: 376009576 L^6 M Q^2 \rb^{17}-37265310 L^8 M
   \rb^{17} - 270 L^{10} \rb^{16} \\
&&
	\quad -\: 65374750 L^4 q^6 Q^6
   \rb^{16}+26140620 L^4 Q^6 \rb^{16}+162689328 L^4 q^4 Q^6
   \rb^{16} \\
&&
	\quad -\: 121673662 L^4 q^2 Q^6 \rb^{16}+181051678 L^6 Q^4
   \rb^{16}+375464447 L^6 q^4 Q^4 \rb^{16} \\
&&
	\quad -\: 574144200 L^6 q^2 Q^4
   \rb^{16}+156598704 L^8 M^2 \rb^{16}+194083179 L^8 Q^2
   \rb^{16} \\
&&
	\quad +\: 448976600 L^6 M^2 Q^2 \rb^{16}-413388705 L^8 q^2 Q^2
   \rb^{16} \\
&&
	\quad -\: 868729408 L^6 M^2 q^2 Q^2 \rb^{16} - 828832702 L^6 M q^4 Q^4
   \rb^{15} \\
&&
	\quad +\: 1289586764 L^6 M q^2 Q^4 \rb^{15}-413125572 L^6 M Q^4
   \rb^{15} - 164124288 L^8 M^3 \rb^{15} \\
&&
	\quad +\: 1872247664 L^8 M q^2 Q^2
   \rb^{15}-923443576 L^8 M Q^2 \rb^{15}-51573060 L^{10} M
   \rb^{15} \\
&&
	\quad +\: 103990738 L^6 Q^6 \rb^{14}-180762075 L^6 q^6 Q^6
   \rb^{14}+520816619 L^6 q^4 Q^6 \rb^{14} \\
&&
	\quad -\: 437564430 L^6 q^2 Q^6
   \rb^{14}+448139600 L^8 Q^4 \rb^{14}+764987253 L^8 q^4 Q^4
   \rb^{14} \\
&&
	\quad -\: 1336458556 L^8 q^2 Q^4 \rb^{14}+210390300 L^{10} M^2
   \rb^{14} \\
&&
	\quad +\: 318061452 L^{10} Q^2 \rb^{14}  + 1078074080 L^8 M^2 Q^2
   \rb^{14}-686947758 L^{10} q^2 Q^2 \rb^{14} \\
&&
	\quad -\: 2096427664 L^8 M^2 q^2
   Q^2 \rb^{14}-1664686178 L^8 M q^4 Q^4 \rb^{13} \\
&&
	\quad +\: 2949083520 L^8 M q^2
   Q^4 \rb^{13}  - 1006115840 L^8 M Q^4 \rb^{13} \\
&&
	\quad -\: 214486200 L^{10} M^3
   \rb^{13}+3037465556 L^{10} M q^2 Q^2 \rb^{13}  \\
&&
	\quad -\:  1476426472 L^{10} M
   Q^2 \rb^{13}  - 45416340 L^{12} M \rb^{13}+254054936 L^8 Q^6
   \rb^{12}  \\
&&
	\quad -\:  289670365 L^8 q^6 Q^6 \rb^{12}+1006383325 L^8 q^4 Q^6
   \rb^{12}  - 970022524 L^8 q^2 Q^6 \rb^{12}  \\
&&
	\quad +\:  724890081 L^{10} Q^4
   \rb^{12}+988450000 L^{10} q^4 Q^4 \rb^{12}  - 2042026856 L^{10} q^2 Q^4
   \rb^{12} \\
&&
	\quad +\: 177812424 L^{12} M^2 \rb^{12}+348607368 L^{12} Q^2
   \rb^{12}+1684701956 L^{10} M^2 Q^2 \rb^{12} \\
&&
	\quad -\: 761936628 L^{12} q^2 Q^2
   \rb^{12}-3328419484 L^{10} M^2 q^2 Q^2 \rb^{12} \\
&&
	\quad -\: 2129775984 L^{10} M
   q^4 Q^4 \rb^{11}  + 4439274664 L^{10} M q^2 Q^4 \rb^{11} \\
&&
	\quad -\: 1602386146
   L^{10} M Q^4 \rb^{11} - 173959488 L^{12} M^3 \rb^{11} \\
&&
	\quad +\: 3310562932
   L^{12} M q^2 Q^2 \rb^{11} - 1579573916 L^{12} M Q^2 \rb^{11} - 24858720
   L^{14} M \rb^{11} \\
&&
	\quad +\: 406828359 L^{10} Q^6
   \rb^{10}-279974030 L^{10} q^6 Q^6 \rb^{10} + 1241111574 L^{10} q^4 Q^6
   \rb^{10} \\
&&
	\quad -\: 1412587877 L^{10} q^2 Q^6 \rb^{10}+786307448 L^{12} Q^4
   \rb^{10}+817825051 L^{12} q^4 Q^4 \rb^{10} \\
&&
	\quad -\: 2099387082 L^{12} q^2 Q^4
   \rb^{10}+90918288 L^{14} M^2 \rb^{10}+254628672 L^{14} Q^2
   \rb^{10} \\
&&
	\quad +\: 1760865424 L^{12} M^2 Q^2 \rb^{10}-562374600 L^{14} q^2 Q^2
   \rb^{10} \\
&&
	\quad -\: 3569449736 L^{12} M^2 q^2 Q^2 \rb^{10} - 1751821658 L^{12} M
   q^4 Q^4 \rb^9 \\
&&
	\quad +\: 4509487672 L^{12} M q^2 Q^4 \rb^9-1712343000 L^{12} M
   Q^4 \rb^9-82401696 L^{14} M^3 \rb^9 \\
&&
	\quad +\: 2413874764 L^{14} M q^2 Q^2
   \rb^9-1126727336 L^{14} M Q^2 \rb^9-7735680 L^{16} M
   \rb^9 \\
&&
	\quad +\: 437700080 L^{12} Q^6 \rb^8-158502105 L^{12} q^6 Q^6
   \rb^8+985006123 L^{12} q^4 Q^6 \rb^8 \\
&&
	\quad -\: 1383307218 L^{12} q^2 Q^6
   \rb^8+569425072 L^{14} Q^4 \rb^8+417700446 L^{14} q^4 Q^4
   \rb^8 \\
&&
	\quad -\: 1444733120 L^{14} q^2 Q^4 \rb^8+24429888 L^{16} M^2
   \rb^8+119331840 L^{16} Q^2 \rb^8 \\
&&
	\quad +\: 1226423392 L^{14} M^2 Q^2
   \rb^8-266027232 L^{16} q^2 Q^2 \rb^8 \\
&&
	\quad -\: 2573503448 L^{14} M^2 q^2 Q^2
   \rb^8 - 893119756 L^{14} M q^4 Q^4 \rb^7 \\
&&
	\quad +\: 3075499116 L^{14} M q^2 Q^4
   \rb^7-1222204072 L^{14} M Q^4 \rb^7  - 17917056 L^{16} M^3
   \rb^7 \\
&&
	\quad +\: 1133034928 L^{16} M q^2 Q^2 \rb^7-515896256 L^{16} M Q^2
   \rb^7-1048320 L^{18} M \rb^7 \\
&&
	\quad +\: 314796400 L^{14} Q^6 \rb^6-45372320
   L^{14} q^6 Q^6 \rb^6+484772678 L^{14} q^4 Q^6 \rb^6 \\
&&
	\quad -\: 905414312 L^{14}
   q^2 Q^6 \rb^6+264921408 L^{16} Q^4 \rb^6+116964608 L^{16} q^4 Q^4
   \rb^6 \\
&&
	\quad -\: 640056544 L^{16} q^2 Q^4 \rb^6+1764864 L^{18} M^2
   \rb^6+32538624 L^{18} Q^2 \rb^6 \\
&&
	\quad +\: 547952320 L^{16} M^2 Q^2
   \rb^6-73153536 L^{18} q^2 Q^2 \rb^6 \\
&&
	\quad -\: 1199712224 L^{16} M^2 q^2 Q^2
   \rb^6  - 251037696 L^{16} M q^4 Q^4 \rb^5 \\
&&
	\quad +\: 1354526192 L^{16} M q^2 Q^4
   \rb^5-560676032 L^{16} M Q^4 \rb^5+663552 L^{18} M^3
   \rb^5 \\
&&
	\quad +\: 310307328 L^{18} M q^2 Q^2 \rb^5-137488896 L^{18} M Q^2
   \rb^5+145589568 L^{16} Q^6 \rb^4 \\
&&
	\quad -\: 2784640 L^{16} q^6 Q^6
   \rb^4+131514624 L^{16} q^4 Q^6 \rb^4-380682880 L^{16} q^2 Q^6
   \rb^4 \\
&&
	\quad +\: 71781120 L^{18} Q^4 \rb^4+11623680 L^{18} q^4 Q^4
   \rb^4-165414912 L^{18} q^2 Q^4 \rb^4 \\
&&
	\quad -\: 368640 L^{20} M^2
   \rb^4+3932160 L^{20} Q^2 \rb^4+142394880 L^{18} M^2 Q^2
   \rb^4 \\
&&
	\quad -\: 8908800 L^{20} q^2 Q^2 \rb^4-327595008 L^{18} M^2 q^2 Q^2
   \rb^4-25551360 L^{18} M q^4 Q^4 \rb^3 \\
&&
	\quad +\: 349112832 L^{18} M
   q^2 Q^4 \rb^3-149848320 L^{18} M Q^4 \rb^3+737280 L^{20} M^3
   \rb^3 \\
&&
	\quad +\: 37754880 L^{20} M q^2 Q^2 \rb^3-16244736 L^{20} M Q^2
   \rb^3+39242496 L^{18} Q^6 \rb^2 \\
&&
	\quad +\: 998400 L^{18} q^6 Q^6
   \rb^2+12867840 L^{18} q^4 Q^6 \rb^2-93103104 L^{18} q^2 Q^6
   \rb^2 \\
&&
	\quad +\: 8626176 L^{20} Q^4 \rb^2-921600 L^{20} q^4 Q^4
   \rb^2-18984960 L^{20} q^2 Q^4 \rb^2 \\
&&
	\quad +\: 16392192 L^{20} M^2 Q^2
   \rb^2-39874560 L^{20} M^2 q^2 Q^2 \rb^2+1843200 L^{20} M q^4 Q^4
   \rb \\
&&
	\quad +\: 40089600 L^{20} M q^2 Q^4 \rb-17768448 L^{20} M Q^4
   \rb+4694016 L^{20} Q^6 \\
&&
	\quad -\: 921600 L^{20} q^4 Q^6-10076160 L^{20} q^2 Q^6 \big),
\end{IEEEeqnarray*}
\par \vspace{-6pt} \begin{IEEEeqnarray*}{rCl}
F^{\mathcal{K}}_{r \lpow{4}} &=& -135 E^6 \left(15 L^4-82 \rb^2 L^2+31 \rb^4\right)
   \rb^{22} \\
&&
	-\: 6 E^5 q Q \big(102400 L^{12}+517632 \rb^2
   L^{10}+1054368 \rb^4 L^8+1089768 \rb^6 L^6 \\
&&
	\quad +\: 582321 \rb^8
   L^4+150810 \rb^{10} L^2+1841 \rb^{12}\big) \rb^{13} \\
&&
	+\: 3
   E^4 \big(3105 \rb^{16}+354 M \rb^{15}-2025 L^2
   \rb^{14}+125863 q^2 Q^2 \rb^{14}-35481 Q^2 \rb^{14} \\
&&
	\quad +\: 166050 L^2
   M \rb^{13}-4725 L^4 \rb^{12}-602319 L^2 Q^2 \rb^{12}+2708274
   L^2 q^2 Q^2 \rb^{12} \\
&&
	\quad +\: 727074 L^4 M \rb^{11}+405 L^6
   \rb^{10}-2881275 L^4 Q^2 \rb^{10}+11820387 L^4 q^2 Q^2
   \rb^{10} \\
&&
	\quad +\: 1380102 L^6 M \rb^9-6737805 L^6 Q^2 \rb^8+25055452
   L^6 q^2 Q^2 \rb^8 \\
&&
	\quad +\: 1358532 L^8 M \rb^7-8837032 L^8 Q^2
   \rb^6+29373604 L^8 q^2 Q^2 \rb^6 \\
&&
	\quad +\: 674208 L^{10} M \rb^5-6654432
   L^{10} Q^2 \rb^4+19576608 L^{10} q^2 Q^2 \rb^4 \\
&&
	\quad +\: 134400 L^{12} M
   \rb^3-2699520 L^{12} Q^2 \rb^2+6958080 L^{12} q^2 Q^2
   \rb^2 \\
&&
	\quad -\: 458752 L^{14} Q^2+1024000 L^{14} q^2 Q^2\big) \rb^{10} \\
&&
	-\: 6
   E^3 q Q \big(-3925 \rb^{16}+41574 M \rb^{15}-255943 L^2
   \rb^{14}+254925 q^2 Q^2 \rb^{14} \\
&&
	\quad -\: 128957 Q^2 \rb^{14}+1062854
   L^2 M \rb^{13}-1328823 L^4 \rb^{12}-1969647 L^2 Q^2
   \rb^{12} \\
&&
	\quad +\: 4378224 L^2 q^2 Q^2 \rb^{12}+4957278 L^4 M
   \rb^{11}-3212301 L^6 \rb^{10} \\
&&
	\quad -\: 9149151 L^4 Q^2
   \rb^{10}+19748544 L^4 q^2 Q^2 \rb^{10}+10822898 L^6 M
   \rb^9 \\
&&
	\quad -\: 4321320 L^8 \rb^8-21004165 L^6 Q^2 \rb^8+43653868 L^6
   q^2 Q^2 \rb^8 \\
&&
	\quad +\: 13080420 L^8 M \rb^7-3317664 L^{10}
   \rb^6-27171768 L^8 Q^2 \rb^6 \\
&&
	\quad +\: 54001655 L^8 q^2 Q^2
   \rb^6+8998896 L^{10} M \rb^5-1367360 L^{12} \rb^4 \\
&&
	\quad -\: 20243616
   L^{10} Q^2 \rb^4+38358312 L^{10} q^2 Q^2 \rb^4+3313280 L^{12} M
   \rb^3 \\
&&
	\quad -\: 235520 L^{14} \rb^2-8140736 L^{12} Q^2 \rb^2+14682880
   L^{12} q^2 Q^2 \rb^2+507904 L^{14} M \rb \\
&&
	\quad -\: 1373184 L^{14} Q^2+2355200
   L^{14} q^2 Q^2 \big) \rb^9 \\
&&
	-\: E^2 \big(6075 \rb^{20}+4608 M
   \rb^{19}+8505 L^2 \rb^{18}-49752 M^2 \rb^{18}+478596 q^2 Q^2
   \rb^{18}  \\
&&
	\quad -\: 173820 Q^2 \rb^{18}-1533432 M q^2 Q^2 \rb^{17}+460656
   M Q^2 \rb^{17}+816336 L^2 M \rb^{17} \\
&&
	\quad -\: 1215 L^4 \rb^{16}-2497542
   q^4 Q^4 \rb^{16}+2120292 q^2 Q^4 \rb^{16}-245262 Q^4
   \rb^{16} \\
&&
	\quad -\: 1868220 L^2 M^2 \rb^{16}-3299742 L^2 Q^2
   \rb^{16}+10700304 L^2 q^2 Q^2 \rb^{16} \\
&&
	\quad -\: 29484816 L^2 M q^2 Q^2
   \rb^{15}+8451444 L^2 M Q^2 \rb^{15}+4786884 L^4 M
   \rb^{15} \\
&&
	\quad -\: 3645 L^6 \rb^{14}-38826517 L^2 q^4 Q^4
   \rb^{14}-4066843 L^2 Q^4 \rb^{14} \\
&&
	\quad +\: 31659224 L^2 q^2 Q^4
   \rb^{14}-10390068 L^4 M^2 \rb^{14}-19219482 L^4 Q^2
   \rb^{14} \\
&&
	\quad +\: 60064308 L^4 q^2 Q^2 \rb^{14}-154136880 L^4 M q^2 Q^2
   \rb^{13}+47859444 L^4 M Q^2 \rb^{13} \\
&&
	\quad +\: 12179196 L^6 M
   \rb^{13}-22800693 L^4 Q^4 \rb^{12}-176561379 L^4 q^4 Q^4
   \rb^{12} \\
&&
	\quad +\: 153234768 L^4 q^2 Q^4 \rb^{12}-25811172 L^6 M^2
   \rb^{12}-56206914 L^6 Q^2 \rb^{12} \\
&&
	\quad +\: 166914204 L^6 q^2 Q^2
   \rb^{12}-403677096 L^6 M q^2 Q^2 \rb^{11}+135948852 L^6 M Q^2
   \rb^{11} \\
&&
	\quad +\: 16833564 L^8 M \rb^{11}-65405457 L^6 Q^4
   \rb^{10}-395889990 L^6 q^4 Q^4 \rb^{10} \\
&&
	\quad +\: 376611312 L^6 q^2 Q^4
   \rb^{10}-34602948 L^8 M^2 \rb^{10}-95792802 L^8 Q^2
   \rb^{10} \\
&&
	\quad +\: 269613144 L^8 q^2 Q^2 \rb^{10}-621132576 L^8 M q^2 Q^2
   \rb^9+224862588 L^8 M Q^2 \rb^9 \\
&&
	\quad +\: 13184964 L^{10} M
   \rb^9-110042297 L^8 Q^4 \rb^8-497842361 L^8 q^4 Q^4
   \rb^8 \\
&&
	\quad +\: 538891516 L^8 q^2 Q^4 \rb^8-25718544 L^{10} M^2
   \rb^8-99847608 L^{10} Q^2 \rb^8 \\
&&
	\quad +\: 265971444 L^{10} q^2 Q^2
   \rb^8-590550024 L^{10} M q^2 Q^2 \rb^7+227372808 L^{10} M Q^2
   \rb^7 \\
&&
	\quad +\: 5517360 L^{12} M \rb^7-113581920 L^{10} Q^4
   \rb^6-358700535 L^{10} q^4 Q^4 \rb^6 \\
&&
	\quad +\: 468025608 L^{10} q^2 Q^4
   \rb^6-9599328 L^{12} M^2 \rb^6-62971104 L^{12} Q^2
   \rb^6 \\
&&
	\quad +\: 158661264 L^{12} q^2 Q^2 \rb^6-343702776 L^{12} M q^2 Q^2
   \rb^5+139082832 L^{12} M Q^2 \rb^5 \\
&&
	\quad +\: 961920 L^{14} M
   \rb^5-71062800 L^{12} Q^4 \rb^4-137853360 L^{12} q^4 Q^4
   \rb^4 \\
&&
	\quad +\: 242715936 L^{12} q^2 Q^4 \rb^4-1059840 L^{14} M^2
   \rb^4-22145280 L^{14} Q^2 \rb^4 \\
&&
	\quad +\: 52757760 L^{14} q^2 Q^2
   \rb^4-112906368 L^{14} M q^2 Q^2 \rb^3+47435136 L^{14} M Q^2
   \rb^3 \\
&&
	\quad -\: 24820096 L^{14} Q^4 \rb^2-21136000 L^{14} q^4 Q^4
   \rb^2+68776448 L^{14} q^2 Q^4 \rb^2 \\
&&
	\quad +\: 184320 L^{16} M^2
   \rb^2-3342336 L^{16} Q^2 \rb^2+7526400 L^{16} q^2 Q^2
   \rb^2 \\
&&
	\quad -\: 16112640 L^{16} M q^2 Q^2 \rb+6942720 L^{16} M Q^2
   \rb-3723264 L^{16} Q^4 \\
&&
	\quad +\: 460800 L^{16} q^4 Q^4+8110080 L^{16} q^2 Q^4\big)
   \rb^6 \\
&&
	-\: 2 E q Q \big(6117 \rb^{20}-111150 M
   \rb^{19}+311079 L^2 \rb^{18}+216372 M^2 \rb^{18} \\
&&
	\quad -\: 520955 q^2
   Q^2 \rb^{18}+333703 Q^2 \rb^{18}+1357072 M q^2 Q^2
   \rb^{17}-845696 M Q^2 \rb^{17} \\
&&
	\quad -\: 2812044 L^2 M \rb^{17}+2009481
   L^4 \rb^{16}+915945 q^4 Q^4 \rb^{16}-1267778 q^2 Q^4
   \rb^{16} \\
&&
	\quad +\: 396409 Q^4 \rb^{16}+4602252 L^2 M^2 \rb^{16}+5571760
   L^2 Q^2 \rb^{16} \\
&&
	\quad -\: 9734497 L^2 q^2 Q^2 \rb^{16}+23610232 L^2 M q^2 Q^2
   \rb^{15}-13791712 L^2 M Q^2 \rb^{15} \\
&&
	\quad -\: 16044576 L^4 M
   \rb^{15}+6067539 L^6 \rb^{14}+13414185 L^2 q^4 Q^4
   \rb^{14} \\
&&
	\quad +\: 6060405 L^2 Q^4 \rb^{14}-18869382 L^2 q^2 Q^4
   \rb^{14}+24947196 L^4 M^2 \rb^{14} \\
&&
	\quad +\: 30903192 L^4 Q^2
   \rb^{14}-55053717 L^4 q^2 Q^2 \rb^{14}+127966134 L^4 M q^2 Q^2
   \rb^{13} \\
&&
	\quad -\: 74792556 L^4 M Q^2 \rb^{13}-43920420 L^6 M
   \rb^{13}+10511046 L^8 \rb^{12} \\
&&
	\quad +\: 32671953 L^4 Q^4
   \rb^{12}+60860775 L^4 q^4 Q^4 \rb^{12}-95126325 L^4 q^2 Q^4
   \rb^{12} \\
&&
	\quad +\: 65169228 L^6 M^2 \rb^{12}+87626078 L^6 Q^2
   \rb^{12}-157101589 L^6 q^2 Q^2 \rb^{12} \\
&&
	\quad +\: 353456414 L^6 M q^2 Q^2
   \rb^{11}-207076180 L^6 M Q^2 \rb^{11}-69000786 L^8 M
   \rb^{11} \\
&&
	\quad +\: 11071026 L^{10} \rb^{10}+91263815 L^6 Q^4
   \rb^{10}+136601970 L^6 q^4 Q^4 \rb^{10} \\
&&
	\quad -\: 247247791 L^6 q^2 Q^4
   \rb^{10}+96999048 L^8 M^2 \rb^{10}+145928177 L^8 Q^2
   \rb^{10} \\
&&
	\quad -\: 262068230 L^8 q^2 Q^2 \rb^{10}+575469110 L^8 M q^2 Q^2
   \rb^9-336392276 L^8 M Q^2 \rb^9 \\
&&
	\quad -\: 65724624 L^{10} M
   \rb^9+7031640 L^{12} \rb^8+150442386 L^8 Q^4 \rb^8 \\
&&
	\quad +\: 171627570
   L^8 q^4 Q^4 \rb^8-379051677 L^8 q^2 Q^4 \rb^8+86517144 L^{10} M^2
   \rb^8 \\
&&
	\quad +\: 149310786 L^{10} Q^2 \rb^8-267904446 L^{10} q^2 Q^2
   \rb^8+578480178 L^{10} M q^2 Q^2 \rb^7 \\
&&
	\quad -\: 335545980 L^{10} M Q^2
   \rb^7-37606560 L^{12} M \rb^7+2484960 L^{14} \rb^6 \\
&&
	\quad +\: 152716608
   L^{10} Q^4 \rb^6+122546115 L^{10} q^4 Q^4 \rb^6-357973743 L^{10} q^2
   Q^4 \rb^6 \\
&&
	\quad +\: 45591480 L^{12} M^2 \rb^6+92720032 L^{12} Q^2
   \rb^6-165986750 L^{12} q^2 Q^2 \rb^6 \\
&&
	\quad +\: 354784780 L^{12} M q^2 Q^2
   \rb^5-203065328 L^{12} M Q^2 \rb^5-11912640 L^{14} M
   \rb^5 \\
&&
	\quad +\: 376320 L^{16} \rb^4+94213696 L^{12} Q^4 \rb^4+45539100
   L^{12} q^4 Q^4 \rb^4 \\
&&
	\quad -\: 205581200 L^{12} q^2 Q^4 \rb^4+12979968 L^{14}
   M^2 \rb^4+32176480 L^{14} Q^2 \rb^4 \\
&&
	\quad -\: 57425440 L^{14} q^2 Q^2
   \rb^4+122208832 L^{14} M q^2 Q^2 \rb^3-68659840 L^{14} M Q^2
   \rb^3 \\
&&
	\quad -\: 1606656 L^{16} M \rb^3+32508192 L^{14} Q^4 \rb^2+5952960
   L^{14} q^4 Q^4 \rb^2 \\
&&
	\quad -\: 66075264 L^{14} q^2 Q^4 \rb^2+1514496 L^{16}
   M^2 \rb^2+4800000 L^{16} Q^2 \rb^2 \\
&&
	\quad -\: 8536320 L^{16} q^2 Q^2
   \rb^2+18178560 L^{16} M q^2 Q^2 \rb-9977856 L^{16} M Q^2
   \rb \\
&&
	\quad +\: 4824576 L^{16} Q^4-480000 L^{16} q^4 Q^4-9139200 L^{16} q^2 Q^4\big)
   \rb^5 \\
&&
	+\: \big(675 \rb^{24}+4410 M \rb^{23}+2160 L^2
   \rb^{22}-39600 M^2 \rb^{22}+115227 q^2 Q^2 \rb^{22} \\
&&
	\quad -\: 67197 Q^2
   \rb^{22}+56160 M^3 \rb^{21}-667672 M q^2 Q^2 \rb^{21}+355376 M
   Q^2 \rb^{21} \\
&&
	\quad +\: 316620 L^2 M \rb^{21}+2430 L^4 \rb^{20}-587260
   q^4 Q^4 \rb^{20}+735120 q^2 Q^4 \rb^{20} \\
&&
	\quad -\: 187940 Q^4
   \rb^{20}-1484604 L^2 M^2 \rb^{20}-1430352 L^2 Q^2
   \rb^{20}-455824 M^2 Q^2 \rb^{20} \\
&&
	\quad +\: 2833833 L^2 q^2 Q^2
   \rb^{20}+895280 M^2 q^2 Q^2 \rb^{20}+1411256 M q^4 Q^4
   \rb^{19} \\
&&
	\quad -\: 1776112 M q^2 Q^4 \rb^{19}+454008 M Q^4
   \rb^{19}+1685448 L^2 M^3 \rb^{19} \\
&&
	\quad -\: 14546000 L^2 M q^2 Q^2
   \rb^{19}+7329928 L^2 M Q^2 \rb^{19}+2294820 L^4 M
   \rb^{19} \\
&&
	\quad +\: 1080 L^6 \rb^{18}+501540 q^6 Q^6 \rb^{18}-1073644 q^4
   Q^6 \rb^{18}+692172 q^2 Q^6 \rb^{18} \\
&&
	\quad -\: 120068 Q^6
   \rb^{18}-10072039 L^2 q^4 Q^4 \rb^{18}-3569853 L^2 Q^4
   \rb^{18}  \\
&&
	\quad +\:  12893588 L^2 q^2 Q^4 \rb^{18}-10117656 L^4 M^2
   \rb^{18}-9912042 L^4 Q^2 \rb^{18} \\
&&
	\quad -\: 9130652 L^2 M^2 Q^2
   \rb^{18}+20420511 L^4 q^2 Q^2 \rb^{18}+18000100 L^2 M^2 q^2 Q^2
   \rb^{18} \\
&&
	\quad +\: 23162154 L^2 M q^4 Q^4 \rb^{17}-30157168 L^2 M q^2 Q^4
   \rb^{17}+8418646 L^2 M Q^4 \rb^{17} \\
&&
	\quad +\: 11036592 L^4 M^3
   \rb^{17}-98521416 L^4 M q^2 Q^2 \rb^{17}+49340940 L^4 M Q^2
   \rb^{17} \\
&&
	\quad +\: 7417080 L^6 M \rb^{17}+135 L^8 \rb^{16}+7040165 L^2
   q^6 Q^6 \rb^{16}-15864231 L^2 q^4 Q^6 \rb^{16} \\
&&
	\quad -\: 2137341 L^2 Q^6
   \rb^{16}+10900607 L^2 q^2 Q^6 \rb^{16}-23698128 L^4 Q^4
   \rb^{16} \\
&&
	\quad -\: 56264808 L^4 q^4 Q^4 \rb^{16}+79043190 L^4 q^2 Q^4
   \rb^{16}-31698072 L^6 M^2 \rb^{16} \\
&&
	\quad -\: 35182752 L^6 Q^2
   \rb^{16}-59942928 L^4 M^2 Q^2 \rb^{16}+74145261 L^6 q^2 Q^2
   \rb^{16} \\
&&
	\quad +\: 116331564 L^4 M^2 q^2 Q^2 \rb^{16}+125883672 L^4 M q^4 Q^4
   \rb^{15} \\
&&
	\quad -\: 180246336 L^4 M q^2 Q^4 \rb^{15} + 54779628 L^4 M Q^4
   \rb^{15}+33719184 L^6 M^3 \rb^{15} \\
&&
	\quad -\: 342471956 L^6 M q^2 Q^2
   \rb^{15}  + 170248816 L^6 M Q^2 \rb^{15}+13419090 L^8 M
   \rb^{15} \\
&&
	\quad +\: 31686225 L^4 q^6 Q^6 \rb^{14}  - 13783656 L^4 Q^6
   \rb^{14}-81531882 L^4 q^4 Q^6 \rb^{14} \\
&&
	\quad +\: 62596737 L^4 q^2 Q^6
   \rb^{14} - 82079698 L^6 Q^4 \rb^{14}-159932735 L^6 q^4 Q^4
   \rb^{14} \\
&&
	\quad +\: 254716494 L^6 q^2 Q^4 \rb^{14}  - 55696320 L^8 M^2
   \rb^{14}-74548785 L^8 Q^2 \rb^{14} \\
&&
	\quad -\: 201783416 L^6 M^2 Q^2
   \rb^{14} + 159829962 L^8 q^2 Q^2 \rb^{14} \\
&&
	\quad +\: 390092596 L^6 M^2 q^2 Q^2
   \rb^{14}+350770294 L^6 M q^4 Q^4 \rb^{13}   \\
&&
	\quad - \: 568301012 L^6 M q^2 Q^4
   \rb^{13}  + 186253644 L^6 M Q^4 \rb^{13}+57715200 L^8 M^3
   \rb^{13}   \\
&&
	\quad -\:  715146268 L^8 M q^2 Q^2 \rb^{13} + 350887688 L^8 M Q^2
   \rb^{13}+14570460 L^{10} M \rb^{13}  \\
&&
	\quad -\:  46895866 L^6 Q^6
   \rb^{12} + 70609410 L^6 q^6 Q^6 \rb^{12} \\
&&
	\quad -\: 217064585 L^6 q^4 Q^6
   \rb^{12}    +  190779597 L^6 q^2 Q^6 \rb^{12} - 171019308 L^8 Q^4
   \rb^{12} \\
&&
	\quad -\: 265596491 L^8 q^4 Q^4 \rb^{12} + 496902142 L^8 q^2 Q^4
   \rb^{12}  - 58236300 L^{10} M^2 \rb^{12} \\
&&
	\quad -\: 100276872 L^{10} Q^2
   \rb^{12}  - 405598576 L^8 M^2 Q^2 \rb^{12}  + 218061390 L^{10} q^2 Q^2
   \rb^{12} \\
&&
	\quad +\: 792120188 L^8 M^2 q^2 Q^2 \rb^{12} + 574503078 L^8 M q^4 Q^4
   \rb^{11}  \\
&&
	\quad -\:  1088345924 L^8 M q^2 Q^4 \rb^{11}+381311264 L^8 M Q^4
   \rb^{11}  + 58190760 L^{10} M^3 \rb^{11}   \\
&&
	\quad -\:  953694804 L^{10} M q^2 Q^2
   \rb^{11}+459363816 L^{10} M Q^2 \rb^{11} + 9470880 L^{12} M
   \rb^{11}  \\
&&
	\quad -\:  96470388 L^8 Q^6 \rb^{10}+87628105 L^8 q^6 Q^6
   \rb^{10}  - 340876605 L^8 q^4 Q^6 \rb^{10} \\
&&
	\quad +\: 352873195 L^8 q^2 Q^6
   \rb^{10}-227137305 L^{10} Q^4 \rb^{10}-268853358 L^{10} q^4 Q^4
   \rb^{10} \\
&&
	\quad +\: 620816598 L^{10} q^2 Q^4 \rb^{10}-35630424 L^{12} M^2
   \rb^{10}-86806752 L^{12} Q^2 \rb^{10} \\
&&
	\quad -\: 517574604 L^{10} M^2 Q^2
   \rb^{10}+191078160 L^{12} q^2 Q^2 \rb^{10} \\
&&
	\quad +\: 1034499432 L^{10} M^2 q^2
   Q^2 \rb^{10} +\: 576600936 L^{10} M q^4 Q^4 \rb^9 \\
&&
	\quad -\: 1339557588 L^{10} M
   q^2 Q^4 \rb^9+497960478 L^{10} M Q^4 \rb^9 + 33377328 L^{12} M^3
   \rb^9 \\
&&
	\quad -\: 822714868 L^{12} M q^2 Q^2 \rb^9+387239708 L^{12} M Q^2
   \rb^9 + 3409200 L^{14} M \rb^9 \\
&&
	\quad -\: 126860433 L^{10} Q^6
   \rb^8+60917535 L^{10} q^6 Q^6 \rb^8 - 328530204 L^{10} q^4 Q^6
   \rb^8 \\
&&
	\quad +\: 417974592 L^{10} q^2 Q^6 \rb^8-194638808 L^{12} Q^4
   \rb^8-162756277 L^{12} q^4 Q^4 \rb^8 \\
&&
	\quad +\: 502209060 L^{12} q^2 Q^4
   \rb^8-11293056 L^{14} M^2 \rb^8-47027712 L^{14} Q^2
   \rb^8 \\
&&
	\quad -\: 424999312 L^{12} M^2 Q^2 \rb^8+104628144 L^{14} q^2 Q^2
   \rb^8 \\
&&
	\quad +\: 879562328 L^{12} M^2 q^2 Q^2 \rb^8 + 347823026 L^{12} M q^4 Q^4
   \rb^7 \\
&&
	\quad -\: 1071487612 L^{12} M q^2 Q^4 \rb^7+419819436 L^{12} M Q^4
   \rb^7+8949312 L^{14} M^3 \rb^7 \\
&&
	\quad -\: 446095928 L^{14} M q^2 Q^2
   \rb^7+204396160 L^{14} M Q^2 \rb^7+524160 L^{16} M
   \rb^7 \\
&&
	\quad -\: 107832056 L^{12} Q^6 \rb^6+21062280 L^{12} q^6 Q^6
   \rb^6-190432201 L^{12} q^4 Q^6 \rb^6 \\
&&
	\quad +\: 320089812 L^{12} q^2 Q^6
   \rb^6-104560848 L^{14} Q^4 \rb^6-53022544 L^{14} q^4 Q^4
   \rb^6 \\
&&
	\quad +\: 255371888 L^{14} q^2 Q^4 \rb^6-1043712 L^{16} M^2
   \rb^6-14548992 L^{16} Q^2 \rb^6 \\
&&
	\quad -\: 218337728 L^{14} M^2 Q^2
   \rb^6+32679168 L^{16} q^2 Q^2 \rb^6+472732336 L^{14} M^2 q^2 Q^2
   \rb^6 \\
&&
	\quad +\: 113591328 L^{14} M q^4 Q^4 \rb^5-540812632 L^{14} M q^2 Q^4
   \rb^5 \\
&&
	\quad +\: 221997808 L^{14} M Q^4 \rb^5 - 9216 L^{16} M^3
   \rb^5-138635904 L^{16} M q^2 Q^2 \rb^5 \\
&&
	\quad +\: 61637376 L^{16} M Q^2
   \rb^5-57533136 L^{14} Q^6 \rb^4+1829120 L^{14} q^6 Q^6
   \rb^4 \\
&&
	\quad -\: 59753232 L^{14} q^4 Q^6 \rb^4+153702272 L^{14} q^2 Q^6
   \rb^4-32116608 L^{16} Q^4 \rb^4 \\
&&
	\quad -\: 6215040 L^{16} q^4 Q^4
   \rb^4+74401536 L^{16} q^2 Q^4 \rb^4+184320 L^{18} M^2
   \rb^4 \\
&&
	\quad -\: 1966080 L^{18} Q^2 \rb^4-64025856 L^{16} M^2 Q^2
   \rb^4+4454400 L^{18} q^2 Q^2 \rb^4 \\
&&
	\quad +\: 146352384 L^{16} M^2 q^2 Q^2
   \rb^4+13582080 L^{16} M q^4 Q^4 \rb^3 \\
&&
	\quad -\: 157017216 L^{16} M q^2 Q^4
   \rb^3+67150464 L^{16} M Q^4 \rb^3-368640 L^{18} M^3
   \rb^3 \\
&&
	\quad -\: 18877440 L^{18} M q^2 Q^2 \rb^3+8122368 L^{18} M Q^2
   \rb^3-17567616 L^{16} Q^6 \rb^2 \\
&&
	\quad -\: 499200 L^{16} q^6 Q^6
   \rb^2-6837120 L^{16} q^4 Q^6 \rb^2+42143232 L^{16} q^2 Q^6
   \rb^2 \\
&&
	\quad -\: 4313088 L^{18} Q^4 \rb^2+460800 L^{18} q^4 Q^4
   \rb^2+9492480 L^{18} q^2 Q^4 \rb^2 \\
&&
	\quad -\: 8196096 L^{18} M^2 Q^2
   \rb^2+19937280 L^{18} M^2 q^2 Q^2 \rb^2-921600 L^{18} M q^4 Q^4
   \rb \\
&&
	\quad -\: 20044800 L^{18} M q^2 Q^4 \rb+8884224 L^{18} M Q^4
   \rb-2347008 L^{18} Q^6 \\
&&
	\quad +\: 460800 L^{18} q^4 Q^6+5038080 L^{18} q^2 Q^6\big)
   \rb^2,
\end{IEEEeqnarray*}

\begin{equation}
F_{\theta \lpow{4}} = 0,
\end{equation}

\begin{equation}
F_{\phi \lpow{4}} = \frac{\rbdot}{120 \pi  L \rb^{13} \left(L^2+\rb^2\right)^{9/2}} (F^{\mathcal{E}}_{r \lpow{4}} \mathcal{E} + F^{\mathcal{K}}_{r \lpow{4}} \mathcal{K}),
\end{equation}
where
\par \vspace{-6pt} \begin{IEEEeqnarray*}{rCl}
F^{\mathcal{E}}_{\phi \lpow{4}} &=&  -135 E^4 \left(43 L^4-82 \rb^2 L^2+3
   \rb^4\right) \rb^{18} \\
&&
	+\: 60 E^3 L^2 q Q \big(32768
   L^{12}+161792 \rb^2 L^{10}+320224 \rb^4 L^8+318848 \rb^6
   L^6 \\
&&
	\quad +\: 161653 \rb^8 L^4+35426 \rb^{10} L^2+269 \rb^{12}\big)
   \rb^7 \\
&&
	+\: 6 E^2 \big(135
   \rb^{18}-1755 L^2 \rb^{16}+1610 q^2 Q^2 \rb^{16}-170 Q^2
   \rb^{16}+10650 L^2 M \rb^{15} \\
&&
	\quad -\: 1755 L^4 \rb^{14}+29422 L^2 Q^2
   \rb^{14}-92200 L^2 q^2 Q^2 \rb^{14}-21390 L^4 M \rb^{13} \\
&&
	\quad +\: 135
   L^6 \rb^{12}+507662 L^4 Q^2 \rb^{12}-2269025 L^4 q^2 Q^2
   \rb^{12}-232230 L^6 M \rb^{11} \\
&&
	\quad +\: 2521482 L^6 Q^2
   \rb^{10}-10796050 L^6 q^2 Q^2 \rb^{10}-544410 L^8 M
   \rb^9 \\
&&
	\quad +\: 6086820 L^8 Q^2 \rb^8-23759035 L^8 q^2 Q^2 \rb^8-591900
   L^{10} M \rb^7 \\
&&
	\quad +\: 8210640 L^{10} Q^2 \rb^6-28531640 L^{10} q^2 Q^2
   \rb^6-312960 L^{12} M \rb^5 \\
&&
	\quad +\: 6347072 L^{12} Q^2 \rb^4-19319200
   L^{12} q^2 Q^2 \rb^4-65280 L^{14} M \rb^3 \\
&&
	\quad +\: 2638592 L^{14} Q^2
   \rb^2-6932480 L^{14} q^2 Q^2 \rb^2+458752 L^{16} Q^2 \\
&&
	\quad -\: 1024000 L^{16}
   q^2 Q^2\big) \rb^4 \\
&&
	-\: 4 E q Q \big(2880 M \rb^{17}+30000 L^2
   \rb^{16}+5220 q^2 Q^2 \rb^{16}-3060 Q^2 \rb^{16} \\
&&
	\quad -\: 7956 L^2 M
   \rb^{15}  + 780630 L^4 \rb^{14}+96659 L^2 Q^2 \rb^{14}-250655 L^2
   q^2 Q^2 \rb^{14} \\
&&
	\quad -\: 1617216 L^4 M \rb^{13} + 4230810 L^6
   \rb^{12}+2341049 L^4 Q^2 \rb^{12} \\
&&
	\quad -\: 5330480 L^4 q^2 Q^2
   \rb^{12}-9600216 L^6 M \rb^{11}+10649970 L^8 \rb^{10} \\
&&
	\quad +\: 12479569
   L^6 Q^2 \rb^{10}-25840225 L^6 q^2 Q^2 \rb^{10}-24498720 L^8 M
   \rb^9 \\
&&
	\quad +\: 14747310 L^{10} \rb^8+31062135 L^8 Q^2 \rb^8-58843050
   L^8 q^2 Q^2 \rb^8 \\
&&
	\quad -\: 33686580 L^{10} M \rb^7+11611440 L^{12}
   \rb^6+42602360 L^{10} Q^2 \rb^6 \\
&&
	\quad -\: 73864550 L^{10} q^2 Q^2
   \rb^6-26140416 L^{12} M \rb^5+4894080 L^{14} \rb^4 \\
&&
	\quad +\: 33272224
   L^{12} Q^2 \rb^4-52783600 L^{12} q^2 Q^2 \rb^4-10826496 L^{14} M
   \rb^3 \\
&&
	\quad +\: 860160 L^{16} \rb^2+13926784 L^{14} Q^2 \rb^2-20204800
   L^{14} q^2 Q^2 \rb^2 \\
&&
	\quad -\: 1867776 L^{16} M \rb+2433024 L^{16} Q^2-3225600
   L^{16} q^2 Q^2\big) \rb^3 \\
&&
	+\: \big(-405 \rb^{22}+135 L^2 \rb^{20}+4320 M^2 \rb^{20}-2460 q^2 Q^2
   \rb^{20}+1020 Q^2 \rb^{20} \\
&&
	\quad +\: 17040 M q^2 Q^2 \rb^{19}-8400 M Q^2
   \rb^{19}+36360 L^2 M \rb^{19}+1755 L^4 \rb^{18} \\
&&
	\quad +\: 13752 q^4 Q^4
   \rb^{18}-18864 q^2 Q^4 \rb^{18}+5112 Q^4 \rb^{18}-47844 L^2
   M^2 \rb^{18} \\
&&
	\quad -\: 304254 L^2 Q^2 \rb^{18}+712086 L^2 q^2 Q^2
   \rb^{18}-1346780 L^2 M q^2 Q^2 \rb^{17} \\
&&
	\quad +\: 700012
   L^2 M Q^2 \rb^{17}+927180 L^4 M \rb^{17}+1485 L^6
   \rb^{16}-278353 L^2 q^4 Q^4 \rb^{16} \\
&&
	\quad -\: 335473 L^2 Q^4
   \rb^{16}+636306 L^2 q^2 Q^4 \rb^{16}-1861164 L^4 M^2
   \rb^{16} \\
&&
	\quad -\: 4676154 L^4 Q^2 \rb^{16}+12140046 L^4 q^2 Q^2
   \rb^{16}-26210120 L^4 M q^2 Q^2 \rb^{15} \\
&&
	\quad +\: 11627788 L^4 M Q^2
   \rb^{15}+5097780 L^6 M \rb^{15}+270 L^8 \rb^{14}-5653339 L^4
   Q^4 \rb^{14} \\
&&
	\quad -\: 7807693 L^4 q^4 Q^4 \rb^{14}+14911296 L^4 q^2 Q^4
   \rb^{14}-10362204 L^6 M^2 \rb^{14} \\
&&
	\quad -\: 25490994 L^6 Q^2
   \rb^{14}+67132326 L^6 q^2 Q^2 \rb^{14}-148624780 L^6 M q^2 Q^2
   \rb^{13} \\
&&
	\quad +\: 63404228 L^6 M Q^2 \rb^{13}+12914820 L^8 M
   \rb^{13}-31286291 L^6 Q^4 \rb^{12} \\
&&
	\quad -\: 39298730 L^6 q^4 Q^4
   \rb^{12}+86555580 L^6 q^2 Q^4 \rb^{12}-25568820 L^8 M^2
   \rb^{12} \\
&&
	\quad -\: 72000090 L^8 Q^2 \rb^{12}+187923870 L^8 q^2 Q^2
   \rb^{12}-418882860 L^8 M q^2 Q^2 \rb^{11} \\
&&
	\quad +\: 176148876 L^8 M Q^2
   \rb^{11}+17928180 L^{10} M \rb^{11}-88446081 L^8 Q^4
   \rb^{10} \\
&&
	\quad -\: 89608695 L^8 q^4 Q^4 \rb^{10}+243191400 L^8 q^2 Q^4
   \rb^{10}-33567120 L^{10} M^2 \rb^{10} \\
&&
	\quad -\: 119826600 L^{10} Q^2
   \rb^{10}+306787500 L^{10} q^2 Q^2 \rb^{10}-684871640 L^{10} M q^2
   Q^2 \rb^9 \\
&&
	\quad +\: 286492480 L^{10} M Q^2 \rb^9+14133600 L^{12} M
   \rb^9-146647408 L^{10} Q^4 \rb^8 \\
&&
	\quad -\: 109473943 L^{10} q^4 Q^4
   \rb^8+391930626 L^{10} q^2 Q^4 \rb^8-23754384 L^{12} M^2
   \rb^8 \\
&&
	\quad -\: 122599104 L^{12} Q^2 \rb^8+306058296 L^{12} q^2 Q^2
   \rb^8-683280460 L^{12} M q^2 Q^2 \rb^7 \\
&&
	\quad +\: 285529832 L^{12} M Q^2
   \rb^7+5961600 L^{14} M \rb^7-149211728 L^{12} Q^4 \rb^6 \\
&&
	\quad -\: 72365318 L^{12} q^4 Q^4
   \rb^6+382585176 L^{12} q^2 Q^4 \rb^6-7700544 L^{14} M^2
   \rb^6 \\
&&
	\quad -\: 76114944 L^{14} Q^2 \rb^6+184519776 L^{14} q^2 Q^2
   \rb^6-411961840 L^{14} M q^2 Q^2 \rb^5 \\
&&
	\quad +\: 172336448 L^{14} M Q^2
   \rb^5+1048320 L^{16} M \rb^5-92058944 L^{14} Q^4
   \rb^4 \\
&&
	\quad -\: 21986048 L^{14} q^4 Q^4 \rb^4+224292096 L^{14} q^2 Q^4
   \rb^4-129024 L^{16} M^2 \rb^4 \\
&&
	\quad -\: 26394624 L^{16} Q^2
   \rb^4+61940736 L^{16} q^2 Q^2 \rb^4-138385920 L^{16} M q^2 Q^2 \rb^3 \\
&&
	\quad +\: 58015488
   L^{16} M Q^2 \rb^3-31716096 L^{16} Q^4 \rb^2-364800 L^{16}
   q^4 Q^4 \rb^2 \\
&&
	\quad +\: 72821760 L^{16} q^2 Q^4 \rb^2+368640 L^{18} M^2
   \rb^2-3932160 L^{18} Q^2 \rb^2 \\
&&
	\quad +\: 8908800 L^{18} q^2 Q^2
   \rb^2-19937280 L^{18} M q^2 Q^2 \rb+8380416 L^{18} M Q^2
   \rb \\
&&
	\quad -\: 4694016 L^{18} Q^4+921600 L^{18} q^4 Q^4+10076160 L^{18} q^2 Q^4 \big),
\end{IEEEeqnarray*}
\par \vspace{-6pt} \begin{IEEEeqnarray*}{rCl}
F^{\mathcal{K}}_{\phi \lpow{4}} &=& 135 E^4 \left(\rb^2-15 L^2\right) \left(3 \rb^2-L^2\right)
   \rb^{18} \\
&&
	-\: 120 E^3 L^2 q Q \big(8192 L^{10}+33280 \rb^2
   L^8+51320 \rb^4 L^6+36199 \rb^6 L^4 \\
&&
	\quad +\: 10592 \rb^8 L^2+337
   \rb^{10}\big) \rb^9 \\
&&
	-\: 6 E^2 \big(135 \rb^{16}-810 L^2
   \rb^{14}+1610 q^2 Q^2 \rb^{14}-170 Q^2 \rb^{14}+4800 L^2 M
   \rb^{13} \\
&&
	\quad -\: 945 L^4 \rb^{12}+22427 L^2 Q^2 \rb^{12}-80705 L^2 q^2
   Q^2 \rb^{12}-20190 L^4 M \rb^{11} \\
&&
	\quad +\: 296896 L^4 Q^2
   \rb^{10}-1341430 L^4 q^2 Q^2 \rb^{10}-115260 L^6 M
   \rb^9+1181215 L^6 Q^2 \rb^8 \\
&&
	\quad -\: 4948565 L^6 q^2 Q^2 \rb^8-185550
   L^8 M \rb^7+2223220 L^8 Q^2 \rb^6 \\
&&
	\quad -\: 8234490 L^8 q^2 Q^2
   \rb^6-127920 L^{10} M \rb^5+2205520 L^{10} Q^2 \rb^4 \\
&&
	\quad -\: 7042640
   L^{10} q^2 Q^2 \rb^4-32640 L^{12} M \rb^3+1118592 L^{12} Q^2
   \rb^2 \\
&&
	\quad -\: 3018240 L^{12} q^2 Q^2 \rb^2+229376 L^{14} Q^2-512000 L^{14}
   q^2 Q^2\big) \rb^6 \\
&&
	+\: 4 E q Q \big(2880 M \rb^{15}+26625 L^2
   \rb^{14}+5220 q^2 Q^2 \rb^{14}-3060 Q^2 \rb^{14} \\
&&
	\quad -\: 26496 L^2 M
   \rb^{13}+469980 L^4 \rb^{12}+81869 L^2 Q^2 \rb^{12}-205265 L^2
   q^2 Q^2 \rb^{12} \\
&&
	\quad -\: 1019778 L^4 M \rb^{11}+2019645 L^6
   \rb^{10}+1406452 L^4 Q^2 \rb^{10} \\
&&
	\quad -\: 3146275 L^4 q^2 Q^2
   \rb^{10}-4643940 L^6 M \rb^9+3949650 L^8 \rb^8+5935025 L^6 Q^2
   \rb^8 \\
&&
	\quad -\: 11944775 L^6 q^2 Q^2 \rb^8-9083730 L^8 M \rb^7+4014000
   L^{10} \rb^6 \\
&&
	\quad +\: 11462630 L^8 Q^2 \rb^6-20801825 L^8 q^2 Q^2
   \rb^6-9092400 L^{10} M \rb^5 \\
&&
	\quad +\: 2070720 L^{12} \rb^4+11531560
   L^{10} Q^2 \rb^4-18862600 L^{10} q^2 Q^2 \rb^4 \\
&&
	\quad -\: 4596096 L^{12} M
   \rb^3+430080 L^{14} \rb^2+5898944 L^{12} Q^2 \rb^2-8691200
   L^{12} q^2 Q^2 \rb^2 \\
&&
	\quad -\: 933888 L^{14} M \rb+1216512 L^{14} Q^2-1612800
   L^{14} q^2 Q^2\big) \rb^5 \\
&&
	+\: \big(405 \rb^{20}+675 L^2
   \rb^{18}-4320 M^2 \rb^{18}+2460 q^2 Q^2 \rb^{18}-1020 Q^2
   \rb^{18} \\
&&
	\quad -\: 17040 M q^2 Q^2 \rb^{17}+8400 M Q^2 \rb^{17}-33660
   L^2 M \rb^{17}+135 L^4 \rb^{16} \\
&&
	\quad -\: 13752 q^4 Q^4 \rb^{16}+18864
   q^2 Q^4 \rb^{16}-5112 Q^4 \rb^{16}+50004 L^2 M^2
   \rb^{16} \\
&&
	\quad +\: 216204 L^2 Q^2 \rb^{16}-520716 L^2 q^2 Q^2
   \rb^{16}+1019300 L^2 M q^2 Q^2 \rb^{15} \\
&&
	\quad -\: 506452 L^2 M Q^2
   \rb^{15}-566100 L^4 M \rb^{15}-135 L^6 \rb^{14}+252229 L^2 q^4
   Q^4 \rb^{14} \\
&&
	\quad +\: 241429 L^2 Q^4 \rb^{14}-508938 L^2 q^2 Q^4
   \rb^{14}+1146852 L^4 M^2 \rb^{14} \\
&&
	\quad +\: 2738532 L^4 Q^2
   \rb^{14}-7171158 L^4 q^2 Q^2 \rb^{14}+15646900 L^4 M q^2 Q^2
   \rb^{13} \\
&&
	\quad -\: 6844472 L^4 M Q^2 \rb^{13}-2443860 L^6 M
   \rb^{13}+3331811 L^4 Q^4 \rb^{12} \\
&&
	\quad +\: 4688825 L^4 q^4 Q^4
   \rb^{12}-9010740 L^4 q^2 Q^4 \rb^{12}+4955940 L^6 M^2
   \rb^{12} \\
&&
	\quad +\: 12379860 L^6 Q^2 \rb^{12}-32628840 L^6 q^2 Q^2
   \rb^{12}+72544040 L^6 M q^2 Q^2 \rb^{11} \\
&&
	\quad -\: 30716848 L^6 M Q^2
   \rb^{11}-4797900 L^8 M \rb^{11}+15224623 L^6 Q^4
   \rb^{10} \\
&&
	\quad +\: 18284320 L^6 q^4 Q^4 \rb^{10}-42365070 L^6 q^2 Q^4
   \rb^{10}+9332460 L^8 M^2 \rb^{10} \\
&&
	\quad +\: 28465560 L^8 Q^2
   \rb^{10}-73843530 L^8 q^2 Q^2 \rb^{10}+164824460 L^8 M q^2 Q^2
   \rb^9 \\
&&
	\quad -\: 69047080 L^8 M Q^2 \rb^9-4884480 L^{10} M \rb^9+34937905
   L^8 Q^4 \rb^8 \\
&&
	\quad +\: 31525405 L^8 q^4 Q^4 \rb^8-95323560 L^8 q^2 Q^4
   \rb^8+8703000 L^{10} M^2 \rb^8 \\
&&
	\quad +\: 37203360 L^{10} Q^2
   \rb^8-94024320 L^{10} q^2 Q^2 \rb^8+209925820 L^{10} M q^2 Q^2
   \rb^7 \\
&&
	\quad -\: 87699740 L^{10} M Q^2 \rb^7-2522160 L^{12} M
   \rb^7+45401768 L^{10} Q^4 \rb^6 \\
&&
	\quad +\: 27067913 L^{10} q^4 Q^4
   \rb^6-118856916 L^{10} q^2 Q^4 \rb^6+3644064 L^{12} M^2
   \rb^6 \\
&&
	\quad +\: 28107264 L^{12} Q^2 \rb^6-68780016 L^{12} q^2 Q^2
   \rb^6+153536600 L^{12} M q^2 Q^2 \rb^5 \\
&&
	\quad -\: 64190992 L^{12} M Q^2
   \rb^5-524160 L^{14} M \rb^5+34060624 L^{12} Q^4 \rb^4 \\
&&
	\quad +\: 10459024
   L^{12} q^4 Q^4 \rb^4-84379968 L^{12} q^2 Q^4 \rb^4+225792 L^{14} M^2
   \rb^4 \\
&&
	\quad +\: 11476992 L^{14} Q^2 \rb^4-27072768 L^{14} q^2 Q^2
   \rb^4+60470400 L^{14} M q^2 Q^2 \rb^3 \\
&&
	\quad -\: 25341312 L^{14} M Q^2
   \rb^3+13804416 L^{14} Q^4 \rb^2+585600 L^{14} q^4 Q^4
   \rb^2 \\
&&
	\quad -\: 32002560 L^{14} q^2 Q^4 \rb^2-184320 L^{16} M^2
   \rb^2+1966080 L^{16} Q^2 \rb^2 \\
&&
	\quad -\: 4454400 L^{16} q^2 Q^2
   \rb^2+9968640 L^{16} M q^2 Q^2 \rb-4190208 L^{16} M Q^2
   \rb \\
&&
	\quad +\: 2347008 L^{16} Q^4-460800 L^{16} q^4 Q^4-5038080 L^{16} q^2 Q^4\big)
   \rb^2.
\end{IEEEeqnarray*}

As in the $f(r)$ case, we were unable to obtain data to verify these results, however, we could do a simple comparison to the \Sch parameters and check that  for $Q=q=0$, one gets the same parameters.  As this is the case, we are confident in the validity of these parameters.  However, as previously stated, this is part of ongoing work into the investigation of the cosmic censorship conjecture.  Therefore, this is only a basis on which to build, we plan to team these results with the retarded field in due time in order to carry out our investigation.


\section{Second Order Self-force}
The original derivation of the MiSaTaQuWa equations, although widely accepted,
required several assumptions. In 2008, Gralla and Wald produced a more rigorous
derivation of the equations of motion that avoided most of the previous
assumptions \cite{Gralla:Wald:2008}. They considered a smooth, one parameter
family of metrics that satisfy Einstein's equations in two regions - the near
zone and far zone. In the far zone, the black hole or massive body can be seen
to shrink down to a world line that is a geodesic of the background space time
with perturbations from the massive body, while in the near zone, the body
remains a fixed size and is only perturbed by the background space time. The
approach required the calculation of the metric perturbation in the far zone for
some gauge as well as a smooth gauge transformation to the near zone that
ensured the background metric is mass centred. Initially these calculations were
done for the Lorentz gauge and were later expanded to encompass more gauges
\cite{Gralla:2011zr}. This more rigorous technique was done in a manner that
would allow perturbations in the mass ratio to infinite order.  However, in these
papers, they developed only the first order equations of motion. 

In the last
year, Gralla, Pound and Detweiller have all independently developed outlines for calculations of the self-force up to the second order in
the mass ratio \cite{Gralla:2012db, Detweiler:2011tt, Pound:2012nt}. In these
methods, they work in the reverse order to \cite{Gralla:Wald:2008,
Gralla:2011zr}, in that they begin with a series expansion of the metric perturbation
in a mass-centred gauge and, by considering smooth gauge transformations, they
compute the metric perturbations in these gauges. The resulting definition for the second order GSF requires a very smooth
effective source, very much like that which we produced in
Sec.~\ref{sec:EffectiveSource}.  In particular, the given expressions, given in
detail in the Appendix of \cite{Gralla:2012db, Pound:2012nt} require the double
covariant derivative of this regularised field. This can be obtained by using
the mode sum approach, discussed in Sec.~\ref{sec: modeSum}, carrying out the
double derivative on the numerically obtained retarded field and analytically
obtained singular field separately, and calculating the resulting regularisation
parameters to regularise the differentiated retarded field. 

To this end we have
calculated these regularisation parameters as a starting point to these second
order calculations. It should be noted that there are many other `ingredients'
required for obtaining the expressions for the second order self-force. Some of
these also involve high order coordinate expansions and regularisation
parameters, making our current work a solid base on which to build towards
obtaining the second order self-force.

\subsection{Mode-sums for the Second Derivative of the Singular Field}
In this section, we derive mode-sum expression for the coordinate expansion of
the second derivative of the singular field and show how they may be used to
derive regularisation parameters for the second derivative of the retarded
field. This calculation follows closely the strategy for computing
regularisation parameters for the self-force. As such, we describe here only the
relevant differences which arise in the second derivative case.

Given our previously calculated coordinate expansions for the singular
field of the form Eq.~\eqref{eqn:fasum}, its second partial derivative naturally takes the form
\begin{equation}
\label{eq:phiabsum}
\Phi_{,ab} \left( r, t, \alpha, \beta \right) = \sum_{n=1}^{\infty} \frac{B_{ab}^{(3 n - 1)}}{ \zrho_0^{2 n + 3}} \epsilon^{n-4},
\end{equation}
where $B_{\text{\fixme{$ab$}}}^{(k)} = b_{\text{\fixme{$ab$}}|a_1 a_2 \dots a_k}(\bar{x}) \Delta x^{a_1} \Delta x^{a_2} \dots \Delta x^{a_k}$.
This is true independently of whether one considers scalar, electromagnetic or
gravitational cases; for simplicity, we describe here the scalar field case and
note that the calculation proceeds in exactly the same way for the
electromagnetic and gravitational cases.

As was the case for self-force regularization, in using Eq.~\eqref{eq:phiabsum}
to derive the regularization parameters, we only need to take the sum to the
appropriate order: $n=1$ for $\Phi_{,ab [-2]}$, $n=2$ for $\Phi_{,ab [-1]}$,
etc. For the self-force, only the first three terms were non-vanishing in the
limit $\Delta x^a \to 0$; in the second derivative case one additional term is
non-vanishing and we must include up to $n=4$ in order to compute a correct
regularised second derivative. Before addressing the mode decomposition, we
first recall the identity used in Eq.~\eqref{eqn: rhoAl},
\begin{equation}
\left(\delta^2 + 1 - u \right)^{p/2} = \sum_{l=0}^{\infty} \mathcal{A}_l^{p/2}(\delta) P_l(u)
\end{equation}
where
\par \vspace{-6pt} \begin{IEEEeqnarray}{rCl}
\mathcal{A}_l^{-1/2} &=& \sqrt{2} \left( \sqrt{\frac{\delta^2}{2}+1} - \frac{\delta}{\sqrt{2}} \right)^{2l+1}, \nonumber \\
\mathcal{A}_l^{p/2} &=& \frac{-1}{(2n-1)\delta} \frac{d \mathcal{A}_l^{p/2-1}}{d\delta}.
\end{IEEEeqnarray}
As in the self-force case, these may be used to derive expressions for
$\zrho_0^{2n+3}$ before taking the limit $\delta \to 0$ (equivalently,
$\Delta r \to 0$). In the present context, the relevant terms are:
\par \vspace{-6pt} \begin{IEEEeqnarray}{rCl}
\mathcal{A}_l^{-5/2} &=& \frac{2 l+1}{3 \delta ^3}-\frac{l (l+1) (2 l+1)}{3 \delta } \fixme{\text{$+$}} \mathcal{O}(\delta^{0}), \nonumber \\
\mathcal{A}_l^{-7/2} &=& \frac{2 l+1}{5 \delta ^5}-\frac{l (l+1) (2 l+1)}{15 \delta ^3}+\frac{(l-1)l (l+1) (l+2) (2 l+1)}{30 \delta }+ \mathcal{O}(\delta^{0}), \nonumber \\
\mathcal{A}_l^{-9/2} &=& \frac{2 l+1}{7 \delta ^7}-\frac{l (l+1) (2 l+1)}{35 \delta ^5} \nonumber \\
  && +\:\frac{(l-1) l (l+1) (l+2) (2 l+1)}{210 \delta ^3}-\frac{(l-2) (l-1) l (l+1) (l+2)(l+3) (2 l+1)}{630 \delta }  \nonumber \\
&& +\: \mathcal{O}(\delta^{0}), \nonumber \\
\mathcal{A}_l^{-11/2} &=& \frac{2 l+1}{9 \delta ^9}-\frac{l (l+1) (2 l+1)}{63 \delta ^7}+\frac{(l-1) l (l+1) (l+2) (2 l+1)}{630 \delta ^5} \nonumber \\
  && -\:\frac{(l-2) (l-1) l (l+1) (l+2) (l+3) (2 l+1)}{5670 \delta ^3} \nonumber \\
&& +\: \frac{(l-3) (l-2) (l-1) l (l+1) (l+2) (l+3) (l+4) (2 l+1)}{22680 \delta } \nonumber \\
  && -\:\frac{(2 l-7) (2 l-5) (2 l-3) (2 l-1) (2 l+1)^2 (2 l+3) (2 l+5) (2l+7) (2 l+9)}{14288400\sqrt{2}}  \nonumber \\
&& +\:  \mathcal{O}(\delta^{0}).
\end{IEEEeqnarray}
for $n=1, 2, 3, 4$, respectively. Using these in Eq.~\eqref{eq:phiabsum} above and
taking the limit $\Delta r \to 0$, we find that all divergent terms vanish. We
may now perform the integrating over the two-sphere as required by the
spherical-harmonic decomposition. Doing so, we obtain regularisation parameters
sufficient to render the sum over $l$ finite. These generally have the following
dependence on $l$:
\begin{gather}
\Phi^l_{,rr\lnpow{2}} = (2l+1)^2 \Phi_{,rr\lnpow{2}}, \quad \Phi^l_{,rr\lnpow{1}} = (2l+1) \Phi_{,rr\lnpow{1}}, \quad \Phi^l_{,rr[0]} = \Phi_{,rr[0]}, \nonumber \\
 \Phi^l_{,rr\lpow{2}} = \frac{\Phi_{,rr\lpow{2}}}{(2l-1)(2l+3)}, \quad
\Phi^l_{,rr\lpow{4}} = \frac{\Phi_{,rr\lpow{4}}}{(2l-3)(2l-1)(2l+3)(2l+5)}, \nonumber \\
\Phi^l_{,rr\lpow{6}} = \frac{\Phi_{,rr\lpow{6}}}{(2l-5)(2l-3)(2l-1)(2l+3)(2l+5)(2l+7)}.
\end{gather}

\subsection{Regularisation Parameters}
In this section, we apply this calculation to the
case of the second radial derivative for a circular geodesic orbit in the
scalar field case. This particular case was chosen as it most simply illustrates
the structure without the unnecessary complexity of having many components. Its
should be noted, however, that the calculation proceeds in exactly the same way
for more generic orbits and for electromagnetic and gravitational cases. We give
the regularisation parameters resulting from doing so electronically.

The regularisation parameters for the second derivative of the scalar field are
then as follows
\begin{equation}
\Phi_{,rr [-2]} = 
\frac{\left(L^2+\rb^2\right) \left(\rb-3 M\right)^{3/2}  \mathcal{E}}{2 \pi  \rb^4 (\rb-2 M)^{5/2}},
\end{equation}
\begin{equation}
\Phi_{,rr [-1]} = 
\frac{\sgn(\Delta r)  (\rb-3 M) \left(L^2 (3 \rb-5 M)+\rb^2 (2 \rb-3 M)\right)}{2 \rb^{7/2} \sqrt{L^2+\rb^2} (\rb-2M)^{5/2}},
\end{equation}
\begin{equation}
\Phi_{,rr [0]} =
\frac{\sqrt{\rb-3 M}}{12 \pi  L^2 \rb^5 \left(L^2+\rb^2\right) (\rb-2 M)^{5/2}} \left( \Phi_{,rr [0]}^{\mathcal{E}} \mathcal{E} + \Phi_{,rr [0]}^{\mathcal{K}} \mathcal{K}\right),
\end{equation}
\begin{IEEEeqnarray}{rCl}
\Phi_{,rr [0]}^{\mathcal{K}} &=& \rb^2 (2 M-\rb) \left(76 L^4 M-48 L^4 \rb+54 L^2 M \rb^2-39 L^2 \rb^3-4 M \rb^4\right) \nonumber \\
&& -\: 4 E^2 M \rb^5 \left(2 L^2+\rb^2\right), \nonumber \\
\Phi_{,rr [0]}^{\mathcal{E}} &=& 4 E^2 M \rb^3 \left(4 L^4+7 L^2 \rb^2+\rb^4\right) \nonumber \\
 && -\:(2 M-\rb) \left(8 L^6 M+76 L^4 M \rb^2-39 L^4 \rb^3+28 L^2 M \rb^4-21 L^2 \rb^5-4 M \rb^6\right). \nonumber
\end{IEEEeqnarray}

\subsection{Example}
As a demonstration of the feasibility of this approach, we now consider the case
of a scalar particle on a circular geodesic orbit of the Schwarzschild
spacetime. In this case, the retarded field may be computed using frequency
domain methods \cite{Detweiler:Messaritaki:Whiting:2002}. This allows us to
obtain accurate values for the spherical-harmonic modes of the retarded field
and its derivative on the worldline (note that the value obtained depends on
whether the worldline is approached from inside or outside in the radial
direction). 
The spherical-harmonic decomposed wave equation, previously derived in Eq.~\eqref{eqn: phiLm},
\begin{equation}
\Phi^{lm}_{,rr} = -\frac{2(r-M)}{r(r-2M)} \Phi^{lm}_{,r} - \Big[\frac{\omega^2 r^2}{(r-2M)^2} - \frac{l(l+1)}{r(r-2M)}\Big] \Phi^{lm},
\end{equation}
then gives an algebraic relation for the second radial derivative of the
spherical-harmonic $l$ modes of the retarded field in terms of these. As
expected, these modes diverge with $l$ like $(2l+1)^2$.

\begin{figure}[ht]
\begin{center}
\includegraphics[width=12cm]{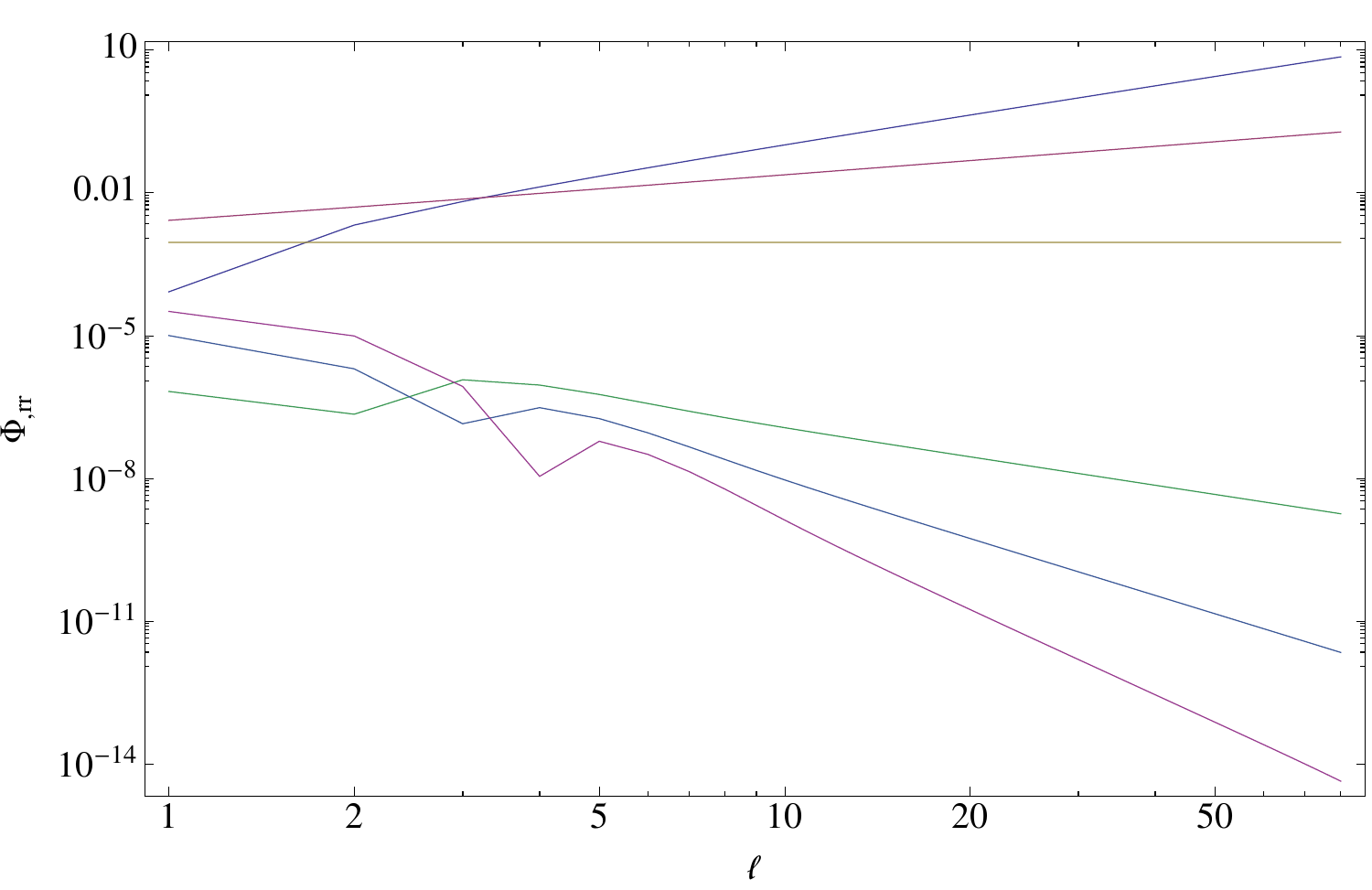}
\caption[Regularisation of the Second Radial Derivative of the Retarded Field]{Regularisation of the second radial derivative of the retarded field
for the case of a scalar particle on a circular geodesic of radius $r_0 = 10M$
in Schwarzschild spacetime.  In decreasing slope the above lines represent the
unregularised second derivative, and the second derivative regularised by subtracting from it in turn
the cumulative sum of $\Phi^l_{,rr[-2]}$, $\Phi^l_{,rr[-1]}$, $\Phi^l_{,rr\lpow{0}}$,
$\Phi^l_{,rr\lpow{2}}$ and $\Phi^l_{,rr\lpow{4}}$.}
\label{fig:phi_rr_modes}
\end{center}
\end{figure}
In Fig.~\ref{fig:phi_rr_modes}, we show the effect of subtracting in turn the
cumulative sums of the regularisation parameters from the second derivative of
the full retarded field. 
The parameters $\Phi_{,rr\lnpow{2}}$,
$\Phi_{,rr\lnpow{1}}$ and $\Phi_{,rr[0]}$ are the analytically derived ones
given above. The parameters $\Phi_{,rr\lpow{2}}$ and $\Phi_{,rr\lpow{4}}$ were
determined through a numerical fit to the data. The resulting rapid convergence
with $l$ enables the calculation of an extremely accurate value for the second
derivative of the retarded field. Summing over $l$, we find $\Phi_{,rr} =
-0.00000287908637(7)$, where the uncertainty in the last digit is estimated by
assuming that the error comes purely from the fact that the sum is only done up
to a finite $l_{\rm max} = 80$. \fixme{It is worth noting that our resulting expression here has fewer digits than our previous calculations due to the nature of this calculation, mainly that we are dealing with a quadratic $l$ dependence of the potential (or more loosely speacking we are taking a double covariant derivative).}



\chapter{Discussion} 


\ifpdf
    \graphicspath{{7/figures/PNG/}{7/figures/PDF/}{7/figures/}}
\else
    \graphicspath{{7/figures/EPS/}{7/figures/}}
\fi


Despite its profound fundamental importance, general relativity is a subject matter that has struggled to find its place in the world of everyday lives.  It was with great excitement in the 1970's, with the development of the global positioning system (GPS), that it was realised that general relativity must be taken into consideration to ensure accuracy.  Now, almost a century after the theory was born, we are on the brink of bringing about another influential application of the theory - as a telescope into the deepest, and certainly darkest, parts of our universe.  Gravitational wave astronomy is almost a reality, and the excitement of that reality has brought a new surge of energy to tackling many of general relativity's long-standing, open problems.

Bringing about a new era in astronomy is not without its challenges, challenges that thousands of scientist are currently working hard to overcome.  On the theory side, owing to the weakness of the signal strain ($10^{\text{\fixme{$-$}}21}$), the requirements for some of the most exciting detections necessitate prior knowledge of the expected waveforms for the gravitational radiation.  To this end, the two body problem is once again taking centre stage amongst relativists, and as a result, the self-force approach is coming under major focus.

The problem with the self-force lies in the singular nature of the field at the particle's position, making it unclear \fixme{as} to how the field \fixme{affects} the motion of the particle, a problem which also exists in electromagnetism.  The MiSaTaWaQu equations were a milestone towards overcoming this issue by correctly identifying the regular component of the field that is responsible for the dynamics.  Detweiler and Whiting also brought about a breakthrough in the understanding of the regularisation procedure, with the introduction of the Detweiler-Whiting singular field - the main focus of this thesis.  The Detweiler-Whiting singular field when subtracted from the retarded field gives the regular field, which, by construction, is \fixme{smooth and} wholly responsible for the self-force.

In calculating the self-force, one requires both the retarded and singular field \fixme{to} determine this regular field.  By concentrating on the singular field, we were able to bring about high order expansions with crucial benefits to the self-force calculation.  Prior to this work, focus on pushing the singular field to higher orders had diminished considerably with the general belief that calculating \fixme{these} terms would be so difficult as to make it infeasible.

In the mode-sum scheme, these high order \fixme{terms had} their greatest impact, allowing for the calculation of high order regularisation parameters.  One of the key ingredients that enabled us to obtain these parameters was the realisation that the rotation, that greatly simplifies \fixme{the resulting} parameters, can be done prior to calculating the singular field.  By readjusting the order in which we carried out the various steps, we were able to greatly reduce the computational stress attached to the high order terms, and complete the successful calculation of the previously unknown regularisation parameters. 

For the gravitational \Sch case  and Kerr space-times, only the first two parameters were previously known, from the original work by Barack and Ori \cite{Barack:Ori:2000, Barack:Ori:2002}.  This lack of higher order parameters resulted in very time-consuming and difficult numerical challenges.  The \Sch scalar and electromagnetic cases had had more success with the third parameter available, however, our high order parameters were warmly welcomed as they sped up calculations in these scenarios (which are still used as toy models to test new techniques).  In all cases, our results reduced the burden on computational resources and introduced an immediate improvement in the accuracy of self-force calculations.  As we were able to produce these results in all geodesic cases in both \Sch space-time and, thus far, for equatorial geodesics in Kerr space-times, this effect was not isolated to one or two situations but had an impact on a wide range of calculations \fixme{\cite{Thornburg:Wardell, Yang:2012bb, Akcay:2012ea, Ottewill:2012aj}}.

The impact of the high order expansions of the singular field is not restricted to the mode-sum approach.  Although the \fixme{mode-sum} has been the most accurate\fixme{,} practical regularisation scheme to date, more recently\fixme{,} new techniques have been introduced that are also showing encouraging results.  One such method is the effective source, independently introduced by Detweiler and Vega \cite{Vega:Detweiler:2008} and Barack and Golbourn \cite{Barack:Golbourn:2007}.  This involves directly solving the homogeneous wave equation for the regular field in the normal neighbourhood of the particle while solving the same wave equation for the retarded field away from the particle.  The two regions are united either by the world function (Detweiler, Vega) or a boundary condition (Barck, Golbourn).  The result is a regular field fully derived from an approximated singular field, easily obtainable from our singular field expansion, resulting in a very smooth effective source in both \Sch and Kerr.

As the effective source stems from the fact that the source must be evaluated in an extended region around the world-line, numerical evaluation can be time consuming, in particular, when using high order expansions such as the ones produced in this thesis.  Existing calculations have settled on expansions of $\epsilon^2$ to be a particular `sweet spot' for these calculations - up to this order the increase in complexity of the singular field `source' is rewarded with an increase in accuracy.  At the current state of the art, expansions above this order slow the calculations down to such a degree that the extra orders offer more of an hindrance than a help.  Nevertheless, our expansions have still \fixme{proved useful} - in the gravitational \Sch case and Kerr cases (equatorial plane), as the $\epsilon^2$ terms were previously unknown, therefore our expansions up to these terms were immediately desirable.

One of the applications of our results in the effective source approach is in the $m$-mode scheme developed by Barack and Golbourn \cite{Barack:Golbourn:Sago:2007}.  This scheme decomposes the retarded field and effective source into azimuthal modes, making it more suited to the Kerr space-time as it conserves the axial symmetry.  By carrying out these calculations with an effective source accurate to $\epsilon^2$, one can obtain an expression for the self-force in $m$-modes.  We have now \fixme{introduced a new} method of using our higher terms of the singular field to obtain `$m$-mode' parameters, which we have shown lead to a faster convergence of the $m$-mode sum.

There has also been recent interest in the self-force in its application to non-geodesic motion, due to the role it appears to play in the cosmic censorship conjecture.  The cosmic censorship conjecture, first proposed by Penrose in 1969 \cite{Penrose:1969}, suggests that singularities produced from gravitational collapse will always be `hidden' behind an horizon.  The \rn space-time, however, naturally gives rise to the concept of a naked singularity if the magnitude of the charge of the black hole is greater than that of its mass.  One can, therefore, infer that for the cosmic conjecture to be correct, we have an upper limit on the charge of the black hole.  Similarly the Kerr-Newman space-time imposes an upper limit on the charge and spin of the black hole.  This seems to suggest, if one can increase a black hole's charge or spin past these thresholds, one would force a naked singularity.  To this end, researchers have been trying to evolve such scenarios by attempting to \emph{overspin} or \emph{overcharge} a black hole by saturating it with suitable particles \cite{Wald:1974, Hubeny:1998, Jacobson:2009, Zimmerman:2010, Barausse:2011}.  

Wald was the first to attempt such a scenario by considering an extremal black hole (at the threshold) in Kerr-Newman \cite{Wald:1974}, and firing test particles at the black hole that carry the required charge or spin which, when swallowed by the black hole, would push its charge over the threshold.  However, Wald found that an extremal black hole would not absorb such particles, in fact it seemed to repel them.  Hubeny carried out a similar scenario in \rn space-time \cite{Hubeny:1998}, but considered a near extremal black hole.  The simulation was a success in the sense that the black hole did capture the required particles and became overcharged.  However, the self-force was not considered in these scenarios so they were not completely conclusive.  Jacobson and Sotiriou similarly managed to overspin a black hole in Kerr space-time \cite{Jacobson:2009}, but again\fixme{,} these calculations were carried out without the self-force correction.  It is only recently \cite{Zimmerman:2010}, that these calculations are being carried out with the self-force.  Barausse and collaborators \cite{Barausse:2011} recently considered similar scenarios, and found the self-force reduces the size of the parameter space of possible `overcharging' particles, however\fixme{,} they could not come to a definitive conclusion.  

All of these developments have led researchers to believe that it is perhaps the self-force that protects these particles from being swallowed by the black hole, suggesting its role as the cosmic censor.  This is leading to a push for more accurate self-force calculations for non-geodesic motion.  These expressions are more complicated than those of geodesic motion, making the regularisation parameters even more crucial.  To this end, we have calculated the regularisation parameters for generic motion in a spherically symmetric space-time, as well as those for more specialised cases\fixme{,} such as generic motion in \Sch or radial infall in a spherically symmetric space-time, \fixme{and that of} a charged particle in \rn space-time.  All of these have been in the scalar case as they are considered stepping stones towards such calculations in higher spins.  We have calculated the parameters and checked them in various toy models, and so are confident, as we have all the necessary tools, that we will shortly produce the required parameters for the higher spins, which in turn can be used to investigate the nature of the self-force in such scenarios.

One of the main motivations for the current work is the application of self-force in predicting the necessary wave-forms for gravitational wave detection.  The more accurate the predicted wave-form, the higher the possibility of detection.  To date, the self-force, which arises in a perturbation expansion in the mass ratio of two bodies, has only ever been calculated to first order.  Recent developments have now made available an outline for calculating the second order self force  \cite{Detweiler:2011tt, Gralla:2012db, Pound:2012nt}.  These calculations, amongst other necessities, require regularisation parameters for derivatives of high order expansions of the singular field as well as the ability to calculate other coordinate expansions.  To this end\fixme{,} we have calculated the required second order derivative regularisation parameters of the singular field, and are confident that we can build on these in calculating other high order expansions required for the second-order calculation.

In summary, the work carried out in this thesis on the singular field has led to many applications in the self-force problem.  Our regularisation parameters have dramatically increased the accuracy of current self-force calculations.  We have also made the desired smooth effective source available, and offered a new method to use the higher terms to increase the convergence of the self-force within the $m$-mode scheme.  All of these were accomplished in scalar, electromagnetic and gravitational cases in both \Sch and Kerr space-time.  We were also able to offer an application in the more fundamental elements of general relativity, in assisting in the investigation of self-force's role in the cosmic censorship conjecture.  Finally, we showed how our results have an importance in the very exciting and ongoing work towards obtaining the first ever second order self force calculation.  Both applications in cosmic censorship and second order are still ongoing and offer further work.  We have made only baby steps in these areas, but we are optimistic about our future results.  To this end, we have enjoyed the journey so far and look forward to its exciting continuation.



\appendix


\chapter{Coordinate Expansions in \Sch Space-Time} \label{sec:CoordinateExpansions} 


\ifpdf
    \graphicspath{{9_backmatter/AppendixCoordinate/figures/PNG/}{9_backmatter/AppendixCoordinate/figures/PDF/}{9_backmatter/AppendixCoordinate/figures/}}
\else
    \graphicspath{{9_backmatter/AppendixA/figures/EPS/}{9_backmatter/AppendixCoordinate/figures/}}
\fi


In this Appendix we give coordinate expansions of the key quantities appearing in the singular field,
Eqs.~\eqref{eq:PhiS}, \eqref{eq:AS} and \eqref{eq:hS}.  Using spherical symmetry, time translation and reversal invariance, any regular biscalar respecting the symmetries of the space-time may be written as,
\begin{equation}
\sum_{i,j,k=0}^{i+j+2 k\le9} \sigma_{ijk} (t'-t)^i (r'-r)^j (1-\cos \gamma)^k
 +\mathcal{O}(\epsilon^{10}),
\end{equation}
which we will use in sections \ref{sec: ASWF} and \ref{sec: AVV} to produce high order expansions of both Synge's world function and the Van Vleck determinant.


\section{Synge World function} \label{sec: ASWF}

Letting $\cos \gamma = \cos \theta \cos \theta' - \sin \theta \sin \theta' \cos(\phi - \phi')$
so that $2(1-\cos\gamma) = \Delta w_1^2 + \Delta w_2^2$, the expansion of the
world function to the order required in this paper is
\begin{align} \label{eqn: sigma expansion}
\sigma(x&,x') = \sum_{i,j,k=0}^{i+j+2 k\le9} \sigma_{ijk} (t'-t)^i (r'-r)^j (1-\cos \gamma)^k
 +\mathcal{O}(\epsilon^{10}),
\end{align}
where the non-zero coefficients are
\begin{gather} \label{eqn: sigmaExpandedFully}
\sigma_{001}=r^2, \quad
\sigma_{002}=\frac{M r}{3},\quad
\sigma_{003}=\frac{1}{90} M (9 r-2 M),\nonumber \\
\sigma_{004}=\frac{M \left(14 M^2-19 M r+30 r^2\right)}{840 r}, \quad
\sigma_{011}=r,\quad
\sigma_{012}=\frac{M}{6},\quad
\sigma_{013}=\frac{M}{20},\nonumber \\
\sigma_{014}=\frac{1}{840} M \left(15-\frac{7 M^2}{r^2}\right),\quad
\sigma_{020}=-\frac{r}{4 M-2 r},\quad
\sigma_{021}=\frac{M}{12 M-6 r},\nonumber \\
\sigma_{022}=\frac{M (r-M)}{60 r (2 M-r)},\quad
\sigma_{023}=\frac{M \left(-14 M^2+6 M r+3 r^2\right)}{840 r^2 (2 M-r)},\quad
\sigma_{030}=-\frac{M}{2 (r-2 M)^2},\nonumber \\
\sigma_{031}=\frac{M}{12 (r-2 M)^2},\quad
\sigma_{032}=\frac{M \left(2 M^2-2 M r+r^2\right)}{120 r^2 (r-2 M)^2},\nonumber \\
\sigma_{033}=\frac{M \left(56 M^3-54 M^2 r+12 M r^2+3 r^3\right)}{1680 r^3 (r-2 M)^2},\quad
\sigma_{040}=-\frac{M (M-8 r)}{24 r (r-2 M)^3},\nonumber \\
\sigma_{041}=\frac{M \left(5 M^2-3 M r-6 r^2\right)}{120 r^2 (r-2 M)^3},\quad
\sigma_{042}=\frac{M \left(42 M^3-70 M^2 r+39 M r^2-16 r^3\right)}{3360 r^3 (r-2 M)^3},\nonumber \\
\sigma_{050}=-\frac{M \left(M^2-2 M r+6 r^2\right)}{24 r^2 (r-2 M)^4},\quad
\sigma_{051}=\frac{M \left(20 M^3-31 M^2 r+12 M r^2+8 r^3\right)}{240 r^3 (r-2 M)^4},\nonumber \\
\sigma_{052}=\frac{M \left(7 M^4-21 M^3 r+23 M^2 r^2-11 M r^3+5 r^4\right)}{1680 r^4 (r-2 M)^4},\nonumber \\
\sigma_{060}=-\frac{M \left(35 M^3-86 M^2 r+86 M r^2-144 r^3\right)}{720 r^3 (r-2 M)^5},\nonumber \\
\sigma_{061}=\frac{M \left(245 M^4-532 M^3 r+421 M^2 r^2-120 M r^3-40 r^4\right)}{1680 r^4 (r-2 M)^5},\nonumber\\
\sigma_{070}=\frac{M \left(-15 M^4+44 M^3 r-54 M^2 r^2+36 M r^3-40 r^4\right)}{240 r^4 (r-2 M)^6},\nonumber\\
\sigma_{071}=\frac{M \left(280 M^5-763 M^4 r+832 M^3 r^2-444 M^2 r^3+100 M r^4+20 r^5\right)}{1120 r^5 (r-2 M)^6},\nonumber \\
\sigma_{080}=\frac{M \left(385 M^5-1316 M^4 r+1928 M^3 r^2-1576 M^2 r^3+788 M r^4-640 r^5\right)}{4480 r^5 (2 M-r)^7},\nonumber \\
\sigma_{090}=\frac{M}{40320 r^6 (r-2 M)^8} \big(-5005 M^6+19558 M^5 r-33394 M^4 r^2+32584 M^3 r^3 \nonumber \\
-19960 M^2 r^4+7984 M r^5-5040 r^6\big), \nonumber \\
\sigma_{200}=\frac{M}{r}-\frac{1}{2}, \quad
\sigma_{201}=\frac{M (r-2 M)}{6 r^2},\quad
\sigma_{202}=\frac{M \left(10 M^2-11 M r+3 r^2\right)}{60 r^3},\nonumber \\
\sigma_{203}=\frac{M \left(-92 M^3+142 M^2 r-78 M r^2+15 r^3\right)}{840 r^4},\quad
\sigma_{210}=-\frac{M}{2 r^2},\nonumber \\
\sigma_{211}=\frac{M (4 M-r)}{12 r^3},\quad
\sigma_{212}=-\frac{M \left(30 M^2-22 M r+3 r^2\right)}{120 r^4},\nonumber \\
\sigma_{213}=\frac{M \left(368 M^3-426 M^2 r+156 M r^2-15 r^3\right)}{1680 r^5},\quad
\sigma_{220}=\frac{M (5 M-4 r)}{12 r^3 (2 M-r)},\nonumber \\
\sigma_{221}=-\frac{M \left(23 M^2-20 M r+3 r^2\right)}{60 r^4 (2 M-r)},\nonumber \\
\sigma_{222}=\frac{M \left(686 M^3-802 M^2 r+281 M r^2-24 r^3\right)}{1680 r^5 (2 M-r)},\quad
\sigma_{230}=-\frac{M (M-r)^2}{4 r^4 (r-2 M)^2},\nonumber \\
\sigma_{231}=\frac{M \left(24 M^3-55 M^2 r+32 M r^2-4 r^3\right)}{120 r^5 (r-2 M)^2},\nonumber \\
\sigma_{232}=\frac{M \left(-630 M^4+1234 M^3 r-832 M^2 r^2+218 M r^3-15 r^4\right)}{1680 r^6 (r-2 M)^2},\nonumber \\
\sigma_{240}=-\frac{M \left(M^3-78 M^2 r+116 M r^2-48 r^3\right)}{240 r^5 (r-2 M)^3},\nonumber \\
\sigma_{241}=-\frac{M \left(553 M^4-269 M^3 r-444 M^2 r^2+322 M r^3-40 r^4\right)}{1680 r^6 (r-2 M)^3},\nonumber \\
\sigma_{250}=\frac{M \left(75 M^4-84 M^3 r-51 M^2 r^2+104 M r^3-40 r^4\right)}{240 r^6 (r-2 M)^4},\nonumber \\
\sigma_{251}=\frac{M \left(-4396 M^5+8227 M^4 r-4760 M^3 r^2+350 M^2 r^3+412 M r^4-60 r^5\right)}{3360 r^7 (r-2 M)^4},\nonumber \\
\sigma_{260}=\frac{M \left(2317 M^5-5560 M^4 r+4220 M^3 r^2-146 M^2 r^3-1254 M r^4+480 r^5\right)}{3360 r^7 (r-2 M)^5},\nonumber \\
\sigma_{270}=\frac{M}{3360 r^8 (r-2 M)^6} \big(3759 M^6-12402 M^5 r+16015 M^4 r^2-9336 M^3 r^3\nonumber \\ +\:1347 M^2 r^4 
+1048 M r^5-420 r^6\big),\nonumber \\
\sigma_{400}=\frac{M^2 (2 M-r)}{24 r^5}, \quad
\sigma_{401}=-\frac{M^2 \left(54 M^2-49 M r+11 r^2\right)}{360 r^6},\nonumber\\
\sigma_{402}=\frac{M^2 \left(1956 M^3-2702 M^2 r+1228 M r^2-183 r^3\right)}{10080 r^7},\quad
\sigma_{410}=\frac{M^2 (2 r-5 M)}{24 r^6},\nonumber \\
\sigma_{411}=\frac{M^2 \left(324 M^2-245 M r+44 r^2\right)}{720 r^7},\nonumber \\
\sigma_{412}=\frac{M^2 \left(-3423 M^3+4053 M^2 r-1535 M r^2+183 r^3\right)}{5040 r^8},\nonumber \\
\sigma_{420}=\frac{M^2 \left(429 M^2-394 M r+86 r^2\right)}{720 r^7 (2 M-r)},\nonumber \\
\sigma_{421}=\frac{M^2 \left(-7509 M^3+9170 M^2 r-3575 M r^2+438 r^3\right)}{5040 r^8 (2 M-r)},\nonumber \\
\sigma_{430}=\frac{M^2 \left(-301 M^3+460 M^2 r-228 M r^2+36 r^3\right)}{240 r^8 (r-2 M)^2},\nonumber \\
\sigma_{431}=\frac{M^2 \left(35472 M^4-63269 M^3 r+40912 M^2 r^2-11240 M r^3+1088 r^4\right)}{10080 r^9 (r-2 M)^2},\nonumber \\
\sigma_{440}=-\frac{M^2 \left(42507 M^4-94876 M^3 r+77976 M^2 r^2-27740 M r^3+3546 r^4\right)}{20160 r^9 (r-2 M)^3},\nonumber \\
\sigma_{450}=\frac{M^2}{20160 r^{10} (r-2 M)^4}  \big(-57987 M^5+178306 M^4 r-214952 M^3 r^2 \nonumber \\
+\: 126744 M^2 r^3 -36328 M r^4+3992 r^5\big),\nonumber \\
\sigma_{600}=\frac{M^3 \left(26 M^2-25 M r+6 r^2\right)}{720 r^9},\nonumber \\
\sigma_{601}=\frac{M^3 \left(-1818 M^3+2475 M^2 r-1115 M r^2+166 r^3\right)}{15120 r^{10}},\nonumber \\
\sigma_{610}=-\frac{M^3 \left(117 M^2-100 M r+21 r^2\right)}{720 r^{10}},\nonumber \\
\sigma_{611}=\frac{M^3 \left(18180 M^3-22275 M^2 r+8920 M r^2-1162 r^3\right)}{30240 r^{11}},\nonumber \\
\sigma_{620}=\frac{M^3 \left(23931 M^3-31560 M^2 r+13652 M r^2-1930 r^3\right)}{30240 r^{11} (2 M-r)},\nonumber \\
\sigma_{630}=\frac{M^3 \left(-27687 M^4+51002 M^3 r-34745 M^2 r^2+10344 M r^3-1131 r^4\right)}{10080 r^{12} (r-2 M)^2},\nonumber \\
\sigma_{800}=\frac{M^4 \left(978 M^3-1393 M^2 r+660 M r^2-104 r^3\right)}{40320 r^{13}}, \nonumber \\
\sigma_{810}=\frac{M^4 \left(-6357 M^3+8358 M^2 r-3630 M r^2+520 r^3\right)}{40320 r^{14}}.
\end{gather}


\section{Van Vleck determinant} \label{sec: AVV}
Inserting the above expansion for $\sigma(x,x')$ into the definition of the Van Vleck determinant,
Eq.~\eqref{eq:vanVleckDefinition}, gives
\begin{align}
\Delta^{1/2}(x,x')= 1 + \sum_{i,j,k=0}^{i+j+2 k\le7} \Delta^{1/2}_{ijk} (t'-t)^i (r'-r)^j (1-\cos \gamma)^k
 +\mathcal{O}(\epsilon^{\text{\fixme{8}}}),
\end{align}
where the non-zero coefficients are
\begin{gather}
\Delta^{1/2}_{002}=\frac{M^2}{15 r^2}, \quad
\Delta^{1/2}_{003}=\frac{M^2 (27 r-34 M)}{378 r^3}, \quad
\Delta^{1/2}_{012}=-\frac{M^2}{15 r^3},  \nonumber \\
\Delta^{1/2}_{013}=\frac{M^2 (17 M-9 r)}{126 r^4}, \quad
\Delta^{1/2}_{021}=\frac{M^2}{60 M r^3-30 r^4}, \quad
\Delta^{1/2}_{022}=\frac{M^2 (322 M-177 r)}{2520 r^4 (2 M-r)},  \nonumber \\
\Delta^{1/2}_{031}=\frac{M^2 (2 r-3 M)}{30 r^4 (r-2 M)^2}, \quad
\Delta^{1/2}_{032}=\frac{M^2 \left(-308 M^2+332 M r-93 r^2\right)}{1260 r^5 (r-2 M)^2},  \nonumber \\
\Delta^{1/2}_{040}=\frac{M^2}{60 r^4 (r-2 M)^2}, \quad
\Delta^{1/2}_{041}=-\frac{M^2 \left(910 M^2-1268 M r+459 r^2\right)}{5040 r^5 (r-2 M)^3},  \nonumber \\
\Delta^{1/2}_{050}=\frac{M^2 (4 M-3 r)}{60 r^5 (r-2 M)^3}, \quad
\Delta^{1/2}_{051}=\frac{M^2 \left(-1190 M^3+2664 M^2 r-2035 M r^2+537 r^3\right)}{5040 r^6 (r-2 M)^4}, \nonumber \\
\Delta^{1/2}_{060}=\frac{M^2 \left(5432 M^2-7720 M r+2943 r^2\right)}{30240 r^6 (r-2 M)^4}, \nonumber \\
\Delta^{1/2}_{070}=\frac{M^2 \left(1036 M^3-2120 M^2 r+1524 M r^2-393 r^3\right)}{2520 r^7 (r-2 M)^5}, \quad
\Delta^{1/2}_{201}=\frac{M^2 (r-2 M)}{30 r^5}, \nonumber \\
\Delta^{1/2}_{202}=\frac{M^2 \left(460 M^2-428 M r+99 r^2\right)}{2520 r^6}, \quad
\Delta^{1/2}_{211}=\frac{M^2 (5 M-2 r)}{30 r^6}, \nonumber \\
\Delta^{1/2}_{212}=-\frac{M^2 \left(690 M^2-535 M r+99 r^2\right)}{1260 r^7}, \quad
\Delta^{1/2}_{220}=-\frac{M^2}{30 r^6}, \nonumber \\
\Delta^{1/2}_{221}=-\frac{M^2 \left(443 M^2-412 M r+90 r^2\right)}{1260 r^7 (2 M-r)}, \quad
\Delta^{1/2}_{230}=\frac{M^2}{10 r^7}, \nonumber \\
\Delta^{1/2}_{231}=\frac{M^2 \left(161 M^3-464 M^2 r+317 M r^2-60 r^3\right)}{1260 r^8 (r-2 M)^2}, \nonumber \\
\Delta^{1/2}_{240}=\frac{M^2 \left(-3526 M^2+3746 M r-981 r^2\right)}{5040 r^8 (r-2 M)^2}, \nonumber \\
\Delta^{1/2}_{250}=-\frac{M^2 \left(9392 M^3-15628 M^2 r+8631 M r^2-1572 r^3\right)}{5040 r^9 (r-2 M)^3},  \nonumber \\
\Delta^{1/2}_{400}=\frac{M^2 (r-2 M)^2}{60 r^8}, \quad
\Delta^{1/2}_{401}=\frac{M^2 \left(-1300 M^3+1682 M^2 r-714 M r^2+99 r^3\right)}{5040 r^9}, \nonumber \\
\Delta^{1/2}_{410}=-\frac{M^2 \left(16 M^2-14 M r+3 r^2\right)}{60 r^9}, \nonumber \\
\Delta^{1/2}_{411}=\frac{M^2 \left(5850 M^3-6728 M^2 r+2499 M r^2-297 r^3\right)}{5040 r^{10}}, \nonumber \\
\Delta^{1/2}_{420}=\frac{M^2 \left(5648 M^2-4800 M r+981 r^2\right)}{10080 r^{10}},  \nonumber \\
\Delta^{1/2}_{430}=-\frac{M^2 \left(2020 M^2-1872 M r+393 r^2\right)}{2520 r^{11}}, \quad
\Delta^{1/2}_{600}=\frac{M^3 (199 M-89 r) (r-2 M)^2}{7560 r^{12}}, \nonumber \\
\Delta^{1/2}_{610}=\frac{M^3 \left(-3184 M^3+4224 M^2 r-1850 M r^2+267 r^3\right)}{5040 r^{13}}.
\end{gather}


\section{Expansions of an arbitrary point on the world line about $x^\ab$}

The four velocity of a general geodesic orbit taken to lie in the equatorial plane is given by  the standard expressions~\cite{Chandrasekhar}
\begin{gather} \label{eqn: ualpha}
\dot{t}(\tau) = \frac{E r(\tau)}{r(\tau) - 2M}, \quad \quad
\dot{r}(\tau) = \sqrt{E^2 - \left(1 - \frac{2M}{r(\tau)}\right) \left(1 + \frac{L^2}{r(\tau)^2}\right)}, \nonumber \\
\dot{\theta}(\tau) = 0, \quad \quad
\dot{\phi}(\tau) = \frac{L}{r(\tau)^2}.
\end{gather}
It is straightforward to calculate the higher order proper time derivatives of these expressions and evaluate both the four velocity and its higher derivatives at $x^{\bar{a}}$, giving, for example,   
\begin{gather}
\dot{t}_0 = \frac{E r_0}{r_0 - 2M},
\quad \quad \dot{r}_0 = \sqrt{E^2 - \left(1 - \frac{2M}{r_0}\right) \left(1 + \frac{L^2}{r_0^2}\right)}, 
\quad \quad \dot{\theta}_0 = 0, 
\quad \quad \dot{\phi}_0 = \frac{L}{r_0^2}, 
\quad \nonumber\\ 
\ddot{t}_0 = -\frac{2 E M \dot{r}_0}{\left(r_0-2 M\right)^2}, 
\quad \quad \ddot{r}_0 = \frac{L^2 r_0-M r_0^2-3 L^2 M}{r_0^4}, 
\quad \quad \ddot{\theta}_0 = 0, 
\quad \quad \ddot{\phi}_0 = -\frac{2 L \dot{r}_0}{r_0^3}, 
\quad \nonumber \\
\dddot{t}_0 = \frac{2 E M \left[2 \left(E^2-1\right) r_0^4-r_0^2 \left(3 L^2+2 M^2\right)+9 L^2 M r_0+5 M r_0^3-6 L^2 M^2\right]}{r_0^4 \left(r_0-2 M\right)^3},
\quad \nonumber \\
\dddot{r}_0 = \frac{\dot{r}_0 \left(-3 L^2 r_0+2 M r_0^2 + 12 L^2 M\right)}{r_0^5},
\quad \nonumber \\
\dddot{\theta}_0 = 0,
\quad \quad \dddot{\phi}_0 = \frac{2 L \left[3 \left(E^2-1\right) r_0^3-4 L^2 r_0+7 M r_0^2+9 L^2 M\right]}{r_0^7}.
\label{eqn: Four Velocity Bar}
\end{gather}
Combining Eq.~\eqref{eqn: Four Velocity Bar} with Eq.~\eqref{eqn: xtilde}, we can express $x^{a'}$ in terms of $x^{\bar{a}}$ and $\Delta \tau$:
\begin{gather} 
t' = \frac{E r_0}{r_0-2 M} \Delta\tau -\frac{E M \dot{r}_0}{\left(r_0-2 M\right)^2}  \Delta\tau^2 + \cdots, \nonumber \\
r' = r_0+ \dot{r}_0  \Delta\tau - \frac{ \left(-L^2 r_0+M r_0^2+3 L^2 M\right)}{2 r_0^4} \Delta\tau ^2 + \cdots, 
\nonumber \\
\theta' = \frac{\pi}{2}, 
\quad \quad \phi' = \frac{L}{r_0^2} \Delta\tau - \frac{L \dot{r}_0}{r_0^3} \Delta\tau^2 + \cdots.
\label{Anna eqn: xprimealpha}
\end{gather}
It is also straightforward to obtain $\delta x^{a'}$, in terms of $\Delta x^{a}$,
$x^{\bar{a}}$ and $\Delta \tau$ by noting that $\delta x^{a'} = x^{a'} - \Delta x^a - x^{\bar{a}}$.
Finally, we can calculate $u^{a'}$ in terms of $x^{\bar{a}}$ and $\Delta \tau$ by inserting $r'$ from Eq.~\eqref{Anna eqn: xprimealpha} into our equations for the four velocity, Eq.~\eqref{eqn: ualpha}
\par \vspace{-6pt} \begin{IEEEeqnarray}{rCl}
u^{t'} &=& \frac{E r_0}{r_0-2 M} - \frac{2 E M r  \rbdot}{\left(r_0-2 M\right)^2} \Delta \tau 
	\nonumber \\
&&
	+\: \frac{E M}{r_0^4 \left(2 M-r_0\right)^3} \big\{6 L^2 M^2 - 9 L^2 M r_0 + r_0^2 \big[3 L^2+2 M^2 - 2 \left( E^2 - 1 
	\right) r_0^2 \nonumber \\
&&	
	\quad -\: 5 M r_0\big]\big\} \Delta \tau ^2 + \cdots, 
	\nonumber \\
u^{r'} &=& \rbdot + \frac{r_0 \left(L^2-M r_0\right)-3 L^2 M}{r_0^4} \Delta \tau  + \frac{\rbdot 
	\left(12 L^2 M -3 L^2 r_0+2 M r_0^2\right)}{2 r_0^5} \Delta \tau ^2 + \cdots, \nonumber 
	\\
u^{\theta'} &=& 0, \nonumber \\
u^{\phi'} &=& \frac{L}{r_0^2} - \frac{2 L \rbdot}{r_0^3} \Delta \tau + \frac{L \left[3 \left( E^2 - 1 
	\right) r_0^3-4 L^2 r_0+7 M r_0^2+9 L^2 M\right]}{r_0^7} \Delta \tau ^2 + \cdots.
\label{eqn: ualphaprime}
\end{IEEEeqnarray}


\section{Expansions of retarded and advanced points}
Taking $\Delta \tau$ to have leading order $\epsilon$, the same leading order of our $\Delta x$ terms, we can further expand it in orders of $\epsilon$, giving
\begin{equation} \label{eqn: deltatau}
\Delta \tau = \tau_1 \epsilon + \tau_2 \epsilon^2 + \tau_3 \epsilon^3 + \tau_4 \epsilon^4 + \cdots.
\end{equation}
Substituting $\delta x^{a'}$ obtained from Eq.~\eqref{Anna eqn: xprimealpha} and $\Delta \tau$ from Eq.~\eqref{eqn: deltatau} into $\sigma (x, x')$, Eq.~\eqref{eqn: nullseparated}, gives $\sigma (x, x')$ as a function of $\Delta x^{a}$, $x^{\ab}$ and the $\tau_n$'s:
\par \vspace{-6pt} \begin{IEEEeqnarray}{rCl}
\sigma(x,x') &=& \frac{1}{2} \Bigg[\frac{\left(2 M - \bar{r} \right) \Delta t^2}{ \bar{r}} + \frac{ 
	\bar{r} \left(\Delta r-2 \rbdot \tau _1 \right) \Delta r}{\bar{r}-2 m}+\bar{r}^2 \left(\Delta 
	\theta^2 + \Delta \phi ^2\right) -\big(2 L \Delta \phi +\tau _1 \nonumber \\
&&
	-\: 2 E \Delta t \big) \tau _1 \Big] + \frac{1}{2} \Bigg\{ \frac{2}{\bar{r}} \left(\frac{E M \Delta 
	t}{\bar{r}-2 M}- L \Delta \phi \right) \tau _1 \Delta r +\frac{M \left(\rbdot \tau _1 -\Delta 
	r\right) \Delta r^2}{\left(\bar{r}-2 M\right)^2} \nonumber \\
&&
	+\:\bar{r} \left[\left(\rbdot \tau _1 + \Delta r\right) \left(\Delta \theta ^2+\Delta \phi ^2 
	\right) -\frac{2 \rbdot  \tau _2 \Delta r}{\bar{r}-2 M}\right] -\frac{M \left(\rbdot \tau _1 + 
	\Delta r \right) \Delta t^2 }{\bar{r}^2} \nonumber \\
&&
	-\: 2 \left(L \Delta \phi -E \Delta t +\tau _1\right) \tau _2 \Bigg\} + \cdots.
\end{IEEEeqnarray}
If we now specify that $x^{a'}$ coincides with the point where the world line intersects with the light cone of $x$, we can use the equation $\sigma (x, x') = 0$ to solve for the $\tau_n$'s in terms of $\Delta x^{a}$ and $x^{\ab}$.  This gives us
\begin{align}
\tau_1 =& E \Delta t - L \Delta \phi + \frac{ \bar{r} \rbdot \Delta r}{2 M-\bar{r}} \pm \zrho, \nonumber \\
\tau_2 =& \frac{\pm 1}{8 \zrho } \Bigg( \frac{\Delta r}{M^2 \bar{r}} \big\{L^2 \left[\Delta r^2+4 
	M^2 \left(\Delta \theta ^2+3 \Delta \phi ^2\right)\right]-4  L M^2  (E \Delta t\pm 2 \zrho 
	) \Delta \phi  \nonumber \\
&
	-\: 2 E M^2 \big(3 E \Delta t \pm  2 \zrho\big ) \Delta t \big\} + \frac{1}{M 
	\bar{r}^2} \big\{4 L M^2 \rbdot \Delta t^2 \Delta \phi -4 M^2 
	\big[\rbdot (E \Delta t\pm \zrho )\nonumber \\
&
	+\: 2 \Delta r \big] \Delta t^2 + L^2 \Delta r^3 \big\} + \frac{ \Delta r }{M^2 \left(2 M-\bar{r}\right)} \big[4  L M^2 \left(E \Delta t-2 \rbdot 
	\Delta r \right) \Delta \phi \nonumber \\
&
	+\:8 E^2 M^3 \bar{r} \left(\Delta \theta ^2+\Delta \phi ^2\right) 
	+ L^2 \Delta r^2 - 2 E M^2 (3 E \Delta t\pm 2 \zrho ) \Delta t \big] \nonumber \\
&
	-\:\frac{ \Delta r^2}{M \left(\bar{r}-2 M 
	\right) ^3} \Big\{ \left(\bar{r}-2 M\right) \big[4 L M^2 \rbdot \Delta \phi + 4 M^2 \big(E 
	\rbdot \Delta t \mp  \rbdot \zrho + E^2 \Delta r\big)\nonumber \\
&
	-\:  L^2 \Delta r \big] + 8 E^2 M^3 \Delta r \Big\} - 
	\frac{4 L^2 M \Delta r \Delta t^2}{\bar{r}^4} - 4 \bar{r} \left(\Delta \theta ^2+\Delta \phi ^2\right) \big[\left(E^2-2\right) \Delta r \nonumber \\
&
	-\: 
	\rbdot \left( E \Delta t- L \Delta \phi \pm \zrho \right)\big]\Bigg), 
\label{eqn: taus}
\end{align}
with the higher order terms following in the same manner.

Using our equations for $\Delta \tau$, Eqs.~\eqref{eqn: deltatau} and \eqref{eqn: taus}, we rewrite $x^{a'}$ (and consequently $\delta x^{a'}$), the four velocity, $u^{a'}$, $\Delta^{\frac{1}{2}} (x, x')$ and $\sigma_{a'}$  (Eqs.~\eqref{Anna eqn: xprimealpha}, \eqref{eqn: ualphaprime}, ~\eqref{eqn: Van Vleck 2} and \eqref{eqn: sigmaalphaprime} respectively), in terms of $\Delta x^{a}$ and $x^{\ab}$.  


\section{Bivector of Parallel Transport}

To calculate the bivector of parallel transport, $g^a{}_{b'} (x,x')$, we first write it
in terms of a coordinate expansion about $x$,
\begin{equation} \label{eqn: bivector expansion}
g^a{}_{b'}(x,x') = \delta^a{}_{b'} + G^a{}_{bc}(x) \delta x^{c'} + G^a{}_{bcd} (x) \delta x^{c'} \delta x^{d'} + G^a{}_{bcde}(x) \delta x^{c'} \delta x^{d'} \delta x^{e'} + \dots,
\end{equation}
where the coefficients $G^a{}_{b...}(x)$ are functions of $x^a$ written in terms of $\Delta x^a$ and $x^{\bar{a}}$.  Calculating $g^a{}_{b',c'} (x,x')$ is straight forward:
\begin{align} \label{eqn: gabc}
g^a{}_{b',c'} (x,x') =& G^a{}_{bc}(x) + 2 G^a{}_{bcd}(x) \delta x^{d'} + 3 G^a{}_{bcde} (x) \delta 
	x^{d'} \delta x^{e'}  \nonumber \\
&
	+\: 4 G^a{}_{bcde}(x) \delta x^{d'} \delta x^{e'} \delta x^{f'} + \cdots.
\end{align}
Using the identity $g^a{}_{b';c'} \sigma^{c'} = g^a{}_{b',c'} \sigma^{c'} - \Gamma^{d'}_{b'c'} g^{a}_{d'} \sigma^{c'} = 0 $ with Eqs.~\eqref{eqn: sigmaaprime}, \eqref{eqn: gabc}, \eqref{eqn: bivector expansion}, and our expression for $\delta x^{a'}$ (obtained from the previous section), one can calculate the above coefficients and hence obtain the bivector of parallel transport, $g^a{}_{b'} (x,x')$, in terms of $\Delta x^a$ and $x^{\bar{a}}$.


\section{Scalar Singular Field}

Combining Eqs.~\eqref{eq:PhiS}, \eqref{eqn: GgGen} and \eqref{eq: U}, the scalar singular field can be written as

\begin{equation} \label{eq: PhiS Simp}
\Phi^{\rm \sing}(x) =  \frac{q}{2} \Bigg[ \frac{\Delta^{\tfrac{1}{2}}(x,x')}{\sigma_{c'}(x,x') u^{c'}(x')} \Bigg]_{x'=x_{\rm \ret}}^{x'=x_{\rm \adv}} + \frac{q}{2} \int_{\tau_{\rm \ret}}^{\tau_{\rm \adv}} V(x,x(\tau')) d\tau.
\end{equation}
We already have everything required for the first term here, which gives the direct part of the
scalar singular field.  It should be noted that $x'=x_{\rm \ret}$ and $x'=x_{\rm \adv}$
are the equivalent of setting $\pm \rho = -\rho$ and $\pm \rho = +\rho$ respectively when
substituting $\tau_n$, Eq.~\eqref{eqn: taus}, into $\Delta^{1/2}$, $\sigma_{a'}$ and $u^{a'}$.

In the scalar case, Eq.~\eqref{eq:V} for the scalar tail part becomes
\begin{equation} \label{eqn: VScalar}
V (x, x') = \sum^{\infty}_{\num=0} V_\num (x, x') \sigma^\num (x, x').
\end{equation}
To calculate coordinate expansions of the $V_\num$, first we require a coordinate expansion for $V_0$ about $x$ of the form,
\begin{equation}
V_0 (x, x') = v_{0}(x) + v_{0a} (x) \delta x^{a'} + v_{0ab} (x) \delta x^{a'} \delta x^{b'} + v_{0abc} (x) \delta x^{a'} \delta x^{b'} \delta x^{c'} + \cdots.
\end{equation}
The `initial condition' described by Eq.~\eqref{eq:recursionV0} and derived from Eq.~\eqref{eqn: VDerived}, in the scalar case, then becomes
\begin{equation}
2 \sigma^{;a'} V_{0;a'} - 2 V_0 \Delta^{-\tfrac{1}{2}} \sigma^{;a'} \Delta^{\tfrac{1}{2}}{}_{;a'} + 2 V_0 + \left(\square' - m^2 -\xi R \right) \Delta^{\tfrac{1}{2}} = 0,
\end{equation}
and from this, it is quite simple to read off expressions for the coefficients $v_{0a...}$.  Once we have $V_0$ to the desired order, we compute a coordinate expansion for $V_\num$ ($\num > 0$) of the form 
\begin{equation}
V_\num (x, x') = v_{\num }(x) + v_{\num a} (x) \delta x^{a'} + v_{\num ab} (x) \delta x^{a'} \delta x^{b'} + v_{\num abc} (x) \delta x^{a'} \delta x^{b'} \delta x^{c'} + \cdots.
\end{equation}
The recursion relation for $V_\num$, Eq.~\eqref{eq:recursionVn} in the scalar case is then
\begin{equation}
2 \num \sigma^{;a'} V_{\num ;a'} - 2 \num V_\num \Delta^{-\tfrac{1}{2}} \sigma^{;a'} \Delta^{\tfrac{1}{2}}{}_{;a'} + 2 \num \left( \num + 1\right) V_\num + \left(\square' - m^2 -\xi R \right) V_{\num-1} = 0,
\end{equation}
from which we can obtain expressions for the coefficients, $v_{\num a...}$. Here, the
number of terms which must be computed is determined by the accuracy to which
we require the singular field. For the present calculation, we require up to $v_{0\,abcde}$,
$v_{1\,abc}$ and $v_{2\,a}$.

Once we have $V_\num$ to the required $\num$, using Eqs.~\eqref{eqn:sigma} and \eqref{eqn: VScalar}, along with our expression for $\delta x^{a'}$ obtained from Eq.~\eqref{Anna eqn: xprimealpha}, we get $V(x,x')$ in terms of $\Delta \tau$, $\Delta x^a$ and $x^{\bar{a}}$.  This can be easily integrated over $\tau$ as required by Eq.~\eqref{eq: PhiS Simp}.  Our final expression for $\Phi^{\sing} (x)$ is then obtained by using Eqs.~\eqref{eqn: deltatau} and \eqref{eqn: taus} to remove the $\Delta \tau$ dependence.  As before $\tau_{\rm \ret}$ and $\tau_{\rm \adv}$ are obtained by allowing $\pm \rho = -\rho$ and $\pm \rho = +\rho$, respectively.


\section{Electromagnetic Singular Field}

For the electromagnetic singular field, we use Eqs.~\eqref{eq:AS}, \eqref{eqn: GgGen} and \eqref{eq: U} to give
\begin{equation} \label{eqn: A Simp}
A_{a}^{\rm \sing} = \frac{e}{2} \Bigg[\frac{\Delta^{\tfrac{1}{2}}(x,x') g_{aa'}(x,x')u^{a'}(x')}{\sigma_{c'}(x,x') u^{c'}(x')} \Bigg]_{x'=x_{\rm \ret}}^{x'=x_{\rm \adv}} + \frac{e}{2} \int_{\tau_{\rm \ret}}^{\tau_{\rm \adv}} V_{aa'}(x,z(\tau)) u^{a'}(x') d\tau,
\end{equation}
where $V_{aa'} (x, z(\tau'))$ is given by Eq.~\eqref{eq:V},
\begin{equation}
V^{aa'} (x, x') = \sum^{\infty}_{\num=0} V^{aa'}_\num (x, x') \sigma^\num (x, x'),
\end{equation}
and the relevant metrics at $x$ and $x'$ can be used to lower indices.  We require a coordinate expansion of $V^{aa'}_{0}$ of the form,
\begin{equation} \label{eqn: v0 EM}
V^{aa'}_0 (x, x') = v^{aa'}_{0}(x) + v^{aa'}_{0}{}_{b} (x) \delta x^{b'} + v^{aa'}_{0}{}_{bc} (x) \delta x^{b'} \delta x^{c'} + v^{aa'}_{0}{}_{bcd} (x) \delta x^{b'} \delta x^{c'} \delta x^{d'} + \cdots.
\end{equation}
Substituting this into the initial condition in Eq.~\eqref{eq:recursionV0} and derived from Eq.~\eqref{eqn: VDerived}, which in the electromagnetic case is
\begin{equation}
2 \sigma^{;b'} V^{aa'}_{0}{}_{;b'} - 2 V^{aa'}_0 \Delta^{-\tfrac{1}{2}} \sigma^{;b'} \Delta^{\tfrac{1}{2}}{}_{;b'} + 2 V^{aa'}_0 + \left(\delta^{a'}{}_{b'}\square' - R^{a'}{}_{b'} \right) \left(\Delta^{\tfrac{1}{2}} g^{ab'}\right) = 0,
\end{equation}
the coefficients of Eq.~\eqref{eqn: v0 EM}, $v^{aa'}_{0}{}_{b \cdots}$, can easily be recursively obtained. It should be noted that the covariant derivatives require the appropriate Christoffel symbols, Eq.~\eqref{eqn: christoffel}, which can be obtained from the suitable metric at $x'$.
Next, we construct coordinate expansions for the $V^{aa'}_{\num}$.  These have the form
\begin{equation}\label{eqn: vn EM}
V^{aa'}_\num (x, x') = v^{aa'}_{\num}(x) + v^{aa'}_{\num}{}_{b} (x) \delta x^{b'} + v^{aa'}_{\num}{}_{bc} (x) \delta x^{b'} \delta x^{c'} + v^{aa'}_{\num}{}_{bcd} (x) \delta x^{b'} \delta x^{c'} \delta x^{d'} + \cdots.
\end{equation}
Substituting Eq.~\eqref{eqn: vn EM} into the recursion relation \eqref{eq:recursionVn}, which for the electromagnetic case becomes
\begin{equation}
2 \num \sigma^{;b'} V^{aa'}_{\num}{}_{;b'} - 2 \num V^{aa'}_\num \Delta^{-\tfrac{1}{2}} \sigma^{;b'} \Delta^{\tfrac{1}{2}}{}_{;b'} + 2 \num \left( \num + 1\right) V^{aa'}_\num + \left(\delta^{a'}{}_{b'}\square' - R^{a'}{}_{b'} \right)V^{ab'}_{\num-1} = 0,
\end{equation}
we can recursively solve for the coefficients of Eq.~\eqref{eqn: vn EM}, $v^{aa'}_{\num}{}_{b \cdots}$.  Once we have $V^{aa'}_{\num}$ to the required $\num$, we carry out the same remaining steps as in the scalar case and use Eq.~\eqref{eqn: A Simp} to calculate the electromagnetic singular field.


\section{Gravitational Singular Field}

In the gravitational case, Eqs.~\eqref{eq:hS}, \eqref{eqn: GgGen} and \eqref{eq: U} give
\begin{align} \label{eqn: h Simp}
\hb_{a b}^{\rm \sing} =& 2 \mu \Bigg[\frac{\Delta^{\tfrac{1}{2}}(x,x') g_{a' (a} g_{b) b'}(x,x') u^{a'}
	(x') u^{b'}(x') }{\sigma_{c'}(x,x') u^{c'}(x')} \Bigg]_{x'=x_{\rm \ret}}^{x'=x_{\rm \adv}} 
	\nonumber \\
&
	+\: 2 \mu \int_{\tau_{\rm \ret}}^{\tau_{\rm \adv}} V_{aba'b'}(x,z(\tau')) u^{a'}(x') u^{b'}(x') 
	d\tau,
\end{align}
where $V_{aba'b'}(x,z(\tau'))$ is given by Eq.~\eqref{eq:V}.  For the gravitational case, this is
\begin{equation}
V^{aba'b'} (x, x') = \sum^{\infty}_{\num=0} V^{aba'b'}_\num (x, x') \sigma^\num (x, x'),
\end{equation}
where the appropriate metric at $x$ or $x'$ can be used to lower indices.  The coordinate expansion for $V^{aba'b'}_0 (x, x') $ is of the form
\begin{align} \label{eqn: v0 G}
V^{aba'b'}_0 (x, x') =& v^{aba'b'}_{0}(x) + v^{aba'b'}_{0}{}_{c} (x) \delta x^{c'} + v^{aba'b'}_{0}{}_{cd} 
	(x) \delta x^{c'} \delta x^{d'} \nonumber \\
&
	+\: v^{aba'b'}_{0}{}_{cde} (x) \delta x^{c'} \delta x^{d'} \delta x^{e'} + \cdots.
\end{align}
We replace $V^{aba'b'}_0 (x, x')$ with Eq~\eqref{eqn: v0 G} in the initial condition described by Eq.~\eqref{eq:recursionV0} and derived from Eq.~\eqref{eqn: VDerived}, which for the gravitational case is
\begin{align}
2 \sigma^{;c'} V^{aba'b'}_{0}{}_{;c'} -& 2 V^{aba'b'}_0 \Delta^{-\tfrac{1}{2}} \sigma^{;c'} \Delta ^{ 
	\tfrac{ 1}{2}}{}_{;c'} + 2 V^{aba'b'}_0 \nonumber \\
&
	+\: \left(\delta^{a'}{}_{c'}\delta^{b'}{}_{d'}\square' + 2 C^{a'}{}_{c'}{}^{b'}{}_{d'} \right) \left( 
	\Delta ^{\tfrac{1}{2}} g^{c'(a}g^{b) d'}\right) = 0.
\end{align}
This equation may be used to recursively solve for the coefficients of Eq.~\eqref{eqn: v0 G}, $v^{aba'b'}_{0}{}_{c \cdots}$.
Next, the coordinate expansion of $V^{aba'b'}_\num (x, x')$ for $\num > 0$ has the form,
\begin{align}\label{eqn: vn G}
V^{aba'b'}_\num (x, x') =& v^{aba'b'}_{\num}{}_{0}(x) + v^{aba'b'}_{\num}{}_{c} (x) \delta x^{c'} + 
	v^{ aba'b'}_{\num}{}_{cd} (x) \delta x^{c'} \delta x^{d'} \nonumber \\
&
	+\: v^{aba'b'}_{\num}{}_{cde} (x) \delta x^{c'} \delta x^{d'} \delta x^{e'} + \cdots.
\end{align}
Substituting this into the recursion relation of Eq.~\eqref{eq:recursionVn}, which for the gravitational case has the form
\begin{align}
2 \num \sigma^{;c'} V^{aba'b'}_{\num}{}_{;c'} -& 2 \num V^{aba'b'}_\num \Delta^{-\tfrac{1}{2}} 
	\sigma^{;c'} \Delta^{\tfrac{1}{2}}{}_{;c'} + 2 \num \left( \num + 1\right) V^{aba'b'}_\num 
	\nonumber \\
&
	+\: \left( \delta^{a'}{}_{c'}\delta^{b'}{}_{d'}\square' + 2 C^{a'}{}_{c'}{}^{b'}{}_{d'} \right)  V ^{ 
	abc'd'} _{\num-1} = 0,
\end{align}
we can recursively solve for the coefficiens of Eq.~\eqref{eqn: vn G}, $v^{aba'b'}_{\num}{}_{c\cdots}$.
As in the previous two cases, once we have $V^{aba'b'}_\num (x, x') $ for the required $\num$,
 it is straightforward to calculate the singular field using Eq.~\eqref{eqn: h Simp}.


\chapter{Covariant Bitensor Expansions*} \label{sec:ExpansionCoefficients}


\ifpdf
    \graphicspath{{9_backmatter/AppendixCovariant/figures/PNG/}{9_backmatter/AppendixCovariant/figures/PDF/}{9_backmatter/AppendixCovariant/figures/}}
\else
    \graphicspath{{9_backmatter/AppendixCovariant/figures/EPS/}{9_backmatter/AppendixCovariant/figures/}}
\fi


In this Appendix, we give covariant expansions for the \fixme{bitensors} appearing in the formal expression for the
singular field, Eq.~\eqref{eq:SingularField}. These are given in terms of the biscalars
$\sbar \equiv (g^{\alphab \betab} + u^{\alphab} u^{\betab}) \sigma_{\alphab} \sigma_{\betab}$,
(the projection of $\sigma_{\bar{a}}(x,\xb)$ orthogonal to the worldline), and
$\bar{r} = \sigma_{\alphab} u^{\alphab}$ (the projection of $\sigma_{\bar{a}}(x,\xb)$ along the worldline).
In writing the coefficients, we use the notation $[T_{a_1 \cdots a_n}]_{(k)}$ to denote the term of
order $\epsilon^k$ in the expansion of the tensor $T_{a_1 \cdots a_n}$, so that
\begin{equation}
T_{a_1 \cdots a_n} = \sum_{k=0}^\infty [T_{a_1 \cdots a_n}]_{(k)} \epsilon^k.
\end{equation}


\section{Advanced and retarded points}
Eq.~\eqref{eq:SingularField} for the singular field includes bitensors at points $x'$ on the
world-line between the advanced and retarded points of $x$. We consolidate this dependance
to a single arbitrary point, $\xb$, on the world by expanding the dependence on $x'$ about $\xb$.
Denoting the proper distance along the world-line between $x_{\rm \adv}/x_{\rm \ret}$ and $\xb$ by
$\Delta \tau$, we may write the expansion of this distance in powers of $\epsilon$ as
\begin{gather*}
\Dtau{1} = \rbar \pm \sbar, \quad \Dtau{2} = 0, \quad
\Dtau{3} = \mp\frac{(\rbar \pm \sbar)^2 R_{u \sigma u \sigma}}{6\sbar}, \nonumber \\
\Dtau{4} = \mp\frac{(\rbar \pm \sbar)^2 \big((\rbar \pm \sbar) R_{u \sigma u \sigma ;u}- R_{u \sigma u \sigma ;\sigma}\big)}{24\sbar},
\end{gather*}
\par \vspace{-6pt} \begin{IEEEeqnarray}{rCl}
\Dtau{5} &=& \mp\frac{(\rbar \pm \sbar)^2}{360 \sbar^3} \Big\{ 5 R_{u \sigma u \sigma} R_{u 
	\sigma u \sigma} (\rbar \mp 3 \sbar) (\rbar \pm \sbar) + \sbar^2 \Big[ (\rbar \pm 
	\sbar)^2 \Big(3 R_{u \sigma u \sigma; u u} \nonumber \\
&& \quad \qquad
	+\: 4 R_{u \sigma u \ab} R_{u \sigma u}{}^{\ab}\Big) - (\rbar \pm \sbar) \Big(3 R_{u \sigma 
	u \sigma; u \sigma} - 16 R_{u \sigma u \ab} R_{u \sigma \sigma}{}^{\ab}\Big) + 3 R_{u 
	\sigma u \sigma; \sigma \sigma} \nonumber \\
&& \qquad
	+\: 4 R_{u \sigma \sigma \ab} R_{u \sigma \sigma}{}^{\ab} \Big] \Big\}, \nonumber \\
\Dtau{6} &=& \pm\frac{(\rbar \pm \sbar)^2}{4320 \sbar^3} \Big( 30 R_{u \sigma u \sigma} 
	\Big[R_{u \sigma u \sigma ; \sigma} (\rbar \mp 3 \sbar) (\rbar \pm \sbar) - R_{u \sigma u 
	\sigma; u} (\rbar \mp 4 \sbar) (\rbar \pm \sbar)^2 \Big] \nonumber \\ 
&& \quad
	+\: \sbar^2 \Big\{6 R_{u \sigma u \sigma; \sigma \sigma \sigma} + 36 R_{u \sigma \sigma 
	\ab; \sigma} R_{u \sigma \sigma}{}^{\ab} - 2 (\rbar \pm \sbar) \Big[3 R_{u \sigma u 
	\sigma; u \sigma \sigma} \nonumber \\
&& \quad \qquad
	-\: 36 R_{u \sigma \sigma \ab; \sigma} R_{u \sigma u}{}^{\ab} 
	- R_{u \sigma \sigma}{}^{\ab} (16 R_{u \sigma u \ab; \sigma} + 5 R_{u \sigma u \sigma; 
	\ab} - 10 R_{u \sigma \sigma \ab; u})\Big] \nonumber \\
&& \qquad
        + 2 (\rbar \pm \sbar)^2 \Big[ 3 R_{u \sigma u 
	\sigma; u u \sigma} - 30 R_{u \sigma u \ab; u} R_{u \sigma \sigma}{}^{\ab} + R_{u \sigma u}{}^{\ab} (13 R_{u 
	\sigma u \ab; \sigma} + 5 R_{u \sigma u \sigma; \ab} \nonumber \\ 
&& \qquad \qquad
        -\: 25 R_{u \sigma \sigma \ab; u}) 
	\Big] - 6  (\rbar \pm \sbar)^3 (R_{u \sigma u \sigma; u u u} + 6 R_{u \sigma u \ab; u} R_{u 
	\sigma u}{}^{\ab}) \Big\} \Big).
\end{IEEEeqnarray}


\section{Advanced and retarded distances}
Taking two derivatives of the world function, we obtain a bitensor that has the covariant expansion
\par \vspace{-6pt} \begin{IEEEeqnarray}{rCl}
\sigma_{ab} &=& g_{ab}
  - \tfrac13 R_{acbd} \sigma^c \sigma^d
  + \tfrac{1}{12} R_{acbd;e} \sigma^c \sigma^d \sigma^e
  + \Big(
      \tfrac{1}{45} R_{acpd} R^p{}_{ebf}
    + \tfrac{1}{60} R_{acbd;ef}
    \Big) \sigma^c \sigma^d \sigma^e \sigma^f
\nonumber \\ &&
  + \Big(
      \tfrac{1}{120} R_{acpd;e} R^p{}_{fbg}
    + \tfrac{1}{120} R_{acpd} R^p{}_{ebf;g}
    + \tfrac{1}{360} R_{acbd;efg}
    \Big) \sigma^c \sigma^d \sigma^e \sigma^f \sigma^g
\nonumber \\ &&
  - \Big(
      \tfrac{2}{945} R_{acpd} R^p{}_{eqf} R^q{}_{gbh}
    + \tfrac{1}{504} R_{acpd;ef} R^p{}_{gbh}
    + \tfrac{17}{5040} R_{acpd;e} R^p{}_{fbg;h}
\nonumber \\ && \quad
    + \tfrac{1}{504} R_{acpd} R^p{}_{ebf;gh}
    + \tfrac{1}{2520} R_{acbd;efgh}
    \Big) \sigma^c \sigma^d \sigma^e \sigma^f \sigma^g \sigma^h
\nonumber \\ &&
  + \Big(
      \tfrac{17}{20160} R_{acpd;e} R^p{}_{fqg} R^q{}_{hbi}
    + \tfrac{29}{30240} R_{acpd} R^p{}_{eqf;g} R^q{}_{hbi}
    + \tfrac{11}{30240} R_{acpd;efg} R^p{}_{hbi}
\nonumber \\ && \quad
    + \tfrac{17}{20160} R_{acpd;ef} R^p{}_{gbh;i}
    + \tfrac{17}{20160} R_{acpd} R^p{}_{eqf} R^q{}_{gbh;i}
    + \tfrac{17}{20160} R_{acpd;e} R^p{}_{fbg;hi}
\nonumber \\ && \quad
    + \tfrac{11}{30240} R_{acpd} R^p{}_{ebf;ghi}
    + \tfrac{1}{20160} R_{acbd;efghi}
    \Big) \sigma^c \sigma^d \sigma^e \sigma^f \sigma^g \sigma^h \sigma^i.
\end{IEEEeqnarray}
For the singular field, we require the expansion of $[\sigma_{a'}u^{a'}](z_\pm,x)$. Writing
$[\sigma_{a'}u^{a'}](\tau) = [\sigma_{a'}u^{a'}](z(\tau), x)$, expanding the dependence on $\tau$
about $\xb$ (using the method of Sec.~\ref{sec:Covariant} and making use of the above expansion
of $\sigma_{ab}$) and evaluating at $\tau = \tau_\pm$,
we obtain the coefficients of the expansion of
$[\sigma_{a'}u^{a'}]_\pm \equiv [\sigma_{a'}u^{a'}](z_\pm, x)$ about $\xb$.
They are:
%
\par \vspace{-6pt} \begin{IEEEeqnarray}{rCl}
r_{(1)} &=& \sbar, \qquad \qquad \qquad
r_{(3)} =
 - \frac{\rbar^2 - \sbar^2}{6\sbar} R_{u \sigma u \sigma}, \nonumber \\
r_{(4)} &=&
 \frac{\rbar\pm\sbar}{24 \sbar} \Big[R_{u \sigma u \sigma \sigma} (\rbar \mp \sbar)
   - R_{u \sigma u \sigma u} (\rbar \pm \sbar) (\rbar \mp 2 \sbar)\bigg], \nonumber \\
r_{(5)} &=& -\frac{1}{360 \sbar^3}
\Big\{
\sbar^2 \Big[(\rbar^2 - \sbar^2) \big(
    3 R_{u \sigma u \sigma \sigma \sigma}
    + 4 R_{u \sigma \sigma}{}^{\ab} R_{u \sigma \sigma \ab}
  \big)
  \nonumber \\ && \qquad
    -\: (\rbar \pm \sbar)^2 (\rbar \mp 2 \sbar) \big(
      3  R_{u \sigma u \sigma \sigma u}
- 16 R_{u \sigma \sigma}{}^\ab R_{u \sigma u \ab}
  \big)
  \nonumber \\ && \qquad
+\: (\rbar \pm \sbar)^3 (\rbar \mp 3 \sbar) \big(
      3 R_{u \sigma u \sigma u u}
    + 4 R_{u \sigma u}{}^\ab R_{u \sigma u \ab}
    \big)
  \Big]
+  5 \Big[(\rbar^2 - \sbar^2) R_{u \sigma u \sigma}\Big]^2
\Big\},
\nonumber \\
r_{(6)} &=& \frac{1}{4320 \sbar^3}
\Big(
\sbar^2 \Big\{
  6 (\rbar^2 - \sbar^2) \big(
      R_{u \sigma u \sigma \sigma \sigma \sigma}
    + 6 R_{u \sigma \sigma}{}^{\ab}{}_{\sigma} R_{u \sigma \sigma \ab}
  \big)\nonumber \\ && \qquad
  -\: 6 (\rbar \pm \sbar)^4  (\rbar \mp 4 \sbar) \Big(
            R_{u \sigma u \sigma u u u}
+ 6 R_{u \sigma u \ab u} R_{u \sigma u}{}^{\ab}
      \Big)
  \nonumber \\ && \qquad
      -\:  2 (\rbar \pm \sbar)^2 (\rbar \mp 2 \sbar)
    \Big[
        3 R_{u \sigma u \sigma u \sigma \sigma}
      - 36 R_{u \sigma \sigma \ab \sigma} R_{u \sigma u}{}^{\ab}
- R_{u \sigma \sigma}{}^{\ab} (16 R_{u \sigma u \ab \sigma}
        \nonumber \\ && \qquad \qquad
        +\: 5 R_{u \sigma u \sigma \ab}
        - 10 R_{u \sigma \sigma \ab u})
    \Big]
  + 2 (\rbar \pm \sbar)^3  (\rbar \mp 3 \sbar) \Big[
          3 R_{u \sigma u \sigma u u \sigma}
\nonumber \\ && \qquad
  -\: 30 R_{u \sigma u \ab u} R_{u \sigma \sigma}{}^{\ab}
        + R_{u \sigma u}{}^{\ab} (13 R_{u \sigma u \ab \sigma} + 5 R_{u \sigma u \sigma \ab}
          - 25 R_{u \sigma \sigma \ab u})
      \Big]
  \Big\}
\nonumber \\ && \quad
  +  30 R_{u \sigma u \sigma} \Big[
   (\rbar^2 - \sbar^2)^2 R_{u \sigma u \sigma \sigma}
  -(\rbar\pm\sbar)^3(\rbar^2 \mp 3\rbar \sbar+4\sbar^2) R_{u \sigma u \sigma u}
  \Big]
\Big).
\end{IEEEeqnarray}


\section{Van Vleck Determinant}
The Van Vleck determinant has the covariant expansion
\par \vspace{-6pt} \begin{IEEEeqnarray}{rCl}
\Delta^{1/2}(x,x') &=&
  1
  + \frac{1}{12} R_{ab} \sigma^a \sigma^b
  - \frac{1}{24} R_{ab;c} \sigma^a \sigma^b \sigma^c
  + \Big(
      \frac{1}{360} R_{paqb} R^p{}_c{}^q{}_d
    + \frac{1}{288} R_{ab} R_{cd}
\nonumber \\ && \quad
    +\: \frac{1}{80}  R_{ab;cd}
    \Big)  \sigma^a \sigma^b \sigma^c \sigma^d
  - \Big(
      \frac{1}{360} R_{paqb} R^p{}_c{}^q{}_{d;e}
    + \frac{1}{288} R_{ab} R_{cd;e}
\nonumber \\ && \quad
    +\: \frac{1}{360}  R_{ab;cde}
    \Big)  \sigma^a \sigma^b \sigma^c \sigma^d \sigma^e
  + \Big(
      \frac{1}{1260} R_{paqb} R^p{}_c{}^q{}_{d;ef}
\nonumber \\ && \quad
    +\: \frac{1}{1344} R_{paqb;c} R^p{}_d{}^q{}_{e;f}
+ \frac{1}{5670} R_{paqb} R_{rc}{}^p{}_d R^q{}_e{}^r{}_f
    + \frac{1}{4320} R_{paqb} R^p{}_c{}^q{}_d R_{ef}
\nonumber \\ && \quad
        + \frac{1}{10368} R_{ab} R_{cd} R_{ef}
+ \frac{1}{1152} R_{ab;c} R_{de;f}
    + \frac{1}{960} R_{ab} R_{cd;ef}
\nonumber \\ && \quad
        + \frac{1}{2016}  R_{ab;cdef}
    \Big)  \sigma^a \sigma^b \sigma^c \sigma^d \sigma^e \sigma^f
 - \Big(
      \frac{1}{6048} R_{paqb} R^p{}_c{}^q{}_{d;efg}
\nonumber \\ && \quad
        +\: \frac{1}{2240} R_{paqb;c} R^p{}_d{}^q{}_{e;fg}
    + \frac{1}{3780} R_{paqb} R_{rc}{}^p{}_d R^q{}_e{}^r{}_{f;g}
\nonumber \\ && \quad
    + \frac{1}{4320} R_{paqb} R^p{}_c{}^q{}_{d;e} R_{fg}
    + \frac{1}{8640} R_{paqb} R^p{}_c{}^q{}_d R_{ef;g}
    + \frac{1}{6912} R_{ab} R_{cd} R_{ef;g}
\nonumber \\ && \quad 
    +\: \frac{1}{1920} R_{ab;c} R_{de;fg}
    + \frac{1}{4320} R_{ab} R_{cd;efg}
\nonumber \\ && \quad
    + \frac{1}{13440}  R_{ab;cdefg}
    \Big)  \sigma^a \sigma^b \sigma^c \sigma^d \sigma^e \sigma^f \sigma^g.
\end{IEEEeqnarray}
Writing $\Delta^{1/2}(\tau) = \Delta^{1/2}(z(\tau), x)$, expanding the dependence on $\tau$
about $\xb$ (using the method of Sec.~\ref{sec:Covariant}) and evaluating at $\tau = \tau_\pm$,
we obtain the coefficients in the expansion of
$\Delta^{1/2}_\pm \equiv \Delta^{1/2}(z_\pm, x)$ about $\xb$. Specialized to the vacuum case,
they are:
\begin{equation*}
\Delta^{1/2}_{(0)} = 1,\quad
\Delta^{1/2}_{(1)} = 0,\quad
\Delta^{1/2}_{(2)} = 0,\quad
\Delta^{1/2}_{(3)} = 0,
\end{equation*}
\par \vspace{-6pt} \begin{IEEEeqnarray}{rCl}
\Delta^{1/2}_{(4)} &=& \frac{1}{360}
\Big[
    C_{\sigma \ab \sigma \bb}
  + 2 (\rbar \pm \sbar) C_{u (\ab |\sigma| \bb)}
  + (\rbar \pm \sbar)^2 C_{u \ab u \bb}
\Big]\Big[
    C_\sigma{}^\ab{}_\sigma{}^\bb
  + 2 (\rbar \pm \sbar) C_u{}^\ab{}_\sigma{}^\bb
\nonumber \\ && \quad
  + (\rbar \pm \sbar)^2 C_u{}^\ab{}_u{}^\bb
\Big],
\nonumber \\
\Delta^{1/2}_{(5)} &=& \frac{1}{360}
\Big[
    C_{\sigma \ab \sigma \bb}
  + 2 (\rbar \pm \sbar) C_{u (\ab |\sigma| \bb)}
  + (\rbar \pm \sbar)^2 C_{u \ab u \bb}
\Big]
\Big[
  (\rbar \pm \sbar)
  \Big(
      C_\sigma{}^{\ab}{}_\sigma{}^{\bb}{}_u
    - 2 C_u{}^{\ab}{}_\sigma{}^{\bb}{}_\sigma
  \Big)
\nonumber \\ &&
\qquad
- (\rbar \pm \sbar)^2
  \Big(
      C_u{}^\ab{}_u{}^\bb{}_\sigma
    - 2 C_u{}^\ab{}_\sigma{}^\bb{}_u
  \Big)
+ (\rbar \pm \sbar)^3 C_u{}^\ab{}_u{}^\bb{}_u
- C_\sigma{}^\ab{}_\sigma{}^\bb{}_\sigma
\Big].
\end{IEEEeqnarray}


\section{Bivector of parallel transport}
The derivative of the bivector of parallel transport has the covariant expansion
\par \vspace{-6pt} \begin{IEEEeqnarray}{rCl}
g_a{}^{a'} g_{a' b ; c}(x,x') &=&
  - \frac12 R_{bacd} \sigma^d
  + \frac16 R_{bacd;e} \sigma^d \sigma^e
  - \frac{1}{24} \Big(
      R_{bapd} R^p{}_{ecf}
    + R_{bacd;ef}
    \Big) \sigma^d \sigma^e \sigma^f
\nonumber \\ &&
  + \Big(
      \frac{1}{60} R_{bapd} R^p{}_{ecf;g}
    + \frac{7}{360} R_{bapd;e} R^p{}_{fcg}
    + \frac{1}{120} R_{bacd;efg}
    \Big) \sigma^d \sigma^e \sigma^f \sigma^g
\nonumber \\ &&
  - \Big(
      \frac{1}{240} R_{bapd} R^p{}_{ecf;gh}
      \frac{1}{120} R_{bapd;e} R^p{}_{fcg;h}
    + \frac{1}{180} R_{bapd;ef} R^p{}_{gch}
\nonumber \\ &&
    \qquad + \frac{1}{240} R_{bapd} R^p{}_{eqf} R^q{}_{gch}
    + \frac{1}{720} R_{bacd;efgh}
    \Big) \sigma^d \sigma^e \sigma^f \sigma^g \sigma^h.
\end{IEEEeqnarray}
For the singular field, we require the expansion of $g_{aa'}u^{a'}(z_\pm,x)$. Writing
$[g_{aa'}u^{a'}](\tau) = [g_{aa'}u^{a'}](z(\tau), x)$, expanding the dependence on $\tau$
about $\xb$ (using the method of Sec.~\ref{sec:Covariant} and making use of the above expansion
of the bivector of parallel transport) and evaluating at $\tau = \tau_\pm$,
we obtain the coefficients of the expansion of
$[g_{aa'}u^{a'}]_\pm \equiv g_{aa'}u^{a'}(z_\pm, x)$ about $\xb$. They are:
\par \vspace{-6pt} \begin{IEEEeqnarray}{rCl}
\Big[g_{aa'} u^{a'}\Big]_{(0)} &=& g_{a\ab}u^\ab, \qquad
\Big[g_{aa'} u^{a'}\Big]_{(1)} = 0, \qquad
\Big[g_{aa'} u^{a'}\Big]_{(2)} = -\frac12 (\rbar \pm \sbar) g_{a}{}^\ab R_{u \sigma u \ab}, \nonumber \\
\Big[g_{aa'} u^{a'}\Big]_{(3)} &=& \frac16 (\rbar \pm \sbar) g_{a}{}^\ab \left[R_{u \sigma u \ab;\sigma}-(\rbar \pm \sbar) R_{u \sigma u \ab;u}\right], \nonumber \\
\Big[g_{aa'} u^{a'}\Big]_{(4)} &=& \pm \frac{1}{24 \sbar} g_{a}{}^\ab (\rbar \pm \sbar)
  \Big[
    2 (\rbar \pm \sbar) R_{u \sigma u \ab} R_{u \sigma u \sigma}
    - \sbar \Big(
        R_{u \sigma u \ab ; \sigma \sigma}
      + R_{u \ab \sigma \bb} R_{u \sigma \sigma}{}^\bb
\nonumber \\ && \qquad
    +\: (\rbar \pm \sbar)^2 R_{u \sigma u a ; u u}
    + (\rbar \pm \sbar) \Big\{
        R_{u \sigma u}{}^\bb ( 2 R_{u \ab \sigma \bb} + 3 R_{u \sigma \ab \bb})
      - R_{u \sigma u \ab ; u \sigma}
\nonumber \\ && \quad \qquad
      + R_{u \ab u}{}^{\bb} \left[
        R_{u \sigma \sigma \bb}
      + (\rbar \pm \sbar) R_{u \sigma u \bb}
      \right]
    \Big\}
    \Big)
  \Big], \nonumber \\
\Big[g_{aa'} u^{a'}\Big]_{(5)} &=& \pm \frac{1}{2160 \sbar} g_{a}{}^\ab (\rbar \pm \sbar)
  \Big(
    \sbar \Big\{
      18 \Big[
          R_{u \sigma u a ; \sigma \sigma  \sigma }
        -  R_{u \sigma u a ; u \sigma \sigma } (\rbar \pm \sbar)
        \nonumber \\ && \qquad \quad
       +\: R_{u \sigma u a ; u u \sigma } (\rbar \pm \sbar)^2
 - R_{u \sigma u a ; u u u} (\rbar \pm \sbar)^3 \Big]
    + 6 R_{u \sigma  \sigma b} \Big[
        7 R_{u a \sigma }{}^{b}{}_{ ;\sigma }
\nonumber \\ && \qquad \quad
        +\: (\rbar \pm \sbar) ( 3 R_{u a u}{}^{b}{}_{ ;\sigma }
- 4 R_{u a \sigma }{}^{b}{}_{;u}
        + 4 R_{u \sigma u a}{}^{;b})
      - 8 R_{u a u}{}^{b}{}_{;u}(\rbar \pm \sbar)^2\Big]
\nonumber \\ &&  \qquad
          +\: 9 R_{u a \sigma b} \Big[
        4 R_{u \sigma  \sigma }{}^{b}{}_{ ;\sigma }
+  (\rbar \pm \sbar) (5 R_{u \sigma u}{}^{b}{}_{ ;\sigma }
        - 3 R_{u \sigma  \sigma }{}^{b}{}_{;u})
\nonumber \\ && \quad \qquad
    -\: 6 R_{u \sigma u}{}^{b}{}_{;u} (\rbar \pm \sbar)^2\Big]
+ 2 R_{u \sigma u b} \Big[
        21  (\rbar \pm \sbar) (2 R_{u a \sigma }{}^{b}{}_{ ;\sigma }
        + 3 R_{u \sigma a}{}^{b}{}_{ ;\sigma } )
\nonumber \\ && \quad \qquad
             +\:  (\rbar \pm \sbar)^2 ( 17 R_{u a u}{}^{b}{}_{; \sigma } 
- 32 R_{u a \sigma }{}^{b}{}_{;u}
        - 36 R_{u \sigma a}{}^{b}{}_{;u}
        + 16 R_{u \sigma u a}{}^{;b})
\nonumber \\ && \quad \qquad
          -\: 21 R_{u a u}{}^{b}{}_{;u} (\rbar \pm \sbar)^3 \Big]
+ 9 R_{u a u b} \Big[
        4 R_{u \sigma  \sigma }{}^{b}{}_{ ;\sigma } (\rbar \pm \sbar)
\nonumber \\ && \quad \qquad
            +\: 3  (\rbar \pm \sbar)^2 (R_{u \sigma u}{}^{b}{}_{ ;\sigma }
        - R_{u \sigma  \sigma }{}^{b}{}_{;u})
- 4 R_{u \sigma u}{}^{b}{}_{;u} (\rbar \pm \sbar)^3\Big]
\nonumber \\ && \qquad
        +  54 (\rbar \pm \sbar) R_{u \sigma a b} \Big[
         R_{u \sigma u}{}^{b}{}_{; \sigma }
     - 2 R_{u \sigma u}{}^{b}{}_{;u} (\rbar \pm \sbar)\Big]
    \Big\}
\nonumber \\ && \qquad
    +\: 15 (\rbar \pm \sbar) \Big\{
       4 R_{u \sigma u \sigma } \Big[
          2  (\rbar \pm \sbar) R_{u \sigma u a; u}
        -   R_{u \sigma u a ;\sigma }\Big]
\nonumber \\ && \quad \qquad
     + 3 R_{u \sigma u a} \Big[
            R_{u \sigma u \sigma ; u} (\rbar \pm \sbar)
        -   R_{u \sigma u \sigma ; \sigma }\Big]
     \Big\}
  \Big).
\end{IEEEeqnarray}


\section{Scalar tail}
The scalar tail bitensor, $V(x,x')$, may be expanded in a covariant series by writing it in the
form of a Hadamard series,
\begin{equation}
V(x,x') = V_0 (x,x') + V_1 (x,x') \sigma(x,x') + \cdots,
\end{equation}
and expanding each of the Hadamard coefficients $V_0(x,x'), V_1(x,x'), \cdots$ in a covariant Taylor
series,
\begin{align*}
V_{0} &= v_{0} - \tfrac{1}{2} v_{0\,;c} \sigma^c
+\tfrac{1}{2}   v_{0\,cd}  \sigma^c\sigma^d +\tfrac{1}{6} \left(  - \tfrac{3}{2} v_{0\,(cd;e)} + \tfrac{1}{4} v_{0\,;(cde)} \right) \sigma^c\sigma^d\sigma^e + \cdots,\\
V_{1} &= v_{1} - \tfrac{1}{2} v_{1\,;c}  \sigma^c + \cdots.
\end{align*}
The series coefficients required to obtain the expansion of the singular field to
$\mathcal{O}(\epsilon^4)$ [$V(x,x')$ to $\mathcal{O}(\epsilon^3)$] are given by
\par \vspace{-6pt} \begin{IEEEeqnarray}{rCl}
v_{0} &=& \tfrac{1}{2} \left((\xi- \tfrac{1}{6}) R + m^2 \right), \nonumber \\
v_{0\,}{}^{cd} &=& -\tfrac{1}{180}R_{pqr}{}^{c}R^{pqrd}-\tfrac{1}{180}R^{c}{}_{p}{}^{d} {}_{q} R^{pq} 
	+ \tfrac{1}{90}R^{c}{}_{p}R^{dp} -\tfrac{1}{120}\Box R^{cd}+\tfrac{1}{12}(\xi-\tfrac{1}{6}) R 
	R^{cd} \nonumber \\
&&
	+\: \tfrac{1}{6}\xi-\tfrac{1}{40})R^{;cd}+\tfrac{1}{12}m^2 R^{cd}, \nonumber \\
v_1 &=& \tfrac{1}{720}R_{pqrs}R^{pqrs} -\tfrac{1}{720}R_{pq}R^{pq} +\tfrac{1}{8}(\xi - \tfrac{1}
	{6})^2 R^2-\tfrac{1}{24}(\xi - \tfrac{1}{5})\Box R+\tfrac{1}{4}m^2(\xi - \tfrac{1}{6}) R \nonumber \\
&&	
	+\: 
	\tfrac{1}{8}m^4. \nonumber
\end{IEEEeqnarray}
For the singular field, we require the expansion of 
$\int_{\tau_{\rm \ret}}^{\tau_{\rm \adv}} V d\tau'$. Writing $V(\tau) = V(z(\tau), x)$ and expanding the dependence on $\tau$
about $\xb$ (using the method of Sec.~\ref{sec:Covariant} and making use of the above expansion
of $V$), we obtain an expansion in powers of $\Delta \tau$ that can be trivially integrated
between $\tau = \tau_-$ and $\tau = \tau_+$. Specialized to the vacuum case,
the required expansion coefficients are then:
\begin{gather*}
\Big[ \int_{\tau_{\rm \ret}}^{\tau_{\rm \adv}} V d\tau' \Big]_{(1)} = 0, \quad
\Big[ \int_{\tau_{\rm \ret}}^{\tau_{\rm \adv}} V d\tau' \Big]_{(2)} = 0, \\
\Big[ \int_{\tau_{\rm \ret}}^{\tau_{\rm \adv}} V d\tau' \Big]_{(3)} = 0, \quad
\Big[ \int_{\tau_{\rm \ret}}^{\tau_{\rm \adv}} V d\tau' \Big]_{(4)} = 0.
\end{gather*}


\section{Electromagnetic tail}
The electromagnetic tail bitensor, $V_{ab'}(x,x')$, may be expanded in a covariant series by
writing it in the form of a Hadamard series,
\begin{equation}
V_{ab'}(x,x') = g_{b'}{}^{b} [ V_{0 ab} (x,x') + V_{1 ab} (x,x') \sigma(x,x') + \cdots],
\end{equation}
and expanding each of the Hadamard coefficients, $V_{0 ab}(x,x'), V_{1 ab}(x,x'), \cdots$, in a
covariant Taylor series,
\begin{align*}
V_{0\,ab} &= v_{0\,(ab)} + \left( - \tfrac{1}{2} v_{0\,(ab);c} + v_{0\,[ab]c}\right) \sigma^c
+\tfrac{1}{2} \left(   v_{0\,(ab)cd} - v_{0\,[ab](c;d)}\right) \sigma^c\sigma^d\\
&\qquad +\tfrac{1}{6} \left(  - \tfrac{3}{2} v_{0\,(ab)(cd;e)} + \tfrac{1}{4} v_{0\,(ab);(cde)} + v_{0\,[ab]cde}\right) \sigma^c\sigma^d\sigma^e + \cdots,\\
V_{1\,ab} &= v_{1\,(ab)} + \left( - \tfrac{1}{2} v_{1\,(ab);c} + v_{1\,[ab]c}\right) \sigma^c + \cdots.
\end{align*}
The series coefficients required to obtain the expansion of the singular field to
$\mathcal{O}(\epsilon^4)$ [$V_{ab'}(x,x')$ to $\mathcal{O}(\epsilon^3)$] are given by
\begin{align*}
v_{0\,(ab)} &= \tfrac{1}{2} R_{ab}-\tfrac{1}{12}R g_{ab},\\
v_{0\,[ab]}{}^{c} &= \tfrac{1}{6} R^c{}_{[b;a]},\\
v_{0\,(ab)}{}^{cd} &= \tfrac{1}{6} R_{ab}{}^{;(cd)}+\tfrac{1}{12}R_{ab}R^{cd}+\tfrac{1}{12}R_{(a}
	{}^{pqc}R_{b)pq}{}^{d} + g_{ab}\Big(-\tfrac{1}{180}R_{pqr}{}^{c}R^{pqrd}\nonumber \\
& \quad
	-\:\tfrac{1}{180}R^{c}
	{}_{p}{}^{d}{}_{q}R^{pq} + \tfrac{1}{90}R^{c}{}_{p}R^{dp} -\tfrac{1}{72}RR^{cd}-\tfrac{1}{40}R^{;cd}-\tfrac{1}{120} 
	\Box R^{cd}\Big),\\
v_{0\,[ab]}{}^{cde} &= -\tfrac{3}{20}R^{(c}{}_{[a;b]}{}^{de)} -\tfrac{1}{12}R^{(c}{}_{[a;b]}R^{de)} 
	-\tfrac{1}{20}R_{[a}{}^{pq(c}R_{b]pq}{}^{d;e)} -\tfrac{1}{30}R_{ab}{}^{p(c}{}_{;q}R_{p}{}^{de)q} 
	\nonumber \\
&
	+\: \tfrac{1}{60}R_{abp}{}^{(c;d}R^{e)p}+\tfrac{1}{20}R_{abp}{}^{(c}R^{de);p} - \tfrac{1}{20} 
	R_{ab}{}^{p(c}R_{p}{}^{d;e)},
\end{align*}
and
\begin{align*}
v_{1\,(ab)} =& -\tfrac{1}{48}R_{apqr}R_{b}{}^{pqr} +\tfrac{1}{8}R_{ap}R_{b}{}^{p} -\tfrac{1}{24} 
	RR_{ab} -\tfrac{1}{24}\Box R_{ab} \\
& 
	+\: g_{ab} \left(\tfrac{1}{720}R_{pqrs}R^{pqrs} -\tfrac{1}{720}R_{pq}R^{pq} +\tfrac{1}{288} 
	R^2 + \tfrac{1}{120}\Box R\right),\\
v_{1\,[ab]}{}^{c} =&  \tfrac{1}{240}R_{[a}{}^{pqr}R_{b]pqr}{}^{;c} + \tfrac{1}{24}R_{[a}{}^{pqc} 
	R_{b]p;q} + \tfrac{1}{120}R_{[a}{}^{pqc}R_{b]q;p} - \tfrac{1}{120}R^{c}{}_{pq[a}R^{pq}{}_{;b]}\\
& 
	+\: \tfrac{1}{24}R^{pc}{}_{;[a}R_{b]p}+ \tfrac{1}{24}R^{p}{}_{[a}R_{b]}{}^{c}{}_{;p} - \tfrac{1}
	{360} R^{pc}R_{p[a;b]}- \tfrac{1}{24}R^{p}{}_{[a}R_{b]p}{}^{;c}+ \tfrac{1}{72}R R^c{}_{[a;b]}\\
& 
	+\: \tfrac{1}{120}R^c{}_{[a;b]p}{}^p - \tfrac{1}{540}R_{abpq;r}R^{pqrc}- \tfrac{1}{540} 
	R_{abpq;r} R^{rqpc}- \tfrac{1}{360}R_{abp}{}^{c}{}_{;q}R^{pq} \\
& 
	+\: \tfrac{1}{120}R_{abpq}R^{cp;q}- \tfrac{1}{120}R_{ab}{}^{pc}R_{;p}.
\end{align*}
These are the same as those given by Brown and Ottewill \cite{Brown:Ottewill:1986} with the exception
of $v_{1\,[ab]}{}^{c}$, where we have corrected a sign error in one of their terms and
combined another two terms into a single term.

For the singular field, we require the expansion of 
$\int_{\tau_{\rm \ret}}^{\tau_{\rm \adv}} V_{ab'} u^{b'} d\tau'$. Writing
$[V_{ab'} u^{b'}](\tau) = [V_{ab'} u^{b'}](z(\tau), x)$ and expanding the dependence on $\tau$
about $\xb$ (using the method of Sec.~\ref{sec:Covariant} and making use of the above expansion
of $V_{ab'}$), we obtain an expansion in powers of $\Delta \tau$ that can be trivially integrated
between $\tau = \tau_-$ and $\tau = \tau_+$. Specialized to the vacuum case,
the required expansion coefficients are then:
\par \vspace{-6pt} \begin{IEEEeqnarray}{rCl}
\Big[ \int_{\tau_{\rm \ret}}^{\tau_{\rm \adv}} V_{ab'} u^{b'} d\tau' \Big]_{(1)} &=& 0, \qquad \qquad
\Big[ \int_{\tau_{\rm \ret}}^{\tau_{\rm \adv}} V_{ab'} u^{b'} d\tau' \Big]_{(2)} = 0, \nonumber \\
\Big[ \int_{\tau_{\rm \ret}}^{\tau_{\rm \adv}} V_{ab'} u^{b'} d\tau' \Big]_{(3)} &=&
  \frac{1}{1152} (\bar{r} \pm \bar{s}) g_{a}{}^{\ab}
  \Big\{
      16 R_{ubac} R_{u}{}^{b}{}_{u}{}^{c} (\rbar \pm \sbar)^2
    + 48 R_{u}{}^{b}{}_{\sigma}{}^{c} R_{\sigma c a b}
\nonumber \\ &&
    +\: 24  (\rbar \pm \sbar) \Big[
        R_{u}{}^{b}{}_{\sigma}{}^{c} R_{ucab}
      + R_{u}{}^{b}{}_{u}{}^{c} R_{\sigma b a c}
\nonumber \\ &&
          +\: (\rbar \mp 2 \sbar) R_{bcde} R^{bcde}u_{a}
    \Big]
  \Big\},
\nonumber \\
\Big[ \int_{\tau_{\rm \ret}}^{\tau_{\rm \adv}} V_{ab'} u^{b'} d\tau' \Big]_{(4)} &=&
  \frac{1}{51840} (\bar{r} \pm \bar{s}) g_{a}{}^{\ab}
  \Big\{
      144 \Big[
          2  R_{ua\sigma}{}^{b;c} R_{\sigma b\sigma c}
        - 6  R_{\sigma}{}^{b}{}_{a}{}^{c}{}_{; \sigma} R_{uc\sigma b}
\nonumber \\ &&
        - \: 9  R_{u}{}^{b}{}_{\sigma}{}^{c}{}_{; \sigma} R_{\sigma cab}
        +  R_{\sigma}{}^{bcd}{}_{; \sigma} R_{\sigma bcd} u_{a}
        \Big]
\nonumber \\ &&
            -\: 4 (\rbar \pm \sbar) \Big[
      18 \Big(
          6  R_{\sigma}{}^{b}{}_{a}{}^{c}{}_{; \sigma} R_{ubuc}
- 2  R_{ua\sigma}{}^{b;c} R_{ub\sigma c}
\nonumber \\ &&
                +\: 9  R_{u}{}^{b}{}_{\sigma}{}^{c}{}_{; \sigma} R_{ucab}
        +  R_{uc\sigma b} (
            6  R_{u}{}^{b}{}_{a}{}^{c}{}_{; \sigma}
         - 2  R_{ua\sigma}{}^{b;c}
- 9  R_{\sigma}{}^{b}{}_{a}{}^{c}{}_{;u})
 \nonumber \\ &&
               +\: 9  R_{u}{}^{b}{}_{u}{}^{c}{}_{; \sigma} R_{\sigma bac}
        - 2  R_{uau}{}^{b;c} R_{\sigma b\sigma c}
        - 6  R_{u}{}^{b}{}_{\sigma}{}^{c}{}_{;u} R_{\sigma cab}
        \Big)
\nonumber \\ &&
        +\:  (\rbar \mp 2 \sbar) \Big(27  R_{u}{}^{bcd}{}_{; \sigma} R_{abcd}
        + 18  R_{a}{}^{bcd}{}_{; \sigma} R_{ubcd}
        + 4  R_{ua}{}^{bc;d} (R_{\sigma bcd}
\nonumber \\ &&
        -  R_{\sigma dbc})\Big)
     - 3 \Big(6  R_{\sigma}{}^{bcd}{}_{; \sigma} R_{ubcd}
        + 6 ( R_{u}{}^{bcd}{}_{; \sigma}
        -   R_{\sigma}{}^{bcd}{}_{;u}) R_{\sigma bcd}
\nonumber \\ &&
        +\:  R^{bcde}{}_{; \sigma} R_{bcde} (\rbar \mp 2 \sbar)
        \Big) u_{a}
      \Big]
    + (\rbar \pm \sbar)^2 \Big[
        48 \Big(
            R_{ubuc} (
                2  R_{ua\sigma}{}^{b;c}
\nonumber \\ &&
              -\: 6  R_{u}{}^{b}{}_{a}{}^{c}{}_{; \sigma}
              + 9  R_{\sigma}{}^{b}{}_{a}{}^{c}{}_{;u}
              )
        + 6 R_{u}{}^{b}{}_{\sigma}{}^{c}{}_{;u} R_{ucab}
        + 9  R_{u}{}^{b}{}_{a}{}^{c}{}_{;u} R_{uc\sigma b}
\nonumber \\ &&
       +\: 2  R_{uau}{}^{b;c} (
            R_{ub\sigma c}
          + R_{uc\sigma b}
          )
        + 6  R_{u}{}^{b}{}_{u}{}^{c}{}_{;u} R_{\sigma bac}
        \Big)
     +  (\rbar 
\nonumber \\ &&
	        \mp \: 3 \sbar) \Big(
            R_{ubcd} (
             27 R_{a}{}^{bcd}{}_{;u}
           - 4  R_{ua}{}^{bc;d}
           )
        + 4 R_{ua}{}^{bc;d} R_{udbc}
        \Big)
\nonumber \\ &&
      +\: 3 \Big(
          16 (
              R_{u}{}^{bcd}{}_{;\sigma}
            - R_{\sigma}{}^{bcd}{}_{;u}) R_{ubcd}
            - R^{bcde}{}_{;u} R_{bcde} (\rbar \mp 3 \sbar)
        \Big) u_{a}
\nonumber \\ &&
      +\: 6 R_{u}{}^{bcd}{}_{;u} \Big(
          3 R_{abcd} (\rbar \mp 3 \sbar)
        - 8 R_{\sigma bcd} u_{a}
        \Big)
      - 432  R_{u}{}^{b}{}_{u}{}^{c}{}_{; \sigma} R_{ubac}
      \Big]
\nonumber \\ &&
    +\: 36 (\rbar \pm \sbar)^3 \Big[
        6  R_{u}{}^{b}{}_{u}{}^{c}{}_{;u} R_{ubac}
      + R_{ubuc} (2  R_{uau}{}^{b;c} + 9  R_{u}{}^{b}{}_{a}{}^{c}{}_{;u})
\nonumber \\ &&
      - R_{u}{}^{bcd}{}_{;u} R_{ubcd} u_{a}\Big]
  \Big\}.
\end{IEEEeqnarray}


\section{Gravitational tail}
The gravitational tail bitensor, $V_{aa'bb'}(x,x')$, may be expanded in a covariant series by
writing it in the form of a Hadamard series,
\begin{equation}
V_{aa'bb'}(x,x') = g_{a'}{}^{c} g_{b'}{}^{d} [ V_{0 acbd} (x,x') + V_{1 acbd} (x,x') \sigma(x,x') + \cdots],
\end{equation}
and expanding each of the Hadamard coefficients, $V_{0 acbd}(x,x'), V_{1 acbd}(x,x'), \cdots$, in a
covariant Taylor series,
\begin{align*}
V_{0\,AB} &= v_{0\,(AB)} + \left( - \tfrac{1}{2} v_{0\,(AB);e} + v_{0\,[AB]e}\right) \sigma^e
+\tfrac{1}{2} \left(   v_{0\,(AB)ef} - v_{0\,[AB](e;f)}\right) \sigma^e\sigma^f\\
&\qquad +\tfrac{1}{6} \left(  - \tfrac{3}{2} v_{0\,(AB)(ef;g)} + \tfrac{1}{4} v_{0\,(AB);(efg)} + v_{0\,[AB]efg}\right) \sigma^e\sigma^f\sigma^g + \cdots,\\
V_{1\,AB} &= v_{1\,(AB)} + \left( - \tfrac{1}{2} v_{1\,(AB);c} + v_{1\,[AB]e}\right) \sigma^e+ \cdots.
\end{align*}
The required coefficients for $V_{0\,AB}$ are
\par \vspace{-6pt} \begin{IEEEeqnarray}{rCl}
   v_{0\,(AB)}&=&v_{0\,(\overline{ab}\,\overline{cd})}=-C_{acbd},
\label{eq:v01}\\
   v_{0\,[AB]}{}^{e}&=&0,
\label{eq:v02}\\
   v_{0\,(AB)}{}^{ef}&=&v_{0\,(\overline{ab}\,\overline{cd})}{}^{ef} 
\nonumber \\&=&
      -\frac{1}{3}C_{acbd}{}^{;(ef)}
      -\frac{1}{6}C_{ac}{}^{p(e}C_{bdp}{}^{f)}
      +\frac{1}{6}g_{ac}C_b{}^{pq(e}
         C_{dpq}{}^{f)}
      \nonumber \\ &&
        -\frac{1}{720} {\Pi}_{abcd} g^{ef} C^{pq{r}s}
         C_{pq{r}s},
\label{eq:v03}
\end{IEEEeqnarray}
and
\begin{eqnarray}
   v_{0\,[AB]}{}^{ef{g}}&=&v_{0\,[\overline{ab}\,\overline{cd}]}{}^{ef{g}}=
      \frac{1}{10}g_{ac}C_{[b}{^{pq(e;f}}
         C_{d]pq}{}^{g)}
      +\frac{1}{15}g_{ac}C_{bdpe;q}
         C^p{_f}{^q}{_{g}},
\label{eq:v04}
\end{eqnarray}
where
\begin{equation}
{\Pi}_{abcd} = {1 \over 2} g_{ac}
g_{bd} + {1 \over 2} g_{ad} g_{bc} +
{\kappa}g_{ab} g_{cd}.
\end{equation}
In Eqs.\ (\ref{eq:v01}) -- (\ref{eq:v04}), the right hand sides are
understood to be symmetrized on the index pairs $(ab)$ and $(cd)$.
The required coefficients for $V_{1\,AB}$ are
\par \vspace{-6pt} \begin{IEEEeqnarray}{rCl}
   v_{1\, (AB)} = v_{1\,(\overline{ab}\,\overline{cd})}&=&
      \frac{1}{12}{\Box}C_{acbd}
      +\frac{1}{2}C_a{^p}{_b}{^q}C_{cpdq}
      +\frac{1}{24}C_{ac}{^{pq}}C_{bdpq}
      -\frac{1}{96}g_{ac}g_{bd}C^{pqrs}C_{pqrs}
\nonumber \\ &&
      +\: \frac{1}{720}{\Pi}_{abcd} C^{pqrs}
         C_{pqrs},
\label{eq:v11}
\end{IEEEeqnarray}
and
\begin{align}
v_{1\,[AB]}{}^{e}=&v_{1\,[\overline{ab}\,\overline{cd}]}{}^{e} \nonumber \\
=&
\frac{1}{12} \left( C_{a}{}^{p}{}_b{}^q C_{cpdq}{}^{;e}-C_{c}{}^{p}{}_d{}^q C_{apbq}{}^{;e} \right)
+\frac{1}{12} \left( C_{a}{}^{p}{}^e{}^q C_{bcdp;q} - C_{c}{}^{p}{}^e{}^q C_{dabp;q} \right)
\nonumber \\ &
+\: \frac{1}{90} g_{ac} C^e{}^{pqr}C_{bdpq;r},
\label{eq:v12}
\end{align}
where again there is implicit symmetrization on the index pairs
($ab$) and ($cd$).
When ${\kappa}=-1/2$, Eq. (\ref{eq:v11}) agrees
with Eq.\ (A23) of Allen, Folacci and Ottewill \cite{Allen:1987bn} 
specialized to the vacuum case. Our expressions also agree with
Anderson, Flanagan and Ottewill \cite{Anderson:Flanagan:Ottewill:2004}, but we write
them here in a slightly more compact form. Note that the
expressions (\protect{\ref{eq:v01}}) -- (\protect{\ref{eq:v04}})
and (\ref{eq:v11}) -- (\ref{eq:v12})
are all traceless on the index pair $(cd)$, aside from the
terms involving the  
tensor ${\Pi}_{abcd}$.  This means that performing
a trace reversal on the index pair $(cd)$ is equivalent to
changing the value of ${\kappa}$ from $0$ to $-1/2$. For the calculation of the gravitational singular field, we require an
expansion of the trace reversed singular field and so we choose $\kappa = 0$.



\renewcommand{\bibname}{References} 

\bibliographystyle{apsrev2}
\lhead{}
\bibliography{references} 

\begin{thebibliography}{124}
\expandafter\ifx\csname natexlab\endcsname\relax\def\natexlab#1{#1}\fi
\expandafter\ifx\csname bibnamefont\endcsname\relax
  \def\bibnamefont#1{#1}\fi
\expandafter\ifx\csname bibfnamefont\endcsname\relax
  \def\bibfnamefont#1{#1}\fi
\expandafter\ifx\csname citenamefont\endcsname\relax
  \def\citenamefont#1{#1}\fi
\expandafter\ifx\csname url\endcsname\relax
  \def\url#1{\texttt{#1}}\fi
\expandafter\ifx\csname urlprefix\endcsname\relax\def\urlprefix{URL }\fi
\providecommand{\bibinfo}[2]{#2}
\providecommand{\eprint}[2][]{\url{#2}}

\bibitem[{\citenamefont{Glashow}(1961)}]{Glenshow:1961}
\bibinfo{author}{\bibfnamefont{S.~L.} \bibnamefont{Glashow}},
  \bibinfo{journal}{Nucl. Phys.} \textbf{\bibinfo{volume}{22}},
  \bibinfo{pages}{579} (\bibinfo{year}{1961}).

\bibitem[{\citenamefont{Weinberg}(1967)}]{Weinberg:1967}
\bibinfo{author}{\bibfnamefont{S.}~\bibnamefont{Weinberg}},
  \bibinfo{journal}{Phys.Rev.Lett} \textbf{\bibinfo{volume}{19}},
  \bibinfo{pages}{1264} (\bibinfo{year}{1967}),
  \eprint{http://link.aps.org/doi/10.1103/PhysRevLett.19.1264}.

\bibitem[{\citenamefont{Salam}(1968)}]{Salam:1964}
\bibinfo{author}{\bibfnamefont{A.}~\bibnamefont{Salam}},
  \bibinfo{journal}{``Elementary Particle Physics'', N. Svartholm, ed. (Nobel
  Symposium No. 8, Almquist and Wiksell, Stockholm)} p. \bibinfo{pages}{367}
  (\bibinfo{year}{1968}).

\bibitem[{\citenamefont{Higgs}(1964)}]{Higgs:1964}
\bibinfo{author}{\bibfnamefont{P.~W.} \bibnamefont{Higgs}},
  \bibinfo{journal}{Physics Letters} \textbf{\bibinfo{volume}{12}},
  \bibinfo{pages}{132} (\bibinfo{year}{1964}).

\bibitem[{\citenamefont{Einstein}(1916)}]{Einstein:1916}
\bibinfo{author}{\bibfnamefont{A.}~\bibnamefont{Einstein}},
  \bibinfo{journal}{Annalen der Physik} \textbf{\bibinfo{volume}{7}},
  \bibinfo{pages}{769} (\bibinfo{year}{1916}).

\bibitem[{\citenamefont{Dyson et~al.}(1920)\citenamefont{Dyson, Eddington, and
  Davidson}}]{Eddington:1919}
\bibinfo{author}{\bibfnamefont{F.~W.} \bibnamefont{Dyson}},
  \bibinfo{author}{\bibfnamefont{A.~S.} \bibnamefont{Eddington}},
  \bibnamefont{and} \bibinfo{author}{\bibfnamefont{C.}~\bibnamefont{Davidson}},
  \bibinfo{journal}{Philos. Trans. Royal Soc. London}
  \textbf{\bibinfo{volume}{220A}}, \bibinfo{pages}{291} (\bibinfo{year}{1920}).

\bibitem[{\citenamefont{Pound and Rebka~Jr.}(1959)}]{Pound:1959}
\bibinfo{author}{\bibfnamefont{R.~V.} \bibnamefont{Pound}} \bibnamefont{and}
  \bibinfo{author}{\bibfnamefont{G.~A.} \bibnamefont{Rebka~Jr.}},
  \bibinfo{journal}{Phys. Rev. Lett.} \textbf{\bibinfo{volume}{3}},
  \bibinfo{pages}{439} (\bibinfo{year}{1959}).

\bibitem[{\citenamefont{Walsh et~al.}(1979)\citenamefont{Walsh, Carswell, and
  Weymann}}]{Walsh:1979}
\bibinfo{author}{\bibfnamefont{D.}~\bibnamefont{Walsh}},
  \bibinfo{author}{\bibfnamefont{R.~F.} \bibnamefont{Carswell}},
  \bibnamefont{and} \bibinfo{author}{\bibfnamefont{R.~J.}
  \bibnamefont{Weymann}}, \bibinfo{journal}{Nature}
  \textbf{\bibinfo{volume}{279}}, \bibinfo{pages}{381} (\bibinfo{year}{1979}).

\bibitem[{\citenamefont{Shapiro}(1964)}]{Shapiro:1964}
\bibinfo{author}{\bibfnamefont{I.~I.} \bibnamefont{Shapiro}},
  \bibinfo{journal}{Phys.Rev. Lett.} \textbf{\bibinfo{volume}{13}},
  \bibinfo{pages}{789} (\bibinfo{year}{1964}).

\bibitem[{\citenamefont{Weisberg et~al.}(1981)\citenamefont{Weisberg, Taylor,
  and Fowler}}]{Taylor:1981}
\bibinfo{author}{\bibfnamefont{J.~M.} \bibnamefont{Weisberg}},
  \bibinfo{author}{\bibfnamefont{J.~H.} \bibnamefont{Taylor}},
  \bibnamefont{and} \bibinfo{author}{\bibfnamefont{L.~A.}
  \bibnamefont{Fowler}}, \bibinfo{journal}{Scientific American}
  \textbf{\bibinfo{volume}{245}}, \bibinfo{pages}{74} (\bibinfo{year}{1981}).

\bibitem[{\citenamefont{Damour}(1983)}]{Damour:1983}
\bibinfo{author}{\bibfnamefont{T.}~\bibnamefont{Damour}},
  \bibinfo{journal}{Phys. Rev. Lett.} \textbf{\bibinfo{volume}{51}},
  \bibinfo{pages}{1019} (\bibinfo{year}{1983}),
  \urlprefix\url{http://link.aps.org/doi/10.1103/PhysRevLett.51.1019}.

\bibitem[{\citenamefont{Hewish et~al.}(1968)\citenamefont{Hewish, Bell,
  Pilkington, Scott, and Collins}}]{Hewish:1968}
\bibinfo{author}{\bibfnamefont{A.}~\bibnamefont{Hewish}},
  \bibinfo{author}{\bibfnamefont{S.}~\bibnamefont{Bell}},
  \bibinfo{author}{\bibfnamefont{J.~D.~H.} \bibnamefont{Pilkington}},
  \bibinfo{author}{\bibfnamefont{P.}~\bibnamefont{Scott}}, \bibnamefont{and}
  \bibinfo{author}{\bibfnamefont{R.~A.} \bibnamefont{Collins}},
  \bibinfo{journal}{Nature} \textbf{\bibinfo{volume}{217}},
  \bibinfo{pages}{709} (\bibinfo{year}{1968}).

\bibitem[{\citenamefont{Klebesadel et~al.}(1973)\citenamefont{Klebesadel,
  Strong, and Olson}}]{Klebesadel:1973}
\bibinfo{author}{\bibfnamefont{R.}~\bibnamefont{Klebesadel}},
  \bibinfo{author}{\bibfnamefont{I.}~\bibnamefont{Strong}}, \bibnamefont{and}
  \bibinfo{author}{\bibfnamefont{R.}~\bibnamefont{Olson}},
  \bibinfo{journal}{Astophys. J.} \textbf{\bibinfo{volume}{182}},
  \bibinfo{pages}{L85} (\bibinfo{year}{1973}).

\bibitem[{LIG()}]{LIGO}
\bibinfo{howpublished}{\url{http://www.ligo.caltech.edu}}.

\bibitem[{VIR()}]{VIRGO}
\bibinfo{howpublished}{\url{http://www.virgo.infn.it}}.

\bibitem[{GEO()}]{GEO}
\bibinfo{howpublished}{\url{http://geo600.aei.mpg.de}}.

\bibitem[{TAM()}]{TAMA}
\bibinfo{howpublished}{\url{http://tamago.mtk.nao.ac.jp/}}.

\bibitem[{Adv()}]{AdvancedLigo}
\bibinfo{howpublished}{\url{https://www.advancedligo.mit.edu/science.html}}.

\bibitem[{\citenamefont{Pitkin et~al.}(2011)\citenamefont{Pitkin, Reid, Rowan,
  and Hough}}]{Pitkin:2011}
\bibinfo{author}{\bibfnamefont{M.}~\bibnamefont{Pitkin}},
  \bibinfo{author}{\bibfnamefont{S.}~\bibnamefont{Reid}},
  \bibinfo{author}{\bibfnamefont{S.}~\bibnamefont{Rowan}}, \bibnamefont{and}
  \bibinfo{author}{\bibfnamefont{J.}~\bibnamefont{Hough}},
  \bibinfo{journal}{Living Reviews in Relativity}  (\bibinfo{year}{2011}).

\bibitem[{NGO()}]{NGO}
\bibinfo{howpublished}{\url{http://www.elisa-ngo.org/}}.

\bibitem[{\citenamefont{Andersson}(2011)}]{Anderson:2011}
\bibinfo{author}{\bibfnamefont{N.}~\bibnamefont{Andersson}},
  \bibinfo{journal}{Progress in Particle and Nuclear Physics}
  \textbf{\bibinfo{volume}{66}}, \bibinfo{pages}{239 } (\bibinfo{year}{2011}),
  ISSN \bibinfo{issn}{0146-6410}.

\bibitem[{LIS()}]{LISAPath}
\bibinfo{howpublished}{\url{http://www.rssd.esa.int/index.php?project=LISAPATHFINDER}}.

\bibitem[{\citenamefont{Eisenhauer et~al.}(2005)\citenamefont{Eisenhauer,
  Genzel, Alexander, Abuter, Paumard et~al.}}]{Eisenhauer:2005}
\bibinfo{author}{\bibfnamefont{F.}~\bibnamefont{Eisenhauer}},
  \bibinfo{author}{\bibfnamefont{R.}~\bibnamefont{Genzel}},
  \bibinfo{author}{\bibfnamefont{T.}~\bibnamefont{Alexander}},
  \bibinfo{author}{\bibfnamefont{R.}~\bibnamefont{Abuter}},
  \bibinfo{author}{\bibfnamefont{T.}~\bibnamefont{Paumard}},
  \bibnamefont{et~al.}, \bibinfo{journal}{Astrophys.J.}
  \textbf{\bibinfo{volume}{628}}, \bibinfo{pages}{246} (\bibinfo{year}{2005}),
  \eprint{astro-ph/0502129}.

\bibitem[{\citenamefont{Gillessen et~al.}(2009)\citenamefont{Gillessen,
  Eisenhauer, Trippe, Alexander, Genzel et~al.}}]{Gillessen:2009}
\bibinfo{author}{\bibfnamefont{S.}~\bibnamefont{Gillessen}},
  \bibinfo{author}{\bibfnamefont{F.}~\bibnamefont{Eisenhauer}},
  \bibinfo{author}{\bibfnamefont{S.}~\bibnamefont{Trippe}},
  \bibinfo{author}{\bibfnamefont{T.}~\bibnamefont{Alexander}},
  \bibinfo{author}{\bibfnamefont{R.}~\bibnamefont{Genzel}},
  \bibnamefont{et~al.}, \bibinfo{journal}{Astrophys.J.}
  \textbf{\bibinfo{volume}{692}}, \bibinfo{pages}{1075} (\bibinfo{year}{2009}),
  \eprint{0810.4674}.

\bibitem[{\citenamefont{Barack}(2009)}]{Barack:2009}
\bibinfo{author}{\bibfnamefont{L.}~\bibnamefont{Barack}},
  \bibinfo{journal}{Classical Quantum Gravity} \textbf{\bibinfo{volume}{26}},
  \bibinfo{pages}{213001} (\bibinfo{year}{2009}), \eprint{arXiv:0908.1664}.

\bibitem[{\citenamefont{Farrell et~al.}(2009)\citenamefont{Farrell, Webb,
  Barret, Godet, and Rodrigues}}]{Farrell:2009}
\bibinfo{author}{\bibfnamefont{S.~A.} \bibnamefont{Farrell}},
  \bibinfo{author}{\bibfnamefont{N.~A.} \bibnamefont{Webb}},
  \bibinfo{author}{\bibfnamefont{D.}~\bibnamefont{Barret}},
  \bibinfo{author}{\bibfnamefont{O.}~\bibnamefont{Godet}}, \bibnamefont{and}
  \bibinfo{author}{\bibfnamefont{J.~M.} \bibnamefont{Rodrigues}},
  \bibinfo{journal}{Nature} \textbf{\bibinfo{volume}{460}}, \bibinfo{pages}{73}
  (\bibinfo{year}{2009}).

\bibitem[{\citenamefont{Farrell et~al.}(2012)\citenamefont{Farrell, Servillat,
  Pforr, Maccarone, Knigge, Godet, Maraston, Webb, Barret, Godet
  et~al.}}]{Farrell:2012}
\bibinfo{author}{\bibfnamefont{S.~A.} \bibnamefont{Farrell}},
  \bibinfo{author}{\bibfnamefont{M.}~\bibnamefont{Servillat}},
  \bibinfo{author}{\bibfnamefont{J.}~\bibnamefont{Pforr}},
  \bibinfo{author}{\bibfnamefont{T.}~\bibnamefont{Maccarone}},
  \bibinfo{author}{\bibfnamefont{C.}~\bibnamefont{Knigge}},
  \bibinfo{author}{\bibfnamefont{O.}~\bibnamefont{Godet}},
  \bibinfo{author}{\bibfnamefont{C.}~\bibnamefont{Maraston}},
  \bibinfo{author}{\bibfnamefont{N.~A.} \bibnamefont{Webb}},
  \bibinfo{author}{\bibfnamefont{D.}~\bibnamefont{Barret}},
  \bibinfo{author}{\bibfnamefont{O.}~\bibnamefont{Godet}},
  \bibnamefont{et~al.}, \bibinfo{journal}{Astrophysical Journal Letters}
  \textbf{\bibinfo{volume}{747}}, \bibinfo{pages}{L13} (\bibinfo{year}{2012}).

\bibitem[{\citenamefont{Amaro-Seoane et~al.}(2007)\citenamefont{Amaro-Seoane,
  Gair, Freitag, Miller, Mandel, Cutler, and Babak}}]{AmaroSeoane:2007}
\bibinfo{author}{\bibfnamefont{P.}~\bibnamefont{Amaro-Seoane}},
  \bibinfo{author}{\bibfnamefont{J.~R.} \bibnamefont{Gair}},
  \bibinfo{author}{\bibfnamefont{M.}~\bibnamefont{Freitag}},
  \bibinfo{author}{\bibfnamefont{M.~C.} \bibnamefont{Miller}},
  \bibinfo{author}{\bibfnamefont{I.}~\bibnamefont{Mandel}},
  \bibinfo{author}{\bibfnamefont{C.~J.} \bibnamefont{Cutler}},
  \bibnamefont{and} \bibinfo{author}{\bibfnamefont{S.}~\bibnamefont{Babak}},
  \bibinfo{journal}{Classical and Quantum Gravity}
  \textbf{\bibinfo{volume}{24}}, \bibinfo{pages}{R113} (\bibinfo{year}{2007}),
  \urlprefix\url{http://stacks.iop.org/0264-9381/24/i=17/a=R01}.

\bibitem[{\citenamefont{Konstantinidis
  et~al.}(2012)\citenamefont{Konstantinidis, Amaro-Seoane, and
  Kokkotas}}]{Knostantinidis:2012}
\bibinfo{author}{\bibfnamefont{S.}~\bibnamefont{Konstantinidis}},
  \bibinfo{author}{\bibfnamefont{P.}~\bibnamefont{Amaro-Seoane}},
  \bibnamefont{and} \bibinfo{author}{\bibfnamefont{K.~D.}
  \bibnamefont{Kokkotas}} (\bibinfo{year}{2012}), \eprint{arXiv:1108.5175}.

\bibitem[{\citenamefont{Smith et~al.}(2013)\citenamefont{Smith, Mandel, and
  Vechhio}}]{Smith:2013}
\bibinfo{author}{\bibfnamefont{R.}~\bibnamefont{Smith}},
  \bibinfo{author}{\bibfnamefont{I.}~\bibnamefont{Mandel}}, \bibnamefont{and}
  \bibinfo{author}{\bibfnamefont{A.}~\bibnamefont{Vechhio}}
  (\bibinfo{year}{2013}), \eprint{1302.6049}.

\bibitem[{\citenamefont{Blanchet et~al.}(2010)\citenamefont{Blanchet,
  Detweiler, Le~Tiec, and Whiting}}]{Blanchet:2010}
\bibinfo{author}{\bibfnamefont{L.}~\bibnamefont{Blanchet}},
  \bibinfo{author}{\bibfnamefont{S.}~\bibnamefont{Detweiler}},
  \bibinfo{author}{\bibfnamefont{A.}~\bibnamefont{Le~Tiec}}, \bibnamefont{and}
  \bibinfo{author}{\bibfnamefont{B.}~\bibnamefont{Whiting}},
  \bibinfo{journal}{Phys. Rev. D} \textbf{\bibinfo{volume}{81}},
  \bibinfo{pages}{084033} (\bibinfo{year}{2010}).

\bibitem[{\citenamefont{Baker et~al.}(2007)\citenamefont{Baker, van Meter,
  McWilliams, Centrella, and Kelly}}]{Baker:2007}
\bibinfo{author}{\bibfnamefont{J.~G.} \bibnamefont{Baker}},
  \bibinfo{author}{\bibfnamefont{J.~R.} \bibnamefont{van Meter}},
  \bibinfo{author}{\bibfnamefont{S.~T.} \bibnamefont{McWilliams}},
  \bibinfo{author}{\bibfnamefont{J.}~\bibnamefont{Centrella}},
  \bibnamefont{and} \bibinfo{author}{\bibfnamefont{B.~J.} \bibnamefont{Kelly}},
  \bibinfo{journal}{Phys. Rev. Lett.} \textbf{\bibinfo{volume}{99}},
  \bibinfo{pages}{181101} (\bibinfo{year}{2007}).

\bibitem[{\citenamefont{Boyle et~al.}(2007)\citenamefont{Boyle, Brown, Kidder,
  Mroue, Pfeiffer et~al.}}]{Boyle:2007ft}
\bibinfo{author}{\bibfnamefont{M.}~\bibnamefont{Boyle}},
  \bibinfo{author}{\bibfnamefont{D.~A.} \bibnamefont{Brown}},
  \bibinfo{author}{\bibfnamefont{L.~E.} \bibnamefont{Kidder}},
  \bibinfo{author}{\bibfnamefont{A.~H.} \bibnamefont{Mroue}},
  \bibinfo{author}{\bibfnamefont{H.~P.} \bibnamefont{Pfeiffer}},
  \bibnamefont{et~al.}, \bibinfo{journal}{Phys.Rev.}
  \textbf{\bibinfo{volume}{D76}}, \bibinfo{pages}{124038}
  (\bibinfo{year}{2007}), \eprint{0710.0158}.

\bibitem[{\citenamefont{Hannam et~al.}(2008)\citenamefont{Hannam, Husa,
  Gonzalez, Sperhake, and Brugmann}}]{Hannam:2008}
\bibinfo{author}{\bibfnamefont{M.}~\bibnamefont{Hannam}},
  \bibinfo{author}{\bibfnamefont{S.}~\bibnamefont{Husa}},
  \bibinfo{author}{\bibfnamefont{J.~A.} \bibnamefont{Gonzalez}},
  \bibinfo{author}{\bibfnamefont{U.}~\bibnamefont{Sperhake}}, \bibnamefont{and}
  \bibinfo{author}{\bibfnamefont{B.}~\bibnamefont{Brugmann}},
  \bibinfo{journal}{Phys. Rev. D} \textbf{\bibinfo{volume}{77}},
  \bibinfo{pages}{044020} (\bibinfo{year}{2008}).

\bibitem[{\citenamefont{Le~Tiec et~al.}(2012)\citenamefont{Le~Tiec, Barausse,
  and Buonanno}}]{LeTiec:2012}
\bibinfo{author}{\bibfnamefont{A.}~\bibnamefont{Le~Tiec}},
  \bibinfo{author}{\bibfnamefont{E.}~\bibnamefont{Barausse}}, \bibnamefont{and}
  \bibinfo{author}{\bibfnamefont{A.}~\bibnamefont{Buonanno}},
  \bibinfo{journal}{Phys.Rev.Lett.} \textbf{\bibinfo{volume}{108}},
  \bibinfo{pages}{131103} (\bibinfo{year}{2012}), \eprint{1111.5609}.

\bibitem[{\citenamefont{Le~Tiec et~al.}(2011)\citenamefont{Le~Tiec, Mroue,
  Barack, Buonanno, Pfeiffer et~al.}}]{LeTiec:2011}
\bibinfo{author}{\bibfnamefont{A.}~\bibnamefont{Le~Tiec}},
  \bibinfo{author}{\bibfnamefont{A.~H.} \bibnamefont{Mroue}},
  \bibinfo{author}{\bibfnamefont{L.}~\bibnamefont{Barack}},
  \bibinfo{author}{\bibfnamefont{A.}~\bibnamefont{Buonanno}},
  \bibinfo{author}{\bibfnamefont{H.~P.} \bibnamefont{Pfeiffer}},
  \bibnamefont{et~al.}, \bibinfo{journal}{Phys.Rev.Lett.}
  \textbf{\bibinfo{volume}{107}}, \bibinfo{pages}{141101}
  (\bibinfo{year}{2011}), \eprint{1106.3278}.

\bibitem[{\citenamefont{Dirac}(1938)}]{Dirac-1938}
\bibinfo{author}{\bibfnamefont{P.~A.~M.} \bibnamefont{Dirac}},
  \bibinfo{journal}{Proc. R. Soc. Lond. A} \textbf{\bibinfo{volume}{167}},
  \bibinfo{pages}{148} (\bibinfo{year}{1938}).

\bibitem[{\citenamefont{DeWitt and Brehme}(1960)}]{DeWitt:1960}
\bibinfo{author}{\bibfnamefont{B.~S.} \bibnamefont{DeWitt}} \bibnamefont{and}
  \bibinfo{author}{\bibfnamefont{R.~W.} \bibnamefont{Brehme}},
  \bibinfo{journal}{Ann. Phys.} \textbf{\bibinfo{volume}{9}},
  \bibinfo{pages}{220} (\bibinfo{year}{1960}).

\bibitem[{\citenamefont{Hobbs}(1968)}]{Hobbs:1968a}
\bibinfo{author}{\bibfnamefont{J.~M.} \bibnamefont{Hobbs}},
  \bibinfo{journal}{Ann. Phys.} \textbf{\bibinfo{volume}{47}},
  \bibinfo{pages}{141} (\bibinfo{year}{1968}).

\bibitem[{\citenamefont{Mino et~al.}(1997)\citenamefont{Mino, Sasaki, and
  Tanaka}}]{Mino:Sasaki:Tanaka:1996}
\bibinfo{author}{\bibfnamefont{Y.}~\bibnamefont{Mino}},
  \bibinfo{author}{\bibfnamefont{M.}~\bibnamefont{Sasaki}}, \bibnamefont{and}
  \bibinfo{author}{\bibfnamefont{T.}~\bibnamefont{Tanaka}},
  \bibinfo{journal}{Phys. Rev.} \textbf{\bibinfo{volume}{D55}},
  \bibinfo{pages}{3457} (\bibinfo{year}{1997}), \eprint{gr-qc/9606018}.

\bibitem[{\citenamefont{Quinn and Wald}(1997)}]{Quinn:Wald:1997}
\bibinfo{author}{\bibfnamefont{T.~C.} \bibnamefont{Quinn}} \bibnamefont{and}
  \bibinfo{author}{\bibfnamefont{R.~M.} \bibnamefont{Wald}},
  \bibinfo{journal}{Phys. Rev.} \textbf{\bibinfo{volume}{D56}},
  \bibinfo{pages}{3381} (\bibinfo{year}{1997}), \eprint{gr-qc/9610053}.

\bibitem[{\citenamefont{Barack and Ori}(2000)}]{Barack:Ori:2000}
\bibinfo{author}{\bibfnamefont{L.}~\bibnamefont{Barack}} \bibnamefont{and}
  \bibinfo{author}{\bibfnamefont{A.}~\bibnamefont{Ori}},
  \bibinfo{journal}{Phys.Rev.} \textbf{\bibinfo{volume}{D61}},
  \bibinfo{pages}{061502} (\bibinfo{year}{2000}), \eprint{gr-qc/9912010}.

\bibitem[{\citenamefont{Quinn}(2000)}]{Quinn:2000}
\bibinfo{author}{\bibfnamefont{T.~C.} \bibnamefont{Quinn}},
  \bibinfo{journal}{Phys. Rev.} \textbf{\bibinfo{volume}{D62}},
  \bibinfo{pages}{064029} (\bibinfo{year}{2000}), \eprint{gr-qc/0005030}.

\bibitem[{\citenamefont{Poisson et~al.}(2011)\citenamefont{Poisson, Pound, and
  Vega}}]{Poisson:2003}
\bibinfo{author}{\bibfnamefont{E.}~\bibnamefont{Poisson}},
  \bibinfo{author}{\bibfnamefont{A.}~\bibnamefont{Pound}}, \bibnamefont{and}
  \bibinfo{author}{\bibfnamefont{I.}~\bibnamefont{Vega}},
  \bibinfo{journal}{Living Rev. Relativity} \textbf{\bibinfo{volume}{14}},
  \bibinfo{pages}{7} (\bibinfo{year}{2011}), \eprint{arXiv:1102.0529v3}.

\bibitem[{\citenamefont{Detweiler}(2005)}]{Detweiler:2005}
\bibinfo{author}{\bibfnamefont{S.}~\bibnamefont{Detweiler}},
  \bibinfo{journal}{Class. Quantum Grav.} \textbf{\bibinfo{volume}{22}},
  \bibinfo{pages}{S681} (\bibinfo{year}{2005}), \eprint{gr-qc/0501004}.

\bibitem[{Poi(1998)}]{Poisson:Wiseman:1998}
\emph{\bibinfo{title}{{Suggestion at the 1st Capra ranch meeting on radiation
  reaction}}} (\bibinfo{year}{1998}).

\bibitem[{\citenamefont{Anderson and Hu}(2004)}]{Anderson:2003}
\bibinfo{author}{\bibfnamefont{P.~R.} \bibnamefont{Anderson}} \bibnamefont{and}
  \bibinfo{author}{\bibfnamefont{B.~L.} \bibnamefont{Hu}},
  \bibinfo{journal}{Phys. Rev.} \textbf{\bibinfo{volume}{D69}},
  \bibinfo{pages}{064039} (\bibinfo{year}{2004}), \eprint{gr-qc/0308034}.

\bibitem[{\citenamefont{Anderson and Wiseman}(2005)}]{Anderson:Wiseman:2005}
\bibinfo{author}{\bibfnamefont{W.~G.} \bibnamefont{Anderson}} \bibnamefont{and}
  \bibinfo{author}{\bibfnamefont{A.~G.} \bibnamefont{Wiseman}},
  \bibinfo{journal}{Class. Quantum Grav.} \textbf{\bibinfo{volume}{22}},
  \bibinfo{pages}{S783} (\bibinfo{year}{2005}), \eprint{gr-qc/0506136}.

\bibitem[{\citenamefont{Ottewill and Wardell}(2008)}]{Ottewill:Wardell:2008}
\bibinfo{author}{\bibfnamefont{A.~C.} \bibnamefont{Ottewill}} \bibnamefont{and}
  \bibinfo{author}{\bibfnamefont{B.}~\bibnamefont{Wardell}},
  \bibinfo{journal}{Phys. Rev.} \textbf{\bibinfo{volume}{D77}},
  \bibinfo{pages}{104002} (\bibinfo{year}{2008}), \eprint{arXiv:0711.2469}.

\bibitem[{\citenamefont{Ottewill and Wardell}(2009)}]{Ottewill:Wardell:2009}
\bibinfo{author}{\bibfnamefont{A.~C.} \bibnamefont{Ottewill}} \bibnamefont{and}
  \bibinfo{author}{\bibfnamefont{B.}~\bibnamefont{Wardell}},
  \bibinfo{journal}{Phys. Rev.} \textbf{\bibinfo{volume}{D79}},
  \bibinfo{pages}{024031} (\bibinfo{year}{2009}), \eprint{arXiv:0810.1961}.

\bibitem[{\citenamefont{Casals et~al.}(2009)\citenamefont{Casals, Dolan,
  Ottewill, and Wardell}}]{Casals:Dolan:Ottewill:Wardell:2009}
\bibinfo{author}{\bibfnamefont{M.}~\bibnamefont{Casals}},
  \bibinfo{author}{\bibfnamefont{S.~R.} \bibnamefont{Dolan}},
  \bibinfo{author}{\bibfnamefont{A.~C.} \bibnamefont{Ottewill}},
  \bibnamefont{and} \bibinfo{author}{\bibfnamefont{B.}~\bibnamefont{Wardell}},
  \bibinfo{journal}{Phys. Rev. D} \textbf{\bibinfo{volume}{79}},
  \bibinfo{pages}{124043} (\bibinfo{year}{2009}), \eprint{arXiv:0903.0395}.

\bibitem[{\citenamefont{Casals et~al.}(2013)\citenamefont{Casals, Dolan,
  Ottewill, and Wardell}}]{Casals:2013}
\bibinfo{author}{\bibfnamefont{M.}~\bibnamefont{Casals}},
  \bibinfo{author}{\bibfnamefont{S.}~\bibnamefont{Dolan}},
  \bibinfo{author}{\bibfnamefont{A.~C.} \bibnamefont{Ottewill}},
  \bibnamefont{and} \bibinfo{author}{\bibfnamefont{B.}~\bibnamefont{Wardell}}
  (\bibinfo{year}{2013}), \eprint{1306.0884}.

\bibitem[{\citenamefont{Barack et~al.}(2007)\citenamefont{Barack, Golbourn, and
  Sago}}]{Barack:Golbourn:Sago:2007}
\bibinfo{author}{\bibfnamefont{L.}~\bibnamefont{Barack}},
  \bibinfo{author}{\bibfnamefont{D.~A.} \bibnamefont{Golbourn}},
  \bibnamefont{and} \bibinfo{author}{\bibfnamefont{N.}~\bibnamefont{Sago}},
  \bibinfo{journal}{Phys.Rev.} \textbf{\bibinfo{volume}{D76}},
  \bibinfo{pages}{124036} (\bibinfo{year}{2007}), \eprint{0709.4588}.

\bibitem[{\citenamefont{Barack and Golbourn}(2007)}]{Barack:Golbourn:2007}
\bibinfo{author}{\bibfnamefont{L.}~\bibnamefont{Barack}} \bibnamefont{and}
  \bibinfo{author}{\bibfnamefont{D.~A.} \bibnamefont{Golbourn}},
  \bibinfo{journal}{Phys.Rev.} \textbf{\bibinfo{volume}{D76}},
  \bibinfo{pages}{044020} (\bibinfo{year}{2007}), \eprint{0705.3620}.

\bibitem[{\citenamefont{Vega and Detweiler}(2008)}]{Vega:Detweiler:2008}
\bibinfo{author}{\bibfnamefont{I.}~\bibnamefont{Vega}} \bibnamefont{and}
  \bibinfo{author}{\bibfnamefont{S.}~\bibnamefont{Detweiler}},
  \bibinfo{journal}{Phys. Rev. D} \textbf{\bibinfo{volume}{77}},
  \bibinfo{pages}{084008} (\bibinfo{year}{2008}).

\bibitem[{\citenamefont{Rosenthal}(2005)}]{Rosenthal:2005ju}
\bibinfo{author}{\bibfnamefont{E.}~\bibnamefont{Rosenthal}},
  \bibinfo{journal}{Class.Quant.Grav.} \textbf{\bibinfo{volume}{22}},
  \bibinfo{pages}{S859} (\bibinfo{year}{2005}), \eprint{gr-qc/0501046}.

\bibitem[{\citenamefont{Rosenthal}(2006)}]{Rosenthal:2006iy}
\bibinfo{author}{\bibfnamefont{E.}~\bibnamefont{Rosenthal}},
  \bibinfo{journal}{Phys.Rev.} \textbf{\bibinfo{volume}{D74}},
  \bibinfo{pages}{084018} (\bibinfo{year}{2006}), \eprint{gr-qc/0609069}.

\bibitem[{\citenamefont{Detweiler}(2012)}]{Detweiler:2011tt}
\bibinfo{author}{\bibfnamefont{S.}~\bibnamefont{Detweiler}},
  \bibinfo{journal}{Phys.Rev.} \textbf{\bibinfo{volume}{D85}},
  \bibinfo{pages}{044048} (\bibinfo{year}{2012}), \eprint{arXiv:1107.2098}.

\bibitem[{\citenamefont{Pound}(2012)}]{Pound:2012nt}
\bibinfo{author}{\bibfnamefont{A.}~\bibnamefont{Pound}} (\bibinfo{year}{2012}),
  \eprint{arXiv:1201.5089}.

\bibitem[{\citenamefont{Gralla}(2012)}]{Gralla:2012db}
\bibinfo{author}{\bibfnamefont{S.~E.} \bibnamefont{Gralla}}
  (\bibinfo{year}{2012}), \eprint{arXiv:1203.3189}.

\bibitem[{\citenamefont{Barack et~al.}(2002)\citenamefont{Barack, Mino, Nakano,
  Ori, and Sasaki}}]{Barack:Mino:Nakano:Ori:Sasaki:2001}
\bibinfo{author}{\bibfnamefont{L.}~\bibnamefont{Barack}},
  \bibinfo{author}{\bibfnamefont{Y.}~\bibnamefont{Mino}},
  \bibinfo{author}{\bibfnamefont{H.}~\bibnamefont{Nakano}},
  \bibinfo{author}{\bibfnamefont{A.}~\bibnamefont{Ori}}, \bibnamefont{and}
  \bibinfo{author}{\bibfnamefont{M.}~\bibnamefont{Sasaki}},
  \bibinfo{journal}{Phys. Rev. Lett.} \textbf{\bibinfo{volume}{88}},
  \bibinfo{pages}{091101} (\bibinfo{year}{2002}), \eprint{gr-qc/0111001}.

\bibitem[{\citenamefont{Barack and Ori}(2002)}]{Barack:Ori:2002}
\bibinfo{author}{\bibfnamefont{L.}~\bibnamefont{Barack}} \bibnamefont{and}
  \bibinfo{author}{\bibfnamefont{A.}~\bibnamefont{Ori}},
  \bibinfo{journal}{Phys. Rev. D} \textbf{\bibinfo{volume}{66}},
  \bibinfo{pages}{084022} (\bibinfo{year}{2002}), \eprint{gr-qc/0204093}.

\bibitem[{\citenamefont{Barack}(2001)}]{Barack:2001}
\bibinfo{author}{\bibfnamefont{L.}~\bibnamefont{Barack}},
  \bibinfo{journal}{Phys. Rev. D} \textbf{\bibinfo{volume}{64}},
  \bibinfo{pages}{084021} (\bibinfo{year}{2001}).

\bibitem[{\citenamefont{Mino and Sasaki}(2002)}]{Mino:Nakano:Sasaki:2002}
\bibinfo{author}{\bibfnamefont{N.}~\bibnamefont{Mino}} \bibnamefont{and}
  \bibinfo{author}{\bibnamefont{Sasaki}}, \bibinfo{journal}{Prog. Theor. Phys.}
  \textbf{\bibinfo{volume}{108}}, \bibinfo{pages}{1039} (\bibinfo{year}{2002}),
  \eprint{gr-qc/0111074}.

\bibitem[{\citenamefont{Detweiler and Whiting}(2003)}]{Detweiler-Whiting-2003}
\bibinfo{author}{\bibfnamefont{S.}~\bibnamefont{Detweiler}} \bibnamefont{and}
  \bibinfo{author}{\bibfnamefont{B.~F.} \bibnamefont{Whiting}},
  \bibinfo{journal}{Phys. Rev. D} \textbf{\bibinfo{volume}{67}},
  \bibinfo{pages}{024025} (\bibinfo{year}{2003}).

\bibitem[{\citenamefont{Detweiler et~al.}(2003)\citenamefont{Detweiler,
  Messaritaki, and Whiting}}]{Detweiler:Messaritaki:Whiting:2002}
\bibinfo{author}{\bibfnamefont{S.}~\bibnamefont{Detweiler}},
  \bibinfo{author}{\bibfnamefont{E.}~\bibnamefont{Messaritaki}},
  \bibnamefont{and} \bibinfo{author}{\bibfnamefont{B.~F.}
  \bibnamefont{Whiting}}, \bibinfo{journal}{Phys. Rev.}
  \textbf{\bibinfo{volume}{D67}}, \bibinfo{pages}{104016}
  (\bibinfo{year}{2003}), \eprint{gr-qc/0205079}.

\bibitem[{\citenamefont{Haas}(2011)}]{Haas:2011bt}
\bibinfo{author}{\bibfnamefont{R.}~\bibnamefont{Haas}} (\bibinfo{year}{2011}),
  \eprint{arXiv:1112.3707}.

\bibitem[{\citenamefont{Barack and Sago}(2010)}]{Barack:Sago:2010}
\bibinfo{author}{\bibfnamefont{L.}~\bibnamefont{Barack}} \bibnamefont{and}
  \bibinfo{author}{\bibfnamefont{N.}~\bibnamefont{Sago}},
  \bibinfo{journal}{Phys. Rev.} \textbf{\bibinfo{volume}{D81}},
  \bibinfo{pages}{084021} (\bibinfo{year}{2010}), \eprint{arXiv:1002.2386}.

\bibitem[{\citenamefont{Warburton and Barack}(2011)}]{Warburton:Barack:2010}
\bibinfo{author}{\bibfnamefont{N.}~\bibnamefont{Warburton}} \bibnamefont{and}
  \bibinfo{author}{\bibfnamefont{L.}~\bibnamefont{Barack}},
  \bibinfo{journal}{Phys. Rev.} \textbf{\bibinfo{volume}{D83}},
  \bibinfo{pages}{124038} (\bibinfo{year}{2011}), \eprint{arXiv:1103.0287}.

\bibitem[{Bar()}]{BarryWardell.net}
\bibinfo{howpublished}{\url{http://www.annaheffernan.com}, \\
  \url{http://www.barrywardell.net/research/code}}.

\bibitem[{\citenamefont{Misner et~al.}(1973)\citenamefont{Misner, Thorne, and
  Wheeler}}]{Misner:Thorne:Wheeler:1974}
\bibinfo{author}{\bibfnamefont{C.~W.} \bibnamefont{Misner}},
  \bibinfo{author}{\bibfnamefont{K.~S.} \bibnamefont{Thorne}},
  \bibnamefont{and} \bibinfo{author}{\bibfnamefont{J.~A.}
  \bibnamefont{Wheeler}}, \emph{\bibinfo{title}{{Gravitation}}}
  (\bibinfo{publisher}{{Freeman}}, \bibinfo{address}{San Francisco},
  \bibinfo{year}{1973}).

\bibitem[{\citenamefont{Haas and Poisson}(2006)}]{Haas:Poisson:2006}
\bibinfo{author}{\bibfnamefont{R.}~\bibnamefont{Haas}} \bibnamefont{and}
  \bibinfo{author}{\bibfnamefont{E.}~\bibnamefont{Poisson}},
  \bibinfo{journal}{Phys. Rev.} \textbf{\bibinfo{volume}{D74}},
  \bibinfo{pages}{044009} (\bibinfo{year}{2006}), \eprint{gr-qc/0605077}.

\bibitem[{\citenamefont{Chandrasekhar}(1992)}]{Chandrasekhar}
\bibinfo{author}{\bibfnamefont{S.}~\bibnamefont{Chandrasekhar}},
  \emph{\bibinfo{title}{The Mathematical Theory of Black Holes}}
  (\bibinfo{publisher}{Oxford University Press}, \bibinfo{year}{1992}).

\bibitem[{\citenamefont{Synge}(1960)}]{Synge}
\bibinfo{author}{\bibfnamefont{J.}~\bibnamefont{Synge}},
  \emph{\bibinfo{title}{Relativity: The General Theory}}
  (\bibinfo{publisher}{North-Holland}, \bibinfo{address}{Amsterdam},
  \bibinfo{year}{1960}).

\bibitem[{\citenamefont{Schwarzschild}(1916)}]{Schwarzschild:1916}
\bibinfo{author}{\bibfnamefont{K.}~\bibnamefont{Schwarzschild}},
  \bibinfo{journal}{Sitzungsber.Preuss.Akad.Wiss.Berlin (Math.Phys.)} pp.
  \bibinfo{pages}{189--196} (\bibinfo{year}{1916}),
  \eprint{arXiv:physics/9905030}.

\bibitem[{\citenamefont{Birkhoff}(1923)}]{Birkhoff:1923}
\bibinfo{author}{\bibfnamefont{G.~D.} \bibnamefont{Birkhoff}},
  \emph{\bibinfo{title}{Relativity and Modern Physics}}
  (\bibinfo{publisher}{Harvard University Press}, \bibinfo{address}{Cambridge,
  Massachusettts}, \bibinfo{year}{1923}).

\bibitem[{\citenamefont{{Wolfram Research, Inc.}}(2008)}]{Mathematica}
\bibinfo{author}{\bibnamefont{{Wolfram Research, Inc.}}},
  \emph{\bibinfo{title}{Mathematica}} (\bibinfo{publisher}{Wolfram Research,
  Inc.}, \bibinfo{address}{Champaign, Illinois}, \bibinfo{year}{2008}),
  \bibinfo{edition}{{Version 8.0}} ed.

\bibitem[{\citenamefont{Hawking and Ellis}(1973)}]{Hawking:Ellis}
\bibinfo{author}{\bibfnamefont{S.~W.} \bibnamefont{Hawking}} \bibnamefont{and}
  \bibinfo{author}{\bibfnamefont{G.~F.~R.} \bibnamefont{Ellis}},
  \emph{\bibinfo{title}{The Large-Scale Structure of Spacetime}}
  (\bibinfo{publisher}{Cambridge University Press}, \bibinfo{year}{1973}).

\bibitem[{\citenamefont{Penrose}(1969)}]{Penrose:1969}
\bibinfo{author}{\bibfnamefont{R.}~\bibnamefont{Penrose}},
  \bibinfo{journal}{Riv. Nuovo Cim.} \textbf{\bibinfo{volume}{D79}},
  \bibinfo{pages}{252} (\bibinfo{year}{1969}).

\bibitem[{\citenamefont{DeWitt}(1965)}]{DeWitt:1965jb}
\bibinfo{author}{\bibfnamefont{B.~S.} \bibnamefont{DeWitt}}
  (\bibinfo{year}{1965}).

\bibitem[{\citenamefont{D\'ecanini and Folacci}(2006)}]{Decanini:Folacci:2005a}
\bibinfo{author}{\bibfnamefont{Y.}~\bibnamefont{D\'ecanini}} \bibnamefont{and}
  \bibinfo{author}{\bibfnamefont{A.}~\bibnamefont{Folacci}},
  \bibinfo{journal}{Phys. Rev.} \textbf{\bibinfo{volume}{D73}},
  \bibinfo{pages}{044027} (\bibinfo{year}{2006}), \eprint{gr-qc/0511115}.

\bibitem[{GRT()}]{GRTensor}
\bibinfo{howpublished}{\url{http://www.grtensor.org}}.

\bibitem[{xTe()}]{xTensorOnline}
\bibinfo{howpublished}{\url{http://metric.iem.csic.es/Martin-Garcia/xAct/}}.

\bibitem[{\citenamefont{Barack and Burko}(2000)}]{Barack:Burko:2000}
\bibinfo{author}{\bibfnamefont{L.}~\bibnamefont{Barack}} \bibnamefont{and}
  \bibinfo{author}{\bibfnamefont{L.~M.} \bibnamefont{Burko}},
  \bibinfo{journal}{Phys. Rev.} \textbf{\bibinfo{volume}{D62}},
  \bibinfo{pages}{084040} (\bibinfo{year}{2000}), \eprint{gr-qc/0007033}.

\bibitem[{\citenamefont{Burko}(2000)}]{Burko:2000b}
\bibinfo{author}{\bibfnamefont{L.~M.} \bibnamefont{Burko}},
  \bibinfo{journal}{Phys. Rev. Lett.} \textbf{\bibinfo{volume}{84}},
  \bibinfo{pages}{4529} (\bibinfo{year}{2000}).

\bibitem[{\citenamefont{Diaz-Rivera et~al.}(2004)\citenamefont{Diaz-Rivera,
  Messaritaki, Whiting, and Detweiler}}]{DiazRivera:2004}
\bibinfo{author}{\bibfnamefont{L.~M.} \bibnamefont{Diaz-Rivera}},
  \bibinfo{author}{\bibfnamefont{E.}~\bibnamefont{Messaritaki}},
  \bibinfo{author}{\bibfnamefont{B.~F.} \bibnamefont{Whiting}},
  \bibnamefont{and} \bibinfo{author}{\bibfnamefont{S.~L.}
  \bibnamefont{Detweiler}}, \bibinfo{journal}{Phys. Rev.}
  \textbf{\bibinfo{volume}{D70}}, \bibinfo{pages}{124018}
  (\bibinfo{year}{2004}), \eprint{gr-qc/0410011}.

\bibitem[{\citenamefont{Haas}(2007)}]{Haas:2007}
\bibinfo{author}{\bibfnamefont{R.}~\bibnamefont{Haas}}, \bibinfo{journal}{Phys.
  Rev.} \textbf{\bibinfo{volume}{D75}}, \bibinfo{pages}{124011}
  (\bibinfo{year}{2007}), \eprint{arXiv:0704.0797}.

\bibitem[{\citenamefont{Canizares and
  Sopuerta}(2009)}]{Canizares:Sopuerta:2009}
\bibinfo{author}{\bibfnamefont{P.}~\bibnamefont{Canizares}} \bibnamefont{and}
  \bibinfo{author}{\bibfnamefont{C.~F.} \bibnamefont{Sopuerta}},
  \bibinfo{journal}{Phys. Rev.} \textbf{\bibinfo{volume}{D79}},
  \bibinfo{pages}{084020} (\bibinfo{year}{2009}), \eprint{arXiv:0903.0505}.

\bibitem[{\citenamefont{Canizares et~al.}(2010)\citenamefont{Canizares,
  Sopuerta, and Jaramillo}}]{Canizares:Sopuerta:Jaramillo:2010}
\bibinfo{author}{\bibfnamefont{P.}~\bibnamefont{Canizares}},
  \bibinfo{author}{\bibfnamefont{C.~F.} \bibnamefont{Sopuerta}},
  \bibnamefont{and} \bibinfo{author}{\bibfnamefont{J.~L.}
  \bibnamefont{Jaramillo}}, \bibinfo{journal}{Phys. Rev.}
  \textbf{\bibinfo{volume}{D82}}, \bibinfo{pages}{044023}
  (\bibinfo{year}{2010}), \eprint{arXiv:1006.3201}.

\bibitem[{\citenamefont{Barack and Sago}(2007)}]{Barack:Sago:2007}
\bibinfo{author}{\bibfnamefont{L.}~\bibnamefont{Barack}} \bibnamefont{and}
  \bibinfo{author}{\bibfnamefont{N.}~\bibnamefont{Sago}},
  \bibinfo{journal}{Phys. Rev.} \textbf{\bibinfo{volume}{D75}},
  \bibinfo{pages}{064021} (\bibinfo{year}{2007}), \eprint{gr-qc/0701069}.

\bibitem[{\citenamefont{Barack and Lousto}(2002)}]{Barack:Lousto:2002}
\bibinfo{author}{\bibfnamefont{L.}~\bibnamefont{Barack}} \bibnamefont{and}
  \bibinfo{author}{\bibfnamefont{C.~O.} \bibnamefont{Lousto}},
  \bibinfo{journal}{Phys. Rev.} \textbf{\bibinfo{volume}{D66}},
  \bibinfo{pages}{061502} (\bibinfo{year}{2002}), \eprint{gr-qc/0205043}.

\bibitem[{\citenamefont{Sago et~al.}(2008)\citenamefont{Sago, Barack, and
  Detweiler}}]{Sago:Barack:Detweiler:2008}
\bibinfo{author}{\bibfnamefont{N.}~\bibnamefont{Sago}},
  \bibinfo{author}{\bibfnamefont{L.}~\bibnamefont{Barack}}, \bibnamefont{and}
  \bibinfo{author}{\bibfnamefont{S.}~\bibnamefont{Detweiler}},
  \bibinfo{journal}{Phys. Rev. D} \textbf{\bibinfo{volume}{78}},
  \bibinfo{pages}{124024} (\bibinfo{year}{2008}), \eprint{arXiv:0810.2530}.

\bibitem[{\citenamefont{Detweiler}(2008)}]{Detweiler:2008}
\bibinfo{author}{\bibfnamefont{S.}~\bibnamefont{Detweiler}},
  \bibinfo{journal}{Phys. Rev. D} \textbf{\bibinfo{volume}{77}},
  \bibinfo{pages}{124026} (\bibinfo{year}{2008}), \eprint{arXiv:0804:3529}.

\bibitem[{\citenamefont{Sago}(2009)}]{Sago:2009}
\bibinfo{author}{\bibfnamefont{N.}~\bibnamefont{Sago}},
  \bibinfo{journal}{Classical and Quantum Gravity}
  \textbf{\bibinfo{volume}{26}}, \bibinfo{pages}{094025}
  (\bibinfo{year}{2009}).

\bibitem[{\citenamefont{Warburton and Barack}(2010)}]{Warburton:Barack:2009}
\bibinfo{author}{\bibfnamefont{N.}~\bibnamefont{Warburton}} \bibnamefont{and}
  \bibinfo{author}{\bibfnamefont{L.}~\bibnamefont{Barack}},
  \bibinfo{journal}{Phys. Rev.} \textbf{\bibinfo{volume}{D81}},
  \bibinfo{pages}{084039} (\bibinfo{year}{2010}), \eprint{arXiv:1003.1860}.

\bibitem[{\citenamefont{Thornburg}(2010)}]{Thornburg:2010}
\bibinfo{author}{\bibfnamefont{J.}~\bibnamefont{Thornburg}}
  (\bibinfo{year}{2010}), \eprint{arXiv:1006.3788}.

\bibitem[{\citenamefont{Warburton et~al.}(2012)\citenamefont{Warburton, Akcay,
  Barack, Gair, and Sago}}]{Warburton:2011fk}
\bibinfo{author}{\bibfnamefont{N.}~\bibnamefont{Warburton}},
  \bibinfo{author}{\bibfnamefont{S.}~\bibnamefont{Akcay}},
  \bibinfo{author}{\bibfnamefont{L.}~\bibnamefont{Barack}},
  \bibinfo{author}{\bibfnamefont{J.~R.} \bibnamefont{Gair}}, \bibnamefont{and}
  \bibinfo{author}{\bibfnamefont{N.}~\bibnamefont{Sago}},
  \bibinfo{journal}{Phys.Rev.} \textbf{\bibinfo{volume}{D85}},
  \bibinfo{pages}{061501} (\bibinfo{year}{2012}), \eprint{arXiv:1111.6908}.

\bibitem[{\citenamefont{Hopper and Evans}(2010)}]{Hopper:2010uv}
\bibinfo{author}{\bibfnamefont{S.}~\bibnamefont{Hopper}} \bibnamefont{and}
  \bibinfo{author}{\bibfnamefont{C.~R.} \bibnamefont{Evans}},
  \bibinfo{journal}{Phys.Rev.} \textbf{\bibinfo{volume}{D82}},
  \bibinfo{pages}{084010} (\bibinfo{year}{2010}), \eprint{arXiv:1006.4907}.

\bibitem[{\citenamefont{Keidl et~al.}(2010)\citenamefont{Keidl, Shah, Friedman,
  Kim, and Price}}]{Keidl:2010pm}
\bibinfo{author}{\bibfnamefont{T.~S.} \bibnamefont{Keidl}},
  \bibinfo{author}{\bibfnamefont{A.~G.} \bibnamefont{Shah}},
  \bibinfo{author}{\bibfnamefont{J.~L.} \bibnamefont{Friedman}},
  \bibinfo{author}{\bibfnamefont{D.-H.} \bibnamefont{Kim}}, \bibnamefont{and}
  \bibinfo{author}{\bibfnamefont{L.~R.} \bibnamefont{Price}},
  \bibinfo{journal}{Phys.Rev.} \textbf{\bibinfo{volume}{D82}},
  \bibinfo{pages}{124012} (\bibinfo{year}{2010}), \eprint{arXiv:1004.2276}.

\bibitem[{\citenamefont{Shah et~al.}(2011)\citenamefont{Shah, Keidl, Friedman,
  Kim, and Price}}]{Shah:2010bi}
\bibinfo{author}{\bibfnamefont{A.~G.} \bibnamefont{Shah}},
  \bibinfo{author}{\bibfnamefont{T.~S.} \bibnamefont{Keidl}},
  \bibinfo{author}{\bibfnamefont{J.~L.} \bibnamefont{Friedman}},
  \bibinfo{author}{\bibfnamefont{D.-H.} \bibnamefont{Kim}}, \bibnamefont{and}
  \bibinfo{author}{\bibfnamefont{L.~R.} \bibnamefont{Price}},
  \bibinfo{journal}{Phys.Rev.} \textbf{\bibinfo{volume}{D83}},
  \bibinfo{pages}{064018} (\bibinfo{year}{2011}), \eprint{arXiv:1009.4876}.

\bibitem[{\citenamefont{Abramowitz and Stegun}(1972)}]{Abramowitz:Stegun}
\bibinfo{author}{\bibfnamefont{M.}~\bibnamefont{Abramowitz}} \bibnamefont{and}
  \bibinfo{author}{\bibfnamefont{I.~A.} \bibnamefont{Stegun}},
  \emph{\bibinfo{title}{Handbook of Mathematical Functions}}
  (\bibinfo{publisher}{Dover}, \bibinfo{address}{New Yord},
  \bibinfo{year}{1972}).

\bibitem[{\citenamefont{Barack and Sago}(2011)}]{Barack:2011ed}
\bibinfo{author}{\bibfnamefont{L.}~\bibnamefont{Barack}} \bibnamefont{and}
  \bibinfo{author}{\bibfnamefont{N.}~\bibnamefont{Sago}},
  \bibinfo{journal}{Phys.Rev.} \textbf{\bibinfo{volume}{D83}},
  \bibinfo{pages}{084023} (\bibinfo{year}{2011}), \eprint{arXiv:1101.3331}.

\bibitem[{\citenamefont{Wheeler}(1955)}]{Wheeler:1955}
\bibinfo{author}{\bibfnamefont{J.~A.} \bibnamefont{Wheeler}},
  \bibinfo{journal}{Phys. Rev.} \textbf{\bibinfo{volume}{97}},
  \bibinfo{pages}{511} (\bibinfo{year}{1955}).

\bibitem[{\citenamefont{Heffernan et~al.}(2012)\citenamefont{Heffernan, Nolan,
  Ottewill, and Wardell}}]{Heffernan:2012b}
\bibinfo{author}{\bibfnamefont{A.}~\bibnamefont{Heffernan}},
  \bibinfo{author}{\bibfnamefont{P.}~\bibnamefont{Nolan}},
  \bibinfo{author}{\bibfnamefont{A.}~\bibnamefont{Ottewill}}, \bibnamefont{and}
  \bibinfo{author}{\bibfnamefont{B.}~\bibnamefont{Wardell}},
  \bibinfo{journal}{(in preparation)}  (\bibinfo{year}{2012}).

\bibitem[{\citenamefont{Dolan and Barack}(2011)}]{Dolan:Barack}
\bibinfo{author}{\bibfnamefont{S.~R.} \bibnamefont{Dolan}} \bibnamefont{and}
  \bibinfo{author}{\bibfnamefont{L.}~\bibnamefont{Barack}},
  \bibinfo{journal}{Phys.Rev.} \textbf{\bibinfo{volume}{D83}},
  \bibinfo{pages}{024019} (\bibinfo{year}{2011}), \eprint{1010.5255}.

\bibitem[{\citenamefont{Dolan et~al.}(2011)\citenamefont{Dolan, Barack, and
  Wardell}}]{Dolan:Barack:Wardell}
\bibinfo{author}{\bibfnamefont{S.~R.} \bibnamefont{Dolan}},
  \bibinfo{author}{\bibfnamefont{L.}~\bibnamefont{Barack}}, \bibnamefont{and}
  \bibinfo{author}{\bibfnamefont{B.}~\bibnamefont{Wardell}},
  \bibinfo{journal}{Phys.Rev.} \textbf{\bibinfo{volume}{D84}},
  \bibinfo{pages}{084001} (\bibinfo{year}{2011}), \eprint{1107.0012}.

\bibitem[{\citenamefont{Wardell et~al.}(2012)\citenamefont{Wardell, Vega,
  Thornburg, and Diener}}]{Diener:Vega:Wardell:Detweiler}
\bibinfo{author}{\bibfnamefont{B.}~\bibnamefont{Wardell}},
  \bibinfo{author}{\bibfnamefont{I.}~\bibnamefont{Vega}},
  \bibinfo{author}{\bibfnamefont{J.}~\bibnamefont{Thornburg}},
  \bibnamefont{and} \bibinfo{author}{\bibfnamefont{P.}~\bibnamefont{Diener}},
  \bibinfo{journal}{Phys.Rev.} \textbf{\bibinfo{volume}{D85}},
  \bibinfo{pages}{104044} (\bibinfo{year}{2012}), \eprint{1112.6355}.

\bibitem[{\citenamefont{Thornburg and Wardell}(2012)}]{Thornburg:Wardell}
\bibinfo{author}{\bibfnamefont{J.}~\bibnamefont{Thornburg}} \bibnamefont{and}
  \bibinfo{author}{\bibfnamefont{B.}~\bibnamefont{Wardell}}
  (\bibinfo{year}{2012}), \bibinfo{note}{in preparation}.

\bibitem[{m-m()}]{m-mode-online}
\bibinfo{howpublished}{\url{http://www.annaheffernan.com}, \\
  \url{http://www.barrywardell.net/research/code/covariantseries}}.

\bibitem[{\citenamefont{Hubeny}(1999)}]{Hubeny:1998}
\bibinfo{author}{\bibfnamefont{V.~E.} \bibnamefont{Hubeny}},
  \bibinfo{journal}{Phys.Rev.} \textbf{\bibinfo{volume}{D59}},
  \bibinfo{pages}{064013} (\bibinfo{year}{1999}), \eprint{gr-qc/9808043}.

\bibitem[{\citenamefont{Zimmerman et~al.}(2012)\citenamefont{Zimmerman, Vega,
  Poisson, and Haas}}]{Zimmerman:2010}
\bibinfo{author}{\bibfnamefont{P.}~\bibnamefont{Zimmerman}},
  \bibinfo{author}{\bibfnamefont{I.}~\bibnamefont{Vega}},
  \bibinfo{author}{\bibfnamefont{E.}~\bibnamefont{Poisson}}, \bibnamefont{and}
  \bibinfo{author}{\bibfnamefont{R.}~\bibnamefont{Haas}}
  (\bibinfo{year}{2012}), \eprint{1211.3889}.

\bibitem[{\citenamefont{{Wald}}(1974)}]{Wald:1974}
\bibinfo{author}{\bibfnamefont{R.}~\bibnamefont{{Wald}}},
  \bibinfo{journal}{Annals of Physics} \textbf{\bibinfo{volume}{82}},
  \bibinfo{pages}{548} (\bibinfo{year}{1974}).

\bibitem[{\citenamefont{Jacobson and Sotiriou}(2009)}]{Jacobson:2009}
\bibinfo{author}{\bibfnamefont{T.}~\bibnamefont{Jacobson}} \bibnamefont{and}
  \bibinfo{author}{\bibfnamefont{T.~P.} \bibnamefont{Sotiriou}},
  \bibinfo{journal}{Phys.Rev.Lett.} \textbf{\bibinfo{volume}{103}},
  \bibinfo{pages}{141101} (\bibinfo{year}{2009}), \eprint{0907.4146}.

\bibitem[{\citenamefont{Barausse et~al.}(2011)\citenamefont{Barausse, Cardoso,
  and Khanna}}]{Barausse:2011}
\bibinfo{author}{\bibfnamefont{E.}~\bibnamefont{Barausse}},
  \bibinfo{author}{\bibfnamefont{V.}~\bibnamefont{Cardoso}}, \bibnamefont{and}
  \bibinfo{author}{\bibfnamefont{G.}~\bibnamefont{Khanna}},
  \bibinfo{journal}{Phys.Rev.} \textbf{\bibinfo{volume}{D84}},
  \bibinfo{pages}{104006} (\bibinfo{year}{2011}), \eprint{1106.1692}.

\bibitem[{\citenamefont{Haas et~al.}()\citenamefont{Haas, Heffernan, Ottewill,
  and Wardell}}]{Haas:Heffernan:2012}
\bibinfo{author}{\bibfnamefont{R.}~\bibnamefont{Haas}},
  \bibinfo{author}{\bibfnamefont{A.}~\bibnamefont{Heffernan}},
  \bibinfo{author}{\bibfnamefont{A.}~\bibnamefont{Ottewill}}, \bibnamefont{and}
  \bibinfo{author}{\bibfnamefont{B.}~\bibnamefont{Wardell}}, \bibinfo{note}{in
  preparation}.

\bibitem[{\citenamefont{Casals et~al.}(2012)\citenamefont{Casals, Poisson, and
  Vega}}]{Casals:Poisson:Vega:2012}
\bibinfo{author}{\bibfnamefont{M.}~\bibnamefont{Casals}},
  \bibinfo{author}{\bibfnamefont{E.}~\bibnamefont{Poisson}}, \bibnamefont{and}
  \bibinfo{author}{\bibfnamefont{I.}~\bibnamefont{Vega}},
  \bibinfo{journal}{Phys.Rev.} \textbf{\bibinfo{volume}{D86}},
  \bibinfo{pages}{064033} (\bibinfo{year}{2012}), \eprint{1206.3772}.

\bibitem[{\citenamefont{Gralla and Wald}(2008)}]{Gralla:Wald:2008}
\bibinfo{author}{\bibfnamefont{S.~E.} \bibnamefont{Gralla}} \bibnamefont{and}
  \bibinfo{author}{\bibfnamefont{R.~M.} \bibnamefont{Wald}},
  \bibinfo{journal}{Class. Quantum Grav.} \textbf{\bibinfo{volume}{25}},
  \bibinfo{pages}{205009} (\bibinfo{year}{2008}), \eprint{arXiv:0806.3293}.

\bibitem[{\citenamefont{Gralla}(2011)}]{Gralla:2011zr}
\bibinfo{author}{\bibfnamefont{S.~E.} \bibnamefont{Gralla}},
  \bibinfo{journal}{Phys.Rev.} \textbf{\bibinfo{volume}{D84}},
  \bibinfo{pages}{084050} (\bibinfo{year}{2011}), \eprint{arXiv:1104.5635}.

\bibitem[{\citenamefont{Yang et~al.}(2012)\citenamefont{Yang, Miao, and
  Chen}}]{Yang:2012bb}
\bibinfo{author}{\bibfnamefont{H.}~\bibnamefont{Yang}},
  \bibinfo{author}{\bibfnamefont{H.}~\bibnamefont{Miao}}, \bibnamefont{and}
  \bibinfo{author}{\bibfnamefont{Y.}~\bibnamefont{Chen}}
  (\bibinfo{year}{2012}), \eprint{1211.5410}.

\bibitem[{\citenamefont{Akcay et~al.}(2012)\citenamefont{Akcay, Barack, Damour,
  and Sago}}]{Akcay:2012ea}
\bibinfo{author}{\bibfnamefont{S.}~\bibnamefont{Akcay}},
  \bibinfo{author}{\bibfnamefont{L.}~\bibnamefont{Barack}},
  \bibinfo{author}{\bibfnamefont{T.}~\bibnamefont{Damour}}, \bibnamefont{and}
  \bibinfo{author}{\bibfnamefont{N.}~\bibnamefont{Sago}},
  \bibinfo{journal}{Phys.Rev.} \textbf{\bibinfo{volume}{D86}},
  \bibinfo{pages}{104041} (\bibinfo{year}{2012}), \eprint{1209.0964}.

\bibitem[{\citenamefont{Ottewill and Taylor}(2012)}]{Ottewill:2012aj}
\bibinfo{author}{\bibfnamefont{A.~C.} \bibnamefont{Ottewill}} \bibnamefont{and}
  \bibinfo{author}{\bibfnamefont{P.}~\bibnamefont{Taylor}},
  \bibinfo{journal}{Phys.Rev.} \textbf{\bibinfo{volume}{D86}},
  \bibinfo{pages}{024036} (\bibinfo{year}{2012}), \eprint{1205.5587}.

\bibitem[{\citenamefont{Brown and Ottewill}(1986)}]{Brown:Ottewill:1986}
\bibinfo{author}{\bibfnamefont{M.~R.} \bibnamefont{Brown}} \bibnamefont{and}
  \bibinfo{author}{\bibfnamefont{A.~C.} \bibnamefont{Ottewill}},
  \bibinfo{journal}{Phys. Rev.} \textbf{\bibinfo{volume}{D34}},
  \bibinfo{pages}{1776} (\bibinfo{year}{1986}).

\bibitem[{\citenamefont{Allen et~al.}(1988)\citenamefont{Allen, Folacci, and
  Ottewill}}]{Allen:1987bn}
\bibinfo{author}{\bibfnamefont{B.}~\bibnamefont{Allen}},
  \bibinfo{author}{\bibfnamefont{A.}~\bibnamefont{Folacci}}, \bibnamefont{and}
  \bibinfo{author}{\bibfnamefont{A.}~\bibnamefont{Ottewill}},
  \bibinfo{journal}{Phys.Rev.} \textbf{\bibinfo{volume}{D38}},
  \bibinfo{pages}{1069} (\bibinfo{year}{1988}).

\bibitem[{\citenamefont{Anderson et~al.}(2005)\citenamefont{Anderson, Flanagan,
  and Ottewill}}]{Anderson:Flanagan:Ottewill:2004}
\bibinfo{author}{\bibfnamefont{W.~G.} \bibnamefont{Anderson}},
  \bibinfo{author}{\bibfnamefont{E.~E.} \bibnamefont{Flanagan}},
  \bibnamefont{and} \bibinfo{author}{\bibfnamefont{A.~C.}
  \bibnamefont{Ottewill}}, \bibinfo{journal}{Phys. Rev.}
  \textbf{\bibinfo{volume}{D71}}, \bibinfo{pages}{024036}
  (\bibinfo{year}{2005}), \eprint{gr-qc/0412009}.

\end{thebibliography}


\end{document}